\documentclass{amsarte}

\usepackage{latexsym,pifont}
\usepackage{epsfig}
\usepackage{rotating}

\usepackage{ifpdf}
\ifpdf
\usepackage{epstopdf}
\usepackage{hyperref}
\else
\usepackage[hypertex]{hyperref}
\fi

\usepackage{amsfonts,amsma th,amssymb,mathrsfs,extarrows,MnSymbol}
\usepackage[all]{xy}

\usepackage{tikz}
\usetikzlibrary{matrix,arrows}

\usepackage[makeroom]{cancel}

\usepackage[toc,page]{appendix}
\newcommand{\stoptocwriting}{%
  \addtocontents{toc}{\protect\setcounter{tocdepth}{-5}}}
\newcommand{\resumetocwriting}{%
  \addtocontents{toc}{\protect\setcounter{tocdepth}{\arabic{tocdepth}}}}

\newcommand{\alxydim}[2]{\begin{aligned}\xymatrix#1{#2}\end{aligned}}

\newcommand{\brem}{\begin{Rem}}
\newcommand{\erem}{\end{Rem}\medskip}
\newcommand{\beg}{\begin{Eg}}
\newcommand{\eeg}{\end{Eg}}
\newcommand{\bedef}{\begin{Def}}
\newcommand{\exdef}{\begin{flushright}$\diamond$\end{flushright}
\end{Def}\vskip0.1cm}
\newcommand{\berop}{\begin{Prop}}
\newcommand{\eerop}{\end{Prop}}
\newcommand{\belem}{\begin{Lem}}
\newcommand{\elem}{\end{Lem}}
\newcommand{\bethe}{\begin{Thm}}
\newcommand{\ethe}{\end{Thm}}
\newcommand{\becor}{\begin{Cor}}
\newcommand{\ecor}{\end{Cor}}
\newcommand{\beroof}{\noindent\begin{proof}}
\newcommand{\eroof}{\end{proof}}
\newcommand{\becon}{\begin{Conv}}
\newcommand{\econ}{\begin{flushright}$\checkmark$\end{flushright}\end{Conv}}
\newcommand{\befact}{\begin{Fact}}
\newcommand{\efact}{\begin{flushright}$\checkmark$\end{flushright}\end{Fact}}
\newcommand{\bequest}{\begin{Quest}}
\newcommand{\equest}{\end{Quest}}
\newcommand{\brob}{\begin{Prob}}
\newcommand{\erob}{\end{Prob}}
\newcommand{\becj}{\begin{conj}}
\newcommand{\ecj}{\begin{flushright}$\boxtimes$\end{flushright}\end{conj}}

\newcommand{\barr}{\begin{array}}
\newcommand{\earr}{\end{array}}
\newcommand{\ben}{\begin{enumerate}}
\newcommand{\een}{\end{enumerate}}
\newcommand{\bit}{\begin{itemize}}
\newcommand{\eit}{\end{itemize}}

\newcommand{\qq}{\begin{eqnarray}}
\newcommand{\qqq}{\end{eqnarray}}

\newcommand{\nn}{\nonumber}

\newcommand{\ovl}[1]{\overline{#1}}
\newcommand{\unl}[1]{\underline{#1}}

\newcommand{\Reqref}[1]{Eq.\,\eqref{#1}}
\newcommand{\Rcite}[1]{Ref.\,\cite{#1}}
\newcommand{\Rxcite}[2]{Ref.\,\cite[#1]{#2}}

\newcommand\void[1]{}

\newcommand{\tx}[1]{\textrm{#1}} 

\newcommand{\gt}[1]{\mathfrak{#1}}

\def\cA{\mathcal{A}}
\def\cB{\mathcal{B}}
\def\cC{\mathcal{C}}
\def\cD{\mathcal{D}}

\def\cF{\mathcal{F}}
\def\cG{\mathcal{G}}

\def\cI{\mathcal{I}}
\def\cJ{\mathcal{J}}
\def\cK{\mathcal{K}}
\def\ceL{\mathcal{L}}
\def\cM{\mathcal{M}}
\def\cN{\mathcal{N}}
\def\cO{\mathcal{O}}

\def\cS{\mathcal{S}}
\def\cT{\mathcal{T}}

\def\cV{\mathcal{V}}
\def\cW{\mathcal{W}}

\def\cZ{\mathcal{Z}}

\def\xcA{\mathscr{A}}
\def\xcB{\mathscr{B}}
\def\xcC{\mathscr{C}}
\def\xcD{\mathscr{D}}

\def\xcF{\mathscr{F}}

\def\xcH{\mathscr{H}}
\def\xcI{\mathscr{I}}

\def\xcL{\mathscr{L}}
\def\xcM{\mathscr{M}}

\def\xcO{\mathscr{O}}

\def\xcT{\mathscr{T}}

\def\xcV{\mathscr{V}}

\def\t{\mathbf{t}}

\def\bC{{\mathbb{C}}}
\def\bD{{\mathbb{D}}}

\def\bH{{\mathbb{H}}}

\def\bN{{\mathbb{N}}}

\def\bP{{\mathbb{P}}}

\def\bR{{\mathbb{R}}}
\def\bS{{\mathbb{S}}}
\def\bT{{\mathbb{T}}}

\def\bZ{{\mathbb{Z}}}

\def\a{\alpha}
\def\b{\beta}
\def\g{\gamma}
\def\G{\Gamma}
\def\d{\delta}
\def\D{\Delta}
\def\ep{\epsilon}
\def\vep{\varepsilon}

\def\th{\theta}

\def\k{\kappa}

\def\la{\lambda}
\def\La{\Lambda}
\def\om{\omega}
\def\Om{\Omega}

\def\si{\sigma}
\def\Si{\Sigma}

\def\t{\tau}

\def\z{\zeta}

\def\agt{\gt{a}}

\def\Bgt{\gt{B}}

\def\dgt{\gt{d}}

\def\ggt{\gt{g}}

\def\tgt{\gt{t}}

\newcommand{\sfa}{{\mathsf a}}

\newcommand{\sfd}{{\mathsf d}}

\newcommand{\sfe}{{\mathsf e}}

\newcommand{\sfi}{{\mathsf i}}

\newcommand{\sfk}{{\mathsf k}}

\newcommand{\sfL}{{\mathsf L}}

\newcommand{\sfN}{{\mathsf N}}

\newcommand{\sfP}{{\mathsf P}}

\newcommand{\sfT}{{\mathsf T}}

\newcommand{\sfY}{{\mathsf Y}}

\newcommand{\txA}{{\rm A}}

\newcommand{\txB}{{\rm B}}

\newcommand{\ee}{{\rm e}}

\newcommand{\txF}{{\rm F}}
\newcommand{\txg}{{\rm g}}
\newcommand{\txG}{{\rm G}}

\newcommand{\txH}{{\rm H}}

\newcommand{\txK}{{\rm K}}

\newcommand{\txm}{{\rm m}}

\newcommand{\txp}{{\rm p}}
\newcommand{\txP}{{\rm P}}

\newcommand{\txW}{{\rm W}}

\newcommand{\Gx}{{\rm G}}

\def\id{{\rm id}}
\newcommand{\pr}{{\rm pr}}

\def\too{\longrightarrow}

\def\Hom{{\rm Hom}}

\def\1morf{1{\rm -Mor}}
\def\2morf{2{\rm -Mor}}
\def\dim{{\rm dim}}
\def\im{{\rm im}}
\def\ker{{\rm ker}}

\def\End{{\rm End}}

\newcommand{\Id}{{\rm Id}}
\def\Inv{{\rm Inv}}

\def\bgrb{\gt{BGrb}}

\newcommand{\Set}{{\rm {\bf Set}}}

\newcommand{\sMan}{{\rm {\bf sMan}}}

\newcommand{\pLie}[1]{\,{-\hspace{-8pt}\xcL}_{#1}}
\def\p{\partial}

\def\con{\righthalfcup}

\def\emb{\hookrightarrow}

\def\curv{{\rm curv}}
\def\Hol{{\rm Hol}}

\def\bd1{{\boldsymbol{1}}}
\def\brd0{{\boldsymbol{0}}}

\def\rk{{\rm rk}}
\def\det{{\rm det}}
\def\tr{{\rm tr}}
\def\diag{\textrm{diag}}

\def\Ad{{\rm Ad}}

\def\Cliff{{\rm Cliff}}

\def\ggtk{\widehat{\gt{g}}_\sfk}
\newcommand{\faff}[1]{P^{\sfk}_{+}(#1)}

\newcommand{\uj}{{\rm U}(1)}
\newcommand{\sug}{{\rm SU}(2)}

\def\x{\times}
\def\ox{\otimes}
\def\bigox{\bigotimes}

\def\rx{\rtimes}

\def\ract{\vartriangleleft}
\def\lact{\vartriangleright}
\def\must{\stackrel{!}{=}}

\def\rstr{\mathord{\restriction}}

\newcommand{\corr}[1]{\left\langle #1 \right\rangle}

\newcommand{\ups}[1]{{}^{\tx{\tiny $#1$}}\hspace{-1pt}}

\newtheorem{Thm}{Theorem}
\newtheorem{Prop}[Thm]{Proposition}
\newtheorem{Lem}[Thm]{Lemma}
\newtheorem{conj}{Conjecture}
\newtheorem{Cor}[Thm]{Corollary}
\theoremstyle{definition}
\newtheorem{Rem}[Thm]{Remark}
\newtheorem{Def}[Thm]{Definition}
\newtheorem{Eg}[Thm]{Example}
\newtheorem{Conv}[Thm]{Convention}
\newtheorem{Fact}[Thm]{Fact}
\newtheorem{Quest}[Thm]{Question}
\newtheorem{Prob}[Thm]{Problem}

\numberwithin{equation}{section} 

\newcount\hour\newcount\minute
        \hour=\time \divide\hour by60 \minute=\time
        {\multiply\hour by60 \global\advance\minute by-\hour}
        \edef\militarytime{\number\hour:\ifnum\minute<10 0\fi\number\minute}

\begin{document}

\title{On symmetric simplicial (super)string backgrounds,\\ (super-)WZW defect fusion and the Chern--Simons theory}

\author{Rafa\l ~R.\ ~Suszek}
\address{R.R.S.:\ Katedra Metod Matematycznych Fizyki,\ Wydzia\l ~Fizyki
Uniwersytetu Warszawskiego,\ ul.\ Pasteura 5,\ PL-02-093 Warszawa,
Poland} \email{suszek@fuw.edu.pl}

\begin{abstract}
The super-$\si$-model of dynamics of the super-charged loop in an ambient supermanifold in the presence of worldsheet defects of arbitrary topology is formalised within Gaw\c{e}dzki's higher-cohomological approach,\ drawing inspiration from the precursor \Rcite{Runkel:2008gr}.\ A distinguished class of the corresponding backgrounds (supertargets with additional bicategorial supergeometric data),\ organised into simplicial hierarchies,\ is considered.\ To these,\ configurational (super)symmetry of the bulk field theory is lifted coherently,\ whereby the notion of a maximally (super)symmetric background,\ and in particular that of a simplicial Lie background,\ arises as the target structure requisite for the definition of the super-$\si$-model with defects fully transmissive to the currents of the bulk (super)symmetry.\ The formal concepts are illustrated in two settings of physical relevance:\ that of the WZW $\si$-model of the bosonic string in a compact simple 1-connected Lie group and that of the GS super-$\si$-model of the superstring in the Minkowski super-space.\ In the former setting,\ the structure of the background is fixed through a combination of simplicial,\ symmetry(-reducibility) and cohomological arguments,\ and a novel link between fusion of the maximally symmetric WZW defects of Fuchs {\it et al.} and the 3$d$ CS theory with timelike Wilson lines with fixed holonomy is established.\ Moreover,\ a purely geometric interpretation of the Verlinde fusion rules is proposed.\ In the latter setting,\ a multiplicative structure compatible with supersymmetry is shown to exist on the GS super-1-gerbe of hep-th/1706.05682,\ and subsequently used in a novel construction of a class of maximally (rigidly) supersymmetric bi-branes whose elementary fusion is also studied.
\end{abstract}

\begin{flushright}
{\it In memory of K${}^\dagger$}\vspace{30pt}
\end{flushright}


\maketitle

\tableofcontents

\newpage

\part*{Introduction}

Recent years have witnessed an essential rethinking and generalisation of the notion of a physical symmetry and the attendant upsurge of interest in its field-theoretic realisations furnished by currents flowing through codimension $\,>1\,$ submanifolds of the physical spacetime (beside the usual codimension-1 spacelike slices),\ {\it cf.},\ {\it e.g.},\ Refs.\,\cite{Freed:2006yc,Freed:2006ya,Gaiotto:2014kfa} and the references within.\ The ensuing research has led to the emergence of a plethora of novel notions and mechanisms such as `higher-form',\ `higher-group' or `higher-categorial' symmetries,\ and -- upon an ingenious incorporation of the good old gauge principle -- also `non-invertible' symmetries,\ {\it cf.}\ \Rcite{Bhardwaj:2022yxj}.\ These are now collectively beginning to supplant the `ancient' (sub-)concept of (0-form) symmetry in its r\^ole of an organising principle in the investigation and solving of a field theory.\  A common feature shared by (many of) these developments is the emergence of `defects' of various dimensionality decorated by some `background' data (often (higher-)cohomological or (higher-)categorial (or both) in nature),\ whose insertion in spacetime effects symmetry transformations (resp.\ `twisting') on the degrees of freedom of the field theory of interest,\ with the quantum-mechanical correlators depending solely on the homotopy class of the insertion,\ whence the qualifier `topological' given to these objects.\smallskip

In low-dimensional field theories of a topological character (such as,\ {\it e.g.},\ the 3$d$ Chern--Simons theory of Refs.\,\cite{Witten:1988hf,Witten:1989wf}) and in those modelling the geometrodynamics of extended distributions of charge (such as,\ {\it e.g.},\ the 2$d$ $\si$-model of the bosonic string),\ the interplay between -- on the one hand -- `defects' and -- on the other hand -- symmetries,\ both rigid and gauged,\ or even dualities between theories,\ has been known for a long time,\ {\it cf.},\ {\it e.g.},\ Refs.\,\cite{Dixon:1985jw,Dixon:1986jc,Petkova:2000ip,Bachas:2001vj,Graham:2003nc,Frohlich:2004ef,Bachas:2004sy,Frohlich:2006ch,Schweigert:2007wd,Fuchs:2007fw,Brunner:2007qu,Runkel:2008gr,Sarkissian:2008dq,Frohlich:2009gb,Davydov:2011kb,Suszek:2011hg,Suszek:2012ddg,Carqueville:2012st,Suszek:2013,Novak:2015ela,Runkel:2020zgg},\ and oftentimes associated,\ in the classical lagrangean formulation,\ with the appearance of Cheeger--Simons(-type) differential characters ({\it cf.}\ \Rcite{Cheeger:1985}) localised on codimension $\,\geq 1\,$ submanifolds of the spacetime of the field theory.\ In the present work,\ and in an upcoming one dedicated to the study of `non-invertible' maximally symmetric defects in the WZW $\si$-model,\ we undertake the task of defending the claim that the higher-geometric and -cohomological approach to the 2$d$ $\si$-model of charged-loop dynamics in the presence of worldsheet defects,\ as formalised in all generality in \Rcite{Runkel:2008gr},\ still has much to offer by way of identification and elucidation of the nontrivial information on the underlying field theory encoded in its category of defects with fusion,\ and so also by way of inspiration and as a testing ground -- structurally rich yet rigorously tractable -- for the more general field-theoretic context.\ As such,\ the work can be regarded as a continuation of the line of research started in the series of papers \cite{Runkel:2008gr,Runkel:2009sp,Suszek:2011hg,Suszek:2012ddg,Suszek:2013},\ and its extension to the $\bZ/2\bZ$-graded setting motivated by the deep nature of the results obtained and anticipated in the un-graded one and based on the recent developments \cite{Suszek:2017xlw,Suszek:2019cum,Suszek:2018bvx,Suszek:2018ugf,Suszek:2020xcu,Suszek:2020rev,Suszek:2021hjh},\ both -- the continuation and the extension -- to be constructively inscribed herewith in the renaissance of the idea of defect-mediated state correspondences in field theories,\ invoked in the opening paragraph.\smallskip

\begin{figure}[hbt]

$$
 \raisebox{-50pt}{\begin{picture}(0,195)
  \put(-160,-5){\scalebox{0.5}{\includegraphics{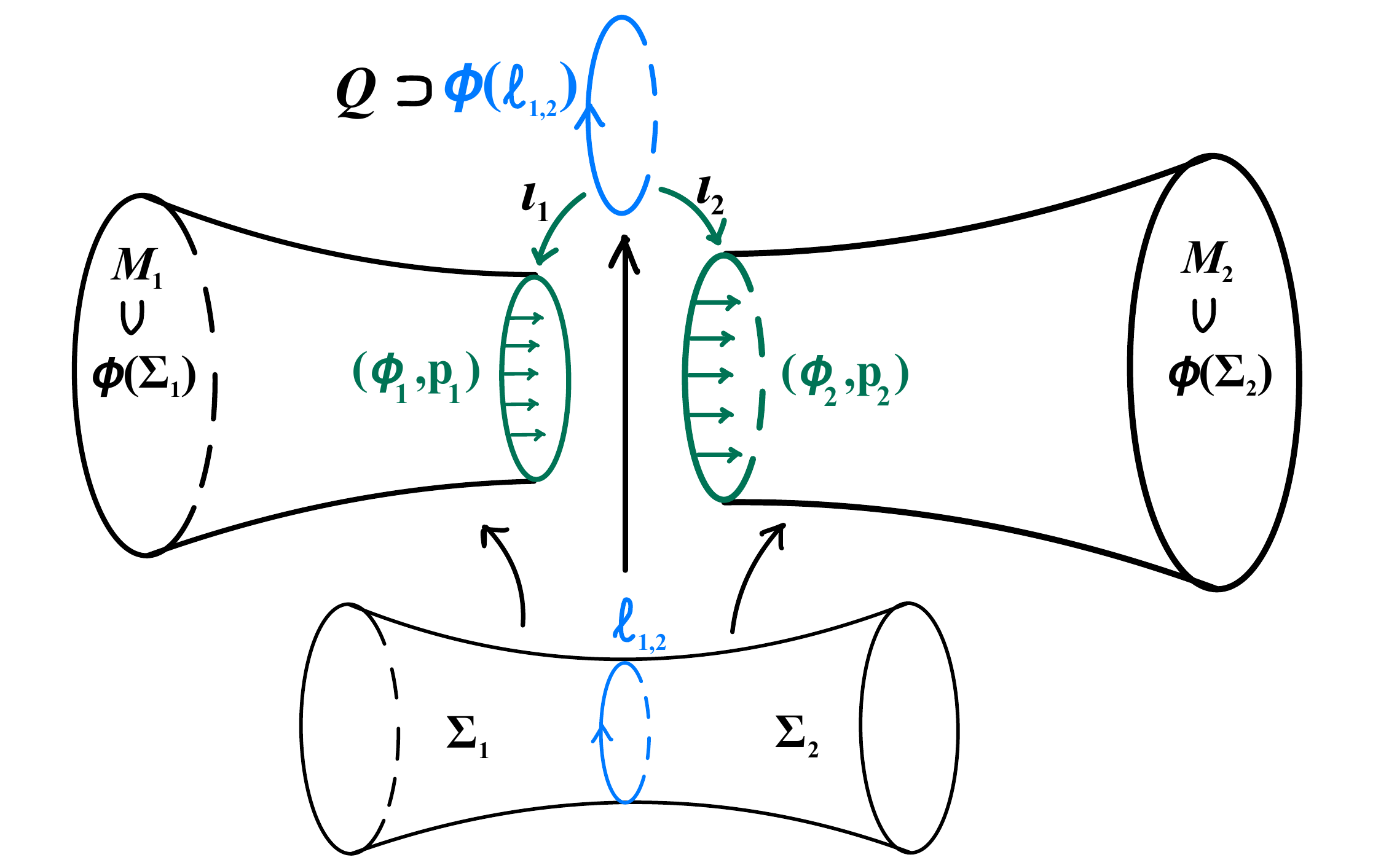}}}
  \end{picture}}
$$

\caption{A correspondence,\ engendered by a circular defect $\,\ell_{1,2}$,\ between the bulk states $\,(\phi_1,\txp_1)\,$ and $\,(\phi_2,\txp_2)\,$ from the respective phases $\,[\Si_A,M_A],\ A\in\{1,2\}\,$ of the field theory separated by $\,\ell_{1,2}$.} \label{fig:defect-corresp}
\end{figure}

\begin{figure}[hbt]

$$
 \raisebox{-50pt}{\begin{picture}(0,280)
  \put(-130,-5){\scalebox{0.5}{\includegraphics{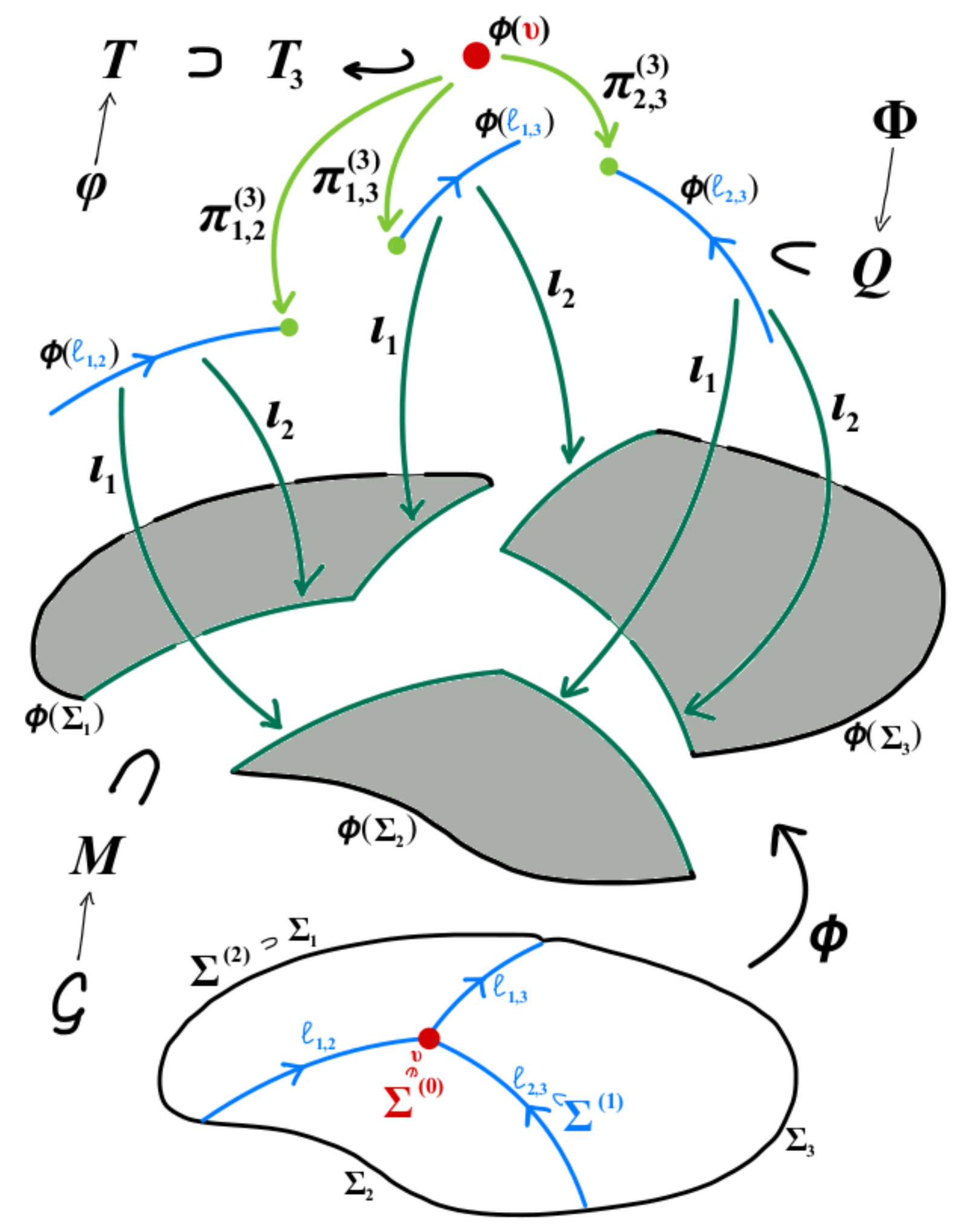}}}
  \end{picture}}
$$

\caption{A decomposition of the worldsheet $\,\Si\,$ into submanifolds $\,\Si^{(k)}\subset\Si\,$ of the respective dimensions $\,k\in\{0,1,2\}$,\ mapped by the field $\,\phi\in[\Si,M\sqcup Q\sqcup T]\,$ of the $\si$-model into the respective components of the composite target $\,M\sqcup Q\sqcup T\,$ endowed with distinguished 0-cells ($\cG$),\ 1-cells ($\Phi$) and 2-cells ($\varphi$) of the bicategory $\,\bgrb^\nabla(M\sqcup Q\sqcup T)$.\ The green color indicates boundaries of the submanifolds (dark green for boundaries of 2$d$ domains and light green for those of the 1$d$ components of the defect graph) at which correspondences arise that generalise,\ in an obvious manner to be formalised presently,\ that of Fig.\,\ref{fig:defect-corresp}.} \label{fig:bicat-sheet}
\end{figure}

\begin{figure}[hbt]

$$
 \raisebox{-50pt}{\begin{picture}(0,130)
  \put(-280,0){\scalebox{0.33}{\includegraphics{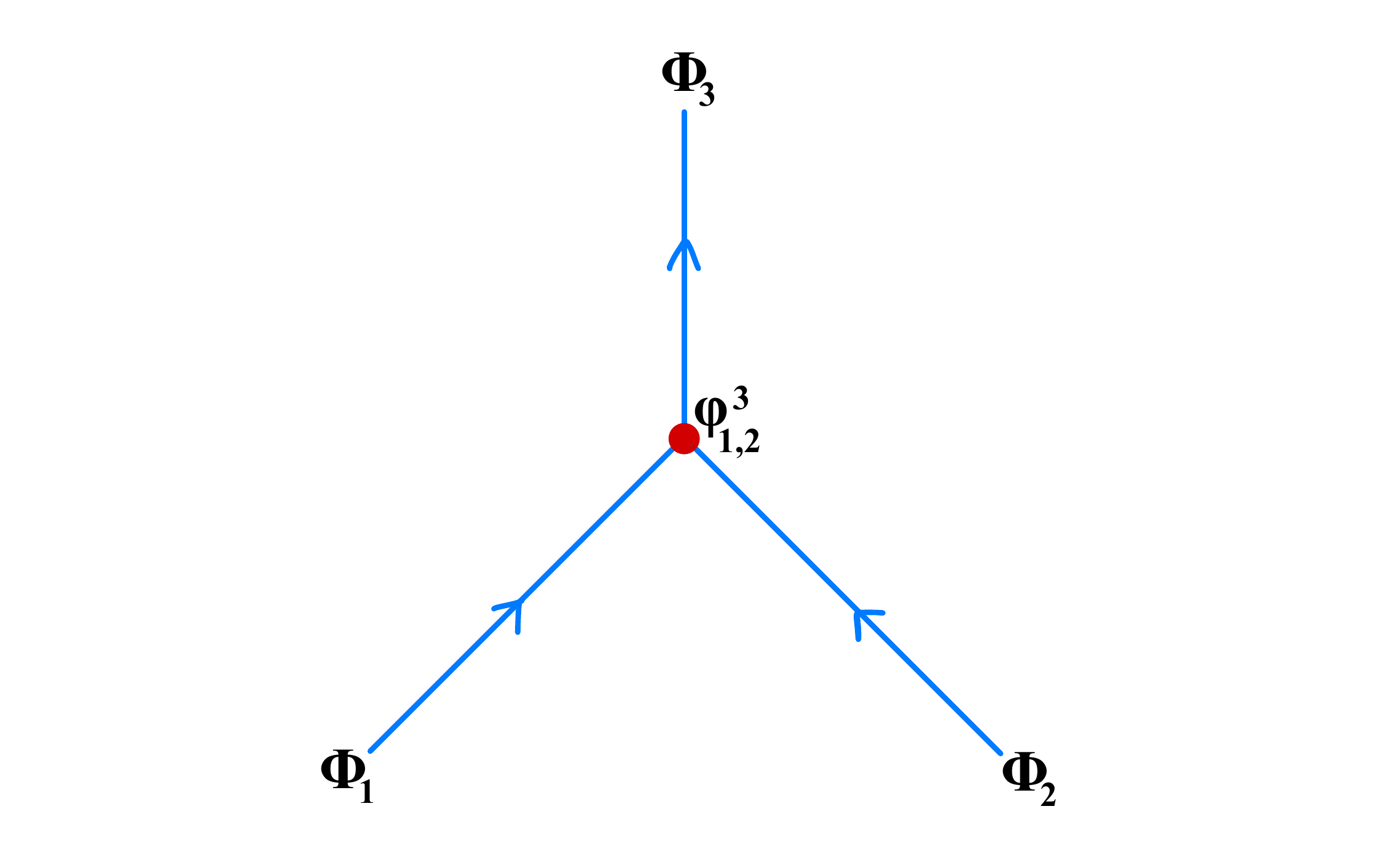}}}\put(-180,-5){(a)}
  \put(-110,0){\scalebox{0.33}{\includegraphics{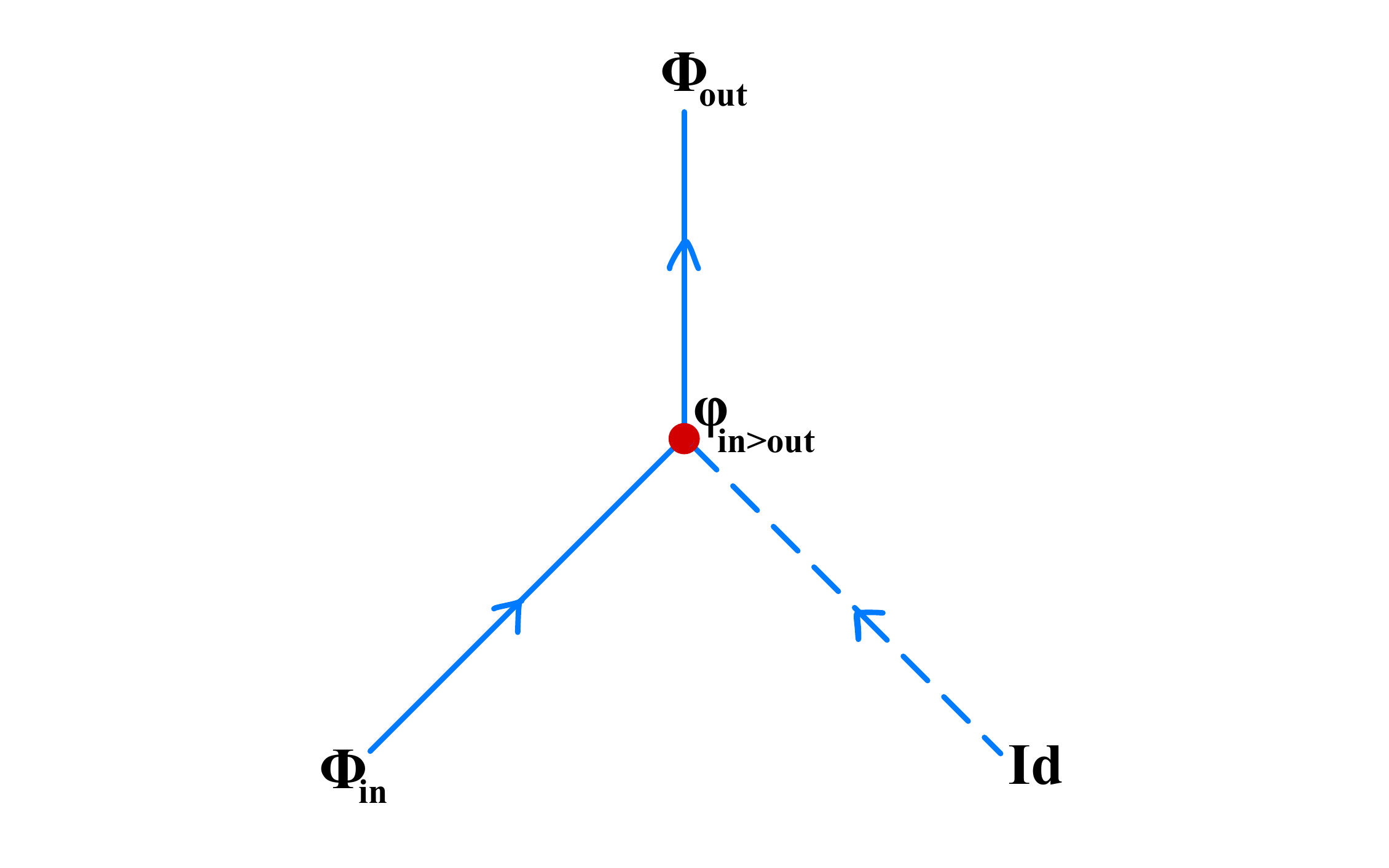}}}\put(-10,-5){(b)}
  \put(65,-5){\scalebox{0.33}{\includegraphics{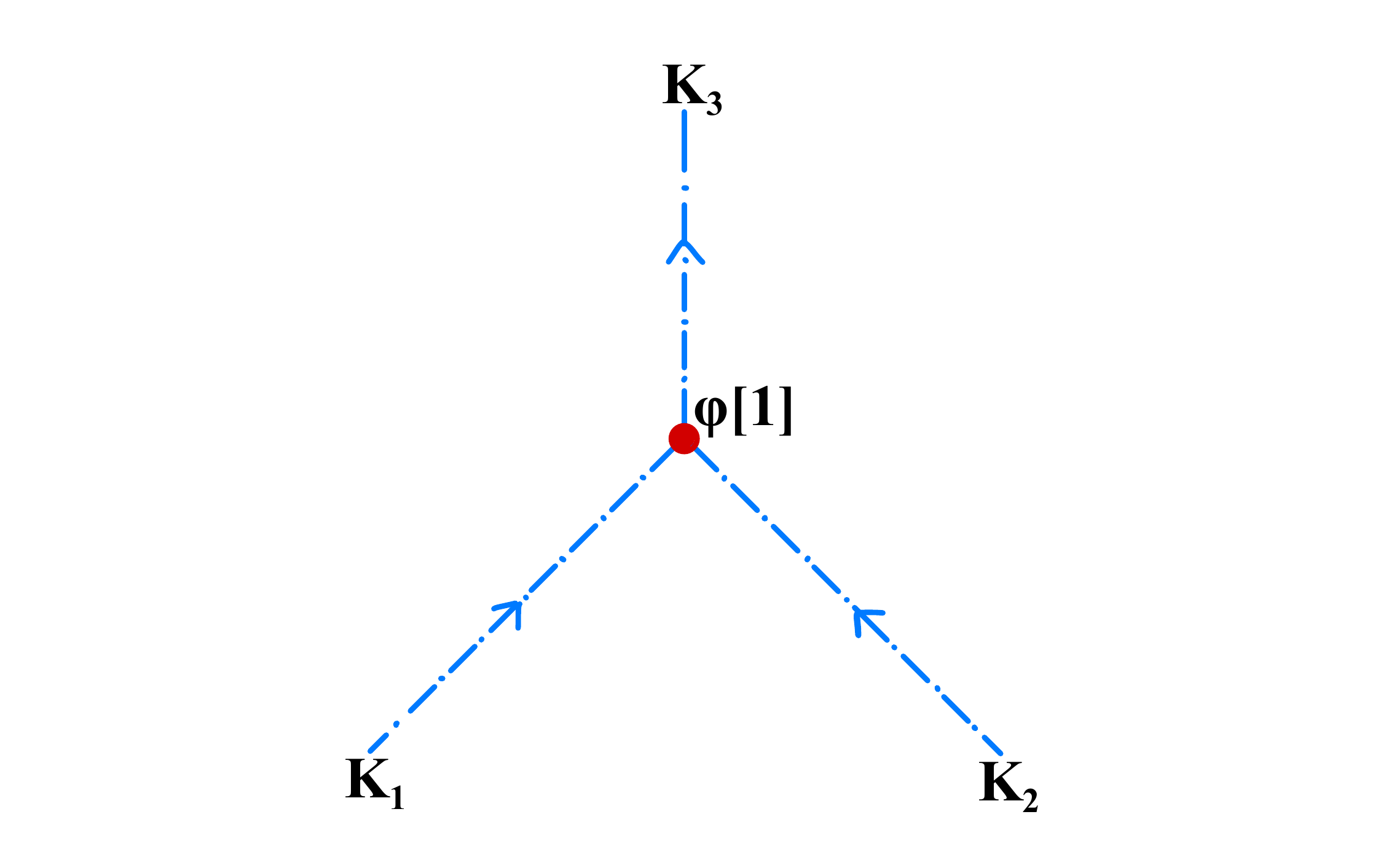}}}\put(165,-5){(c)}
  \end{picture}}
$$

\caption{The zoo of gerbe-theoretic 2-cells and 0-defects.\ (a) A generic fusion 0-defect with a 2-cell $\,\varphi_{1,2}^3\ :\ (\pi_{2,3}^{(3)\,*}\Phi_2\ox\id)\circ\pi_{1,2}^{(3)\,*}\Phi_1\xLongrightarrow{\cong}\pi_{1,3}^{(3)\,*}\Phi_3$.\ (b) A (1-)defect-changing 0-defect with a 2-cell $\,\varphi_{{\rm in}>{\rm out}}\ :\ \pi_{1,2}^{(3)\,*}\Phi_{\rm in}\xLongrightarrow{\cong}\pi_{1,3}^{(3)\,*}\Phi_{\rm out}$.\ (Here,\ $\,\pi_{2,3}^{(3)}\,$ maps to $\,M\subset Q$.) (c) A candidate `genuine' 1-form defect with a 2-cell $\,\varphi_{1,2}^3\ :\ (\pi_{2,3}^{(3)\,*}\txK_2\ox\id)\circ\pi_{1,2}^{(3)\,*}\txK_1\xLongrightarrow{\cong}\pi_{1,3}^{(3)\,*}\txK_3\,$ for some automorphisms $\,\txK_A,\ A\in\{1,2,3\}\,$ of the bulk 0-cell which trivialise under transgression.\ (Here,\ $\,\pi_{1,2}^{(3)},\ \pi_{2,3}^{(3)}\,$ and $\,\pi_{1,3}^{(3)}\,$ map to $\,Q\subset T_3$.)} \label{fig:Chan-Paton-def}
\end{figure}

The basic idea that motivates interest in a rigorous formulation of the 2$d$ $\si$-model with the topological Wess--Zumino term in the presence of defects,\ and -- \`a la fois -- anchors such interest in the broader context indicated above,\ is illustrated in Fig.\,\ref{fig:defect-corresp}:\ A circular (spacelike) defect line $\,\ell_{1,2}\,$ separating the two `domains' $\,\Si_A,\ A\in\{1,2\}\,$ of the two-dimensional spacetime $\,\Si$,\ inhabited by the respective `phases' $\,[\Si_A,M]\,$ of the (bulk) $\si$-model with the target space $\,M\,$ (potentially of the type $\,M=M_1\sqcup M_2\,$ and with $\,\Si_A\,$ constrained to map to $\,M_A$),\ sets in a natural correspondence those pairs of the limiting (classical) field configurations $\,\phi_A\,$ in either phase,\ taken with the respective (kinetic-)momenta $\,\txp_A\,$ determining their `evolution' into the interior of the domains,\ which jointly define an admissible field configuration $\,\phi\in[\Si,M\sqcup Q\,]$ of the $\si$-model \emph{in the presence of the defect line},\ the latter being,\ in general,\ mapped by (the restriction of) $\,\phi\,$ to another manifold $\,Q$,\ sometimes suggestively termed the correspondence space.
In a field theory whose quantum description is governed by Segal's axioms of Refs.\,\cite{Segal:1987sk,Segal:2002},\ incorporation of line defects necessitates admission of their intersections (or junctions),\ coming with their own targets $\,\bigsqcup_{n\geq 3}\,T_n\equiv T\,$ (labelled by the valence of the junctions),\ and so,\ altogether,\ we are led to enquire as to the data which ought to be pulled back to the defect lines and junctions of an arbitrary defect graph $\,\G\,$ embedded in $\,\Si\,$ from the respective targets in order to ensure the well-definedness of the ensuing field theory on the $\G$-decorated spacetime $\,\Si\,$ once the phases of the bulk $\si$-model have been defined in terms of a metric on $\,M\,$ and a geometrisation of an integral class $\,[\txH]\in H^3(M,2\pi\bZ)\,$ of the Kalb--Ramond (torsion) field $\,\txH$,\ known as the gerbe,\ {\it cf.}\ Refs.\,\cite{Gawedzki:1987ak} and \cite{Murray:1994db} for the original hypercohomological and higher-geometric descriptions,\ respectively.\ Drawing inspiration from the study of a class of circular defects and the associated targets reported in \Rcite{Fuchs:2007fw},\ the question was asked and answered in all generality in \Rcite{Runkel:2008gr}:\ Consistency of the $\si$-model for $\,(\Si,\G)\,$ was demonstrated to call for the full bicategory $\,\bgrb^\nabla(M\sqcup Q\sqcup T)\,$ of gerbes with connective structure over the composite target space $\,M\sqcup Q\sqcup T$,\ whose distinguished $k$-cells are to be pulled back to the $(2-k)$-dimensional submanifolds $\,\Si^{(k)}\subset\Si\,$ (with $\,k\in\{0,1,2\}$) into which the embedded defect graph $\,\G\,$ decomposes the spacetime $\,\Si\,$ of the theory,\ as illustrated in Fig.\,\ref{fig:bicat-sheet}.
The results of the detailed worldsheet analysis presented {\it ibidem} helped to concretise and formalise the raw idea of a defect/symmetry correspondence in \Rcite{Suszek:2011hg}:\ The gluing law for the kinetic momenta of the states in correspondence,\ enforced at the defect line $\,\ell_{1,2}\,$ by the so-called Defect Gluing Condition (DGC),\ was shown to define an isotropic subspace in the product of the spaces of states of the two phases separated by $\,\ell_{1,2}\,$ (taken with the difference of the pullbacks of the symplectic forms that the first-order formalism of Refs.\,\cite{Gawedzki:1972ms,Kijowski:1973gi,Kijowski:1974mp,Kijowski:1976ze,Szczyrba:1976,Kijowski:1979dj} associates with the bulk Dirac--Feynman amplitudes for the phases),\ and data of the 1-cell of $\,\bgrb^\nabla(M\sqcup Q\sqcup T)\,$ assigned to the defect line were found to give rise to a linear map between the prequantum bundles associated to the two phases {\it via} Gaw\c{e}dzki's cohomological transgression of \Rcite{Gawedzki:1987ak}.\ The study also shed light upon the nature of the generalisation of the `ancient' concept of symmetry of the $\si$-model (or,\ more generally,\ of a duality between such field theories) furnished by defect lines decorated by the higher-categorial target-space data:\ The latter need not induce a bijective correspondence between the spaces of states of the two phases,\ let alone map the respective hamiltonians into one another -- these are the distinctive features of the very special topological defects discussed at great length in the original study \cite{Runkel:2008gr} ({\it cf.}\ with the general considerations in \Rcite{Bhardwaj:2022yxj}),\ whereas the capacity of the new construct is most vividly demonstrated by the boundary defects of \Rcite{Runkel:2008gr} which separate our favourite $\si$-model from the empty one and in this manner model the insertion of an arbitrary worldsheet boundary.\ In fact,\ the story does not end there:\ Defect junctions decorated with data of the 2-cells of $\,\bgrb^\nabla(M\sqcup Q\sqcup T)\,$ naturally take the r\^ole of zero-dimensional defects (the `non-genuine' ones in the language of \Rcite{Bhardwaj:2022yxj}),\ and with the help of the identity defect they can be made to act on the Chan--Paton(-type) degrees of freedom of the defect-twisted sector.\ Actually,\ one might even try to squeeze the framework for `genuine' 1-form symmetries by considering defect junctions between defect lines carrying data of the 1-cells of $\,\bgrb^\nabla(M\sqcup Q\sqcup T)\,$ mapping to the identity automorphism of the prequantum bundle under transgression,\ {\it cf.}\ Fig.\,\ref{fig:Chan-Paton-def}.

With the bicategorial content of the $\si$-model with defects firmly established,\ it became possible to actually \emph{employ},\ in \Rcite{Runkel:2008gr},\ data of the $Z_\si$-equivariant structures on the bulk gerbe for discrete-symmetry groups $\,Z_\si\,$ of the bulk theory,\ requisite for the gauging of $\,Z_\si\,$ (aka orbifolding resp.\ orientifolding) and established in Refs.\,\cite{Gawedzki:2003pm,Gawedzki:2007uz,Gawedzki:2008um},\ in the construction of defect networks \emph{implementing} the discrete symmetry on bulk states resp.\ giving rise to the so-called $Z_\si$-twisted sector of the $\si$-model with the orbifold $\,M//Z_\si\,$ as the target space (in the spirit of Refs.\,\cite{Dixon:1985jw,Dixon:1986jc}).\ The construction involved fusion (through defect junctions and the attendant bicategorial 2-cells) of the 1-cells attached to the inherently topological discrete-symmetry defect lines,\ from which the cohomological Moore--Seiberg data of \Rcite{Moore:1988qv} of the relevant simple-current sector were extracted in \Rcite{Runkel:2008gr} in the setting of the Wess--Zumino--Witten $\si$-model of \Rcite{Witten:1983ar}.\ An adaptation of the idea to the substantially more structured case of a continuous gauge symmetry (investigated in the bicategorial framework in Refs.\,\cite{Gawedzki:2010rn,Gawedzki:2012fu}) subsequently led the Author,\ in a study reported in Refs.\,\cite{Suszek:2012ddg,Suszek:2013},\ to a reinterpretation (and hence an independent justification) of a full-fledged $\txG_\si$-equivariant structure on the cells of $\,\bgrb^\nabla(M\sqcup Q\sqcup T)$,\ defining a $\si$-model with a continuous rigid symmetry $\,\txG_\si\,$ (and with $\txG_\si$-symmetric defects),\ as data of a topological gauge-symmetry defect implementing the gauging of $\,\txG_\si\,$ in what may be viewed as a generalisation of the worldsheet-orbifold scheme of Refs.\,\cite{Dixon:1985jw,Dixon:1986jc}.\ The bicategorial framework derived in \Rcite{Runkel:2008gr} was also applied in the study of T-duality presented in \Rcite{Sarkissian:2008dq} ({\it cf.}\ also \Rcite{Suszek:2012ddg}),\ and -- finally -- in the analysis of elementary ({\it i.e.},\ ternary) fusion of the maximally symmetric defects of the WZW $\si$-model with the target $\,{\rm SU}(2)\,$ ({\it cf.}\ \Rcite{Fuchs:2007fw}) reported in \Rcite{Runkel:2009sp},\ which pointed towards a novel geometric interpretation of the Verlinde fusion rules of the chiral fusion ring of the $\si$-model in terms of the relevant elementary-fusion 2-cells.\smallskip

The above recapitulation sets the scene for the study documented herein,\ and delineates its broader context.\ The study proposes,\ in its Part \ref{p:genstr},\ a complete framework for the lagrangean description of the dynamics of the (Nambu--Goto--)Green--Schwarz type ({\it cf.}\ Refs.\,\cite{deAzcarraga:1982dhu,Green:1983sg,Green:1983wt}) of (super)charged loops in a supermanifold (or a disjoint union thereof) in the presence of an arbitrary defect graph embedded in the 2$d$ spacetime.\ It subsequently examines at great length a particular (general) instantiation of the framework,\ inspired by the analysis and use of the so-called induction of 2-cell data associated with valence $\,>3\,$ defect junctions of topological defect lines from those of elementary (ternary) 2-cells in the un-graded setting of Refs.\,\cite{Runkel:2008gr,Suszek:2012ddg,Suszek:2013},\ to wit,\ simplicial hierarchies of targets and the attendant simplicial gerbes,\ in which much of the higher-geometric analysis localises neatly on the first three levels of the hierarchy (indexed,\ form level 3 upwards,\ by the valence of the corresponding defect junctions):\ the bulk level (0) with its 0-cell (a gerbe),\ the correspondence level (1) with its 1-cell (a gerbe bi-module),\ and the elementary-fusion level (3) with its 2-cell (an elementary fusion 2-isomorphism).\ The final act of the general discussion sees simplicial hierarchy enriched equivariantly by an action of a (Lie) (super)group of rigid symmetries of the dynamics under consideration -- this places us in the context of (super)field theory with maximally symmetric defects,\ {\it i.e.},\ those fully transmissive to the Noether currents flowing in the bulk.\ The ensuing 2$d$ superfield theory with defects is to be realised in Freed's (inner ${\rm Hom}-)$functorial scheme delineated in \Rcite{Freed:1999} ({\it i.e.},\ in particular,\ with the spacetime of the theory kept Gra\ss mann-\emph{even},\ unlike in,\ {\it e.g.},\ Refs.\,\cite{Novak:2015ela,Runkel:2020zgg},\ and in contrast to the superembedding scheme of \Rcite{Sorokin:1999jx}),\ and constitutes a consistent transcription of the proposal of \Rcite{Runkel:2008gr} into the $\bZ/2\bZ$-graded setting of (higher) target supergeometry,\ which allows for a unified treatment of bosonic and fermionic degrees of freedom of the super-charged loop (or superstring) -- propagating across defects or,\ dually,\ twisted by the defect -- within the higher-cohomological paradigm introduced (in the un-graded setting) by Gaw\c{e}dzki in his seminal paper \cite{Gawedzki:1987ak},\ and subsequently rendered higher-geometric and -categorial by Murray in the foundational \Rcite{Murray:1994db} and by his followers.\ The transcription rests heavily -- implicitly and,\ at later stages,\ explicitly -- upon the recent \emph{constructive} results in the theory of (supersymmetric) gerbes on supermanifolds,\ reported in Refs.\,\cite{Suszek:2017xlw,Suszek:2019cum,Suszek:2018bvx,Suszek:2018ugf,Suszek:2020xcu,Suszek:2020rev,Suszek:2021hjh},\ and upon an obvious bicategorial extension of the very general study of gerbes on supermanifolds in \Rcite{Huerta:2020}.\ The ultimate goal here is a systematic incorporation of supersymmetry present in the world-sheet theories of super-$p$-branes of the superstring theory in the lagrangean modelling of their dynamics in the presence of worldvolume defects preserving all or some of the bulk supersymmetry.\ When viewed from this perspective,\ the current analysis remains,\ admittedly,\ incomplete,\ in that it only deals,\ at later stages of its specialisation,\ with the \emph{global} (left) supersymmetry of the super-$\si$-model with defects -- indeed,\ this supersymmetry is consistently built into the description following the original idea of \Rcite{Suszek:2017xlw}.\ Investigation of the fate,\ in the presence of the rigidly supersymmetric defects,\ of the `infinitesimal gauge' supersymmetry of Refs.\,\cite{deAzcarraga:1982dhu,Siegel:1983ke,Siegel:1983hh},\ responsible for the restoration of balance between the two species of degrees of freedom in the vacuum of the bulk theory (aka $\k$-symmetry) and recently placed in the higher-geometric context in Refs.\,\cite{Suszek:2020xcu,Suszek:2020rev,Suszek:2021hjh},\ is,\ on the other hand,\ postponed to a separate future study.\ Upon the latter augmentation,\ the approach to the dynamics of extended distributions of super-charge advocated herein is expected to provide insights into the structure of the relevant superfield theories similar in their breadth and depth to those offered by its predecessor in the un-graded setting,\ as recapitulated succinctly,\ {\it e.g.},\ in the review \Rcite{Suszek:2020rev},\ and so,\ in particular,\ to contribute significantly to the development of the modern study of generalised symmetries invoked in the opening paragraph.\ This expectation,\ further backed up by the novel findings on deep structural relations between the higher geometry behind the fusion of the maximally symmetric defects of the WZW $\si$-model and the non-perturbative content of the (chiral) bulk theory (as encoded in the associated 3$d$ topological Chern--Simons field theory),\ which we present in Part \ref{p:WZWCS},\ becomes the rationale for the study,\ developed in Part \ref{p:smaxym-bib},\ of the very first -- to the best of the Author's knowledge -- examples of higher-supergeometric 1- and 2-cells associated with maximally (rigid-)supersymmetric defects in the Green--Schwarz super-$\si$-model of superstring dynamics in the flat supertarget $\,{\rm sMink}(d,1|D_{d,1})$.\ Thus,\ in both the un-graded and the $\bZ/2\bZ$-graded geometric (target) sceneries,\ we put some flesh on the bones of the formal constructions of Part \ref{p:genstr} by considering explicit examples of (super)string backgrounds for which the general principles are seen or expected to work,\ to wit,\ those whose 1-cells implement the fundamental symmetry of the bulk target -- the right regular translations in the target Lie (super)group.\smallskip 

Let us now be more specific as to the contents of the present paper.\ It starts with a systematic introduction,\ given in Sec.\,\ref{sec:wrldshtbic},\ of all the geometric ingredients -- topological,\ tensorial and,\ ultimately,\ also higher-geometric -- of the definition of the super-$\si$-model of the Green--Schwarz type on the mapping supermanifold for -- as the domain -- a Gra\ss mann-even closed 2$d$ spacetime (the worldsheet) with an embedded defect graph $\,\G\,$ and -- as the codomain -- a disjoint union of supermanifolds (the target super-space) related by a family of supermanifold morphisms reflecting the simplicial decomposition of the worldsheet induced by $\,\G$.\ The latter comes with a ditinguished family of 0-cells (gerbes),\ 1-cells (gerbe bi-modules) and 2-cells (gerbe 2-isomorphisms) of the bicategory of gerbes with connective structure over the target super-space.\ The introduction,\ culminating in Definition \ref{def:ssimod},\ is augmented with a detailed discussion,\ initiated in Sec.\,\ref{sub:simpltargsp} and continued through Sec.\,\ref{sub:grbmultissi},\ of a special example of the target structures going into the definition in which data associated with defect junctions of arbitrary valence are induced from those defining the elementary (ternary) junctions,\ as they often do,\ through relative homotopy moves of defects junctions (resolutions and fusions),\ in the motivating case of topological defects implementing symmetries studied in \Rcite{Runkel:2008gr,Suszek:2012ddg},\ and in which the existence of the trivial (identity) defect is neatly encoded by certain `upstream' mappings (the so-called degeneracy maps).\ The special example is provided by the simplicial superstring backgrounds of Def.\,\ref{simplssbckgrndcompl} whose face maps \emph{induce} the maps $\,T\too Q\,$ filtering through $\,T_3$,\ whence the induction.\ Their relevance to the present context essentially hinges on their elementary (combinatorial) properties stated in Props.\,\ref{prop:simplicial-target} and \ref{prop:DJI-descent}.\ In the next Sec.\,\ref{sec:susygrb} (global) supersymmetry is incorporated in the three-tier architecture of the super-$\si$-models with defects introduced previously (and in particular -- lifted from the super-target to the higher-geometric structure over it,\ and thus essentially categorified) and the notion of a maximally (super)symmetric background is introduced.\ Furthermore,\ a special construction of a supersymmetric background with explicit (built-in) supersymmetry is presented (in Def.\,\ref{def:supergerbetc}),\ which is at the heart of a programme,\ developed by the Author in a series of papers \cite{Suszek:2017xlw,Suszek:2019cum,Suszek:2018bvx,Suszek:2018ugf,Suszek:2020xcu,Suszek:2020rev,Suszek:2021hjh},\ of (\emph{constructive}) supergeometrisation of the classes in the Cartan--Eilenberg cohomology of supersymmetry Lie superalgebras defining super-$\si$-models for the so-called super-$p$-branes of superstring theory in homogeneous spaces $\,\txG/\txH\,$ of supersymmetry Lie supergroups $\,\txG\,$ (as in Refs.\,\cite{deAzcarraga:1982dhu,Green:1983wt,Green:1983sg,Achucarro:1987nc,Metsaev:1998it,Zhou:1999sm,Arutyunov:2008if,Gomis:2008jt,Fre:2008qc,DAuria:2008vov}.\ The supergeometrisation uses stepwise application (inspired by the works of de Azc\'arraga {\it et al.},\ {\it cf.},\ {\it e.g.},\ \Rcite{Chryssomalakos:2000xd}) of the bijective correspondence between classes in the second cohomology group $\,H^2(\ggt,\agt)\,$ of a Lie superalgebra $\,\ggt\,$ with values in its trivial supercommutative module $\,\agt\,$ and equivalence classes of supercentral extensions of $\,\ggt\,$ through $\,\agt$,\ and can be understood -- morally (and prior to descent from $\,\txG\,$ to $\,\txG/\txH\,$ {\it via} $\txH$-equivariantisation,\ as dictated by a straightforward adaptation the gauge principle of Refs.\,\cite{Gawedzki:2010rn,Gawedzki:2012fu,Suszek:2012ddg}) -- as internalisation of the geometric construction due to Murray {\it et al.} from Refs.\,\cite{Murray:1994db,Murray:1999ew,Stevenson:2000wj,Stevenson:2001grb2,Carey:2002,Johnson:2003,Carey:2004xt} in the category of Lie supergroups.\ The general Part \ref{p:genstr} concludes with a harmoniuos marriage,\ staged in Sec.\,\ref{sec:smultcat},\ between the principle of simplicial hierarchy and the principle of (maximal) (super)symmetry,\ blessed with the principle of (full) reducibility,\ whereby the fomer discussion winds up over a disjoint union of full orbits of the (super)symmetry group embedded in a simplicial $\txG$-supermanifold.\ Upon categorification,\ after \Rcite{Carey:2004xt},\ of the binary operation on the (super)group (whose appearance in this context,\ alongside local trivialisations of the bulk gerbe over (reduced-)supersymmetry orbits within the bulk target super-space,\ is carefully explained {\it ibidem}) -- generalised suitably in Def.\,\ref{def:genmultstr} and rendered compatible with supersymmetry in Def.\,\ref{def:susy-mult-str},\ with hindsight,\ so as to allow for a unified treatment of the un-graded and $\bZ/2\bZ$-graded examples studied afterwards -- this last construction acquires a natural model,\ given by the Lie backgrounds introduced in Sec.\,\ref{sub:WZWorbs}.

The next two parts of the paper provide an extensive illustration of the general considerations of Part \ref{p:genstr}.\ Thus,\ first,\ in Part \ref{p:WZWCS},\ the un-graded maximally symmetric WZW simplicial backround is discussed with the first two rungs of the simplicial hierarchy given by a compact simple 1-connected Lie group $\,\txG\,$ with (the Cartan--Killing metric and) the Cartan 3-form and the associated integer tensor power of the basic gerbe of \Rcite{Meinrenken:2002} (the bulk) and the disjoint sum of the maximally symmetric bi-branes of \Rcite{Fuchs:2007fw} (the correspondence),\ respectively.\ The bulk tensorial data are fixed by the requirement of a non-anomalous conformal symmetry of the ensuing 2$d$ bulk field theory with the given target,\ {\it cf.}\ \Rcite{Witten:1983ar},\ and those of the bi-brane (and of the associated brane) are shown to be determined uniquely by the demand that the ensuing DGC be preserved by the loop-group extension of the left-right regular translational (resp.\ vector,\ {\it i.e.},\ adjoint for the brane) symmetry of the target (Prop.\,\ref{prop:maxym-curv-existuniq} resp.\ \ref{prop:maxym-curv-bdry-existuniq}).\ Next,\ a proposal for the component of the target associated with junctions of the maximally symmetric defects is given in Def.\,\ref{def:fus-2-iso},\ based on the previous work reported in Refs.\,\cite{Runkel:2008gr,Runkel:2009sp} and on the long chain of observations and inferences laid out in Secs.\,\ref{sub:maxym-def-junct} ,\ \ref{sec:cs} and \ref{sub:fus-2iso}.\ A distinguished place among them is occupied by Thm.\,\ref{thm:ker-Om},\ one of the main concrete results of the paper,\ which asserts that the necessary condition (tensorial) for the existence of the maximally symmetric fusion 2-cells is satisfied \emph{precisely} on the disjoint union of orbits of an appropriate lift of the left-right regular action to the (simplicial) nerve of the right-regular action groupoid of $\,\txG\,$ on itself deduced from the mixed simpliciality/symmetry(/reducibility) analysis of Sec.\,\ref{sub:maxym-def-junct}. The theorem exploits an original interpretation of the said condition (on $\,T_n$) in terms of a partially symplectically reduced -- \`a la Alekseev and Malkin ({\it cf.}\ \Rcite{Alekseev:1993rj}) -- presymplectic form of the 3$d$ group-$\txG$ Chern--Simons theory on the (timelike) cylinder $\,\bR\x\bS^2\,$ over the 2-sphere punctured by a collection of ($n$) vertical (timelike) Wilson lines with holonomies from the same conjugacy classes as the (incoming) ones entering the definition of the component maximally symmetric bi-branes under fusion (Prop.\,\ref{prop:Om-AM-vs-RS}).\ The remarkable new link between the 2$d$ WZW $\si$-model and the 3$d$ CS theory thus established extends nontrivially the hitherto evidence on the deep structural relationship between the two studied extensively in the past ({\it cf.}\ Refs.\,\cite{Witten:1988hf,Gawedzki:1989rr,Witten:1991mm,Gawedzki:1999bq,Gawedzki:2001rm}) and forming the basis of the functorial quantisation scheme for the WZW RCFT advanced in \Rcite{Felder:1999mq,Fuchs:2002cm}.\ Furthermore,\ it harmonises nicely with the general findings of \Rcite{Suszek:2012ddg} on the transgression (into symmetry intertwiners) of the fusion 2-cells in the presence of a continuous bulk symmetry preserved at the defect.\ All this,\ put in conjunction with the results of \Rcite{Frohlich:2006ch} on the relation between the so-called Verlinde fusion rules of the WZW $\si$-model and junctions of the maximally symmetric defects (in the TFT-based quantum-mechanical description of the WZW RCFT),\ leads -- through a gerbe-theoretic analysis of Sec.\,\ref{sec:WZW-target} making ample use of the multiplicative structure on the basic gerbe -- to a very natural conjectural identification of the (higher-)geometric structure encoding the said fusion rules (Conjecture \ref{conj:2-iso-vs-Verlinde}) and to a definition of a bicategorial fusing matrix of the maximally symmetric WZW background (Def.\,\ref{def:maxym-WZW-ibi-mat}),\ anticipated to be intimately related to the corresponding component of the Moore--Seiberg data of the RCFT ({\it cf.}\ \Rcite{Moore:1988qv}),\ as -- indeed -- it was proven to be in the simple-current sector in \Rcite{Runkel:2008gr}.

Part \ref{p:smaxym-bib} provides an account of a pioneering,\ and hence -- in particular -- largely non-systematic exploration of the bicategorial structure associated with the Green--Schwarz super-$\si$-model of dynamics of the superstring in $\,{\rm sMink}(d,1|D_{d,1})\,$ (where $\,D_{d,1}\,$ is the dimension of the relevant Majorana-spinor representation of the Clifford algebra $\,\Cliff(\bR^{d,1})$) in the presence of (candidate) maximally symmetric defects.\ Drawing inspiration from the long-known structural affinity between the 2$d$ superfield theory and the bosonic WZW $\si$-model,\ noted already in \Rcite{Henneaux:1984mh} and more recently elaborated in \Rcite{Suszek:2019cum},\ we adapt the definition of the bicategory for the maximally symmetric defect from Part \ref{p:WZWCS}.\ In so doing,\ we take as the starting point the \emph{explicit} construction,\ proposed in \Rcite{Suszek:2017xlw},\ of the Green--Schwarz super-1-gerbe $\,\cG_{\rm GS}\,$ (recalled in Sec.\,\ref{sec:GSWZW}),\ which (super)geometrises,\ in the hands-on and manifestly supersymmetric procedure \`a la de Azc\'arraga mentioned above,\ the Cartan--Eilenberg super-3-cocycle on the bulk Lie supergroup $\,{\rm sMink}(d,1|D_{d,1})$.\ In the first step,\ made in Sec.\,\ref{sec:multGSs1g},\ we prove the existence and explicitly reconstruct -- essentially as a corollary to the non-obvious Prop.\,\ref{prop:GS-mult-4coc} -- a supersymmetric generalised multiplicative structure on $\,\cG_{\rm GS}\,$ (Thm.\,\ref{thm:mult-str-GS}),\ another important result of the present paper.\ Next,\ in Sec.\,\ref{sec:susyGbrs},\ we identify,\ in a semi-systematic manner (drawing heavily on the un-graded intuition and the findings of \Rcite{Suszek:2021hjh},\ anticipated in \Rcite{Suszek:2020rev} and concerning the behaviour of $\,\cG_{\rm GS}\,$ in restriction to the vacuum of the super-$\si$-model in the dual,\ purely topological Hughes--Polchinsky formulation of \Rcite{Hughes:1986dn},\ {\it cf.}\ also Refs.\,\cite{Gauntlett:1989qe,Suszek:2020xcu}) three classes of (rigidly) supersymmetric $\cG_{\rm GS}$-branes (Props.\,\ref{prop:AdGbr},\ \ref{prop:sstring-GS-brane} and \ref{prop:spoint-GS-brane}).\ Upon coupling to the multiplicative structure on $\,\cG_{\rm GS}\,$ derived previously,\ these yield -- in a manner fully analogous to the one originally devised in \Rcite{Fuchs:2007fw} -- the corresponding three classes of maximally (rigidly) supersymmetric $\cG_{\rm GS}$-bi-branes of Thm.\,\ref{thm:maxym-GS-bib}.\ The case study in the $\bZ/2\bZ$-graded setting closes with a detailed analysis -- carried out on all three levels:\ equivariantly geometric,\ tensorial and higher-geometric in Sec.\,\ref{sec:sibib} -- of the elementary fusion of the newly found species of super-bi-brane,\ summarised in Thm.\,\ref{thm:sfusion}.\smallskip

The general considerations and constructions presented in Part \ref{p:genstr} and the detailed discussion,\ in Parts \ref{p:WZWCS} and \ref{p:smaxym-bib},\ of the highly symmetric field theories of the (super-)WZW type in which they are seen to be of relevance leave us with a number of interesting questions and open several avenues of prospective research,\ which we indicate in the closing section of (the main text of) the paper,\ and to which we hope to return in near future.\bigskip

\noindent{\bf Acknowledgements:}  The work reported herein arises at the nexus of the many ideas and themes discussed with,\ learned from and worked upon jointly with the late Professor Krzysztof Gaw\c{e}dzki ($\ast\,$ 02.07.1947 -- $\,\dagger\,$ 21.01.2022) over the many years of profoundly inspiring and congenial interactions.\ The Author considers this work to be a humble tribute to the grateful memory of K -- a distinguished member of the Warsaw School of Mathematical Physics established by Professor Krzysztof Maurin,\ a great Friend and Teacher,\ and -- quite simply -- a noble Man.

The origins of the present work date back to the period 2007-2012 of a collaboration between the Author and Professor Ingo Runkel from Universit\"at Hamburg.\ Discussions with Professor Runkel about various aspects of the un-graded component of the work are gratefully acknowledeged herewith.

\part{Simplicial $\si$-models with (super)symmetry -- general structures}\label{p:genstr}

In this opening part,\ we give a formal definition of the two-dimensional lagrangean superfield theory describing the dynamics of a (super-)charged loop in (the body of) an ambient supermanifold in the presence of an arbitrary defect embedded in the two-dimensional spacetime.\ The definition is a fairly straightforward generalisation of the one derived by Runkel and the Author in the setting of the bosonic two-dimensional $\si$-model with the Wess--Zumino term in \Rcite{Runkel:2008gr}.\ Upon setting up the general bicategorial structure over the configuration (target) space of the superfield theory,\ we discuss at some length a particularly natural instantiation of the abstract definition,\ inspired by the study of topological defects in Refs.\,\cite{Runkel:2008gr,Suszek:2011hg,Suszek:2012ddg},\ in which the inherently simplicial relations between elements of the decomposition of the spacetime of the theory induced by the embedded defect graph are faithfully encoded in a hierarchical structure of the target space and the higher-geometric object over it,\ leading to induction of the data for defect junctions of valence higher than 3 from those of the elementary (ternary) ones.\ Subsequent coherent incorporation of the rigid (super)symmetry of the bulk theory into the simplicial picture leads to the emergence of distinguished maximally symmetric backgrounds,\ the main subject of this study.\ Final specialisation to the most elementary highly symmetric setting of a (target) Lie-(super)group (super)geometry then leads us to a systematic construction of a large class of defects compatible with the (super)symmetry of the phases of the 2d theory separated by them.\ The construction employs a suitable categorification of the constitutive binary operation on the target (super)group,\ and provides novel insights into the original 2d (super)field theory which base on the ensuing igher-geometric realisation of its fundamental rigid symmetry induced by the binary operation -- these we consider subsequently in the next two parts.

\section{From decorated (super)string worldsheets to target bicategories}\label{sec:wrldshtbic}

As our aim here is to put the study of supersymmetric defects of superstring theory in the higher-geometric framework developed in Refs.\,\cite{Suszek:2017xlw,Suszek:2019cum,Suszek:2018bvx,Suszek:2018ugf,Suszek:2020xcu,Suszek:2020rev,Suszek:2021hjh},\ we begin by adapting the formalism laid out in \Rcite{Runkel:2008gr} ({\it cf.}\ also Refs.\,\cite{Runkel:2009sp,Suszek:2011hg,Suszek:2012ddg},\ and,\ in particular,\ the pioneering \Rcite{Fuchs:2007fw}) for the study of multi-phase bosonic two-dimensional $\si$-models to the supersymmetric $\bZ/2\bZ$-graded setting of interest.

\subsection{Super-$\si$-models with defects}\label{sub:ssigmod-def}

In the absence of defects,\ {\bf superfield configurations} of the super-$\si$-model over a two-dimensional (purely Gra\ss mann-even) oriented closed manifold $\,\Si$,\ termed the {\bf worldsheet},\ that describes propagation of a (closed) superstring in (the body of) a supermanifold $\,M$,\ termed the ({\bf super}){\bf target},\ are defined to be `elements' of the mapping supermanifold 
\qq\nn
[\Si,M]\equiv\unl\Hom{}_\sMan(\Si,M):=\Hom_\sMan(\Si\x-,M)\ :\ \sMan\too\Set\,,
\qqq
evaluated on the family $\,\{\bR^{0|N}\}^{N\in\bN}\,$ of Gra\ss mann-odd hyperplanes\footnote{Accordingly,\ whenever we consider restrictions (or extensions) of superfield configurations in the present paper,\ they are to be understood `functorially'.},\ as proposed and elucidated by Freed in \Rcite{Freed:1999}.\ The (super)target is assumed to carry an even metric tensor (degenerate in the Gra\ss mann-odd coordinate directions) and a de Rham-closed even super-3-form field.\ In the presence of defects,\ the worldsheet and the supertarget are equipped with more structure,\ and field configurations have to satisfy additional conditions.\ Let us start by defining a worldsheet with defects.

\bedef\label{def:worldsheet}
An {\bf oriented worldsheet with defects} (to be referred to simply as a {\bf worldsheet} henceforth) is a pair $\,(\Si,d)\,$ composed of a closed oriented two-dimensional manifold $\,\Sigma\,$ and a distinguished decomposition $\,d\ :\ \Sigma\xrightarrow{\ \cong\ }\Sigma^{(2)} \cup \Sigma^{(1)} \cup \Sigma^{(0)}\,$ into submanifolds subject to the following conditions:
\ben
\item $\,\Sigma^{(k)}\,$ is an oriented $k$-dimensional submanifold of $\,\Sigma$,\ with,\ in particular,\ the orientation of $\,\Sigma^{(2)}\,$ induced by that of $\,\Sigma$;
\item $\,\Sigma^{(k)} \cap \Sigma^{(l)} = \emptyset\,$ for $\,k \neq l$,\ and $\,\Sigma^{(0)} \cup \Sigma^{(1)}\equiv\G\,$ is a closed subset in $\,\Sigma\,$ (sometimes called the {\bf defect quiver}).
\een
The connected components of the defect quiver of dimension 1 ({\it i.e.},\ those from $\,\Si^{(1)}$) are called {\bf defect lines},\ and those of dimension 0 ({\it i.e.},\ from $\,\Si^{(0)}$) are called {\bf defect junctions}.
\exdef

\brem
Note that any open subset $\,U\subset\Si\,$ of a worldsheet with defects $\,(\Si,d)\,$ gives rise to a worldsheet with defects $\,(U,d\rstr_U)$.\ Note also that condition (2) is nontrivial in that it ensures that the three unions: $\,\Sigma^{(0)}$,\ $\Sigma^{(0)} \cup \Sigma^{(1)}$,\ and $\,\Sigma^{(0)} \cup \Sigma^{(1)} \cup \Sigma^{(2)}\,$ are closed in $\,\Sigma$,\ which,\ in turn,\ implies that $\,\overline{\Sigma^{(k)}} \setminus \Sigma^{(k)}\,$ is contained in the union of lower dimensional components.
\erem

Although we shall not make much use of the extra structure in the remainder of the present paper,\ we add,\ for completeness,\ that a {\bf morphism} between two worldsheets with defects: $\,(\Sigma_1,d_1)\,$ and $\,(\Sigma_2,d_2)\,$ is a smooth orientation-preserving map $\,f\ :\ \Sigma_1\too\Sigma_2\,$ which is a diffeomorphism onto its image and compatible with the respective decompositions.\ This means that for $\,k\in\{0,1,2\}$,\ we have $\,f(\Sigma^{(k)}_1) \subset\Sigma_2^{(k)}\,$ and $\,f\rstr_{\Sigma_1^{(k)}}\,$ is orientation-preserving.

\brem 
The definition of the super-$\si$-model with defects makes ample use of disjoint sums of supermanifolds of various superdimensions.\ For the sake of brevity,\ we shall call any such disjoint union a {\bf super-space} in what follows.\erem

The {\bf target super-space} of a super-$\si$-model with defects over a worldsheet $\,(\Si,d)\,$ is a super-space $\,\xcT\,$ with a disjoint-sum decomposition $\,\xcT = M \sqcup Q \sqcup T\,$ and further structure.\ A {\bf superfield configuration} of the super-$\si$-model is an `element' of the mapping super-space $\,[\Sigma,\xcT]\,$ (with an obvious meaning generalising the standard notion of the mapping supermanifold),\ which we shall denote as $\,\xi$,\ subject to certain coherence conditions.\ We shall describe the additional structure on the target super-space and the conditions imposed on a superfield configuration in parallel below.

\subsubsection*{The super-space $\,M$}

\noindent {\bf (T1)} The {\bf bulk target super-space} $\,M\,$ is a super-space equipped with a Gra\ss mann-even  metric tensor $\,\txg\,$ (typically degenerate in the Gra\ss mann-odd coordinate directions) and a de Rham-closed even super-3-form $\,\txH$,\ 
\qq\label{eq:Hclos}
\sfd\txH=0\,.
\qqq
Just to underline:\ We do not require that $\,M\,$ be connected or that all connected components of $\,M\,$ have the same superdimension.\ The same holds for the super-spaces $\,Q\,$ and $\,T\,$ to be defined below.

\begin{figure}[hbt]

$$
 \raisebox{-50pt}{\begin{picture}(-50,165)
  \put(-205,-5){\scalebox{0.6}{\includegraphics{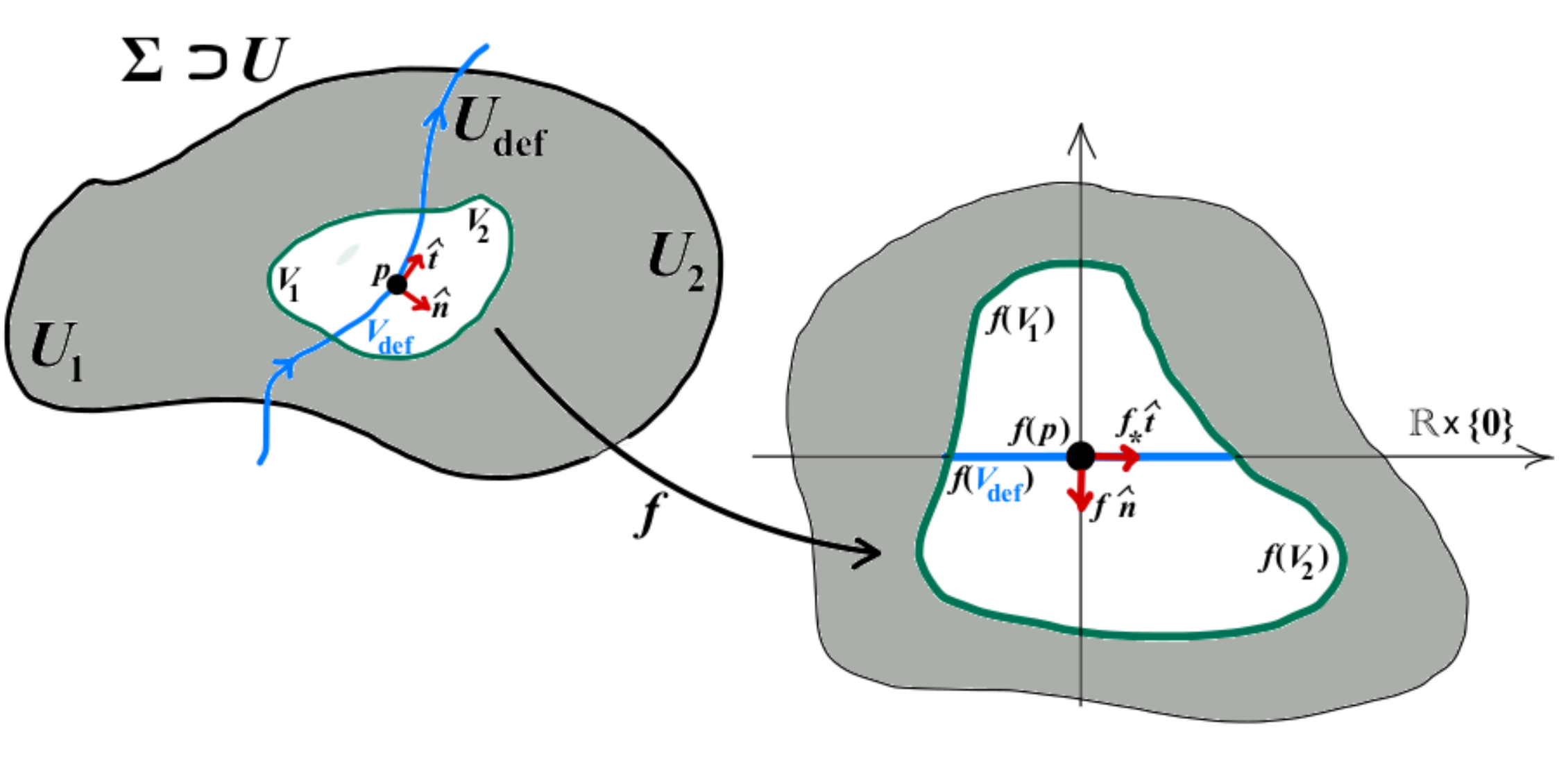}}}
  \end{picture}}
$$

\caption{The splitting of an open neighbourhood $\,U\subset\Si\,$ by a defect line,\ with the attendant model furnished by a diffeomorphism $\,f$.} \label{fig:defect-split-patch-mod}
\end{figure}

To describe the {\bf extension property} of a superfield configuration,\ we need some notation to distinguish the `values' of the superfield taken on a little patch to the `left' and to the `right' of a defect line.\ Thus,\ let $\,(\Sigma,d),\ d\ :\ \Si\xrightarrow{\ \cong\ }\Sigma^{(2)} \cup \Sigma^{(1)} \cup \Sigma^{(0)}\,$ be a worldsheet and let $\,U\,$ be an open set in $\,\Sigma$.\ We say that $\,U\,$ {\bf is split by} $\,\Sigma^{(1)}\,$ {\bf with decomposition} $\,U = U_1 \cup U_{\rm def} \cup U_2\,$ if the following conditions hold:
\begin{itemize}
\item $\,U_1, U_2\,$ are open and $\,U_{\rm def} = \Sigma^{(1)} \cap U\,$ is non-empty.\ The three sets:\ $\,U_1,\ U_2\,$ and $\,U_{\rm def}\,$ are mutually disjoint, and $\,U_1 \cup U_{\rm def}\,$ and $\,U_2 \cup U_{\rm def}\,$ are path-connected.
\item Each point $\,p \in U_{\rm def}\,$ has a neighbourhood $\,V\,$ which is diffeomorphic to an open subset of $\,\bR^2\,$ such that $\,U_{\rm def}\,$ gets mapped to the axis $\,\bR\x\{0\}\,$ in such a manner that the orientation is preserved,\ $\,U_1\,$ gets mapped to the upper half plane and $\,U_2\,$ to the lower half plane,\ also preserving the orientation.\ {\it Cf.}\ Fig.\,\ref{fig:defect-split-patch-mod} for an illustration.
\end{itemize}
Intuitively,\ looking along the defect line in the direction of its orientation,\ $\,U_1\,$ lies to the left of the defect,\ and $\,U_2\,$ to its right.\ The conditions above ensure that the decomposition $\,U_1 \cup U_{\rm def} \cup U_2\,$ is unique.

\medskip\noindent
{\bf (F1)} The restriction of $\,\xi\,$ to $\,\Sigma^{(2)}\,$ is an `element' of $\,[\Si^{(2)},M]$.\ For $\,p \in
\Sigma^{(1)}$,\ let $\,U\,$ be an open neighbourhood of $\,p\,$ which is split by $\,\Sigma^{(1)}\,$ with the decomposition $\,U = U_1 \cup U_{\rm def} \cup U_2$.\ Then,\ $\,\xi\rstr_{U_1}\,$ has an extension to $\,U_1 \cup U_{\rm def}\,$ (in $\,[U_1 \cup U_{\rm def},M]$),\ and $\,\xi\rstr_{U_2}\,$ has an extension to $\,U_2 \cup U_{\rm def}\,$ (in $\,[U_2 \cup U_{\rm def},M]$).

\medskip

Condition (F1) allows to compare the `value' of $\,\xi\,$ and of its differential on either side of a defect line by taking limits approaching a point on $\,\Sigma^{(1)}\,$ from the corresponding side.\ Below,\ we shall write the extension of $\,\xi\rstr_{U_A}\,$ to $\,U_A \cup U_{\rm def}\,$ as $\,\xi_{|A}$, $A\in\{1,2\}$,\ and we shall usually do so without explicitly specifying a choice of $\,U$.

\subsubsection*{The super-space $\,Q$}

\noindent {\bf (T2)} The {\bf bi-brane superworldvolume} (or the {\bf correspondence super-space}) $\,Q\,$ is a super-space equipped with an even super-2-form $\,\omega\,$ and two maps $\,\iota_1,\iota_2\ :\ Q \too M\,$ such that
\qq\label{eq:om-triv-HH}
\D_Q\txH = -\sfd\om\,,
\qqq
where
\qq\label{eq:DelQ}
\D_Q:=\iota_2^*-\iota_1^*\,.
\qqq

\medskip\noindent
{\bf (F2)} The restriction of $\,\xi\,$ to $\,\Sigma^{(1)}\,$ is an `element' of $\,[\Si^{(1)},Q]$.\ It satisfies the conditions\footnote{It is to be understood that the conditions are satisfied identically for \emph{all} $\,N\in\bN\,$ labelling the Gra\ss mann-odd hyperplane completion of $\,\Si$.\label{foot:restrext}}
\qq\nn
\xi_{|A}\rstr_{U_{\rm def}}=\iota_A \circ \xi\rstr_{U_{\rm def}}\,,\qquad A\in\{1,2\}\,.
\qqq

\medskip

Condition (F2) constrains the possible `jumps' that the superfield $\,\xi\,$ may have across a defect line.\ There is also a second condition which constrains the change in the derivative of the superfield normal to the defect line.\ For this,\ we need to introduce a linear map $\,N\,$ (for `normal') which rotates a tangent vector on $\,\Sigma\,$ into one orthogonal to it with respect to the euclidean metric $\,\d^{(2,0)}_{\rm E}=\d_{ij}\,\sfd\si^i\ox\sfd\si^j\,$ (in local worldsheet coordinates $\,\{\si^i\}^{i\in\{1,2\}}$),
\qq\label{eq:normal-map-def}
N\ :\ \G(\sfT\Sigma)\too\G(\sfT\Sigma)\ :\ v(\cdot)=v^i(\cdot)\,\tfrac{\p\ }{\p\si^i}\equiv v^i(\cdot)\,\p_i\longmapsto v^i(\cdot)\,\ep_{ij}\,\d^{jk}\,\p_k\,,
\qqq
where $\,\ep^{ij}=\ep_{ij}=\ep_{[ij]}\,$ is the antisymmetric tensor in two dimensions,\ with the normalisation $\,\ep_{12}=1$.\ The map $\,N\,$ squares to $\,N \circ N = -\id_{\G(\sfT\Sigma)}\,$ and satisfies $\,\d^{(2,0)}_{\rm E}(v,Nv)=0\,$ and $\,\d^{(2,0)}_{\rm E}(Nv,Nv) =\d^{(2,0)}_{\rm E}(v,v)$.\ Note also that the pair $\,(v,Nv)\,$ defines the standard orientation $\,[(\p_1,\p_2)]\,$ of the euclidean worldsheet for any non-zero vector $\,v\in\sfT_p\Si$.

\medskip\noindent
{\bf (F3)} -- The {\bf Defect Gluing Condition} (DGC):\ Let $\,t \in\G(\sfT\Sigma^{(1)})\,$ be the vector field tangent to $\,\Si^{(1)}$.\ Then, 
\qq\label{eq:DGC}
\txp_{|2}\circ\iota_{2*}-\txp_{|1}\circ\iota_{1*}= \om_\xi(\cdot,\xi_*t) \,,
\qqq
with
\qq\nn
\txp_{|A}=\sqrt{\det\,(\xi_{|A}^*\txg)}\cdot\txg_{\xi_{|2}}\bigl(\cdot,\xi_{|A\,*}(Nt)\bigr)\,,\qquad A\in\{1,2\}\,.
\qqq
Here,\ $\,\xi_{|A\,*}(Nt)\,$ is understood as the limit of $\,\xi_{*}(Nt)\,$ in which $\,x\in\Sigma^{(2)}\,$ approaches the defect line from the appropriate side (namely,\ inside $\,U_A\,$ in the notation of condition (F1)).\ This condition was derived in \Rcite{Runkel:2008gr} by varying the $\si$-model action functional\footnote{In \Rcite{Runkel:2008gr},\ the so-called Polyakov formulation of the $\si$-model was considered,\ but this only affects the metric (`kinetic') term and it is straightforward to adapt the calculation in \Rxcite{App.\,A.2}{Runkel:2008gr} to the definition of the action functional given in \Reqref{eq:ssimod}.} and demanding that the `boundary' terms of the variation,\ localised at $\,\Si^{(1)}\,$ vanish.\ It was subsequently shown,\ in \Rcite{Suszek:2011hg},\ to define an isotropic subspace in the two-phase space of states of the $\si$-model (the $\,\txp_{|A}\,$ acquiring the interpretation of the limiting values of the component of the kinetic momentum normal to the defect line on either side of the latter -- clearly, the momentum suffers a discontinuity controlled by $\,\om$),\ from which an explicit correspondence between defects and dualities of the $\si$-model was inferred.\ The correspondence is at the core of the construction of the gauge-symmetry defect laid out in \Rcite{Suszek:2012ddg}.

\subsubsection*{The super-space $\,T$}

\begin{figure}[hbt]

$$
 \raisebox{-50pt}{\begin{picture}(-50,233)
  \put(-210,-5){\scalebox{0.6}{\includegraphics{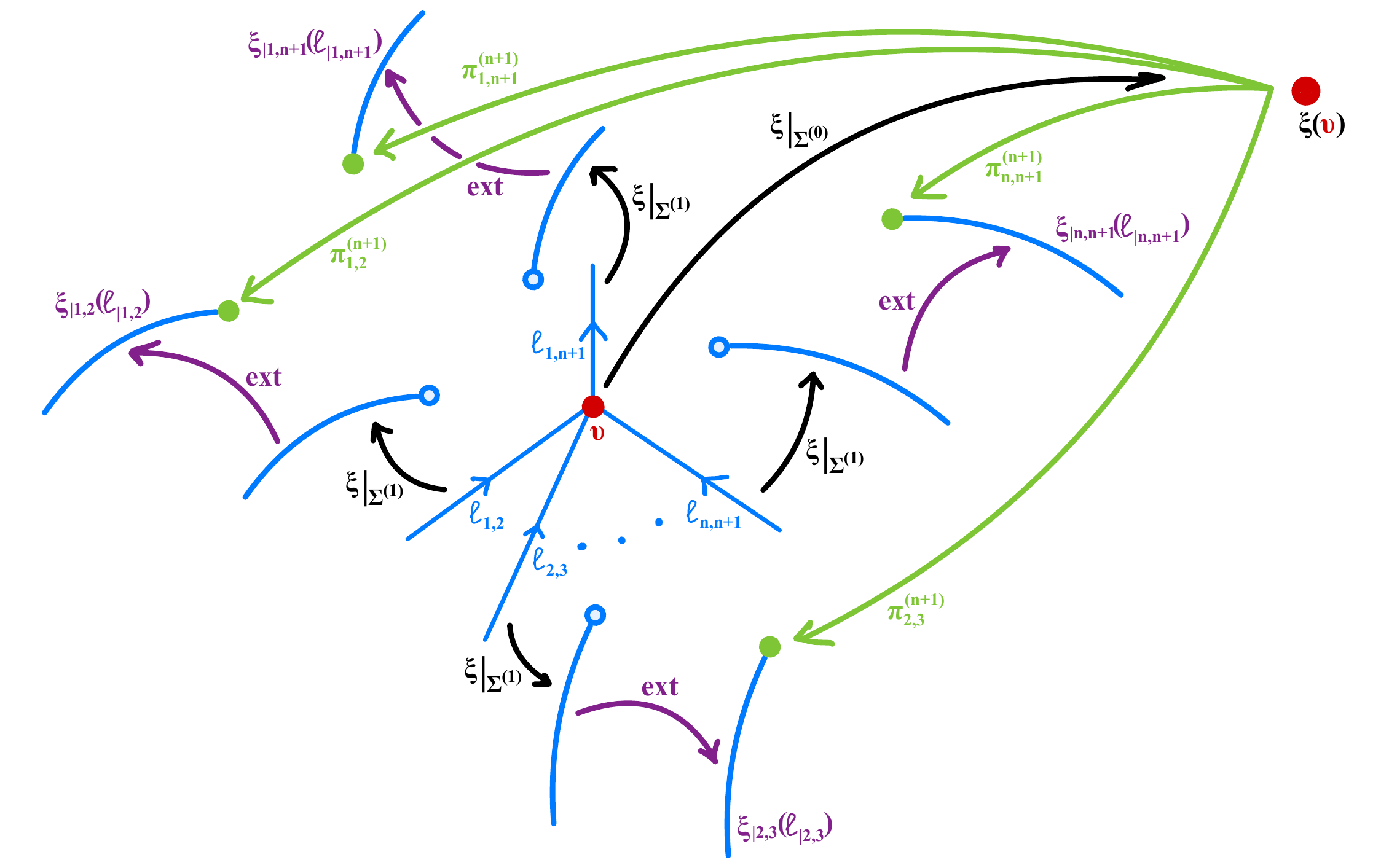}}}
  \end{picture}}
$$

\caption{Extensions (marked in violet) of the (super)field configuration to the closures $\,\ell_{|k,k+1}\,$ and $\,\ell_{|1,n+1}\,$ of the defect lines converging at a given defect junction $\,v\,$ of valence $\,n+1$.} \label{fig:n-ext-config}
\end{figure}

The super-space $\,T\,$ is associated with junctions of defect lines,\ that is,\ it serves as the (super)target for $\,\xi\,$ restricted to $\,\Sigma^{(0)}$.\ We shall only consider junctions which have $\,n\,$ defect lines oriented towards the junction and one line oriented away from it.\ For the general discussion with arbitrary orientations,\ we refer the reader to\footnote{In \Rcite{Runkel:2008gr},\ no orientation was assigned to defect junctions.\ From the point of view of this article,\ this means that all junctions in \Rcite{Runkel:2008gr} implicitly carry orientation `$+$'.} \Rxcite{Sect.\,2.5}{Runkel:2008gr}.\ In the present restricted setting,\ each point $\,x \in \Sigma^{(0)}\,$ has a neighbourhood which is diffeomorphic to the two worldsheets shown in Fig.\,\ref{fig:n-ext-config}.\ The super-space $\,T\,$ has components $\,T_{n,1},\ n\in\bZ_{\ge0}\,$ to which a junction point $\,x\,$ with $\,n+1\,$ defect lines is mapped ({\it cf.}\ Fig.\,\ref{fig:bicat-sheet} for an illustration of the case $\,n=2$).\ We call $\,T\,$ (resp.\ $\,T_{n,1}$) the {\bf inter-bi-brane superworldvolume} (resp.\ the {\bf $(n,1)$-valent component inter-bi-brane superworldvolume}).

\medskip\noindent
{\bf (T3)} The super-space $\,T\,$ has a disjoint-sum decomposition $\,T =\bigsqcup_{n=2}^\infty T_{n,1}$.\ Each component $\,T_{n,1}\,$ is equipped with maps $\,\pi_{1,n+1}^{(n+1)}\,$ and $\,\pi_{k,k+1}^{(n+1)},\ k\in\ovl{1,n}\,$ from $\,T_{n,1}\,$ to $\,Q$.\ These maps satisfy the compatibility conditions
\qq 
\raisebox{2.6em}{\xymatrix@C=1em@R=1em{
& T_{n,1} \ar[dl]_{\pi_{k-1,k}^{(n+1)}} \ar[dr]^{\pi_{k,k+1}^{(n+1)}}
\\
Q \ar[dr]_{\iota_2} && Q \ar[dl]^{\iota_1} \\
&M
}}\,,\qquad  k\in\ovl{2,n}\,, \qquad \text{ and } \qquad
\raisebox{2.6em}{\xymatrix@C=1em@R=1em{
& T_{n,1} \ar[dl]_{\pi_{1,n+1}^{(n+1)}} \ar[dr]^{\pi_{1,2}^{(n+1)}}
\\
Q \ar[dr]_{\iota_1} && Q \ar[dl]^{\iota_1} \\
&M
}}\qquad ,\qquad
\raisebox{2.6em}{\xymatrix@C=1em@R=1em{
& T_{n,1} \ar[dl]_{\pi_{1,n+1}^{(n+1)}} \ar[dr]^{\pi_{n,n+1}^{(n+1)}}
\\
Q \ar[dr]_{\iota_2} && Q \ar[dl]^{\iota_2} \\
&M
}}\,.\cr
\label{eq:def-jun-extend}
\qqq
The pullbacks of the 2-form $\,\omega\,$ from $\,Q\,$ satisfy
the {\bf Defect-Junction Identity} (DJI)
\qq \label{eq:def-jun-id}
  \D_{T_{n,1}}\om=0\,,
\qqq
where
\qq\nn
\D_{T_{n,1}}:=\sum_{k=1}^n\,\pi_{k,k+1}^{(n+1)\,*}-\pi_{1,n+1}^{(n+1)\,*}\,.
\qqq

\medskip

Condition \eqref{eq:def-jun-id} was derived in \Rxcite{Sect.\,2.7}{Runkel:2008gr} from the requirement of (higher) gauge invariance of the $\si$-model action functional in the presence of defect junctions combined with the variational principle for the action functional.\ The said requirement translates into the necessary condition for the existence of a certain 2-isomorphism between 1-isomorphisms of 1-gerbes -- this will be restated in the supergemoetric setting (on the basis of \Rcite{Runkel:2008gr}) in Sec.\,\ref{sub:grbmultissi}.

Condition \eqref{eq:def-jun-extend} merely encodes the compatibility of the different extensions of the super-$\si$-model superfield $\,\xi$,\ as we now describe.\ In a small neighbourhood $\,V\,$ around a point $\,v\in \Sigma^{(0)}$,\ as shown in Fig.\,\ref{fig:n-ext-config},\ the restriction of $\,\xi\,$ to each connected component $\,I\,$ of $\,\Sigma^{(1)}
\cap V\,$ (an open interval) has an extension to $\,I \cup \{v\}$.\ We denote these extensions by $\,\xi_{|k,k+1}\,$ and
$\,\xi_{|1,n+1}$,\ with a labelling as shown in Fig.\,\ref{fig:n-ext-config}.\ For a field configuration close to a junction point,\ we require that these extensions allow a consistent extension of $\,\xi\,$ in each of the wedges $\,1,2,\ldots,n+1\,$ in Fig.\,\ref{fig:n-ext-config} to $\,v$.

\medskip\noindent
{\bf (F4)} Let $\,v\in \Sigma^{(0)}\,$ have a neighbourhood as shown in Fig.\,\ref{fig:n-ext-config}.\ The mapping $\,\xi\,$ satisfies\footnote{The seemingly artificial notation should be read along the lines of footnote \ref{foot:restrext}.} $\,\xi\rstr_{\{v\}}\in[\{v\},T_{n,1}]$,\ and $\,\pi_{1,n+1}^{(n+1)} \circ \xi\rstr_{\{v\}} = \xi_{|1,n+1}\rstr_{\{v\}}\,$ and
$\,\pi_{k,k+1}^{(n+1)} \circ \xi\rstr_{\{v\}} = \xi_{|k,k+1}\rstr_{\{v\}},\ k\in\ovl{1,n}$.

\medskip

Consider for example the defect line labelled $\,\ell_{1,2}\,$ in Fig.\,\ref{fig:n-ext-config}.\ The restriction $\,\xi_{|1,2}\rstr_{\{v\}}$,\ mapping to $\,Q$,\ is obtained as the limit of $\,\xi\rstr_{\ell_{1,2}}\,$ in which a point $\,q\in\ell_{1,2}\,$ approaches $\,v$.\ Since $\,\iota_2 \circ\xi\rstr_{\ell_{1,2}} =\xi_{|2}\rstr_{\ell_{1,2}}$,\ mapping to $\,M$,\
is,\ itself,\ the limit of $\,\xi\rstr_{p_2}$,\ restricted to the wedge labelled $\,p_2$,\ in which $\,p\in p_2\,$ approaches $\,q$,\ we see that the limit $\,p\to v$,\ taken in the wedge $\,p_2$,\ gives $\,\iota_2 \circ \xi_{|1,2}\rstr_{\{v\}} = \iota_2 \circ \pi_{1,2}^{(3)} \circ \xi\rstr_{\{v\}}$.\ If we repeat this procedure with respect to the defect line labelled $\,\ell_{2,3}$,\ we obtain the limiting value $\,\iota_1 \circ \pi_{2,3}^{(3)} \circ\xi\rstr_{\{v\}}$.\ These two values agree in virtue of condition \eqref{eq:def-jun-extend},\ and this is the reason to impose the condition in the first place.

\medskip

Let us package the above description in two definitions:\ 
\bedef\label{def:targetss}
A {\bf target super-space} $\,\xcT\,$ {\bf for a multi-phase super-$\si$-model with defects} (or a {\bf target super-space} for brevity) is a super-space $\,\xcT\,$ with a disjoint-sum decomposition $\,\xcT = M \sqcup Q \sqcup T\,$ having the properties described in (T1), (T2) and (T3).
\exdef 
\noindent and
\bedef
A {\bf superfield configuration over a worldsheet with defects} $\,(\Sigma,d)\,$ {\bf of the multi-phase super-$\si$-model with the target super-space} $\,\xcT\,$ (or a {\bf superfield configuration} for brevity) is an `element' $\,\xi\in[\Sigma,\xcT]\,$ such that $\,\xi\rstr_{\Sigma^{(2)}}\in[\Si^{(2)},M]$,\ $\xi\rstr_{\Sigma^{(1)}}\in[\Sigma^{(1)},Q]\,$ and $\,\xi\rstr_{\Sigma^{(0)}}\in[\Sigma^{(0)},T]$,\ with the restrictions subject to the conditions (F1),\ (F2),\ (F3) and (F4).\ We shall collectively denote the entirety of superfield configurations over a worldsheet with defects $\,(\Sigma,d)\,$ as $\,\cF(\Sigma,d)$.
\exdef

We now come to the superfield equations (of evolution).\ The {\bf Dirac--Feynman amplitude of the multi-phase super-$\si$-model over a worldsheet with defects} $\,(\Sigma,d)\,$ (to be given in Sec.\,\ref{sub:grbmultissi}) is a `functional'\footnote{It makes sense to think of it as a (nested) family of functionals labelled by the set $\,\bN\ni N\,$ of superdimensions $\,(0,N)\,$ of the Gra\ss mann-odd hyperplanes $\,\bR^{0|N}\,$ mentioned earlier.} $\,\cA_{\rm DF}\equiv\ee^{\sfi\,S}\,$ on $\,\cF(\Sigma,d)$,\ typically expressed in terms of an action `functional' $\,S$.\ The corresponding superfield equations are obtained by varying the DF amplitude $\,\cA_{\rm DF}\,$ along an arbitrary path in $\,\cF(\Sigma,d)$;\ in the case of the multi-phase (super-)$\si$-model (in the Polyakov formulation) with a purely Gra\ss mann-even target (super-)space,\ this calculation was carried out in \Rxcite{App.\,A.2}{Runkel:2008gr},\ and it is straightforward to repeat the computation in the $\bZ/2\bZ$-graded setting (in the spirit of \Rcite{Freed:1999}).\ The result is that $\,\xi \in \cF(\Sigma,d)\,$ is a critical `point' of the `functional' $\,\cA_{\rm DF}\,$ iff the restriction of $\,\xi\,$ to $\,\Sigma^{(2)}\,$ satisfies the following `differential equation'\footnote{Once again,\ we are dealing with a (nested) family of differential equations.}
\qq\label{eq:sfield-eqs}\qquad\qquad
\d^{ij}\,\bigl[\txg\bigl(\cdot,\p_i\bigl(\sqrt{\det\,(\xi^*\txg)}\,(\xi_*\p_j)^a\bigr)\,\tfrac{\p\ }{\p\xi^a}\bigr)-\sqrt{\det\,(\xi^*\txg)}\,\G\bigl(\cdot,\xi_*\p_i,\xi_*\p_j\bigr)-\txH\bigl(\cdot,\xi_*\p_i,\xi_*N\p_j\bigr)\bigr]=0\,,
\qqq
expressed in the local (super)coordinates $\,\xi^a\,$ on $\,M$,\ and in terms of the Christoffel form $\,\G\,$ of $\,\txg$,\ the latter evaluating on a triple $\,(\cV_1,\cV_2,\cV_3)\,$ of (Gra\ss mann-even) vector fields on $\,M\,$ as
\qq\nn
-2\G(\cV_1,\cV_2,\cV_3)=\cV_1^a\,\p_a\txg_{bc}\,\cV_2^b\,\cV_3^c-\cV_2^b\,\p_b\txg_{ac}\,\cV_1^a\,\cV_3^c-\cV_3^c\,\p_c\txg_{ba}\,\cV_2^b\,\cV_1^a\,.
\qqq

\subsubsection*{World-sheets with boundary}

We shall briefly describe,\ along the lines of \Rxcite{Sect.\,2.4}{Runkel:2008gr},\ how worldsheets with boundary can be understood as a special case of worldsheets with defect lines.\ In the decomposition $\,\xcT = M
\sqcup Q\,$ (we have $\,T = \emptyset\,$ in this case), choose $\,M = N \sqcup\bR^{0|0}$.\ The two maps $\,\iota_A\ :\ Q \too M\,$ are such that 
\qq\nn
\alxydim{}{Q \ar[rr]^{\iota_1} \ar[rd]_{\unl\iota{}_1} & & M \\ & N \ar[ur]_{\jmath_N}}
\qqq
and
\qq\nn
\alxydim{}{Q \ar[rr]^{\iota_2} \ar@{-->}[rd] & & M \\ & \bR^{0|0} \ar[ur]_{\jmath_{\bR^{0|0}}}}\,,
\qqq
where $\,\jmath_X,\ X\in\{N,\bR^{0|0}\}\,$ are the canonical inclusions,\ and the dotted arrow represents the unique mapping of $\,Q\,$ to the terminal object in $\,\sMan$.\ Thus, in a small neighbourhood $\,U\,$ of a point $\,p \in \Sigma^{(1)}\,$ with a decomposition $\,U = U_1 \cup U_{\rm def} \cup U_2\,$ as before,\ the restriction of the superfield $\,\xi\,$ to $\,U_2\,$ has $\,\bR^{0|0}\,$ as the codomain.\ If we delete all the regions of the worldsheet which are mapped to $\,\bR^{0|0}$,\ we obtain a worldsheet with a boundary.\ Let us write $\,Q_\p\,$ instead of $\,Q\,$ under such circumstances,\ also later on when describing boundary conditions.\ The super-2-form on $\,Q_\p\,$ will be denoted as $\,\om_\p$.\ It satisfies
\qq\label{eq:om-b-triv-H}
\iota_1^* \txH=\sfd\om_\p\,.
\qqq
The DGC \eqref{eq:DGC} turns into
\qq\label{eq:DGC-b}
\txp_{|1}\circ\iota_{1*}=-\om_{\p\,\xi}(\cdot,\xi_*t) \,
,
\qqq
which is imposed for all $\,t\,$ tangent to the worldsheet boundary.\ The super-space $\,Q_\p\,$ is called the {\bf D-brane worldvolume},\ and $\,\om_\p\,$ is referred to as the {\bf D-brane curvature}.

Typically, one takes $\,Q\equiv Q_\p\,$ to be a sub-supermanifold of $\,N\,$ in the boundary setting.\ When discussing boundaries in the following,\ this is the situation that we shall (often) look at.\ We shall not make explicit the component $\,\bR^{0|0}\,$ of the target super-space,\ nor shall we (always) spell out the embedding map $\,\iota_1$.

\subsection{Simplicial target super-spaces}\label{sub:simpltargsp}

Our definition of the target super-space does not presuppose the existence of mappings between components of the inter-bi-brane superworldvolume associated with defect junctions of different valence which would be consistent with the compatibility conditions \eqref{eq:def-jun-extend} or which would imply the latter for components of higher valence once they are satisfied for components of lower valence,\ and,\ indeed,\ nothing forces us to impose such relations generically.\ Neither does it reflect (automatically) the physically quite natural possibility to enhance any pre-existing defect through adjunction of an arbitrary number of identity defect lines (either intersecting the defect or not) across which the superfield continues smoothly.\ On the other hand,\ inspection of the prototypical poly-phase scenario,\ based on the concept of the pair groupoid,\ reveals a natural alternative.\ Indeed,\ consider a bi-brane worldvolume (assumed un-graded for the sake of simplicity) in the form of a submanifold $\,Q\subset M^{\x 2}\,$ in the cartesian square of the bulk target space $\,M\,$ and equipped with the obvious bi-brane maps $\,\iota_A=\pr_A\ :\ Q\too M\,$ given by (the restrictions of) the canonical projections (to $\,Q$).\ It is now completely natural to take as the worldvolume $\,T_{2,1}\,$ of the `elementary' inter-bi-brane the intersection $\,T_{2,1}=\bigcap_{i\in\{1,2,3\}}\,\widehat\pr{}_i^{(3)\,-1}(Q)\subset M^{\x 3}\,$ of the preimages of the correspondence space $\,Q\,$ under the maps $\,\widehat\pr{}_i^{(3)}\ :\ M^{\x 3}\too M^{\x 2},\ i\in\{1,2,3\}\,$ of which the $i$-th one projects \emph{out} the $i$-th factor,\ {\it i.e.},\ $\,(\widehat\pr{}_1^{(3)},\widehat\pr{}_2^{(3)},\widehat\pr{}_3^{(3)})(x_1,x_2,x_3)=((x_2,x_3),(x_1,x_3),(x_1,x_2))$,\ with the said projections as the inter-bi-brane maps,\ $\,(\pi^{(3)}_{1,2},\pi^{(3)}_{2,3},\pi^{(3)}_{1,3})=(\widehat\pr{}_3^{(3)},\widehat\pr{}_1^{(3)},\widehat\pr{}_2^{(3)})$.\ Proceeding recursively,\ we may now take $\,T_{n+1,1}\subset M^{\x n+2}\,$ in the form $\,T_{n+1,1}=\bigcap_{i\in\ovl{1,n+2}}\,\widehat\pr{}_i^{(n+2)\,-1}(T_{n,1})\,$ where $\,\widehat\pr{}_i^{(n+1)}\ :\ T_{n+1,1}\too T_{n,1}\ :\ (x_1,x_2,\ldots,x_{n+2})\longmapsto(x_1,x_2,\ldots,x_{i-1},x_{i+1},x_{i+2},\ldots,x_{n+2})$.\ In order to work out the inter-bi-brane maps,\ we first change the representation of inter-bi-brane worldvolumes conveniently by referring to the worldsheet-defect picture:\ The $(n+1)$-tuple $\,(x_1,x_2,\ldots,x_{n+1})\in T_{n,1}\,$ encodes the raw data of the discontinuities of the $\si$-model field in the $\,n+1\,$ phases meeting at the defect junction of valence $\,(n,1)$,\ and these data can be equivalently transcribed as data of the $n$ consecutive jumps of the field at the $n$ incoming defect lines written as $\,((x_1,x_2),(x_2,x_3),\ldots,(x_n,x_{n+1}))$,\ which reflects the obvious identification $\,M^{\x n+1}\equiv M^{\x 2}\x_M M^{\x 2}\x_M\cdots\x_M M^{\x 2}\,$ ($n$ copies).\ This new representation enables us to view the maps $\,\widehat\pr{}_i^{(n+1)}\,$ from a different angle as the fibred product is none other than the submanifold of \emph{composable} morphisms in the $n$-fold cartesian power of the morphism manifold $\,M^{\x 2}\,$ of the pair groupoid of $\,M$.\ In this picture,\ the map $\,\widehat\pr{}_i^{(n+1)}\,$ corresponds to the (post-)composition of the $i$-th morphism with the $(i-1)$-st one,\ $\,((x_1,x_2),(x_2,x_3),\ldots,(x_n,x_{n+1}))\longmapsto((x_1,x_2),(x_2,x_3),\ldots,(x_{i-2},x_{i-1}),(x_i,x_{i+1})\circ(x_{i-1},x_i),(x_{i+1},x_{i+2}),(x_{i+2},x_{i+3}),\ldots,(x_n,x_{n+1}))$,\ which models the process of obtaining the defect junction through `fusion' of the pair $\,(\ell_i,\ell_{i+1})\,$ of defect lines at a trivalent defect junction an infinitesimal distance $\,\ep>0\,$ away from the original one of valence $\,(n,1)\,$ and subsequently passing to the limit $\,\ep\to 0$,\ {\it cf.}\ [Figure:Ind] -- a most natural scenario in the setting of the 2d Conformal Field Theory (with the so-called topological defects).\ The inter-bi-brane maps are now simply the canonical projections to the $n$ cartesian factors $\,M^{\x 2}$,\ augmented by the `total-composition' map which yields the pair $\,(x_1,x_{n+1})\equiv(x_n,x_{n+1})\circ(x_{n-1},x_n)\circ\cdots\circ(x_1,x_2)$,\ and it is easy to see how these can be recovered from various \emph{equivalent} fusion `cascades',\ modelled by concatenations of the descent maps $\,T_{k,1}\too T_{k-1,1},\ k\in\ovl{n,3}\,$ and the $\,\iota_A$.\ Such an {\bf induction scheme} was originally conceived in \Rcite{Runkel:2008gr} and later elaborated in \Rcite{Suszek:2011hg}.\ As for the identity-defect enhancement,\ it is readily achieved in this setting by augmentation of the target-space structure $\,\xcT\,$ above $\,M\,$ by the $M$-fibred powers of the bulk target space and by all possible (multiple) extensions of its components by the $M$-fibred product with the target space $\,M\,$ (to be marked symbolically by tildas on the enhanced worldvolumes),\ which enables us to consider defect-line and defect-junction data of the special form $\,((x_1,x_2),(x_2,x_3),\ldots,(x_{k-1},x_k),(x_k,x_k),(x_k,x_{k+1}),(x_{k+1},x_{k+2}),\ldots,(x_n,x_{n+1}))\,$ lying in the image of obvious embeddings $\,M\too\widetilde Q,\ \widetilde Q\too\widetilde T{}_{2,1}\,$ and $\,\widetilde T{}_{k-1,k}\too\widetilde T{}_{k,1}$.\ The ensuing two families of maps going `upwards' and 'downwards' in the (enhanced) hierarchy of manifolds $\,(M,\widetilde Q,\widetilde T{}_{2,1},\widetilde T{}_{3,1},\ldots)\,$ satisfy some obvious identities expressing equivalences of distinguished pairs of paths in the hierarchy traced by their superpositions.\ Below,\ we formalise the resultant structure with view to its physically important implementation outside the model setting just described.\medskip

A useful abstraction of the properties of the above model target super-space,\ and a point of departure for our subsequent considerations is specified in
\bedef\label{def:simplicial}
Let $\,\cC\,$ be a category with the set of objects $\,{\rm Ob}(\cC)$.\ A {\bf simplicial object} $\,(X_\bullet,d^{(\bullet)}_\cdot,s^{(\bullet)}_\cdot)\,$ {\bf in} $\,\cC\,$ is a collection of objects $\,X_n\in{\rm Ob}(\cC),\ n \in\bN$,\ together with distinguished morphisms:\ the {\bf face maps} $\,d^{(n)}_i\in\Hom_\cC(X_n,X_{n-1})\,$ and the {\bf degeneracy maps} $\,s^{(n)}_i\in\Hom_\cC(X_n,X_{n+1})$,\ defined for all $\,0\leq i\leq n\,$ and satisfying the {\bf simplicial identities}:
\qq\nn
  d^{(n-1)}_i \circ d^{(n)}_j &=& d^{(n-1)}_{j-1} \circ d^{(n)}_i\,,\quad i<j\,,\cr\cr
  s^{(n+1)}_i \circ s^{(n)}_j &=& s^{(n+1)}_{j+1} \circ s^{(n)}_i\,,\quad i \le j\,,\cr\cr
  d^{(n+1)}_i \circ s^{(n)}_j &=& \begin{cases}
  s^{(n-1)}_{j-1} \circ d^{(n)}_i & \text{if} \quad  i<j\,, \\
  \id_{X_n} & \text{if} \quad i=j ~\text{or}~ i=j+1\,, \\
  s^{(n-1)}_j \circ d^{(n)}_{i-1} & \tx{if} \quad i>j+1\,.
  \end{cases}
\qqq
A simplicial object in the category $\,\Set\,$ (resp.\ $\,{\rm {\bf Top}},\ ({\rm {\bf s}}){\rm {\bf Man}}\,$ {\it etc.}) is termed a {\bf simplicial set} (resp.\ {\bf space},\ ({\bf super}){\bf manifold} {\it etc.}).
\exdef 
\noindent A fundamental class of examples,\ generalising straightforwardly the model target space just described and thus of immediate relevance to us,\ is provided by nerves of categories ({\it cf.}\ \Rcite{Segal:1968}).
\bedef
Let $\,\cC\,$ be a small category with the set of objects $\,\mathrm{Ob}(\cC)$,\ the set of morphisms
$\,\mathrm{Mor}(\cC)\,$ and structure maps:\ $\,s\ :\ \mathrm{Mor}(\cC) \too \mathrm{Ob}(\cC)\,$ (the source map\footnote{One should not confuse the source map $\,s\,$ with a degeneracy map $\,s_i$.}),\ $\,t\ :\ \mathrm{Mor}(\cC) \too \mathrm{Ob}(\cC)\,$ (the target map) and $\,\id_\cdot\ :\ \mathrm{Ob}(\cC)\too\mathrm{Mor}(\cC)\,$ (the identity map).\ The {\bf nerve of} $\,\cC\,$ is the simplicial set $\,\sfN_\bullet(\cC)\,$ with the following data:\ $\,\sfN_0(\cC) = \mathrm{Ob}(\cC)\,$ and,\ for $\,n\geq 1$,
\qq\nn
  \sfN_n(\cC) = \{\ (f_1,f_2,\ldots,f_n) \in \mathrm{Mor}(\cC)^{\times n}
  \quad |\quad t(f_i) = s(f_{i+1})\ \} \,,
\qqq
{\it i.e.},\ $\,\sfN_n(\cC)\,$ is the set of all $n$-tuples of composable morphisms (note that in this ordering convention $\,f_i\,$ and $\,f_{i+1}\,$ are composable if $\,f_{i+1} \circ f_i\,$ makes sense).\ The degeneracy maps are:\ $\,s_0(a) = \id_a\,$ for $\,a\in\mathrm{Ob}(\cC)$,\ and $\,s_i^{(n)}(f_1,f_2,\ldots,f_n) = (f_1,f_2,\ldots,f_i,\id_{t(f_i)},f_{i+1},\ldots,f_n)\,$ for $\,n\ge 1$.\ The face maps are:\ $\,d_0(f) = t(f)\,$ and $\,d_1(f) = s(f)\,$ for $\,f\in\mathrm{Mor}(\cC)$,\ and,\ for $\,n \ge 2$,
\qq\nn
  d_i^{(n)}(f_1,f_2,\ldots,f_n) = \begin{cases}
  (f_2,f_3,\ldots,f_n) & \text{for} \quad   i=0\,,\\
  (f_1,f_2,\ldots,f_{i+1} \circ f_i,\ldots,f_n) & \text{for} \quad   0<i<n\,,\\
  (f_1,f_2,\ldots,f_{n-1}) & \text{for} \quad   i=n\,.
  \end{cases}\,,
\qqq
\exdef
\noindent That simplicial super-spaces provide a physically interesting abstraction of the properties of the nerve of the pair groupoid of the bulk target space,\ discussed earlier,\ and thus become distinguished candidates for target super-spaces is a direct consequence of the following
\berop\label{prop:simplicial-target}
Adopt the notation of Def.\,\ref{def:simplicial}.\ Let $\,\cC\,$ be a category and let $\,(X_\bullet,d^{(\bullet)}_\cdot,s^{(\bullet)}_\cdot)\,$ be a simplicial object in $\,\cC$.\ For any $\,n\geq 2\,$ there are (generically,\ {\it i.e.},\ without any further relations between the face maps) precisely $\,n+1\,$ inequivalent ways to obtain morphisms $\,X_n\too X_1\,$ through composition of the face maps $\,d^{(k)}_i,\ i\in\ovl{0,k},\ k\in\ovl{2,n}\,$ and the morphisms thus induced admit presentations
\qq\nn
\pi^{(n+1)}_{k,k+1}&=&d^{(2)}_2\circ d^{(3)}_2\circ\cdots\circ d^{(n-k+1)}_2\circ d^{(n-k+2)}_0\circ d^{(n-k+3)}_0\circ\cdots\circ  d^{(n-1)}_0\circ d^{(n)}_0\,,\qquad k\in\ovl{1,n}\,,\\ \label{eq:ind-ibb-maps}\\
\pi^{(n+1)}_{1,n+1}&=&d^{(2)}_1\circ d^{(3)}_1\circ\cdots\circ d^{(n)}_1\,.\nonumber
\qqq
These satisfy the compatibility conditions
\qq\nn
\raisebox{2.6em}{\xymatrix@C=1em@R=1em{
& X_n \ar[dl]_{\pi_{k-1,k}^{(n+1)}} \ar[dr]^{\pi_{k,k+1}^{(n+1)}}
\\
X_1 \ar[dr]_{d^{(1)}_0} && X_1 \ar[dl]^{d^{(1)}_1} \\
& X_0
}}\,,\qquad  k\in\ovl{2,n}\,, \qquad \text{ and } \qquad
\raisebox{2.6em}{\xymatrix@C=1em@R=1em{
& X_n \ar[dl]_{\pi_{1,n+1}^{(n+1)}} \ar[dr]^{\pi_{1,2}^{(n+1)}}
\\
X_1 \ar[dr]_{d^{(1)}_1} && X_1 \ar[dl]^{d^{(1)}_1} \\
& X_0
}}\qquad ,\qquad
\raisebox{2.6em}{\xymatrix@C=1em@R=1em{
& X_n \ar[dl]_{\pi_{1,n+1}^{(n+1)}} \ar[dr]^{\pi_{n,n+1}^{(n+1)}}
\\
X_1 \ar[dr]_{d^{(1)}_0} && X_1 \ar[dl]^{d^{(1)}_0} \\
& X_0
}}\,,
\qqq
and - for $\,(i,j)\in\{(1,2),(2,3),(1,3)\},\ k\in\{0,1\}\,$ and $\,l-1,m\in\ovl{0,n}\,$
\qq\nn
&\pi^{(3)}_{i,j}\circ s^{(1)}_k=\left\{ \barr{cl} \iota_1 & \tx{if}\quad (i,j,k)=(1,2,0) \\ \iota_2 & \tx{if}\quad (i,j,k)=(2,3,1) \\ \id_{X_1} & \tx{otherwise} \earr \right.\,,&\cr\cr\cr
&\pi^{(n+2)}_{1,n+2}\circ s^{(n)}_i=\pi^{(n+1)}_{1,n+1}\,,\qquad\qquad\pi^{(n+2)}_{l,l+1}\circ s^{(n)}_m=\left\{ \barr{cl} \pi^{(n+1)}_{l-1,l} & \tx{if}\quad l\geq m+2 \\ \iota_1\circ\pi^{(n+1)}_{l,l+1} & \tx{if}\quad l=m+1 \\ \pi^{(n+1)}_{l,l+1} & \tx{if}\quad l\leq m \\ \iota_2\circ\pi^{(n+1)}_{n,n+1} & \tx{if}\quad (l,m)=(n+1,n) \earr \right.\,,&
\qqq
where $\,\iota_A=s^{(0)}_0\circ d^{(1)}_{2-A}\ :\ X_1\circlearrowleft\,,\ A\in\{1,2\}$.
\eerop
\beroof
{\it Cf.}\ App.\,\ref{app:proof-simpl}.
\eroof
\noindent Before we conclude that simplicial super-spaces may,\ indeed,\ consistently model target super-spaces of multi-phase super-$\si$-models,\ we need to discuss the interpretation and fate of the identities satisfied by the various tensors supported on $\,X_0\,$ and $\,X_1$.\ To this end,\ we now set up a natural framework\footnote{The underlying idea is an extension of the standard concept of relative cohomology,\ {\it cf.}\ the discussion in \Rcite{Suszek:2012ddg} in the present setting.} for the description of the de Rham cohomology of a simplicial super-space.

\bedef\label{def:relcohom}
Let $\,(X_\bullet,d^{(\bullet)}_\cdot,s^{(\bullet)}_\cdot)\,$ be a simplicial super-space.\ The {\bf simplicial de Rham cohomology of} $\,(X_\bullet,d^{(\bullet)}_\cdot,s^{(\bullet)}_\cdot)\,$ is the total cohomology of the semi-bounded bicomplex
\qq\nn
\alxydim{@C=1cm@R=1cm}{ & \vdots \ar[d]^(.3){\D^{(p)}_{n-1}} & \\ \cdots \ar[r]^(.3){\sfd^{p-1}_{(X_n)}} & \Om^p(X_n) \ar[r]^(.7){\sfd^p_{(X_n)}} \ar[d]^(.7){\D^{(p)}_n} & \cdots  \\ & \vdots & }\,,\qquad p,n\in\bN
\qqq
with the de Rham exterior derivatives as the horizontal coboundary operators and with the vertical coboundary operators
\qq\nn
\D^{(p)}_n\equiv\sum_{i=0}^{n+1}\,(-1)^i\,d^{(n+1)\,*}_i\ :\ \Om^p(X_n)\too\Om^p(X_{n+1})\,,\qquad(p,n)\in\bN^{\x 2}\,,
\qqq
{\it i.e.},\ the cohomology of the diagonal cochain complex (a.k.a.\ the {\bf simplicial de Rham complex})
\qq\nn
\bigl(\cA^\bullet(X_\bullet),\cD^\bullet\bigr)\qquad :\qquad \cA^0(X_\bullet)\xrightarrow{\ \cD_0\ }\cA^1(X_\bullet)\xrightarrow{\ \cD_1\ }\cdots\xrightarrow{\ \cD_{q-1}\ }\cA^q(X_\bullet)\xrightarrow{\ \cD_q\ }\cdots
\qqq
with the $q$-cochain groups
\qq\nn
\cA^q(X_\bullet)=\bigoplus_{p=0}^q\,A^{p,q-p}(X_\bullet)\,,\qquad\qquad A^{p,q-p}(X_\bullet)\equiv\Om^p(X_{q-p})\,,\qquad\qquad q\in\bN
\qqq
related by the couboundary operators
\qq\nn
\cD^q\ :\ \cA^q(X_\bullet)\too\cA^{q+1}(X_\bullet)\,,\qquad\qquad\cD^q\rstr_{A^{p,q-p}(X_\bullet)}=(-1)^{q-p+1}\,\sfd^p_{(X_{q-p})}+\D^{(p)}_{q-p}\,.
\qqq
\exdef
\brem\label{rem:simpldRcohom}
The definition is a standard one for a pair of families of commuting coboundary operators,\ and so the only nontrivial fact which renders it meaningful is the following elementary property of the weighted-pullback operators (written for arbitrary $\,p,n\in\bN$):\ $\,\D^{(p)}_{n+1}\circ\D^{(p)}_n=0$,\ whose verification uses the simplicial identities.\ The de Rham cohomology of simplicial manifolds was originally considered by Dupont in \Rcite{Dupont:1976} ({\it cf.}\ also \Rcite{Dupont:1988}).
\erem
\noindent The naturality of the cohomological framework proposed is put on display by the simple observation:\ If we were to think -- in conformity with Prop.\,\ref{prop:simplicial-target} -- of the bulk target super-space $\,M$,\ the bi-brane superworldvolume $\,Q\,$ and the component inter-bi-brane superworldvolume $\,T_{2,1}\,$ as the first three elements of a simplicial super-space $\,(X_\bullet,d^{(\bullet)}_\cdot,s^{(\bullet)}_\cdot)$,\ {\it i.e.},\ if we were to set $\,(X_0,X_1,X_2)\equiv(M,Q,T_{2,1})\,$ with the previously suggested identification of the corresponding face maps $\,(d^{(1)}_0,d^{(1)}_1)\equiv(\iota_2,\iota_1)\,$ and $\,(d^{(2)}_0,d^{(2)}_1,d^{(2)}_2)\equiv(\pi^{(3)}_{2,3},\pi^{(3)}_{1,3},\pi^{(3)}_{1,2})$,\ then the identities \eqref{eq:Hclos},\ \eqref{eq:om-triv-HH} and \eqref{eq:def-jun-id} satisfied by the triple $\,(\txH,\om,0,0)\in\Om^3(X_0)\oplus\Om^2(X_1)\oplus\Om^1(X_2)\oplus\Om^0(X_3)\,$ would make it a 3-cocycle in the relative cohomology of $\,(X_\bullet,d^{(\bullet)}_\cdot,s^{(\bullet)}_\cdot)$.\ Of course,\ we should then still have to impose the DJI's coming from defect junctions of valence $\,(n,1),\ n>2$,\ but it turns out that the simplicial identities would come to our rescue once more,\ as clarified in
\berop\label{prop:DJI-descent}
Let $\,(X_\bullet,d^{(\bullet)}_\cdot,s^{(\bullet)}_\cdot)\,$ be a simplicial super-space with the maps $\,\pi^{(n+1)}_{1,n+1},\pi^{(n+1)}_{k,k+1}\ :\ X_n\too X_1,\ k\in\ovl{1,n},\ n\geq 2\,$ as in \Reqref{eq:ind-ibb-maps},\ and let $\,\om\in\Om^2(X_1)$.\ Denote $\,\D_{X_n}=\sum_{k=1}^n\,\pi^{(n+1)\,*}_{k,k+1}-\pi^{(n+1)\,*}_{1,n+1}$.\ The following implication holds true:
\qq\nn
\om\in\ker\,\D^{(2)}_1\qquad\Longrightarrow\qquad\forall_{n\geq 2}\ :\ \D_{X_n}\om=0\,.
\qqq
\eerop
\beroof
The proof proceeds by induction.\ Thus,\ first,\ we deal with the case $\,n=3$.\ Invoking the observations and results of (the proof of) Prop.\,\ref{prop:simplicial-target},\ in conjunction with the assumption and the simplicial identities,\ we compute
\qq\nn
&&\bigl(\pi^{(4)\,*}_{1,2}+\pi^{(4)\,*}_{2,3}+\pi^{(4)\,*}_{3,4}-\pi^{(4)\,*}_{1,4}\bigr)\om=d^{(3)\,*}_3\circ\bigl(\D^{(2)}_1+\pi^{(3)\,*}_{1,3}\bigr)\om+\bigl(\pi^{(4)\,*}_{3,4}-\pi^{(4)\,*}_{1,4}\bigr)\om\cr\cr
&=&d^{(3)\,*}_3\circ\pi^{(3)\,*}_{1,3}\om+\bigl(\pi^{(4)\,*}_{3,4}-\pi^{(4)\,*}_{1,4}\bigr)\om\equiv\bigl(d^{(2)}_1\circ d^{(3)}_3\bigr)^*\om+\bigl(d^{(2)}_0\circ d^{(3)}_1-d^{(2)}_1\circ d^{(3)}_1\bigr)^*\om=d^{(3)\,*}_1\circ\D^{(2)}_1\om=0\,,
\qqq
as desired.

Next,\ we posit the induction hypothesis for $\,n\equiv N>2$,\ {\it i.e.},\ assume that $\,\D_{X_N}\om=0$,\ and use it,\ alongside the original assumption,\ to prove the induction step.\ We obtain,\ upon repeated application of the simplicial identities to the aforementioned results,
\qq\nn
&&\D_{X_{N+1}}\om=d^{(N+1)\,*}_{N+1}\circ\bigl(\D_{X_N}+\pi^{(N+1)\,*}_{1,N+1}\bigr)\om+\bigl(\pi^{(N+2)\,*}_{N+1,N+2}-\pi^{(N+2)\,*}_{1,N+2}\bigr)\om\cr\cr
&=&\bigl(d^{(2)}_2\circ d^{(3)}_1\circ d^{(4)}_1\circ\cdots\circ d^{(N+1)}_1\bigr)^*\om+\bigl(d^{(2)}_0\circ d^{(3)}_1\circ d^{(4)}_1\circ\cdots\circ d^{(N+1)}_1-d^{(2)}_1\circ d^{(3)}_1\circ\cdots\circ d^{(N+1)}_1\bigr)^*\om\cr\cr
&=&\bigl(d^{(3)}_1\circ d^{(4)}_1\circ\cdots\circ d^{(N+1)}_1\bigr)^*\circ\D^{(2)}_1\om=0\,,
\qqq
which concludes the proof.
\eroof

At this stage,\ there seems to be no imprint of the extra structure modelling the physically relevant identity defect other than the very natural coherence identity in Prop.\,\ref{prop:simplicial-target} (involving the degeneracy maps).\ We read off the additional information from
\berop
Let $\,(X_\bullet,d^{(\bullet)}_\cdot,s^{(\bullet)}_\cdot)\,$ be a simplicial super-space and let $\,(\cA^\bullet(X_\bullet),\cD^\bullet)\,$ be the corresponding simplicial de Rham complex of Def.\,\ref{def:relcohom}.\ The family of maps
\qq\nn
\nabla_r\ :\ \cA^r(X_\bullet)\too\cA^{r-1}(X_\bullet)\ :\ \bigl(\om_0,\om_1,\ldots,\om_r\bigr)\longmapsto\bigl(\widetilde\D{}_{(0)}^{r-1}\om_0,\widetilde\D{}_{(1)}^{r-2}\om_1,\ldots,\widetilde\D{}_{(r-1)}^0\om_{r-1}\bigr)\,,\qquad r\in\bN^\x
\qqq
induced by the degeneracy maps $\,s^{(\bullet)}_\cdot\,$ of the simplicial super-space as
\qq\nn
\widetilde\D{}_{(p)}^{r-p-1}\ :\ \Om^p(X_{r-p})\too\Om^p(X_{r-p-1})\ :\ \om_p\longmapsto\sum_{k=0}^{r-p-1}\,(-1)^k\,s^{(r-p-1)\,*}_k\om_p\,,\qquad p\in\ovl{0,r-1}
\qqq
defines a chain complex
\qq\nn
\bigl(\cB_\bullet(X_\bullet),\nabla_\bullet\bigr)\,,\qquad\cB_r(X_\bullet)\equiv\cA^r(X_\bullet)\,.
\qqq
Its boundary operators canonically determine cohomology maps
\qq\nn
[\nabla_r]\ :\ H^r\bigl(\cA^\bullet(X_\bullet),\cD^\bullet\bigr)\too H^{r-1}\bigl(\cA^\bullet(X_\bullet),\cD^\bullet\bigr)\,.
\qqq
In what follows,\ we shall refer to the image $\,[\nabla_r]([\Om_r])\in H^{r-1}(\cA^\bullet(X_\bullet),\cD^\bullet)\,$ of the class $\,[\Om_r]\,$ of a simplicial de Rham $r$-cocycle $\,\Om_r\in\cA^r(X_\bullet)\,$ by the name of the {\bf degeneracy shadow of} $\,[\Om_r]$.
\eerop
\beroof
The proof of the first part of the statement hinges upon the identity:\ $\,\widetilde\D{}_{(p)}^n\circ\widetilde\D{}_{(p)}^{n+1}=0$,\ analogous to (and checked similarly as) that encountered in Rem.\,\ref{rem:simpldRcohom}.

The second part of the statement follows directly from the `anticommutation' relations $\,\widetilde\D{}_{(p)}^{n+1}\circ\D^{(p)}_{n+1}=-\D^{(p)}_n\circ\widetilde\D{}_{(p)}^{n}\,,\ p,n\in\bN\,$ proven with the help of the identity $\,(d^{(n+2)}_{k+1}-d^{(n+2)}_k)\circ s^{(n+1)}_k=0\,$ which follows from the simplicial identities,\ and from the equality $\,\widetilde\D{}_{(p)}^0\circ\D^{(p)}_0=0$,\ implied by a specialisation of the latter identity.\ Jointly,\ these yield $\,\nabla_r\circ\cD^{r-1}=-\cD^{r-2}\circ\nabla_{r-1}$.
\eroof
\noindent In the setting of the super-$\si$-model with a simplicial target super-space $\,(X_\bullet,d^{(\bullet)}_\cdot,s^{(\bullet)}_\cdot)$,\ the above gives us the degeneracy shadow $\,(s^{(0)\,*}_0\om,0,0)\in\cA^2(X_\bullet)\,$ of the previously considered simplicial de Rham 3-cocycle 
\qq\label{eq:simpl3coc}
\Om\equiv(\txH,\om,0,0)\in\cA^3(X_\bullet)\,.
\qqq

Altogether,\ motivated by our findings,\ and with some hindsight,\ we give
\bedef\label{def:simpl-target-sspace}
A {\bf simplicial target super-space for a multi-phase super-$\si$-model with defects} (or a {\bf simplicial target super-space} for brevity) is a target super-space $\,\xcT=M\sqcup Q\sqcup T\,$  in the sense of Def.\,\ref{def:targetss} for which there exists a simplicial super-space $\,(X_\bullet,d^{(\bullet)}_\cdot,s^{(\bullet)}_\cdot)\,$ such that each connected component of $\,M\,$ is a sub-supermanifold of $\,X_0$,\ each connected component of $\,Q\,$ is a sub-supermanifold of $\,X_1\equiv Q\,$ and,\ for every $\,n\geq 2$,\ each connected component  of $\,T_{n,1}\,$ is a sub-supermanifold of $\,X_n$,\ and such that the structural maps are induced from the face maps of the simplicial super-space as {\it per}
\qq\nn
\bigl(\iota_1,\iota_2\bigr):=\bigl(d^{(1)}_1,d^{(1)}_0\bigr)\,,\qquad\qquad\bigl(\pi^{(3)}_{1,2},\pi^{(3)}_{2,3},\pi^{(3)}_{1,3}\bigr):=\bigl(d^{(2)}_2,d^{(2)}_0,d^{(2)}_1\bigr)
\qqq
and -- for $\,n>2\,$ --
\qq\nn
\pi^{(n+1)}_{k,k+1}&=&d^{(2)}_2\circ d^{(3)}_2\circ\cdots\circ d^{(n-k+1)}_2\circ d^{(n-k+2)}_0\circ d^{(n-k+3)}_0\circ d^{(n)}_0\,,\qquad k\in\ovl{1,n}\,,\cr\cr
\pi^{(n+1)}_{1,n+1}&=&d^{(2)}_1\circ d^{(3)}_1\circ\cdots\circ d^{(n)}_1\,.
\qqq
Its tensorial data then compose a simplicial de Rham 3-cocycle $\,\Om\equiv(\txH,\om,0,0)$.
\exdef

\brem
The construction of the cohomological bicomplex that captures the skew-tensorial data of a consistent target super-space clearly requires less structure than is present in the context in hand,\ to wit,\ only that of an incomplete simplicial space (a family of spaces with face maps),\ or even a truncation thereof at some fixed level $\,n\in\bZ_{\ge 2}$.\ Any triple $\,(M,Q,T_{n,1})\,$ with properties (T1),\ (T2) and (T3) specified in Sec.\,\ref{sub:ssigmod-def} gives rise to such a (semi-bounded) bicomplex owing to the compatibility conditions \eqref{eq:def-jun-extend}.\ The ensuing cohomology can be understood as the dual of the relative (in the sense of \Rxcite{Sec.\,7}{Bott:1982}) homology of $\,(M,Q,T_{n,1})\,$ (suggested by the homological decomposition of $\,X(\Si)\subset\cT$).\ The latter was elaborated in \Rcite{Suszek:2012ddg} and subsequently employed in a discussion of the algebroidal structure on the set of (infinitesimal) generators of global symmetries of the $\si$-model,\ whereas the former was instrumental in the cohomological classification of the corresponding gauge anomalies carried out in Refs.\,\cite{Gawedzki:2010rn,Gawedzki:2012fu}. 
\erem

\brem
For an elementary exposition on simplicial methods,\ {\it cf.},\ {\it e.g.},\ \Rcite{Friedman:2008}.\ Groupoids and their nerves are covered neatly in Refs.\,\cite{Segal:1968,Moerdijk:2002}.
\erem

Presently,\ we shall discuss at length an important example of a simplicial target super-space,\ but in the meantime we take a closer look at the higher-geometric structures requisite for a rigorous definition of the Wess--Zumino (topological) term of the super-$\si$-model,\ and subsequently incorporate (global) (super)symmetry into the picture,\ also in the simplicial setting.

\subsection{A handful of rudimentary facts about 1-gerbes}
\label{sub:rudi-gerbe}

A natural mathematical framework for an in-depth study of the two-dimensional (super-)$\si$-model is the theory of 1-gerbes advanced by Murray and Stevenson in Refs.\,\cite{Murray:1994db,Murray:1999ew,Stevenson:2000wj} ({\it cf.}\ also a more recent review \cite{Murray:2007ps}) and providing us with a convenient geometrisation of the (hyper-)cohomological description pioneered by Alvarez in \Rcite{Alvarez:1984es} and,\ in particular,\ Gaw\c{e}dzki in \Rcite{Gawedzki:1987ak}.\ Our use of the theory in the present article is of a twofold nature:\ For one thing,\ we need it to write out a consistent definition of the action `functional' of the theory for a generic (defect-decorated) worldsheet,\ which -- as it happens -- opens a direct avenue to the quantum theory and hence gives rise to the expectation that some important results obtained in this framework in the classical r\'egime carry over to the quantum r\'egime.\ For another thing,\ we invoke certain specific constructs,\ such as 1-gerbe trivialisations, bi-modules, multiplicative structures {\it etc.},\ some of which are distinctive of the Lie (super)group-theoretic setting of interest,\ in order to formulate and (partly) solve the existence problem for the geometric structures postulated to describe what we shall call,\ in analogy with the un-graded precedent,\ maximally supersymmetric defects.\ In both cases,\ one may take the general discussion all the way to its logical conclusion without making an explicit use of the underlying extensive differential-(super)geometric and cohomological formalism,\ developing it,\ instead,\ along the lines of an essentially categorial approach that features 1-gerbes as objects on which a number of natural operations (such as,\ {\it e.g.},\ the tensor product and pullback) can be carried out,\ and which can be mapped into one another by means of morphisms,\ the latter being themselves related by higher-order morphisms.\ This approach finds its formalisation in the notion of the weak monoidal 2-category (with duality) of abelian bundle 1-gerbes with connection over a given base supermanifold, introduced,\ in the un-graded setting,\ in \Rcite{Stevenson:2000wj},\ further developed in \Rcite{Waldorf:2007mm},\ and finally reconstructed in the $\bZ/2\bZ$-graded setting in \Rcite{Huerta:2020} (the author explicitly treats the $0$-cells,\ but it is clear how the reconstruction extends to 1- and 2-cells) after having been rather extensively used by the Author in the context of superstring theory.\ The weak 2-category (implicitly) organises the subsequent brief recapitulation of the relevant elements of the general theory.\ On the other hand,\ identification of the relevant gerbe-theoretic structures for the concrete tensorial data of the (Green--Schwarz) super-$\si$-model with the bulk target super-space given by the super-Minkowski space requires concretisation of the general categorial definitions and relations,\ for which we need geometrisations of the abstract concepts.\ These we recall in a concise manner below,\ urging the interested Reader to consult the rich literature on the subject (cited in the Introduction) for details.\ In fact,\ incorporating supersymmetry in the higher-(super)geometric picture,\ as required by the underlying physical model,\ calls for a substantial refinement of the general discussion reflected in the choice of the cohomology to be geometrised.\ This we relegate to a separate Sec.\,\ref{sec:susygrb},\ focusing on generic structures in the present one. \medskip

The basic notion of gerbe theory is that of an {\bf abelian bundle 1-gerbe with connection} $\,\cG\,$ (to be referred to as a {\bf 1-gerbe} for brevity) over a supermanifold $\,M$,\ termed the {\bf base} of $\,\cG$,\ which is to be understood as a geometric structure that realises a class in the cohomology group $\,H^3(|M|;2\pi\bZ)$,\ termed the {\bf Dixmier--Douady class} of $\,\cG\,$ and represented by an even de Rham super-3-cocycle on $\,M\,$ which we call the {\bf curvature} of $\,\cG\,$ and denote by $\,\curv(\cG)$.\ As such,\ a 1-gerbe is represented by the class $\,[b]_{\rm D}\,$ of a 2-cocycle of its local data in the (Beilinson--)Deligne hypercohomology group $\,\bH^2(M,\cD(2)^\bullet)$,\ for the Deligne complex $\,\cD(2)^\bullet\ :\ \cO^{\bC\,\x}_{M,0}\xrightarrow{\ -\sfi\,\sfd\log
\ }\unl{\Om^1}(M)^\bC_0\xrightarrow{\ \ \sfd\ \ }\unl{\Om^2}(M)^\bC_0\,$ of differential sheaves over $\,M$,\ {\it cf.},\ {\it e.g.},\ Refs.\,\cite{Brylinski:1993ab,Murray:1994db,Huerta:2020},\ with the following components:\ the subsheaf $\,\cO^{\bC\,\x}_{M,0}\,$ of Gra\ss mann-even invertible elements of the complexification $\,\cO_M^\bC=\cO_M\ox\bC\,$ of the structure sheaf $\,\cO_M\,$ of $\,M$;\ the subsheaves $\,\unl{\Om^k}(M)^\bC_0\,$ of Gra\ss mann-even elements of the complexifications $\,\unl{\Om^k}(M)^\bC=\unl{\Om^k}(M)\ox\bC\,$ of the respective sheaves $\,\unl{\Om^k}(M)\,$ of $k$-forms on $\,M$.\ The geometric realisation of a class $\,[b]_{\rm D}\,$ with the curvature $\,\txH\in\Om^3(M)_0\,$ takes the form of a septuple 
\qq\nn
\cG\doteq\bigl(\sfY M,\pi_{\sfY M},\txB,L,\pi_L,\cA_L,\mu_L\bigr)
\qqq 
composed of a surjective submersion $\,\pi_{\sfY M}\ :\ \sfY M\too M\,$ endowed with a Gra\ss mann-even primitive (termed the {\bf curving}) $\,\txB\in\Om^2(\sfY M)_0\,$ such that $\,\sfd\txB=\pi_{\sfY M}^*\txH$,\ and of a principal $\bC^\x$-bundle $\,\pi_L\ :\ L\too\sfY^{[2]}M\,$ ({\it cf.}\ App.\,\ref{app:convs}) with a principal $\bC^\x$-connection super-1-form (termed the {\bf connection}) $\,\cA_L\in\Om^1(L)_0\,$ such that $\,\sfd\cA_L=\pi_L^*(\pr_2^*-\pr_1^*)\txB$,\ and a groupoid structure on its fibres,\ {\it i.e.},\ a distinguished connection-preserving principal $\bC^\x$-bundle isomorphism $\,\mu_L\ :\ \pr_{1,2}^*L\ox\pr_{2,3}^*L\xrightarrow{\ \cong\ }\pr_{1,3}^*L\,$ over $\,\sfY^{[3]}M$,\ subject to an associativity constraint over $\,\sfY^{[4]}M\,$ ({\it cf.}, {\it e.g.},\ \Rcite{Waldorf:2007mm}).\ A distinguished class of gerbes is composed of those whose curvature admits a global primitive (on $\,M$).\ A member of the class with the primitive $\,\om\in \Om^2(M)_0\,$ shall be denoted by $\,\cI_\om$,\ and we have $\,\curv(\cI_\om)=\sfd\om$.\ For these,\ we may take $\,\cI_\om=(M,\id_M,\om,M\x\bC^\x,\pr_1,\pr_2^*\vartheta,\bd1)$,\ with $\,\vartheta\in\Om^1(\bC^\x)\,$ the Maurer--Cartan 1-form on the Lie group $\,\bC^\x$,\ and with the trivial groupoid structure $\,\bd1\,$ given by group multiplication in the fibre.

There are a number of natural operations on 1-gerbes, such as,\ {\it e.g.},\ the ({\bf tensor}) {\bf product} $\,\cG_1\ox\cG_2\,$ of any two 1-gerbes $\,\cG_A,\ A=1,2\,$ over the same base $\,M$,\ and the {\bf pullback} $\,f^*\cG\,$ of a 1-gerbe $\,\cG\,$ over $\,M\,$ along an arbitrary supermanifold map $\,f\ :\ N\too M$.\ The latter is the 1-gerbe $\,f^*\cG=(f^*\sfY M,\pi_{f^*\sfY M},\widehat f{}^*\txB,\widehat f{}^{[2]\,*}L,\pi_{\widehat f{}^{[2]\,*}L},\widehat{\widehat f}{}^{[2]\,*}\cA_L,\widehat f{}^{[3]\,*}\mu_L)\,$ defined in terms of pullback surjective submersions: 
\qq\nn
\alxydim{@C=2.cm@R=1.5cm}{ \sfY_f N\equiv f^*\sfY M \ar[r]^{\qquad\widehat f} \ar[d]_{\pi_{f^*\sfY M}} & \sfY M \ar[d]^{\pi_{\sfY M}} \\ N \ar[r]_{f} & M }\qquad\quad{\rm and}\qquad\quad\alxydim{@C=2.cm@R=1.5cm}{ \widehat f{}^{[2]\,*}L \ar[r]^{\widehat{\widehat f}{}^{[2]}} \ar[d]_{\pi_{\widehat f{}^{[2]\,*}L}} & L \ar[d]^{\pi_L} \\ \sfY_f^{[2]} N \ar[r]_{\widehat f{}^{[2]}} & \sfY^{[2]}M }\,.
\qqq
Under these operations,\ the curvature behaves as follows:\ $\,\curv(\cG_1\ox\cG_2)=\curv(\cG_1)+\curv(\cG_2)\,$ and $\,\curv\bigl(f^*\cG\bigr)=f^*\curv(\cG)$,\ and for trivial gerbes over $\,M\,$ we find the identities $\,\cI_{\om_1}\ox\cI_{\om_2}=\cI_{\om_1+\om_2}\,$ and $\,f^*\cI_\om=\cI_{f^*\om}$.
~\smallskip

The geometric realisation of cohomological equivalence of 1-gerbes $\,\cG_A,\ A=1,2\,$ (of the same curvature) over the common base,\ $\,[b_1]_{\rm D}=[b_2]_{\rm D}$,\ goes under the name of a \emph{1-isomorphism} $\,\Phi\,$ between $\,\cG_1\,$ and $\,\cG_2\,$ and is denoted as $\,\Phi\ :\ \cG_1\xrightarrow{\ \cong\ }\cG_2$.\ Thus,\ a 1-isomorphism is represented by a 1-cochain $\,p\,$ in the Deligne hypercohomology introduced previously.\ It is now easy to see that ({\it cf.}\ Refs.\,\cite{Gajer:1996,Johnson:2003})
\berop\label{prop:1-isoclof1-grb}
The set of 1-isomorphism classes of 1-gerbes of a given curvature over $\,M\,$ is a torsor under an action of the group $\,\cW^3(M;\txH=0)\cong H^2(M,\uj)\,$ of {\bf flat gerbes} over $\,M\,$ ({\it i.e.},\ those with a vanishing curvature).
\eerop 
\noindent The action referred to above as well as the binary operation in $\,\cW^3(M;\txH=0)\,$ are induced by the tensor product as $\,\cW^3(M;0)\x\bH^2(M,\cD(2)^\bullet)\too\bH^2(M,\cD(2)^\bullet)\ :\ ([b_0]_{\rm D},[b]_{\rm D})\longmapsto[b_0]_{\rm D}\ox[b]_{\rm D}$.\ Given geometrisations $\,\cG_A=(\sfY_A M,\pi_{\sfY_A M},\txB_A,L_A,\pi_{L_A},\cA_{L_A},\mu_{L_A}),\ A\in\{1,2\}\,$ of the two 1-gerbes related by the 1-isomorphism $\,\Phi$,\ we obtain the geometric realisation of the latter in the form of a sextuple 
\qq\nn
\Phi\doteq\bigl(\sfY\sfY_{1,2}M,\pi_{\sfY\sfY_{1,2}M},E,\pi_E,\cA_E,\a_E\bigr)
\qqq
composed of a surjective submersion $\,\pi_{\sfY\sfY_{1,2}M}\ :\ \sfY\sfY_{1,2}M\too\sfY_1 M\x_M\sfY_2 M\equiv\sfY_{1,2}M\,$ and of a principal $\bC^\x$-bundle $\,\pi_E\ :\ E\too\sfY\sfY_{1,2}M\,$ over it,\ endowed with a principal $\bC^\x$-connection 1-form $\,\cA_E\in\Om^1(E)_0\,$ such that $\,\sfd\cA_E=\pi_E^*\pi_{\sfY\sfY_{1,2}M}^*(\pr_2^*\txB_2-\pr_1^*\txB_1)$,\ and with a distinguished connection-preserving principal $\bC^\x$-bundle isomorphism $\,\a_E\ :\ \pi_{\sfY\sfY_{1,2}M}^{\x 2\,*}\pr_{1,3}^*L_1\ox\pr_2^*E\xrightarrow{\ \cong\ }\pr_1^*E\ox\pi_{\sfY\sfY_{1,2}M}^{\x 2\,*}\pr_{2,4}^*L_2\,$ over $\,\sfY^{[2]}\sfY_{1,2}M\equiv\sfY\sfY_{1,2}M\x_M\sfY\sfY_{1,2}M$,\ subject to a further coherence constraint involving the two groupoid structures $\,\mu_{L_A}$,\ {\it cf.}, {\it e.g.},\ \Rcite{Waldorf:2007mm}.

Amidst 1-isomorphisms between 1-gerbes over a given base $\,M$,\ there are special types which play an important r\^ole in what follows.\ These are {\bf trivialisations},\ of the form
\qq\nn
\cT_\om\ :\ \cG\xrightarrow{\ \cong\ }\cI_\om\,,
\qqq
with $\,\sfd\om=\curv(\cG)$,\ and,\ for an arbitrary pair $\,(\cG_1,\cG_2)\,$ of 1-gerbes,\ the associated {\bf $(\cG_1,\cG_2)$-bi-modules},\ defined as
\qq\nn
\Phi_\om\ :\ \cG_1\xrightarrow{\ \cong\ }\cG_2\ox\cI_\om\,,
\qqq
with $\,\sfd\om=\curv(\cG_1)-\curv(\cG_2)$.\ Whenever the two 1-gerbes involved in the latter definition are pullbacks $\,\cG_A=\iota_A^*\cG,\ A\in\{1,2\}\,$ of a single 1-gerbe over a manifold $\,M\,$ to another manifold $\,Q\,$ along the respective smooth maps $\,\iota_A\ :\ Q\too M$,\ we should think of the corresponding pair $\,(\cG,\Phi_\om)\,$ as a
geometric realisation of the class $\,[\curv(\cG)\oplus\om]\,$ of the 3-cocycle $\,\curv(\cG)\oplus\om\,$ in the relative de Rham cohomology of the pair $\,(Q,M)$.

As in the case of 1-gerbes,\ we have several natural operations on 1-isomorphisms,\ including the {\bf tensor product} $\,\Phi_{1,3}\ox\Phi_{2,4}\,$ of any two 1-isomorphisms $\,\Phi_{A,B}\ :\ \cG_A\xrightarrow{\ \cong\ }\cG_B,\ (A,B)=(1,3),(2,4)$,\ with $\,\Phi_{1,3}\ox\Phi_{2,4}\ :\ \cG_1\ox\cG_2\xrightarrow{\ \cong\ }\cG_3\ox \cG_4$,\ the {\bf composition} $\,\Phi_{2,3}\circ\Phi_{1,2}\,$ of any two 1-isomorphisms $\,\Phi_{A,B}\ :\ \cG_A\xrightarrow{\ \cong\ }\cG_B,\ (A,B)=(1,2),(2,3)$,\ with $\,\Phi_{2,3}\circ\Phi_{1,2}\ :\ \cG_1\xrightarrow{\ \cong\ }\cG_3$,\ and the {\bf pullback} $\,f^*\Phi\,$ of a 1-isomorphism $\,\Phi\ :\ \cG_1\xrightarrow{\ \cong\ }\cG_2\,$ between gerbes $\,\cG_A,\ A=1,2\,$ over $\,M\,$ along an arbitrary supermanifold map $\,f\ :\ N\too M$,\ with $\,f^*\Phi\ :\ f^*\cG_1\xrightarrow{\ \cong\ }f^*\cG_2$.

Besides 1-{\it iso}morphisms,\ there also exist non-invertible 1-morphisms between 1-gerbes,\ with the underlying structure of a vector bundle of a rank strictly greater than 1.\ We shall not have a need for these in the present paper (but {\it cf.},\ {\it e.g.},\ \Rcite{Gawedzki:2004tu}).~\smallskip

On the lowest rung of the cohomology ladder,\ we find geometric objects that realise cohomological equivalences between 1-isomorphisms,\ represented by 0-cochains in the Deligne hypercohomology.\ Thus,\ for any two 1-isomorphisms $\,\Phi_{1,2}^A,\ A=1,2\,$ between the same two gerbes $\,\cG_B,\ B=1,2$,\ with equivalent Deligne data $\,[p_2-p_1]_{\rm D}=0$,\ we have a \emph{2-isomorphism} $\,\varphi\,$ between $\,\Phi_{1,2}^1\,$ and $\,\Phi_{1,2}^2$,\ denoted as $\,\varphi\ :\ \Phi_{1,2}^1\xLongrightarrow{\ \cong\ }\Phi_{1,2}^2$,\ or -- in the extended notation -- as
\qq\nn
\alxydim{}{\cG_1 \ar@/^1.6pc/[rrr]^{\Phi_{1,2}^1}="5"
\ar@/_1.6pc/[rrr]_{\Phi_{1,2}^2}="6"
\ar@{=>}"5"+(0,-4);"6"+(0,4)|{\varphi} & & & \cG_2}\,.
\qqq
Clearly,\ $\,\varphi\,$ is represented by a 0-cochain $\,f\,$ in the Deligne hypercohomology based on $\,\cD(2)^\bullet$,\ and inequivalent 2-isomorphisms between given two 1-isomorphisms are labelled by classes $\,[f_2-f_1]_{\rm D}\,$ of 0-cocycles in the Deligne hypercohomology group $\,\bH^0(M,\cD(2)^\bullet)$.\ Given geometrisations $\,\Phi^A_{1,2}=(\sfY^A\sfY_{1,2}M,\pi_{\sfY^A\sfY_{1,2}M},E_A,\pi_{E_A},$ $\cA_{E_A},\a_{E_A}),\ A\in\{1,2\}\,$ of the two 1-isomorphisms related by the 2-isomorphism $\,\varphi$,\ we obtain the geometric realisation of the latter in the form of a triple
\qq\nn
\varphi\doteq\bigl(\sfY\sfY^{1,2}\sfY_{1,2}M,\pi_{\sfY\sfY^{1,2}\sfY_{1,2}M},\b\bigr)
\qqq
composed of a surjective submersion $\,\pi_{\sfY\sfY^{1,2}\sfY_{1,2}M}\ :\ \sfY\sfY^{1,2}\sfY_{1,2}M\too\sfY^1\sfY_{1,2}M\x_{\sfY_{1,2}M}\sfY^2\sfY_{1,2}M\equiv\sfY^{1,2}\sfY_{1,2}M$,\ and,\ over it,\ a distinguished connection-preserving principal $\bC^\x$-bundle isomorphism $\,\b\ :\ \pi_{\sfY\sfY^{1,2}\sfY_{1,2}M}^*\pr_1^*E_1\xrightarrow{\ \cong\ }\pi_{\sfY\sfY^{1,2}\sfY_{1,2}M}^*\pr_2^*E_2$,\ subject to a further coherence constraint involving the two isomorphisms $\,\a_{E_A}$,\ {\it cf.}, {\it e.g.},\ \Rcite{Waldorf:2007mm}.\ We have the important ({\it cf.}\ Refs.\,\cite{Gajer:1996,Johnson:2003})
\berop\label{prop:2-isoclof1-isos}
The set of 2-isomorphism classes of 1-isomorphisms between a given pair of 1-gerbes over $\,M\,$ is a torsor under an action of the group $\,H^1(M,\uj)$.
\eerop 

Once again,\ we can subject the geometric objects in hand to a variety of natural operations.\ There is,\ {\it e.g.},\ the {\bf tensor product} $\,\varphi_{1,3}\ox\varphi_{2,4}\,$ of any two 2-isomorphisms $\,\varphi_{A,B}\ :\ \Phi_A\xLongrightarrow{\ \cong\ }\Phi_B,\ (A,B)\in\{(1,3),(2,4)\}$,\ with $\,\varphi_{1,3}\ox\varphi_{2,4}\ :\ \Phi_1\ox\Phi_2 \xLongrightarrow{\ \cong\ }\Phi_3\ox\Phi_4$,\ the {\bf horizontal composition} $\,\varphi_{2,3}\circ\varphi_{1,2}\,$ of any two 2-isomorphisms $\,\varphi_{A,B}\ :\ \Phi_{A,B}^1\xLongrightarrow{\ \cong\ }\Phi_{A,B}^2,\ (A,B)\in\{(1,2),(2,3)\}\,$ between 1-isomorphisms $\,\Phi_{A,B}^C\ :\ \cG_A\xrightarrow{\ \cong\ }\cG_B,\ C\in\{1,2\}\,$ between 1-gerbes $\,\cG_A\,$ over $\,M$,\ with
\qq\nn
\alxydim{}{\cG_1
\ar@/^1.6pc/[rrr]^{\Phi_{2,3}^1\circ\Phi_{1,2}^1}="5"
\ar@/_1.6pc/[rrr]_{\Phi_{2,3}^2\circ\Phi_{1,2}^2}="6"
\ar@{=>}"5"+(0,-4);"6"+(0,4)|{\varphi_{2,3}\circ\varphi_{1,2}} & & &
\cG_3}\ \equiv\ \alxydim{}{\cG_1
\ar@/^1.6pc/[rrr]^{\Phi_{1,2}^1}="5"
\ar@/_1.6pc/[rrr]_{\Phi_{1,2}^2}="6"
\ar@{=>}"5"+(0,-4);"6"+(0,4)|{\varphi_{1,2}} & & & \cG_2
\ar@/^1.6pc/[rrr]^{\Phi_{2,3}^1}="7"
\ar@/_1.6pc/[rrr]_{\Phi_{2,3}^2}="8"
\ar@{=>}"7"+(0,-4);"8"+(0,4)|{\varphi_{2,3}} & & & \cG_3}\,,
\qqq
the {\bf vertical composition} $\,\varphi_{2,3}\bullet\varphi_{1,2}\,$ of any two 2-isomorphisms $\,\varphi_{A,B}\ :\ \Phi_A\xLongrightarrow{\ \cong\ }\Phi_B,\ (A,B)\in\{(1,2),(2,3)\}\,$ between 1-isomorphisms $\,\Phi_A\ :\ \cG_1\xrightarrow{\ \cong\ }\cG_2,\ A\in\{1,2,3\}\,$ between 1-gerbes $\,\cG_1\,$ and $\,\cG_2\,$ over $\,M$,\ with
\qq\nn
\alxydim{}{\cG_1 \ar@/^1.6pc/[rrr]^{\Phi_1}="5"
\ar@/_1.6pc/[rrr]_{\Phi_3}="6"
\ar@{=>}"5"+(0,-4);"6"+(0,4)|{\varphi_{2,3}\bullet\varphi_{1,2}} & &
& \cG_2}\ \equiv\ \alxydim{}{\cG_1
\ar@/^3pc/[rrr]^{\Phi_1}="5" \ar[rrr]|{\Phi_2}="6"
\ar@/_3pc/[rrr]_{\Phi_3}="7"
\ar@{=>}"5"+(0,-4);"6"+(0,4)|{\varphi_{1,2}}
\ar@{=>}"6"+(0,-4);"7"+(0,4)|{\varphi_{2,3}} & & & \cG_2}\,,
\qqq
and the {\bf pullback} $\,f^*\varphi\,$ of a 2-isomorphism $\,\varphi\ :\ \Phi_{1,2}^1\xLongrightarrow{\ \cong\ }\Phi_{1,2}^2\,$ between 1-isomorphisms $\,\Phi_{1,2}^B\ :\ \cG_1\xrightarrow{\ \cong\ }\cG_2,\ B\in\{1,2\}\,$ between 1-gerbes $\,\cG_A,\ A\in\{1,2\}\,$ over $\,M\,$ along an arbitrary supermanifold map $\,f\ :\ N\too M$,\ with
\qq\nn
\alxydim{}{f^*\cG_1 \ar@/^1.6pc/[rrr]^{f^*\Phi_{1,2}^1}="5"
\ar@/_1.6pc/[rrr]_{f^*\Phi_{1,2}^2}="6"
\ar@{=>}"5"+(0,-4);"6"+(0,4)|{f^*\varphi} & & & f^*\cG_2}\,.
\qqq
~\smallskip

Prior to concluding this section,\ let us remark that 1-gerbes over a given supermanifold $\,M$,\ together with the attendant 1-morphisms and 2-isomorphisms,\ compose a weak monoidal 2-category (with duality),\ termed the {\bf weak 2-category of abelian bundle gerbes with connection over} $\,M\,$ and denoted as $\,\bgrb^\nabla(M)$,\ {\it cf.}\ Refs.\,\cite{Stevenson:2000wj,Waldorf:2007mm} (the structure carries over to the $\bZ/2\bZ$-graded setting).

\subsection{The gerbe theory for the multi-phase super-$\si$-model}\label{sub:grbmultissi}

The physical significance of the weak 2-category $\,\bgrb^\nabla(\xcT)\,$ (understood in a natural manner) over the target super-space $\,\xcT=M\sqcup Q\sqcup T\,$ of the super-$\si$-model on a generic worldsheet with defects $\,(\Si,d)\,$ stems from the observation that 0-, 1- and 2-cells of the 2-category provide just the cohomological data needed to define the topological term in the action `functional' of the super-$\si$-model in the presence of the
self-intersecting defect embedded by $\,d\,$ in $\,\Si$.\ This was substantiated in all generality in \Rcite{Runkel:2008gr},\ which is also where the said action functional was derived from elementary worldsheet invariance analysis,\ and the findings reported in that paper retain their validity in the $\bZ/2\bZ$-graded setting.\ In order to appreciate this observation,\ we need to recall the basic gerbe-theoretic structure behind the two-dimensional (super)field theory under
consideration.\ To this end,\ we introduce (adapting the nomenclature introduced in Refs.\,\cite{Fuchs:2007fw,Runkel:2008gr},\ {\it cf.}\ also Refs.\,\cite{Runkel:2009sp,Suszek:2011hg,Suszek:2012ddg,Gawedzki:2012fu,Suszek:2013})
\bedef\label{def:str-bgrnd}
A {\bf superstring background} is a triple $\,\Bgt=(\cM,\cB,\cJ)\,$ composed of the following geometric structures:
\bit
\item the {\bf target} $\,\cM=(M,\txg,\cG)\,$ consisting of a super-space $\,M\,$ (the {\bf bulk target super-space}) with a Gra\ss mann-even metric tensor $\,\txg\,$ (typically degenerate in the Gra\ss mann-odd coordinate directions) and a bundle gerbe $\,\cG\,$ (with connection) of curvature $\,\curv(\cG)=\txH\in\Om^3(M)_0$;
\item the {\bf $\cG$-bi-brane} $\,\cB=(Q,\iota_1,\iota_2,\om,\Phi)\,$ consisting of a super-space $\,Q$,\ termed the {\bf $\cG$-bi-brane worldvolume},\ with a Gra\ss mann-even super-2-form $\,\om$,\ termed
the {\bf $\cG$-bi-brane curvature},\ and a pair of supermanifold maps $\,\iota_A\ :\ Q\too M,\ A\in\{1,2\}$,\ and of a $(\iota_1^*\cG,\iota_2^*\cG)$-bi-module
\qq\nn
\Phi\ :\ \iota_1^*\cG\xrightarrow{\ \cong\ }\iota_2^*\cG\ox\cI_\om\,,
\qqq
to be called the {\bf correspondence bi-module};
\item the {\bf $(\cG,\cB)$-inter-bi-brane}\footnote{As indicated in Sec.\,\ref{sub:ssigmod-def},\ we restrict our attention here to the background structure necessary to describe a distinguished class of worldsheet defects,\ to wit,\ those with defect junctions with a single outgoing defect line.\ This affects our choice of inter-bi-brane data,\ {\it cf.}\ Refs.\,\cite{Runkel:2008gr,Suszek:2011hg,Suszek:2012ddg} for a more general description.} $\,\cJ=\bigsqcup_{n=2}^\infty\,(T_{n,1},\pi_{1,2}^{(n+1)},\pi_{2,3}^{(n+1)},\ldots,\pi_{n,n+1}^{(n+1)},\pi_{1,n+1}^{(n+1)},\varphi_{n+1})\,$ consisting of a super-space $\,T=\bigsqcup_{n=2}^\infty\,T_{n,1}$,\ termed the {\bf $(\cG,\cB)$-inter-bi-brane worldvolume},\ with a collection of maps $\,\pi_{k,k+1}^{(n+1)},\pi_{1,n}^{(n+1)}\ :\ T_{n,1}\too Q\,$ for each component worldvolume,\ subject to the constraints \eqref{eq:def-jun-extend},\ and a distinguished 1-gerbe 2-isomorphism ($\Phi^\vee\,$ is the dual of $\,\Phi$)
\qq\label{diag:2iso}
\xy (50,0)*{\bullet}="G3"+(5,4)*{\bigl(\iota_1\circ\pi_{3,4}^{(n+1)}\bigr)^*\cG\ox
\cI_{\pi_{1,2}^{(n+1)\,*}\om+\pi_{2,3}^{(n+1)\,*}\om}};
(25,-20)*{\bullet}="G2"+(-20,0)*{\bigl(\iota_1\circ\pi_{2,3}^{(n+1)}\bigr)^*\cG\ox
\cI_{\pi_{1,2}^{(n+1)\,*}\om}}; (75,-20)*{\ \vdots}="dots"; (85,-20)*{\,,};
(35,-40)*{\bullet}="G1"+(-10,-4)*{\bigl(\iota_1\circ\pi_{1,2}^{(n+1)}\bigr)^*\cG};
(65,-40)*{\bullet}="G1add"+(18,-4)*{\bigl(\iota_1\circ\pi_{1,n+1}^{(n+1)}\bigr)^*\cG\ox
\cI_{\D_{T_{n,1}}\om}}; (50,-40)*{}="id"; \ar@{->}|{\pi_{2,3}^{(n+1)\,*}
\Phi\ox\id_{\cI_{\pi_{1,2}^{(n+1)\,*}\om}}\hspace{45pt}} "G2";"G3"
\ar@{->}|{\hspace{75pt}\pi_{3,4}^{(n+1)\,*}
\Phi\ox\id_{\cI_{\pi_{1,2}^{(n+1)\,*}\om+\pi_{2,3}^{(n+1)\,*}\om}}} "G3";"dots"
\ar@{->}|{\pi_{1,2}^{(n+1)\,*}\Phi} "G1";"G2"
\ar@{->}|{\pi_{1,n+1}^{(n+1)\,*}\Phi^\vee\ox\id_{\cI_{\D_{T_{n,1}}\om}}}
"dots";"G1add" \ar@{=}|{\id_{(\iota_1\circ\pi_{1,2}^{(n+1)})^*\cG}}
"G1"+(2,0);"G1add"+(-2,0) \ar@{=>}|{\varphi_{n+1}}
"G3"+(0,-3);"id"+(0,+3)
\endxy
\qqq
to be called the ({\bf component}) {\bf fusion 2-isomorphism}.
\eit
\exdef
\brem\label{rem:elemfusion}
In order to make the definition of the inter-bi-brane ever so slightly more tractable,\ and,\ simultaneously,\ to pave the way to subsequent considerations,\ we write out the relevant 2-isomorphism for the elementary (trivalent) vertex corresponding to $\,n=2\,$ that we shall call the {\bf elementary fusion 2-isomorphism}.\ We have
{\small\qq\hspace{-1.5cm}
\alxydim{@C=10em@R=6em}{\bigl(\iota_2\circ\pi_{1,2}\bigr)^*\cG\ox\cI_{\pi_{1,2}^*\om}\equiv\bigl(\iota_1\circ\pi_{2,3}\bigr)^*\cG\ox\cI_{\pi_{1,2}^*\om}
\ar[r]^{\pi_{2,3}^*\Phi\ox\id_{\cI_{\pi_{1,2}^*\om}}\qquad\qquad}
& \bigl(\iota_2\circ\pi_{2,3}\bigr)^*\cG\ox\cI_{\pi_{1,3}^*\om}\ox\cI_{\D_{T_{2,1}}\om}\equiv\bigl(\iota_2\circ\pi_{1,3}\bigr)^*\cG\ox\cI_{\pi_{1,3}^*\om}\ox\cI_{\D_{T_{2,1}}\om}
\ar[d]^{\pi_{1,3}^*\Phi^\vee\ox\id_{\cI_{\D_{T_{2,1}}\om}}} \ar@{=>}[dl]|{\varphi_3} \\
\bigl(\iota_1\circ\pi_{1,2}\bigr)^*\cG \ar[u]^{\pi_{1,2}^*\Phi}
\ar@{=}[r]|{\ \id_{(\iota_1\circ\pi_{1,2})^*\cG}\ }
& \bigl(\iota_1\circ\pi_{1,3}\bigr)^*\cG\ox\cI_{\D_{T_{2,1}}\om} }\cr\label{diag:elem-ibb}
\qqq} 
\erem

The existence of the structure of a simplicial super-space underlying the target super-space of the super-$\si$-model,\ with the cohomological constraints \eqref{eq:def-jun-extend} on the component inter-bi-brane super-worldvolumes $\,T_{n,1}\,$ associated with defect junctions of higher valence $\,n+1>3\,$ ensured by the `master' DJI satisfied over  the `elementary' super-worldvolume $\,T_{2,1}$,\ immediately suggests a distinguished higher-geometric sub-structure over such a target super-space which we describe below.\ For that,\ we need
\bedef\label{def:FBT}
Fix $\,n\in\bN\setminus\{0,1\}$.\ A \textbf{full binary tree with $\,n\,$ leaves} is a connected acyclic (undirected) graph,\ with a distinguished node of degree 2 (termed the {\bf root}),\ and all other nodes of degree either 3 (the internal nodes) or 1 (the external nodes,\ termed the {\bf leaves}),\ there being $\,n\,$ of the latter,\ and with a distinguished ordering of every pair of nodes that are connected to an arbitrary node of degree greater than 1 and lie farther from the root than the node -- these nodes are termed the {\bf children of the node} and the ordering distinguishes a {\bf left child} from a {\bf right child}.\ We shall denote the set of all such trees as
\qq\nn
{\rm FBT}_n=\{\ \tx{full binary trees with $\,n\,$ leaves}\ \}
\qqq
and distinsuish,\ for later reference,\ two of its elements:
\qq
\tau_n^{\rm L}\quad\doteq\hspace{-10pt}
\xy (50,10)*{\bullet}="R";
(45,5)*{\bullet}="V12_n-1";
(35,-5)*{\bullet}="V123";
(30,-10)*{\bullet}="V12";
(25,-15)*{\bullet}="V1"+(0,-4)*{1};
(35,-15)*{\bullet}="V2"+(0,-4)*{2};
(45,-15)*{\bullet}="V3"+(0,-4)*{3};
(65,-15)*{\bullet}="Vn-1"+(0,-4)*{n-1};
(75,-15)*{\bullet}="Vn"+(0,-4)*{n}; 
\ar@{-} "R";"V12_n-1" \ar@{-}
"R";"Vn" \ar@{-} "V12_n-1";"Vn-1" \ar@{.} "V12_n-1";"V123" \ar@{-}
"V123";"V3" \ar@{-} "V123";"V12" \ar@{-} "V12";"V2" \ar@{-}
"V12";"V1"
\endxy\qquad\,,\qquad
\tau_n^{\rm R}\quad\doteq\hspace{-10pt}
\xy (50,10)*{\bullet}="R";
(55,5)*{\bullet}="V23_n";
(65,-5)*{\bullet}="Vn-2n-1n";
(70,-10)*{\bullet}="Vn-1n";
(25,-15)*{\bullet}="V1"+(0,-4)*{1};
(35,-15)*{\bullet}="V2"+(0,-4)*{2};
(55,-15)*{\bullet}="Vn-2"+(0,-4)*{n-2};
(65,-15)*{\bullet}="Vn-1"+(0,-4)*{n-1};
(75,-15)*{\bullet}="Vn"+(0,-4)*{n}; 
\ar@{-} "R";"V23_n" \ar@{-}
"R";"V1" \ar@{-} "V23_n";"V2" \ar@{.} "V23_n";"Vn-2n-1n" \ar@{-}
"Vn-2n-1n";"Vn-2" \ar@{-} "Vn-2n-1n";"Vn-1n" \ar@{-} "Vn-1n";"Vn-1" \ar@{-}
"Vn-1n";"Vn"
\endxy\,.\cr \label{eq:LRtreen}
\qqq
A {\bf level of the} ({\bf full binary}) {\bf tree} is the subset of its nodes at a fixed distance from the root,\ measured by the number of edges on the path from the root to that node.\ In particular,\ the root is the unique element at level 0.\ The {\bf height of the} ({\bf full binary}) {\bf tree} is the number of edges between the root and any node at the highest level.

To every full binary tree with $\,n\,$ leaves $\,\t\in{\rm FBT}_n\,$ of height $\,H\,$ we associate its {\bf level-ordered presentation} given by an embedding of the graph $\,\t\,$ in $\,\bR^{\x 2}\,$ obtained in the following manner:
\bit
\item[-] the root of $\,\t\,$ is placed at $\,(0,0)$;\ 
\item[-] the nodes of degree 3 at level $\,L>0\,$ are placed on the horizontal line $\,\bR\x\{-L\}\,$ so that the first coordinate of the left child of a given node at level $\,L-1\,$ is greater than that of the right child,\ and no node is embedded between them on that line;
\item[-] the leaves of $\,\t\,$ are placed on the horizontal line $\,\bR\x\{-H\}\,$ in such a manner that for any pair of children of some node above them the first coordinate of the left child is greater than that of the right child,\ and no node is embedded between them on that line -- this placement determines a numbering of the leaves in the direction of the increasing first coordinate,\ with the range $\,\ovl{1,n}$;
\item[-] a child of a node is linked to the latter by a segment of the straight line passing through the node and the child.
\eit

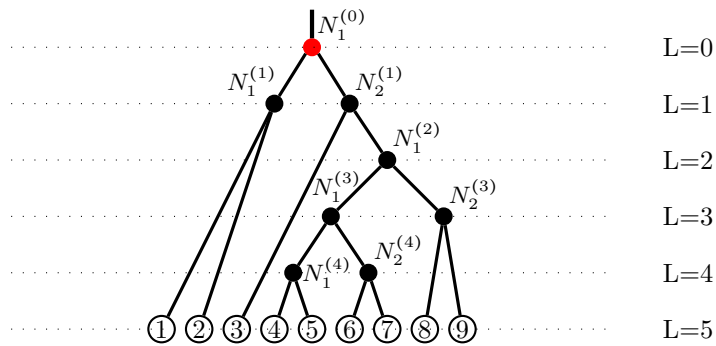
\begin{figure}[h!]
\begin{tikzpicture}
[root/.style={circle,draw=red!100,fill=red!100,thick,inner sep=0pt,minimum size=6pt}, node/.style={circle,draw=black!100,fill=black!100,thick,inner sep=0pt,minimum size=6pt}, leaf/.style={circle,draw=black!100,fill=white!100,thick,inner sep=0pt,minimum size=10pt}]

\node[root] (R) at (0,0) {};
\node at (0.4,0.3) {{\small $N^{(0)}_1$}};
\node[leaf] (L1) at (-2,-3.75) {1};
\node[leaf] (L2) at (-1.5,-3.75) {2};
\node[leaf] (L3) at (-1.0,-3.75) {3};
\node[leaf] (L4) at (-0.5,-3.75) {4};
\node[leaf] (L5) at (0.0,-3.75) {5};
\node[leaf] (L6) at (0.5,-3.75) {6};
\node[leaf] (L7) at (1.0,-3.75) {7};
\node[leaf] (L8) at (1.5,-3.75) {8};
\node[leaf] (L9) at (2.0,-3.75) {9};
\node[node] (N10) at (-0.25,-3.0) {};
\node at (0.2,-3) {{\small $N^{(4)}_1$}};
\node[node] (N11) at (0.75,-3.0) {};
\node at (1.15,-2.7) {{\small $N^{(4)}_2$}};
\node[node] (N12) at (0.25,-2.25) {};
\node at (0.325,-1.85) {{\small $N^{(3)}_1$}};
\node[node] (N13) at (1.75,-2.25) {};
\node at (2.15,-1.95) {{\small $N^{(3)}_2$}};
\node[node] (N14) at (1.0,-1.5) {};
\node at (1.4,-1.2) {{\small $N^{(2)}_1$}};
\node[node] (N15) at (-0.5,-0.75) {};
\node at (-0.8,-0.45) {{\small $N^{(1)}_1$}};
\node[node] (N16) at (0.5,-0.75) {};
\node at (0.9,-0.45) {{\small $N^{(1)}_2$}};

\draw[-,ultra thick] (R.north) -- (0,0.5);
\draw[-,very thick] (L1) -- (N15);
\draw[-,very thick] (L2) -- (N15);
\draw[-,very thick] (L3) -- (N16);
\draw[-,very thick] (L4) -- (N10);
\draw[-,very thick] (L5) -- (N10);
\draw[-,very thick] (L6) -- (N11);
\draw[-,very thick] (L7) -- (N11);
\draw[-,very thick] (N10) -- (N12);
\draw[-,very thick] (N11) -- (N12);
\draw[-,very thick] (L8) -- (N13);
\draw[-,very thick] (L9) -- (N13);
\draw[-,very thick] (N12) -- (N14);
\draw[-,very thick] (N13) -- (N14);
\draw[-,very thick] (N14) -- (N16);
\draw[-,very thick] (N15) -- (R.south west);
\draw[-,very thick] (N16) -- (R.south east);

\node at (5,0) {L=0};
\node at (5,-0.75) {L=1};
\node at (5,-1.5) {L=2};
\node at (5,-2.25) {L=3};
\node at (5,-3.0) {L=4};
\node at (5,-3.75) {L=5};

\draw[loosely dotted,-] (-4,0) -- (4,0);
\draw[loosely dotted,-] (-4,-0.75) -- (4,-0.75);
\draw[loosely dotted,-] (-4,-1.5) -- (4,-1.5);
\draw[loosely dotted,-] (-4,-2.25) -- (4,-2.25);
\draw[loosely dotted,-] (-4,-3) -- (4,-3);
\draw[loosely dotted,-] (-4,-3.75) -- (4,-3.75);
\end{tikzpicture}
\caption{The planting of a level-ordered presentation of a full binary tree $\,\t_*\,$ with 9 leaves (of height 5).\ The induced numbering of the leaves is shown. The dotted lines sweep the levels of the tree (indexed on the right). The nodes of degree 3 are labelled for later reference.} \label{fig:plantLOpresFBT9}
\end{figure}

Finally,\ the {\bf planting} of the level-ordered presentation of the given full binary tree with $\,n\,$ leaves $\,\t\in{\rm FBT}_n\,$ is the graph $\,\widehat\t\,$ obtained from that presentation through attachment of a single vertical edge $\,\{0\}\x[0,1]\,$ to the (embedded) root of $\,\t$,\ {\it cf.}\ Fig.\,\ref{fig:plantLOpresFBT9}.
\exdef
\noindent The link between full binary trees and simplicial target super-spaces employs the pictorial representation of the face maps of a simplicial object given in App.\,\ref{app:proof-simpl} and is established in 
\bedef\label{def:trees-vs-simpltargets}
Let $\,\cC\,$ be a category and let $\,(X_\bullet,d^{(\bullet)}_\cdot,s^{(\bullet)}_\cdot)\,$ be a simplicial object in $\,\cC$.\ Given a full binary tree with $\,n>2\,$ leaves $\,\t\in{\rm FBT}_n\,$ of height $\,H(>1)$,\ the {\bf simplicial descent pattern for} $\,\t\,$ is the following collection $\,\xcD(\t)\equiv\{\xcD(\t)_L\}_{L\in\ovl{0,H-1}}$,\ indexed by the set $\,\ovl{0,H-1}\,$ of root and branch levels,\ of the following sequences of morphisms:\ Let $\,N_L\,$ be the number of nodes of degree 3 at level $\,L\in\ovl{0,H}\,$ in $\,\widehat\t$,\ so that,\ in particular,\ $\,(N_0,N_H)=(1,0)$.\ These numbers determine the numbers $\,\La_L=n-\sum_{l=L+1}^{H-1}\,N_l-2N_L\,$ of free lines intersecting the respective horizontal lines $\,\bR\x\{-L\}$ in (the planting of) a level-ordered presentation of $\,\t$,\ so that,\ in particular,\ $\,\La_0=0$.\ The free lines at level $\,L\,$ originate from the respective leaves with numbers $\,0\leq n^{(L)}_1<n^{(L)}_2<\cdots<n^{(L)}_{\La_L}\leq n$.\ To every node of degree 3 we associate the corresponding superposition $\,d^{(n)}_{k_3,k_4,\ldots,k_n}\equiv d^{(3)}_{k_3}\circ d^{(4)}_{k_4}\circ\cdots\circ d^{(n)}_{k_n}\ :\ X_n\too X_2\,$ of the face maps of the simplicial object (with $\,0\leq k_m\leq m,\ m\in\ovl{3,n}$) through application of the scheme of the recursive ternary resolution laid out in App.\,\ref{app:proof-simpl} ({\it cf.},\ in particular,\ Fig.\,\ref{fig:higherface}).\ Thus,\ there arises,\ at a given level $\,L$,\ a collection of $\,N_L\,$ morphisms $\,X_n\too X_2\,$ that we augment with the collection of morphisms $\,\pi^{(n+1)}_{n^{(L)}_m,n^{(L)}_m+1}\ :\ X_n\too X_1,\ m\in\ovl{1,\La_L}\,$ associated with the free lines at that level.\ The $\,N_L+\La_L\,$ morphisms obtained in this manner are arranged,\ at each level $\,L\in\ovl{0,H-1}\,$,\ into a sequence in which the position of the morphism (from left to right) is determined by the order of appearance of the corresponding intersection as the line $\,\bR\x\{-L\}\,$ traverses the level-ordered presentation of $\,\t\,$ in the direction of the increasing first coordinate.
\exdef

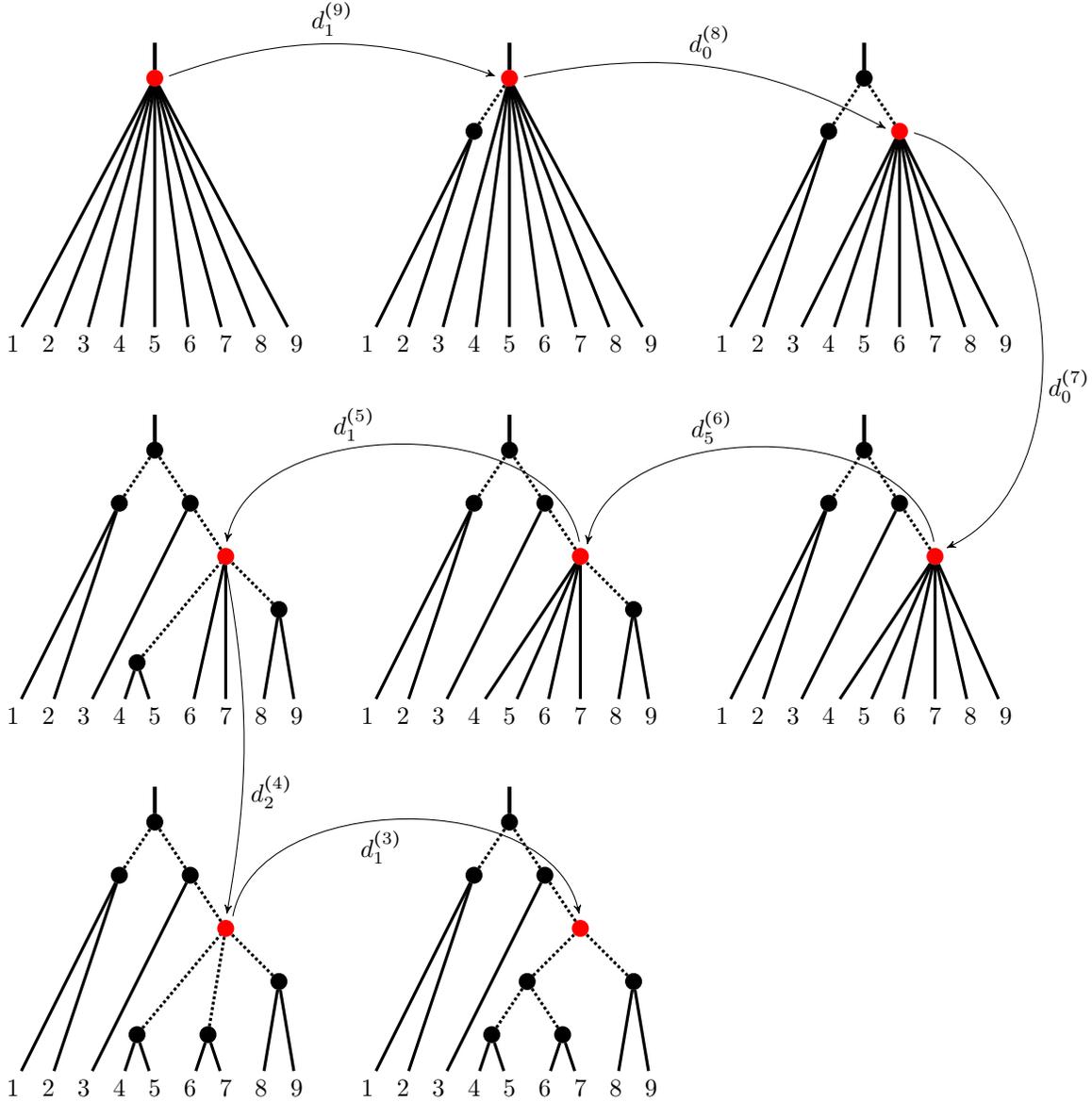
\begin{figure}[h!]
\begin{tikzpicture}
[root/.style={circle,draw=red!100,fill=red!100,thick,inner sep=0pt,minimum size=6pt}, node/.style={circle,draw=black!100,fill=black!100,thick,inner sep=0pt,minimum size=6pt}]

\node[root] (R) at (0,0) {};
\node (L01) at (-2,-3.75) {1};
\node (L02) at (-1.5,-3.75) {2};
\node (L03) at (-1.0,-3.75) {3};
\node (L04) at (-0.5,-3.75) {4};
\node (L05) at (0.0,-3.75) {5};
\node (L06) at (0.5,-3.75) {6};
\node (L07) at (1.0,-3.75) {7};
\node (L08) at (1.5,-3.75) {8};
\node (L09) at (2.0,-3.75) {9};

\draw[-,ultra thick] (R.north) -- (0,0.5);
\draw[-,very thick] (L01) -- (R);
\draw[-,very thick] (L02) -- (R);
\draw[-,very thick] (L03) -- (R);
\draw[-,very thick] (L04) -- (R);
\draw[-,very thick] (L05) -- (R);
\draw[-,very thick] (L06) -- (R);
\draw[-,very thick] (L07) -- (R);
\draw[-,very thick] (L08) -- (R);
\draw[-,very thick] (L09) -- (R);

\node[root] (N101) at (5,0) {};
\node[node] (N111) at (4.5,-0.75) {};
\node (L11) at (3,-3.75) {1};
\node (L12) at (3.5,-3.75) {2};
\node (L13) at (4.0,-3.75) {3};
\node (L14) at (4.5,-3.75) {4};
\node (L15) at (5.0,-3.75) {5};
\node (L16) at (5.5,-3.75) {6};
\node (L17) at (6.0,-3.75) {7};
\node (L18) at (6.5,-3.75) {8};
\node (L19) at (7.0,-3.75) {9};

\draw[-,ultra thick] (N101.north) -- (5,0.5);
\draw[-,very thick] (L11) -- (N111);
\draw[-,very thick] (L12) -- (N111);
\draw[densely dotted,very thick] (N111) -- (N101);
\draw[-,very thick] (L13) -- (N101);
\draw[-,very thick] (L14) -- (N101);
\draw[-,very thick] (L15) -- (N101);
\draw[-,very thick] (L16) -- (N101);
\draw[-,very thick] (L17) -- (N101);
\draw[-,very thick] (L18) -- (N101);
\draw[-,very thick] (L19) -- (N101);

\draw[->,bend left=20,shorten <=2.5pt,shorten >=2.5pt,>=stealth'] (R.east) to node [above] {$d^{(9)}_1$} (N101.west);

\node[node] (N201) at (10,0) {};
\node[node] (N211) at (9.5,-0.75) {};
\node[root] (N212) at (10.5,-0.75) {};
\node (L21) at (8,-3.75) {1};
\node (L22) at (8.5,-3.75) {2};
\node (L23) at (9.0,-3.75) {3};
\node (L24) at (9.5,-3.75) {4};
\node (L25) at (10.0,-3.75) {5};
\node (L26) at (10.5,-3.75) {6};
\node (L27) at (11.0,-3.75) {7};
\node (L28) at (11.5,-3.75) {8};
\node (L29) at (12.0,-3.75) {9};

\draw[-,ultra thick] (N201.north) -- (10,0.5);
\draw[-,very thick] (L21) -- (N211);
\draw[-,very thick] (L22) -- (N211);
\draw[densely dotted,very thick] (N211) -- (N201);
\draw[-,very thick] (L23) -- (N212);
\draw[-,very thick] (L24) -- (N212);
\draw[-,very thick] (L25) -- (N212);
\draw[-,very thick] (L26) -- (N212);
\draw[-,very thick] (L27) -- (N212);
\draw[-,very thick] (L28) -- (N212);
\draw[-,very thick] (L29) -- (N212);
\draw[densely dotted,very thick] (N212) -- (N201);

\draw[->,bend left=20,shorten <=2.5pt,shorten >=2.5pt,>=stealth'] (N101.east) to node [above] {$d^{(8)}_0$} (N212.west);

\node[node] (N301) at (10,-5.25) {};
\node[node] (N311) at (9.5,-6) {};
\node[node] (N312) at (10.5,-6) {};
\node[root] (N321) at (11,-6.75) {};
\node (L31) at (8,-9) {1};
\node (L32) at (8.5,-9) {2};
\node (L33) at (9.0,-9) {3};
\node (L34) at (9.5,-9) {4};
\node (L35) at (10.0,-9) {5};
\node (L36) at (10.5,-9) {6};
\node (L37) at (11.0,-9) {7};
\node (L38) at (11.5,-9) {8};
\node (L39) at (12.0,-9) {9};

\draw[-,ultra thick] (N301.north) -- (10,-4.75);
\draw[-,very thick] (L31) -- (N311);
\draw[-,very thick] (L32) -- (N311);
\draw[densely dotted,very thick] (N311) -- (N301);
\draw[-,very thick] (L33) -- (N312);
\draw[-,very thick] (L34) -- (N321);
\draw[-,very thick] (L35) -- (N321);
\draw[-,very thick] (L36) -- (N321);
\draw[-,very thick] (L37) -- (N321);
\draw[-,very thick] (L38) -- (N321);
\draw[-,very thick] (L39) -- (N321);
\draw[densely dotted,very thick] (N312) -- (N301);
\draw[densely dotted,very thick] (N321) -- (N312);

\draw[->,bend left=70,shorten <=2.5pt,shorten >=2.5pt,>=stealth'] (N212.east) to node [pos=0.6,right] {$d^{(7)}_0$} (N321.north east);

\node[node] (N401) at (5,-5.25) {};
\node[node] (N411) at (4.5,-6) {};
\node[node] (N412) at (5.5,-6) {};
\node[root] (N421) at (6,-6.75) {};
\node[node] (N432) at (6.75,-7.5) {};
\node (L41) at (3,-9) {1};
\node (L42) at (3.5,-9) {2};
\node (L43) at (4.0,-9) {3};
\node (L44) at (4.5,-9) {4};
\node (L45) at (5.0,-9) {5};
\node (L46) at (5.5,-9) {6};
\node (L47) at (6.0,-9) {7};
\node (L48) at (6.5,-9) {8};
\node (L49) at (7.0,-9) {9};

\draw[-,ultra thick] (N401.north) -- (5,-4.75);
\draw[-,very thick] (L41) -- (N411);
\draw[-,very thick] (L42) -- (N411);
\draw[densely dotted,very thick] (N411) -- (N401);
\draw[-,very thick] (L43) -- (N412);
\draw[-,very thick] (L44) -- (N421);
\draw[-,very thick] (L45) -- (N421);
\draw[-,very thick] (L46) -- (N421);
\draw[-,very thick] (L47) -- (N421);
\draw[-,very thick] (L48) -- (N432);
\draw[-,very thick] (L49) -- (N432);
\draw[densely dotted,very thick] (N412) -- (N401);
\draw[densely dotted,very thick] (N421) -- (N412);
\draw[densely dotted,very thick] (N432) -- (N421);

\draw[->,bend right=80,shorten <=2.5pt,shorten >=2.5pt,>=stealth'] (N321.north) to node [pos=0.6,above] {$d^{(6)}_5$} (N421.north east);

\node[node] (N501) at (0,-5.25) {};
\node[node] (N511) at (-0.5,-6) {};
\node[node] (N512) at (0.5,-6) {};
\node[root] (N521) at (1,-6.75) {};
\node[node] (N532) at (1.75,-7.5) {};
\node[node] (N531) at (-0.25,-8.25) {};
\node (L51) at (-2,-9) {1};
\node (L52) at (-1.5,-9) {2};
\node (L53) at (-1.0,-9) {3};
\node (L54) at (-0.5,-9) {4};
\node (L55) at (0.0,-9) {5};
\node (L56) at (0.5,-9) {6};
\node (L57) at (1.0,-9) {7};
\node (L58) at (1.5,-9) {8};
\node (L59) at (2.0,-9) {9};

\draw[-,ultra thick] (N501.north) -- (0,-4.75);
\draw[-,very thick] (L51) -- (N511);
\draw[-,very thick] (L52) -- (N511);
\draw[densely dotted,very thick] (N511) -- (N501);
\draw[-,very thick] (L53) -- (N512);
\draw[-,very thick] (L54) -- (N531);
\draw[-,very thick] (L55) -- (N531);
\draw[-,very thick] (L56) -- (N521);
\draw[-,very thick] (L57) -- (N521);
\draw[-,very thick] (L58) -- (N532);
\draw[-,very thick] (L59) -- (N532);
\draw[densely dotted,very thick] (N512) -- (N501);
\draw[densely dotted,very thick] (N521) -- (N512);
\draw[densely dotted,very thick] (N532) -- (N521);
\draw[densely dotted,very thick] (N531) -- (N521);

\draw[->,bend right=80,shorten <=2.5pt,shorten >=2.5pt,>=stealth'] (N421.north) to node [pos=0.6,above] {$d^{(5)}_1$} (N521.north);

\node[node] (N601) at (0,-10.5) {};
\node[node] (N611) at (-0.5,-11.25) {};
\node[node] (N612) at (0.5,-11.25) {};
\node[root] (N621) at (1,-12) {};
\node[node] (N632) at (1.75,-12.75) {};
\node[node] (N631) at (-0.25,-13.5) {};
\node[node] (N642) at (0.75,-13.5) {};
\node (L61) at (-2,-14.25) {1};
\node (L62) at (-1.5,-14.25) {2};
\node (L63) at (-1.0,-14.25) {3};
\node (L64) at (-0.5,-14.25) {4};
\node (L65) at (0.0,-14.25) {5};
\node (L66) at (0.5,-14.25) {6};
\node (L67) at (1.0,-14.25) {7};
\node (L68) at (1.5,-14.25) {8};
\node (L69) at (2.0,-14.25) {9};

\draw[-,ultra thick] (N601.north) -- (0,-10);
\draw[-,very thick] (L61) -- (N611);
\draw[-,very thick] (L62) -- (N611);
\draw[densely dotted,very thick] (N611) -- (N601);
\draw[-,very thick] (L63) -- (N612);
\draw[-,very thick] (L64) -- (N631);
\draw[-,very thick] (L65) -- (N631);
\draw[-,very thick] (L66) -- (N642);
\draw[-,very thick] (L67) -- (N642);
\draw[-,very thick] (L68) -- (N632);
\draw[-,very thick] (L69) -- (N632);
\draw[densely dotted,very thick] (N612) -- (N601);
\draw[densely dotted,very thick] (N621) -- (N612);
\draw[densely dotted,very thick] (N632) -- (N621);
\draw[densely dotted,very thick] (N631) -- (N621);
\draw[densely dotted,very thick] (N642) -- (N621);

\draw[->,bend left=10,shorten <=2.5pt,shorten >=2.5pt,>=stealth'] (N521.south) to node [pos=0.65,right] {$d^{(4)}_2$} (N621.north);

\node[node] (N701) at (5,-10.5) {};
\node[node] (N711) at (4.5,-11.25) {};
\node[node] (N712) at (5.5,-11.25) {};
\node[root] (N721) at (6,-12) {};
\node[node] (N732) at (6.75,-12.75) {};
\node[node] (N731) at (5.25,-12.75) {};
\node[node] (N741) at (4.75,-13.5) {};
\node[node] (N742) at (5.75,-13.5) {};
\node (L71) at (3,-14.25) {1};
\node (L72) at (3.5,-14.25) {2};
\node (L73) at (4.0,-14.25) {3};
\node (L74) at (4.5,-14.25) {4};
\node (L75) at (5.0,-14.25) {5};
\node (L76) at (5.5,-14.25) {6};
\node (L77) at (6.0,-14.25) {7};
\node (L78) at (6.5,-14.25) {8};
\node (L79) at (7.0,-14.25) {9};

\draw[-,ultra thick] (N701.north) -- (5,-10);
\draw[-,very thick] (L71) -- (N711);
\draw[-,very thick] (L72) -- (N711);
\draw[densely dotted,very thick] (N711) -- (N701);
\draw[-,very thick] (L73) -- (N712);
\draw[-,very thick] (L74) -- (N741);
\draw[-,very thick] (L75) -- (N741);
\draw[-,very thick] (L76) -- (N742);
\draw[-,very thick] (L77) -- (N742);
\draw[-,very thick] (L78) -- (N732);
\draw[-,very thick] (L79) -- (N732);
\draw[densely dotted,very thick] (N712) -- (N701);
\draw[densely dotted,very thick] (N721) -- (N712);
\draw[densely dotted,very thick] (N732) -- (N721);
\draw[densely dotted,very thick] (N731) -- (N721);
\draw[densely dotted,very thick] (N741) -- (N731);
\draw[densely dotted,very thick] (N742) -- (N731);

\draw[->,bend left=79,shorten <=2.5pt,shorten >=2.5pt,>=stealth'] (N621.north east) to node [pos=0.45,below] {$d^{(3)}_1$} (N721.north);
\end{tikzpicture}
\caption{A recursive ternary resolution that extracts the map $\,X_9\too X_2\,$ associated to the node $\,N^{(2)}_1$.\ Nodes of intermediate valences obtained at the consecutive stages of the cascade of the face maps are marked in red.} \label{fig:plantLOpresFBT9-RTR}
\end{figure}

\brem
We illustrate the contents of the definition by writing out the simplicial descent pattern for the full binary tree $\,\t_*\,$ depicted in Fig.\,\ref{fig:plantLOpresFBT9}.\ Consider,\ {\it e.g.},\ the node $\,N^{(2)}_1\,$ in Fig.\,\ref{fig:plantLOpresFBT9}.\ The recursive ternary resolution of the relevant graph,\ worked out in detail in Fig.\,\ref{fig:plantLOpresFBT9-RTR},\ 
yields
\qq\nn
N^{(2)}_1\qquad :\qquad d^{(3)}_1\circ d^{(4)}_2\circ d^{(5)}_1\circ d^{(6)}_5\circ d^{(7)}_0\circ d^{(8)}_0\circ d^{(9)}_1\equiv d^{(9)}_{1,2,1,5,0,0,1}\,.
\qqq
Analogously,\ we derive
\qq\nn
N^{(0)}_1\qquad &:&\qquad d^{(3)}_2\circ d^{(4)}_3\circ d^{(5)}_3\circ d^{(6)}_3\circ d^{(7)}_5\circ d^{(8)}_7\circ d^{(9)}_1\equiv d^{(9)}_{2,3,3,3,5,7,1}\,,\cr\cr
N^{(1)}_1\qquad &:&\qquad d^{(3)}_3\circ d^{(4)}_3\circ d^{(5)}_4\circ d^{(6)}_4\circ d^{(7)}_4\circ d^{(8)}_6\circ d^{(9)}_8\equiv d^{(9)}_{3,3,4,4,4,6,8}\,,\cr\cr
N^{(1)}_2\qquad &:&\qquad d^{(3)}_2\circ d^{(4)}_2\circ d^{(5)}_3\circ d^{(6)}_2\circ d^{(7)}_6\circ d^{(8)}_0\circ d^{(9)}_1\equiv d^{(9)}_{2,2,3,2,6,0,1}\,,\cr\cr
N^{(3)}_1\qquad &:&\qquad d^{(3)}_2\circ d^{(4)}_4\circ d^{(5)}_1\circ d^{(6)}_0\circ d^{(7)}_6\circ d^{(8)}_0\circ d^{(9)}_1\equiv d^{(9)}_{2,4,1,0,6,0,1}\,,\cr\cr
N^{(3)}_2\qquad &:&\qquad d^{(3)}_0\circ d^{(4)}_1\circ d^{(5)}_2\circ d^{(6)}_1\circ d^{(7)}_0\circ d^{(8)}_0\circ d^{(9)}_1\equiv d^{(9)}_{0,1,2,1,0,0,1}\,,\cr\cr
N^{(4)}_1\qquad &:&\qquad d^{(3)}_0\circ d^{(4)}_4\circ d^{(5)}_3\circ d^{(6)}_0\circ d^{(7)}_0\circ d^{(8)}_7\circ d^{(9)}_1\equiv d^{(9)}_{0,4,3,0,0,7,1}\,,\cr\cr
N^{(4)}_2\qquad &:&\qquad d^{(3)}_0\circ d^{(4)}_4\circ d^{(5)}_1\circ d^{(6)}_0\circ d^{(7)}_0\circ d^{(8)}_7\circ d^{(9)}_1\equiv d^{(9)}_{0,4,1,0,0,7,1}\,.
\qqq
Taking into account the positioning and labels on the free lines of the planting of Fig.\,\ref{fig:plantLOpresFBT9},\ we ultimately arrive at the complete simplicial descent pattern
\qq\nn
\xcD(\t_*)&=&\bigl\{\ \bigl(d^{(9)}_{2,3,3,3,5,7,1}\bigr)\equiv\xcD(\t_*)_0,\bigl(d^{(9)}_{3,3,4,4,4,6,8},d^{(9)}_{2,2,3,2,6,0,1}\bigr)\equiv\xcD(\t_*)_1,\cr\cr
&&\hspace{-0.4cm}\bigl(\pi^{(10)}_{1,2},\pi^{(10)}_{2,3},\pi^{(10)}_{3,4},d^{(9)}_{1,2,1,5,0,0,1}\bigr)\equiv\xcD(\t_*)_2,\bigl(\pi^{(10)}_{1,2},\pi^{(10)}_{2,3},\pi^{(10)}_{3,4},d^{(9)}_{2,4,1,0,6,0,1},d^{(9)}_{0,1,2,1,0,0,1}\bigr)\equiv\xcD(\t_*)_3,\cr\cr
&&\hspace{0.3cm}\bigl(\pi^{(10)}_{1,2},\pi^{(10)}_{2,3},\pi^{(10)}_{3,4},d^{(9)}_{0,4,3,0,0,7,1},d^{(9)}_{0,4,1,0,0,7,1},\pi^{(10)}_{8,9},\pi^{(10)}_{9,10}\bigr)\equiv\xcD(\t_*)_4\ \bigr\}\,.
\qqq
\erem
\noindent The last two definitions enable us to finally formulate
\bedef\label{simplssbckgrndcompl}
A {\bf simplicial superstring background} is a superstring background with a simplicial target super-space,\ as described in Def.\,\ref{def:simpl-target-sspace}.\ We call it {\bf descent-complete} if it has the following property:\ For every $\,n>2\,$ and every full binary tree $\,\t\in{\rm FBT}_n$,\ the component inter-bi-brane super-worldvolume $\,T_{n,1}\,$ contains the common intersection $\,T_{n,1}^{\rm int}\,$ of the preimages of the component inter-bi-brane super-worldvolume $\,T_{2,1}\,$ under the maps $\,d^{(n)}_{k_3,k_4,\ldots,k_n}\ :\ X_n\too X_2\,$ of the simplicial descent pattern $\,\xcD(\t)\,$  for $\,\t\,$ composed of the face maps of the underlying simplicial super-space as in Def.\,\ref{def:trees-vs-simpltargets} and the preimages of the bi-brane super-worldvolume $\,Q\,$ under the maps $\,\pi^{(n+1)}_{n^{(L)}_m,n^{(L)}_m+1}\,$ of the same simplicial descent pattern,\ so that the trees determine,\ over the intersection,\ the corresponding component fusion 2-isomorphisms $\,\varphi_{n+1}[\t]\,$ in terms of the elementary fusion 2-isomorphism $\,\varphi_3\,$ and the 1-gerbe 2-isomorphism 
\qq\nn
\alxydim{@C=10em@R=6em}{\bigl(\iota_2\circ\pi_{1,2}\bigr)^*\cG\ox\cI_{\pi_{1,2}^*\om}
\ar[r]^{\pi_{2,3}^*\Phi\ox\id_{\cI_{\pi_{1,2}^*\om}}\qquad\qquad} \ar@{=>}[dr]|{\widetilde\varphi{}_3}
& \bigl(\iota_2\circ\pi_{2,3}\bigr)^*\cG\ox\cI_{\pi_{1,3}^*\om}\ox\cI_{\D_{T_{2,1}}\om} \\
\bigl(\iota_1\circ\pi_{1,2}\bigr)^*\cG \ar[u]^{\pi_{1,2}^*\Phi}
\ar@{=}[r]|{\ \id_{(\iota_1\circ\pi_{1,2})^*\cG}\ } 
& \bigl(\iota_1\circ\pi_{1,3}\bigr)^*\cG\ox\cI_{\D_{T_{2,1}}\om} \ar[u]_{\pi_{1,3}^*\Phi\ox\id_{\cI_{\D_{T_{2,1}}\om}}} }
\qqq
canonically induced by it as {\it per}
\qq\nn
\varphi_{n+1}[\t]\rstr_{T_{n,1}^{\rm int}}=\xcD(\t)_0(\varphi_3)\bullet\xcD(\t)_1(\widetilde\varphi{}_3)\bullet\xcD(\t)_2(\widetilde\varphi{}_3)\bullet\cdots\bullet\xcD(\t)_{H-1}(\widetilde\varphi{}_3)\,,
\qqq
where
\bit
\item $\,\xcD(\t)_0(\varphi_3)\,$ is the pullback of $\,\varphi_3\,$ by the unique element of $\,\xcD(\t)_0$;
\item $\,\xcD(\t)_L(\widetilde\varphi{}_3),\ L\in\ovl{1,H-1}\,$ is the horizontal composition of the identity 2-isomorphisms on the pullbacks of $\,\Phi\,$ along the maps $\,T_{n,1}\too Q\,$ belonging to $\,\xcD(\t)_L\,$ (the pullbacks being corrected,\ whenever necessary,\ by appropriate identity 1-isomorphisms on trivial 1-gerbes coming from the higher-level bi-modules) with the pullbacks of $\,\widetilde\varphi{}_3\,$ along the maps $\,T_{n,1}\too T_{2,1}\,$ belonging to $\,\xcD(\t)_k$,\ the horizontal compositions being performed in the order of the appearance of the relevant maps of the two types in the sequence $\,\xcD(\t)_L$.
\eit
\exdef
\noindent We shall spend some time discussing an important example of a superstring background of the type introduced above in later sections.\ Meanwhile,\ we recall the (super)field-theoretic r\^ole of the higher-geometric structure and,\ in the next section,\ incorporate target super-space (super)symmetry in the higher-geometric picture.\ We begin with the long-expected
\bedef\label{def:ssimod}
The {\bf two-dimensional Green--Schwarz}({\bf -type}) {\bf super-$\si$-model} (or the {\bf GS super-$\si$-model} for brevity) with background $\,\Bgt\,$ as in Def.\,\ref{def:str-bgrnd} on a worldsheet with defects $\,(\Si,d)\,$ as in Def.\,\ref{def:worldsheet} is a theory of superfield configurations $\,\xi\,$ from $\,\cF(\Si,d)\,$ determined by the principle of least action for the Dirac--Feynman amplitude with the (action) `functional'
\qq\label{eq:ssimod}
S_\si[(\xi\,\vert\,\G)]=\int_\Si\,\sqrt{\det\,\bigl(\xi^*\txg\bigr)}+S_{\rm WZ}[(\xi\,\vert\,\G)]\,,
\qqq
where the so-called `topological',\ or Wess--Zumino term
\qq\nn
S_{\rm WZ}[(\xi\,\vert\,\G)]=-\sfi\,\log\Hol_{\cG,\Phi,(\varphi_\cdot)}(\xi\,\vert\,\G)
\qqq
is given by the {\bf decorated-surface holonomy} $\,\Hol_{\cG,\Phi,(\varphi_\cdot)}(\xi\,\vert\,\G)\,$ for the superfield configuration $\,\xi$,\ derived in the un-graded setting in \Rxcite{Sect.\,2}{Runkel:2008gr} and transcribed {\it verbatim} into the $\bZ/2\bZ$-graded setting.
\exdef

The r\^ole of the gerbe-theoretic structures in the modelling of the dynamics of linear distributions of energy and topological charge (resp.\ super-charge) does not end with the rigorous definition of the DF amplitude.\ As was demonstrated in detail in the un-graded mono-phase setting by Gaw\c{e}dzki in the seminal paper \cite{Gawedzki:1987ak},\ and subsequently worked out for the un-graded multi-phase setting by the Author in Refs.\,\cite{Suszek:2011hg,Suszek:2012ddg} ({\it cf.}\ also Refs.\,\cite{Suszek:2011,Suszek:2013}),\ the 1-gerbe associated with a single `phase' of the two-dimensional field theory canonically defines,\ through the so-called transgression,\ a pre-quantum bundle of the theory in that phase,\ whereas a bi-brane attached to a defect line between phases gives rise to a correspondence\footnote{The interpretation of the inter-bi-brane in this paradigm is a little more involved,\ and so shall be omitted here.\ The interested Reader is referred to the original papers \cite{Suszek:2011hg,Suszek:2012ddg}.} between global sections of the mono-phase pre-quantum bundles (to be thought of as quantum state spaces upon choosing a polarisation) which,\ under certain circumstances (discussed at length in \Rcite{Suszek:2011}),\ becomes an equivalence (or {\it duality}) of the quantised mono-phase theories (aka a {\it quantomorphism}).\ This coherent quantum-mechanical interpretation of the higher geometry behind the multi-phase $\si$-model (which is further enhanced in the presence of global (target-space) symmetries and,\ in particular,\ in the context of their gauging) provides the fundamental motivation for its further study in the un-graded setting and the discussion of its analogon in the $\bZ/2\bZ$-graded environment.\ In either case,\ it is customary to assume the presence of the action of a supersymmetry Lie supergroup,\ and to demand invariance of the ensuing super-$\si$-model (or,\ more concretely,\ of its DF amplitude) under the induced action on the superfields.\ Here,\ supersymmetry features as a global (target super-space) symmetry of the superfield theory,\ and so its consistent incorporation in the higher-geometric framework calls for a discussion of a gerbe-theoretic rendering of global symmetries in general.\ It also raises the question of compatibility of the defect with the global symmetries of the bulk phases separated by it.\ These are the issues that we review in the next section.

\section{Enter global (super)symmetry}\label{sec:susygrb}

Global symmetries of the charged geometrodynamics of the (super-)$\si$-model are induced,\ in a natural manner,\ from those automorphisms of the target super-space which preserve the bulk metric tensor and lift to isomorphisms of the gerbe-theoretic structure.\ In the presence of a nontrivial Gra\ss mann-odd component of the structure sheaf of the target super-space,\ used to model fermionic degrees of freedom ({\it e.g.},\ in the effective field theory of excitations of the embedded worldsheet),\ it is customary to assume that the symmetries form a Lie supergroup with a nontrivial Gra\ss mann-odd component of the tangent Lie superalgebra that generates supersymmetry transformations mapping into one another components of the superfield of different statistics.\ Accordingly,\ in what follows,\ we assume given a Lie supergroup $\,\txG_{\rm s}=(|\txG_{\rm s}|,\ggt_{\rm s})$,\ to be thought of as a group object in $\,\sMan\,$ or,\ equivalently,\ after Kostant\footnote{{\it Cf.}\ also \Rcite{Koszul:1982} and,\ in particular,\ \Rcite{Carmeli:2011} for a very lucid introduction into the formalism.} \cite{Kostant:1975},\ as a super-Harish--Chandra pair composed of a (body) Lie group $\,|\txG_{\rm s}|\,$ and a Lie superalgebra $\,\ggt_{\rm s}=\ggt_{\rm s}^{(0)}\oplus\ggt_{\rm s}^{(1)}\,$ with the Gra\ss mann-even Lie subalgebra given by the tangent Lie algebra of the body Lie group,\ $\,\ggt_{\rm s}^{(0)}\equiv{\rm Lie}\,|\txG_{\rm s}|$,\ and the latter group realised on $\,\ggt_{\rm s}\,$ through an extension of the standard adjoint action of $\,|\txG_{\rm s}|\,$ on $\,{\rm Lie}\,|\txG_{\rm s}|$,\ together with a {\bf coherent} action 
\qq\nn
\xcT\la\equiv T\la\sqcup Q\la\sqcup M\la\equiv\bigsqcup_{n=2}^\infty\,T_{n,1}\la\sqcup Q\la\sqcup M\la\ :\ \txG_{\rm s}\x\xcT\too\xcT\,,
\qqq
by which we mean a (left) action in the sense of \Rxcite{Def.\,8.1.2}{Carmeli:2011} intertwined by the structure maps of the background,\ {\it i.e.},\ such that 
\qq
&\iota_A\circ Q\la=M\la\circ(\id_{\txG_{\rm s}}\x\iota_A)\,,\qquad A\in\{1,2\}\,,&\label{eq:bibinter}\\ \cr\cr
&\left\{ \barr{l} \pi^{(n+1)}_{k,k+1}\circ T_{n,1}\la=Q\la\circ\bigl(\id_{\txG_{\rm s}}\x\pi^{(n+1)}_{k,k+1}\bigr)\,,\qquad k\in\ovl{1,n} \\ \\
\pi^{(n+1)}_{1,n+1}\circ T_{n,1}\la=Q\la\circ\bigl(\id_{\txG_{\rm s}}\x\pi^{(n+1)}_{1,n+1}\bigr) \earr\right.\,,\qquad n\in\bN_{\geq 2}\,.&\label{eq:ibbinter}
\qqq

Next,\ we consider invariance of the superstring background under $\,\xcT\la$.\ As we intend to give a description of rather special backgrounds,\ namely,\ those (maximally) compatible with the supersymmetry present (in the form of the above coherent action),\ we shall impose the constraints of invariance \emph{uniformly} on \emph{all} components of the background \emph{at the same time},\ rather than distinguishing the bulk component and then considering various reduction schemes for the bulk supersymmetry group.\ The absence of points (in the standard sense of the word) in a non-even supermanifold makes the issue of invariance slightly more involved than in the un-graded setting and leaves us with two convenient formulations that find a practical application in what follows.\ Below,\ we detail the first one,\ which is closest to the standard notion known from the un-graded setting.\ The discussion of the other one,\ making ample use of the notion of an $\cS$-point in a supermanifold,\ is postponed to Sec.\,\ref{sec:GSWZW} in which circumstances arise that make its application natural and,\ indeed,\ convenient.\ Meanwhile,\ we restrict our thinking of `points' in the relevant supermanifolds $\,\txG_{\rm s}\,$ and (those composing) $\,\xcT\,$ to the respective bodies and,\ accordingly,\ split the description of the property of interest into a global one for the body (which may have an arbitrary topology) with the help of the induced $|\txG_{\rm s}|$-action: 
\qq\nn
|\xcT\la|_g\equiv\xcT\la\circ\bigl(\widehat g\x\id_\xcT\bigr)\ :\ \xcT\equiv\bR^{0|0}\x\xcT\too\txG_{\rm s}\x\xcT\too\xcT\,,\qquad g\in|\txG_{\rm s}|\,,
\qqq
using the topological points $\,\{\widehat g\ :\ \bR^{0|0}\too\txG_{\rm s}\}_{g\in|\txG_{\rm s}|}\equiv|\txG_{\rm s}|$,\ and an infinitesimal one for the contractible soul with the help of the Lie derivative along fundamental supervector fields induced by $\,\cT\la\,$ and associated with vectors in $\,\ggt_{\rm s}$.\ As for the latter,\ it proves convenient\footnote{This mode of description becomes indispensable in the analysis of \emph{gauge} symmetries,\ {\it cf.}\ \Rxcite{Sec.\,7}{Suszek:2020xcu}.} -- as shall become clear later on -- to work with Gra\ss mann-even supervector fields only,\ and so we present the said vector fields as sections $\,\cK_\cdot\equiv\cK_\cdot^M\sqcup\cK_\cdot^Q\sqcup\bigsqcup_{n=2}^\infty\,\cK_\cdot^{T_{n,1}}\in\G(\cO_{\ggt_{\rm s}}\ox\cT\xcT)\equiv\G(\cO_{\ggt_{\rm s}}\ox{\rm sDer}(\cO_\xcT))\,$ of a subsheaf $\,\cO_{\ggt_{\rm s}}\ox\cT\xcT\,$ of the tangent sheaf $\,\cT(\ggt_{\rm s}\x\xcT)\equiv{\rm sDer}(\cO_{\ggt_{\rm s}}\ox\cO_\xcT)\,$ over the super-\emph{space} $\,\ggt_{\rm s}\x\xcT\,$ that are linear in the global generators $\,\{X^a\}^{a\in\ovl{1,D}},\ D\equiv\dim\,\ggt_{\rm s}\,$ of the structure sheaf $\,\cO_{\ggt_{\rm s}}$ (supercoordinates) on $\,\ggt_{\rm s}\,$ associated with a Gra\ss mann-homogeneous basis $\,\{t_a\}_{a\in\ovl{1,D}}\,$ of the Lie superalgebra $\,\ggt_{\rm s}$,\ of the respective Gra\ss mann parities $\,|X^a|=|t_a|$.\ Thus,\ we consider vector fields on $\,\ggt_{\rm s}\x\xcT\,$ from the integrable superdistribution $\,\cT\xcT\,$ within $\,\cT(\ggt_{\rm s}\x\xcT)\,$ ({\it i.e.},`along' the second cartesian factor),\ of the form $\,\cK_X\equiv X^a\,\cK_{t_a}\equiv X^a\,\cK_a$,\ expressed in terms of the `standard' fundamental vector fields\footnote{We put an extra minus sign in the definition in order to ensure that $\,t_a\longmapsto\cK_a\,$ extends to a Lie-superalgebra morphism rather than \emph{anti}-morphism.} $\,\cK_a=-(t_a\ox\id_{\cO_\xcT})\circ\xcT\la^*\in{\rm sDer}(\cO_\xcT)$,\ {\it cf.}\ \Rxcite{Def.\,8.2.1}{Carmeli:2011}.\ In what follows,\ we use the shorthand notation $\,X\in\ggt_{\rm s}$.\ When taking Lie derivatives of (local) covariant tensor fields on $\,\xcT$,\ viewed as a subclass of tensor fields on the product super-space $\,\ggt_{\rm s}\x\xcT$,\ in the setting of \emph{global} supersymmetry,\ the resulting (local) covariant tensor fields are to be evaluated on vector fields from the aforementioned superdistribution $\,\cT\xcT\,$ exclusively,\ which we shall not indicate explicitly for the sake of transparency of the notation.\ This may be understood formally as restricting to its integral leaves within $\,\ggt_{\rm s}\x\xcT$,\ and practically boils down to treating the supercoordinate sections $\,X^a\,$ as `constant'\footnote{This is to be contrasted with the standard treatment of gauge (super)symmetry,\ {\it e.g.},\ in the construction of equivariant structures on the higher-geometric objects.}.\ After these preparations,\ necessitated by peculiarities of supergeometry,\ we finally come to the conditions of $\txG_{\rm s}$-invariance.\ Thus,\ a covariant tensor $\,T\,$ on $\,\xcT\,$ is called {\bf $\txG_{\rm s}$-invariant} if it satisfies the following conditions
\qq\nn
\left\{ \barr{c}
|\xcT\la|_g^*T=T\,,\qquad g\in|\txG_{\rm s}|\cr\cr
\pLie{\cK_X}\Om=0\,,\qquad X\in\ggt_{\rm s} \earr\right.\,.
\qqq

In the superfield-theoretic setting in hand,\ the above conditions are imposed on the tensorial components of the background $\,\Bgt$.\ Thus,\ we presuppose that the metric tensor on the bulk target super-space and both curvatures:\ that of the bulk 1-gerbe and that of the attendant super-bi-brane,\ composing the simplicial de Rham 3-cocycle $\,\Om\,$ of \Reqref{eq:simpl3coc},\ satisfy
\qq\label{eq:bckgrndinv}
\left\{ \barr{c}
\bigl(|M\la|_g^*\txg,|\xcT\la|_g^*\Om\bigr)=(\txg,\Om)\,,\qquad g\in|\txG_{\rm s}|\cr\cr\bigl(\pLie{\cK^M_X}\txg,\pLie{\cK_X}\Om\bigr)=(0,0)\,,\qquad X\in\ggt_{\rm s} \earr\right.\,,
\qqq
where -- just to reemphasise -- the bottom identities are understood to obtain upon restriction to (the integral foliation of) the superdistribution $\,\cT\xcT\,$ in $\,\cT(\ggt_{\rm s}\x\xcT)$.\ In this way,\ we arrive at
\bedef
Let $\,\txG_{\rm s}\,$ be a Lie supergroup,\ as described above.\ A target super-space $\,\xcT\,$ for a multi-phase super-$\si$-model with defects,\ understood as in Def.\,\ref{def:targetss} ,\ is called a {\bf target $\txG_{\rm s}$-super-space for a multi-phase super-$\si$-model with defects} (or a {\bf target $\txG_{\rm s}$-super-space} for short) if it admits a coherent action $\,\xcT\la\,$ of $\,\txG_{\rm s}\,$ which satisfies the equivariance conditions \eqref{eq:bibinter} and \eqref{eq:ibbinter},\ and if the tensors $\,(\txg,\Om)\,$ are $\txG_{\rm s}$-invariant in the sense of \Reqref{eq:bckgrndinv}.
\exdef
\noindent An obvious specialisation of the last definition of interest to us is given in
\bedef\label{def:simpltarGsspace}
Let $\,\txG_{\rm s}\,$ be a Lie supergroup,\ as described above.\ A {\bf simplicial target $\txG_{\rm s}$-super-space for a multi-phase super-$\si$-model with defects} (or a {\bf simplicial target $\txG_{\rm s}$-super-space} for short) is a simplicial target super-space,\ as introduced in Def.\,\ref{def:simpl-target-sspace},\ whose underlying simplicial super-space is a simplicial object in the category of $\txG_{\rm s}$-super-spaces,\ so that,\ in particular,\ its face maps and degeneracy maps are $\txG_{\rm s}$-equivariant,\ and if the tensors $\,(\txg,\Om)\,$ are $\txG_{\rm s}$-invariant in the sense of  \Reqref{eq:bckgrndinv}.
\exdef

The assumed coherence of the action of the supersymmetry Lie supergroup $\,\txG_{\rm s}\,$ on the simplicial target super-space enables us to consistently restrict the associated simplicial de Rham complex of Def.\,\ref{def:relcohom} to $\txG_{\rm s}$-invariant forms,\ whereupon the {\bf $\txG_{\rm s}$-invariant simplicial de Rham complex} arises.\ The de Rham 3-cocycle $\,\Om\,$ from the above definitions is a cochain from this subcomplex.\ As was recalled in Secs.\,\ref{sub:rudi-gerbe} and \ref{sub:grbmultissi},\ the 1-gerbe,\ the bi-brane and the inter-bi-brane of the super-$\si$-model provide a geometrisation of the class $\,[\Om]\,$ in the (integral) simplicial de Rham cohomology,\ and it is now natural to look for a refinement of their construction based on a class in the {\bf $\txG_{\rm s}$-invariant simplicial de Rham cohomology}.\ Unlike in the case of compact Lie groups,\ the refinement is not trivial {\it a priori} as there is no counterpart of the Cartan--Eilenberg theorem in the $\bZ/2\bZ$-graded setting.

Let us,\ consequently,\ lift $\txG_{\rm s}$-invariance to the higher-geometric components of the background.\ For that,\ we need,\ first,\ to introduce another piece of notation:\ Given,\ for some $\,k\in\{0,1,2\}$,\ a $k$-cell $\,\Psi\,$ in $\,\bgrb^\nabla(\xcT)\,$ that geometrises (the class of) a $(2-k)$-cochain $\,\psi\,$ in the Deligne hypercohomology of $\,\xcT$,\ we denote by $\,\pLie{\cK_X}\Psi\,$ the $k$-cell in the (weak) 2-category of abelian bundle gerbes with connection over the integral foliation of $\,\cT\xcT\,$ in $\,\ggt_{\rm s}\x\xcT\,$ that provides a geometric realisation of the $(2-k)$-cochain $\,\pLie{\cK_X}\psi\,$ in the Deligne hypercohomology of that foliation.\ It is now straightforward to spell out the $\txG_{\rm s}$-invariance conditions.\ Thus,\ we call the bulk 1-gerbe $\,\cG\,$ {\bf supersymmetric},\ or {\bf $\txG_{\rm s}$-invariant},\ if there exist two families of 1-gerbe 1-isomorphisms:
\qq\nn
\La_g\ :\ |M\la|_g^*\cG\xrightarrow{\ \cong\ }\cG\,,\qquad g\in|\txG_{\rm s}|\qquad\quad{\rm and}\qquad\quad K_X\ :\ \pLie{\cK^M_X}\cG\xrightarrow{\ \cong\ }\cI_0\,,\qquad X\in\ggt_{\rm s}\,.
\qqq
Given such a 1-gerbe $\,\cG$,\ we call the $\cG$-bi-brane $\,\cB\,$ {\bf maximally supersymmetric} if (condition \eqref{eq:bibinter} is satisfied and) there exist the corresponding families of 1-gerbe 2-isomorphisms:
\qq\nn
\alxydim{@C=7em@R=6em}{ |Q\la|_g^*\iota_1^*\cG\equiv\iota_1^*|M\la|_g^*\cG
\ar[r]^{|Q\la|_g^*\Phi\hspace{4em}} \ar[d]_{\iota_1^*\La_g} & |Q\la|_g^*\iota_2^*\cG\ox|Q\la|_g^*\cI_\om\equiv\iota_2^*|M\la|_g^*\cG\ox\cI_{|Q\la|_g^*\om} \ar[d]^{\iota_2^*\La_g\ox\id_{\cI_\om}} \ar@{=>}[dl]|{\la_g} \\
\iota_1^*\cG \ar[r]_{\Phi} & \iota_2^*\cG\ox\cI_\om }
\qqq 
and
\qq\nn
\alxydim{@C=7em@R=6em}{ \pLie{\cK^Q_X}\iota_1^*\cG\equiv\iota_1^*\pLie{\cK^M_X}\cG
\ar[r]^{\pLie{\cK^Q_X}\Phi\hspace{2em}} \ar[d]_{\iota_1^*K_X} & \pLie{\cK^Q_X}\bigl(\iota_2^*\cG\ox\cI_\om\bigr)\equiv\iota_2^*\pLie{\cK^M_X}\cG \ar[d]^{\iota_2^*K_X} \ar@{=>}[dl]|{\k_X} \\
\cI_0 \ar@{=}[r]_{\id_{\cI_0}} & \cI_0 }\,.
\qqq 
Note that the above canonically induce 2-isomorphisms
\qq\nn
\alxydim{@C=7em@R=6em}{ |Q\la|_g^*\iota_1^*\cG\equiv\iota_1^*|M\la|_g^*\cG
\ar[d]_{\iota_1^*\La_g}  \ar@{=>}[dr]|{\widetilde\la{}_g} & |Q\la|_g^*\iota_2^*\cG\ox|Q\la|_g^*\cI_\om\equiv\iota_2^*|M\la|_g^*\cG\ox\cI_{|Q\la|_g^*\om} \ar[l]_{|Q\la|_g^*\Phi^\vee\hspace{4em}} \ar[d]^{\iota_2^*\La_g\ox\id_{\cI_{|Q\la|_g^*\om}}} \\ \iota_1^*\cG & \iota_2^*\cG\ox\cI_\om \ar[l]^{\Phi^\vee} }
\qqq 
and
\qq\nn
\alxydim{@C=7em@R=6em}{ \pLie{\cK^Q_X}\iota_1^*\cG\equiv\iota_1^*\pLie{\cK^M_X}\cG \ar[d]_{\iota_1^*K_X} \ar@{=>}[dr]|{\widetilde\k{}_X} & \pLie{\cK^Q_X}\bigl(\iota_2^*\cG\ox\cI_\om\bigr)\equiv\iota_2^*\pLie{\cK^M_X}\cG \ar[d]^{\iota_2^*K_X} \ar[l]_{\qquad\pLie{\cK^Q_X}\Phi^\vee\hspace{2em}} \\ \cI_0 \ar@{=}[r]_{\id_{\cI_0}} & \cI_0 }\,.
\qqq 
Finally,\ for $\,\cG\,$ and $\,\cB\,$ as above,\ we call the inter-$(\cG,\cB)$-bi-brane $\,\cJ\,$ {\bf maximally supersymmetric} if (condition \eqref{eq:ibbinter} is satisfied and) the following identities (with readily reconstructible domains of the identity 2-isomorphisms,\ left out for the sake of transparency) hold true for all $\,n\in\bN_{\geq 2}$:
\qq\nn
\id\circ|T_{n,1}\la|_g^*\varphi_{n+1}=\bigl(\varphi_{n+1}\circ\id\bigr)\bullet\bigl(\id\circ\pi_{1,2}^*\la_g\bigr)\bullet\bigl(\id\circ\pi_{2,3}^*\la_g\circ\id\bigr)\bullet\cdots\bullet\bigl(\id\circ\pi_{n,n+1}^*\la_g\circ\id\bigr)\bullet\bigl(\pi_{1,n+1}^*\widetilde\la{}_g\circ\id\bigr)
\qqq
and
\qq\nn
\id\circ\pLie{\cK^{T_{n,1}}_X}\varphi_{n+1}=\bigl(\id\circ\pi_{1,2}^*\k_X\bigr)\bullet\bigl(\id\circ\pi_{2,3}^*\k_X\circ\id\bigr)\bullet\cdots\bullet\bigl(\id\circ\pi_{n,n+1}^*\k_X\circ\id\bigr)\bullet\bigl(\pi_{1,n+1}^*\widetilde\k{}_X\circ\id\bigr)\,.
\qqq
Whenever a superstring background consists of a supersymmetric target equipped with a $\txG_{\rm s}$-invariant metric tensor and a $\txG_{\rm s}$-invariant 1-gerbe,\ and of a maximally supersymmetric bi-brane and of an inter-bi-brane with the same property,\ we speak of a {\bf maximally supersymmetric superstring background} (or a {\bf maximally supersymmetric background} for brevity).
~\medskip

Prior to concluding the present section in which the conceptual foundation of the paper is completed,\ we discuss special circumstances in which maximal supersymmetry of the superstring background is attained.\ This we do in all generality,\ in anticipation of the concrete findings of the subsequent investigation of a particular background with the bulk supergeometry $\,{\rm sMink}(d,1|D_{d,1})$.\ We begin with
\bedef\label{def:supergerbetc}
Let $\,\Bgt\,$ be a superstring background as in Def.\,\ref{def:str-bgrnd} and assume that the underlying super-space $\,\xcT=M\sqcup Q\sqcup T\,$ is endowed with a coherent action $\,\xcT\la\equiv T\la\sqcup Q\la\sqcup M\la\,$ of a (supersymmetry) Lie supergroup $\,\txG_{\rm s}\,$ as described above.\ We call $\,\Bgt\,$ a {\bf superstring super-background} (or a {\bf super-background} for brevity) if the following conditions are satisfied:
\bit
\item[(sBM)] There exist lifts of the action $\,M\la$:\ $\,\sfY M\la\,$ to $\,\sfY M\,$ and,\ then,\ that of $\,\sfY^{[2]}M\la\equiv(\sfY M\la\circ\pr_{1,2},\sfY M\la\circ\pr_{1,3})$ ({\it cf.}\ App.\,\ref{app:convs}) to $\,L$,
\qq\nn
\alxydim{@C=2.cm@R=1.5cm}{ \txG_{\rm s}\x\sfY M \ar[r]^{\sfY M\la} \ar[d]_{\id_{\txG_{\rm s}}\x\pi_{\sfY M}} & \sfY M \ar[d]^{\pi_{\sfY M}} \\ \txG_{\rm s}\x M \ar[r]_{M\la} & M }\qquad\quad{\rm and}\qquad\quad\alxydim{@C=2.cm@R=1.5cm}{ \txG_{\rm s}\x L \ar[r]^{L\la} \ar[d]_{\id_{\txG_{\rm s}}\x\pi_L} & L \ar[d]^{\pi_L} \\ \txG_{\rm s}\x\sfY^{[2]}M \ar[r]_{\sfY^{[2]}M\la} & \sfY^{[2]}M }\,,
\qqq
with respect to which the curving $\,\txB\,$ and the connection $\,\cA_L\,$ are $\txG_{\rm s}$-invariant (as defined above),\ and the groupoid structure $\,\mu_L\,$ is $\txG_{\rm s}$-equivariant (in the usual sense).\ We call a 1-gerbe with these properties a {\bf super-1-gerbe}.
\item[(sBQ)] For given lifts of the action $\,Q\la\,$ to the $\,\sfY_A Q\equiv\iota_A^*\sfY M$,
\qq\nn
\alxydim{@C=2.cm@R=1.5cm}{ \txG_{\rm s}\x\sfY_A Q \ar[r]^{\sfY_A Q\la} \ar[d]_{\id_{\txG_{\rm s}}\x\pi_{\sfY_A Q}} & \sfY_A Q \ar[d]^{\pi_{\sfY_A Q}} \\ \txG_{\rm s}\x Q \ar[r]_{Q\la} & Q }\,,
\qqq
there exists a lift of the product action $\,\sfY_{1,2}Q\la\equiv(\sfY_1 Q\la\circ\pr_{1,2},\sfY_2 Q\la\circ\pr_{1,3})\,$ ({\it cf.}\ App.\,\ref{app:convs}) to $\,\sfY\sfY_{1,2}Q$,\ and,\ then,\ that of the latter to $\,E$,
\qq\nn
\alxydim{@C=2.cm@R=1.5cm}{ \txG_{\rm s}\x\sfY\sfY_{1,2}Q \ar[r]^{\sfY\sfY_{1,2}Q\la} \ar[d]_{\id_{\txG_{\rm s}}\x\pi_{\sfY\sfY_{1,2}Q}} & \sfY\sfY_{1,2}Q \ar[d]^{\pi_{\sfY\sfY_{1,2}Q}} \\ \txG_{\rm s}\x\sfY_{1,2}Q \ar[r]_{\sfY_{1,2}Q\la} & \sfY_{1,2}Q }\qquad\quad{\rm and}\qquad\quad\alxydim{@C=2.cm@R=1.5cm}{ \txG_{\rm s}\x E \ar[r]^{E\la} \ar[d]_{\id_{\txG_{\rm s}}\x\pi_E} & E \ar[d]^{\pi_E} \\ \txG_{\rm s}\x\sfY\sfY_{1,2}Q \ar[r]_{\sfY\sfY_{1,2}Q\la} & \sfY\sfY_{1,2}Q }\,,
\qqq
with respect to which the connection $\,\cA_E\,$ is $\txG_{\rm s}$-invariant and the isomorphism $\,\a_E\,$ is $\txG_{\rm s}$-equivariant.\ Here,\ we speak of a {\bf super-$\cG$-bi-brane}.
\item[(sBT)] For given lifts of $\,T\la\,$ to the pullbacks of the total spaces of all relevant principal $\bC^\x$-bundles,\ the isomorphism $\,\b\,$ is $\txG_{\rm s}$-equivariant.\ We thus end up with a {\bf super-$(\cG,\cB)$-inter-bi-brane}.
\eit
\exdef

\noindent As we shall argue,\ a super-background is an example of a maximally supersymmetric background defined previously.\ We give a sketch of a demonstration below,\ focusing on $\txG_{\rm s}$-invariance under the induced action $\,|\xcT\la|\,$ of the body Lie group $\,\txG_{\rm s}\,$ and leaving the tangential action of the Lie superalgebra $\,\ggt_{\rm s}\,$ as an exercise to the Reader.\ We start our analysis with the 1-gerbe $\,\cG=(\sfY M,\pi_{\sfY M},\txB,L,\pi_L,\cA_L,\mu_L)\,$ and systematically reconstruct the corresponding pullback 1-gerbe $\,|M\la|_g^*\cG\,$ for an arbitrary $\,g\in|\txG_{\rm s}|$.\ The existence of $\,\sfY M\la\,$ enables us to choose the surjective submersion $\,|M\la|_g^*\sfY M\,$ of $\,|M\la|_g^*\cG\,$ as\footnote{The standard model of the pullback surjective submersion given by $\,M{}_{|M\la|_g}\hspace{-1pt}\x_{\pi_{\sfY M}}\hspace{-1pt}\sfY M\,$ is surjectively submersed by our $\,|M\la|_g^*\sfY M\equiv\sfY M\,$ in virtue of the universal property of the pullback.}   
\qq\nn
\alxydim{@C=3.cm@R=1.5cm}{ |M\la|_g^*\sfY M:=\sfY M \ar[r]^{\qquad\widehat{|M\la|_g}\equiv|\sfY M\la|_g} \ar[d]_{\pi_{|M\la|_g^*\sfY M}:=\pi_{\sfY M}} & \sfY M \ar[d]^{\pi_{\sfY M}} \\ M \ar[r]_{|M\la|_g} & M }\,,
\qqq
and the identification of the curving $\,\widehat{|M\la|_g}{}^*\txB\equiv|\sfY M\la|_g^*\txB=\txB\,$ follows.\ Next,\ we use the existence of $\,L\la\,$ (a connection-preserving bundle automorphism that covers $\,M\la$) to take 
\qq\nn
\alxydim{@C=3.cm@R=1.5cm}{ \widehat{|M\la|_g}{}^{[2]\,*}L=L \ar[r]^{\qquad\widehat{|\sfY^{[2]}M\la|_g}\equiv|L\la|_g} \ar[d]_{\pi_{\widehat{|M\la|_g}{}^{[2]\,*}L}:=\pi_L} & L \ar[d]^{\pi_L} \\ \sfY^{[2]}M \ar[r]_{|\sfY^{[2]}M\la|_g} & \sfY^{[2]}M }\,,
\qqq
whereupon we obtain $\,\widehat{|\sfY^{[2]}M\la|_g}{}^*\cA_L\equiv|L\la|_g^*\cA_L=\cA_L$.\ Finally,\ the assumed $\txG_{\rm s}$-equivariance of $\,\mu_L\,$ permits us to write $\,\widehat{|M\la|_g}{}^{[3]\,*}\mu_L=\mu_L$,\ thereby completing a convenient definition of the pullback 1-gerbe
\qq\nn
|M\la|_g^*\cG&\equiv&\bigl(|M\la|_g^*\sfY M,\pi_{|M\la|_g^*\sfY M},\widehat{|M\la|_g}{}^*\txB,\widehat{|M\la|_g}{}^{[2]\,*}L,\pi_{\widehat{|M\la|_g}{}^{[2]\,*}L},\widehat{|\sfY^{[2]}M\la|_g}{}^*\cA_L\equiv|L\la|_g^*\cA_L,\widehat{|M\la|_g}{}^{[3]\,*}\mu_L\bigr)\cr\cr
&\equiv&\cG\,.
\qqq
Therefore,\ we may consistently set $\,\La_g:=\id_\cG\,,\qquad g\in|\txG_{\rm s}|$.

Passing to the $\cG$-bi-brane $\,\cB=(Q,\iota_1,\iota_2,\om,\Phi)$,\ we first convince ourselves that we have $\,|Q\la|_g^*\iota_A^*\cG\equiv\iota_A^*\cG,\ A\in\{1,2\}$,\ with,\ {\it e.g.},\ $\,\iota_A^*\cG=(\sfY_A Q\equiv Q\hspace{3pt}{}_{\iota_A}\hspace{-3pt}\x_{\pi_{\sfY M}}\sfY M,\pr_1,\pr_2^*\txB,\sfY_A^{[2]}Q\hspace{3pt}{}_{\pr_{2,4}}\hspace{-3pt}\x_{\pi_L}L,\pi_L\circ\pr_2,\pr_2^*\cA_L,\pr_{2,4,6}^*\mu_L)$,\ and so it remains to pin down the pullback 1-isomorphism $\,|Q\la|_g^*\Phi\,$ in a judiciously chosen manner.\ To this end,\ we lift $\,Q\la\,$ to the $\,\sfY_A Q\,$ as {\it per} $\,\sfY_A Q\la\equiv(Q\la\circ\pr_{1,2},\sfY M\la\circ\pr_{1,3})\,$ ({\it cf.}\ App.\,\ref{app:convs}),\ and,\ then,\ choose
\qq\nn
\alxydim{@C=3.5cm@R=1.5cm}{ \bigl(|\sfY_1 Q\la|_g\x|\sfY_2 Q\la|_g\bigr)^*E=E \ar[r]^{\qquad\qquad\quad\widehat{|\sfY_1 Q\la|_g\x|\sfY_2 Q\la|_g}\equiv|E\la|_g} \ar[d]_{\pi_{(|\sfY_1 Q\la|_g\x|\sfY_2 Q\la|_g)^*E}:=\pi_E} & E \ar[d]^{\pi_E} \\ \sfY\sfY_{1,2}Q\equiv\sfY_{1,2}Q \ar[r]_{|\sfY_1 Q\la|_g\x|\sfY_2 Q\la|_g} & \sfY_{1,2}Q }\,,
\qqq
only to find $\,\widehat{|\sfY_1 Q\la|_g\x|\sfY_2 Q\la|_g}{}^*\cA_E\equiv|E\la|_g^*\cA_E=\cA_E$.\ Taking into account the assumed $\txG_{\rm s}$-equivariance of $\,\a_E$,\ we thus establish the anticipated identity
\qq\nn
|Q\la|_g^*\Phi&\equiv&\bigl(|Q\la|_g^*\sfY_1 Q\x_Q|Q\la|_g^*\sfY_2 Q,\pi_{|Q\la|_g^*\sfY_1 Q\x_Q|Q\la|_g^*\sfY_2 Q},\bigl(|\sfY_1 Q\la|_g\x|\sfY_2 Q\la|_g\bigr)^*E,\pi_{(|\sfY_1 Q\la|_g\x|\sfY_2 Q\la|_g)^*E},\cr\cr
&&\widehat{|\sfY_1 Q\la|_g\x|\sfY_2 Q\la|_g}{}^*\cA_E,\bigl(\widehat{|\sfY_1 Q\la|_g\x|\sfY_2 Q\la|_g}\x\widehat{|\sfY_1 Q\la|_g\x|\sfY_2 Q\la|_g}\bigr)^*\a_E\bigr)\equiv\Phi\,,
\qqq
and so also\footnote{Here,\ we are being a little sloppy and (intentionally) \emph{not} taking into account the existence of nontrivial left- and right-unit 2-cells in the weak 2-category $\,\bgrb^\nabla(Q)$.\ We rectify this in Sec.\,\ref{sec:multGSs1g}.} $\,\la_g\equiv\id_\Phi$.\ The $\txG_{\rm s}$-equivariance of the isomorphism of the 1-gerbe 2-isomorphisms $\,\varphi_{n+1}$,\ in conjunction with the triviality of the 1-isomorphisms $\,\La_g\,$ and 2-isomorphisms $\,\la_g\,$ just established,\ yields the identities
\qq\nn
|T_{n,1}\la|_g^*\varphi_{n+1}=\varphi_{n+1}\,,
\qqq
valid for all $\,n\in\bN_{\geq 2}$,\ and thus ensures maximal supersymmetry of the super-inter-$(\cG,\cB)$-bi-brane.\ This concludes (the sketch of) our proof of the claim.

\section{Simplicial (super)symmetric Lie backgrounds}\label{sec:smultcat}

Considerations of supersymmetry distinguish a class of superstring backgrounds,\ to wit,\ those whose target super-spaces are disjoint sums of complete orbits of the supersymmetry group,\ the criterion of distinction here being `reducibility' (and `elementarity') with respect to the action of the (super)group.\ We are thus led to focus our attention on (super)string geometrodynamics on homogeneous spaces of Lie supergroups (as bulk target super-spaces).\ As was shown convincingly in the works \cite{Gawedzki:2010rn,Gawedzki:2012fu,Suszek:2012ddg,Suszek:2011,Suszek:2013} of Gaw\c{e}dzki,\ Waldorf and the Author,\ the higher-geometric structure requisite for the formulation of the geometrodynamics on a homogeneous space $\,\txG/\txH\,$ associated with a \emph{nontrivial} subgroup $\,\txH\subset\txG\,$ of the supersymmetry group $\,\txG\,$ arises through equivariantisation of the given  higher-geometric structure over $\,\txG\,$ with respect to the action of $\,\txH\,$ on $\,\txG$,\ and so it is well justified to focus further on superstring backgrounds with bulk targets given by Lie supergroups first,\ which is what we do in this paper.\ Thus,\ we fix a Lie supergroup $\,\txG\,$ for the remainder of our discussion,\ together with its binary operation
\qq\label{eq:binary}
\txm\ :\ \txG\x\txG\too\txG\,,
\qqq 
In this setting,\ the supersymmetry group $\,\txG_{\rm s}\,$ is either the product Lie supergroup $\,\txG\x\txG\,$ (the un-graded case) with the (left) action 
\qq\label{eq:Gact-bulk}
\ell\wp\equiv\ell\circ\bigl(\id_\txG\x\wp\bigr)\circ\bigl(\id_\txG\x\t\circ\bigl(\Inv\x\id_\txG\bigr)\bigr)\ :\ \txG^{\x 2}\x\txG\too\txG
\qqq
written in terms of the standard left ($\ell$,\ with the second cartesian factor acted upon) and right ($\wp$,\ with the first cartesian factor acted upon) regular actions $\,\ell\equiv\txm\equiv\wp\,$ and the canonical transposition $\,\t\ :\ \txG_{(1)}\x\txG_{(2)}\too\txG_{(2)}\x\txG_{(1)}\,$ of the cartesian factors,\ or its left factor $\,\txG\x\bR^{0|0}\equiv\txG\,$ (the $\bZ/2\bZ$-graded case\footnote{What makes the situation quite a bit more complicated in this case is the presence of an additional `infinitesimal' right \emph{local} supersymmetry,\ known as $\k$-symmetry,\ {\it cf.}\ the original \Rcite{deAzcarraga:1982dhu} for an account of its discovery,\ \Rcite{Siegel:1983hh} for that of its re-discovery,\ and Refs.\,\cite{Suszek:2020xcu,Suszek:2020rev,Suszek:2021hjh} for an in-depth explanation of its geometric nature.\ The symmetry is crucial for the supersymmetric balance in the vacuum of the theory,\ but we shall not deal with it in the present paper,\ postponing its treatment to a future study -- hence the simplification.}) with the (left) action 
\qq\label{eq:Gact-bulk-grad}
\ell\equiv\ell\wp\circ\bigl(\id_\txG\x\widehat e\x\id_\txG\bigr).\ 
\qqq
The best studied (super-)$\si$-models of the corresponding dynamics are collectively referred to as the Wess--Zumino--Witten (WZW) (super-)$\si$-models in the literature.

\subsection{Group-homomorphicity categorified}\label{sub:homocat}

The WZW bulk target super-space combines the structure of a supermanifold and that of a supergroup,\ the two being related by the supermanifold map \eqref{eq:binary},\ and so a class of superbackgrounds becomes distinguished,\ namely those whose associated defects implement the binary operation.\ Fix a 1-gerbe $\,\cG\,$ over $\,\txG\,$ that defines the bulk dynamics.\ Following the general philosophy recalled in Sec.\,\ref{sub:grbmultissi},\ we then,\ in particular,\ look for a sub-supermanifold $\,Q\subset\txG\x\txG\,$ which supports a 1-gerbe 1-isomorphism $\,\pr_1^*\cG\rstr_Q\cong\txm^*\cG\rstr_Q\ox\cI_\om\,$ for some $\,\om\in\Om^2(Q)$,\ or,\ equivalently,\ one on which we have\label{pgref:guidelines} $\,\txm^*\cG\rstr_Q\cong\bigl(\pr_1^*\cG\ox\pr_2^*\cG\bigr)\rstr_Q\ox\bigl(\pr_2^*\cG^*\rstr_Q\ox\cI_{-\om}\bigr)$.\ Such a reformulation of the original problem emphasises the significance of the condition of compatibility of the higher-geometric structure $\,\cG\,$ with the binary operation on its base that generalises the condition of homomorphicity encountered in the study of functions on $\,\txG$.\ Indeed,\ if we had $\,\txm^*\cG\cong\pr_1^*\cG\ox\pr_2^*\cG$,\ a solution to the above problem could be chosen in the form $\,Q=\txG\x D\,$ for an \emph{arbitrary} $\,D\subset\txG\,$ that supports a trivialisation $\,\cG\rstr_D\cong\cI_{\om_D}$.\ It would then suffice to take $\,\om=-\pr_2^*\om_D$.\ The rigidity of the former assumption,\ precluding the much desired application of the argument in the familiar setting of the WZW $\si$-model with $\,\txG\,$ compact and 1-connected ({\it cf.}\ below),\ is readily circumvented by allowing a trivial but not necessarily zero correction 
\qq\label{eq:multcorr}
\cI_\varrho\cong\txm^*\cG^*\ox\pr_1^*\cG\ox\pr_2^*\cG\,,
\qqq 
which does not invalidate the conclusion qualitatively but merely calls for a quantitative redefinition $\,-\pr_2^*\om_D\longmapsto-\pr_2^*\om_D+\varrho\,$ of the bi-brane curvature $\,\om$.\ This,\ then,\ is the sort of additional structure on the bulk 1-gerbe $\,\cG\,$ over $\,\txG\,$ that we consider in some detail below,\ with view to subsequent applications in the the study of maximally (super)symmetric WZW defects.\medskip

The point of departure in our discussion is the {\bf Polyakov--Wiegmann}({\bf --type}) {\bf identity}  
\qq\label{eq:PolWiegH}
\txm^*\txH=\pr_1^*\txH+\pr_2^*\txH-\sfd\varrho
\qqq
that has to be satisfied by the curvature $\,\txH\,$ of the 1-gerbe $\,\cG$.\ Following \Rcite{Gawedzki:2009jj},\ we develop its geometrisation over the nerve of the category
\qq\nn
\alxydim{@C=2cm@R=1.5cm}{ \mathrm{Mor}\bigl(\txG^{\rm op}\bigr)\x\mathrm{Mor}\bigl(\txG^{\rm op}\bigr)\equiv\txG\x\txG \ar[r]^{\qquad\quad\circ=\txm\circ\t_\txG} & \mathrm{Mor}\bigl(\txG^{\rm op}\bigr) \equiv\txG \ar@{-->}@<0.5ex>[r]^s \ar@{-->}@<-0.5ex>[r]_t & \bR^{0|0}\equiv\mathrm{Ob}\bigl(\txG^{\rm op}\bigr) \ar@/_1.5pc/[l]_{\Id_\cdot=\widehat e} } 
\qqq
opposite to the category canonically associated with the Lie supergroup $\,\txG$.\ It has the terminal supermanifold $\,\bR^{0|0}\,$ as its object supermanifold and the Lie supergroup itself as the morphism supermanifold.\ The source and target maps are both given by the unique morphism $\,\alxydim{}{\txG \ar@{-->}[r] & \bR^{0|0}}\,$ into the terminal supermanifold,\ the identity morphism is the topological unit in $\,\txG$,\ and composition of morphisms is the binary operation in {\it reverse} order ($\t_\txG\ :\ \txG\x\txG\circlearrowleft\,$ is the transposition,\ with $\,\pr_1\circ\t_\txG=\pr_2,\ \pr_2\circ\t_\txG=\pr_1$).\ In order to spell out the algebraic information encoded by the nerve $\,\sfN_\bullet\txG^{\rm op}$,\ we pass to the $\cS$-point picture and write out the structure for a fixed $\,\cS\,$ and the associated $\cS$-points $\,{\rm Hom}_\sMan(\cS,\txG)\equiv\txG(\cS)$,\ denoting the mapping $\,{\rm Hom}_\sMan(\cS,\txm)\ :\ \txG(\cS)^{\x 2}\too\txG(\cS)\,$ induced by $\,\txm\,$ simply by a dot.\ Thus,\ an $\cS$-point in the morphism supermanifold is $\,g\in\txG(\cS)$,\ and the source and target maps are both represented by $\,g\longmapsto \bullet$,\ whereas the identity morphism associates the unit $\,e\equiv e_{\txG(\cS)}\in\txG(\cS)\,$ to the (unique) $\cS$-point $\,\bullet\in\bR^{0|0}(\cS)$.\ In this notation,\ composition of morphisms is simply given by $\,(g_1,g_2)\longmapsto g_2\cdot g_1$.\ Now,\ the nerve consists,\ in the $\cS$-point picture,\ of spaces $\,\sfN_n\txG^{\rm op}(\cS) =\txG^{\times n}(\cS)\,$ with the face and degeneracy maps given (for any $\,n$) by
\qq
\barr{ll}\displaystyle
  d^{(n)}_i(\cS)(g_1,g_2,\ldots,g_n)=
  \begin{cases} (g_2,\ldots,g_n) & \text{if} \quad i=0 \cr
(g_1,g_2,\ldots,g_{i-1},g_i\cdot g_{i+1},g_{i+2},\ldots, g_n) & \text{if} \quad  0<i<n\,, \cr
  (g_1,g_2,\ldots,g_{n-1}) & \text{if} \quad i=n \end{cases}
  \label{eq:GG-face-deg}\\[1.5em] \displaystyle\\
  s^{(n)}_i(\cS)(g_1,g_2,\ldots,g_n)= (g_1,g_2,\ldots,g_{i},e,g_{i+1},\ldots,g_n)\,,
\earr
\qqq
and the non-trivial simplicial identities reflect the group axioms.

At this stage,\ a specialisation of the construction of Def.\,\ref{def:relcohom} gives us the cohomology of relevance to our subsequent considerations.\ Thus,\ we obtain the cochain bicomplex (here,\ $\,\txG^{\x 0}\equiv\bR^{0|0}$)
\qq\label{diag:bicG}
\alxydim{@C=2.cm@R=1.cm}{ & \vdots \ar[d]_(.3){\D_{\txG}^{(n-1;p)}} & \\ \cdots \ar[r]^(.3){\sfd^{(p-1;n)}_{\rm dR}\equiv\sfd_{(\txG^{\x n})}} & \Om^p\bigl(\txG^{\x n}\bigr) \ar[r]^(.7){\sfd^{(p;n)}_{\rm dR}\equiv\sfd_{(\txG^{\x n})}} \ar[d]_(.7){\D_{\txG}^{(n;p)}} & \cdots  \\ & \cdots & }\,,\qquad p,n\in\bN
\qqq
with the vertical coboundary operators
\qq\nn
\D_{\txG}^{(n;p)}\equiv\sum_{i=0}^{n+1}\,(-1)^i\,d^{(n+1)\,*}_i\ :\ \Om^p\bigl(\txG^{\x n}\bigr)\too\Om^p\bigl(\txG^{\x n+1}\bigr)\,,\qquad(n,p)\in\bN^{\x 2}\,.
\qqq
On its diagonal,\ we find the cochain complex 
\qq\nn
\bigl(\cA^\bullet\bigl(\sfN_\bullet\txG^{\rm op}\bigr),\cD^{(\bullet)}_{\txG}\bigr)\qquad :\qquad \cA^0\bigl(\sfN_\bullet\txG^{\rm op}\bigr)\xrightarrow{\ \cD_{\txG}^{(0)}\ }\cA^1\bigl(\sfN_\bullet\txG^{\rm op}\bigr)\xrightarrow{\ \cD^{(1)}_{\txG}\ }\cdots\xrightarrow{\ \cD^{(q-1)}_{\txG}\ }\cA^q\bigl(\sfN_\bullet\txG^{\rm op}\bigr)\xrightarrow{\ \cD^{(q)}_{\txG}\ }\cdots
\qqq
with the $q$-cochain groups
\qq\nn
\cA^q\bigl(\sfN_\bullet\txG^{\rm op}\bigr)=\bigoplus_{p=0}^{q}\,A^{p,q-p}\,,\qquad\qquad A^{p,q-p}\equiv\Om^p\bigl(\txG^{\x q-p}\bigr)\,,\qquad\qquad q\in\bN^\x
\qqq
related by the couboundary operators
\qq\nn
\cD_{\txG}^{(q)}\ :\ \cA^q\bigl(\sfN_\bullet\txG^{\rm op}\bigr)\too\cA^{q+1}\bigl(\sfN_\bullet\txG^{\rm op}\bigr)\,,\qquad\qquad\cD_{\txG}^{(q)}\rstr_{A^{p,q-p}}=(-1)^{q-p+1}\,\sfd^{(p;q-p)}_{\rm dR}+\,\D_\txG^{(q-p;p)}\,,
\qqq
and it is the ensuing total cohomology of the bicomplex that captures the structure of the Polyakov--Wiegmann identity and,\ as such,\ becomes the starting point of the geometrisation/categorification that we are after.\ Prior to giving higher-geometric flesh to the last statement,\ we fix one more

\noindent We are now finally ready to formulate
\bedef\label{def:genmultstr}
Let $\,\txG\,$ be a Lie supergroup,\ let $\,\cG=(\sfY\txG,\pi_{\sfY\txG},\txB,L,\pi_L,\txA,\mu_L)\,$ be a 1-gerbe of curvature $\,\txH\equiv{\rm curv}\,(\cG)\in Z^3_{\rm dR}(\txG)$,\ and suppose there exist:\ a 2-form $\,\varrho\in\Om^2(\txG^{\x 2})$,\ a 1-form $\,\vartheta\in\Om^1(\txG^{\x 3})\,$ and a 0-form $\,\varphi\in\Om^0(\txG^{\x 4})\,$ such that $\,\cZ_\cG\equiv(0,\txH,\varrho,\vartheta,\varphi)\in{\rm Ker}\,\cD^{(4)}_\txG$.\ A {\bf generalised multiplicative structure on} $\,\cG\,$ associated with the 4-cocycle $\,\cZ_\cG\,$ is a pair $\,(\cM,\a)\,$ that consists of a 1-gerbe 1-isomorphism
\qq\nn
\cM\ :\  d_2^{(2)\,*}\cG\ox d_0^{(2)\,*}\cG\xrightarrow{\ \cong\ } d_1^{(2)\,*}\cG\ox\cI_\varrho
\qqq
and a 1-gerbe 2-isomorphism ({\it cf.}\ App.\,\ref{app:convs}) 
\qq\label{diag:aleph}
\xy (50,0)*{\bullet}="G12"+(0,4)*{\cG^{(3)}_1\ox\cG^{(3)}_2\ox\cG^{(3)}_3};
(25,-20)*{\bullet}="G1r1"+(-15,0)*{\cG^{(3)}_{12}\ox\cG^{(3)}_3\ox\cI_{ d_3^{(3)\,*}\varrho}\quad}; (75,-20)*{\bullet}="G2om"+(16,0)*{\quad\cG^{(3)}_1\ox\cG^{(3)}_{23}\ox
\cI_{ d_0^{(3)\,*}\varrho}}; (35,-40)*{\bullet}="G2or1"+(-12,-4)*{\cG^{(3)}_{123}
\ox\cI_{ d_3^{(3)\,*}\varrho+ d_1^{(3)\,*}\varrho}};
(65,-40)*{\bullet}="G2or2"+(14,-4)*{\cG^{(3)}_{123}\ox\cI_{ d_0^{(3)\,*}\varrho+
 d_2^{(3)\,*}\varrho}}; \ar@{->}|{ d_3^{(3)\,*}\cM\ox\id_{\cG^{(3)}_3}} "G12";"G1r1"
\ar@{->}|{\id_{\cG^{(3)}_1}\ox d_0^{(3)\,*}\cM} "G12";"G2om"
\ar@{->}|{ d_1^{(3)\,*}\cM\ox\id_{\cI_{ d_3^{(3)\,*}\varrho}}} "G1r1";"G2or1"
\ar@{->}|{ d_2^{(3)\,*}\cM\ox\id_{\cI_{ d_0^{(3)\,*}\varrho}}} "G2om";"G2or2"
\ar@{->}|{\id_{\cG^{(3)}_{123}}\ox\cJ_{-\vartheta}} "G2or2";"G2or1" \ar@{=>}|{\ \a\ } "G1r1"+(3,0);"G2om"+(-3,0)
\endxy\,,
\qqq
written in terms of the trivial 0-gerbe $\,\cJ_{-\vartheta}=(\txG^{\x 3}\x\bC^\x,\pr_1,-\vartheta)\ :\ \cI_{ d_0^{(3)\,*}\varrho+d_2^{(3)\,*}\varrho}\xrightarrow{\ \cong\ }\cI_{ d_3^{(3)\,*}\varrho+ d_1^{(3)\,*}\varrho}\,$ over $\,\txG^{\x 3}\,$ and subject to the coherence constraint expressed by the commutative diagram 
{\tiny\qq\nn
\xy (50,0)*{(\cM^{(4)}_{123,4}\ox\id_{\cI_{\varrho^{(4)}_{1,2}+\varrho^{(4)}_{12,3}}})\circ(\cM^{(4)}_{12,3}\ox\id_{\cG^{(4)}_4\ox\cI_{\varrho^{(4)}_{1,2}}})\circ(\cM^{(4)}_{1,2}\ox\id_{\cG^{(4)}_3\ox\cG^{(4)}_4})}="G12";
(0,-30)*{\barr{c} (\id_{\cG^{(4)}_{1234}}\ox\cJ_{- d_4^{(4)\,*}\vartheta})\circ(\cM^{(4)}_{123,4}\ox\id_{\cI_{\varrho^{(4)}_{2,3}+\varrho^{(4)}_{1,23}}}) \\ \circ(\cM^{(4)}_{1,23}\ox\id_{\cG^{(4)}_4\ox\cI_{\varrho^{(4)}_{2,3}}})\circ(\id_{\cG^{(4)}_1}\ox\cM^{(4)}_{2,3}\ox\id_{\cG^{(4)}_4}) \earr}="G1r1"; 
(100,-30)*{\barr{c} (\id_{\cG^{(4)}_{1234}}\ox\cJ_{- d_1^{(4)\,*}\vartheta})\circ(\cM^{(4)}_{12,34}\ox\id_{\cI_{\varrho^{(4)}_{3,4}+\varrho^{(4)}_{1,2}}}) \\ \circ(\cM^{(4)}_{1,2}\ox\id_{\cG^{(4)}_{34}\ox\cI_{\varrho^{(4)}_{3,4}}})\circ(\id_{\cG^{(4)}_1\ox\cG^{(4)}_2}\ox\cM^{(4)}_{3,4}) \earr}="G2om"; 
(0,-70)*{\barr{c} (\id_{\cG^{(4)}_{1234}}\ox\cJ_{- d_4^{(4)\,*}\vartheta- d_2^{(4)\,*}\vartheta})\circ(\cM^{(4)}_{1,234}\ox\id_{\cI_{\varrho^{(4)}_{2,3}+\varrho^{(4)}_{23,4}}})\\ \circ(\id_{\cG^{(4)}_1\ox\cI_{\varrho^{(4)}_{2,3}}}\ox\cM^{(4)}_{23,4})\circ(\id_{\cG^{(4)}_1}\ox\cM^{(4)}_{2,3}\ox\id_{\cG^{(4)}_4}) \earr}="G2or1";
(100,-70)*{\barr{c} (\id_{\cG^{(4)}_{1234}}\ox\cJ_{- d_1^{(4)\,*}\vartheta- d_3^{(4)\,*}\vartheta})\circ(\cM^{(4)}_{1,234}\ox\id_{\cI_{\varrho^{(4)}_{3,4}+\varrho^{(4)}_{2,34}}}) \\ \circ(\id_{\cG^{(4)}_1\ox\cI_{\varrho^{(4)}_{3,4}}}\ox\cM^{(4)}_{2,34})\circ(\id_{\cG^{(4)}_1\ox\cG^{(4)}_2}\ox\cM^{(4)}_{3,4}) \earr}="G2or2"; 
(50,-100)*{(\id_{\cG^{(4)}_{1234}}\ox\cJ_{- d_4^{(4)\,*}\vartheta- d_2^{(4)\,*}\vartheta-d^{(4)\,*}_0\vartheta})\circ(\cM^{(4)}_{1,234}\ox\id_{\cI_{\varrho^{(4)}_{3,4}+\varrho^{(4)}_{2,34}}}) \circ(\id_{\cG^{(4)}_1\ox\cI_{\varrho^{(4)}_{3,4}}}\ox\cM^{(4)}_{2,34})\circ(\id_{\cG^{(4)}_1\ox\cG^{(4)}_2}\ox\cM^{(4)}_{3,4})}="Gfin";
\ar@{=>}|{\id\circ( d_4^{(4)\,*}\a\ox\id)} "G12";"G1r1"
\ar@{=>}|{d^{(4)\,*}_1\a\circ\id} "G12";"G2om"
\ar@{=>}|{\id\circ d_2^{(4)\,*}\a\circ\id} "G1r1";"G2or1"
\ar@{=>}|{\id\circ d_3^{(4)\,*}\a\circ\id} "G2om";"G2or2"
\ar@{=>}|{\id\circ(\id\ox d^{(4)\,*}_0\a)} "G2or1";"Gfin" 
\ar@{=>}|{(\id\ox\cK_{-\varphi})\circ\id} "G2or2";"Gfin" 
\endxy\,,
\qqq}
\qq\label{diag:cohmult}
\qqq
of 1-gerbe 2-isomorphisms over $\,\txG^{\x 4}$,\ written in terms of the trivial $(-1)$-gerbe
\qq\nn
\cK_{-\varphi}\ :\ \cJ_{- d_1^{(4)\,*}\vartheta- d_3^{(4)\,*}\vartheta}\xrightarrow{\ \cong\ }\cJ_{- d_4^{(4)\,*}\vartheta- d_2^{(4)\,*}\vartheta-d^{(4)\,*}_0\vartheta}\,.
\qqq
Whenever $\,\varphi\equiv 0$,\ we call the corresponding generalised multiplicative structure {\bf 0-flat},\ and 
if $\,(\vartheta,\varphi)=(0,0)$,\ we speak of a ({\bf standard}) {\bf multiplicative structure}.

A 1-gerbe $\,\cG\,$ that admits a generalised (resp.\ 0-flat,\ resp.\ standard) multiplicative structure shall be called {\bf generalised-multiplicative} (resp.\ {\bf 0-flat generalised-multiplicative},\ resp.\ {\bf multiplicative}).
\exdef

\subsection{WZW (super)targets as $\txG_{\rm s}$-orbits}\label{sub:WZWorbs}

The foregoing discussion identifies the nerve $\,(\unl\sfN{}_\bullet\txG\equiv\sfN_\bullet(\txG\rx\txG),\unl d{}^{(\bullet)}_\cdot,\unl s{}^{(\bullet)}_\cdot)\,$ of the action groupoid
\qq\nn
\alxydim{@C=2.5cm@R=1.5cm}{ \mathrm{Mor}\bigl(\txG\rx\txG\bigr)\equiv\txG\x\txG \ar@<0.5ex>[r]^{s=\pr_1\equiv\unl d{}^{(1)}_1} \ar@<-0.5ex>[r]_{t=\wp\equiv\unl d{}^{(1)}_0} & \txG\equiv\mathrm{Ob}\bigl(\txG\rx\txG\bigr) \ar@/_2pc/[l]_{\Id_\cdot=\id_\txG\x\widehat e\equiv\unl s{}^{(0)}_0} }\,,
\qqq  
as the simplicial supermanifold in which to look for the simplicial WZW target super-space whose associated higher-geometric data -- containing,\ in particular,\ a 1-gerbe over $\,M=\txG\equiv\unl\sfN{}_0\txG\,$ and the composite 1-isomorphism $\,(\cM\ox\id)\circ(\id\ox\pr_2^*\cT_D^{-1}\ox\id)\ :\ \pr_1^*\cG\equiv\pr_1^*\cG\ox\cI_{\pr_2^*\om_D}\ox\cI_{-\pr_2^*\om_D}\xrightarrow{\ \cong\ }\txm^*\cG\rstr_{\txG\x D}\ox\cI_\om\,$ over $\,Q=\txG\x D\subset\txG^{\x 2}\equiv\unl\sfN{}_1\txG\,$ (for some disjoint union $\,D\,$ of $\txG\la$-orbits within $\,\txG$) -- determine a multi-phase super-$\si$-model with a defect realising the basic algebraic structure \eqref{eq:binary} on the bulk WZW target.\ The lowest rung of the nerve comes with one of the previously distinguished $\txG_{\rm s}$-actions 
\qq\nn
(\txG_{\rm s},\txG\la\}\in\bigl\{\bigl(\txG^{\x 2},\ell\wp\bigr),(\txG,\ell)\bigr\}\,,
\qqq
from which we readily induce the structure of a simplicial $\txG_{\rm s}$-supermanifold on $\,\unl\sfN{}_\bullet\txG\,$ by declaring its face maps to be $\txG_{\rm s}$-equivariant.\ Once this is achieved,\ we may decompose the latter into complete $\txG_{\rm s}$-orbits and subsequently look for suitable higher-geometric structures compatible with the supersymmetry over these supergeometries.\ The first part of this task shall be completed below.\medskip

Let us,\ first,\ determine the equivariant lift of $\,\txG\la\,$ from $\,\txG\equiv\unl\sfN{}_0\txG\,$ to the higher components $\,\unl\sfN{}_n\txG\equiv\sfN_n(\txG\rx\txG)\equiv\sfN_{n+1}(\txG^{\rm op}),\ n\in\bN^\x\,$ of the nerve\footnote{In order to convince ourselves that this is,\ in fact,\ the nerve of the said action groupoid,\ it suffices to pass,\ in the $\cS$-point picture,\ from the standard representation $\,((g,h_1),(g\cdot h_1,h_2),\ldots,(g\cdot h_1\cdot\cdots\cdot h_{n-1},h_n))\,$ of an $n$-tuple of composable morphisms of $\,\txG\rx\txG\,$ (with $\,g,h_i\in\txG,\ i\in\ovl{1,n}$) to the equivalent one $\,(g,h_1,h_2,\ldots,h_n)$.} along its face maps $\,\unl d{}^{(n)}_i\equiv d^{(n+1)}_{i+1}$,\ turning also the degeneracy maps $\,\unl s{}^{(n)}_i\equiv s^{(n+1)}_{i+1},\ i\in\ovl{0,n}\,$ into $\txG_{\rm s}$-equivariant mappings.\ For this purpose,\ introduce the canonical diagonal embedding $\,\D\ :\ \txG\too\txG\x\txG\,$ determined by the identities $\,\pr_1\circ\D=\id_\txG=\pr_2\circ\D$,\ and subsequently use it to define the conjugation (the adjoint action) $\,\Ad\equiv\ell\wp\circ(\D\x\id_\txG)\ :\ \txG\x\txG\too(\txG\x\txG)\x\txG\too\txG$.\  We then readily establish,\ for $\,(\txG_{\rm s},\txG\la)=(\txG\x\txG,\ell\wp)$,\ the following form of the $\txG_{\rm s}$-action at level $\,n\geq 1$:
\qq\nn
\txG\la^{(n)}\equiv\ell\wp^{(n)}=\bigl(\ell\wp\circ\pr_{1,2,3},\Ad\circ\pr_{1,2,4},\Ad\circ\pr_{1,2,5},\ldots,\Ad\circ\pr_{1,2,n+3}\bigr)\ :\ \bigl(\txG\x\txG\bigr)\x\txG^{\x n+1}\too\txG^{\x n+1}\,.
\qqq
For $\,(\txG_{\rm s},\txG\la)=(\txG,\ell)$,\ on the other hand,\ we obtain
\qq\label{eq:lactn}
\txG\la^{(n)}\equiv\ell^{(n)}=\ell\x\id_{\txG^{\x n}}\ :\ \txG\x\txG^{\x n+1}\too\txG^{\x n+1}\,,\qquad n\geq 1\,.
\qqq

It is now natural to investigate the decomposition of the newly found simplicial supermanifold $\,\unl\sfN{}_\bullet\txG\,$ into `irreducible' sub-$\txG_{\rm s}$-supermanifolds,\ {\it i.e.},\ orbits of the action $\,\txG\la^{(\bullet)}$.\ In the case of $\,(\txG_{\rm s},\txG\la)=(\txG,\ell)$,\ of relevance,\ {\it e.g.},\ to the study of the super-$\si$-model on the Lie supergroup $\,{\rm sMink}(d,1|D_{d,1})$,\ this investigation trivialises,\ but its fully fledged application in the superfield-theoretic setting calls,\ first,\ for the incorporation of the `infinitesimal gauged right' supersymmetry known as the $\k$-symmetry into the picture,\ and so we leave it out of the present discussion.\ Instead,\ we take a closer look at orbits of the action $\,\ell\wp^{(\bullet)}\,$ of $\,\txG\x\txG\,$ on components of $\,\unl\sfN{}_\bullet\txG\,$ for a Lie group $\,\txG$.\ We treat the spaces $\,\unl\sfN{}_n\txG\,$ for the first few values of $n$ explicitly,\ and,\ thus prepared,\ eventually give the general structure.

\bit
\item[$\bullet$] $n=0$:\ An element $\,(x,y) \in \txG\x\txG\,$ acts on $\,\unl\sfN{}_0\txG\equiv\txG\ni g\,$ as $\,\ell\wp^{(0)}_{(x,y)}(g) = x\cdot g\cdot y^{-1}$,\ so that $\,\unl\sfN{}_0\txG\,$ is a single orbit.
\item[$\bullet$] $n=1$:\ $\,(x,y)\,$ acts on $\,\unl\sfN{}_1 \txG\ni (g,h)\,$ as $\,\ell\wp^{(1)}_{(x,y)}(g,h) = (x\cdot g\cdot y^{-1}, \Ad_y(h))$,\ and so we find the decomposition into orbits 
\qq\nn
\unl\sfN{}_1 \txG = \bigsqcup_{\la\in \sfk\,\xcA_{\rm W}(\ggt)}\, \txG
\times \xcC_\la \,,\qquad\qquad\xcC_\la
=\bigl\{\ \Ad_x(\ee_\la) \ \vert \ x\in\txG \
\bigr\}\,,\qquad \ee_\la=\ee^{2\pi\sfi\frac{\la}{\sfk}}\,,
\qqq
labelled -- with hindsight -- by vectors in the $\sfk$-rescaled Weyl alcove
\qq\nn
\xcA_W(\ggt)=\bigl\{\ \la\in\tgt \quad\big\vert\quad\langle\la
,\th\rangle\leq 1\quad\land\quad\langle\la,\a_i\rangle\geq 0\,,\ i
=\ovl{1,{\rm rank}\,\ggt} \ \bigr\}\,,
\qqq
the latter being defined in terms of the simple roots $\a_i\,$ and of the longest root $\,\th\,$ of $\,\ggt$,\ and in terms of the canonical pairing $\,\corr{\cdot,\cdot}\,$ between roots and weights -- the scaling,\ inconsequential at this stage,\ prepares us for the subsequent use of the $\sfk$-rescaled Cartan--Killing metric on $\,\txG\,$ (affording,\ {\it i.a.},\ the identification between $\,\ggt\,$ and its dual).
\item[$\bullet$] $n=2$:\ $\,(x,y)\,$ acts on $\,\unl\sfN{}_2 \txG\ni (g,h_1,h_2)\,$ as $\,\ell\wp^{(2)}_{(x,y)}(g,h_1,h_2) = (x\cdot g \cdot y^{-1},\Ad_y(h_1),\Ad_y(h_2))$.\ Fixing $\,y\,$ and varying $\,x\,$ over $\,\txG$,\ we obtain,\ again,\ the entire $\,\txG\,$ as the first cartesian factor of a single orbit.\ Let $\,\la_i \in\sfk\,\xcA_{\rm W}(\ggt)$, $i=1,2\,$ be such that $\,h_i \in\xcC_{\la_i}$,\ and choose some $\,x,w \in \txG\,$ so that $\,h_1=\Ad_x(\ee_{\la_1})\,$ and $\,h_2 = \Ad_{x\cdot w}(\ee_{\la_2})$.\ Then,\ the $\txG\x\txG$-orbit of $\,(g,h_1,h_2)\,$ coincides with that of $\,(g,\ee_{\la_1},\Ad_w(\ee_{\la_2}))$.\ Any two such orbits for $\,w,w' \in \txG\,$ are equal iff $\,w' = u\cdot w\cdot v\,$ for some $\,u \in \cS_{\la_1}\,$ and $\,v \in \cS_{\la_2}$,\ where
\qq\nn
\cS_{\la_i}:=\{\ g\in\txG \quad\vert\quad \Ad_g(\ee_{\la_i})=\ee_{\la_i} \ \}
\qqq
is the $\Ad$-stabiliser of $\,\ee_{\la_i}$.\ This relation defines the double coset $\,\cS_{\la_1}\backslash \txG/\cS_{\la_2}$.\ Altogether,\ then,\ we obtain the decomposition
\qq\nn
  \unl\sfN{}_2\txG = \bigsqcup_{(\la_1,\la_2)\in \sfk\,\xcA_{\rm W}(\ggt)^{\x 2}}\,
  \bigsqcup_{[w] \in \cS_{\la_1}\backslash \txG/\cS_{\la_2}}\, \txG
  \times T_{(\la_1,\la_2)}^{[w]}  \,,
\qqq
where
\qq\nn
  T_{(\la_1,\la_2)}^{[w]} = \left\{\ \left(\Ad_y(\ee_{\la_1}),
  \Ad_{y\cdot w}(\ee_{\la_2})\right) \quad\big|\quad y \in \txG \
  \right\} \subset \xcC_{\la_1}  \times \xcC_{\la_2}\,.
\qqq
Define a map
\qq\nn
\rho_{(\lambda_1,\lambda_2)}\ :\ \cS_{\la_1} \backslash \txG /
\cS_{\la_2} \too \sfk\,\xcA_{\rm W}(\ggt)
\qqq
by taking $\,\rho_{(\lambda_1,\lambda_2)}([w])\,$ to be the unique element of the $\sfk$-rescaled Weyl alcove such that $\,\ee_{\la_1}\cdot \Ad_w(\ee_{\la_2}) \in\xcC_{\rho_{(\lambda_1,\lambda_2)}([w])}$.\ Fixing $\,[w]\in\rho_{(\lambda_1,\lambda_2)}^{-1}(\{\la\})\,$ and applying the face maps $\,\unl d{}^{(2)}_\cdot\ :\ \unl\sfN{}_2 \txG \too \unl\sfN{}_1 \txG$
({\it i.e.},\ the inter-bi-brane maps for $\,T_{2,1}$),\ we find
\qq \label{eq:T12-facemaps}
\alxydim{@C=2cm@R=1cm}{
&& (g\cdot h_1,h_2) \in \txG \times \xcC_{\la_2} \\
\txG\x T_{(\la_1,\la_2)}^{[w]} \ni (g,h_1,h_2) \ar[urr]^{\unl d{}^{(2)}_0} \ar[rr]^{\unl d{}^{(2)}_1}
\ar[drr]^{\unl d{}^{(2)}_2}
&& (g,h_1\cdot h_2) \in \txG \times \xcC_\la \\
&& (g,h_1) \in \txG \times \xcC_{\la_1}
}\,,
\qqq
and so $\,T_{(\la_1,\la_2)}^{[w]}\,$ sits in the intersection 
\qq\nn
T_{\la_1,\la_2}^\la:=\pr_{2,3}\bigl(\unl d{}^{(2)\,-1}_2\bigl(\txG\x\xcC_{\la_1}\bigr)\cap\unl d{}^{(2)\,-1}_0\bigl(\txG\x\xcC_{\la_2}\bigr)\cap\unl d{}^{(2)\,-1}_1\bigl(\txG\x\xcC_\la\bigr)\bigr)\equiv\bigl(\xcC_{\la_1}\x\xcC_{\la_1}\bigr)\cap\txm^{-1}(\xcC_\la)\supset T_{(\la_1,\la_2)}^{[w]}\,.
\qqq

With view to our subsequent analyses,\ we introduce two more maps on $\,\txG^{\x 2}$,\ to wit,
\qq\label{eq:W-and-La}
[W]\ :\ \txG^{\x 2}\too\bigsqcup_{(\la_1,\la_2)\in \sfk\,\xcA_{\rm W}(\ggt)^{\x 2}}\,\cS_{\la_1}\backslash\txG/\cS_{\la_2}\,,
\qquad\qquad\La\ :\ \txG^{\x 2}\too \sfk\,\xcA_{\rm W}(\ggt)
\qqq
with the defining properties
\qq\nn
(g_1,g_2)\in\xcC_{\la_1}\x\xcC_{\la_2}\ \Longrightarrow\ \left(\
(g_1,g_2)\in T_{(\la_1,\la_2)}^{[W](g_1,g_2)} \qquad \land\qquad g_1\cdot g_2
\in\xcC_{\La(g_1,g_2)} \ \right)\,.
\qqq
We have the obvious identity
\qq\nn
\rho_{(\lambda_1,\lambda_2)}\circ[W]\vert_{\xcC_{\la_1}\x
\xcC_{\la_2}}=\La\vert_{\xcC_{\la_1}\x \xcC_{\la_2}}\,.
\qqq
Note that for $\,\sug\,$ the class $\,[w]\,$ is uniquely fixed by $\,\la=\rho_{(\lambda_1,\lambda_2)}([w])$,\ so that $\,T_{(\la_1,\la_2)}^\la\equiv T_{(\la_1,\la_2)}^{[w]}\,$ is a single orbit.\ 
\item[$\bullet$] $n>2$:\ This case is treated analogously to that of $\,n=2$.\ For $\,\la_1,\la_2,\ldots,\la_n \in \sfk\,\xcA_{\rm W}(\ggt)\,$ and $\,w_2,w_3,\ldots,w_n \in \txG$,\ we define
\qq\label{eq:Tlawe}
T_{(\la_1,\la_2,\ldots,\la_n)}^{[w_2,w_3,\ldots,w_n]}=\left\{\
\left(\Ad_y(\ee_{\la_1}),\Ad_{y\cdot w_2}(\ee_{\la_2}),
\ldots,\Ad_{y\cdot w_{n}}(\ee_{\la_{n}})\right) \quad\vert\quad
y\in\txG\ \right\} \,,
\qqq
where $\,[w_2,w_3,\ldots,w_n]\,$ denotes an equivalence class of tuples $\,(w_2,w_3,\ldots,w_n)\in\txG^{\x n-1}$,\ with two tuples $\,(w_2,w_3,\ldots,w_n)\,$ and $\,(w_2',w_3',\ldots,w_n')\,$ considered equivalent iff $\,T_{(\la_1,\la_2,\ldots,\la_n)}^{[w_2,w_3,\ldots,w_n]}= T_{(\la_1,\la_2,\ldots,\la_n)}^{[w_2',w_3',\ldots,w_n']}$.\ In other words,\ $\,[w_2,w_3,\ldots,w_n] \in \cS_{\la_1}\backslash\x_{i=2}^{n}\,\left(\txG/\cS_{\la_i}\right)$,\ where the left action of $\,\txG\,$ (defining the left coset) is the diagonal left multiplication:\ $\bigl(\,g,\bigl(w_2\cdot\cS_{\la_2},w_3\cdot\cS_{\la_3}, \ldots,w_n\cdot\cS_{\la_n}\bigr)\bigr)\longmapsto\left(g\cdot w_2\cdot\cS_{\la_2} ,g\cdot w_3\cdot\cS_{\la_3},\ldots,g\cdot w_n\cdot\cS_{\la_n} \right)$.\ The decomposition of $\,\unl\sfN{}_n\txG\,$ into orbits can now be written as
\qq\nn
\unl\sfN{}_n\txG = \bigsqcup_{\overrightarrow\la\in\sfk\,\xcA_{\rm W}
(\ggt)^{\x n}}\,\bigsqcup_{[\overrightarrow w]\in\cS_{\la_1}
\backslash\x_{i=2}^{n}\,\left(\txG/\cS_{\la_i}\right)}\,\txG\times
T_{\overrightarrow\la}^{[\overrightarrow w]}\,.
\qqq
By construction,\ for a given $\,\overrightarrow\la:=(\la_1,\la_2,\ldots,\la_n)\,$ and every class $\,[\overrightarrow w]:=[w_2,w_3,\ldots,w_n]$,\ there exists a unique $\,\la\in\sfk\,\xcA_{\rm W}(\ggt)\,$ such that the orbit $\,\txG \times T_{\overrightarrow\la}^{[\overrightarrow w]}\,$ gets mapped to $\,\txG \times \xcC_\la\,$ on multiplication in the second factor.\ As previously,\ this gives rise to a map
\qq\nn
\rho_{\overrightarrow \lambda}\ :\ \cS_{\la_1}\backslash
\x_{i=2}^n\,\left(\txG/\cS_{\la_i}\right)\too\sfk\,\xcA_{\rm
W}(\ggt)
\qqq
with the defining property:
\qq \label{eq:mu[w]-def}
  \la \equiv \rho_{\overrightarrow \lambda}([\overrightarrow w])
  \quad \text{ is such that } \quad
  \ee_{\la_1} \cdot \Ad_{w_2}(\ee_{\la_2}) \cdot \cdots \cdot
  \Ad_{w_n}(\ee_{\la_n}) \in \xcC_\la \,,
\qqq
and we land in
\qq\label{eq:Tlala}
T_{\overrightarrow\la}^\la:=\x_{i=1}^n\,\xcC_{\la_i}\cap\txm_{1,2,\ldots,
n}^{-1}(\xcC_\la)\supset T_{\overrightarrow\la}^{[\overrightarrow w]}\,.
\qqq
\eit
By way of a closing remark,\ we note that the orbits $\,\txG\x T_{\overrightarrow\la}^{[\overrightarrow w]}\subset\unl\sfN{}_n\txG\,$ for $\,n>2\,$ come with a number of smooth ($\txG$-equivariant) maps to the `elementary' geometries $\,\txG\x T_{(\la_1,\la_2)}^{[w]}$.\ All these maps act as the identity $\,\id_\txG\,$ on the spectator factor $\,\txG$,\ and so we focus on their nontrivial cartesian factors supported on the $\,T_{\overrightarrow\la}^{[\overrightarrow w]}$.\ A map from the class that we have in mind is labelled by arbitrary integers $\,1\leq i\leq j< k\leq l\leq n\,$ and takes the following general form
\qq\label{eq:proj-nibb-elemibb}
\pi_{\overrightarrow\la}^{[\overrightarrow w]}\bigl(i,j\vert k,l\bigr)\ &:&\ T_{\overrightarrow\la}^{[
\overrightarrow w]}\too\bigsqcup_{(\la,\mu)\in\faff{\ggt}^{\x 2}}\,\bigsqcup_{[w]\in\cS_\la\backslash\txG/
\cS_\mu}\,T_{(\la,\mu)}^{[w]}\\\cr
&:&\ (h_1,h_2,\ldots,h_n)\longmapsto(h_i\cdot h_{i+1}\cdot\cdots
\cdot h_j,h_k\cdot h_{k+1}\cdot\cdots\cdot h_l)\,,\nonumber
\qqq
with the understanding that $\,\pi_{\overrightarrow\la}^{[\overrightarrow w]}(i,i\vert k,l>k)\equiv(\pr_i,\txm^{(n)}_{k\,k+1\ldots l}),\pi_{\overrightarrow\la}^{[\overrightarrow w]}(i,j>i\vert k,k)\equiv(\txm^{(n)}_{i\,i+1\ldots j},\pr_k)\,$ and $\,\pi_{\overrightarrow\la}^{[\overrightarrow w]}(i,i\vert j,j)\equiv(\pr_i,\pr_j)$.\ The maps are restrictions (which we have not made explicit,\ for the sake of transparency) of the corresponding (bi-)multiplications $\,\txG^{\x n}\too\txG^{\x 2}\,$ which we denote,\ accordingly,\ as
\qq\label{eq:mult-of-pi}
\txm^{(n)}\bigl(i,j\vert k,l\bigr)\ :\ \txG^{\x n}\too\txG^{\x 2}\,,\qquad\qquad\txm^{(n)}\bigl(i,j\vert k,l\bigr)\rstr_{T_{\overrightarrow\la}^{[\overrightarrow w]}}\equiv\pi_{\overrightarrow\la}^{[
\overrightarrow w]}\bigl(i,j\vert k,l\bigr)\,.
\qqq
Upon picking up an arbitrary $n$-tuple $\,(h_1,h_2,\ldots,h_n)\in T_{\overrightarrow\la}^{[\overrightarrow w]}$,\ define
\qq\nn
[W]_{\overrightarrow\la}^{[\overrightarrow w]}(i,j\vert k,l):=[W]\circ\pi_{\overrightarrow\la}^{[\overrightarrow w]}\bigl(i,j\vert k,l\bigr)(h_1,h_2,\ldots,h_n)
\qqq
and
\qq\nn
\La_{\overrightarrow\la}^{[\overrightarrow w]}(i,j):=\La\circ\pi_{\overrightarrow\la}^{[\overrightarrow w]}\bigl(i,j-1\vert j,j\bigr)(h_1,h_2,\ldots,h_n)\,,
\qqq
the two being manifestly independent of the choice of the $n$-tuple.\ We then readily see that
\qq\nn
\pi_{\overrightarrow\la}^{[\overrightarrow w]}(i,j\vert k,l)\left(
T_{\overrightarrow\la}^{[\overrightarrow w]}\right)=T_{(
\La_{\overrightarrow\la}^{[\overrightarrow w]}(i,j),
\La_{\overrightarrow\la}^{[\overrightarrow w]}(k,l))}^{[W
]_{\overrightarrow\la}^{[\overrightarrow w]}(i,j\vert k,l)}\,.
\qqq
The last observation shall be used in Sec.\,\ref{sub:fus-2iso} to induce cohomological structure on the $\,\cT_n\,$ from that existing on $\,\cT_2\,$ (under additional hypotheses). 

The above structure shall be reflected,\ in its entirety,\ in the construction of higher-geometric objects attached to a maximally (super)symmetric defect.\ Hence,\ we shall have more to say about its naturality later on,\ when we come to discuss the gerbe-theoretic components of the simplicial WZW background in the presence of the defects.\ But before we get there,\ we conclude this section with a refinement of our former treatment of the geometrisation of the Polyakov--Wiegmann identity.\medskip

To this end,\ we employ $\,\txG\la\,$ to induce,\ in the previously considered manner,\ the (pointwise) action of the body Lie group $\,|\txG_{\rm s}|\,$ on $\,\unl\sfN{}_\bullet\txG\,$ by the automorphisms
\qq\nn
|\ell\wp^{(n)}|_{(g,h)}\equiv\ell\wp^{(n)}\circ\bigl(\widehat g\x\widehat h\x\id_\txG\bigr)\ :\ \txG\equiv\bR^{0|0}\x\bR^{0|0}\x\txG\too\bigl(\txG\x\txG\bigr)\x\txG\too\txG\,,
\qqq
labelled by $\,(g,h)\in|\txG\x\txG|\equiv|\txG|\x|\txG|$,\ and
\qq\nn
|\ell^{(n)}|_g\equiv\ell^{(n)}\circ\bigl(\widehat g\x\id_\txG\bigr)\ :\ \txG\equiv\bR^{0|0}\x\txG\too\txG\x\txG\too\txG\,,
\qqq
labelled by $\,g\in|\txG|$,\ respectively,\ alongside the standard action of the respective Lie superalgebras by superderivations.\ On the object supermanifold $\,\txG\,$ of $\,\txG\rx\txG$,\ the latter is implemented by the fundamental vector fields of $\,\txG\la^{(0)}$,\ {\it i.e.},
\qq\nn
\ggt\oplus\ggt\ni(X,Y)\longmapsto\cK^\txG_{(X,Y)}=X^A\,R_A-Y^A\,L_A
\qqq
for $\,(\txG_{\rm s},\txG\la)=(\txG^{\x 2},\ell\wp)\,$ with $\,\ggt_s=\ggt\oplus\ggt$,\ and 
\qq\nn
\ggt\ni X\longmapsto\cK^\txG_X=X^A\,R_A\,,
\qqq
for $\,(\txG_{\rm s},\txG\la)=(\txG,\ell)\,$ with $\,\ggt_s=\ggt$,\ where $\,L_A=(\id_{\cO_\txG}\ox t_A)\circ\txm^*\,$ and $\,R_A=\bigl(t_A\ox\id_{\cO_\txG}\bigr)\circ\txm^*\,$ are the basis right- and left-invariant vector fields,\ respectively.\ These lift to the higher rungs of the simplicial supermanifold (with $\,\cO_{\txG^{\x n+1}}\equiv\cO_\txG^{\ox n+1}$) as
\qq\nn
\ggt\oplus\ggt\ni(X,Y)\longmapsto\cK^{\txG^{\x n+1}}_{(X,Y)}=\cK^\txG_{(X,Y)}\ox\id_{\cO_\txG}^{\ox n}+Y^A\,\sum_{k=1}^n\,\id_{\cO_\txG}^{\ox k}\ox\bigl(R_A-L_A\bigr)\ox\id_{\cO_\txG}^{\ox n-k}
\qqq
and
\qq\nn
\ggt\ni X\longmapsto\cK^{\txG^{\x n+1}}_X=\cK^\txG_X\ox\id_{\cO_\txG}^{\ox n}\,,
\qqq
respectively.

At this stage,\ we may meaningfully restrict each row of the bicomplex \eqref{diag:bicG} that captures (the de Rham representative of) the characteristic class of the (generalised) multiplicative structure on the 1-gerbe to the subcomplex composed of $\txG_{\rm s}$-invariant super-differential forms on the corresponding component supermanifold $\,\sfN_n\txG^{\rm op}$,
\qq\nn
\bigl(\Om^\bullet\bigl(\sfN_n\txG^{\rm op}\bigr)^{\txG\la^{(n-1)}},\sfd_{\rm dR}^{(\bullet)}\bigr)\qquad :\qquad \Om^0\bigl(\txG^{\x n}\bigr)^{\txG\la^{(n-1)}}\xrightarrow{\ \sfd\ }\Om^1\bigl(\txG^{\x n}\bigr)^{\txG\la^{(n-1)}}\xrightarrow{\ \sfd\ }\cdots\xrightarrow{\ \sfd\ }\Om^p\bigl(\txG^{\x n}\bigr)^{\txG\la^{(n-1)}}\xrightarrow{\ \sfd\ }\cdots\,.
\qqq
Here,\ $\,\txG\la^{(-1)}\equiv\pr_2\ :\ \txG_{\rm s}\x\bR^{0|0}\too\bR^{0|0}\,$ and $\,\Om^0(\txG)^{\txG\la^{(0)}}\,$ are just (locally) constant sections of the structure sheaf $\,\cO_\txG\,$ of $\,\txG$.\ Note that the cohomology of the subcomplex for $\,(\txG_{\rm s},\txG\la)=(\txG,\ell)\,$ and $\,n=1\,$ is the {\bf Cartan--Eilenberg cohomology of} $\,\txG\,$ that we shall stumble upon presently,\ and so denote as $\,{\rm CaE}^\bullet(\txG)\equiv H^\bullet_{\rm dR}(\txG,\bR)^\ell\equiv H^\bullet_{\rm dR}(\txG,\bR)^{{\rm L}(\txG)}$.\ The restriction cannot be extended to the entire bicomplex in a manner that defines a $\txG_{\rm s}$-invariant refinement of the associated cohomology because of the presence of the extra structural (degeneracy and face) map at each level of $\,\sfN_\bullet\txG^{\rm op}\,$ (relative to $\,\unl\sfN{}_\bullet\txG$) which is readily seen to be non-equivariant with respect to the $\txG_{\rm s}$-actions present.\ Nevertheless,\ it makes perfect sense to consider

\bedef\label{def:susy-mult-str}
Adopt the hitherto notation.\ A generalised multiplicative structure $\,(\cM,\a)\,$ on a Cartan--Eilenberg super-1-gerbe $\,\cG=(\sfY\txG,\pi_{\sfY\txG},\txB,L,\pi_L,\txA,\mu_L)\,$ over a Lie supergroup $\,\txG\,$ associated with a 4-cocycle $\,\cZ_\cG\,$ is termed $\txG\la^{(\bullet)}-${\bf su\-per\-sym\-met\-ric} if the tensorial data $\,\cZ_\cG\,$ are ($\txG\la^{(\bullet)}-$)LI,\ the structural 1-isomorphism $\,\cM\,$ is a $\txG\la^{(2)}$-supersymmetric 1-gerbe 1-isomorphism,\ and the 2-isomorphism $\,\a\,$ is a $\txG\la^{(3)}$-supersymmetric 1-gerbe 2-isomorphism.
\exdef

\noindent The physical rationale behind the requirement that the generalised multiplicative structure be $\txG\la^{(\bullet)}$-su\-per\-sym\-met\-ric shall become clear in the next section in which we turn to un-graded physics with view to developing an intuition as to how the categorification of the Lie supergroup structure discussed above can be employed in the construction of the multi-phase WZW super-$\si$-model with the maximally supersymmetric defect implementing the right action of the Lie supergroup $\,\txG\,$ on itself.

\part{The maximally symmetric WZW defects and the CS theory}\label{p:WZWCS}

Below,\ we place the abstract considerations and constructions of the previous part in the concrete and well-studied context of the Wess--Zumino--Witten $\si$-model of \Rcite{Witten:1983ar},\ with the bulk target space given by a compact simple 1-connected Lie group.\ The conceptual framework set up previously paves the way to an elaboration of a programme,\ pioneered by Gaw\c{e}dzki in the seminal \Rcite{Gawedzki:1987ak} and continued and extended by His students and collaborators,\ and in particular by Fuchs,\ Schweigert and Waldorf in \Rcite{Fuchs:2007fw},\ and by Runkel and the Author in Refs.\,\cite{Runkel:2008gr,Runkel:2009sp},\ of decoding the constitutive algebraic data of the underlying 2d (quantum) Rational Conformal Field Theory from the higher-geometric and -categorial gerbe-theoretic constructs which arise in a rigorous approach to the definition and application of the topological Wess--Zumino term in the model's Dirac--Feynman amplitude,\ and to its extension to the situation in which defects compatible with the rich bi-chiral bulk symmetry enter the field-theoretic picture.\ In the course of the present study,\ we identify -- conjecturally yet with a solid partial evidence to lean upon -- new elements of the said higher-geometric `code' through a detailed analysis,\ organised by the general considerations of Part \ref{p:genstr},\ of the geometric and gerbe-theoretic data of the maximally symmetric inter-bi-brane (and the associated fusion 2-isomorhisms),\ which,\ remarkably but understandably (from the point of view of the RCFT and its functorial quantisation),\ lands us in the setting of the 3d topological gauge field theory of Chern and Simons coupled to a Wilson line defect.\ It is also to be noted that structures discussed hereunder provide a very concrete realisation of the idea of representing symmetries of a field theory by defects,\ the symmetry in question being the configurational symmetry induced from the right regular action of the target Lie group on itself.

\section{The geometry of the un-graded maximally symmetric WZW defect}\label{sec:wzw}

As emphasised at the beginning of the closing section of the general Part \ref{p:genstr},\ the category of Lie (super)groups and their homogeneous spaces is a rich source of examples of highly symmetric geometries.\ The corresponding bosonic two-dimensional field theories,\ with the said geometries as targets,\ have long been known under the name of bosonic (gauged) Wess--Zumino--Witten (WZW) $\si$-models and explored extensively since their inception in Refs.\,\cite{Witten:1983ar,Knizhnik:1984nr,Gepner:1986wi} (the pure Lie-group geometry) and Refs.\,\cite{Goddard:1984vk,Goddard:1986ee,Bardakci:1987ee,Gawedzki:1988nj}.\ While the ones on homogeneous spaces are quite heavily structured as field theories ({\it cf.},\ in particular,\ \Rcite{Gawedzki:1988nj}),\ the models of loop dynamics in group manifolds admit a fairly straightforward description in terms of elementary constructs of Cartan's differential calculus and their higher geometrisations.\ Among other things,\ the issue of (maximal) symmetry of the associated backgrounds has been investigated at great length over the years,\ and shall be recapitulated and elaborated below with view to illustrating the former abstract definitions and to developing intuitions to be invoked in the $\bZ/2\bZ$-graded setting of interest.

Thus,\ in this section,\ we discuss the $\txG\x\txG$-invariant simplicial submanifold $\,\cT_\bullet\,$ of $\,\unl\sfN{}_\bullet \txG\,$ which can serve as the target space of the WZW $\si$-model with a maximally symmetric defect,\ starting with the target-space of an empty defect quiver $\,\G=\emptyset\,$ and then passing to the bi-brane worldvolume for the maximally symmetric defect line.\ This paves the way to a proposal for the worldvolume of the maximally symmetric inter-bi-brane associated with junctions of an arbitrary number of such defect lines,\ which we provide towards the end of the section.\ In its derivation,\ we invoke the three-dimensional Chern--Simons theory that has long been known to encode the structure of the topological sector of the WZW $\si$-model in the presence of the maximally symmetric boundary defect. Further concrete and circumstantial evidence for our proposal shall be provided in the next section in which we introduce the full-fledged higher-geometric structure for the defect.\medskip

Prior to launching the technical analysis of the reference field theory with defects,\ we note that its standard formulation,\ while being essentially equivalent to the Nambu--Goto one introduced in Def.\,\ref{def:ssimod},\ differs from the latter in that it presupposes the existence of a metric $\,\g\,$ on the worldsheet $\,\Si\,$ and employs the metric term in the Polyakov form: 
\qq\nn
S_{\si,{\rm metr}}^{\rm P}[\xi]=-\tfrac{1}{2}\,\Vert\sfd\xi\Vert_{(\g,\txg)}^2\,,\qquad\qquad\Vert\sfd\xi\Vert_{(\g,\txg)}^2=\int_\Si\,\bigl(\wedge\ox\txg_\xi\bigr)\bigl(\sfd\xi,\bigl(\star_\g\ox\id_{\xi^*\sfT\xi(\Si)}\bigr)(\sfd\xi)\bigr)
\qqq
which computes the norm squared of the tangent operator $\,\sfd\xi=\sfd\si^i\ox\xi_*\p_i\in\G(\sfT^*\Si\ox_{\Si,\bR}\xi^*\sfT\xi(\Si))\,$ (thus,\ the latter has the $\sfT\xi(\Si)$-legs contracted with the help of the target-space metric and the $\sfT^*\Si$-legs,\ of which the second one has been transformed by the Hodge operator $\,\star_\g\,$ for the worldsheet metric $\,\g$,\ wedged to yield a top-degree form on $\,\Si$).\ In what follows,\ we denote the counterpart of $\,\cF(\Si,d)\,$ for the $\si$-model in the Polyakov formulation as $\,\cF(\Si,d;\g)$,\ and use the symbol $\,\cF_{\rm eom}(\Si,d;\g)\,$ to refer to those field configurations which satisfy the ensuing field equations.

The reformulation of the field theory entails a parallel restatement of the field-theoretic description of the defects.\ Thus,\ we now consider a linear map (written in the previously introduced notation)
\qq\nn
\widetilde N\ :\ \G(\sfT\Si)\too\G(\sfT\Si)\ :\ v(\cdot)=v^i(\cdot)\,\p_i\longmapsto-\sqrt{|\det\,\g_\cdot|}\,v^k(\cdot)\,\bigl(\g_\cdot^{-1}\bigr)^{ij}\,\ep_{jk}\,\p_i
\qqq
which squares to $\,\widetilde N\circ\widetilde N=\id_{\G(\sfT\Si)}\,$ and satisfies $\,g\bigl(v,\widetilde Nv\bigr)=0\,$ and $\,\g\bigl(\widetilde Nv,\widetilde Nv\bigr)=-\g(v,v)$.\ We also have $\,v\con\widetilde Nv\con\sqrt{|\det\,\g|}\,\sfd\si^1\wedge\sfd\si^2=\g(v,v)$,\ so that the pair $\,(\widetilde Nv,\g(v,v)^{-1}\,v)\,$ defines the orientation of the minkowskian worldsheet (equivalent to $\,(\p_1,\p_2)$) for any $\,v\,$ which is  not light-like.\ Given the above and for an arbitrary $\,p\in \Sigma^{(1)}$,\ $\,t \in \sfT_p \Sigma^{(1)}\,$ and $\,V\in\sfT_{\xi(p)}Q$,\ the DGC takes the form
\qq\label{eq:DGC-WZW}
\txg_{\xi_{|2}(p)}(\iota_{2*}V,\xi_{|2\,*}(p)(\widetilde Nt))-\txg_{\xi_{|1}(p)}(\iota_{1*}V,\xi_{|1\,*}(p)(\widetilde Nt))= \om_{\xi(p)}(V,\xi_*(p)t)\,,
\qqq
with $\,\xi_{|A\,*}(p)(\widetilde Nt)\,$ understood as the limit of $\,\xi_{*}(x)(\widetilde Nt)\,$ as $\,x \in \Sigma^{(2)}\,$ approaches $\,p\,$ from the appropriate side (namely, inside $\,U_A\,$ in the notation
of condition (F1)).\ This condition was derived in \Rcite{Runkel:2008gr} by varying the $\si$-model action
functional\footnote{In \Rcite{Runkel:2008gr},\ a euclidean world-sheet metric was used,\ but this only affects the metric (`kinetic') term and it is straightforward to adapt the calculation in \Rxcite{App.\,A.2}{Runkel:2008gr} to the present definition of the action functional.}.\medskip

\subsection{The geometry of the bulk WZW
$\si$-model}\label{sub:WZW-target-space}

The target space of the WZW $\si$-model is the group manifold of the compact simple 1-connected Lie group $\,\Gx$,\ on which we take,\ as tensorial data of the background,\ the Cartan--Killing metric
\qq\label{eq:g-coord}
\txg\equiv\txg_{\rm CK}^{(\sfk)}=-\tfrac{\sfk}{4\pi}\,\bigl(\id_{\Om^1(\txG)^{\ox 2}}\ox\tr_\ggt\bigr)\bigl(\theta_{\rm L}\ox\theta_{\rm L}\bigr)=
\tfrac{\sfk}{8\pi}\,\d_{AB}\,\theta_{\rm L}^A\ox\theta_{\rm L}^B\,,\qquad\sfk\in\bR_{>0}\,,
\qqq
and the $\sfk$-fold multiple
\qq\nn
\txH^{(\sfk)}_{\rm C}=\sfk\,\txH_{\rm C}
\qqq
of the representative
\qq\label{eq:H-coord}
\txH_{\rm C}=\tfrac{1}{12\pi}\,\tr_\ggt\bigl(\theta_{\rm L}\wedge\theta_{\rm L}
\wedge\theta_{\rm L}\bigr)=-\tfrac{1}{48\pi}\,f_{ABC}\,\theta_{\rm L}^A\wedge\theta_{\rm L}^B\wedge\theta_{\rm L}^C
\qqq
of the generating class in the cohomology group $\,H^3_{\rm dR}(\txG,2\pi\bZ)\cong2\pi\bZ\,$ of a compact simple 1-connected Lie group $\,\txG$,\ known as the Cartan 3-form and written in terms of the standard left-invariant Maurer--Cartan 1-form $\,\theta_{\rm L}\equiv\theta_{\rm L}^A\ox t_A\in\Om^1(\Gx)\ox\ggt\,$ on $\,\txG\,$ and the trace\footnote{Whenever we are dealing with a $\ggt$-valued element of a quotient $\,\xcI\,$ of the tensor algebra of $\,\Om^1(X)\,$ on a manifold $\,X$,\ we abuse the notation mildly and write $\,\tr_\ggt\,$ instead of $\,\id_\xcI\ox\tr_\ggt\,$ in order to unburden our formul\ae.} $\,\tr_\ggt\,$ on $\,\ggt\,$ which we normalise so that the relation $\,\tr_\ggt(t_A\,t_B)=-\frac{1}{2}\,\d_{AB}\,$ obtains for the generators $\,t_A\,$ of $\,\ggt$,\ the latter satisfying the defining commutation relations $\,[t_A,t_B]=f_{ABC}\,t_C$,\ with $\,f_{ABC}\,$ the structure constants of $\,\ggt\,$ (this yields the standard matrix trace for,\ {\it e.g.},\ $\,\Gx={\rm SU}(2)$).\ In fact,\ the two tensors admit an equivalent definition in terms of the right-invariant Maurer--Cartan 1-form:
\qq\label{eq:WZW-L-is-R}
\txg_{\rm CK}^{(\sfk)}\equiv-\tfrac{\sfk}{4\pi}\,\bigl(\id_{\Om^1(\txG)^{\ox 2}}\ox\tr_\ggt\bigr)\bigl(\theta_{\rm R}\ox\theta_{\rm R}\bigr)\,,\qquad\qquad\txH_{\rm C}^{(\sfk)}\equiv\tfrac{\sfk}{12\pi}\,\tr_\ggt\bigl(\theta_{\rm R}\wedge\theta_{\rm R}\wedge\theta_{\rm R}\bigr)\,,
\qqq
and so we conclude that\vspace{12pt}
\begin{center}
\fbox{\begin{minipage}[c]{9.5cm} 
\noindent {\it The bulk target space of the WZW $\si$-model,}
\qq\nn
M=\cT_0=\unl\sfN{}_0\txG\,,
\qqq 
{\it is a single orbit of the action}
\qq\nn
\ell\wp^{(0)}\ :\ (\txG\x\txG)\x\unl\sfN{}_0\txG\too\unl\sfN{}_0\txG\,,
\qqq
{\it endowed with $\ell\wp$-invariant tensorial data $\,(\txg_{\rm CK}^{(\sfk)},\txH_{\rm C}^{(\sfk)})$,}
\qq\nn
\forall_{(x,y)\in\txG\x\txG}\ :\
\bigl(\ \ell\wp_{(x,y)}^*\txg_{\rm CK}^{(\sfk)}=\txg_{\rm CK}^{(\sfk)}\quad\land\quad\ell\wp_{(x,y)}^*\txH_{\rm C}^{(\sfk)}=\txH_{\rm C}^{(\sfk)}\ \bigr)\,.
\qqq
\end{minipage}}
\end{center}
~\vspace{12pt}

The normalisation of the metric tensor and of the associated metric term of the action functional \emph{relative to} the topological term,\ parametrised by the so-called {\bf level} $\,\sfk$,\ is determined by the requirement of a non-anomalous conformal symmetry of the quantised theory,\ {\it cf.}\ \Rcite{Witten:1983ar}.\ The quantisation of the level in the latter term ({\it i.e.},\ the restriction $\,\sfk\in\bN^\x\subset\bR_{>0}\,$ of the domain of the level),\ on the other hand,\ follows from topological considerations (and so could be skipped at this stage):\ It ensures that the Cartan 3-form has periods in $\,2\pi\bZ\,$ and becomes a representative of the generator class in $\,H^3(\txG,2\pi\bZ)\cong 2\pi\bZ$.\ Its cohomological nontriviality,\ thus constrained,\ leads to the appearance of the associated gerbe in the lagrangean description of the $\si$-model,\  {\it cf.}\ \Rcite{Gawedzki:1987ak}.\ The relevant differential-geometric structure realising the de Rham class $\,[\txH_{\rm C}^{(\sfk)}]\,$ shall be indicated in Sec.\,\ref{sub:WZW-target-bulk}.\ Here,\ we note that the normalisation of the Cartan--Killing metric fixes the size of the group manifold and confine ourselves to the ensuing field equations.\medskip

Let us put the WZW $\si$-model,\ with $\,M=\txG\,$ and the tensorial data introduced above,\ on the cylinder $\,\Si\cong\bR\x\bS^1\,$ with rectilinear coordinates $\,\{\si^i\}^{i\in\{1,2\}}\,$ and the minkowskian metric $\,\g\equiv\eta=-\sfd\si^1\ox\sfd\si^1+\sfd\si^2\ox\sfd\si^2$,\ and write
\qq\label{eq:Convgdg}
g^{-1}\,\p_i g:=\p_i\con g^*\theta_{\rm L}\,,\qquad\qquad\p_i g\,g^{-1}:=\p_i\con g^*\theta_{\rm R}
\qqq
for an arbitrary $\,g\in[\Si,\txG]\,$ in what follows.\ By varying the DF amplitude of the ensuing bulk WZW $\si$-model,\ we obtain the equations for the embedding field $\,\xi\equiv g\in[\Si,\txG]\,$ in the compact form $\,(\eta^{ij}+\ep^{ij})\,\p_i\bigl(g^{-1}\,\p_j g\bigr)=0$,\ or,\ using the light-cone coordinates $\,\si^\pm=\si^2\pm\si^1\,$ and the corresponding derivatives $\,\p_\pm=\tfrac{\p\ }{\p\si^\pm}$,
\qq\nn
\p_+\bigl(g^{-1}\,\p_-g\bigr)=0\,.
\qqq
A general solution to this equation factorises as
\qq\label{eq:solution-bulk-EOM}
g(\si)=g_L(\si^+)\cdot g_R(\si^-)^{-1}\,,
\qqq
with $\,g_L\,$ and $\,g_R\,$ arbitrary $\Gx$-valued maps on $\,\bR\,$ with equal monodromies,\ {\it cf.}\ \Rcite{Gawedzki:2001rm}.

In addition to the conformal symmetry,\ that is a symmetry engendered by arbitrary worldsheet coordinate transformations
\qq\label{eq:conf-sym}
\si^\pm\longmapsto f_\pm(\si^\pm)
\qqq
under orientation-preserving diffeomorphisms $\,f_\pm\,$ of the circle (a remnant of the combined Weyl and diffeomorphism invariance of $\,\cF_{\rm eom}(\Si,d;\g)\,$ preserving the conformal class of the
minkowskian gauge $\,\g=\eta$),\ the bulk theory enjoys a level-$\sfk$ $\,\ggtk^{(L)}\oplus \ggtk^{(R)}\,$ Ka\v c--Moody symmetry,\ realised on fields through Poisson brackets with the chiral currents
\qq\nn
J_L(\si)=\tfrac{\sfk}{2\pi}\,g(\si)\,\p_+g(\si)^{-1}\,,\qquad
\qquad J_R(\si)=\tfrac{\sfk}{2\pi}\,g(\si)^{-1}\,\p_-g(\si)\,.
\qqq
Note that the currents become functions of the respective light-cone coordinates $\,\si^\pm\,$ upon using the equations of motion of the $\si$-model.\ The Ka\v c--Moody symmetry is an infinitesimal counterpart of\medskip

\noindent {\it Loop-group invariance:} In the WZW $\si$-model on an (oriented) worldsheet $\,\Si\,$ without defects,\ {\it i.e.},\ for $\,\Si^{(1)}\cup\Si^{(0)}=\emptyset\,$ and $\,d=\id_\Si$,\ the loop-group action
\qq
\sfL\ell\wp\ :\ (\sfL\txG\x\sfL\txG)\x[\Si,\txG]\too[\Si,\txG]\ :\ \left( (h_L,h_R),g\right)\longmapsto h_L^+\cdot g\cdot(h_R^-)^{-1}\,,\cr \label{eq:loop-sym}
\qqq
defined in terms of the maps $\,h_{L/R}^\pm:=h_{L/R}\circ\pi_\pm\,$ with $\,\pi_\pm(\si):=\si^\pm$,\ and lifting the geometric action $\,\ell\wp\,$ of \Reqref{eq:Gact-bulk} to the mapping space $\,[\Si,\txG]$,\ preserves $\,\cF_{\rm eom}(\Si,d;\g)$.

\subsection{The (untwisted) maximally symmetric WZW boundary}
\label{sub:maxym-b-def}

One possible criterion of characterisation of boundary conditions,\ or more general defect conditions satisfied by a field configuration of the WZW $\si$-model consists in specifying the scheme of reduction of the symmetries of the bulk model that occurs in the presence of the defect.\ In this brief review,\ we shall be concerned with those conformal defects of the WZW $\si$-model that preserve the maximal amount of the bulk symmetry,\ the latter being encoded in Eqs.\,\eqref{eq:conf-sym} and \eqref{eq:loop-sym}.\ Infinitesimally,\ the symmetry generators (currents) will span at least one copy of the Ka\v c--Moody algebra $\,\ggtk\subset\ggtk^{(L)}\oplus\ggtk^{(R)}$.\ Since the chiral components of the stress-energy tensor of the WZW $\si$-model can be written as squares of the respective chiral Ka\v c--Moody currents (up to a multiplicative constant),\ this ensures that a copy of the Virasoro algebra $\,{\rm Vir}\,$ embedded diagonally in the bulk symmetry algebra $\,{\rm Vir}^{(L)} \oplus{\rm Vir}^{(R)}\,$ and corresponding to worldsheet diffeomorphisms that leave the defect unchanged is preserved.\ In the boundary case,\ this leads us to consider the so-called maximally symmetric\footnote{Here, we consider solely the \emph{untwisted} maximally symmetric D-brane,\ distinguished by the diagonal character of the embedding $\,\ggtk\subset\ggtk^{(L)}\oplus\ggtk^{(R)}$,\ which corresponds to the reduction of the isometry group $\,\txG\x\txG\,$ of left- and right-regular translations in the target space $\,\txG\,$ to a single copy of $\,\txG\,$ acting by conjugation (the adjoint action) as the isometry group of the D-brane worldvolume.\ In general,\ one finds also one-sided translates of these in the group manifold,\ as well as orbits of the adjoint action twisted by an outer automorphism of the Lie algebra $\,\ggt\,$ as admissible maximally symmetric D-brane worldvolumes,\ {\it cf.}\ \Rcite{Recknagel:1997sb}.} WZW D-brane,\ with the worldvolume given by a sub-collection
\qq\label{eq:Qp-def}\qquad
Q_\p=\bigsqcup_{\la\in\faff{\ggt}}\,\xcC_\la\subset\bigsqcup_{\la\in\sfk\,\xcA_{\rm W}(\ggt)}\,\xcC_\la
\qqq
with labels from a subset $\,\faff{\ggt}\subset\sfk\,\xcA_{\rm W}(\ggt)$,\ to be (re-)established presently through a cohomological analysis.\ Each connected component of the D-brane worldvolume carries the structure of a homogeneous $\txG$-space with respect to the left action
\qq\nn
\ell\wp^\p\equiv\Ad\ :\ \txG\x\xcC_\la\too\xcC_\la\ :\ (x,g)\longmapsto x\cdot g\cdot x^{-1}\equiv\Ad_x(g)\,,
\qqq
which is an explicit geometric realisation of the bulk-boundary symmetry reduction $\,\txG\x\txG\searrow\txG$,\ given by the restriction of $\,\ell\wp\,$ to the diagonal,\ $\,\ell\wp^\p_x(g)\equiv\ell\wp_{(x,x)}(g)$.

In order to fix the D-brane curvature,\ we demand that the latter reduction admit a lift to the configuration space of the boundary WZW $\si$-model.  
\berop\label{prop:maxym-curv-bdry-existuniq}
In the WZW $\si$-model on a worldsheet $\,\Si\cong_d\Si^{(2)}\cup\Si^{(1)}\,$ with a (boundary-type) defect $\,\Si^{(1)}\neq\emptyset=\p\Si^{(1)}\,$ (without junctions) embedded in the worldvolume $\,Q_\p\,$ of the maximally symmetric WZW D-brane of curvature $\,\om_\p^{(\sfk)}$,\ the latter \emph{assumed $\ell\wp^\p$-invariant},\ the sets $\,\cF(\Si,d;\eta)\,$ and $\,\cF_{\rm eom}(\Si,d;\eta)\,$ admit a left action of $\,\sfL\txG\,$ of the form
\qq
\sfL\ell\wp^\p\ :\ \sfL\txG\x\cF_{({\rm eom})}(\Si,d;\eta)\too\cF_{({\rm eom})}(\Si,d;\eta)\ :\ (h,g)\longmapsto h^+\cdot g\cdot
(h^-)^{-1}=:\ups{h}g\cr \label{eq:LG-trafo-bdry}
\qqq
and the chiral currents of the bulk Ka\v c--Moody symmetry glue at the boundary as dictated by the identity
\qq\label{eq:bchirglueWZW}
(J_R-J_L)\vert_{\p\Si}=0
\qqq
iff the curvature restricts to the connected components of $\,Q_\p\,$ as
\qq\label{eq:om-WZW-b}
\om_{\p,\la}^{(\sfk)}\equiv\om_\p^{(\sfk)}\vert_{\xcC_\la}=\tfrac{\sfk}{8\pi}\,\tr_\ggt\left(\theta_{\rm L}
\wedge\left(\tfrac{\id_\ggt+\sfT_e\Ad_\cdot}{\id_\ggt-\sfT_e\Ad_\cdot}\right)\circ\theta_{\rm L}\right)\,.
\qqq
\eerop
\beroof
{\it Cf.}\ App.\,\ref{app:proof1}. \eroof

\brem\label{rem:Llift-cond-b-nontriv}
It deserves to be emphasised that \Reqref{eq:om-b-triv-H} in conjunction with the condition that $\,\om_\p^{(\sfk)}\,$ be $\ell\wp^\p$-invariant does not fix the D-brane curvature uniquely,\ and so the additional requirement stated in the thesis of the proposition plays a nontrivial r\^ole.\ Indeed,\ the former two only determine $\,\om_\p^{(\sfk)}\,$ up to an $\ell\wp^\p$-invariant 2-cocycle on each connected component $\,\xcC_\la\,$ of the D-brane worldvolume.\ Each such 2-cocycle defines a class in the second cohomology group $\,H^2_{\rm dR}(\xcC_\la;\bR)^{\Ad_\cdot}\,$ of the complex of $\Ad_\cdot$-invariant forms on the conjugacy class $\,\xcC_\la$.\ From the assumed compactness and connectedness of $\,\txG$,\ we infer -- upon invoking Theorem 2.3 of \Rcite{Chevalley:1948} (see also,\ {\it e.g.},\ Theorem I of \Rxcite{Chap.\,IV, \S\,1, Sec.\,4.3}{Greub:1973g}) -- the existence of an isomorphism between the said cohomology group and the standard second de Rham cohomology group,\ $\,H^2_{\rm dR}(\xcC_\la;\bR)^{\Ad_\cdot}\cong H^2_{\rm dR}(\xcC_\la;\bR)$.\ The de Rham groups of (non-singular) conjugacy classes are not trivial in general,\ hence we have a whole wealth of cohomologically non-equivalent solutions to the problem in hand prior to imposing the requirement of the existence of a loop-group lift of the symmetry-reduction scheme. \erem

\subsection{The maximally symmetric WZW defect}\label{sub:maxym-nb-def}

We continue our discussion of the maximally symmetric defects of the WZW $\si$-model,\ passing now to the non-boundary case and focusing on the defect implementing the group multiplication in the manner discussed earlier.\ The corresponding geometric structure was first identified in \Rcite{Fuchs:2007fw}.\ Its existence follows directly from the old observation:\ The Cartan 3-form \eqref{eq:H-coord} satisfies the Polyakov--Wiegmann identity\footnote{The identity was first shown and used in \Rcite{Polyakov:1984et}.} $\,\txm^*\txH_{\rm C}=\pr_1^*\txH_{\rm C}+\pr_2^*\txH_{\rm C}-\sfd\varrho\,$ ({\it cf.}\ \Reqref{eq:PolWiegH}) in which 
\qq\label{eq:PolWieg-2}
\varrho=\tfrac{1}{4\pi}\,\tr_\ggt\bigl(\pr_1^*\th_{\rm L}\wedge\pr_2^*\th_{\rm R}\bigr)\in\Om^2\bigl(\txG^{\x 2}\bigr)
\qqq
is such that the familar identity
\qq\label{eq:D22Grho}
\D^{(2;2)}_\txG\varrho=0
\qqq
holds true.\ Accordingly,\ we may adduce the reasoning from the beginning of Sec.\,\ref{sub:homocat} in conjunction with the old results quoted in the previous section and tentatively write the worldvolume of the maximally symmetric WZW bi-brane in the form 
\qq\label{eq:maxym-bib-wvol}
Q=\txG\x Q_\p\,,
\qqq
postponing the discussion of the existence of the requisite higher-geometric structure to Sec.\,\ref{sec:WZW-target}.\ Note that each of its connected components,\ $\,Q_\la=\txG\x\xcC_\la$,\ is endowed with the structure of a homogeneous $\txG\x\txG$-space with respect to the left action
\qq\label{eq:Gact-bib}
\ell\wp^{(1)}\ :\ (\txG\x\txG)\x Q_\la\too Q_\la\ :\
\bigl((x,y),(g,h)\bigr)\longmapsto\left(x\cdot g\cdot
y^{-1},\Ad_y(h)\right)\,.
\qqq
We encountered it previously in Sec.\,\ref{sub:WZWorbs}.\ It is $\txG\x\txG$-equivariantly isomorphic with the biconjugacy class $\,\xcB_\la=\{\ \left(x\cdot\Ad_y(\ee_\la),x\right) \ \vert \ x,y\in \txG \ \}$,\ originally postulated as a component worldvolume in \Rcite{Fuchs:2007fw}.\ The isomorphism is explicitly given by the map $\,mu_\la\ :\ Q_\la\too\xcB_\la\ :\ (g,h)\longmapsto(g\cdot h,g)$,\ and it is readily seen to intertwine the left $\txG\x\txG$-action $\,\D\ell\wp\ :\ (\txG\x\txG)\x\xcB_\la\too\xcB_\la\ :\ ((x,y)(g_1,g_2))\longmapsto(x\cdot g_1\cdot y^{-1},x\cdot g_2\cdot y^{-1})\,$ on $\,\xcB_\la\,$ with $\,\ell\wp^{(1)}$,\ {\it i.e.},\ $\,\mu_\la\circ\ell\wp^{(1)}_{(x,y)}=\D\ell\wp_{(x,y)}\circ\mu_\la$.\ The advantage of our model of the bi-brane worldvolume is the ability to incorporate it into the simplicial framework laid out in Sec.\,\ref{sub:simpltargsp} and \ref{sec:smultcat},\ which neatly organises our subsequent search for a candidate structure to be associated with defect-line intersections.

The bi-brane maps for the bi-brane worldvolume of \Reqref{eq:maxym-bib-wvol} are
\qq\label{eq:maxym-nb-bib-maps}
\iota_1=\pr_1\,,\qquad\qquad\iota_2=\txm\,,
\qqq
{\it i.e.},\ restrictions of the degeneracy maps of the nerve $\,\unl\sfN{}_\bullet\txG\,$ at level 1.\ As discussed earlier,\ they intertwine the $\txG\x\txG$-action $\,\ell\wp^{(1)}\,$ structure on $\,Q\,$ with the action $\,\ell\wp\,$ on $\,\txG$,
\qq\label{eq:intertwact-iota}
\iota_\a\circ\ell\wp^{(1)}_{(x,y)}=\ell\wp_{(x,y)}\circ\iota_\a\,.
\qqq
~\medskip

In our search for the bi-brane curvature,\ we are guided by an analogon of the symmetry(-lift) principle employed in the boundary setting.\ We thus arrive at 
\berop\label{prop:maxym-curv-existuniq}
In the WZW $\si$-model on a worldsheet $\,\Si\cong_d\Si^{(2)}\cup\Si^{(1)}\,$ with a defect $\,\Si^{(1)}\neq\emptyset=\p\Si^{(1)}\,$ (without junctions) embedded in the worldvolume $\,Q\,$ of the maximally symmetric WZW bi-brane of curvature $\,\om^{(\sfk)}$,\ the latter \emph{assumed $\ell\wp^{(1)}$-invariant},\ the sets $\,\cF(\Si,d;\eta)\,$ and $\,\cF_{\rm eom}(\Si,d;\eta)$,\ with elements represented by pairs $\,(\xi\vert_{\Si^{(2)}},\xi\vert_{\Si^{(1)}})\equiv\left(g_{\rm b},(g,h)\right)$,\ admit a left action of $\,\sfL\txG\x\sfL\txG\,$ of the form
\qq
\sfL\ell\wp^{(1)}\ &:&\ (\sfL\txG\x\sfL\txG)\x\cF_{({\rm eom})}(\Si,d;\eta)\too\cF_{({\rm eom})}(\Si,d;\eta)\cr
&&\label{eq:LGLG-trafo}\\
&:&\ \left((h_L,h_R),\left(g_{\rm b},(g,h)\right)\right)\longmapsto\left(h_L^+\cdot g_{\rm b}\cdot(h_R^-)^{-1},\left(h_L^+\cdot g\cdot(h_R^-)^{-1},\Ad_{h_R^-}(h) \right)\right) \nonumber
\qqq
and the chiral currents of the bulk Ka\v c--Moody symmetry are continuous across the defect as expressed by the identities
\qq\label{eq:current-continuity-cond}
(J_{L\,1}-J_{L\,2})\vert_\G=0\,,\qquad\qquad(J_{R\,1}-J_{R\,2})\vert_\G=0
\qqq
written for 
\qq\nn
J_{L\,\a}(\si)=\tfrac{\sfk}{2\pi}\,g_{{\rm b}|\a}\,\p_+ g_{{\rm
b}|\a}^{-1}\,,
\qquad\qquad J_{R\,\a}(\si)=\tfrac{\sfk}{2\pi}\,g_{{\rm b}|\a}^{-1}\,\p_-
g_{{\rm b}|\a}
\qqq
with
\qq\label{eq:WZW-field-defext}
g_{{\rm b}|\a}\vert_\G=\iota_\a\circ(g,h)\,,
\qqq
iff the curvature restricts to the connected components of $\,Q\,$ as
\qq\label{eq:om-WZW-nb}
\om_\la^{(\sfk)}\equiv\om^{(\sfk)}\vert_{Q_\la}=\varrho^{(\sfk)}\vert_{Q_\la}-\pr_2^*\om_{\p,\la}^{(\sfk)}\,,
\qqq
where
\qq\label{eq:varrho}
\varrho^{(\sfk)}=\tfrac{\sfk}{4\pi}\,\tr_\ggt\bigl(\pr_1^*\th_{\rm L}\wedge\pr_2^*\th_{\rm R}\bigr)\,.
\qqq
\eerop
\beroof
{\it Cf.}\ App.\,\ref{app:proof2}. \eroof

\brem
The nontriviality of the requirement of the existence of a loop-group lift in the thesis of the proposition can be demonstrated with the help of arguments similar to those adduced in Remark \ref{rem:Llift-cond-b-nontriv}.\ Indeed,\ the compact and connected group $\,\txG\,$ is now replaced by the product group $\,\txG\x\txG\,$ with the same topological properties,\ and so we end up with the de Rham cohomology group $\,H^2_{\rm dR}(\txG\x\xcC_\la;
\bR)\,$ as the structure that enumerates cohomologically inequivalent $\ell\wp^{(1)}$-invariant 2-forms that satisfy
condition \eqref{eq:om-triv-HH}.\ By the K\"unneth Theorem,\ and in virtue of the 1-connectedness of both $\,\txG\,$ and the $\,\xcC_\la$,\ we obtain the result $\,H^2_{\rm dR}(\txG\x\xcC_\la;\bR)\cong H^2_{\rm dR}(\xcC_\la;\bR)$,\
which puts us back in the topological context discussed previously.
\erem

In summary,\vspace{12pt}
\begin{center}
\fbox{\begin{minipage}[c]{9.5cm} 
\noindent {\it The worldvolume of the maximally symmetric WZW bi-brane,}
\qq\nn
  Q=\cT_1\equiv\bigsqcup_{\la\in\faff{\ggt}}\,\txG \times \xcC_\la\subseteq
  \bigsqcup_{\la\in\sfk\,\xcA_{\rm W}(\ggt)}\,\txG \times \xcC_\la
  \subset \unl\sfN{}_1\txG  \,,
\qqq
{\it is a disjoint union of a family of orbits of the action}
\qq\nn
\ell\wp^{(1)}\ :\ (\txG\x\txG)\x\unl\sfN{}_1\txG\too\unl\sfN{}_1\txG\,,
\qqq
{\it endowed with an $\ell\wp^{(1)}$-invariant curvature $\,\om^{(\sfk)}$,}
\qq\nn
\forall_{(x,y)\in\txG\x\txG}\ :\ \ell\wp_{(x,y)}^{(1)\,*}=\om^{(\sfk)}\,,
\qqq
and $\txG\x\txG$-equivariant bi-brane maps
\qq\nn
\iota_1=\unl d{}^{(1)}_1\rstr_{\cT_1}\,,\qquad\qquad\iota_2=\unl d{}^{(1)}_0\rstr_{\cT_1}\,.
\qqq
\end{minipage}}
\end{center}

\subsection{Junctions of the maximally symmetric WZW defect}\label{sub:maxym-def-junct}

Our concrete choice of the boundary data made in Sec.\,\ref{sub:maxym-b-def} leads us to a \emph{proposal} for the worldvolume of the maximally symmetric WZW inter-bi-brane based on the \emph{tentative} identification of the WZW target space under reconstruction as a simplicial $\txG\x\txG$-space.\ Such identification enables us to turn the simplicial crank,\ worked out in all generality in Sec.\,\ref{sub:simpltargsp},\ and yields inter-bi-brane worldvolumes as $\txG\x\txG$-submanifolds (sub-families of $\txG\x\txG$-orbits)
\qq\nn
\cT_n\equiv T_{n,1}\subset\bigsqcup_{\overrightarrow\la\in\sfk\,\xcA_{\rm W}
(\ggt)^{\x n}}\,\bigsqcup_{[\overrightarrow w]\in\cS_{\la_1}
\backslash\x_{i=2}^{n}\,\left(\txG/\cS_{\la_i}\right)}\,\txG\times
T_{\overrightarrow\la}^{[\overrightarrow w]} 
\qqq
in the disjoint unions of $\txG\x\txG$-orbits found in Sec.\,\ref{sub:WZWorbs}.\ These are to be equipped with the inter-bi-brane maps which we read off from Def.\,\ref{def:simpl-target-sspace} and \Reqref{eq:GG-face-deg} in the form
\qq
\pi^{(n+1)}_{1,n+1}(g,h_1,h_2,\ldots,h_n)&=&(g,h_1\cdot h_2\cdot\cdots\cdot
h_n)\,, \cr && \label{eq:WZW-ibb-maps}\\
\pi^{(n+1)}_{k,k+1}(g,h_1,h_2,\ldots,h_n)&=&(g\cdot h_1\cdot h_2\cdot\cdots
\cdot h_{k-1},h_k)\,,\qquad k\in\ovl{1,n}\,.\nonumber
\qqq
Having the above map $\,\cT_n\,$ to the bi-brane worldvolume $\,Q\,$ of Sec.\,\ref{sub:maxym-nb-def} imposes further constraints upon the admissible inter-bi-brane worldvolumes.\ For example,\ in the case of $\,n=2$,\ we see from \Reqref{eq:T12-facemaps} that only those spaces $\,\txG \times T_{(\la_1,\la_2)}^{[w]}\,$ \emph{can} appear in $\,\cT_2\,$ for which $\,\la_1,\la_2,\rho_{(\la_1,\la_2)}([w])\in\faff{\ggt}$.\ In order to describe $\,\cT_2$,\ we introduce,\ for each triple $\,\la_1,\la_2,\la \in\faff{\ggt}$,\ a subset
\qq\nn
  \cF_{(\la_1,\la_2)}^\la
  \subset \cS_{\la_1}\backslash \txG/\cS_{\la_2}
\qqq
with the property
\qq\nn
[w] \in \cF_{(\la_1,\la_2)}^\la\quad\Longrightarrow\quad
\rho_{(\la_1,\la_2)}([w]) = \la\,.
\qqq
We emphasise that the set $\,\cF_{(\la_1,\la_2)}^\la\,$ need \emph{not} consist of all the $\,[w]\,$ satisfying the above
condition,\ in fact it could even be empty.\ A precise definition of $\,\cF_{(\la_1,\la_2)}^\la\,$ will be stated in Section
\ref{sub:fus-2iso} in terms of the existence of a certain gerbe 2-isomorphism.\ Altogether,
\qq\nn
  \cT_2 =  \txG \times \bigsqcup_{(\la_1,\la_2,\la) \in \faff{\ggt}^{\x 3}}\,
  \bigsqcup_{[w]\in \cF_{(\la_1,\la_2)}^\la}\,T_{(\la_1,\la_2)}^{[w]}
  \subset \unl\sfN{}_2\txG  \,.
\qqq
The sets $\,\cF_{(\la_1,\la_2)}^\la\,$ admit an obvious generalisation relevant to the definition of defect junctions of arbitrary valence,\ to be denoted as
\qq\nn
  \cF_{\overrightarrow\la}^\la
  \subset \cS_{\la_1}\backslash
\x_{i=2}^n\,\left(\txG/\cS_{\la_i}\right)\,,
\qqq
with the property
\qq\nn
[\overrightarrow w] \in
\cF_{\overrightarrow\la}^\la\quad\Longrightarrow\quad
\rho_{\overrightarrow\la}([\overrightarrow w]) = \la\,.
\qqq
These enter the definition of the inter-bi-brane worldvolumes
\qq\label{eq:ibb-n-wvol}
\cT_n = \txG\x\bigsqcup_{(\overrightarrow\la,\la)\in\faff{\ggt}^{\x
n+1}}\,\bigsqcup_{[\overrightarrow w]\in
\cF_{\overrightarrow\la}^\la}\,T_{\overrightarrow\la}^{[\overrightarrow
w]}\subset\unl\sfN{}_n\txG\,.
\qqq
As in the case of the $\,\cF_{\la_1,\la_2}^\la$,\ we assume that the set $\,\cF_{\overrightarrow\la}^\la\,$ is non-empty if a certain gerbe 2-isomorphism exists over the manifold $\,\txG\x T_{\overrightarrow\la}^{[\overrightarrow w]}\,$ for some $\,[\overrightarrow w]\in \cF_{\overrightarrow\la}^\la$.\ Let us remark that the $\,\cT_n\,$ can be ($\txG\x\txG$-equivariantly) equivalently described as a disjoint union
\qq\nn
\cT_n \cong \bigsqcup_{(\overrightarrow\la,\la)\in\faff{\ggt}^{\x
n+1}}\,\bigsqcup_{[\overrightarrow w]\in
\cF_{\overrightarrow\la}^\la}\,[Q_{\la_1}
{\,}_{\unl d{}^{(1)}_0}\hspace{-3pt}\x_{\unl d{}^{(1)}_1}Q_{\la_2}{\,}_{\unl d{}^{(1)}_0}\hspace{-3pt}\x_{\unl d{}^{(1)}_1}\cdots{\,}_{\unl d{}^{(1)}_0}\hspace{-3pt}\x_{\unl d{}^{(1)}_1}Q_{\la_n}]_{(\la,[\overrightarrow
w])}
\qqq
of full $\txG\x\txG$-orbits (labelled by the classes $\,[\overrightarrow w]$) contained in the preimages with respect to the map $\,\widetilde\pi{}^{(n+1)}_{1,n+1}:=\pi^{(n+1)}_{1,n+1}\circ\pr_{1,2,4,\ldots 2n}\,$ of the `outgoing' component bi-brane worldvolume $\,Q_\la\,$ within the product of the `incoming' component bi-brane worldvolumes $\,Q_{\la_i},\ i\in\ovl{1,n}\,$ fibred over $\,M\equiv \txG\,$ relative to the bi-brane maps $\,\unl d{}^{(1)}_i\ :\ Q\too M,\ i\in\{0,1\}$.\ The factors of the fibred product are ordered in the manner determined by the ordering of the corresponding defect lines converging at the defect junction.\medskip

In order to verify the above proposal,\ arisen at the conjunction of the simplicial `scheme' and elementary symmetry considerations,\ we must check that it solves the constraints \eqref{eq:def-jun-id},\ understood here as an identity defining the tangent distribution of the connected component $\,\txG\x T_{\overrightarrow\la}^{[\overrightarrow w]}\,$ of $\,\cT_n\,$ within the larger space $\,\txG\x T_{\overrightarrow\la}^\la\,$ with $\,\la={\rho_{\overrightarrow\la}([\overrightarrow w])}\,$ ({\it cf.}\ \Reqref{eq:Tlala}) as the kernel of the 2-form $\,\D_{\cT_n}\om^{(\sfk)}$.\ Just to remind the Reader:\ The identity is a necessary condition for the existence of the 1-gerbe 2-isomorphism $\,\varphi_{n+1}\,$ of Def.\,\ref{def:str-bgrnd},\ to be pulled back,\ along the map $\,X\ :\ \Si^{(0)}\too T$,\ to a junction $\,x_{n+1}\in\Si^{(0)}\,$ of $\,n+1\,$ lines of the maximally symmetric WZW defect.\ Hence,\ its satisfaction is crucial for the consistency of the multi-phase WZW $\si$-model.\ Taking into account the explicit form of the inter-bi-brane maps \eqref{eq:WZW-ibb-maps} in conjunction with the cohomological identity \eqref{eq:D22Grho},\ we obtain the formula
\qq\label{eq:DJI-form}
\D_{\cT_n}\om^{(\sfk)}=\pr_{2,3,\ldots,n+1}^*\Om_n^{(\sfk)}\,,
\qqq
written in terms of the universal 2-form
\qq\label{eq:Omn-def}
\Om_n^{(\sfk)}=\sum_{i=1}^{n-1}\,\txm^{(n)}\bigl(i,i\vert i+1,n\bigr)^*\varrho^{(\sfk)}
+\txm_n^*\om_{\p,\la}^{(\sfk)}-\sum_{j=1}^n\,\pr_j^*\om_{\p,\la_j}^{(\sfk)}\in\Om^2\bigl(T_{\overrightarrow\la}^\la\bigr)
\qqq 
on the space $\,T_{\overrightarrow\la}^\la\,$ of \Reqref{eq:Tlala} with the help of the maps \eqref{eq:mult-of-pi}.\ We readily establish

\berop\label{prop:Omn-in-kern}
Let $\,T_{\overrightarrow\la}^{[\overrightarrow w]}\subset T_{\overrightarrow\la}^\la\,$ be the manifolds defined in
\Reqref{eq:Tlawe} and \Reqref{eq:Tlala},\ respectively,\ and let $\,\Om_n^{(\sfk)}\,$ be the 2-form on $\,T_{\overrightarrow\la}^\la\,$ given in \Reqref{eq:Omn-def}.\ Then,\ for $\,\rho_{\overrightarrow\la}\,$ as in \Reqref{eq:mu[w]-def},\ we have
\qq\nn
\forall_{[w]\in\rho_{\overrightarrow\la}^{-1}(\{\la\})}\ :\
\Om_n^{(\sfk)}\vert_{T_{\overrightarrow\la}^{[\overrightarrow w]}}=0\,,\ i.e.,\qquad\qquad\G(\sfT T_{\overrightarrow\la}^{[\overrightarrow w]})\subset\ker\,\Om_n^{(\sfk)}\,.
\qqq
\eerop
\beroof
{\it Cf.}\ Appendix \ref{app:Omn-in-kern}. \eroof

\noindent The proof of the observation that vector fields on $\,T_{\overrightarrow\la}^\la\,$ tangent to the submanifold $\,T_{\overrightarrow\la}^{[\overrightarrow w]}\,$ annihilate $\,\Om_n^{(\sfk)}\,$ is completely straightforward.\ By contrast,\ it is far from obvious that the said vector fields span the \emph{whole} kernel of the 2-form over the larger spaces $\,T_{\overrightarrow\la}^\la\,$ (composed of full $\txG\x\txG$-orbits),\ so that the
$\,T_{\overrightarrow\la}^{[\overrightarrow w]}\,$ emerge as candidates for elementary inter-bi-brane worldvolumes.\ The choice of the smooth space $\,T_{\overrightarrow\la}^\la\,$ as the point of departure of our search for these elementary worldvolumes here is dictated by the worldsheet interpretation thereof as the connected components of the codomain of the $\si$-model field restricted to a defect junction $\,x_{n+1}\,$ with $n$ incoming defect lines with
field discontinuities from the respective conjugacy classes $\,\xcC_{\la_i}\,$ and a single outgoing one with the limiting value of the field jump given by the product $\,\pr_2\circ\pi_{1,n+1}\circ X(x_{n+1})=(\pr_2\circ\pi_{1,2})\circ X(
x_{n+1})\cdot(\pr_2\circ\pi_{2,3})\circ X(x_{n+1})\cdot\cdots\cdot(\pr_2\circ\pi_{n,n+1})\circ X(x_{n+1})$,\ {\it cf.}\ \Reqref{eq:WZW-ibb-maps}.\ Arguments of symmetry alone (other than `reducibility') leave unanswered the natural question as to whether connected components of the space $\,T_{\overrightarrow\la}^\la\,$ can be regarded as elementary inter-bi-brane worldvolumes or they split further into a disjoint union of the manifolds $\,\txG\x T_{\overrightarrow\la}^{[\overrightarrow w]}\,$ with $\,\rho_{\vec\la}([\overrightarrow w])=\la$.\ As complete $\txG_{\rm s}$-orbits,\ the latter are the \emph{smallest} manifolds on which the 2-isomorphisms $\,\varphi_{n+1}\,$ could exist in principle,\ consistently with the symmetries present.\ The leaves $\,\ceL\,$ of the characteristic foliation of $\,\Om_n^{(\sfk)}\,$ within $\,T_{\overrightarrow\la}^\la\,$ (some of them),\ on the other hand,\ define -- as we shall explain in detail in Sec.\,\ref{sub:fus-2iso} -- the \emph{largest} possible connected components $\,\txG\x\ceL\,$ of the support of the $\,\varphi_{n+1}$.\ It is one of the main results of the present paper that the leaves $\,\ceL\,$ coincide with the minimal manifolds $\,T_{\overrightarrow\la}^{[\overrightarrow w]}$.\ Its proof hinges on a reinterpretation of the 2-form $\,\Om_n^{(\sfk)}\,$ of \Reqref{eq:Omn-def} in what may,\ on the face of it,\ seem as a completely unrelated mathematical context\footnote{We remark,\ in passing,\ that the existence of the 2-form $\,\Om_2^{(\sfk)}\,$ on the submanifold $\,\pr_1^{-1}(\xcC_{\la_1})\cap\pr_1^{-1}(\xcC_{\la_2})\cap\txm^{-1}(\xcC_\la)\subset\txG^2$,\ the latter arising in a natural manner in the context of the \emph{anticipated} `bi-brane fusion',\ was observed,\ as a highly non-trivial `fact',\ already in \Rcite{Fuchs:2007fw},\ with explicit reference to the geometric quantisation of the moduli space of flat connections on a triply punctured Riemann sphere in which this 2-form had been known to play an important r\^ole.},\ to wit,\ the symplectic geometry of the space of connections on a trivial principal $\txG$-bundle over the Riemann sphere $\,\bC\sfP^1\,$ punctured at $n+1$ distinct points,\ admitting a flat connection with fixed holonomies around the punctures.\ The advantage of rephrasing the problem in hand in this new context is that it enables us to identify the kernel of $\,\Om_n^{(\sfk)}\,$ as a simple corollary to the beautiful result of Alekseev and Malkin reported in \Rcite{Alekseev:1993rj} from which -- remarkably -- the 2-form $\,\Om_n^{(\sfk)}\,$ emerges as the pre-symplectic form on a finite-dimensional space obtained through partial symplectic reduction of the phase space of the Chern--Simons theory with the gauge group $\,\txG\,$ on $\,\bR\x\bC\sfP^1$,\ coupled to timelike Wilson lines of fixed holonomy.\ Thus,\ solving the DJI will require additional physical and mathematical structures of an increasing complexity,\ which we introduce gradually in the subsequent sections.\ Our analysis starts,\ out of necessity,\ with a brief reminder on the Chern--Simons topological gauge field theory in the presence of timelike Wilson lines,\ {\it cf.}\ \Rcite{Witten:1988hf},\ and its relation to (the chiral component of) the WZW $\si$-model,\ {\it cf.}\ Refs.\,\cite{Gawedzki:1999bq,Gawedzki:2001rm,Gawedzki:2001ye}.

\subsection{Insights from the Chern--Simons theory} \label{sec:cs}

Our hitherto discussion emphasises the essential r\^ole of the target geometry for the \emph{boundary} WZW defect and its symmetries as the carrier of the nontrivial topological and cohomological data of the non-boundary WZW defect,\ fixing,\ in particular,\ the labelling of the connected components of the corresponding worldvolume by weights of $\,\ggt\,$ and determining the associated summands $\,\om_{\p,\la}^{(\sfk)}\,$ of the degree-2 components $\,\om_\la^{(\sfk)}\,$ of the simplicial curvature $\,\Om^{(\sfk)}=(\txH^{(\sfk)}_{\rm C},\om^{(\sfk)},0,0)$.\ This constatation shall be strengthened in Sec.\,\ref{sec:WZW-target} by our study of the geometrisation of the latter simplicial 3-cocycle,\ and -- in particular -- extended to the defect intersections,\ from which the non-boundary completion of the boundary structure emerges as a mere `spectator'.\ All this suggests that we pause to take a closer look at the boundary WZW $\si$-model in search of hints as to the geometric meaning of the constraints \eqref{eq:def-jun-id} that the worldsheet analysis of \Rcite{Runkel:2008gr} imposes upon the worldvolume of the maximally symmetric inter-bi-brane.\medskip

The boundary WZW $\si$-model was studied at great length in \Rcite{Gawedzki:2001rm} ({\it cf.}\ also \Rcite{Gawedzki:2001ye}),\ where its (pre)symplectic structure $\,\Om_{\p{\rm WZW}}^{(\sfk)}\,$ was demonstrated to decompose,\ in an adapted vertex-IRF-type parametrisation of the chiral components of the extremal configurations,\ into the `dynamical' part $\,\Om^{(\sfk)}_{\sfL\txG}\,$ on the so-called model space $\,\sfL\txG\x\xcA_{\rm W}(\ggt)\,$ of the loop group $\,\sfL\txG$,\ and the 'boundary' part $\,\Om^{(\sfk)}_\p\,$ depending on the boundary data of the configuration.\ In the simplest case of a pair:\ $\,\bR\x\{0\}\,$ and $\,\bR\x\{\pi\}\,$ of connected components of the boundary of a strip worldsheet $\,\Si=\bR\x[0,\pi]\,$ that are presupposed to be mapped to the respective conjugacy classes $\,\xcC_{\la_0}\ni g(\si^1,0)\,$ and $\,\xcC_{\la_\pi}\ni g(\si^1,\pi)\,$ by an extremal configuration $\,g\in[\Si,\txG]$,\ the latter takes the form $\,g(\si)=g_{\rm L}(\si^+)\cdot h_0\cdot g_{\rm L}(-\si^-)^{-1}\equiv g_{\rm L}(\si^+-2\pi)\cdot h_\pi\cdot g_{\rm L}(-\si^-)^{-1}$,\ expressed in terms of an arbitrary twisted loop $\,g_{\rm L}\in[\bR,\txG]\,$ with the monodromy $\,\g=h_\pi\cdot h_0^{-1}\equiv\Ad_{g_0}(\sfe_\mu)$,\ written for $\,(g_0,\mu)\in\txG\x\sfk\,\xcA_{\rm W}(\ggt)$,\ in a conjugacy class with the weight label $\,\mu\,$ fixed by the \emph{fusion} of the boundary data $\,(h_\pi,h_0)\in\pr_1^{-1}(\xcC_{\la_\pi})\cap\pr_2^{-1}(\xcC_{\la_0})\cap\widetilde\txm{}^{-1}(\xcC_\mu)\subset\txG^{\x 2}\,$ under the binary operation $\,\widetilde\txm\equiv\txm\circ(\id_\txG\x\Inv)\,$ whose twist in the second argument accounts for the relative orientation flip of the two components of $\,\p\Si$.\ In this manner,\ the boundary data control the topology of the chiral component of the bulk field in the said parametrisation $\,g_{\rm L}(\si^+)=h(\si^+)\cdot\sfe^{\sfi\,\si^+\frac{\mu}{\sfk}}\cdot g_0^{-1}\,$ with $\,h\in\sfL\txG$.\ In this parametrisation,\ the `boundary' part reads
\qq\nn
\Om^{(\sfk)}_\p(h_\pi,h_0,g_0,\mu)=\tfrac{\sfk}{4\pi}\,\widehat\tr{}_\ggt\bigl(\th_{\rm L}(h_0)\wedge\th_{\rm L}(h_\pi)\bigr)+\sfi\,\widehat\tr{}_\ggt\bigl(\sfd\mu\wedge\th_{\rm L}(g_0)\bigr)+\om_{\p,\la_0}^{(\sfk)}(h_0)-\om_{\p,\la_\pi}^{(\sfk)}(h_\pi)+\om_{\p,\mu}^{(\sfk)}(\g)\,,
\qqq
where $\,\widehat\tr{}_\ggt=\id_{\Om^\bullet(\txG^{\x 3}\x\sfk\,\xcA_{\rm W}(\ggt))}\ox\tr_\ggt$. 

One of the key insights of \Rcite{Gawedzki:2001rm} (inspired by the earlier works \cite{Witten:1988hf,Gawedzki:1989rr,Witten:1991mm,Gawedzki:1999bq} concerned with the (chiral) \emph{bulk} WZW $\si$-model and elaborated in \Rcite{Gawedzki:2001ye}) is the identification of the `boundary'/topological term $\,\Om^{(\sfk)}_\p\,$ in $\,\Om_{\p{\rm WZW}}^{(\sfk)}\,$ as a descendant of the presymplectic form of the three-dimensional Chern--Simons (CS) theory with the structure group $\,\txG\,$ (as defined in Refs.\,\cite{Schwarz:1978cn,Zuckerman:1989cx,Witten:1988hf}) over the cylinder $\,\bR\x\bS^2_{(3)}\,$ with the equitemporal slice given by the Riemann sphere $\,\bS^2_{(3)}\cong\bC\sfP^1\setminus\{P_{\la_\pi},P_{\la_0},P_\mu\}\,$ punctured at the three distinct points.\ At these points,\ a triple of timelike Wilson lines $\,\bR\x\{P_x\},\ x\in\{\la_\pi,\la_0,\mu\}\,$ pierce the sphere.\ The gauge field of the CS theory is coupled covariantly to these field-strength singularities,\ and so it acquires a nontrivial holonomy around the singularity lines,\ further constrained to take values in the respective conjugacy classes $\,\xcC_{\la_\pi},\ \xcC_{\la_0}\,$ and $\,\xcC_\mu$.\ The relevant descent is that to the `almost-physical' phase space of the CS theory with all of the gauge-symmetry group but its finite subgroup $\,\txG\,$ divided out in a Marsden--Weinstein-type reduction.\ The latter was worked out by Alekseev and Malkin in \cite{Alekseev:1993rj} for cylinders $\,\bR\x\Si_{g,n+1}\,$ over Riemann surfaces of an arbitrary genus $\,g\,$ and an arbitrary number $\,n+1\,$ of punctures.\ We shall now recapitulate it briefly for $\,g=0\,$ with view to extracting from it a useful interpretation of the 2-form on the LHS of the DJI for the WZW $\si$-model with a maximally symmetric defect.\ Thus,\ we shall be concerned with the CS theory on the cylinder $\,\cM_{n+1}=\bR\x\bC\sfP^1_{\{\overrightarrow P\}}\,$ over the punctured Riemann sphere $\,\bC\sfP^1_{\{\overrightarrow P\}}=\bC\sfP^1\setminus\{P_i\}_{i\in\ovl{1,n+1}}\,$ in the presence of $\,n\,$ timelike Wilson lines $\,\txW_{\la_i},\ i\in\ovl{1,n}\,$ and a single \emph{anti}-timelike one $\,\txW_{\la_{n+1}}\equiv\txW_\mu$,\ localised at the respective singularity lines $\,\ell_i=\bR\x\{P_i\}\,$ of the field-strength 2-form (with the $\,\ell_{i\neq n+1}\,$ carrying the standard orientation,\ and $\,\ell_{n+1}\,$ -- the opposite one) and having the property that the holonomy of the connection around $\,\ell_i\,$ takes values in the conjugacy class $\,\xcC_{\la_i}\subset\txG$.\medskip

Let us first review elementary facts about the relevant three-dimensional topological field theory.\ The basic field variables of the theory are the $\ggt$-valued connection 1-forms $\,\cA=\cA^A\ox t_A\,$ on (the base of) a trivial principal $\txG$-bundle over $\,\bR\x\bC\sfP^1\,$ which are smooth everywhere except at the excluded lines $\,\ell_i\,$ (where the corresponding curvature is understood to develop singularities),\ and the $\txG$-valued fields $\,\g_i\in[\ell_i,\txG]\,$ supported on the respective singularity lines $\,\ell_i$.\ The action functional takes the form
\qq
S_{\rm CS|W}[\cA,\{\g_i\}]=-\tfrac{\sfk}{4\pi}\int_\cM\,\tr_\ggt\bigl(\cA\wedge\sfd\cA+\tfrac{2}{3}\,\cA\wedge\cA\wedge\cA\bigr)+\sfk\,\sum_{i=1}^{n+1}\,\int_{\ell_i}\,\tr_\ggt\bigl[\la_i\,\bigl(\sfT_e\Ad_{\g_i}\circ\cA+\g_i^*\th_{\rm L}\bigr)\bigr]\cr\label{eq:act-CSW}
\qqq
and is readily verified to be invariant under gauge transformations $\,(\cA,\g_i)\longmapsto(\sfT_e\Ad_\chi\circ\cA-\chi^*\th_{\rm R},\iota_i^*\chi\cdot\g_i)\,$ engendered by maps\footnote{Gauge transformations form a Lie--Fr\'echet group $\,\G(\Ad\,\sfP_\txG)\,$ of global sections of the adjoint bundle (associated with the underlying principal $\txG$-bundle $\,\sfP_\txG$),\ isomorphic with the group $\,{\rm Hom}_\txG(\sfP_\txG,\txG)\,$ of $\txG$-equivariant maps from the total space of the principal $\txG$-bundle to its structure group.\ The existence of a global section of $\,\sfP_\txG\,$ enables us to present gauge transformations as mappings from $\,[\bR\x\bC\sfP^1,\txG]$.} $\,\chi\in[\bR\x\bC\sfP^1,\txG]\,$ and written in terms of the embeddings $\,\iota_i\ :\ \ell_i\emb\cM$.

In the convenient \emph{static gauge} $\,\txA_0=0\,$ of the connection 1-form
\qq\nn
\cA(t,\si)=\txA_0(t,\si)\,\sfd t+\txA(t,\si)\,,\qquad\qquad\txA(t,\si)=\txA_a(t,\si)\,\sfd\si^a
\qqq
(expressed in local coordinates $\,(t,\si)\,$ on $\,\cM_{n+1}$,\ with $\,\si\equiv(\si^1,\si^2)$),\ preserved by the static-gauge
transformations composing the effective gauge group $\,\txG_{\bC\sfP^1}=[\bC\sfP^1,\txG]$,\ the field equations read
\qq\nn
\p_t\txA&=&0\,,\\\cr
\txF(\txA)&=&2\pi\,\bigl(\sum_{i=1}^n\,\sfT_e\Ad_{\g_i}
(\la_i)\,\d_{P_i}-\sfT_e\Ad_{\g_{n+1}}
(\la_{n+1})\,\d_{P_{n+1}}\bigr)\,,\label{eq:curv-source}\\\cr
\p_t\bigl(\sfT_e\Ad_{\g_i}(\la_i)\bigr)&=&0\,,
\qqq
where $\,\d_{P_i}=\d^{(2)}(\si-\si_i)\,\sfd\si^1\wedge\sfd\si^2\,$ ($\si_i\,$ is the value taken by the local coordinates at $\,P_i$) and where $\,\txF(\txA)=\sfd\txA+\txA\wedge\txA\,$ is the curvature of the connection ({\it i.e.},\ the gauge-field strength).\ Thus,\ up to gauge transformations,\ the theory is effectively two-dimensional -- there is no dynamics.\ In particular,\ it has a \emph{finite-dimensional} (physical) phase space
\qq\nn
\xcM_{0,n+1}=\Im_{0,n+1}/\txG_{\bC\sfP^1}
\qqq
defined as the space of orbits of the effective gauge group $\,\txG_{\bC\sfP^1}\,$ within the space $\,\Im_{0,n+1}\,$ of flat connections on a (trivial) principal $\txG$-bundle over the \emph{decorated} Riemann sphere $\,\bC\sfP^1_{\{\overrightarrow P\}}$,\ also known as the {\bf moduli space of flat connections} over $\,\bC\sfP^1_{\{\overrightarrow P\}}$.\ Just to reemphasise:\ The connection 1-forms are assumed smooth everywhere on $\,\bC\sfP^1\,$ except at the non-coincident points $\,P_i,\ i\in \ovl{1,n+1}$,\ where they are constrained to take values (up to gauge transformations) in the respective (co)adjoint orbits\footnote{Recall that we have identified $\,\ggt^*\,$ with $\,\ggt\,$ with the help of the Killing form.}
\qq\nn
\xcO_{\la_i}=\bigl\{\ \tfrac{2\pi\sfi}{\sfk}\,\sfT_e\Ad_x(\la_i) \quad\vert\quad x\in\txG \ \bigr\}\,.
\qqq
The \emph{unphysical} phase space $\,\Im_{0,n+1}\,$ of the gauge theory comes with the presymplectic form
\qq\nn
\Om_{\rm
CS|W}[\cA,\g_i]=\Om_\xcA[\txA]+\sum_{i=1}^n\,\varpi_i(\ovl\g_i)-
\varpi_{n+1}(\ovl\g_{n+1})\,,\qquad\ovl\g_i=\g_i(0)\,,
\qqq
easily derived ({\it e.g.},\ in the first-order formalism of Refs.\,\cite{Gawedzki:1972ms,Kijowski:1973gi,Kijowski:1974mp,Kijowski:1976ze,Szczyrba:1976,Kijowski:1979dj}) in the form of the sum of the canonical Atiyah--Bott form
\qq\nn
\Om_\xcA[\txA]=\tfrac{\sfk}{4\pi}\,\int_{\bC\sfP^1}\,\tr_\ggt(\d
\txA\wedge\d\txA)
\qqq
of \Rcite{Atiyah:1982fa} and the Kirillov--Kostant--Souriau forms 
\qq\nn
\varpi_i=\sfk\,\tr_\ggt(\la_i\,\sfd\th_L)
\qqq
on the respective $\,\xcO_{\la_i}$,\ the latter forms playing a central r\^ole in the geometric quantisation of coadjoint orbits,\ developed in Refs.\,\cite{Kostant:1970,Souriau:1970,Kirillov:1975}.

In order to make contact with the structures appearing in the previous section,\ we should next extend $\,\Om_{\rm CS|W}\,$ (denoting it by $\,\Om_\xcA^{\rm tot}\,$ henceforth) to the space
\qq\nn
\xcA_{0,n+1}^{\rm tot}=\xcA_{0,n+1}\x\x_{i=1}^{n+1}\,\xcO_{\la_i}
\qqq
composed of the space $\,\xcA_{0,n+1}\,$ of \emph{all} connections on a (trivial) principal $\txG$-bundle over the decorated Riemann sphere and of the coadjoint orbits $\,\xcO_{\la_i}\,$ assigned to the punctures.\ It is not hard to demonstrate\footnote{{\it Cf.},\ {\it e.g.},\ the arguments in \Rcite{Sengupta:2002am}.} that the pair $\,(\xcA_{0,n+1}^{\rm tot},\Om_\xcA^{\rm tot})\,$ is actually a \emph{symplectic} space,\ and the gauge group $\,\txG_{\bC\sfP^1}\,$ is readily seen to act on this space in a hamiltonian manner (in the sense of,\ {\it e.g.},\ \Rxcite{Def.\,(3.2.3)}{Woodhouse:1992de}),\ with the momentum
\qq\nn
\mu\ :\ \xcA_{0,n+1}^{\rm tot}\too\ggt^*\ :\
\left(\txA,\sfT_e\Ad_{x_1}(\la_1),\sfT_e\Ad_{x_2}(\la_2),\ldots,\sfT_e\Ad_{x_n}(\la_n)
\right)\longmapsto\mu_{\left(\txA,\sfT_e\Ad_{x_i}(\la_i)\right)}
\qqq
explicitly defined by the formula
\qq\nn
\mu_{\left(\txA,\sfT_e\Ad_{x_i}(\la_i)\right)}(X):=-\tfrac{\sfk}{2\pi}\,
\int_{\bC\sfP^1}\,\tr_\ggt\bigl[X\,\bigl(\txF(\txA)-2\pi\,
\sum_{i=1}^n\,\sfT_e\Ad_{x_i}(\la_i)\,\d_{P_i}+2\pi\,\sfT_e\Ad_{x_{n+1}}
(\la_{n+1})\,\d_{P_{n+1}}\bigr)\bigr]\,,
\qqq
valid for arbitrary $\,X\in\ggt$.\ Hence,\ the unphysical phase space of the Chern--Simons theory on $\,\bC\sfP^1_{\{\overrightarrow P\}}\,$ described earlier coincides with the level set
\qq\nn
\Im_{0,n+1}=\mu^{-1}\bigl(\{0\}\bigr)
\qqq
of the moment map,\ and the \emph{physical} phase space can be obtained through the symplectic reduction of the latter,
\qq\nn
\xcM_{0,n+1}=\mu^{-1}\bigl(\{0\}\bigr)//\txG_{\bC\sfP^1}\,,
\qqq
in the sense of Marsden and Weinstein,\ {\it cf.}\ \Rcite{Marsden:1974sr},\ with respect to the (co)isotropy group $\,\txG_{\bC\sfP^1}\,$ of the zero weight.

The significance of the last result rests upon the fact that it explicitly identifies the kernel of the presymplectic form
$\,\Om_{\rm CS|W}\,$ and thus suggests symplectic reduction as the most natural procedure of establishing the symplectic structure on the finite-dimensional physical phase space of the Chern--Simons theory.\ In this manner,\ we arrive at the construction of Alekseev and Malkin which turns to account the non-trivial topology of the domain $\,\bC\sfP^1_{\{\overrightarrow P\}}\,$ of smoothness of the connection 1-forms so as to \emph{partially} reduce the presymplectic space $\,(\Im_{0,n+1},\Om_\xcA^{\rm tot})\,$ with respect to a judiciously chosen infinite-dimensional subgroup of the gauge group,\ so that at the end of the day there emerges a finite-dimensional quotient equipped with a presymplectic form with a well-defined kernel spanned by generators of the residual gauge symmetry.\ The construction,\ outlined below,\ yields a 2-form related directly to the one appearing on the LHS of the DJI for the maximally symmetric WZW defect,\ given \Reqref{eq:DGC-WZW},\ together with a compact definition of its kernel. 

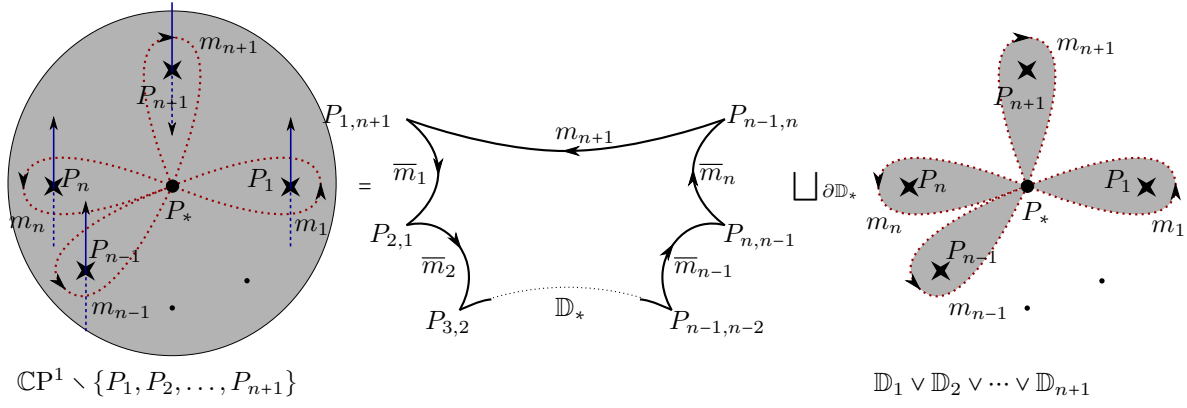
\begin{figure}[hbt]

$$
 \raisebox{-62pt}{\begin{picture}(100,50)
  \put(-10,0){\scalebox{0.25}{\includegraphics{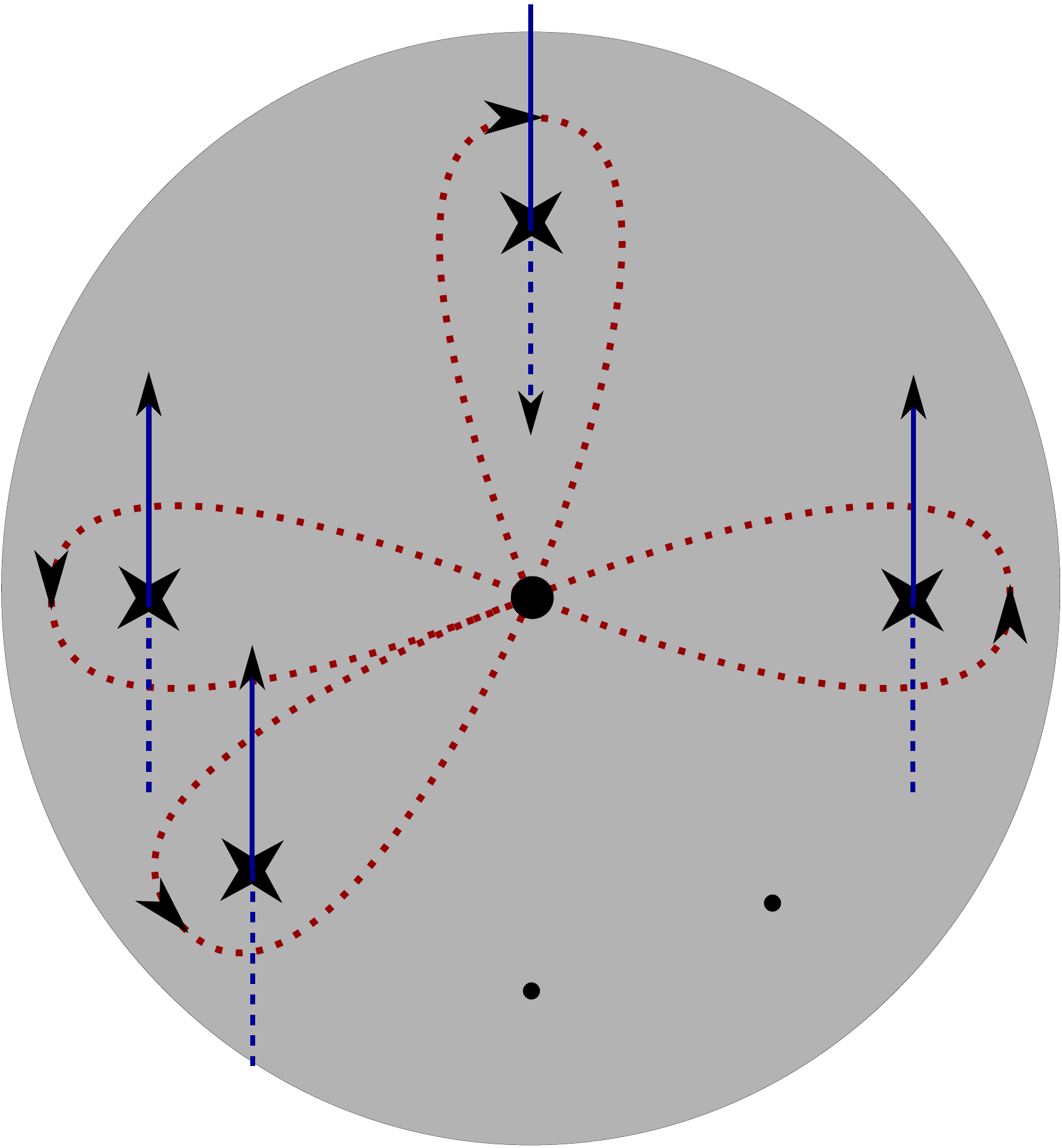}}}
  \end{picture}
  \put(0,0){
     \setlength{\unitlength}{.60pt}\put(-28,-16){
     \put(-155,-5)   { $\bC\txP^1\setminus\{P_1,P_2,\ldots,P_{n+1}\}$ }
     \put(-62,103)    { $P_*$ }
     \put(20,95)    { $m_1$ }
     \put(-160,97)   { $m_n$ }
     \put(-108,43)   { $m_{n-1}$ }
     \put(-40,210)   { $m_{n+1}$ }
     \put(-10,123)    { $P_1$   }
     \put(-128,123)   { $P_n$   }
     \put(-110,78)    { $P_{n-1}$   }
     \put(-80,175)  { $P_{n+1}$   }
     }\setlength{\unitlength}{1pt}}}
  \hspace{20pt}=\hspace{-10pt}
 \raisebox{-45pt}{\begin{picture}(140,50)
  \put(20,0){\scalebox{0.25}{\includegraphics{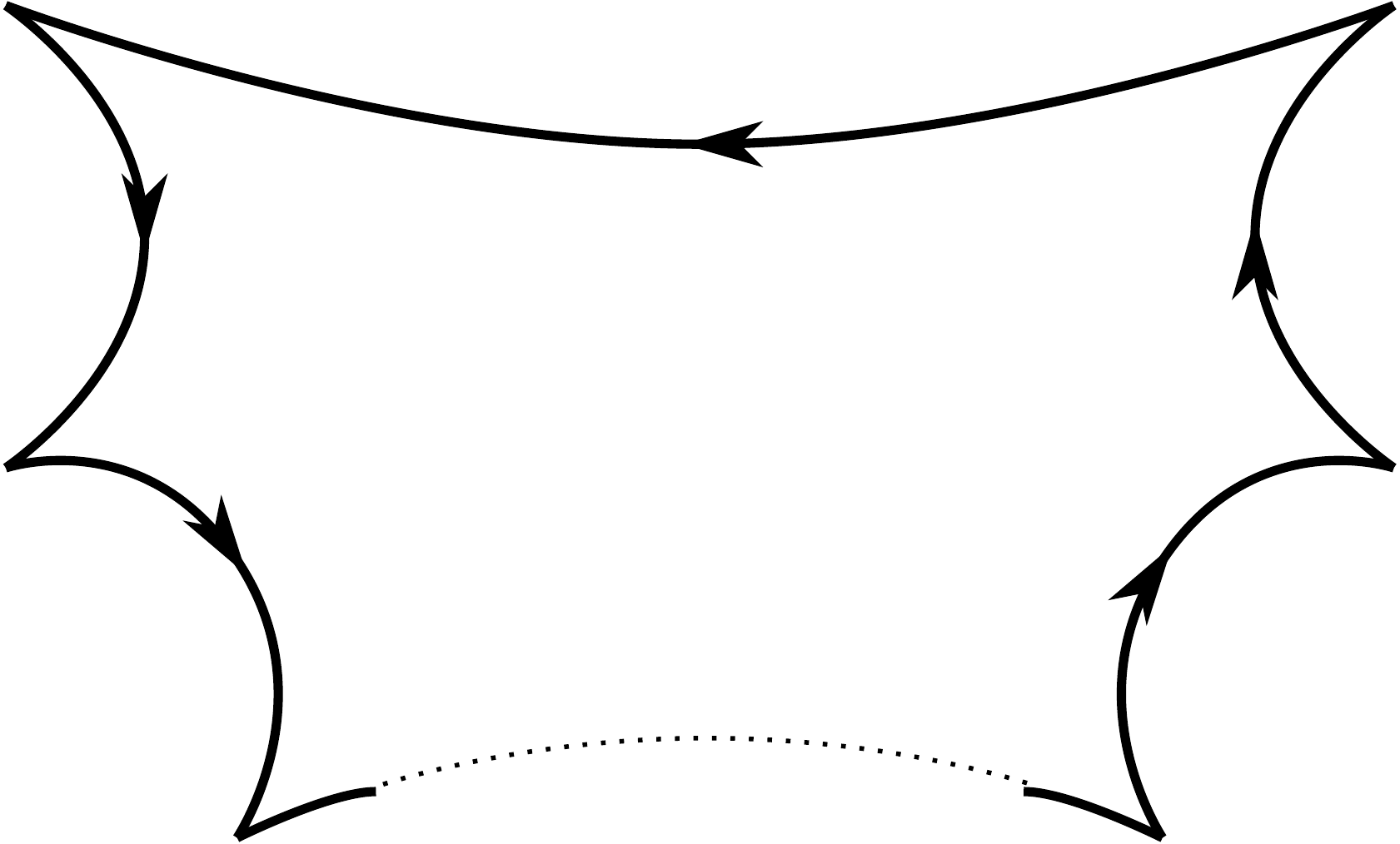}}}
  \end{picture}
  \put(0,0){
     \setlength{\unitlength}{.60pt}\put(-43,-16){
     \put(-68,13)     { $\bD_*$ }
     \put(-68,125)   { $m_{n+1}$ }
     \put(-170,99)   { $\ovl m_1$ }
     \put(23,99)     { $\ovl m_n$ }
     \put(-152,40)   { $\ovl m_2$ }
     \put(7,40)    { $\ovl m_{n-1}$ }
     \put(40,135)     { $P_{n-1,n}$ }
     \put(-215,135)  { $P_{1,n+1}$ }
     \put(38,60)      { $P_{n,n-1}$ }
     \put(-185,60)   { $P_{2,1}$ }
     \put(4,5)     { $P_{n-1,n-2}$ }
     \put(-150,5)    { $P_{3,2}$ }
     }\setlength{\unitlength}{1pt}}}
  \hspace{25pt}\bigsqcup{}_{\p\bD_*}\hspace{0pt}
 \raisebox{-45pt}{\begin{picture}(110,50)
  \put(5,0){\scalebox{0.25}{\includegraphics{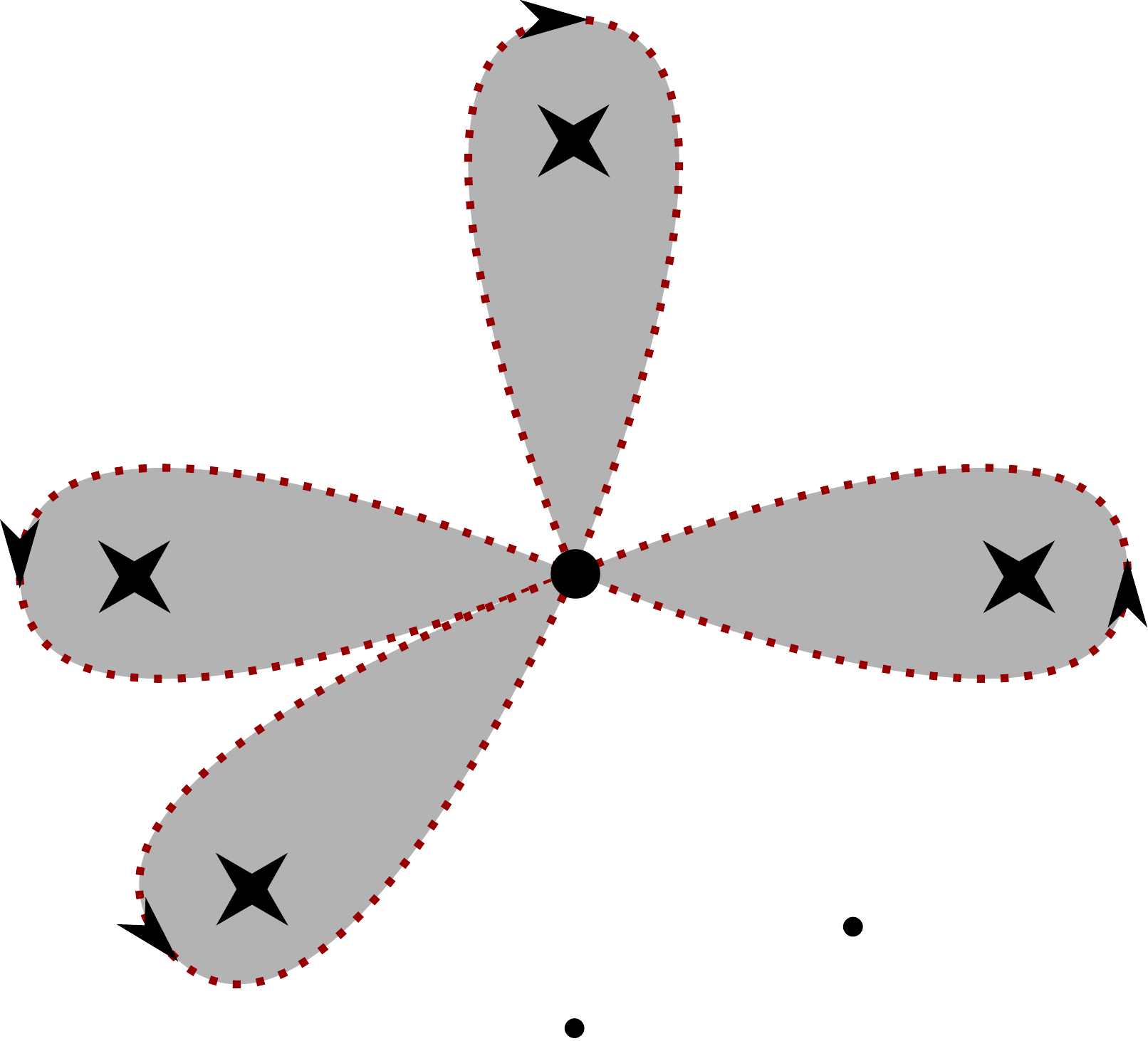}}}
  \end{picture}\put(0,0){
     \setlength{\unitlength}{.60pt}\put(-28,-16){
     \put(-152,-34)   { $\bD_1\vee\bD_2\vee\cdots\vee\bD_{n+1}$ }
     \put(-59,74)    { $P_*$ }
     \put(23,66)    { $m_1$ }
     \put(-157,68)   { $m_n$ }
     \put(-105,14)   { $m_{n-1}$ }
     \put(-37,181)   { $m_{n+1}$ }
     \put(-7,94)    { $P_1$   }
     \put(-125,94)   { $P_n$   }
     \put(-107,49)    { $P_{n-1}$   }
     \put(-77,146)  { $P_{n+1}$   }
     }\setlength{\unitlength}{1pt}}}
$$

\caption{Decomposition of the punctured Riemann sphere $\,\bC\sfP^1_{\{\vec P\}}\,$ into a disjoint union of a bouquet of $n+1$ punctured discs $\,\bD_i\,$ and an un-punctured one,\ $\,\bD_*$,\ sewn (trivially) along the common boundary.\ The decomposition is achieved through incision of $\,\bC\sfP^1_{\{\vec P\}}\,$ along representatives $\,m_i\,$ of the generators of $\,\pi_1(\bC \txP^1_{\{\vec P\}};P_*)$.\ The arrows piercing the
sphere represent the vertical timelike (resp.\ anti-timelike) Wilson lines $\,\txW_{\la_{i\neq n+1}}\,$ (resp.\ $\,\txW_{\la_{n+1}}$) of the gauge field theory.} \label{fig:excision}
\end{figure}

We begin by picking up an arbitrary point $\,P_*\in\bC\sfP^1_{\{\overrightarrow P\}}\,$ together with representatives $\,m_i\,$ of the generators $\,[m_i]\,$ of the based fundamental group $\,\pi_1(\bC\sfP^1_{\{\overrightarrow P\}};P_*)$,\ subject to the single condition
\qq\nn
[m_1\circ m_2\circ\cdots\circ m_n]=[m_{n+1}]
\qqq
(here,\ $\,\circ\,$ stands for the concatenation of loops),\ and subsequently cutting $\,\bC\sfP^1_{\{\overrightarrow P\}}\,$ along the $\,m_i\,$ as in Figure \ref{fig:excision}.
The homological surgery enables us to locally present the connection as a gauge transform of the trivial one,
\qq\nn
\txA\vert_{\bD_*}=-\chi_*^*\th_{\rm R}\,,\qquad\chi_*\in\txG_{\bD_*}\,,
\qqq
over the un-punctured disc,\ and as a gauge transform of a representative $\,\om_i\,$ of the generator class in $\,H^1(\bD_i\setminus\{P_i\},\bZ)\cong\bZ$,
\qq\nn
\txA\vert_{\bD_i}&=&(-1)^{\d_{i,n+1}}\,\tfrac{2\pi\sfi}{\sfk}\,\om_i\ox\sfT_e\Ad_{\chi_i}(\la_i)-\chi_i^*\th_{\rm R}\,,\qquad\chi_i\in\txG_{\bD_i}\,,
\qqq
over the punctured disc $\,\bD_i$.\ The normalisation of $\,\om_i\,$ is such that $\,\int_{C_i}\,\om_i=1\,$ for an arbitrary contour $\,C_i\subset\bD_i\,$ that goes around $\,P_i\,$ once.\ We can now substitute the local presentations of $\,\txA\,$ into the formula for $\,\Om_{\rm CS|W}\,$ (having split the two-dimensional integral into a sum of integrals over the $n+2$ discs beforehand),\ whereby the latter is readily seen to reduce to a finite sum of contour integrals along the loops $\,m_i\,$ and simple contributions from the punctures $\,P_i$.\ At this point,\ it remains to solve the smoothness condition for $\,\txA$,\ imposed at the common boundary of $\,\bD_*\,$ and $\,\bD_1\vee\bD_2\vee\cdots\vee\bD_{n+1}\,$ and relating the boundary values of the restricted gauge transformations $\,\chi_*\,$ and the $\,\chi_i$.\ The thus simplified pre-symplectic form can be expressed in terms of the \emph{transport} (or \emph{monodromy}) operators
\qq\nn
M_i[\chi_i]=P\ee^{\int_{m_i}\,\txA}=\Ad_{\chi_i(P_*)}(\ee_{\la_i})\,,
\qqq
defined with the help of the standard path-ordering operator $\,P\,$ and subject to an obvious constraint
\qq\nn
M_1[\chi_1]\cdot M_2[\chi_2]\cdot\cdots\cdot M_n[\chi_n]=M_{n+1}[\chi_{n+1}]\,.
\qqq
These are manifestly invariant under gauge transformations $\,\chi\,$ from the (infinite-dimensional) subgroup
\qq\nn
\txG_{\bC\sfP^1}(P_*)=\{\ g\in\txG_{\bC\sfP^1} \quad\vert\quad g(
P_*)=e \ \}\subset\txG_{\bC\sfP^1}
\qqq
of the gauge group,\ $\,M_i[\chi\cdot\chi_i]=\Ad_{\chi(P_*)}\left(M_i[\chi_i]\right)\equiv M_i[\chi_i]$,\ and so they coordinatise the \emph{finite-dimensional} partial quotient
\qq
\xcH_{0,n+1}:=\Im_{0,n+1}/\txG_{\bC\sfP^1}(P_*)\equiv\{\ (M_1,M_2,\ldots,M_n,M)\in\x_{i=1}^n\,\xcC_{\la_i}\x
\xcC_{\la_{n+1}}\quad \vert\quad M_1\cdot M_2\cdot\cdots\cdot M_n=M \ \}\,,\cr \label{eq:H0n}
\qqq
embedded in the product manifold
\qq\nn\label{eq:conjoman}
\xcF_{0,n+1}:=\x_{i=1}^n\,\xcC_{\la_i}\x\xcC_{\la_{n+1}}\,.
\qqq
In conformity with the general scheme of symplectic reduction,\ $\,\Om_{\rm CS|W}\,$ descends to the quotient $\,\xcH_{0,n+1}\,$ as a presymplectic form $\,\Om_{\xcF_{0,n+1}}^{(\sfk)}\,$ with a finite-dimensional kernel spanned by the fundamental vector fields engendered by constant gauge transformations $\,M_i\longmapsto\Ad_\chi(M_i),\ \chi\in\txG\,$ from the residual gauge group $\,\txG$.\ Parenthetically,\ let us also point out that the $\txG$-invariance of the descended presymplectic form ensures the independence of the 2-form itself of the choice of the base point $\,P_*$.\ Indeed,\ changing the latter is tantamount to conjugating all transport operators by the same group element.

In order to be able to write out the descended presymplectic form,\ we parameterise $\,\xcF_{0,n+1}\,$ (in a manifestly redundant manner) with the help of the surjective map
\qq\nn
\pi\ &:&\ \txG^{\x n+1}\x T^{\x n+1}\too\xcF_{0,n+1}\cr\cr
&:&\ (u_1,u_2,\ldots,u_n,u,C_1,C_2,\cdots,C_n,C)\longmapsto
\left(\Ad_{u_1}(C_1),\Ad_{u_2}(C_2),\ldots,\Ad_{u_n}(C_n),\Ad_u(C)
\right)\,,
\qqq
written for the $(n+1)$-th cartesian power $\,T^{\x n+1}\,$ of the Cartan subgroup $\,T\subset\txG\,$ of $\,\txG$,\
whereupon we arrive at the definition
\qq\nn
\widetilde\xcH_{0,n+1}:=\pi^{-1}(\xcH_{0,n+1})\,.
\qqq
In this parametrisation,\ the 2-form $\,\Om_{\xcF_{0,n+1}}^{(\sfk)}\,$ of interest reads
\qq\label{eq:Om-part-red}\hspace{.5cm}
\pi^*\Om_{\xcF_{0,n+1}}^{(\sfk)}(u_1,u_2,\ldots,u_n,u,C_1,C_2,\ldots,C_n,C)=\tfrac{\sfk}{4\pi}\,\tr_\ggt\bigg(\sum_{i=1}^n\,\th_L(u_i)\wedge\sfT_e\Ad_{C_i}\circ\th_L(u_i)\cr\cr
-\th_L(u)\wedge\sfT_e\Ad_C\circ\th_L(u)+\sum_{j=1}^{n-1}\,\bigl(K_j^*\th_L\wedge K_{j+1}^*\th_L\bigr)(u_1,u_2,\ldots,u_n,C_1,C_2,\cdots,C_n)\bigg)\,,
\qqq
where
\qq\nn
K_j(u_1,u_2,\ldots,u_n,C_1,C_2,\cdots,C_n)=
\Ad_{u_{n+1-j}}(C_{n+1-j})\cdot\Ad_{u_{n+2-j}}(C_{n+2-j})\cdot\cdots
\cdot\Ad_{u_n}(C_n)\,.
\qqq
As a corollary to Theorem 1 of \Rcite{Alekseev:1993rj} (explicitly stated in \Rxcite{Eq.\,(3.8)}{Alekseev:1993rj}),\ we obtain 

\berop\label{prop:kerOmF}
The kernel of $\,\Om_{\xcF_{0,n+1}}^{(\sfk)}\,$ is spanned by the fundamental vector fields for the left $\txG$-action 
\qq\nn
&&\txG\x\xcH_{0,n+1}\too\xcH_{0,n+1}\cr\cr
&:&\ \left(y,(M_1,M_2,\ldots,M_n,M)\right) \longmapsto\left(\Ad_y(M_1),
\Ad_y(M_2),\ldots,\Ad_y(M_n),\Ad_y(M)\right)\,.
\qqq
\eerop

By now,\ we have all the requisite tools for the solution of the original problem,\ which is the verification of the postulated (preliminary) localisation of the worldvolume of the WZW inter-bi-brane.\ To this end,\ we consider,\ once more,\ the manifold \eqref{eq:H0n} and note that the latter is canonically diffeomorphic to the previously encountered manifold $\,T_{\overrightarrow\la}^\la\,$ of \Reqref{eq:Tlala} as {\it per}
\qq\label{eq:hbar}
\hbar\ :\ T_{\overrightarrow\la}^\la\xrightarrow{\ \cong\ }
\xcH_{0,n+1}\ :\ (h_1,h_2,\ldots,h_n)\longmapsto(h_1,h_2,\ldots,h_n,h_1
\cdot h_2\cdot\cdots\cdot h_n)\,.
\qqq
The isomorphism gives us the fundamental

\berop\label{prop:Om-AM-vs-RS}
Let $\,\Om_n^{(\sfk)}\,$ and $\,\Om_{\xcF_{0,n+1}}^{(\sfk)}\,$ be the 2-forms defined in Eqs.\,\eqref{eq:Omn-def} and \eqref{eq:Om-part-red},\ respectively,\ and let $\,\hbar\,$ be the diffeomorphism of \Reqref{eq:hbar}.\ Then,
\qq\label{eq:OmF-as-n}
\hbar^*\Om_{\xcF_{0,n+1}}^{(\sfk)}=-\Om_n^{(\sfk)}\,.
\qqq
\eerop
\beroof
{\it Cf.}\ Appendix \ref{app:Om-AM-vs-RS}. \eroof 

\noindent The above implies,\ in particular,\ that the kernel of $\,\Om_n^{(\sfk)}\,$ can be obtained through pushforward of $\,\ker\,\Om_{\xcF_{0,n+1}}^{(\sfk)}\,$ along the inverse of $\,\hbar$.\ This yields the all-important

\bethe[The kernel of $\,\Om_n^{(\sfk)}$] \label{thm:ker-Om}
Let $\,T_{\overrightarrow\la}^\la\,$ be the manifold defined in \Reqref{eq:Om-part-red},\ and let the $\,L_A\,$ (resp.\ the $\,R_A$) be the standard left-invariant (resp.\ right-invariant) vector fields on $\,\txG$,\ {\it cf.}\ \Reqref{eq:left-min-right}.\ The kernel of the 2-form $\,\Om_n^{(\sfk)}\in\Om^2(T_{\overrightarrow\la}^\la)\,$ of \Reqref{eq:Omn-def} is spanned by the fundamental vector fields
\qq\nn
\xcV_A(h_1,h_2,\ldots,h_n)=\sum_{i=1}^n\,V_A(h_i)\,,\qquad\qquad
V_A(h_i)=(R_A-L_A)(h_i)
\qqq
of the left action of $\,\txG\,$ on $\,T_{\overrightarrow\la}^\la\,$ given by
\qq\nn
\txG\x T_{\overrightarrow\la}^\la\too T_{\overrightarrow\la}^\la\
:\ \left(y,(h_1,h_2,\ldots,h_n)\right)\longmapsto\left(\Ad_y(h_1),
\Ad_y(h_2),\ldots,\Ad_y(h_n)\right)\,.
\qqq
\ethe
\beroof
The fundamental vector fields for the left $\txG$-action on $\,\xcH_{0,n+1}\,$ push forward along $\,\hbar^{-1}\,$ to the $\,\xcV_A$.\ The thesis now follows directly from that Prop.\,\ref{prop:kerOmF}.
\eroof 
\noindent The theorem (partially) settles the issue of localisation of the elementary inter-bi-brane worldvolumes in that it proves that the $\,\cT_{\overrightarrow\la}^{[\overrightarrow w]}$,\ formerly shown to be the smallest $\txG\x\txG$-spaces that can support the inter-bi-brane data ({\it i.e.},\ the fusion 2-isomorphism $\,\varphi_{n+1}$),\ are in fact also the largest spaces with this property.\ This,\ then,\ confirms the postulated decomposition of the (total) worldvolume of the inter-bi-brane (associated with defect junctions of a given valence) stated in \Reqref{eq:ibb-n-wvol}.\ Further specification of the
worldvolume calls for heavier cohomological tools,\ such as,\ {\it e.g.},\ an explicit construction of $\,\varphi_{n+1}\,$ alluded to previously,\ an idea that we take up in the next section.\\[2pt]

The foregoing analysis,\ including an exposition of the main conceptual points of the Alekseev--Malkin construction (itself embedded in the theory of quasi-hamiltonian $\txG$-spaces,\ {\it cf.}\ \Rcite{Alekseev:1997}),\ leaves us with the intriguing question as to a deeper reason for the existence of a relation between,\ on the one hand,\ the geometry of the phase space of the topological field theory defined by \Reqref{eq:act-CSW} and its symplectic reduction,\ and,\ on the other hand,\ by the gerbe-theoretic structure associated with junctions of the maximally symmetric WZW defect and represented implicitly by the DJI.\ While a full-fledged explanation of this relation falls well beyond the compass of the present paper (and would,\ among other things,\ prerequire an in-depth case study along the lines of \Rcite{Runkel:2009sp}),\ we find it fitting to conclude this section with a heuristic argument that sets up the conceptual framework for further study of the relation.

The upshot of our hitherto analysis is a correspondence between supports of the fusion 2-iso\-morph\-isms $\,\varphi_{n+1}\,$ (and hence also the corresponding component inter-bi-brane 2-isomorphisms and junctions of the maximally symmetric WZW defects) and (some of) leaves of the characteristic foliation of the partially symplectically reduced presymplectic form of the Chern--Simons theory with the Wilson lines $\,\txW_{\la_i},\ i\in\ovl{1,n+1}\,$ on the cylinder $\,\bR\x\bC\sfP^1_{\{\overrightarrow P\}}\,$ over the punctured Riemann sphere.\ The space of these leaves,\ denoted as $\,\xcM_{0,n+1}=\xcH_{0, n+1}/\txG$,\ constitutes the finite-dimensional (classical) space of states of the theory.\ Condition \eqref{eq:curv-source} for the connection which enters the definition of $\,\xcH_{0,n+1}\,$ carries over to the (holomorphically) quantised theory as an operator constraint imposed on \emph{physical} quantum states,\ admitting an interpretation of the (infinitesimal) Ward identity satisfied by the corresponding wave function(al)s.\ As discussed in,\ {\it e.g.},\ \Rxcite{Sec.\,5}{Gawedzki:1999bq},\ this permits to identify states of the Chern--Simons theory on $\,\bR\x\bC\sfP^1_{\{\overrightarrow P\}}\,$ as certain distinguished \emph{intertwiners} of the (diagonal) $\txG$-action on irreducible modules labelled by the weights $\,\la_i$.\ It deserves to be noted that it is precisely the Ward identity that enables to relate sesquilinear combinations of physical wave functions of the Chern--Simons theory to the so-called conformal blocks that compose correlation functions of the WZW model,\ thus establishing a profound correspondence between the two theories.

The connection with maximally symmetric WZW defect junctions suggested by the above collection of facts rests upon the following two related observations:\ First of all,\ as shown in \Rcite{Frohlich:2006ch},\ a junction of maximally symmetric WZW defect lines has a well-defined description in the quantised WZW model iff the Verlinde fusion coefficient $\,\cN_{\la_1,\la_2,\ldots,\la_n}^{\hspace{35pt}\la_{n+1}}\,$ for the weights $\,\la_i\,$ that label the defect lines converging at the junction is non-zero.\ However,\ the said coefficient is known ({\it cf.}\ \Rcite{Witten:1988hf}) to give the dimension of the space of conformal blocks on the Riemann sphere with insertions (at non-coincident points $\,P_i,\ i\in\{1,n+1\}$) of primary operators of the WZW model from the respective integrable highest-weight modules labelled by the $\,\la_i$.\ Thus,\ the quantum junctions are rather straightforwardly linked to the quantum states of the Chern--Simons theory,\ and our findings reported hereinabove indicate that it is reasonable to expect that this link remains visible in the classical r\'egime in the higher-geometric formulation.\ The second observation derives from the study,\ reported in Refs.\,\cite{Suszek:2011hg,Suszek:2012ddg},\ of the r\^ole played by the geometric data associated with conformal defects and their junctions in the canonical description of the $\si$-model,\ both classical and geometrically pre-quantised.\ In this description,\ to a given defect junction of valence $n+1$ one assigns a subspace in the $(n+1)$-th power of the phase space of the $\si$-model by thinking of it as a vertex of a cross-defect splitting-joining interaction of the string (resp.\ of a splitting-joining interaction in the defect-twisted sector).\ As argued in the latter of the two articles,\ in the case of defects that preserve,\ in the sense of a continuity relation for the Noether currents,\ (a subset of) symmetries of the bulk $\si$-model on either side of the defect line,\ the field-space data for a junction define an intertwiner for the action of the preserved symmetry (sub)group on the factors in the product phase space assigned to the junction that correspond to the incoming and outgoing states,\ respectively.\ In the context of the WZW model,\ with its \emph{maximally} symmetric defects,\ this brings us back to the description of states of the Chern--Simons theory in terms of group-$\txG$ intertwiners.

\section{The higher geometry of the WZW background}\label{sec:WZW-target}

The in-depth geometric analysis of the previous section,\ crowned with the argument derived from the study of the topological Chern--Simons theory,\ has provided us with a simplicial $\txG\x\txG$-manifold  
\qq\nn\hspace{-.5cm}
\alxydim{@C=2.cm@R=1cm}{ \cdots \ar@<.75ex>[r]^{\unl d{}_\cdot^{(n+1)}\qquad\qquad\qquad} \ar@<.25ex>[r] \ar@<-.25ex>@{..}[r] \ar@<-.75ex>[r] & \txG\x\bigsqcup_{(\overrightarrow\la,\la)\in\faff{\ggt}^{\x n+1}}\,\bigsqcup_{[\overrightarrow w]\in\cF_{\overrightarrow\la}^\la}\,T_{\overrightarrow\la}^{[\overrightarrow w]} \ar@<.75ex>[d]^{\unl d{}_\cdot^{(n)}\qquad\qquad\qquad} \ar@<.25ex>[d] \ar@<-.25ex>@{..}[d] \ar@<-.75ex>[d] 
\ar@<.75ex>[ddr]^{\pi^{(n+1)}_{\cdot,\cdot}} \ar@<.25ex>[ddr] \ar@<-.25ex>@{..}[ddr] \ar@<-.75ex>[ddr]
& & \\ & \vdots \ar@<.75ex>[d]^{\unl d{}_\cdot^{(3)}\qquad\qquad\qquad} \ar@<.25ex>[d] \ar@<-.25ex>[d] \ar@<-.75ex>[d] & & \\ 
& \txG \times \bigsqcup_{(\la_1,\la_2,\la) \in \faff{\ggt}^{\x 3}}\,\bigsqcup_{[w]\in \cF_{(\la_1,\la_2)}^\la}\,T_{(\la_1,\la_2)}^{[w]} \ar@<.5ex>[r] \ar@<0.ex>[r] \ar@<-.5ex>[r]_{\qquad\qquad\qquad\unl d{}^{(2)}_\cdot\equiv\pi^{(3)}_{\cdot,\cdot}} & \bigsqcup_{\la\in\faff{\ggt}}\,\txG\x\xcC_\la \ar@<.5ex>[r]^{\qquad\quad\unl d{}^{(1)}_1\equiv\iota_1=\pr_1} \ar@<-.5ex>[r]_{\qquad\quad\unl d{}^{(1)}_0\equiv\iota_2=\txm} & \txG}
\qqq
as a tentative model of the target space of the maximally symmetric WZW defect,\ implemeting the binary operation of the underlying Lie group $\,\txG$.\ In this model,\ there are three as yet unspecified objects:\ subsets $\,\faff{\ggt}\subset\sfk\,\xcA_{\rm W}(\ggt)\,$ and $\,\cF_{\overrightarrow\la}^\la\subset\cS_{\la_1}\backslash
\x_{i=2}^n\,(\txG/\cS_{\la_i})$,\ and the submanifolds $\,T_{\overrightarrow\la}^{[\overrightarrow w]}\subset T_{\overrightarrow\la}^\la$.\ These are determined by certain higher-geometric structures over $\,\unl\sfN{}_\bullet\txG\,$ that realise the simplicial de Rham class $\,[(\txH_{\rm C}^{(\sfk)},\om^{(\sfk)},0,0)]\,$ and,\ in so doing,\ complete the definition of the maximally symmetric WZW background.\ Below,\ we specify these structures and discuss their properties with respect to the simplicial structure the $\txG\x\txG$-action on $\,\unl\sfN{}_\bullet\txG$,\ thereby identifying the said objects.

\brem
Following the reasoning from Sec.\,\ref{sub:WZWorbs},\ one is tempted to draw yet another pencil of arrows pointing from $\,\txG\x\bigsqcup_{(\overrightarrow\la,\la)\in\faff{\ggt}^{\x n+1}}\,\bigsqcup_{[\overrightarrow w]\in\cF_{\overrightarrow\la}^\la}\,T_{\overrightarrow\la}^{[\overrightarrow w]}\,$ to $\,\txG \times \bigsqcup_{(\la_1,\la_2,\la) \in \faff{\ggt}^{\x 3}}\,\bigsqcup_{[w]\in \cF_{(\la_1,\la_2)}^\la}\,T_{(\la_1,\la_2)}^{[w]}\,$ and representing (restrictions of) the mappings $\,\id_\txG\x\bigsqcup_{(\overrightarrow\la,\la)\in\faff{\ggt}^{\x n+1}}\,\bigsqcup_{[\overrightarrow w]\in\cF_{\overrightarrow\la}^\la}\,\pi_{\overrightarrow\la}^{[\overrightarrow w]}(i,j\vert k,l)\,$ for arbitrary $\,1\leq i\leq j<k\leq l\leq n$.\ This,\ however,\ would presuppose the existence of certain relations between the \emph{sub}sets $\,\cF_{\overrightarrow\la}^\la\subset\cS_{\la_1}\backslash
\x_{i=2}^n\,(\txG/\cS_{\la_i})\,$ and $\,\cF_{(\mu_1,\mu_2)}^\mu\subset\cS_{\mu_1}\backslash
\txG/\cS_{\mu_2}\,$ that we have not spelled out in Sec.\,\ref{sub:maxym-def-junct}.\ In fact,\ such relations appear quite natural from the point of view of the two-dimensional conformal field theory under consideration,\ and so we come back to them when discussing the fusion 2-isomorphism(s) in Sec.\,\ref{sub:fus-2iso},
\erem

\subsection{The WZW 1-gerbe}\label{sub:WZW-target-bulk}  

We begin with the geometrisation of the bulk de Rham class $\,[\txH_{\rm C}^{(\sfk)}]$,\ {\it i.e.},\ a 1-gerbe over $\,\txG\,$ with curvature given by the $\sfk$-th multiple of the Cartan 3-form of \Reqref{eq:H-coord}.\ The first thing to be noted is that in order for the 1-gerbe to exist we need $\,[\txH_{\rm C}^{(\sfk)}]\in H^3(\txG,2\pi\bZ)\subset H^3(\txG,\bR)$,\ which happens iff $\,\sfk\in\bZ$.\ Taking into account the related normalisation of the metric term in the action functional of the WZW $\si$-model,\ we further restrict $\,\sfk\in\bN$,\ so that the term has the standard sign.\ That the choice of the curvature of the 1-gerbe then determines the class of $\,\cG^{(\sfk)}\,$ uniquely follows from Prop.\,\ref{prop:1-isoclof1-grb} and the isomorphism $\,H^2(\txG,\uj)\cong\bd1\,$ (here,\ and in what follows,\ $\,\bd1\,$ is the trivial group).\ In fact, $\,\cG^{(\sfk)}\,$ is 1-isomorphic with the $\sfk$-th tensor power $\,\cG^{(\sfk)}\cong\cG^{(1)\,\ox\sfk}\,$ of the so-called {\bf basic gerbe}\footnote{Meinrenken's construction of the basic gerbe for a general 1-connected Lie group was preceded by that of \Rcite{Gawedzki:1987ak} for $\,\sug\,$ and that of \Rcite{Chatterjee:1998} for $\,{\rm SU}(N)$.\ The non-simply connected case was worked out in Refs.\ \cite{Gawedzki:2002se,Gawedzki:2003pm}.} $\,\cG^{(1)}\,$ of \Rcite{Meinrenken:2002},\ of curvature $\,\curv(\cG^{(1)})=\txH_{\rm C}$.\ The bi-invariance of the Cartan 3-form then implies,\ by the very same token,\ that the 1-gerbe is $\txG\x\txG$-symmetric, 
\qq\nn
\curv\bigl(\ell\wp_{(x,y)}^*\cG^{(\sfk)}\bigr)=\ell\wp_{(x,y)}^*\curv\bigl(\cG^{(\sfk)}\bigr)=\curv\bigl(\cG^{(\sfk)}\bigr)\qquad\xLongrightarrow{\ \ H^2(\txG,\uj)\cong\bd1\ \ }\qquad\ell\wp_{(x,y)}^*\cG^{(\sfk)}\cong\cG^{(\sfk)}\,.
\qqq 

We may next enquire,\ with hindsight,\ about the existence of a (generalised) multiplicative structure on $\,\cG^{(\sfk)}$,\ a natural higher-cohomological analogon of the concepts of a group homomorphism and that of a multiplicative line bundle over a group manifold,\ introduced in \Rcite{Carey:2004xt} and further elaborated in Refs.\,\cite{Waldorf:2008mult,Gawedzki:2009jj},\ which we recalled and generalised in Def.\,\ref{def:genmultstr} with view to subsequent considerations in the $\bZ/2\bZ$-graded setting.\ In virtue of the Universal Coefficient Theorem,\ in conjunction with the K\"unneth Formula,\ we obtain,\ for $\,\txG\,$ simple and 1-connected,\ $\,H^2(\txG\x\txG,\uj)=\bd1=H^1(\txG\x\txG,\uj)$,\ and so the above identity together with Prop.\,\ref{prop:1-isoclof1-grb} immediately implies the existence of a 1-isomorphism
\qq\nn
\cM\ :\ \cG^{(\sfk)}_1\ox\cG^{(\sfk)}_2\xrightarrow{\ \cong\ }\cG^{(\sfk)}_{12}\ox\cI_{\varrho^{(\sfk)}}\,,\qquad\qquad\varrho^{(\sfk)}=\sfk\,\varrho\,,
\qqq
{\it cf.}\ \Reqref{eq:PolWieg-2},\ which is unique up to a 2-isomorphism by Prop.\,\ref{prop:2-isoclof1-isos}.\ Note that by the previous cohomological arguments and in consequence of the elementary property 
\qq\label{eq:D22Grhok}
\D^{(2;2)}_\txG\varrho^{(\sfk)}=0
\qqq
of the 2-form $\,\varrho^{(\sfk)}$,\ {\it cf.}\ \Reqref{eq:D22Grho},\ a 2-isomorphism \eqref{diag:aleph} with $\,\vartheta=0\,$ always exists on a 1-connected group $\,\txG$.\ As anticipated in Sec.\,\ref{sub:homocat},\ the additional structure $\,(\cM,\a)\,$ plays a central r\^ole in the construction of the maximally symmetric WZW bi-brane proposed in \Rcite{Fuchs:2007fw},\ linking the latter to the long-known construction of the maximally symmetric brane in a manner essentially determined by the guidelines on p.\,\pageref{pgref:guidelines} and symmetry considerations.\ For the sake of a succinct expression of this fact in the subsequent sections,\ we fix below one further piece of notation.

The point of departure is the natural correspondence between,\ on the one hand,\ collections of binary operations on adjacent pairs of cartesian components of $\,\txG^{\x n},\ n\in\bN_{\geq 2}\,$ (or {\bf multiplication patterns}) and,\ on the other hand,\ full binary trees with $n$ leaves of Def.\,\ref{def:FBT}.\ For $\,n\in\bN_{\geq 2}$,\ the correspondence can be neatly encoded in a map
\qq\nn
\mu_n\ :\ {\rm FBT}_n\too 2^{C^\infty(\txG^{\x n},\txG^{\x 2})}
\qqq
that assigns to a given tree $\,\tau\in{\rm FBT}_n\,$ the collection of those smooth maps $\,\txm^{(n)}(i,j\vert j+1,k)\,$ of \Reqref{eq:mult-of-pi} that are represented by its non-unary vertices (or,\ equivalently,\ by pairs of its edges joined by one of its vertices),\ {\it e.g.},
\qq\nn
\mu_3(\tau_3^{\rm L})=\{\txm^{(3)}(1,1\vert 2,2),\txm^{(3)}(1,2\vert 3,3)\}\,,
\qqq
{\it cf.}\ \eqref{eq:LRtreen}.\ Our presentation of the correspondence is fixed by setting
\qq\nn
\mu_2\ :\ {\rm FBT}_2\too 2^{C^\infty(\txG^{\x 2},\txG^{\x 2})}\ :\ \t^{\rm L}_2\longmapsto\id_{\txG^{\x 2}}\,.
\qqq
In parallel,\ we may also define (for $\,n\geq 2$) a map
\qq\nn
\xcM_n\ :\ {\rm FBT}_n\too 1\tx{-}{\rm Iso}\left(\bgrb^\nabla(
\txG^{\x n})\right)
\qqq
that assigns to a given tree $\,\tau\in{\rm FBT}_n\,$ a sequence of pullbacks,\ along elements of $\,\mu_n(\t)$,\ of the 1-isomorphism $\,\cM\,$ of the multiplicative structure that realises the multiplication pattern represented by $\,\tau\,$ on the relevant tensor product of pullbacks of the 1-gerbe $\,\cG^{(\sfk)}$,\ {\it e.g.},
\qq\nn
\xcM_3(\tau_3^{\rm L})=\cM_{12,3}\circ(\cM_{1,2}\ox\id_{\cG_3})\equiv\txm^{(3)}(1,2\vert 3,3)^*\cM\circ\bigl(\txm^{(3)}(1,1\vert 2,2)^*\cM\ox\id_{\cG_3}\bigr)\,.
\qqq
As a simple consequence of the associativity of the multiplicative structure on $\,\cG$,\ we readily establish
\becor\label{cor:An-map}
For every $\,n\in\bN_{\geq 3}$,\ there exists a map
\qq\nn
\xcA_n\ :\ {\rm FBT}_n\too 2\tx{-}{\rm Iso}\left(\bgrb^\nabla(
\txG^{\x n})\right)
\qqq
such that,\ for a given tree $\,\tau\in{\rm FBT}_n$,
\qq\nn
\xcA_n(\tau)\ :\ \xcM_n(\tau_n^{\rm L})\xLongrightarrow{\ \cong\ }
\xcM_n(\tau)\,.
\qqq
\ecor \noindent The thesis of the corollary will be seen to play an important r\^ole in the derivation of linear relations between various explicit realisations of the inter-bi-brane 2-isomorphism obtained through the so-called induction scheme of \Rxcite{Sec.\,2.8}{Runkel:2008gr}.

\subsection{The maximally symmetric WZW brane}\label{sec:mxymbr}

Let us,\ next,\ consider the gerbe-theoretic data for circular ({\it i.e.},\ non-intersecting) maximally symmetric WZW boundaries,\ that is the data of the maximally symmetric $\cG^{(\sfk)}$-brane.\ While we are ultimately interested in WZW defects exclusively,\ the $\cG^{(\sfk)}$-brane structure can be seen to form part of the gerbe-theoretic data of the $\cG$-bi-brane,\ and so we spend some time recalling its details.\ As argued in \Rxcite{p.\,12}{Runkel:2008gr} ({\it cf.}\ also \ref{sub:maxym-b-def}),\ incorporating worldsheet boundaries into the unified framework of description valid for all defects,\ prerequires choosing the target-space in a rather special form,\ to wit,\ as the manifold $\,M=\Gx\sqcup\{\bullet\}\,$ with $\,\bullet\,$ a point,\ and with -- as previously -- the 1-gerbe $\,\cG^{(\sfk)}\,$ over $\,\Gx$.\ We fix this choice (solely) for the present section.\ 

The presence of topological obstructions to the existence of a higher-geometric structure (the relevant gerbe module) over a generic conjugacy class $\,\xcC_\la\,$ leads to a quantisation of the weight label in an otherwise continuous set of orbits of the adjoint action within $\,\txG$,\ confining the label to the fundamental affine Weyl alcove
\qq\nn
\faff{\ggt}=\sfk\,\xcA_{\rm W}(\ggt)\cap P(\ggt)\ni\la\,,
\qqq
given as the intersection of the weight lattice $\,P(\ggt)\,$ of $\,\ggt\,$ with $\,\sfk\,\xcA_{\rm W}(\ggt)$,\ {\it cf.}\ Refs.\,\cite{Carey:2002,Gawedzki:2002se,Gawedzki:2004tu}.

The defect line $\,\ell\cong\bS^1\,$ is then mapped by the $\si$-model field $\,g\ :\ \ell\too Q_\p\,$ to the disjoint union
\qq\label{eq:maxym-b-wvol}
Q_\p=\bigsqcup_{\la\in\faff{\ggt}}\,\xcC_\la
\qqq
of orbits of the adjoint $\txG$-action,\ the conjugacy classes of Cartan elements $\,\sfe_\la\,$ of \Reqref{eq:Qp-def} labelled by weights from $\,\faff{\ggt}$.\ The conjugacy classes compose the worldvolume of the maximally symmetric $\cG^{(\sfk)}$-brane
\qq\label{eq:bdry-bib-WZW}
\cB_\p=(Q_\p,\iota_1,\iota_2,\om_\p,\Phi_\p)
\qqq
with the following additional data:\ The curvature of the bi-brane is given by the 2-form $\,\om^{(\sfk)}_\p\,$ of \Reqref{eq:om-WZW-b},\ ensuring the existence of the $\sfL\txG$-action on field configurations from $\,\cF(\Si,d;\g)$ subject to boundary conditions $\,X(\Si^{(1)})\subset Q_\p$.\ The bi-brane maps are $\,\iota_\la\equiv\iota_1 \vert_{\xcC_\la}\ :\ \xcC_\la\emb\txG\,$ (the embedding of the conjugacy class $\,\xcC_\la\,$ in the group manifold),\ and the constant map $\,\iota_2=\bullet:Q_\p-->\{\bullet\}$.\ Finally,\ the bi-brane 1-isomorphism is glued up from the
trivialisations
\qq\nn
\cT_\la\equiv\Phi_\p\vert_{\xcC_\la}\ :\
\iota_\la^*\cG^{(\sfk)}\equiv\cG^{(\sfk)}\vert_{\xcC_\la}\xrightarrow{\ \cong\ }
\cI_{\om^{(\sfk)}_{\p,\la}}\,.
\qqq
It is vital to note that -- as was shown for $\,\Gx={\rm SU}(N)\,$ in \Rxcite{Sec.\,8.1}{Gawedzki:2002se},\ and for an arbitrary compact simple 1-connected Lie group $\,\txG\,$ in \Rxcite{Sec.\,5.1}{Gawedzki:2004tu} -- the 1-isomorphisms $\,\cT_\la\,$ exist for $\,\la\in\faff{\ggt}\,$ exclusively,\ and so they single out the subset $\,Q_\p\,$ of conjugacy classes in $\,\txG\,$ which coincides with the set of worldvolumes of stable (untwisted) maximally symmetric D-branes of the WZW model at level $\,\sfk$,\ {\it cf.}, {\it e.g.}, \Rcite{Felder:1999ka}. These are,\ in turn,\ in a one-to-one correspondence with the (untwisted) maximally symmetric boundary states of the associated BCFT ({\it ib.}).\ It is also worth mentioning that the 1-isomorphisms are unique up to a (unique) 2-isomorphism due to the 1-connectedness of the $\,\xcC_\la$.\ They are also $\Ad$-symmetric,\ by the same token.

\subsection{The maximally symmetric WZW bi-brane}\label{sec:mxymbibr}

The species of bi-brane central to our considerations provides the data of the non-boundary maximally symmetric WZW defects,\ implementing jumps by elements of the target Lie group in the sense expressed by formul\ae ~\eqref{eq:maxym-nb-bib-maps} and \eqref{eq:WZW-field-defext}.\ A distinguished class of such defects -- the central-jump defects at which $\,g_{|2}=g_{|1}\cdot z\,$ for $\,z\,$ from the centre $\,Z(\Gx)\,$ of $\,\Gx\,$ -- alongside the attendant bi-brane were considered at length in \Rcite{Runkel:2008gr}.\ (One should also mention,\ parenthetically,\ the gauge-symmetry defect of \Rcite{Suszek:2012ddg} ({\it cf.}\ also \Rcite{Suszek:2013}) which in the current setting would realise a worldsheet-\emph{local} variant of the action $\,\ell\wp\,$ under restriction to a gauge-able subgroup $\,\txG_{\rm gauge}\subset\txG\x\txG\,$ of the bulk rigid-symmetry group  ({\it e.g.},\ the diagonal subgroup $\,\D(\txG)$.) The more general jump defects,\ with the jump from a fixed conjugacy class in $\,\txG$,\ were first discussed in \Rcite{Fuchs:2007fw},\ where the notion of a bi-brane was introduced.\ As emphasised in Sec.\,\ref{sub:maxym-nb-def},\ the $\si$-model field maps the defect line into the collection $\,Q=\Gx\x Q_\p\,$ of full $\txG\x\txG$-orbits,\ given by the cartesian product of the group manifold with the worldvolume of the maximally symmetric $\cG^{(\sfk)}$-brane.\ The ensuing maximally symmetric $\cG^{(\sfk)}$-bi-brane 
\qq\label{eq:bib-WZW}
\cB=\bigl(Q,\unl d{}^{(1)}_1,\unl d{}^{(1)}_0,\om^{(\sfk)},\Phi\bigr)
\qqq
has $\,Q\subset\unl\sfN{}_1\txG\,$ as its worldvolume,\ the 2-form $\,\om^{(\sfk)}\,$ from \Reqref{eq:om-WZW-nb} (fixed by the symmetry argument) as its curvature,\ the restrictions (to $\,Q$) of the lowest-level face maps of the simplicial $\txG\x\txG$-space $\,\unl\sfN{}_\bullet\txG\,$ ({\it i.e.},\ of the source and target maps of $\,\txG\rx\txG$) as the bi-brane maps,
\qq\nn
\iota_1=\unl d{}^{(1)}_1\,,\qquad\qquad\iota_2=\unl d{}^{(1)}_0\,,
\qqq
and the composite 1-isomorphism
\qq\nn
\Phi=\bigl(\cM\ox\id_{\cI_{-\pr_2^*\om^{(\sfk)}_\p}}\bigr)\circ\bigl(
\id_{\unl d{}^{(1)\,*}_1\cG}\ox\pr_2^*\Phi_\p^{-1}\ox\id_{\cI_{-\pr_2^*
\om^{(\sfk)}_\p}}\bigr)
\qqq
as the last piece of its data.\ Here,\ it is understood that $\,\Phi\,$ acts on the gerbe $\,\unl d{}^{(1)\,*}_1\cG^{(\sfk)}\ox\cI_{\pr_2^*\om^{(\sfk)}_\p} \ox\cI_{-\pr_2^*\om^{(\sfk)}_\p}\equiv\unl d{}^{(1)\,*}_1\cG^{(\sfk)}\equiv\iota_1^*\cG^{(\sfk)}$,\ as required by the definition of the bi-brane.

It ought to be underlined that -- once again -- the component bi-brane bi-module
\qq\nn
\Phi_\la:=\Phi\vert_{\txG\x\xcC_\la}=\bigl(\cM\vert_{\txG\x
\xcC_\la}\ox\id_{\cI_{-\pr_2^*\om^{(\sfk)}_{\p,\la}}}\bigr)\circ\bigl(
\id_{\unl d{}^{(1)\,*}_1\cG^{(\sfk)}}\ox\pr_2^*\cT_\la^{-1}\ox\id_{\cI_{-\pr_2^*\om^{(\sfk)}_{\p,
\la}}}\bigr)
\qqq
over $\,\txG\x\xcC_\la\,$ is unique up to a (unique) 2-isomorphism for $\,\txG\,$ a compact simple 1-connected Lie group.\ This,\ in conjunction with the unobstructed existence (and uniqueness up to a unique 2-isomorphism) of the 1-isomorphism $\,\cM\,$ determines the worldvolume of the maximally symmetric WZW bi-brane $\,\cB\,$ in terms of its boundary counterpart.

\subsection{The fusion 2-isomorphism}\label{sub:fus-2iso}

In this last section of our elucidation of the gerbe-theoretic structure associated with the maximally symmetric WZW defect,\ we give the definition and perform a (re)construction of the inter-bi-brane for the corresponding defect junctions of arbitrary valence.\ These defect junctions are particularly important for the WZW model as the operators that represent them in the quantum theory are enumerated by the Verlinde fusion coefficients of the RCFT,\ and so establishing a one-to-one correspondence between these coefficients (or the vector spaces of which they are dimensions) and `inequivalent' (in a suitable sense) 2-isomorphisms would be another piece of strong evidence in support of the claim that the gerbe theory of the \emph{classical} $\si$-model encodes a lot of significant information on the quantum theory.\ The first step towards this goal consists in writing out the very definition of the inter-bi-brane 2-isomorphism\footnote{In the diagrams and formul\ae ~of the present section,\ we are dropping the obvious subscripts from some identity morphisms for the sake of brevity.\ These can be reproduced easily from the context.},\ along the lines of \Rcite{Runkel:2008gr},\ which may subsequently be employed to fix the remaining data of the inter-bi-brane,\ as illustrated in the case study reported in \Rcite{Runkel:2009sp}.

Putting together the results of Sec.\,\ref{sub:maxym-def-junct} and the constraints on the admissible weight labels reviewed above,\ we obtain the inter-bi-brane worldvolume 
\qq\nn
T=\bigsqcup_{n\in\bN_{\geq 2}}\,T_{n,1}\,,\qquad
T_{n,1}=\txG\x\bigsqcup_{(\overrightarrow\la,\la)\in\faff{\ggt}^{\x
n+1}}\,\bigsqcup_{[\overrightarrow
w]\in\cF_{\overrightarrow\la}^\la}\,
T_{\overrightarrow\la}^{[\overrightarrow w]}\,,
\qqq
with the connected components defined in \Reqref{eq:Tlawe}.\ The (component) inter-bi-brane then takes the form
\qq\nn
\cJ_{n,1}=\bigl(T_{n,1},\pi^{(n+1)}_{1,2},\pi^{(n+1)}_{2,3},\ldots,,\pi^{(n+1)}_{n,n+1},\pi^{(n+1)}_{1,n+1},\varphi_{n+1}\bigr)\,,
\qqq
with the inter-bi-brane maps as in \Reqref{eq:WZW-ibb-maps},\ and with the 2-isomorphism over $\,T_{n,1}\,$ defined (after some obvious rearrangements of the 1-isomorphisms involved) by the diagram
\qq\label{diag:maxym-ibb-2-iso} \xy
(50,0)*{\bullet}="m2GGI"+(0,4)*{\tx{\scriptsize$\cG^{(\sfk)}_{123}\ox
\bigox_{k=4}^{n+1}\,\cG^{(\sfk)}_k\ox \cI_{\om^{(\sfk)}_{1,2}+\om^{(\sfk)}_{12,3}-
\sum_{k=4}^{n+1}\,\om^{(\sfk)}_{\p\,k}}$}};
(25,-15)*{\bullet}="mGGI"+(-25,0)*{\tx{\scriptsize$\cG^{(\sfk)}_{12}\ox
\bigox_{k=3}^{n+1}\,\cG^{(\sfk)}_k\ox \cI_{\om^{(\sfk)}_{1,2}-
\sum_{k=3}^{n+1}\,\om^{(\sfk)}_{\p\,k}}$}};
(75,-15)*{\bullet}="mGI"+(22,0)*{\tx{\scriptsize$\cG^{(\sfk)}_{12\cdots n+1}
\ox \cI_{\sum_{j=1}^n\,\om^{(\sfk)}_{12\cdots j,j+1}}$}};
(25,-35)*{\bullet}="GI"+(-18,0)*{\tx{\scriptsize$\bigox_{i=1}^{n
+1}\,\cG^{(\sfk)}_i\ox \cI_{-\sum_{k=2}^{n+1}\,\om^{(\sfk)}_{\p\,k}}$}};
(75,-35)*{\bullet}="GmGI"+(32,0)*{\tx{\scriptsize$\cG^{(\sfk)}_1\ox\cG^{(\sfk)}_{23
\cdots n+1}\ox \cI_{\sum_{j=1}^n\,\om^{(\sfk)}_{12\cdots
j,j+1}-\varrho^{(\sfk)}_{1,23\cdots n+1}}$}};
(35,-50)*{\bullet}="G1"+(-22,-4)*{\tx{\scriptsize$\cG^{(\sfk)}_1\equiv
\cG^{(\sfk)}_1\ox \cI_{\sum_{k=2}^{n+1}\,\om^{(\sfk)}_{\p\,k}}\ox
\cI_{-\sum_{k=2}^{n+1}\,\om^{(\sfk)}_{\p\,k}}$}};
(65,-50)*{\bullet}="G1I"+(5,-4)*{\tx{\scriptsize$\cG^{(\sfk)}_1\ox
\cI_{\Om^{(\sfk)}_{n\,2,3,\ldots,n+1}}$}}; (50,-50)*{}="id";
\ar@{->}|{\id\ox\bigox_{k=2}^{n+1}\,\Phi_{\p\,k}^{-1}\ox\id}
"G1";"GI" \ar@{->}|{\cM_{1,2}\ox\id} "GI";"mGGI"
\ar@{->}|{\cM_{12,3}\ox\id} "mGGI";"m2GGI" \ar@{->}|{\cdots}
"m2GGI";"mGI" \ar@{->}|{\cM_{1,23\cdots n+1}^{-1}\ox\id}
"mGI";"GmGI" \ar@{->}|{\id\ox\Phi_{\p\,{23\cdots n+1}}\ox\id}
"GmGI";"G1I" \ar@{=}|{\ \id\ } "G1"+(2,0);"G1I"+(-2,0) \ar@{=>}|{\
\varphi_{n+1}\ } "m2GGI"+(0,-3);"id"+(0,+3)
\endxy
\qqq
in which $\,\Om^{(\sfk)}_n\,$ is the universal 2-form \eqref{eq:Omn-def},\ obtained here with the help of identity \eqref{eq:D22Grhok},\ and in which the shorthand notation of App.\,\ref{app:convs} have been employed.\ Repeated use of the 2-isomorphism $\,\a\,$ from Diagram \eqref{diag:aleph} (with $\,\vartheta=0$),\ in conjunction with the canonical `death' 2-isomorphism $\,d_{\cM_{1,23\ldots n+1}}\,$ for the 1-isomorphism $\,\cM_{1,23\ldots n+1}\,$ ({\it cf.}\ \Rcite{Waldorf:2007mm}),\ enables us to reduce the upper part of the above diagram as {\it per}
\qq\nn
\xy
(50,0)*{\bullet}="G1G23nI"+(0,4)*{\tx{\tiny$\cG^{(\sfk)}_1\ox\cG^{(\sfk)}_{23\cdots
n+1}\ox \cI_{\sum_{j=1}^n\,\om^{(\sfk)}_{12\cdots j,j+1}-\varrho^{(\sfk)}_{1,23\cdots
n+1}}$}};
(30,-20)*{\bullet}="G12nI"+(-19,0)*{\tx{\tiny$\cG^{(\sfk)}_{12\cdots n+1}\ox
\cI_{\sum_{j=1}^n\,\om^{(\sfk)}_{12\cdots j,j+1}}$}};
(70,-20)*{\bullet}="G1G23nIbis"+(26,0)*{\tx{\tiny$\cG^{(\sfk)}_1\ox\cG^{(\sfk)}_{23\cdots
n}\ox \cI_{\sum_{j=2}^n\,\om^{(\sfk)}_{23\cdots j,j+1}-\om^{(\sfk)}_{\p\,2}}$}};
(30,-40)*{\bullet}="G14G5nI"+(-33,0)*{\tx{\tiny$\cG^{(\sfk)}_{1234}\ox
\bigox_{k=5}^{n+1}\,\cG^{(\sfk)}_k\ox \cI_{\om^{(\sfk)}_{1,2}+\om^{(\sfk)}_{12,3}+\om^{(\sfk)}_{123,4}-
\sum_{k=5}^{n+1}\,\om^{(\sfk)}_{\p\,k}}$}};
(70,-40)*{\bullet}="G1G24G5nI"+(35,0)*{\tx{\tiny$\cG^{(\sfk)}_1\ox\cG^{(\sfk)}_{234}
\ox\bigox_{k=5}^{n+1}\,\cG^{(\sfk)}_k\ox \cI_{\om^{(\sfk)}_{2,3}+\om^{(\sfk)}_{23,4}-\om^{(\sfk)}_{\p\,2}-
\sum_{k=5}^{n+1}\,\om^{(\sfk)}_{\p\,k}}$}};
(30,-60)*{\bullet}="G13G4nI"+(-28,0)*{\tx{\tiny$\cG^{(\sfk)}_{123}\ox
\bigox_{k=4}^{n+1}\,\cG^{(\sfk)}_k\ox \cI_{\om^{(\sfk)}_{1,2}+\om^{(\sfk)}_{12,3}-
\sum_{k=4}^{n+1}\,\om^{(\sfk)}_{\p\,k}}$}};
(70,-60)*{\bullet}="G1G23G4nI"+(30,0)*{\tx{\tiny$\cG^{(\sfk)}_1\ox\cG^{(\sfk)}_{23}
\ox\bigox_{k=4}^{n+1}\,\cG^{(\sfk)}_k\ox \cI_{\om^{(\sfk)}_{2,3}-\om^{(\sfk)}_{\p\,2}-
\sum_{k=4}^{n+1}\,\om^{(\sfk)}_{\p\,k}}$}};
(30,-80)*{\bullet}="G12G3nI"+(-24,0)*{\tx{\tiny$\cG^{(\sfk)}_{12}\ox
\bigox_{k=3}^{n+1}\,\cG^{(\sfk)}_k\ox \cI_{\om^{(\sfk)}_{1,2}-\sum_{k=3}^{n+1}\,
\om^{(\sfk)}_{\p\,k}}$}};
(70,-80)*{\bullet}="G1nI"+(18,0)*{\tx{\tiny$\bigox_{i
=1}^{n+1}\,\cG^{(\sfk)}_k\ox \cI_{-\sum_{k=2}^{n+1}\,\om^{(\sfk)}_{\p\,k}}$}};
(60,-10)*{}="id" \ar@{->}|{\cM_{1,23\cdots n+1}^{-1}\ox\id}
"G12nI";"G1G23nI" \ar@{=}|{\ \id\ }
"G1G23nI"+(2,-2);"G1G23nIbis"+(-2,2) \ar@{->}|{\vdots}
"G14G5nI";"G12nI" \ar@{->}|{\vdots} "G1G24G5nI";"G1G23nIbis"
\ar@{->}|{\cM_{123,4}\ox\id} "G13G4nI";"G14G5nI"
\ar@{->}|{\id\ox\cM_{23,4}\ox\id} "G1G23G4nI";"G1G24G5nI"
\ar@{->}|{\cM_{12,3}\ox\id} "G12G3nI";"G13G4nI"
\ar@{->}|{\id\ox\cM_{2,3}\ox\id} "G1nI";"G1G23G4nI"
\ar@{->}|{\cM_{1,23\cdots n+1}\ox\id} "G1G23nIbis";"G12nI"
\ar@{->}|{\cM_{1,234}\ox\id} "G1G24G5nI";"G14G5nI"
\ar@{->}|{\cM_{1,23}\ox\id} "G1G23G4nI";"G13G4nI"
\ar@{->}|{\cM_{1,2}\ox\id} "G1nI";"G12G3nI"
\ar@{=>}|{\a_{1,2,3}\ox\id} "G12G3nI"+(2,1);"G1G23G4nI"+(-2,-1)
\ar@{=>}|{\a_{1,23,4}\ox\id} "G13G4nI"+(2,1);"G1G24G5nI"+(-2,-1)
\ar@{=>}|{\cdots} "G14G5nI"+(2,1);"G1G23nIbis"+(-2,-1)
\ar@{=>}|{d_{\cM_{1,23\cdots n+1}}} "G12nI"+(2,1);"id"+(-2,-1)
\endxy
\qqq
This permits us to extract from the former diagram the subdiagram
\qq\label{diag:hex-tilde-2iso} \xy
(35,0)*{\bullet}="mGGI"+(-24,4)*{\tx{\scriptsize$\cG^{(\sfk)}_1\ox\cG^{(\sfk)}_{23}\ox
\bigox_{k=4}^{n+1}\,\cG^{(\sfk)}_k\ox \cI_{\om^{(\sfk)}_{2,3}-\om^{(\sfk)}_{\p\,2}-
\sum_{k=4}^{n+1}\,\om^{(\sfk)}_{\p\,k}}$}};
(65,0)*{\bullet}="mGI"+(30,4)*{\tx{\scriptsize$\cG^{(\sfk)}_1\ox\cG^{(\sfk)}_{234}\ox
\bigox_{k=5}^{n+1}\,\cG^{(\sfk)}_k\ox \cI_{\om^{(\sfk)}_{2,3}+\om^{(\sfk)}_{23,4}-\om^{(\sfk)}_{\p\,2}-
\sum_{k=5}^{n+1}\,\om^{(\sfk)}_{\p\,k}}$}};
(25,-20)*{\bullet}="GI"+(-18,0)*{\tx{\scriptsize$\bigox_{i=1}^{n+1}
\,\cG^{(\sfk)}_i\ox \cI_{-\sum_{k=2}^{n+1}\,\om^{(\sfk)}_{\p\,k}}$}};
(75,-20)*{\bullet}="GmGI"+(32,0)*{\tx{\scriptsize$\cG^{(\sfk)}_1\ox\cG^{(\sfk)}_{23
\cdots n+1}\ox \cI_{\sum_{j=1}^n\,\om^{(\sfk)}_{12\cdots j,j+1}-
\varrho^{(\sfk)}_{1,23\cdots n+1}}$}};
(35,-40)*{\bullet}="G1"+(0,-4)*{\tx{\scriptsize$\cG^{(\sfk)}_1$}};
(65,-40)*{\bullet}="G1I"+(5,-4)*{\tx{\scriptsize$\cG^{(\sfk)}_1\ox
\cI_{\Om^{(\sfk)}_{n\,2,3,\ldots,n+1}}$}}; (50,0)*{}="id1"; (50,-40)*{}="id2";
\ar@{->}|{\id\ox\bigox_{k=2}^{n+1}\,\Phi_{\p\,k}^{-1}\ox\id}
"G1";"GI" \ar@{->}|{\id\ox\cM_{2,3}\ox\id} "GI";"mGGI"
\ar@{->}|{\id\ox\cM_{23,4}\ox\id} "mGGI";"mGI" \ar@{->}|{\vdots}
"mGI";"GmGI" \ar@{->}|{\id\ox\Phi_{\p\,{23\cdots n+1}}\ox\id}
"GmGI";"G1I" \ar@{=}|{\ \id\ } "G1"+(2,0);"G1I"+(-2,0) \ar@{=>}|{\
\widetilde\varphi{}_{n+1}\ } "id1"+(0,-3);"id2"+(0,+3)
\endxy
\qqq
with the composite 2-isomorphism
\qq\nn
\widetilde\varphi{}_{n+1}:=\varphi_{n+1}\bullet\d_{n+1}^{-1}
\qqq
that we are after.\ The latter has been expressed in terms of the 2-isomorphism
\qq\nn
\d_{n+1}&:=&\id\circ\bigl[d_{\cM_{1,23\cdots n+1}}\bullet
\bigl(\id\circ(\a_{1,23\cdots n,n+1}\ox\id)\bigr)\bullet\bigl(\id
\circ(\a_{1,23\cdots n-1,n}\ox\id)\bigr)\cr\cr
&&\bullet\cdots\bullet\bigl(\id\circ(\a_{1,2,3}\ox\id)\bigr)\bigr]
\circ\id\,.
\qqq
In order to draw further conclusions from the above reasoning,\ we need the following straightforward

\berop \label{prop:spect-gerbe-iso}
Let $\,\cG,\cG_1\,$ and $\,\cG_2\,$ be 1-gerbes over a common base $\,M$,\ and let
\qq\nn
\Phi_A\ :\ \cG_1\xrightarrow{\ \cong\ }\cG_2\,,\quad A=1,2
\qqq
be two 1-isomorphisms between $\,\cG_1\,$ and $\,\cG_2$.\ There exists a 2-isomorphism
\qq\nn
\alxydim{}{\cG\ox\cG_1 \ar@/^1.6pc/[rrr]^{\id_\cG\ox\Phi_1}="5"
\ar@/_1.6pc/[rrr]_{\id_\cG\ox\Phi_2}="6"
\ar@{=>}"5"+(0,-4);"6"+(0,4)|{\varphi} & & & \cG\ox\cG_2}
\qqq
iff there exists a 2-isomorphism
\qq\nn
\alxydim{}{\cG_1 \ar@/^1.6pc/[rrr]^{\Psi_1}="5"
\ar@/_1.6pc/[rrr]_{\Psi_2}="6" \ar@{=>}"5"+(0,-4);"6"+(0,4)|{\psi}
& & & \cG_2}\,.
\qqq
\eerop
\noindent From the last proposition,\ we immediately infer that the existence of a 2-isomorphism $\,\widetilde\varphi{}_{n+1}\,$ is equivalent to the existence of another 2-isomorphism that we introduce hereunder,\ upon stripping diagram \eqref{diag:hex-tilde-2iso} of the spectator tensor factor $\,\cG^{(\sfk)}_1\,$ and the attendant identity 1-isomorphisms.
\bedef\label{def:fus-2-iso}
Given arbitrary $\,\overrightarrow\la\in\faff{\ggt}^{\x n}\,$ and $\,[\overrightarrow w]\in\cS_{\la_1}\backslash\x_{i=2}^{n}\,(\txG/\cS_{\la_i})\,$ such that $\,\rho_{\overrightarrow\la}([\overrightarrow w])=\la\in\faff{\ggt}$,\ a ({\bf component}) {\bf WZW fusion 2-isomorphism} ({\bf of valence $n+1$}) over $\,T_{\overrightarrow \la}^{[\overrightarrow w]}\,$ is given by the diagram
\qq\label{diag:fus-2-iso}
\alxydim{@C=7em@R=6em}{\bigox_{i=1}^n\,\cG^{(\sfk)}_i
\ar[d]_{\bigox_{i=1}^n\,\pr_i^*\cT_{\la_i}}
\ar[r]^{\xcM_n(\tau_n^{\rm L})\hspace{30pt}}
& \cG^{(\sfk)}_{12\cdots n}\ox\cI_{\sum_{j=2}^n\,\varrho^{(\sfk)}_{12\cdots j-1,j}}
\ar[d]^{\txm_{12\ldots n}^*\cT_{\la} \ox\id} \\
\cI_{\sum_{i=1}^n\,\pr_i^*\om^{(\sfk)}_{\la_i}}
\ar@{=>}[ur]|{\varphi_{\overrightarrow\la}^\la[\overrightarrow w]}
\ar@{=}[r]|{\ \id\ }
& \cI_{\sum_{i=1}^n\,\pr_i^*\om^{(\sfk)}_{\la_i}+\Om^{(\sfk)}_n}}
\qqq
in which all gerbes and morphisms are implicitly restricted to $\,T_{\overrightarrow\la}^{[\overrightarrow w]}$.\ Fusion
2-isomorphisms for $\,n=2$
\qq
\alxydim{@C=7em@R=6em}{\cG^{(\sfk)}_1\ox\cG^{(\sfk)}_2 \ar[d]_{\pr_1^*\cT_{\la_1}\ox
\pr_2^*\cT_{\la_2}} \ar[r]^{\cM\equiv\xcM_2(\t^{\rm L}_2)} &
\cG^{(\sfk)}_{12}\ox \cI_{\varrho^{(\sfk)}_{1,2}} \ar[d]^{\txm^*\cT_\la\ox\id} \\
\cI_{\pr_1^*\om^{(\sfk)}_{\p,\la_1}+\pr_2^*\om^{(\sfk)}_{\p,\la_2}}
\ar@{=>}[ur]|{\varphi_{(\la_1,\la_2)}^\la[w]} \ar@{=}[r]|{\ \id\ }
& \cI_{\varrho^{(\sfk)}_{1,2}+\txm^*\om^{(\sfk)}_{\p,\la}}}
\qqq
shall be called {\bf basic}.

The sets $\,\cF_{\overrightarrow\la}^\la$,\ first mentioned in Sec.\,\ref{sub:maxym-def-junct},\ are characterised by the condition
\qq\nn
[\overrightarrow w]\in\cF_{\overrightarrow\la}^\la\
\Longleftrightarrow\ \tx{there exists a fusion 2-isomorphism }
\varphi_{\overrightarrow\la}^\la [\overrightarrow w]\in 2\tx{-}{\rm
Iso}\left(\bgrb^\nabla(T_{\overrightarrow\la}^{[\overrightarrow w]})
\right)\,.
\qqq
Thus,\ in particular,\ $\,\cF_{\overrightarrow\la}^\la=\emptyset\,$ iff the manifold $\,T_{\overrightarrow\la}^{[\overrightarrow w]}\,$ does not support a fusion 2-isomorphism. 
\exdef

Drawing motivation (in part) from the explicit construction carried
out for $\,\txG=\sug\,$ in \Rcite{Runkel:2009sp},\ we propose
\becj\label{conj:2-iso-vs-Verlinde}
Let $\,\cN_{\la_1\la_2\ldots\la_n}^{\hspace{30pt}\la}\in\bN\,$ be
the structure constants of the Verlinde fusion ring giving the
multiplicity of the chiral sector $\,[\la]\,$ of the WZW $\si$-model
in the fusion product of the chiral sectors $\,[\la_i]$,
\qq\nn
[\la_1]\star[\la_2]\star\cdots\star[\la_n]=\sum_{\la\in\faff{\ggt}}\,
\cN_{\la_1\la_2\ldots\la_n}^{\hspace{30pt}\la}\,[\la]\,.
\qqq
Then,
\qq\nn
\cN_{\la_1\la_2\ldots\la_n}^{\hspace{30pt}\la}=\vert\cF_{(\la_1,
\la_2,\ldots,\la_n)}^\la\vert\,.
\qqq
In other words, the direct summands of type $\,[\la]\,$ in the
decomposition of the fusion product $\,[\la_1]\star[\la_2]\star
\cdots\star[\la_n]\,$ are in a one-to-one
correspondence with the disjoint components $\,T_{\overrightarrow
\la}^{[\overrightarrow w]}\,$ in the decomposition of the manifold
$\,T_{\overrightarrow\la}^\la\,$ into connected $\txG$-submanifolds.  \ecj

We shall next take a closer look at some intricate constructive
implications of Definition \ref{def:fus-2-iso}. As follows from the
general discussion in Sec.\,\ref{sub:rudi-gerbe}, any two fusion
2-isomorphisms on the connected manifold
$\,T_{\overrightarrow\la}^{[\overrightarrow w]}\,$ differ by a phase
factor (a locally constant $\uj$-valued map), and a redefinition
\qq\nn
\varphi_{\overrightarrow\la}^\la[\overrightarrow w]\longmapsto
\varphi_{\overrightarrow\la}^\la[\overrightarrow w]\cdot
c_{\overrightarrow\la}^\la[\overrightarrow w]\,,\qquad
c_{\overrightarrow\la}^\la[\overrightarrow w]\in\uj
\qqq
of a given fusion 2-isomorphism $\,\varphi_{\overrightarrow\la}^\la
[\overrightarrow w]\,$ by an arbitrary phase factor
$\,c_{\overrightarrow\la}^\la[\overrightarrow w]\,$ produces an
admissible fusion 2-isomorphism. This observation, in conjunction
with our earlier discussion of the maps
$\,\pi_{\overrightarrow\la}^{[ \overrightarrow w]}(i,j\vert k,l)\,$
of \Reqref{eq:proj-nibb-elemibb}, prompts us to enquire as to the
possibility of realising in the present setting the induction scheme
put forward in \Rxcite{Sec.\,2.8}{Runkel:2008gr} and further
elaborated in the context of $\si$-model symmetries in
\Rxcite{Rem.\,5.6}{Suszek:2011hg} and
\Rxcite{Sec.\,8.3}{Suszek:2012ddg}. The idea is to use the said maps
and certain distinguished basic fusion 2-isomorphisms $\,\varphi_{(
\la_1,\la_2)}^\la[w]\,$ to derive a fusion 2-isomorphism
$\,\varphi_{\overrightarrow\la}^\la [\overrightarrow w]\,$ through a
systematic construction that we proceed to describe. Let us begin
with
\berop\label{prop:alpha-relns-fus-2-iso}
A fusion 2-isomorphism $\,\varphi_{\overrightarrow\la}^\la[
\overrightarrow w]\,$ exists over $\,T_{\overrightarrow\la}^{[
\overrightarrow w]}\,$ iff there exists a 2-isomorphism
\qq\nn
\alxydim{@C=7em@R=6em}{\bigox_{i=1}^n\,\cG_i
\ar[d]_{\bigox_{i=1}^n\,\pr_i^*\cT_{\la_i}}
\ar[r]^{\xcM_n(\tau)\hspace{30pt}}
& \cG_{12\cdots n}\ox I_{\sum_{j=2}^n\,\varrho_{12\cdots j-1,j}}
\ar[d]^{\txm_{12\ldots n}^*\cT_{\la} \ox\id} \\
I_{\sum_{i=1}^n\,\pr_i^*\om_{\la_i}}
\ar@{=>}[ur]|{\varphi_{\overrightarrow\la}^\la[\overrightarrow
w]_\tau} \ar@{=}[r]|{\ \id\ }
& I_{\sum_{i=1}^n\,\pr_i^*\om_{\la_i}+\Om_n}}
\qqq
for an arbitrary tree $\,\tau\in{\rm PBT}_n$,\ and the two are
related (up to a phase factor) through the equality
\qq\nn
\varphi_{\overrightarrow\la}^\la[\overrightarrow w]_\tau=c\cdot
(\id_{\txm_{12\ldots n}^*\cT_{\la}\ox\id}\circ\xcA_n(\tau))\bullet
\varphi_{\overrightarrow\la}^\la[\overrightarrow w]\,,
\qqq
written for some constant $\,c\in\uj$.
\eerop
\beroof
An immediate consequence of Cor.\,\ref{cor:An-map}. \eroof
\noindent A priori, the proposition affords the possibility of
building a fusion 2-isomorphism over a given manifold
$\,T_{\overrightarrow \la}^{[\overrightarrow w]}\,$ in $\,\vert{\rm
FBT}_n\vert\,$ ways. Indeed, for each (planar) binary tree
$\,\tau\,$ representing a specific multiplication scheme, we may,\ \emph{apparently},\
write the corresponding fusion 2-isomorphism as a composition of
pullbacks along the respective maps $\,\txm^{(n)}_{(i,j\vert j+1,k)}\in
\mu(\tau)\,$ of basic fusion 2-isomorphisms associated with the
binary vertices of $\,\tau$.\ An example of this construction (for
$\,\tau=\tau_n^{\rm L}$) is presented in the diagram below,
\qq\nn\hspace{-.3cm}
{\tiny\alxydim{@C=1em@R=4em}{\bigox_{i=1}^n\,\cG_i
\ar[d]_{\cT_{\la_1}\ox\cT_{\la_2}\ox\id}
\ar[r]^{\cM_{1,2}\ox\id\hspace{25pt}}
& I_{\varrho_{1,2}}\ox\cG_{12}\ox\bigox_{j=3}^n\,\cG_j
\ar[rr]^{\id\ox\cM_{12,3}\ox\id\hspace{15pt}}
\ar[d]^{\id\ox\cT_{\la_{12}}\ox\id}
& & I_{\varrho_{12,3}}\ox\cG_{123}\ox\bigox_{j=4}^n\,\cG_j\ox
I_{\varrho_{1,2}} \ar[rr]^{\cdots}
\ar[ddd]^{\id\ox\cT_{\la_{123}}\ox\id} & & \cG_{12\cdots n}\ox
I_{\sum_{j=2}^n\, \varrho_{12\cdots j-1,j}}
\ar[dddd]^(.6){\cT_{\la_{12\cdots n-1}}\ox\id} \\
I_{\om_{\p,\la_1}+\om_{\p,\la_2}}\ox\bigox_{j=3}^n\,\cG_i \ar@{=}[r]
\ar@{=>}[ur]|{\varphi_{(\la_1,\la_2)}^{\la_{12}}[w_{12}]_{1,2}\ox\id}
\ar[dd]_{\id\ox\cT_{\la_3}\ox\id} &
I_{\varrho_{1,2}+\om_{\p,\la_{12}}}\ox\bigox_{j=
3}^n\,\cG_j & & & & \\ & & & & & \\
I_{\om_{\p,\la_1}+\om_{\p,\la_2}+\om_{\p,\la_3}}\ox
\bigox_{j=4}^n\,\cG_i \ar@{=}[rrr]
\ar@{=>}[uuurrr]|(.4){\id\ox\varphi_{(\la_{12},\la_3)}^{\la_{123}}[w_{123}]_{12,3}\ox\id}
\ar[d]_{\vdots}
& & & I_{\varrho_{1,2}+
\varrho_{12,3}+\om_{\p,\la_{123}}}\ox\bigox_{j=4}^n\,\cG_j &
\ar@{}[dr]|{\ddots} & \\ I_{\sum_{i=1}^n\,\om_{\p,\la_i}}
\ar@{=}[rrrrr] & & & & & I_{\sum_{i=
1}^n\,\om_{\p,\la_i}+\Om_n}}}\,,
\qqq
written over $\,T_{\overrightarrow\la}^{[ \overrightarrow w]}\,$ and
for $\,\la_{12\cdots k}:=\La_{\overrightarrow\la}^{[\overrightarrow
w]}(1,k)\,$ and $\,w_{12\cdots k}:=[W]_{\overrightarrow\la}^{[
\overrightarrow w]}(i,k-1\vert k,k),\ k=1,2,\ldots,n$.\ From the
diagram, we read off the composite fusion 2-isomorphism
\qq\nn
\varphi_{\overrightarrow\la}^\la[\overrightarrow w]\equiv
\varphi_{\overrightarrow \la}^\la[\overrightarrow w]_{\tau_n^{\rm
L}}&=&(\id\ox\varphi_{(\la_{12\cdots n-1},\la_n)}^{\la_{12\cdots n}}
[w_{12\cdots n}]_{12\cdots n-1,n}\ox\id)\cr\cr
&&\bullet\left(\id\circ(\id\ox\varphi_{(\la_{12\cdots n-2},\la_{n-
1})}^{\la_{12\cdots n-1}}[w_{12\cdots n-1}]_{12\cdots n-2,n-1}\ox
\id)\right)\cr\cr
&&\bullet\cdots\bullet\left(\id\circ(\varphi_{(\la_1,\la_2
)}^{\la_{12}}[w_{12}]_{1,2}\ox\id)\right)\,.
\qqq
The above reconstruction procedure may,\ however,\ encounter a topological
obstruction:\ Whenever at least one of the maps
$\,[W]\circ\pi_{\overrightarrow\la}^{[
\overrightarrow w]}(i,j\vert j+1,k)\,$ for $\,\txm^{(n)}(i,j\vert
j+1,k)\in\mu_n(\tau)\,$ sends $\,T_{\overrightarrow\la}^{[
\overrightarrow w]}\,$ outside the corresponding nonempty $\,\cF_{(
\La_{\overrightarrow\la}^{[\overrightarrow w]}(i,j),
\La_{\overrightarrow\la}^{[\overrightarrow w]}(j+1,k))}^{\La_{\overrightarrow\la}^{[
\overrightarrow w]}(i,k)}\subset\cS_{\La_{\overrightarrow\la}^{[\overrightarrow w]}(i,j)}\backslash\txG/
\cS_{\La_{\overrightarrow\la}^{[\overrightarrow w]}(j+1,k)}\,$ with $\,\La_{\overrightarrow\la}^{[
\overrightarrow w]}(i,k)\in\faff{\ggt}$,\ there is no basic fusion 2-isomorphism to pull
back to the vertex. A natural point of departure for the analysis of
the obstruction is an explicit verification of Conjecture
\ref{conj:2-iso-vs-Verlinde} which is available only in the
particular case of $\,\txG=\sug\,$ as of this writing. The expected
correspondence of the gerbe structures under consideration with data
of the RCFT of the WZW $\si$-model leads us to postulate further
\becj\label{conj:indu}
For all $\,\tau\in{\rm FBT}_n\,$ and each $\,\txm^{(n)}(i,j\vert j+1,k
)\in\mu_n(\tau)$,\ the manifold $\,T_{\overrightarrow\la}^{[
\overrightarrow w]}\,$ is mapped by $\,[W]\circ\pi_{\overrightarrow\la}^{[\overrightarrow w]}(i,j\vert j+1,k)\,$ to the set $\,\cF_{(
\La_{\overrightarrow\la}^{[\overrightarrow w]}(i,j),
\La_{\overrightarrow\la}^{[\overrightarrow w]}(j+1,k))}^{\La_{\overrightarrow\la}^{[
\overrightarrow w]}(i,k)}\neq\emptyset$,\ so that the maximally symmetric WZW background is descent-complete in the sense of Def.\,\ref{simplssbckgrndcompl}. \ecj 
\noindent Whenever the obstruction is absent, we have
\berop
Let $\,\tau\in{\rm FBT}_n\,$ be such that for each $\,\txm^{(n)}(i,j
\vert j+1,k)\in\mu_n(\tau)\,$ the manifold $\,T_{\overrightarrow
\la}^{[\overrightarrow w]}\subset T_{\overrightarrow\la}^\la\,$ is mapped by $\,[W]\circ\pi_{\overrightarrow\la}^{[\overrightarrow w]}(i,j\vert j+1,k)\,$ to the set
$\,\cF_{(
\La_{\overrightarrow\la}^{[\overrightarrow w]}(i,j),
\La_{\overrightarrow\la}^{[\overrightarrow w]}(k,l))}^{\La_{\overrightarrow\la}^{[
\overrightarrow w]}(i,k)}\neq\emptyset\,$ with $\,\La_{\overrightarrow\la}^{[
\overrightarrow w]}(i,k)\in\faff{\ggt}$.\ Then, there exists a fusion 2-isomorphism
$\,\varphi_{\overrightarrow\la}^{[\overrightarrow w]}[\tau]\,$ given
by a vertical composition of pullbacks of $\,n\,$ basic fusion
2-isomorphisms from the manifolds $\,T^{[W]_{\overrightarrow\la}^{[\overrightarrow w]}(i,j\vert j+1,k)}_{(
\La_{\overrightarrow\la}^{[\overrightarrow w]}(i,j),
\La_{\overrightarrow\la}^{[\overrightarrow w]}(j+1,k))}\equiv\pi_{\overrightarrow\la}^{[\overrightarrow w]}(i,j\vert j+1,k)(T_{\overrightarrow\la}^{[\overrightarrow w]})\,$ along (all)
the respective maps $\,\txm^{(n)}(i,j\vert j+1,k)\,$ as described in Def.\,\ref{simplssbckgrndcompl}. 
\eerop
\beroof
Obvious.
\eroof

In the next step, we examine the phase factors that relate various
induced fusion 2-isomorphisms on a given connected component of the
inter-bi-brane world-volume. Now,\ any binary tree with $n$ leaves obtained from a
given $n$-ary tree in the recursive ternary resolution detailed in App.\,\ref{app:proof-simpl} can be reduced to the distinguished one
$\,\tau_n^{\rm L}\,$ by successive application of a finite number of
local \emph{associator moves} (the dotted lines represent the rest
of the binary tree which can be arbitrary)
\qq\nn
\xy (60,20)*{}="Up"+(0,4)*{}; (50,10)*{\bullet}="R"+(0,4)*{};
(40,0)*{\bullet}="V12"+(-5,3)*{xy}; (30,-10)*{}="V1"+(0,-4)*{x};
(50,-10)*{}="V2"+(0,-4)*{y}; (70,-10)*{}="V3"+(0,-4)*{z};
(20,-20)*{}="Down_1"+(0,-4)*{}; (60,-20)*{}="Down_2"+(0,-4)*{};
(80,-20)*{}="Down_3"+(0,-4)*{}; \ar@{-} "R";"V12" \ar@{-} "R";"V3"
\ar@{-} "V12";"V1" \ar@{-} "V12";"V2" \ar@{.} "Up";"R" \ar@{.}
"V1";"Down_1" \ar@{.} "V2";"Down_2" \ar@{.} "V3";"Down_3"
\endxy\qquad \alxydim{@C=1.0cm@R=1.5cm}{ \ar@{~>}[r] & }\qquad
\xy (60,20)*{}="Up"+(0,4)*{}; (50,10)*{\bullet}="R"+(0,4)*{};
(60,0)*{\bullet}="V23"+(5,3)*{yz}; (30,-10)*{}="V1"+(0,-4)*{x};
(50,-10)*{}="V2"+(0,-4)*{y}; (70,-10)*{}="V3"+(0,-4)*{z};
(20,-20)*{}="Down_1"+(0,-4)*{}; (60,-20)*{}="Down_2"+(0,-4)*{};
(80,-20)*{}="Down_3"+(0,-4)*{}; \ar@{-} "R";"V23" \ar@{-} "R";"V1"
\ar@{-} "V23";"V2" \ar@{-} "V23";"V3" \ar@{.} "Up";"R" \ar@{.}
"V1";"Down_1" \ar@{.} "V2";"Down_2" \ar@{.} "V3";"Down_3"
\endxy\,.
\qqq
Therefore, as long as we restrict our attention to fusion
2-isomorphisms induced from basic ones, complete information on the
algebraic structure of the phase relations between various induction
schemes (corresponding to various elements of $\,{\rm FBT}_n$) is
contained in the simplest non-basic (induced) fusion 2-isomorphisms,
to wit, those for 4-valent defect junctions. For these, we find $\,{\rm FBT}_3=\{\tau_3^{\rm L},\tau_3^{\rm R}\}$,\ {\it cf.}\ \eqref{eq:LRtreen}.\ Taking into account Prop.\,\ref{prop:alpha-relns-fus-2-iso}, we
arrive at
\berop\label{prop:fusion-phase}
Let $\,\la_1,\la_2,\la_3,\la\in\faff{\ggt}\,$ and $\,[w_2,w_3]\in
\cS_{\la_1}\backslash\left((\txG/\cS_{\la_2})\x(\txG/\cS_{\la_3})
\right)\,$ be such that $\,\emptyset\neq T_{(\la_1,\la_2,\la_3)}^{[w_2,w_3]}\subset T_{(\la_1,\la_2,\la_3)}^\la\,$ is
mapped by $\,[W]\circ
\pi_{(\la_1,\la_2,\la_3)}^{[w_2,w_3]}(i,j\vert j+1,k)\,$ to $\,\cF_{(
\La_{(\la_1,\la_2,\la_3)}^{[w_2,w_3]}(i,j),
\La_{(\la_1,\la_2,\la_3)}^{[w_2,w_3]}(j+1,k))}^{\La_{(\la_1,\la_2,\la_3)}^{[w_2,w_3]}(i,k)}\,$ with $\,\La_{(\la_1,\la_2,\la_3)}^{[w_2,w_3]}(i,k)\in\faff{\ggt}\,$ for every triple $\,(i,j,k)\in\{(1,1,2),
(2,2,3),(1,2,3),(1,1,3)\}$.\ Denote
\qq\nn
&\la_{12}:=\La_{(\la_1,\la_2,\la_3)}^{[w_2,w_3]}(1,2)\,,\qquad\qquad
\la_{23}:=\La_{(\la_1,\la_2,\la_3)}^{[w_2,w_3]}(2,3)\,,&\cr\cr\cr
&[w_{12}]:=[W]_{(\la_1,\la_2,\la_3)}^{[w_2,w_3]}(1,1\vert 2,2)\,,\qquad
\qquad[w_{23}]:=[W]_{(\la_1,\la_2,\la_3)}^{[w_2,w_3]}(2,2\vert 3,3)\,,&
\cr\cr
&[w_{123}^{(1)}]:=[W]_{(\la_1,\la_2,\la_3)}^{[w_2,w_3]}(1,2\vert 3,3)\,,
\qquad\qquad[w_{123}^{(2)}]:=[W]_{(\la_1,\la_2,\la_3)}^{[w_2,w_3]}
(1,1\vert 2,3)\,.&
\qqq
There exists an element
\qq\nn
\Phi_{[w_{123}^{(2)}]\la_{23}[w_{23}],[w_{12}]\la_{12}[w_{123}^{(1
)}]}^{(\la_1,\la_2,\la_3)\ \la}[w_2,w_3]\in\uj\,,
\qqq
to be termed the {\bf fusion phase} in what follows, which satisfies
the relation
\qq\nn
&&\pi_{(\la_1,\la_2,\la_3)}^{[w_2,w_3]}(1,1\vert 2,3)^*\varphi_{(\la_1,\la_{23})}^\la[w_{123}^{(2)}]
\bullet\left(\id_{\id_{I_{\pr_2^*\om_{\la_2}+\pr_3^*\om_{\la_3}}}}
\circ\pi_{(\la_1,\la_2,\la_3)}^{[w_2,w_3]}(2,2\vert 3,3)^*\varphi_{(\la_2,\la_3)}^{\la_{23}}[w_{23}]
\right)\cr\cr
&=&\Phi_{[w_{123}^{(2)}]\la_{23}[w_{23}],[w_{12}]\la_{12}[w_{12
3}^{(1)}]}^{(\la_1,\la_2,\la_3)\ \la}[w_2,w_3]\cdot\pi_{(\la_1,\la_2,\la_3)}^{[w_2,w_3]}(1,2\vert 3,3)^*
\varphi_{(\la_{12},\la_3)}^\la[w_{123}^{(1)}]\cr\cr
&&\bullet\left(
\id_{\id_{I_{\pr_1^*\om_{\la_1}+\pr_2^*\om_{\la_2}}}}\circ\pi_{(\la_1,\la_2,\la_3)}^{[w_2,w_3]}(1,1\vert 2,2)^*\varphi_{(\la_1,\la_2)}^{\la_{12}}[w_{12}]\right)
\qqq
for a given collection $\,\{\varphi_{(\la_1,\la_2)}^{\la_{12}}[
w_{12}],\varphi_{(\la_2,\la_3)}^{\la_{23}}[w_{23}],\varphi_{(
\la_{12},\la_3)}^\la[w_{123}^{(1)}],\varphi_{(\la_1,\la_{23})}^\la[
w_{123}^{(2)}]\}\,$ of basic fusion 2-isomorphisms. The fusion phase
is determined up to redefinitions
\qq\nn
\Phi_{[w_{123}^{(2)}]\la_{23}[w_{23}],[w_{12}]\la_{12}[w_{123}^{(1
)}]}^{(\la_1,\la_2,\la_3)\ \la}[w_2,w_3]\cr\cr\longmapsto
\Phi_{[w_{123}^{(2)}]\la_{23}[w_{23}],[w_{12}]\la_{12}[w_{123}^{(1
)}]}^{(\la_1,\la_2,\la_3)\ \la}[w_2,w_3]\cdot c_{(\la_2,\la_3
)}^{\la_{23}}[w_{23}]^{-1}\cdot c_{(\la_{12},\la_3)}^\la[w_{123}^{(
1)}]\cdot c_{(\la_1,\la_{23})}^\la[w_{123}^{(
2)}]^{-1}\cdot c_{(\la_1,
\la_2)}^{\la_{12}}[w_{12}]
\qqq
that correspond to phase rescalings of the basic fusion
2-isomorphisms.
\eerop
\noindent Our considerations lead us naturally to
\bedef\label{def:maxym-WZW-ibi-mat}
Let $\,\phi:=\{\varphi_{(\la_1,\la_2)}^{[w]}\}_{\la_1,\la_2,\la\in
\faff{\ggt},[w]\in\cF_{(\la_1,\la_2)}^\la}\,$ be a collection of
basic fusion 2-isomorphisms, defined over the respective connected
submanifolds $\,T_{(\la_1,\la_2)}^{[w]}\subset T_{(\la_1,\la_2)}^\la$.\ The {\bf maximally
symmetric WZW inter-bi-brane fusing matrix} for $\,\phi\,$ is the
matrix with entries given by the fusion phases $\,\Phi_{[w_{123}^{(2
)}]\la_{23}[w_{23}],[w_{12}]\la_{12}[w_{123}^{(1)}]}^{(\la_1,\la_2,
\la_3)\ \la}[w_2,w_3]\,$ labelled by $\,\la_1,\la_2,\la_3,\la\in
\faff{\ggt}\,$ and $\,[w_2,w_3]\in\cF_{(\la_1,\la_2,\la_3)}^\la\,$
as well as objects $\,\la_{12},\la_{23},w_{12},w_{23},w_{123}^{(1)},
w_{123}^{(2)}\,$ defined in Prop.\,\ref{prop:fusion-phase}.
\exdef \noindent As a continuation of Conjectures
\ref{conj:2-iso-vs-Verlinde} and \ref{conj:indu},\ we anticipate that the above inter-bi-brane fusing matrix is related directly to the standard fusing matrix $\,F_{[w_{123}^{(2)}]\la_{23}[w_{23}],[w_{12}]\la_{12}[w_{123}^{(1)}]}^{(\la_1,\la_2,\la_3)\ \la}[w_2,w_3]\,$ of
the WZW $\si$-model (written in the notation of \Rcite{Fuchs:2002cm}),\ the latter being regarded as a largely cohomological object,\ {\it cf.}\ \Rxcite{App.\,E}{Moore:1988qv}.\ We hope to return to the question of the precise nature of this relationship in a future study. 

When taken in conjunction with the naturality of the gerbe-theoretic constructions employed (and the underlying relation between the maximally symmetric gerbe (bi)modules resp.\ defects and sectors of the bulk WZW model),\ the remarks made towards the end of Sec.\,\ref{sec:cs} certainly render the above conjectures plausible and the propositions contingent on their veracity highly interesting.\ Let us also note in this context that a potential source of refinement -- suggested by the r\^ole of the 3d Chern--Simons theory in the analysis of Sec.\,\ref{sec:cs} -- of the fundamental correspondence (implicitly posited in Conjecture \ref{conj:2-iso-vs-Verlinde}) between -- on the one hand -- the disjoint unions 
\qq\nn
\cM^\la_{\overrightarrow\la}\equiv\bigsqcup_{[\overrightarrow w]\in\cF^\la_{\overrightarrow\la}}\,T_{\overrightarrow\la}^{[\overrightarrow w]} 
\qqq
of gerbe-theoretically admissible $\txG$-orbits $\,T_{\overrightarrow\la}^{[\overrightarrow w]}\,$ within the `fusion manifolds' $\,T_{\overrightarrow\la}^\la\,$ and -- on the other hand -- the multiplicity spaces\footnote{{\it Cf.},\ {\it e.g.},\ \Rxcite{Eq.\,(7.13)}{Gawedzki:1999bq}.} in the decompositions of the relevant poly-boundary Hilbert spaces into integrable highest-weight modules of the preserved (diagonal) current sub-algebra $\,\ggtk\subset\ggtk^{(L)}\oplus\ggtk^{(R)}\,$ is not expected to provide us with any additional constraints on the admissible `multiplicity labels' $\,[\overrightarrow w]\,$ and thus to determine a proper subset of $\,\cF^\la_{\overrightarrow\la}$.\ Indeed,\ in the light of the canonical relation,\ established in \Rcite{Gawedzki:2001ye},\ between the 3d Chern--Simons theory with the Wilson lines (as before) and the \emph{maximally} gauged WZW model on $\,\txG\,$ (with only \emph{topological} degrees of freedom left),\ and upon taking into account the gerbe-theoretic implementation,\ worked out in Refs.\,\cite{Gawedzki:2010rn,Gawedzki:2012fu},\ of the gauging (of $\,\Ad(\txG)$) in terms of an ($\Ad(\txG)$-)equivariant structure on the gerbe and the attendant bi-modules and fusion 2-isomorphisms,\ such constraints might only come from a restriction to those fusion 2-isomorphisms which admit an $\Ad(\txG)$-equivariant structure.\ However,\ in the latter of the two papers,\ it was demonstrated that (for $\,\txG\,$ connected) there is no obstruction to endowing a maximally symmetric inter-bi-brane with a $\Ad(\txG)$-equivariant structure other than that coming from the bulk gerbe,\ {\it cf.}\ \Rxcite{Cor.\,11.17}{Gawedzki:2012fu},\ and so all inter-bi-branes predicted by our reasoning descend to the topological gauged WZW model.\ In other words,\ the $\,\cM^\la_{\overrightarrow\la}\,$ seem to be the \emph{unique} gerbe-theoretic candidates for the `multiplicity geometries'.\ But that is hardly the end of the story:\ The detailed analysis,\ carried out in \Rcite{Runkel:2008gr},\ of a class of maximally symmetric defects distinguished by the \emph{simple} (0-1) fusion rules of the corresponding chiral sectors of the WZW model,\ namely,\ of the simple-current defects (associated with the degenerate conjugacy classes of central elements of the target Lie group and implementing worldsheet orbifolding of the mono-phase WZW model with respect to the action of the centre of the target Lie group),\ puts some (group-cohomological) flesh on them and corroborates them within the said class.\ Finally,\ the detailed case study,\ reported in \Rcite{Runkel:2009sp},\ of the gerbe-theoretic content of the WZW model with the target Lie group $\,{\rm SU}(2)\,$ and the maximally symmetric (self-intersecting) defect demonstrated the validity of Conjecture \ref{conj:2-iso-vs-Verlinde} in this case.\ Thus motivated,\ we shall return to the task of verifying the Conjectures in a future work.\ And meanwhile,\ we leave the ungraded world and pass to the $\bZ/2\bZ$-graded setting to witness the simplicial machinery at work there for the first time ever.

\part{Candidate maximally supersymmetric defects in the flat GS model}\label{p:smaxym-bib}

In the present part,\ we put the conceptual scheme set up in Part \ref{p:genstr} to work in the supergeometric setting of the much tractable Green--Schwarz super-$\si$-model of the superstring with the bulk target superspace $\,{\rm sMink}(d,1|D_{d,1})$,\ introduced in \Rcite{Green:1983wt},\ which has long been known to share many structural features with the `bosonic' WZW $\si$-model of Part \ref{p:WZWCS},\ {\it cf.}\ \Rcite{Henneaux:1984mh}.\ Accordingly,\ we employ the intuitions developed in the study of the un-graded $\si$-model towards a systematic reconstruction of a class of supersymmetric defects associated with the supersymmetric generalised multiplicative structure on the GS super-1-gerbe of \Rcite{Suszek:2017xlw},\ whose derivation is an independent important result of our investigation.\ The latter is the next logical step in the development of the programme,\ initiated in \Rcite{Suszek:2017xlw} and subsequently elaborated in a variety of directions in Refs.\,\cite{Suszek:2019cum,Suszek:2018bvx,Suszek:2018ugf,Suszek:2020xcu,Suszek:2020rev,Suszek:2021hjh},\ of the (higher-super)geometrisation,\ in the spirit of Murray {\it et al.},\ of the cohomological data of super-$p$-brane super-$\si$-models,\ {\it i.e.},\ of the classes in the Cartan--Eilenberg cohomology of Lie supergroups distinguished by the string-theoretic considerations leading up to the so-called `brane scan' of \Rcite{Achucarro:1987nc},\ and the curved super-$\si$-models related to them through a variant of the {\.I}n{\"o}n{\"u}--Wigner contraction.\ A peculiarity of all these superfield theories is the presence of an infinitesimal right gauge supersymmetry,\ known as the $\k$-symmetry and discovered first by de Azc\'arraga and Lukierski in \Rcite{deAzcarraga:1982dhu} and subsequently (anew) by Siegel in \Rcite{Siegel:1983hh},\ whose geometric nature and a higher geometric signature were elucidated in Refs.\,\cite{Suszek:2020xcu,Suszek:2020rev,Suszek:2021hjh}.\ As the latter hinges on a duality of the original GS super-$\si$-model with the so-called Hughes--Polchinski model of \Rcite{Hughes:1986dn},\ whose adaptation to the setting with (supersymmetric) defects would call for a separate study,\ we content ourselves in what follows with the preliminary construction of candidate maximally supersymmetric bi-branes and postpone the un-circumnavigable task of verifying their $\k$-symmetry as well as the subsequent completion of the bi-categorial scheme laid out in Sec.\,\ref{sec:wrldshtbic} to a future work.\ For similar reasons,\ we are wary of thinking of the structures encountered hereunder as defect realisations of (super)\emph{symmetries} of the 2$d$ superfield theory -- indeed,\ while the instantonic bi-branes of Sec.\,\ref{sub:sbib} do correspond to the translational symmetry in the body,\ and the superstring-like ones can be expected to correspond to $\k$-symmetry,\ the issue certainly requires further scrutiny.

We close this introductory part with a {\it memento}:\ Owing to the topological triviality of the body of the supermanifolds encountered in the analysis to come (and in virtue of the theorem,\ due to Kostant ({\it cf.}\ \Rcite{Kostant:1975}),\ which equates the de Rham cohomology of the supermanifold and that of its body),\ the said analysis is actually nontrivial ({\it i.e.},\ its results are not known {\it a priori}) on the tensorial level,\ whereas the subsequent geometrisations are mere reconstructions done for the sake of bookkeeping.\ This may well obscure the deeper meaning of the tensorial calculus carried out in the \emph{supersymmetric refinement} of the standard de Rham cohomology,\ which was revealed by a beautiful but largely overlooked reasoning due to Rabin and Crane ({\it cf.}\ Refs.\,\cite{Rabin:1984rm,Rabin:1985tv}):\ The refinement encodes the nontrivial topology of the orbifold of the supergeometry with respect to the action of the discrete Rabin--Kosteleck\'y supergroup of \Rcite{Kostelecky:1983qu} (in the usual sense in which the de Rham cohomology encodes the topology (homology) of the manifold),\ and so what we are dealing with is,\ secretly,\ a `proper' higher-geometric analysis over the orbifold,\ which happens to be remarkably compactly packaged in terms of supersymmetry-invariant differential forms.\ With the \emph{added complexity} of the supersymmetric higher geometry thus isolated,\ we should think of the considerations that follow as a judicious first step towards the study of generic supergeometries (with supersymmetry),\ with a body of an arbitrary topology.

\section{The Green--Schwarz super-WZW-model and its super-1-gerbe}\label{sec:GSWZW}

We shall now set the stage for our considerations in the $\bZ/2\bZ$-graded setting,\ which we take to be that of the Green--Schwarz super-$\si$-model  of superstring dynamics in the bulk target with the body given by the standard minkowskian spacetime $\,{\rm Mink}(d,1)$.\ This choice of the target supergeometry -- flat and topologically trivial (in the body) -- is dictated by the intention to separate the truly novel phenomena engendered in the presence of the Gra\ss mann-odd `fibre' over the riemannian `base'\footnote{The rigorous structure behind this heuristic picture is provided by the Gaw\c{e}dzki--Batchelor Theorem of Refs.\,\cite{Gawedzki:1977pb,Batchelor:1979a}.} from the usual cohomological effects of a nontrivial topology of (the body of) the target.

\subsection{The geometry of the bulk GS super-$\si$-model}

The target superspace of the GS super-$\si$-model is the supermanifold 
\qq\nn
\bT\equiv{\rm sMink}(d,1|D_{d,1})\equiv\bR^{d,1|D_{d,1}}\,,
\qqq
with its global generators of the structure sheaf $\,\cO_\bT\,$ (coordinates):\ the Gra\ss mann-even ones $\,\{x^a\}^{a\in\ovl{0,d}}\,$ and the Gra\ss mann-odd ones $\,\{\theta^\a\}^{\a\in\ovl{1,D_{d,1}}}$,\ the latter carrying indices of a Majorana-spinor representation of the group $\,{\rm Spin}(d,1)$,\ the latter being realised,\ in the standard manner,\ within the Clifford algebra $\,\Cliff(\bR^{d,1})\,$ with generators $\,\{\G_a\}_{a\in\ovl{0,d}}\,$ subject to the anticommutation relations $\,\{\G_a,\G_b\}=2\eta_{ab}\,\bd1$,\ written for $\,\eta\equiv(\eta_{ab})_{a,b\in\ovl{0,d}}=\diag(-,\underbrace{+,+,\ldots,+}_{d\ \tx{times}})$.\ In the Majorana-spinor representation of interest,\ we have a skew charge-cojugation matrix $\,C=(C_{\a\b})_{\a,\b\in\ovl{1,D_{d,1}}}\,$ and we further assume that the following symmetry relations
\qq\nn
\ovl\G{}_a^{\rm T}=\ovl\G{}_a\,,\qquad\qquad\ovl\G{}_a\equiv C\,\G_a\,,
\qqq
and the Fierz identity
\qq\label{eq:Fierz}
\eta_{ab}\,\ovl\G{}^a_{\a(\b}\,\ovl\G{}^b_{\g\d)}=0
\qqq
hold true.\ Now,\ the bulk target superspace $\,\bT\,$ is a Lie supergroup,\ {\it i.e.},\ in particular,\ an orbit of the left regular action $\,\ell\equiv\txm$.\ In the global coordinates,\ the binary operation $\,\txm\ :\ \bT\x\bT\too\bT\,$ reads
\qq\nn
\txm\bigl(\bigl(\th_1^\a,x_1^a\bigr),\bigl(\th_2^\b,x_2^b\bigr)\bigr)\equiv\bigl(\th_1^\a,x_1^a\bigr)\cdot\bigl(\th_2^\b,x_2^b\bigr)=\bigl(\th_1^\a+\th_2^\a,x_1^a+x_2^a-\tfrac{1}{2}\,\th_1\,\ovl\G{}^a\,\th_2\bigr)\,.
\qqq
According to the general scheme,\ $\,\ell\,$ determines the left actions
\qq\label{eq:leftreg}\qquad\qquad
\bT\la^{(n)}\equiv\ell^{(n)}\equiv\ell\x\id_{{\rm sMink}(d,1|D_{d,1})^{\x n}}\ :\ \bT\x{\rm sMink}(d,1|D_{d,1})^{\x n+1}\too{\rm sMink}(d,1|D_{d,1})^{\x n+1}
\qqq
that lift to the de Rham complexes $\,(\Om^\bullet({\rm sMink}(d,1|D_{d,1})^{\x n+1}),\sfd_{\rm dR}^{(\bullet)})\,$ over $\,{\rm sMink}(d,1|D_{d,1})^{\x n+1}$.\ Here,\ the existence of global generators of the structure sheaf $\,\cO_\bT\,$ of $\,\bT\,$ (the global coordinates introduced earlier) enables us to rewrite the conditions of left invariance in a uniform and concise manner as
\qq\nn
\forall_{(\vep,t)\in\bT}\ :\ \bT\la_{(\vep,t)}^{(n)\,*}\om=\om\,,
\qqq
where we effectively treat the $\,(\vep,t)\,$ as (if they were) constant -- this is a convenient alternative to the more general approach to invariance described in Sec.\,\ref{sec:susygrb}.\ In particular,\ these conditions distinguish the basis LI super-1-forms on $\,\bT^{\x 1}$:
\qq\nn
\si_{\rm L}^\a(\th,x)=\sfd\theta^\a\,,\qquad\a\in\ovl{1,D_{d,1}}\,,\qquad\qquad\qquad e_{\rm L}^a(\th,x)=\sfd x^a+\tfrac{1}{2}\,\th\,\ovl\G{}^a\,\sfd\th\,,\qquad a\in\ovl{0,d}\,,
\qqq
written in terms of the matrices $\,\ovl\G{}^a\equiv\eta^{-1\,ab}\,\ovl\G{}_b$.\ We note parenthetically that the Lie supergroup admits an equivalent description,\ due to Kostant,\ as the super-Harish--Chandra pair $\,{\rm sMink}(d,1|D_{d,1})=({\rm Mink}(d,1),\gt{smink}(d,1|D_{d,1}))$,\ {\it cf.}\ \Rcite{Kostant:1975}.\ Here,\ 
\qq\nn
&\tgt\equiv\gt{smink}(d,1|D_{d,1})=\gt{smink}(d,1|D_{d,1})^{(0)}\oplus\gt{smink}(d,1|D_{d,1})^{(1)}\equiv\tgt^{(0)}\oplus\tgt^{(1)}\,,&\cr\cr
&\tgt^{(0)}=\bigoplus_{a=0}^d\,\corr{P_a}\qquad\qquad\tgt^{(1)}=\bigoplus_{\a=1}^{D_{d,1}}\,\corr{Q_\a}&
\qqq
is the (supersymmetry) Lie superalgebra of $\,\bT$,\ on which the Lie group $\,{\rm Mink}(d,1)\equiv\bR^{d,1}\,$ acts trivially. 

On the above-introduced supergeometry,\ we have the (odd-degenerate) `metric' 
\qq\nn
\txg=\eta_{ab}\,e_{\rm L}^a\ox e_{\rm L}^b\,,
\qqq
and the LI Green--Schwarz super-3-cocycle 
\qq\nn
\txH_{\rm GS}=\si_{\rm L}\wedge\ovl\G{}_a\,\si_{\rm L}\wedge e_{\rm L}^a\in Z^3_{\rm dR}(\bT)^{\bT\la}_0
\qqq 
of Refs.\,\cite{Green:1983wt,Green:1983sg},\ which together compose the tensorial data of the superbackground.\ Note that the super-3-cocycle can be rewritten entirely in terms of the RI counterparts 
\qq\nn
\si_{\rm R}^\a(\th,x)=\sfd\theta^\a\,,\qquad\a\in\ovl{1,D_{d,1}}\,,\qquad\qquad\qquad e_{\rm R}^a(\th,x)=\sfd x^a-\tfrac{1}{2}\,\th\,\ovl\G{}^a\,\sfd\th\,,\qquad a\in\ovl{0,d}\,,
\qqq
of the previously introduced basis LI super-1-forms as
\qq\label{eq:GS-L-is-R}
\txH_{\rm GS}=\si_{\rm R}\wedge\ovl\G{}_a\,\si_{\rm R}\wedge e_{\rm R}^a\,,
\qqq 
in analogy with the WZW 3-cocycle \eqref{eq:WZW-L-is-R}.\ In fact the analogy goes much further:\ The GS super-3-cocycle \emph{is} the Cartan 3-form on the Lie supergroup $\,\bT$,\ as noticed in \Rcite{Henneaux:1984mh}.\ It is not hard to verify that the GS super-3-cocycle does \emph{not} possess an LI primitive ({\it cf.}\ \Rcite{Suszek:2017xlw}),\ and so it defines a class $\,[\txH_{\rm GS}]\in{\rm CaE}^3(\bT)\,$ whose \emph{concrete} geometrisation was worked out in \Rcite{Suszek:2017xlw}.\ The latter is the central object in the analysis that follows.

\subsection{The higher geometry of the bulk GS superbackground}

The Lie supergroup $\,\bT\,$ carries a distinguished 1-gerbe with curvature $\,\txH_{\rm GS}$,\ which was constructed in \Rcite{Suszek:2017xlw} in what may be regarded as a gerbe-theoretic variant of the ingenious  stepwise extension scheme,\ proposed by de Azc\'arraga {\it et al.} in \Rcite{Chryssomalakos:2000xd},\ for the supersymmetry Lie superalgebra of $\,\bT\,$ associated with the physically distinguished classes in higher ($p+2>2$) Cartan--Eilenberg cohomology groups of $\,\bT$.\ In each step of the procedure,\ the $\bZ/2\bZ$-graded analogon\footnote{{\rm Cf.}\ \Rxcite{App.\,C}{Suszek:2017xlw}.} of the classic one-to-one correspondence between equivalence classes of central extensions of a Lie algebra $\,\ggt\,$ and classes in the second group $\,{\rm CE}^2(\ggt)\,$ of the Chevalley--Eilenberg cohomology of $\,\ggt\,$ is used,\ and for $\,\ggt\equiv\tgt\,$ and $\,p=1$,\ we thus arrive at a {\bf Cartan--Eilenberg super-1-gerbe} (in the sense of \Rxcite{Def.\,5.11}{Suszek:2017xlw})
\qq\nn
\cG_{\rm GS}\equiv\bigl(\sfY\bT\equiv\sfY{\rm sMink}(d,1|D_{d,1}),\pi_{\sfY\bT},\txB_{\rm GS},L,\pi_L,\cA_L,\mu_L\bigr)
\qqq
of curvature $\,\txH_{\rm GS}\equiv{\rm curv}\,(\cG_{\rm GS})$,\ termed the {\bf Green--Schwarz} (GS) {\bf super-1-gerbe} in \Rcite{Suszek:2017xlw}.\ This is an example of a 1-gerbe over a supermanifold -- such structures were studied at length in all generality in \Rcite{Huerta:2020}.\ On the total space of its surjective submersion 
\qq\nn
\pi_{\sfY\bT}\equiv\pr_1\ :\ \sfY\bT\equiv\bT\x\bR^{0|D_{d,1}}\too\bT\,,
\qqq
with its global Gra\ss mann-odd coordinates $\,\{\xi_\a\}_{\a\in\ovl{1,D_{d,1}}}\,$ in the fibre (co-spinors of $\,{\rm Spin}(d,1)$),\ there exists the curving 
\qq\nn
\txB_{\rm GS}=e^{(2)}_\a\wedge\pi_{\sfY\txG}^*\si_{\rm L}^\a\in\Om^2(\sfY\bT)^{\sfY\bT\la}_0\,,\qquad\qquad\sfd\txB_{\rm GS}=\pi_{\sfY\bT}^*\txH_{\rm GS}\,,
\qqq
defined in terms of the LI super-1-forms
\qq\label{eq:LIBGS}
e^{(2)}_\a(\th,x,\xi)=\sfd\xi_\a-\ovl\G{}_{a\,\a\b}\,\th^\b\,\bigl(\sfd x^a+\tfrac{1}{6}\,\theta\,\ovl\G{}^a\,\sfd\th\bigr)
\qqq
that are LI with respect to the left regular action of the Lie supergroup $\,\sfY\bT\,$ on itself,
\qq\nn
\sfY\bT\la\equiv\sfY\ell\equiv\sfY\txm\ :\ \sfY\bT\x\sfY\bT\too\sfY\bT\,,
\qqq
determined by the binary operation $\,\sfY\txm\,$ with the coordinate presentation
\qq\nn
&&\sfY\txm\bigl(\bigl(\th_1^\a,x_1^a,\xi_{1\,\b}\bigr),\bigl(\th_2^\g,x_2^b,\xi_{2\,\d}\bigr)\bigr)\cr\cr
&=&\bigl(\th^\a_1+\th^\a_2,x^a_1+x^a_2-\tfrac{1}{2}\,\th_1\,\ovl\G{}^a\,\theta_2,\xi_{1\,\b}+\xi_{2\,\b}+\ovl\G{}_{b\,\b\g}\,\th_1^\g\,x_2^b-\tfrac{1}{6}\,\bigl(\th_1\,\ovl\G{}_b\,\th_2\bigr)\,\ovl\G{}^b_{\b\g}\,\bigl(2\th_1^\g+\th_2^\g\bigr)\bigr)\,,
\qqq
{\it i.e.},\ we have
\qq\label{eq:curvLI}
\forall_{(\vep,t,\z)\in\sfY\bT}\ :\ \sfY\ell_{(\vep,t,\z)}^*\txB_{\rm GS}=\txB_{\rm GS}\,.
\qqq
It will be convenient to rewrite the curving in the form
\qq\label{eq:curvdec}
\txB_{\rm GS}(\th,x,\xi)=\th\,\ovl\G{}_a\,\sfd\th\wedge\sfd x^a+\sfd\xi_\a\wedge\sfd\th^\a\,.
\qqq
Over the fibred square $\,\sfY^{[2]}\bT\equiv\sfY\bT{}_{\pi_{\sfY\bT}}\hspace{-1pt}\x_{\pi_{\sfY\bT}}\hspace{-1pt}\sfY\bT\equiv\sfY\bT\x_\bT\hspace{-1pt}\sfY\bT$,\ we have the principal $\bC^\x$-bundle 
\qq\nn
\pi_L=\pr_1\ :\ L\equiv\sfY^{[2]}\bT\x\bC^\x\too\sfY^{[2]}\bT
\qqq
with a principal $\bC^\x$-connection super-1-form $\,\cA_{\rm GS}\,$ given,\ in the coordinates $\,((\th,x,\xi_1),(\th,x,\xi_2),z)\equiv(y_1,y_2,z)\in\sfY^{[2]}\bT\x\bC^\x$,\ by the formula
\qq\nn
&\cA_L(y_1,y_2,z)=\vartheta(z)+\txA_L(y_1,y_2)\,,&\cr\cr
&\vartheta(z)=\tfrac{\sfi\,\sfd z}{z}\,,\qquad\qquad\txA_L(y_1,y_2)\equiv\theta^\a\,\sfd\xi_{21\,\a}\,,\qquad\xi_{21}\equiv\xi_2-\xi_1\,.&
\qqq
The above super-1-form is LI with respect to the left regular action of the Lie supergroup $\,L\,$ on itself,
\qq\nn
L\ell\equiv L\txm\ :\ L\x L\too L\,,
\qqq
determined by the binary operation $\,L\txm\,$ with the coordinate presentation 
\qq\nn
L\txm\bigl(\bigl(y^1_{\ 1},y^1_{\ 2},z_1\bigr),\bigl(y^2_{\ 1},y^2_{\ 2},z_2\bigr)\bigr)=\bigl(\sfY\txm\bigl(y^1_{\ 1},y^2_{\ 1}\bigr),\sfY\txm\bigl(y^1_{\ 2},y^2_{\ 2}\bigr),\ee^{\sfi\,\la(y^1_{\ 1},y^1_{\ 2},y^2_{\ 1},y^2_{\ 2})}\cdot z_1\cdot z_2\bigr)
\qqq
written in the shorthand notation $\,y^A_{\ B}\equiv(\theta_A,x_A,\xi^A_B),\ A,B\in\{1,2\}\,$ and in terms of the maps
\qq\nn
\la\bigl(y^1_{\ 1},y^1_{\ 2},y^2_{\ 1},y^2_{\ 2}\bigr):=\theta_1^\a\,\bigl(\xi^2_2-\xi^2_1\bigr)_\a\,,
\qqq
{\it i.e.},\ we have
\qq\nn
\forall_{(\upsilon_1,\upsilon_2,\varsigma)\in\sfY\bT}\ :\ L\ell_{(\upsilon_1,\upsilon_2,\varsigma)}^*\cA_L=\cA_L\,.
\qqq
Finally,\ over the fibred cube $\,\sfY^{[3]}\bT\equiv\sfY\bT\x_\bT\hspace{-1pt}\sfY\bT\x_\bT\hspace{-1pt}\sfY\bT$,\ we find the trivial groupoid structure $\,\mu_L\ :\ \pr_{1,2}^*L\ox\pr_{2,3}^*L\xrightarrow{\ \cong\ }\pr_{1,3}^*L\,$ with the coordinate presentation (written for $\,y_A\equiv(\theta,x,\xi_A),\ A\in\{1,2,3\}$)
\qq\nn
\mu_L\bigl((y_1,y_2,1)\ox(y_2,y_3,z)\bigr)=(y_1,y_3,z)\,.
\qqq
In what follows,\ we shall denote such trivial principal $\bC^\x$-bundle isomorphisms suggestively as
\qq\label{eq:muL}
\mu_L\equiv\bd1\,.
\qqq
The isomorphism lifts to the category of Lie supergroups to which its domain and codomain belong,\ being endowed with the binary operation induced,\ in a natural manner,\ from that on $\,L$.

Just to (re-)emphasise:\ The GS super-1-gerbe is an example of a super-1-gerbe in the sense of Def.\,\ref{def:supergerbetc},\ and so it is maximally supersymmetric by construction,\ {\it cf.}\ Sec.\,\ref{sec:susygrb}.

\section{A multiplicative structure on the Green--Schwarz super-1-gerbe}\label{sec:multGSs1g}

In the next step,\ we equip the GS super-1-gerbe with a supersymmetric multiplicative structure,\ with view to constructing a maximally supersymmetric $\cG_{\rm GS}$-bi-brane in structural analogy with the un-graded case reviewed in Sec.\,\ref{sec:mxymbibr}.\smallskip

We start by checking the relevant identity \eqref{eq:PolWiegH} for the curvature of the GS super-1-gerbe,\ with the understanding that it now ought to be satisfied in the \emph{supersymmetric} de Rham cohomology.\ Thus,\ we compute -- employing the Fierz identity \eqref{eq:Fierz},\ and in the shorthand notation $\,m_A\equiv(\theta_A,x_A),\ A\in\{1,2,3,4\}$,\ to be used throughout our analysis --
\qq\nn
\txm^*\txH_{\rm GS}(m_1,m_2)&=&\pr_1^*\txH_{\rm GS}(m_1,m_2)+\pr_2^*\txH_{\rm GS}(m_1,m_2)+\tfrac{1}{2}\,\sfd\th_2\wedge\ovl\G{}_a\,\sfd\th_2\wedge\th_2\,\ovl\G{}^a\,\sfd\th_1\cr\cr
&&+\sfd\bigl[2\eta_{ab}\,e^a_{\rm L}(m_1)\wedge e_{\rm R}^b(m_2)+2\th_2\,\ovl\G{}_a\,\sfd\th_1\wedge\bigl(e_{\rm L}^a(m_1)+\sfd x_2^a\bigr)\bigr]\,.
\qqq
Note that the identity $\,\sfd(\sfd\th_2\wedge\ovl\G{}_a\,\sfd\th_2\,\th_2^\a\,\ovl\G{}^a_{\a\b})=0$,\ which follows directly from \Reqref{eq:Fierz},\ implies -- in view of the contractibility of $\,{\rm sMink}(d,1|D_{d,1})\,$ -- the existence of a primitive of the 2-form in the last bracket.\ Using the homotopy retraction $\,(\t,\th_2)\longmapsto\t\,\th_2,\ \t\in[0,1]$,\ we readily establish
\qq\nn
\txm^*\txH_{\rm GS}(m_1,m_2)&=&\pr_1^*\txH_{\rm GS}(m_1,m_2)+\pr_2^*\txH_{\rm GS}(m_1,m_2)\cr\cr
&&+\sfd\bigl[2\eta_{ab}\,e^a_{\rm L}(m_1)\wedge e_{\rm R}^b(m_2)+2\th_2\,\ovl\G{}_a\,\sfd\th_1\wedge\bigl(e_{\rm L}^a(m_1)+\sfd x_2^a-\tfrac{1}{6}\,\th_2\,\ovl\G{}^a\,\sfd\th_2\bigr)\bigr]
\qqq
so that,\ altogether,
\qq\nn
\D^{(1;3)}_\bT\txH_{\rm GS}\equiv\pr_2^*\txH_{\rm GS}-\txm^*\txH_{\rm GS}+\pr_1^*\txH_{\rm GS}=\sfd\varrho_{\rm GS}\,,
\qqq
where the super-2-form $\,\varrho_{\rm GS}\in\Om^2(\bT^{\x 2})\,$ is given by the formula
\qq\nn
\varrho_{\rm GS}(m_1,m_2)=-2\eta_{ab}\,e^a_{\rm L}(m_1)\wedge e_{\rm R}^b(m_2)-2\th_2\,\ovl\G{}_a\,\si_{\rm L}(m_1)\wedge\bigl(e_{\rm L}^a(m_1)+\om^a(m_2)\bigr)
\qqq
in which
\qq\nn
\om^a(m_2)=\sfd x_2^a-\tfrac{1}{6}\,\th_2\,\ovl\G{}^a\,\sfd\th_2\,.
\qqq
The crucial property of the above super-2-form is its manifest invariance with respect to the formerly introduced action $\,\ell^{(2)}$,
\qq\label{eq:rhoLI}
\forall_{(\vep,t)\in\bT}\ :\ \ell_{(\vep,t)}^{(2)\,*}\varrho_{\rm GS}=\varrho_{\rm GS}\,.
\qqq
Next,\ we examine the multiplicative properties of $\,\varrho_{\rm GS}$.\ To this end,\ we write out -- invoking the Fierz identity \eqref{eq:Fierz} again (thrice),\ and for arbitrary $\,m_A\equiv(\theta_A,x_A)\in\bT,\ A\in\{1,2,3\}\,$ --
\qq\nn
&&\varrho_{\rm GS}(m_2,m_3)-\varrho_{\rm GS}(m_1\cdot m_2,m_3)+\varrho_{\rm GS}(m_1,m_2\cdot m_3)-\varrho_{\rm GS}(m_1,m_2)\cr\cr
&=&\sfd\bigl[-2\theta_2\,\ovl\G{}_a\,\theta_3\,\bigl(e^a_{\rm L}(m_1)+\tfrac{2}{3}\,\th_2\,\ovl\G{}_a\,\sfd\th_1+\tfrac{1}{3}\,\th_3\,\ovl\G{}_a\,\sfd\th_1\bigr)\bigr]\,,
\qqq
or,\ equivalently,
\qq\label{eq:varhothGS}
\D^{(2;2)}_\bT\varrho_{\rm GS}=\sfd\vartheta_{\rm GS}\,,
\qqq
where the super-1-form $\,\vartheta_{\rm GS}\in\Om^1(\bT^{\x 3})\,$ is given by the formula
\qq\nn
\vartheta_{\rm GS}(m_1,m_2,m_3)=-\tfrac{2}{3}\,\theta_2\,\ovl\G{}_a\,\theta_3\,\bigl(3e^a_{\rm L}(m_1)+2\th_2\,\ovl\G{}^a\,\si_{\rm L}(m_1)+\th_3\,\ovl\G{}^a\,\si_{\rm L}(m_1)\bigr)\,,
\qqq
and so,\ in particular,\ it is LI,
\qq\nn
\forall_{(\vep,t)\in\bT}\ :\ \ell_{(\vep,t)}^{(3)\,*}\vartheta_{\rm GS}=\vartheta_{\rm GS}\,.
\qqq
Finally,\ we establish --through another application of the Fierz identity -- the equality
\qq\nn
\D^{(2;2)}_\bT\vartheta_{\rm GS}=0\,.
\qqq

Thus,\ altogether,\ we arrive at
\berop\label{prop:GS-mult-4coc}
The quintuple $\,\cZ_{\rm GS}\equiv(0,\txH_{\rm GS},\varrho_{\rm GS},\vartheta_{\rm GS},0)\,$ with $\,\txH_{\rm GS},\ \varrho_{\rm GS}\,$ and $\,\vartheta_{\rm GS}\,$ defined above is a simplicial de Rham 4-cocycle over $\,\sfN_\bullet\bT^{\rm op}$,\ {\it i.e.},\ it belongs to $\,{\rm Ker}\,\cD^{(4)}_\bT$.
\eerop
\noindent The last result constitutes the point of departure in a reconstruction of a full-fledged 0-flat supersymmetric generalised multiplicative structure on the GS super-1-gerbe.\ In fact,\ in the cohomologically trivial setting of $\,\bT^\bullet$,\ the existence of $\,\cZ_{\rm GS}\,$ \emph{ensures} the existence of the full-fledged supersymmetric multiplicative structure,\ and so what remains at this stage is a tedious derivation,\ which we append to the following theorem as its constructive proof.

\bethe\label{thm:mult-str-GS}
There exists a $\ell^{(\bullet)}$-supersymmetric 0-flat generalised multiplicative structure $\,(\cM_{\rm GS},\a_{\rm GS})\,$ on the Green--Schwarz super-1-gerbe $\,\cG_{\rm GS}\,$ associated with the 4-cocycle $\,\cZ_{\rm GS}\,$ of Prop.\,\ref{prop:GS-mult-4coc}.
\ethe
\beroof
{\it Cf.}\ App.\,\ref{app:mult-str-GS}.
\eroof
\noindent The multiplicative structure found in the present section can subsequently be employed in a construction of supersymmetric bi-branes of the GS super-$\si$-model of the super-Minkowskian superstring,\ carried out in a far-reaching analogy with its un-graded precursor.\ For that,\ however,\ we need one more ingredient,\ to wit,\ $\cG_{\rm GS}$-branes.\ These we discuss in the next section.

\section{Some supersymmetric $\cG_{\rm GS}$-(bi-)branes}\label{sec:susyGbrs}

Below,\ we present three classes of $\cG_{\rm GS}$-branes and subsequently employ them in the construction of maximally supersymmetric $\cG_{\rm GS}$-bi-branes mimicking the un-graded scheme and taking into account the specific form of the chiral (left) rigid supersymmetry of the GS super-$\si$-model indicated in \Reqref{eq:Gact-bulk-grad}.\ For concreteness,\ we fix the superdimensionality of the bulk target super-space $\,{\rm sMink}(d,1|D_{d,1})\,$ as $\,(d+1|D_{d,1})=(10|32)$.\ Thus,\ we shall,\ first,\ be dealing with quintuples 
\qq\nn
\cD_D=\bigl(D,\iota_D,\bullet_D,\om_D,\cT_D)
\qqq 
composed of a sub-supermanifold 
\qq\nn
\iota_D\ :\ D\emb{\rm sMink}(9,1|32)
\qqq 
mapped canonically into the terminal supermanifold $\,\bR^{0|0}\,$ as $\,\bullet_D\ :\ D-->\bR^{0|0}\,$ and equipped with an action of a Lie sub-supergroup $\,\bD\subset\bT$,\ or {\bf reduced supersymmetry group},\ induced by the left action $\,\ell\,$ of $\,\bT\,$ on itself,\ and of a trivialisation
\qq\label{eq:GSbrane}
\cT_D\ :\ \iota_D^*\cG_{\rm GS}\xrightarrow{\ \cong\ }\cI_{\om_D}\,,
\qqq
that is a super-1-gerbe 1-isomorphism between the restriction $\,\iota_D^*\cG_{\rm GS}\equiv\cG_{\rm GS}\rstr_D\,$ to the worldvolume $\,D\,$ and the trivial super-1-gerbe $\,\cI_{\om_D}\,$ of a curving $\,\om_D\in\Om^2(D)^\bD\,$ supported over $\,D$,\ the curving satisfying the identity $\,\sfd\om_D=\iota_D^*\txH_{\rm GS}$,\ or,\ equivalently,\ providing a primitive for the trivialisation $\,[\iota_D^*\txH_{\rm GS}]\equiv 0\in H^3(D;\bR)^\bD$.\ These shall be combined,\ in the manner reviewed in Sec.\,\ref{sec:mxymbibr},\ with the generalised multiplicative structure on $\,\cG_{\rm GS}\,$ derived in Sec.\,\ref{sec:multGSs1g},\ whereby the sought-after maximally supersymmetric $\cG_{\rm GS}$-bi-modules shall be formed.

We do not,\ as yet,\ know a systematic way of identifying the relevant $\cG_{\rm GS}$-modules,\ and so the structures that we discuss arise at the intersection of a variety of observations and intuitions established in the un-graded setting of the WZW $\si$-model and in the in-depth study of $\k$-symmetry related in Refs.\,\cite{Suszek:2020xcu,Suszek:2020rev} and in particular the case study of \Rcite{Suszek:2021hjh}.\ As the latter right gauge supersymmetry is \emph{not} incorporated in our analysis (but {\it cf.}\ Sec.\,\ref{sub:bisusyGbr}),\ we do not elaborate the analysis to the level attained in the un-graded case,\ restricting ourselves to a preliminary investigation of fusion and leaving the completion of the effort thus initiated to a separate future work.

\subsection{The instantonic adjoint $\cG_{\rm GS}$-brane}\label{sub:AdGbr}

In the light of the structural affinity between the bosonic WZW $\si$-model and the GS super-$\si$-model ({\it cf.}\ \Rcite{Henneaux:1984mh},\ and also \Rcite{Suszek:2019cum}),\ and of the bi-invariance of the GS super-3-cocycle $\,\txH_{\rm GS}$,\ it seems natural to begin our search for $\cG_{\rm GS}$-branes with $\bZ/2\bZ$-graded analogons of the maximally symmetric WZW branes of Sec.\,\ref{sec:mxymbr}.\ Given the peculiar nature of the adjoint action of the Lie supergroup $\,\bT\ni(\vep,y),(\theta,x)\,$ on itself,\ $\,\Ad_{(\vep,y)}(\theta,x)=(\theta,x-\vep\,\ovl\G{}^\cdot\,\th)$,\ it is clear that the corresponding superworldvolumes are $(0|0)$-superdimensional and localise at the topological points 
\qq\label{eq:topoint-emb}
\widehat x{}_*\ :\ \bR^{0|0}\emb{\rm sMink}(9,1|32)\,,\qquad x_*\in\bR^{9,1}
\qqq
in the bulk target super-space $\,{\rm sMink}(9,1|32)$.\ We have $\,\widehat x{}_*^*\txH_{\rm GS}=0$,\ and we readily prove 
\berop\label{prop:AdGbr}
Over every topological point $\,x_*\in\bR^{9,1}\,$ in $\,{\rm sMink}(9,1|32)$,\ embedded as in \Reqref{eq:topoint-emb},\ there exists a flat $(0|0)$-su\-per\-di\-men\-sion\-al $\cG_{\rm GS}$-module
\qq\nn
\bigl(\bD_{x_*},\cT_{-1}^{(x_*)}\bigr)\,,\qquad\qquad\bD_{x_*}\equiv\bR^{0|0}
\qqq
with the 1-isomorphisms 
\qq\nn
\cT_{-1}^{(x_*)}=\bigl(\sfY_{x_*}\bR^{0|0}\x_{\bR^{0|0}}\bR^{0|0}\equiv\sfY_{x_*}\bR^{0|0},\id_{\sfY_{x_*}\bR^{0|0}},\sfY_{x_*}\bR^{0|0}\x\bC^\x,\pr_1,\pr_2^*\vartheta,\bd1\bigr)
\qqq
between the pullback 1-gerbe
\qq\nn
\widehat x{}_*^*\cG_{\rm GS}&=&\bigl(\sfY_{x_*}\bR^{0|0}\equiv\widehat x{}_*^*\sfY\bT:=\bR^{0|0}{}_{\widehat x{}_*}\hspace{-2pt}\x_{\pi_{\sfY\bT}}\hspace{-2pt}\sfY\bT,\pi_{\sfY_{x_*}\bR^{0|0}}\equiv\pr_1,\widehat{\widehat x}{}_*^*\txB_{\rm GS}=0,\widehat{\widehat x}{}_*^{[2]\,*}L=\sfY_{x_*}^{[2]}\bR^{0|0}{}_{\widehat{\widehat x}{}_*^{[2]}}\hspace{-2pt}\x_{\pi_L}\hspace{-2pt}L,\cr\cr
&&\pi_{\widehat{\widehat x}{}_*^{[2]\,*}L}\equiv\pr_1,\widehat{\widehat x}{}_*^{[2]\,*}\cA_L\equiv\pr_2^*\pr_2^*\vartheta,\bd1\bigr)
\qqq
and the trivial 1-gerbe
\qq\nn
\cI_0=\bigl(\bR^{0|0},\id_{\bR^{0|0}},0,\bR^{0|0}\x\bC^\x,\pr_1,\pr_2^*\vartheta,\bd1\bigr)\,.
\qqq
Upon restriction to the topological point $\,x_*$,\ the adjoint action trivialises,\ and so the $\cG_{\rm GS}$-module is (trivially) ($\Ad$-)supersymmetric.
\eerop
\beroof
Obvious.
\eroof

\subsection{A bi-supersymmetric $\cG_{\rm GS}$-brane}\label{sub:bisusyGbr}

The derivation of the next example draws heavily upon the detailed investigation,\ anticipated in \Rcite{Suszek:2020rev} and presented in \Rcite{Suszek:2021hjh},\ of the consequences of the supergerbe-theoretic interpretation of $\k$-symmetry advanced in \Rcite{Suszek:2020xcu}:\ As was demonstrated in the convenient dual (purely topological) Hughes--Polchinsky formulation of the GS super-$\si$-model ({\it cf.}\ Refs.\,\cite{Hughes:1986dn,Gauntlett:1989qe}) employed in \Rcite{Suszek:2021hjh} (after \Rcite{Suszek:2019cum}),\ the supergerbe-theoretic manifestation of the critical\footnote{The criticality is to be understood in the sense defined by the principle of least action for the action functor of the GS super-$\si$-model.} embedding of the worldsheet of the fundamental superstring in the supertarget $\,{\rm sMink}(9,1|32)\,$ is the trivialisation of the ${\rm SO}(9,1)$-covariant lift of the GS super-1-gerbe $\,\cG_{\rm GS}\,$ from $\,{\rm sMink}(9,1|32)\,$ to $\,{\rm sISO}(9,1|32)\,$ over the embedded superstring worldsheet.\ The conclusion hinges on the symmetry analysis of the classical vacuum of the dual super-$\si$-model (or,\ in other words,\ of the critical embedding),\ from which the latter emerges as a torsor of the $\k$-symmetry group,\ itself a distinguished Lie sub-supergroup $\,{\rm sISO}(9,1|32)_{\rm vac}\,$ of the bulk supersymmetry group $\,{\rm sISO}(9,1|32)$.\ Consequently,\ the critically embedded superstring worldsheet determines a Lie sub-superalgebra of the Lie superalgebra $\,\gt{siso}(9,1|32)\,$ (the tangent Lie superalgebra of the $\k$-symmetry group whose fundamental vector fields span the tangent sheaf of the vacuum).\ Below,\ we reverse the logic of the analysis of \Rcite{Suszek:2021hjh} and first look for Lie sub-superalgebras of $\,\gt{smink}(9,1|32)\,$ with a structre analogous to the one of the $\k$-symmetry algebra,\ in which the supersymmetric geometry of the critical superstring is neatly encoded.\ Only upon identifying all such Lie sub-superalgebras in the superdimension $\,(9+1|32)\,$ do we address the question as to which of them determine trivialisations of the bulk super-1-gerbe $\,\cG_{\rm GS}$.

The starting point of the construction carried out in \Rcite{Suszek:2021hjh} is a choice of a projector
\qq\nn
\sfP^{(1)}\in{\rm End}\bigl(\tgt^{(1)}\bigr)\,,\qquad\qquad\sfP^{(1)}\cdot\sfP^{(1)}=\sfP^{(1)}
\qqq
of rank $\,0<R\equiv\rk\,\sfP^{(1)}<32$,\ defined on the Gra\ss mann-odd subspace $\,\gt{smink}(9,1|32)^{(1)}\,$ of the Lie superalgebra $\,\gt{smink}(9,1|32)$,\ with the property
\qq\label{eq:P0P1conj}
\sfP^{(1)\,{\rm T}}\cdot\ovl\G{}^a\cdot\sfP^{(1)}=\sfP^{(0)\,a}_{\ \ \ \ \ b}\,\ovl\G{}^b\cdot\sfP^{(1)}\,,
\qqq
where 
\qq\nn
\sfP^{(0)}\in{\rm End}\bigl(\tgt^{(0)}\bigr)\,,\qquad\qquad\sfP^{(0)}\cdot\sfP^{(0)}=\sfP^{(0)}
\qqq
is a projector acting on the Gra\ss mann-even Lie subalgebra $\,\gt{smink}(9,1|32)^{(0)}\equiv\gt{mink}(9,1)\,$ of the Lie superalgebra $\,\gt{smink}(9,1|32)$.\ In the original analysis,\ the rank of $\,\sfP^{(0)}\,$ was constrained to be 2,\ and its image was required to contain the time direction corresponding to the Lie-(super)algebra index $\,a=0$.\ Below,\ we keep the latter condition (after all,\ we aim to model the worldvolume of an extended object to which a connected component of the boundary of an embedded worldvolume of the fundamental object ({\it e.g.},\ the superstring) is confined) while relaxing the former one.\ More specifically,\ we fix,\ just to keep things simple,\ a subset $\,\ovl{0,p}\subset\ovl{0,9}\,$ such that
\qq\nn
{\rm im}\,\sfP^{(0)\,{\rm T}}\equiv\bigoplus_{\unl a=0}^p\,\corr{P_{\unl a}}\,,\qquad\qquad{\rm ker}\,\sfP^{(0)\,{\rm T}}\equiv\bigoplus_{\widehat a=p+1}^9\,\corr{P_{\widehat a}}\,,
\qqq
and,\ basing on the findings of \Rcite{Suszek:2020xcu} and of the works cited therein,\ assume $\,\sfP^{(1)}\,$ to be of the form
\qq\nn
\sfP^{(1)}=\tfrac{1}{2}\,\bigl(\bd1_{32}+\G^0\,\G^1\,\cdots\,\G^p\bigr)\,.
\qqq
The above is a projector iff $\,p(p+1)\equiv 2\mod 4$,\ and so we end up with the limited spectrum of admissible ranks:
\qq\nn
\rk\,\sfP^{(0)}\equiv p+1\in\{2,3,6,7,10\}\,.
\qqq
Under these circumstances,\ we find
\qq\nn
\sfP^{(1)\,{\rm T}}=C\,\tfrac{1}{2}\,\bigl(\bd1_{32}+(-1)^p\,\G^0\,\G^1\,\cdots\,\G^p\bigr)\,C^{-1}\,,
\qqq
and so
\qq\nn
\sfP^{(1)\,{\rm T}}\cdot\ovl\G{}^{\unl a}\cdot\sfP^{(1)}=\ovl\G{}^{\unl a}\cdot\sfP^{(1)}\,,\qquad\qquad\sfP^{(1)\,{\rm T}}\cdot\ovl\G{}^{\widehat a}\cdot\sfP^{(1)}=\brd0_{32}\,,
\qqq
whence the desired identity \eqref{eq:P0P1conj} ensues.\ Consequently,\ we find the Lie brackets
\qq\nn
&\{(Q\,\sfP^{(1)})_\a,(Q\,\sfP^{(1)})_\b\}=\bigl(\sfP^{(1)\,{\rm T}}\cdot\ovl\G{}^a\cdot\sfP^{(1)}\bigr)_{\a\b}\,P_a=\bigl(\ovl\G{}^a\cdot\sfP^{(1)}\bigr)_{\a\b}\,(P\,\sfP^{(0)})_a\,,&\cr\cr
&[(P\,\sfP^{(0)})_a,(P\,\sfP^{(0)})_b]=0\,,\qquad\qquad[(Q\,\sfP^{(1)})_\a,(P\,\sfP^{(0)})_a]=0\,,&
\qqq
defining a Lie sub-superalgebra of $\,\gt{smink}(9,1|32)$,\ and -- indeed -- a Lie sub-supergroup of $\,{\rm sMink}(9,1|32)\,$ (in the Kostant presentation).\ We shall write
\qq\nn
{\rm im}\,\sfP^{(1)\,{\rm T}}=\corr{Q_\b\,\sfP^{(1)\,\b}_{\ \ \ \ \a}\ \vert\ \a\in\ovl{1,32}}\equiv\bigoplus_{\unl\a=1}^R\,\corr{\breve Q{}_{\unl\a}}
\qqq
and
\qq\nn
{\rm ker}\,\sfP^{(1)\,{\rm T}}=\bigoplus_{\widehat\a= R+1}^{32}\,\corr{\widehat Q{}_{\widehat\a}}\,.
\qqq
With the above,\ we may associate sub-supermanifolds of the supertarget $\,{\rm sMink}(9,1|32)\,$ defined as the leaves of the foliation determined,\ through the super-variant of the (Global) Frobenius Theorem ({\it cf.}\ \Rxcite{Thm.\,6.2.1}{Carmeli:2011}),\ with the involutive superdistribution 
\qq\nn
\xcD_{(p,1|R)}={\rm ker}\,\bigl(\id_\tgt-\sfP\bigr)\circ\th_{\rm L}\,,\qquad\qquad\sfP=\sfP^{(0)}\oplus\sfP^{(1)}\,,
\qqq
written in terms of the LI Maurer--Cartan super-1-form $\,\th_{\rm L}\equiv e^a_{\rm L}\ox P_a+\si_{\rm L}^\a\ox Q_\a\in\Om^1(\bT)^\bT\ox\tgt$.\ The superdistribution is spanned by the LI vector fields associated with the vectors from $\,{\rm im}\,\sfP$.\ Such a definition is the most natural first step towards the discussion of the (\emph{right}) $\k$-symmetry and BPS charge of the ensuing $\cG$-brane supergeometry,\ which,\ however,\ we postpone to a future work.

Fix a point $\,\widehat x{}_*\equiv(\widehat x{}^{\widehat a}_*)^{\widehat a\in\ovl{p+1,9}}\in\bR^{9-p}\equiv\{0\}\x\bR^{9-p}\subset\bR^{9,1}\,$ and let
\qq\label{eq:Dpemb}
\ep_{(p,1|R)}^{(\widehat x{}_*)}\ :\ D_{(p,1|R)}^{(\widehat x{}_*)}\emb{\rm sMink}(9,1|32)
\qqq
be a leaf of the foliation of $\,\xcD_{(p,1|R)}\,$ through $\,\widehat x{}_*$.\ Write 
\qq\nn
\sfP\circ\th_{\rm L}\equiv e{}^{\unl a}_{\rm L}\ox P_{\unl a}+\si{}^{\unl\a}_{\rm L}\ox\breve Q_{\unl a}\,,
\qqq
denoting,\ furthermore,\ the global coordinates on 
\qq\nn
D_{(p,1|R)}^{(\widehat x{}_*)}\cong{\rm sMink}(p,1| R)\equiv\bD_{(p,1|R)}
\qqq
induced by the flows of the tangent superdistribution as $\,\{\breve\th{}^{\unl\a}\}^{\unl\a\in\ovl{1,R}}\,$ (the Gra\ss mann-odd ones) and $\,\{\breve x{}^{\unl a}\}^{\unl a\in\ovl{0,p}}\,$ (the Gra\ss mann-even ones),\ so that the embedding $\,\ep_{(p,1|R)}^{(\widehat x{}_*)}\,$ acquires the coordinate form $\,\ep_{(p,1|R)}^{(\widehat x{}_*)}(\breve\th{}^{\unl\a},\breve x{}^{\unl a})=(\breve{\unl\th}{}^\a,\breve x{}^{\unl a},\widehat x{}^{\widehat b}_*)$,\ written in terms of the spinor $\,\breve{\unl\th}\,$ with components $\,\breve{\unl\th}{}^{\unl\a}=\breve\th{}^{\unl\a}\,$ and $\, \breve{\unl\th}{}^{\widehat\a}=0$,\ and we have the basis LI super-1-forms
\qq\nn
\breve e{}^{\unl a}_{\rm L}\bigl(\breve\th,\breve x\bigr)\equiv\ep_{(p,1|R)}^{(\widehat x{}_*)\,*}e{}^{\unl a}_{\rm L}\bigl(\breve\th,\breve x\bigr)=\sfd\breve x{}^{\unl a}+\tfrac{1}{2}\,\breve\th\,\ovl\g{}^{\unl a}\,\sfd\breve\th\,,\qquad\qquad\breve\si{}^{\unl\a}_{\rm L}\bigl(\breve\th,\breve x\bigr)\equiv\ep_{(p,1|R)}^{(\widehat x{}_*)\,*}\si{}^{\unl\a}_{\rm L}\bigl(\breve\th,\breve x\bigr)=\sfd\breve\th^{\unl\a}\,,
\qqq
where\footnote{Here,\ we regard each $\,\ovl\G{}^{\unl a}\,$ as the matrix of a symmetric bilinear form on the 32-dimensional Majorana spinor module.\ In view of the identity $\,\ovl\G{}^{\unl a}\cdot\sfP^{(1)}=\sfP^{(1)\,{\rm T}}\cdot\ovl\G{}^{\unl a}$,\ the form splits into a direct sum of its restrictions to (the cartesian squares of) the subspaces $\,{\rm im}\,\sfP^{(1)}\,$ and $\,{\rm ker}\,\sfP^{(1)}$,\ and $\,\ovl\g{}^{\unl a}\,$ is the matrix of the former restriction.\ This is to be contrasted with the endomorphism $\,\G{}^{\unl a}\,$ of the said module which becomes block-\emph{anti}diagonal with respect to the direct-sum decomposition of the module into $\,{\rm im}\,\sfP^{(1)}\,$ and $\,{\rm ker}\,\sfP^{(1)}\,$ for odd $\,p$.\ There is,\ in particular,\ no contradiction in the case $\,p=1\,$ between the equality $\,\ovl\g{}^0=-\ovl\g{}^1$,\ crucial to our reasoning,\ and the defining anticommutation relations of the underlying Clifford algebra (note that $\,\sfP^{(1)}\,$ does not commute with $\,C$).}
\qq\nn
\ovl\g{}^{\unl a}\equiv\ovl\G{}^{\unl a}\rstr_{\im\,\sfP^{(1)}}\,.
\qqq
We shall also use the notation
\qq\nn
\ovl\G{}^{\widehat b}_{\widehat\g\unl\d}\equiv\widehat\chi{}^{\widehat b}_{\widehat\g\unl\d}
\qqq
in what follows.\ It is now obvious that 
\qq\label{eq:D1Rorbit}
\ep_{(p,1|R)}^{(\widehat x{}_*)}\bigl(D_{(p,1|R)}^{(\widehat x{}_*)}\bigr)\equiv\ep_{(p,1|R)}^{(\widehat x{}_*)}(0)\ract\bD_{(p,1|R)}
\qqq
is the orbit of the free \emph{right} action of the subgroup 
\qq\nn
\jmath_{(p,1|R)}\ :\ \bD_{(p,1|R)}\emb\bT
\qqq
whose embedding in the supertarget Lie supergroup admits the simple presentation $\,\jmath_{(p,1|R)}(\Theta^{\unl\a},X^{\unl a})=(\Theta^{\unl\a},0,X^{\unl a},0)\,$ in the global Gra\ss mann-odd $\,\{\Theta^{\unl\a}\}^{\unl\a\in\ovl{1, R}}\,$ and Gra\ss mann-even $\,\{X^{\unl a}\}^{\unl a\in\ovl{0,p}}\,$ coordinates on $\,\bD_{(p,1|R)}$,\ and whose action is inherited (through the embedding $\,\jmath_{(p,1|R)}\,$ and restriction) from the right action of $\,\bT\,$ on itself.\ This point of view,\ consistent with our choice of the tangent superdistribution $\,\xcD_{(p,1|R)}$,\ is expected to be central to a future analysis of the $\k$-symmetry present.\ It is not a unique action,\ though,\ and it is the alternative one that is of key relevance to our present considerations.\ The latter is founded on the simple observation:\ The leaf $\,\ep_{(p,1|R)}^{(\widehat x{}_*)}(D_{(p,1|R)}^{(\widehat x{}_*)})\,$ is preserved (indeed,\ even generated) by the very same subgroup $\,\jmath_{(p,1|R)}(\bD_{(p,1|R)})\subset\bT\,$ acting freely from the \emph{left} (also through $\,\jmath_{(p,1|R)}\,$ and restriction),
\qq\nn
\ep_{(p,1|R)}^{(\widehat x{}_*)}\bigl(D_{(p,1|R)}^{(\widehat x{}_*)}\bigr)\equiv\bD_{(p,1|R)}\lact\ep_{(p,1|R)}^{(\widehat x{}_*)}(0)\,.
\qqq
The embedding \eqref{eq:Dpemb} is equivariant with respect to the $\bD_{(p,1|R)}$-actions:\ the regular action $\,\breve\ell\equiv\breve\txm\,$ (resp.\ $\,\breve\wp\equiv\breve\txm$) of $\,\bD_{(p,1|R)}\,$ on its domain,\ defined just like the left (resp.\ right) regular action of $\,\bT\,$ on itself,\ and the induced left (resp.\ right) action on its codomain,\ mediated by the embedding $\,\jmath_{(p,1|R)}$,\ {\it i.e.},\ we have,\ in the coordinate presentation,
\qq\label{eq:inD1act}\qquad\qquad
\ell_{\jmath_{(p,1|R)}(\breve\th,\breve x)}\circ\ep_{(p,1|R)}^{(\widehat x{}_*)}=\ep_{(p,1|R)}^{(\widehat x{}_*)}\circ\breve\ell{}_{(\breve\th,\breve x)}\,,\qquad\qquad\wp_{\jmath_{(p,1|R)}(\breve\th,\breve x)}\circ\ep_{(p,1|R)}^{(\widehat x{}_*)}=\ep_{(p,1|R)}^{(\widehat x{}_*)}\circ\breve\wp{}_{(\breve\th,\breve x)}\,,
\qqq
and so the embedded subgroup $\,\jmath_{(p,1|R)}(\bD_{(p,1|R)})\cong\bD_{(p,1|R)}\,$ becomes a candidate for the residual (global-)supersymmetry group of the $\cG_{\rm GS}$-brane.\ Finally,\ it is worth noting that for the specific choice $\,\widehat x{}_*=0$,\ and for that choice exclusively,\ the corresponding leaf 
\qq\nn
\ep_{(p,1|R)}^{(0)}\bigl(D_{(p,1|R)}^{(0)}\bigr)=\ep_{(p,1|R)}^{(0)}(0)\ract\bD_{(p,1|R)}=\bD_{(p,1|R)}\lact\ep_{(p,1|R)}^{(0)}(0)=\jmath_{(p,1|R)}\bigl(\bD_{(p,1|R)}\bigr)
\qqq
is a Lie sub-supergroup of $\,\bT$,\ the embedding $\,\ep_{(p,1|R)}^{(0)}\,$ being now promoted to the rank of a Lie-supergroup monomorphism.\ With these simple observations in mind,\ we may now proceed with our efforts to trivialise the GS super-1-gerbe over $\,\ep_{(p,1|R)}^{(\widehat x{}_*)}(D_{(p,1|R)}^{(\widehat x{}_*)})\,$ in a $\jmath_{(p,1|R)}(\bD_{(p,1|R)})$-invariant manner,\ or,\ equivalently,\ the pullback gerbe $\,\ep_{(p,1|R)}^{(\widehat x{}_*)\,*}\cG_{\rm GS}\,$ over $\,D_{(p,1|R)}^{(\widehat x{}_*)}\,$ in a (left-)$\bD_{(p,1|R)}$-invariant manner.

Following the latter path,\ we note that the curvature of $\,\cG_{\rm GS}\,$ restricts to $\,D_{(p,1|R)}^{(\widehat x{}_*)}\,$ as
\qq\nn
\ep_{(p,1|R)}^{(\widehat x{}_*)\,*}\txH_{\rm GS}=\breve\si_{\rm L}\wedge\ovl\g{}_{\unl a}\,\breve\si_{\rm L}\wedge\breve e{}^{\unl a}_{\rm L}\,.
\qqq
Thus,\ in the case $\,p=1$,\ in which $\,R_{(p=1)}=\frac{32}{2}=16$,\ we obtain
\qq\nn
\ep_{(1,1|16)}^{(\widehat x{}_*)\,*}\txH_{\rm GS}=\breve\si_{\rm L}\wedge\ovl\g{}_0\,\breve\si_{\rm L}\wedge\breve e{}^0_{\rm L}+\breve\si_{\rm L}\wedge\ovl\g{}_1\,\breve\si_{\rm L}\wedge\breve e{}^1_{\rm L}\,.
\qqq
However,
\qq\label{eq:Cgam0-Cgam1}
\ovl\g{}_0=-\ovl\g{}^0\equiv-\ovl\G{}^0\rstr_{\im\,\sfP^{(1)}}=\ovl\G{}^1\rstr_{\im\,\sfP^{(1)}}\equiv\ovl\g{}^1=\ovl\g{}_1\,,
\qqq
and so
\qq\nn
\ep_{(1,1|16)}^{(\widehat x{}_*)\,*}\txH_{\rm GS}=\sfd\bigl(-2\breve e{}^0_{\rm L}\wedge\breve e{}^1_{\rm L}\bigr)\,,
\qqq
which is none other than the desired trivialisation of the curvature,\ with the manifestly $\bD_{(p,1|R)}$-invariant curving
\qq\label{eq:sstringcurv}
\om_{(1,1|16)}=-2\breve e{}^0_{\rm L}\wedge\breve e{}^1_{\rm L}\in\Om^2\bigl(D_{(1,1|16)}^{(\widehat x{}_*)}\bigr)^{\bD_{(1,1|16)}}\,.
\qqq 
Upon decomposing the sought-after primitive of $\,\ep_{(p,1|R)}^{(\widehat x{}_*)\,*}\txH_{\rm GS}\,$ for $\,p>1\,$ in the basis of LI super-2-forms on $\,\bD_{(p,1|R)}$,\ {\it i.e.},\ writing it out as an $\bR$-linear combination of the super-2-forms $\,\breve e{}^{\unl a}_{\rm L}\wedge\breve e{}^{\unl b}_{\rm L},\breve e{}^{\unl a}_{\rm L}\wedge\breve\si_{\rm L}^{\unl\a}\,$ and $\,\breve\si_{\rm L}^{\unl\a}\wedge\breve\si_{\rm L}^{\unl\b}$,\ we easily convince ourselves that the very same mechanism as the one that ensured success in the previous case dooms us to failure this time (unless we impose additional constraints,\ which might further restrict the Majorana-spinor representation under consideration).\ That said,\ it is only natural to expect that the embedded supergeometries $\,D_{(p,1|R)}^{(\widehat x{}_*)}\,$ for $\,p>1\,$ support supersymmetric trivialisations of the higher GS super-$p$-gerbes,\ and so our analysis is certain to be of use in any future study of supersymmetric boundary (and non-boundary) defects in the GS super-$\si$-models for higher super-$p$-branes carried out along the lines drawn in the present paper.\smallskip

Our hitherto analysis leaves us with the well-posed question of existence of a supersymmetric geometrisation of the tensorial trivialisation $\,\ep_{(1,1|16)}^{(\widehat x{}_*)\,*}\txH_{\rm GS}=\sfd\om_{(1,1|16)}$.\ The answer is given in
\berop\label{prop:sstring-GS-brane}
Over every (topological) hyperbolic plane in $\,{\rm sMink}(9,1|32)\,$ labelled by $\,\widehat x{}_*\in\bR^8\subset\bR^{9,1}\,$ (as above),\ there exists a $\bD_{(1,1|16)}$-supersymmetric $(1+1|16)$-su\-per\-di\-men\-sion\-al $\cG_{\rm GS}$-module
\qq\nn
\bigl(D_{(1,1|16)}^{(\widehat x{}_*)},\cT_{(1,1|16)}^{(\widehat x{}_*)}\bigr)\,,\qquad\qquad D_{(1,1|16)}^{(\widehat x{}_*)}\cong {\rm sMink}(1,1|16)
\qqq
of curvature $\,\om_{(1,1|16)}\,$ as in \Reqref{eq:sstringcurv},\ determined by a 1-isomorphism
\qq\label{eq:hyper-triv}
\cT_{(1,1|16)}^{(\widehat x{}_*)}\ :\ \ep_{(1,1|16)}^{(\widehat x{}_*)\,*}\cG_{\rm GS}\xrightarrow{\ \cong\ }\cI_{\om_{(1,1|16)}}
\qqq
between the restricted GS super-1-gerbe $\,\cG_{\rm GS}\,$ and the trivial super-1-gerbe 
\qq\nn
\cI_{\om_{(1,1|16)}}=\bigl(\sfY_0 D_{(1,1|16)}^{(\widehat x{}_*)}\equiv D_{(1,1|16)}^{(\widehat x{}_*)},\id_{D_{(1,1|16)}^{(\widehat x{}_*)}},\om_{(1,1|16)},D_{(1,1|16)}^{(\widehat x{}_*)}\x\bC^\x,\pr_1,\pr_2^*\vartheta,\bd1\bigr)\,.
\qqq 
\eerop
\beroof
{\it Cf.}\ App.\,\ref{app:sstring-GS-brane}.
\eroof

The existence of the structure \eqref{eq:D1Rorbit} of an orbit of a \emph{right} action of the residual supersymmetry group  $\,\bD_{(1,1|16)}\,$ on the worldvolume $\,D_{(1,1|16)}^{(\widehat x{}_*)}\,$ of the $\cG_{\rm GS}$-brane investigated in the previous section provokes a natural question as to the existence of a lift of the right-handed supersymmetry to the restricted GS super-1-gerbe $\,\ep_{(1,1|16)}^{(\widehat x{}_*)\,*}\cG_{\rm GS}\,$ and -- if the answer is positive -- as to the compatibility of its trivialisation $\,\cT_{(1,1|16)}^{(\widehat x{}_*)}\,$ with this extended supersymmetry.\ We inspect both issues below.

Our analysis begins with the verification of the \emph{right} invariance of the curvature of $\,\ep_{(1,1|16)}^{(\widehat x{}_*)\,*}\cG_{\rm GS}$,\ the latter taking the coordinate form (in the previously introduced notation) $\,\ep_{(1,1|16)}^{(\widehat x{}_*)\,*}\txH_{\rm GS}(\breve m)=\sfd\breve\th\wedge\ovl\g{}_0\,\sfd\breve\th\wedge\sfd\breve x{}_+$.\ Given the coordinate transformations
\qq\label{eq:rsusybr}
\breve\th{}^{\unl\a}\longmapsto\breve\th{}^{\unl\a}+\breve\vep{}^{\unl\a}\,,\qquad\qquad\breve x{}_+\longmapsto\breve x{}_++\breve t{}_+-\tfrac{1}{2}\,\breve\th\,\bigl(\ovl\g{}^0+\ovl\g{}^1\bigr)\,\breve\vep=\breve x{}_++\breve t{}_+
\qqq
under a right translation by $\,\breve c=(\breve\vep,\breve t)\in\bD_{(1,1|16)}$,\ we immediately infer the anticipated property of the curvature super-3-form,
\qq\label{eq:rinvHGSrestr}
\breve\wp{}_{\breve c}^*\ep_{(1,1|16)}^{(\widehat x{}_*)\,*}\txH_{\rm GS}=\ep_{(1,1|16)}^{(\widehat x{}_*)\,*}\txH_{\rm GS}\,.
\qqq
The higher-geometric lift of the above tensorial identity is given in
\berop\label{prop:Rsusy-GS-grb}
The restriction of the Green--Schwarz super-1-gerbe to the super-worldvolume of the $\cG_{\rm GS}$-brane of Prop.\,\ref{prop:sstring-GS-brane} is \emph{right}-$\bD_{(1,1|16)}$-invariant,\ {\it i.e.},\ there exists a family of 1-isomorphisms
\qq\nn
\Phi_{\breve c}\ :\ \wp_{\breve c}^*\ep_{(1,1|16)}^{(\widehat x{}_*)\,*}\cG_{\rm GS}\xrightarrow{\ \cong\ }\ep_{(1,1|16)}^{(\widehat x{}_*)\,*}\cG_{\rm GS}\,,\qquad\qquad\breve c\in\bD_{(1,1|16)}\,.
\qqq
\eerop
\beroof
{\it Cf.}\ App.\,\ref{app:Rsusy-GS-grb}.
\eroof
\noindent Compatibility\footnote{Just to reemphasise:\ The compatibility merely qualifies the 1-gerbe 1-isomorphism of the trivialisation.\ We do \emph{not} have a trivialisation in the category of \emph{right} super-1-gerbes.} of the trivialisation established in Prop.\,\ref{prop:sstring-GS-brane} with the above \emph{right} supersymmetry now becomes a direct consequence of
\berop\label{prop:Rsusy-inv-GS-1-brane}\qquad
There exists a $\bD_{(1,1|16)}$-indexed family of 1-gerbe 2-isomorphisms
\qq\label{diag:R-on-L-GS-1-brane}
\alxydim{@C=3.cm@R=2cm}{\ep_{(1,1|16)}^{(\widehat x{}_*)\,*}\cG_{\rm GS} \ar[r]^{\cT_{(1,1|16)}^{(\widehat x{}_*)}} \ar@{=>}[dr]|{\ \varphi_{\breve c}\ }  & \cI_{\om_{(1,1|16)}} \\ \breve\wp{}_{\breve c}^*\ep_{(1,1|16)}^{(\widehat x{}_*)\,*}\cG_{\rm GS} \ar[u]^{\Phi_{\breve c}} \ar[r]_{\breve\wp{}_{\breve c}^*\cT_{(1,1|16)}^{(\widehat x{}_*)}} & \breve\wp{}_{\breve c}^*\cI_{\om_{(1,1|16)}} \ar[u]_{\Psi_{\breve c}} }\,,\qquad\qquad\breve c\in\bD_{(1,1|16)}
\qqq
written for $\,\Phi_{\breve c}\,$ as in Prop.\,\ref{prop:Rsusy-GS-grb} and for a canonical \emph{right}-$\bD_{(1,1|16)}$-supersymmetric structure
\qq\nn
\Psi_{\breve c}\ :\ \breve\wp{}_{\breve c}^*\cI_{\om_{(1,1|16)}}\xrightarrow{\ \cong\ }\cI_{\om_{(1,1|16)}}\,,\qquad\qquad\breve c\in\bD_{(1,1|16)}
\qqq
on the trivial super-1-gerbe of Prop.\,\ref{prop:sstring-GS-brane}.
\eerop
\beroof
{\it Cf.}\ App.\,\ref{app:Rsusy-inv-GS-1-brane}.
\eroof

\brem
In the light of the findings of Refs.\,\cite{Suszek:2020xcu,Suszek:2021hjh},\ the present study of the \emph{right} extension of global supersymmetry seems to be the first step on a path towards giving a higher-geometric meaning to the $\k$-symmetry of the (left-)$\bD_{(1,1|16)}$-supersymmetric $\cG_{\rm GS}$-brane $\,(D_{(1,1|16)}^{(\widehat x{}_*)},\cT_{(1,1|16)}^{(\widehat x{}_*)})$.\ As such,\ it merits a separate investigation,\ involving,\ in particular,\ a verification of its compatibility with the left supersymmetry.\ We intend to return to these issues in a future work.
\erem

\subsection{An odd superpoint-supersymmetric $\cG_{\rm GS}$-brane}\label{sub:sptGbr}
 
The basic idea underlying the identification and reconstruction of $\cG_{\rm GS}$-modules and the associated $\cG_{\rm GS}$-branes that we pursue in this paper is the reduction of the bulk supersymmetry upon restriction to an embedded sub-supermanifold of the target supermanifold -- itself an orbit of a free action of the reduced supersymmetry group whose Lie superalgebra spans the tangent sheaf of the orbit.\ In this picture,\ the bulk supersymmetry has so far been induced from (the adjoint (left) action of $\,\bT\,$ on itself or) a \emph{chiral} (\emph{left}) action of the mother supersymmetry group on itself,\ and so it is \emph{left} actions of subgroups of the latter that we consider.\ As it turns out,\ they may and -- as demonstrated in the previous section -- sometimes do carry an additional \emph{right} supersymmetry that lifts to the restricted super-1-gerbe.\ Thus,\ we end up with a product Lie sub-supergroup $\,\bD_{(p,1|R)}^{({\rm L})}\x\bD_{(p,1|R)}^{({\rm R})},\ \bD_{(p,1|R)}^{({\rm L})}=\bD_{(p,1|R)}^{({\rm R})}\,$ (resp.\ a direct-sum Lie sub-superalgebra $\,\dgt_{(p,1|R)}^{({\rm L})}\oplus\dgt_{(p,1|R)}^{({\rm R})},\ \dgt_{(p,1|R)}^{({\rm L})}=\dgt_{(p,1|R)}^{({\rm R})}$) of the \emph{bi-chiral} supersymmetry group $\,\bT^{({\rm L})}\x\bT^{({\rm R})}\,$ (resp.\ of the supersymmetry algebra $\,\tgt^{({\rm L})}\oplus\tgt^{({\rm R})}$) of the supertarget $\,\bT\,$ rather than that of its chiral (left) Lie sub-supergroup $\,\bT^{({\rm L})}\,$ (resp.\ Lie sub-superalgebra $\,\tgt^{({\rm L})}$).\ The last observation leads us to contemplate a more general scenario in which we take the bi-chiral supersymmetry group as the point of departure and look for supersymmetric trivialisations of the GS super-$1$-gerbe over sub-supermanifolds of $\,\bT\,$ that arise as orbits of actions of its arbitrary Lie sub-supergroups,\ $\,\bD\subset\bT^{({\rm L})}\x\bT^{({\rm R})}\,$ with the corresponding Lie superalgebras $\,\dgt\subset\tgt^{({\rm L})}\oplus\tgt^{({\rm R})}$,\ not necessarily having the structure of direct products (resp.\ sums) of chiral factors.\ In so doing,\ we take care to choose from among sub-supermanifolds of the supertarget $\,\bT\,$ with a non-vanishing Gra\ss mann-odd component of the structure (and tangent) sheaf,\ and hence proper \emph{super}symmetry which ensures that we do not wind up in the old category of bosonic branes.\ Below,\ we elaborate at some length a simple concrete example in which this scenario is seen to work out,\ thus arriving at a new species of $\cG_{\rm GS}$-brane.

Our analysis begins with a choice of a subspace within the bi-chiral Lie superalgebra $\,\tgt^{({\rm L})}\oplus\tgt^{({\rm R})}\,$ spanned by $\,N\leq 32\,$ linearly independent vector combinations
\qq\nn
\cV_{\unl\a}=\bigl(Q_{\unl\a},Q_{\unl\a}\bigr)\,,\qquad\unl\a\in\ovl{1,N}
\qqq
of the Gra\ss mann-odd generators of the left ($(Q_{\unl\a},0)$) and right ($(0,Q_{\unl\a})$) summands,\ respectively,\ which we may -- as we just have -- choose to be the first $\,N\,$ of the 32 generators of the chiral summands without any loss of generality.\ Geometrically,\ the $\,\cV_{\unl\a}\,$ are simply the coordinate differentations $\,\cV_{\unl\a}(\th,x)=\frac{\vec\p\ }{\p\th^{\unl\a}}$,\ and so they span a purely Gra\ss mann-odd superdistribution 
\qq\nn
\xcD_{(0|N)}=\bigoplus_{\unl\a=1}^N\,\corr{\cV_{\unl\a}}_{\cO_\bT}
\qqq
over $\,\bT$,\ with the direct complement at $\,x\in\bR^{9,1}\,$ spanned by the remaining $\,\frac{\vec\p\ }{\p\th^{\widehat\a}},\ \widehat\a\in\ovl{N+1,32}\,$ and the $\,\frac{\p\ }{\p x^a},\ a\in\ovl{0,9}$.\ The superdistribution is manifestly involutive,\ and hence integrable in virtue of the $\bZ/2\bZ$-graded variant of the Global Frobenius Theorem,\ {\it cf.}\ \Rxcite{Thm.\,6.2.1}{Carmeli:2011}.\ Its integral leaf 
\qq\nn
\d_{(0|N)}^{(x_*)}\ :\ D_{(0|N)}^{(x_*)}\too\bT
\qqq
through $\,x_*\in\bR^{9,1}$,\ with the coordinate presentation $\,\d_{(0|N)}^{(x_*)}(\th)=(\unl\th,x_*)\,$ written in terms of the spinor $\,\unl\th\,$ with components $\,\unl\th{}^{\unl\a}=\th^{\unl\a}\,$ and $\,\unl\th{}^{\widehat\a}=0$,\
is the orbit $\,\d_{(0|N)}^{(x_*)}(D_{(0|N)}^{(x_*)})\equiv\bD_{(0|N)}\lact(0,x_*^a)\,$ of that topological point under the (restricted) `twisted adjoint' action 
\qq\nn
\widetilde\Ad\ :\ \bD_{(0|N)}\x\bT\too\bT\,,
\qqq
with the coordinate presentation
\qq\nn
\widetilde\Ad{}_\vep(\th,x):=\bigl(\tfrac{1}{2}\,\unl\vep,0\bigr)\cdot(\th,x)\cdot\bigl(\tfrac{1}{2}\,\unl\vep,0\bigr)=(\th+\unl\vep,x)\,,
\qqq
of the superpoint (Lie supergroup)
\qq\nn
\bD_{(0|N)}\equiv\bR^{0|N}\,.
\qqq
We have an analogon $\,\widetilde\Ad{}_\vep\circ\d_{(0|N)}^{(x_*)}=\d_{(0|N)}^{(x_*)}\circ\t_\vep\,$ of formul\ae ~\eqref{eq:inD1act},\ expressed in terms of the translational action
\qq\nn
\t\ :\ \bD_{(0|N)}\x\bD_{(0|N)}\too\bD_{(0|N)}
\qqq
of $\,\bD_{(0|N)}\,$ on itself,\ with the coordinate presentation
\qq\nn
\t_\vep(\th)=\th+\vep\,,
\qqq
and so the superpoint acquires the status of the reduced supersymmetry group of $\,D_{(0|N)}^{(x_*)}$.

Upon pullback to $\,D_{(0|N)}^{(x_*)}$,\ we find $\,\d_{(0|N)}^{(x_*)\,*}\si^{\unl\a}_{\rm L}(\th)=\sfd\th^{\unl\a},\ \d_{(0|N)}^{(x_*)\,*}\si^{\widehat\a}_{\rm L}(\th)=0\,$ and $\,\d_{(0|N)}^{(x_*)\,*}e^a_{\rm L}(\th)=\tfrac{1}{2}\,\unl\th\,\ovl\G{}^a\,\sfd\unl\th$,\ and so the curvature of the GS super-1-gerbe trivialises as $\,\d_{(0|N)}^{(x_*)\,*}\txH_{\rm GS}=0\,$ by the Fierz identity.\ Altogether,\ then,\ we anticipate a $\bD_{(0|N)}$-supersymmetric structure
\qq\nn
\Xi_\vep\ :\ \t_\vep^*\d_{(0|N)}^{(x_*)\,*}\cG_{\rm GS}\xrightarrow{\ \cong\ }\d_{(0|N)}^{(x_*)\,*}\cG_{\rm GS}\,,\qquad\qquad\vep\in\bD_{(0|N)}\,,
\qqq
and a trivialisation of $\,\d_{(0|N)}^{(x_*)\,*}\cG_{\rm GS}$\,,
\qq\nn
\cT_{(0|N)}^{(x_*)}\ :\ \d_{(0|N)}^{(x_*)\,*}\cG_{\rm GS}\xrightarrow{\ \cong\ }\cI_0\,,
\qqq
compatible with the former.\ As the (zero) curving of the trivialisation is automatically $\bD_{(0|N)}$-invariant,\ it makes sense to speak of a (left) $\bD_{(0|N)}$--supersymmetric trivialisation here,\ which is the desirable circumstance.\ Still,\ there is no {\it a priori} reason to expect the tensorial data of the pullback super-1-gerbe $\,\d_{(0|N)}^{(x_*)\,*}\cG_{\rm GS}\,$ to be $\bD_{(0|N)}$-invariant.\ Instead,\ a generic $\bD_{(0|N)}$-supersymmetric structure is likely to arise.\ These expectations are corroborated in (the proof of)

\berop\label{prop:spoint-GS-brane}
Over every topological point $\,x_*\in\bR^{9,1}\,$ in $\,{\rm sMink}(9,1|32)$,\ embedded as $\,\widehat x{}_*\ :\ \bR^{0|0}\emb{\rm sMink}(9,1|32)$,\ and for every $\,N\in\ovl{1,32}$,\ there exists a flat $\bD_{(0|N)}$-supersymmetric $(0|N)$-su\-per\-di\-men\-sion\-al $\cG_{\rm GS}$-module
\qq\nn
\bigl(D_{(0|N)}^{(x_*)},\cT_{(0|N)}^{(x_*)}\bigr)\,,\qquad\qquad D_{(0|N)}^{(x_*)}\cong\bR^{0|N}
\qqq
determined by a 1-isomorphism
\qq\nn
\cT_{(0|N)}^{(x_*)}\ :\ \d_{(0|N)}^{(x_*)\,*}\cG_{\rm GS}\xrightarrow{\ \cong\ }\cI_0
\qqq
between the restricted GS super-1-gerbe $\,\cG_{\rm GS}\,$ and the trivial super-1-gerbe over $\,D_{(0|N)}^{(x_*)}$.
\eerop
\beroof
{\it Cf.}\ App.\,\ref{app:spoint-GS-brane}.
\eroof

\subsection{The multiplicative $\cG_{\rm GS}$-bi-branes}\label{sub:sbib}

With the three families of supersymmetric $\cG_{\rm GS}$-branes on our hands,\ we may complete the construction of the associated maximally supersymmetric $\cG_{\rm GS}$-bi-branes.\ Indeed,\ given a $\cG_{\rm GS}$-brane \eqref{eq:GSbrane},\ we define,\ over 
\qq\nn
Q_D\equiv\bT\x D
\qqq 
and for 
\qq\nn
\txm_D\equiv\txm\circ\bigl(\id_\bT\x\iota_D\bigr)\,,\qquad\qquad\cM_D\equiv\bigl(\id_\bT\x\iota_D\bigr)^*\cM\,,\qquad\qquad\varrho_{{\rm GS},D}\equiv\bigl(\id_\bT\x\iota_D\bigr)^*\varrho_{\rm GS}\,,
\qqq
the corresponding (multiplicative) $\ell^{(1)}$-supersymmetric 1-isomorphism
\qq\nn
\Phi_D\ &:&\ \pr_1^*\cG_{\rm GS}\equiv\pr_1^*\cG_{\rm GS}\ox\pr_2^*\cI_{\om_D}\ox\cI_{-\pr_2^*\om_D}\xrightarrow{\ \ \id_{\pr_1^*\cG_{\rm GS}}\ox\pr_2^*\cT_D^{-1}\ox\id_{\cI_{-\pr_2^*\om_D}}\ \ }\pr_1^*\cG_{\rm GS}\ox\pr_2^*\iota_D^*\cG_{\rm GS}\ox\cI_{-\pr_2^*\om_D}\cr\cr
&&\equiv\bigl(\id_\bT\x\iota_D\bigr)^*\bigl(\pr_1^*\cG_{\rm GS}\ox\pr_2^*\cG_{\rm GS}\bigr)\ox\cI_{-\pr_2^*\om_D}\xrightarrow{\ \ \cM_D\ox\id_{\cI_{-\pr_2^*\om_D}}\ \ }\txm_D^*\cG_{\rm GS}\ox\cI_{\om_{Q_D}}\,,
\qqq
with
\qq\nn
\om_{Q_D}\equiv\varrho_{{\rm GS},D}-\pr_2^*\om_D\in\Om^2(Q_D)^\bT\,.
\qqq
The above data give rise to a $\cG_{\rm GS}$-bi-brane
\qq\nn
\cB_D=\bigl(Q_D,\pr_1,\txm_D,\om_{Q_D},\Phi_D\bigr)\,.
\qqq

Thus,\ upon substituting for $\,\cD_D\,$ (and $\,\bD$) the data from the previous three sections,\ we obtain 
\bethe\label{thm:maxym-GS-bib}
There exist three species of the {\bf maximally supersymmetric multiplicative $\cG_{\rm GS}$-bi-module}/{\bf -brane}:
\ben
\item of {\bf type $\,(0|0)\,$} ({\bf instantonic}):
\qq\nn 
\cB_{x_*}\equiv\bigl(\bT\x\bR^{0|0},\pr_1,\txm\circ\bigl(\id_\bT\x\widehat x_*\bigr),0,\Phi_{x_*}\bigr)\,,
\qqq
with
\qq\nn
\Phi_{x_*}=\bigl(\id_\bT\x\widehat x_*\bigr)^*\cM\circ\bigl(\id_{\pr_1^*\cG_{\rm GS}}\ox\pr_2^*\cT_{-1}^{(x_*)\,-1}\bigr)\,;
\qqq
\item of {\bf type $\,(1,1|16)\,$} ({\bf superstring-like}):
\qq\nn 
\cB^{\widehat x{}_*}_{(1,1|16)}\equiv\bigl(\bT\x\bR^{1,1|16},\pr_1,\txm\circ\bigl(\id_\bT\x\ep_{(1,1|16)}^{(\widehat x{}_*)}\bigr),\bigl(\id_\bT\x\ep_{(1,1|16)}^{(\widehat x{}_*)}\bigr)^*\varrho_{\rm GS}-\pr_2^*\om_{1,1|16},\Phi^{\widehat x{}_*}_{(1,1|16)}\bigr)\,,
\qqq
with
\qq\nn
\Phi^{\widehat x{}_*}_{(1,1|16)}=\bigl(\bigl(\id_\bT\x\ep_{(1,1|16)}^{(\widehat x{}_*)}\bigr)^*\cM\ox\id_{\cI_{-\pr_2^*\om_{1,1|16}}}\bigr)\circ\bigl(\id_{\pr_1^*\cG_{\rm GS}}\ox\pr_2^*\cT^{\widehat x{}_*\,-1}_{(1,1|16)}\ox\id_{\cI_{-\pr_2^*\om_{1,1|16}}}\bigr)\,;
\qqq
\item of {\bf type $\,(0|N)\,$} ({\bf superpoint-like}):
\qq\nn 
\cB^{x_*}_{(0|N)}\equiv\bigl(\bT\x\bR^{0|N},\pr_1,\txm\circ\bigl(\id_\bT\x\d_{(0|N)}^{(x_*)}\bigr),\bigl(\id_\bT\x\iota_{(0|N)}^{(x_*)}\bigr)^*\varrho_{\rm GS},\Phi^{x_*}_{(0|N)}\bigr)\,,
\qqq
with
\qq\nn
\Phi^{x_*}_{(0|N)}=\bigl(\id_\bT\x\d_{(0|N)}^{(x_*)}\bigr)^*\cM\circ\bigl(\id_{\pr_1^*\cG_{\rm GS}}\ox\pr_2^*\cT^{(x_*)\,-1}_{(0|N)}\bigr)\,.
\qqq
\een
\ethe
\beroof
The statement follows straightforwardly from Props.\,\ref{prop:AdGbr},\ref{prop:sstring-GS-brane} and \ref{prop:spoint-GS-brane} taken in conjunction with the general considerations of Sec.\,\ref{sub:homocat}.
\eroof

\section{Supersymmetric $\cG_{\rm GS}$-bi-brane fusion and elementary $\cG_{\rm GS}$-inter-bi-branes}\label{sec:sibib}

In this closing section,\ we undertake the task of fusing pairs $\,(\cB_{D_1},\cB_{D_2})\,$ of the supersymmetric $\cG_{\rm GS}$-bi-branes found previously.\ Given the incompleteness,\ emphasised earlier,\ of our symmetry analysis in which $\k$-symmetry has not been taken into account,\ we restrict our considerations to the most elementary structure in the inter-bi-brane hierarchy described in Sec.\,\ref{sub:grbmultissi},\ namely,\ the ternary inter-bi-brane and the relevant elementary fusion 2-isomorphism of Remark \ref{rem:elemfusion}.\ Accordingly,\ we shall look for sub-supermanifolds
\qq\nn
\iota_{T_{D_1,D_2}^{D_3}}\equiv\id_\bT\x\iota_{\cF_{D_1,D_2}^{D_3}}\ :\ T_{D_1,D_2}^{D_3}\equiv\bT\x\cF_{D_1,D_2}^{D_3}\emb\bT\x D_1\x D_2\,,\qquad\qquad\iota_{\cF_{D_1,D_2}^{D_3}}\ :\ \cF_{D_1,D_2}^{D_3}\emb D_1\x D_2\,,
\qqq
further embedded in $\,\bT^{\x 3}\,$ by the mappings
\qq\nn
\jmath_{D_1,D_2}^{D_3}\equiv\jmath_{0,1,2}\circ\iota_{T_{D_1,D_2}^{D_3}}\ :\ T_{D_1,D_2}^{D_3}\emb\bT^{\x 3}\,,\qquad\quad\jmath_{0,1,2}\equiv\id_\bT\x\iota_{D_1,D_2}\equiv\id_\bT\x\iota_{D_1}\x\iota_{D_2}\ :\ \bT\x D_1\x D_2\emb\bT^{\x 3}\,,
\qqq
with the property
\qq\nn
\txm^{D_3}_{D_1,D_2}\bigl(\cF_{D_1,D_2}^{D_3}\bigr)=\iota_{D_3}(D_3)\,,\qquad\qquad\txm^{D_3}_{D_1,D_2}\equiv\txm\circ\bigl(\iota_{D_1}\x\iota_{D_2}\bigr)\circ\iota_{\cF_{D_1,D_2}^{D_3}}\equiv\txm_{D_1,D_2}\circ\iota_{\cF_{D_1,D_2}^{D_3}}\,,
\qqq
so that it makest sense to write
\qq\nn
f_{D_1,D_2}^{D_3}\equiv\iota_{D_3}^{-1}\circ\txm^{D_3}_{D_1,D_2}\equiv\mu^{D_3}_{D_1,D_2}\circ\iota_{\cF_{D_1,D_2}^{D_3}}\ :\ \cF_{D_1,D_2}^{D_3}\too D_3\,,
\qqq
and such that there exists,\ over $\,T_{D_1,D_2}^{D_3}\,$ and for
\qq\nn 
&p_A\ :\ T_{D_1,D_2}^{D_3}\too D_A\,,\qquad A\in\{1,2,3\}\,,&\cr\cr
&p_1:=\pr_2\circ\iota_{T_{D_1,D_2}^{D_3}}\,,\qquad\qquad p_2:=\pr_3\circ\iota_{T_{D_1,D_2}^{D_3}}\,,\qquad\qquad p_3:=f_{D_1,D_2}^{D_3}\circ\pr_2&
\qqq
and
\qq\nn
&p_{0,1},p_{01,2},p_{0,12}\ :\ T_{D_1,D_2}^{D_3}\too\bT^{\x 2}\,,&\cr\cr
&p_{0,1}:=\bigl(\id_\bT\x\iota_{D_1}\bigr)\circ\pr_{1,2}\circ\iota_{T_{D_1,D_2}^{D_3}}\equiv\pr_{1,2}\circ\jmath_{D_1,D_2}^{D_3}\,,&\cr\cr\qquad 
&p_{01,2}:=\bigl(\txm\x\id_\bT\bigr)\circ\jmath_{D_1,D_2}^{D_3}\,,\qquad\qquad p_{0,12}:=\bigl(\id_\bT\x\txm\bigr)\circ\jmath_{D_1,D_2}^{D_3}\,,&
\qqq
as well as for
\qq\nn
\pi^{(3)}_{1,2}\equiv\pr_{1,2}\circ\iota_{T_{D_1,D_2}^{D_3}}\,,\qquad\qquad\pi^{(3)}_{2,3}=\bigl(\txm\circ\bigl(\id_\bT\x\iota_{D_1}\bigr)\x\id_{D_2}\bigr)\circ\iota_{T_{D_1,D_2}^{D_3}}\,,\qquad\qquad\pi^{(3)}_{1,3}\equiv\id_\bT\x f_{D_1,D_2}^{D_3}\,,
\qqq
a ($\ell^{(2)}$-supersymmetric) 1-gerbe 2-isomorphism (we omit the obvious subscripts on the identity 1-isomorphisms for the sake of transparency and explicitly mark the dependence of the 2-isomorphism on the choice of the inequivalent ways of obtaining $\,D_3\,$ from $\,D_1\,$ and $\,D_2\,$ through multiplication)
\qq\nn
\xy (25,0)*{\bullet}="mGGI"+(-24,6)*{\tx{\scriptsize$\barr{c}
\jmath_{D_1,D_2}^{D_3\,*}\bigl(\pr_{1,2}^*\txm^*\cG_{\rm GS}\ox\pr_3^*\cG_{\rm GS}\bigr)\ox\cI_{\pi_{1,2}^{(3)\,*}\om_{Q_{D_1}}-p_2^*\om_{D_2}}\earr$}};
(73,0)*{\bullet}="mGI"+(23,6)*{\tx{\scriptsize$\barr{c} \jmath_{D_1,D_2}^{D_3\,*}d_1^{(3)\,*}\txm^*\cG_{\rm GS}\ox\cI_{\pi_{1,2}^{(3)\,*}\om_{Q_{D_1}}+\pi_{2,3}^{(3)\,*}\om_{Q_{D_2}}} \\ \cong\jmath_{D_1,D_2}^{D_3\,*}d_2^{(3)\,*}\txm^*\cG_{\rm GS}\ox\cI_{\pi_{1,2}^{(3)\,*}\om_{Q_{D_1}}+\pi_{2,3}^{(3)\,*}\om_{Q_{D_2}}} \earr$}};
(12,-30)*{\bullet}="GI"+(-23,0)*{\tx{\scriptsize$\barr{c}\pr_1^*\cG_{\rm GS}\hspace{-0.05cm}
\ox p_1^*\iota_{D_1}^*\cG_{\rm GS}\hspace{-0.05cm}\ox p_2^*\iota_{D_2}^*\cG_{\rm GS}\\\hspace{-0.05cm}\ox
\cI_{-p_1^*\om_{D_1}-p_2^*\om_{D_2}} \\ \\ \hspace{-.05cm}\cong\jmath_{D_1,D_2}^{D_3\,*}\bigl(\pr_1^*\cG_{\rm GS}\hspace{-0.05cm}
\ox\pr_2^*\cG_{\rm GS}\hspace{-0.05cm}\ox\pr_3^*\cG_{\rm GS}\bigr)\\\hspace{-0.05cm}\ox
\cI_{-p_1^*\om_{D_1}-p_2^*\om_{D_2}}\earr$}};
(86,-30)*{\bullet}="GmGI"+(25,0)*{\tx{\scriptsize$\barr{c}\jmath_{D_1,D_2}^{D_3\,*}\bigl(\pr_1^*\cG_{\rm GS}\ox\pr_{2,3}^*\txm^*\cG_{\rm GS}\bigr)\\\hspace{-0.1cm}\ox
\cI_{\pi_{1,2}^{(3)\,*}\om_{Q_{D_1}}+\pi_{2,3}^{(3)\,*}\om_{Q_{D_2}}-p_{0,12}^*\varrho_{\rm GS}}\\ \\ \hspace{-.05cm}\cong\pr_1^*\cG_{\rm GS}\ox p_3^*\iota_{D_3}^*\cG_{\rm GS}\\\hspace{-0.1cm}\ox
\cI_{\pi_{1,2}^{(3)\,*}\om_{Q_{D_1}}+\pi_{2,3}^{(3)\,*}\om_{Q_{D_2}}-p_{0,12}^*\varrho_{\rm GS}} \earr$}};
(25,-60)*{\bullet}="G1"+(-13,-4)*{\tx{\scriptsize$\pr_1^*\cG_{\rm GS}
\hspace{-0.05cm}\cong\pr_1^*\cG_{\rm GS}\hspace{-0.05cm}\ox
p_1^*\cI_{\om_{D_1}} \hspace{-0.08cm}\ox p_2^*\cI_{\om_{D_2}}\hspace{-0.05cm}\ox
\cI_{-p_1^*\om_{D_1}-p_2^*\om_{D_2}}$}};
(73,-60)*{\bullet}="G1I"+(22,-4)*{\tx{\scriptsize$\pr_1^*\cG_{\rm GS}
\ox\cI_{\pi_{1,2}^{(3)\,*}\om_{Q_{D_1}}+\pi_{2,3}^{(3)\,*}\om_{Q_{D_2}}-\pi_{1,3}^{(3)\,*}\om_{Q_{D_3}}}$}}; (40,0)*{}="hup";
(40,-60)*{}="hdown"; \ar@{->}|{\id\ox\pi_{1,2}^{(3)\,*}\pr_2^*\cT_{D_1}^{-1}\ox\pi_{2,3}^{(3)\,*}\pr_2^*\cT_{D_2}^{-1}\ox\id\cong\id\ox p_1^*\cT_{D_1}^{-1}\ox p_2^*\cT_{D_2}^{-1}\ox\id\qquad\qquad\qquad} "G1";"GI"
\ar@{->}|{\pi_{1,2}^{(3)\,*}(\id_\bT\x\iota_{D_1})^*\cM_{\rm GS}\ox\id\cong\jmath_{D_1,D_2}^{D_3\,*}d_3^{(3)\,*}\cM_{\rm GS}\ox\id\qquad\qquad} "GI";"mGGI"
\ar@{->}|{\tx{\scriptsize$\barr{c} \pi_{2,3}^{(3)\,*}(\id_\bT\x\iota_{D_2})^*\cM_{\rm GS}\ox\id \\ \cong\jmath_{D_1,D_2}^{D_3\,*}d_1^{(3)\,*}\cM_{\rm GS}\ox\id \earr$}} "mGGI";"mGI"
\ar@{->}|{\qquad\qquad\pi_{1,3}^{(3)\,*}(\id_\bT\x\iota_{D_3})^*\cM_{\rm GS}^{-1}\ox\id\cong\jmath_{D_1,D_2}^{D_3\,*}d_2^{(3)\,*}\cM_{\rm GS}^{-1}\ox\id}
"mGI";"GmGI" \ar@{->}|{\id\ox\pi_{1,3}^{(3)\,*}\pr_2^*\cT_{D_3}\ox\id\cong\id\ox p_3^*\cT_{D_3}\ox\id} "GmGI";"G1I" \ar@{=}|{\ \id\ } "G1"+(2,0);"G1I"+(-2,0)
\ar@{=>}|{\ \varphi_3[\cF_{D_1,D_2}^{D_3}]\ } "hup"+(9,-3);"hdown"+(9,+3)
\endxy\,.
\qqq
Note the emergence of the familiar consistency condition -- the DJI:\ $\,\pi_{1,2}^{(3)\,*}\om_{Q_{D_1}}+\pi_{2,3}^{(3)\,*}\om_{Q_{D_2}}-\pi_{1,3}^{(3)\,*}\om_{Q_{D_3}}=0$.

With the help of the generalised multiplicative structure \eqref{diag:aleph} on $\,\cG_{\rm GS}$,\ whose existence was established in Sec.\,\ref{sec:multGSs1g},\ we may reduce the task substantially to that of finding the ({\bf generalised}) {\bf fusion 2-isomorphism}
\qq\label{eq:gen-fus-2-iso-gen}\qquad\qquad
\alxydim{@C=4cm@R=2cm}{\jmath_{D_1,D_2}^{D_3\,*}\bigl(\pr_2^*\cG_{\rm GS}\ox\pr_3^*\cG_{\rm GS}\bigr) \ar[d]_{p_1^*\cT_{D_1}\ox p_2^*\cT_{D_2}}
\ar[r]^{\jmath_{D_1,D_2}^{D_3\,*}\pr_{2,3}^*\cM_{\rm GS}} & \jmath_{D_1,D_2}^{D_3\,*}\pr_{2,3}^*\txm^*\cG_{\rm GS}\ox\cI_{\jmath_{D_1,D_2}^{D_3\,*}\pr_{2,3}^*\varrho_{\rm GS}} \ar[d]^{p_3^*\cT_{D_3}\ox\cJ_{-\jmath_{D_1,D_2}^{D_3\,*}\vartheta_{\rm GS}}} \\
\cI_{p_1^*\om_{D_1}+p_2^*\om_{D_2}}
\ar@{=>}[ur]|{\phi[\cF_{D_1,D_2}^{D_3}]} \ar@{=}[r]_{\id} & \cI_{p_3^*\om_{D_3}+\jmath_{D_1,D_2}^{D_3\,*}(\pr_{2,3}^*\varrho_{\rm GS}-\sfd\vartheta_{\rm GS})}}
\qqq
Once the latter has been found,\ we arrive at the elementary (trivalent) $\cG_{\rm GS}$-inter-bi-brane
\qq\nn
\cJ_{2,1}[\cF_{D_1,D_2}^{D_3}]=\bigl(T_{D_1,D_2}^{D_3},\pi_{1,2}^{(3)},\pi_{2,3}^{(3)},\pi_{1,3}^{(3)},\varphi_3[\cF_{D_1,D_2}^{D_3}]\bigr)\,.
\qqq

We shall now derive the structures indicated above in the case of the two non-instantonic types ($(1,1|16)\,$ and $\,(0|N)$) of $\cG_{\rm GS}$-bi-branes of Sec.\,\ref{sub:sbib}.

\subsection{Supergeometric fusion of the supersymmetry orbits}

Following the logic laid out in Refs.\,\cite{Fuchs:2007fw} and \cite{Runkel:2009sp} and recalled in Sec.\,\ref{sec:wzw},\ we first look for (examples of) sub-supermanifolds
\qq\nn
\D^{(x_{*\,1}/x_{*\,2})}_{(p_1/p_2,q_1/q_2|N_1/N_2)}\emb D_1\x D_2\,,\qquad\qquad D_A\equiv D_{(p_A,q_A|N_A)}^{(x_{*\,A})}\in\{D_{(1,1|16)}^{(\widehat x{}_{*\,A})},D_{(0|N_A)}^{(x_{*\,A})}\}\,,\qquad A\in\{1,2\}
\qqq
within the products of the (same-type) supersymmetric $\cG_{\rm GS}$-brane worldvolumes introduced in Secs.\,\ref{sub:bisusyGbr} and \ref{sub:sptGbr},\ whose embeddings 
\qq\nn
(\iota_1\x\iota_2)\bigl(\D^{(x_{*\,1}/x_{*\,2})}_{(p_1/p_2,q_1/q_2|N_1/N_2)}\bigr)\,,\qquad\qquad\iota_A\equiv\iota_{(p_A,q_A|N_A)}^{(x_{*\,A})}\in\{\ep_{(1,1|16)}^{(\widehat x{}_{*\,A})},\d_{(0|N)}^{(x_{*\,A})}\}
\qqq
in the cartesian square of the mother supersymmetry group $\,\bT\,$ produce worldvolumes of new (embedded) supersymmetric $\cG_{\rm GS}$-branes under multiplication,
\qq\nn
\txm\circ(\iota_1\x\iota_2)\bigl(\D^{(x_{*\,1}/x_{*\,2})}_{(p_1/p_2,q_1/q_2|N_1/N_2)}\bigr)\supset\iota_3(D_3)\,,
\qqq
with $\,D_3\equiv D_{(p_3,q_3|N_3)}^{(x_{*\,3})}\,$ (embedded by $\,\iota_3\equiv\iota_{(p_3,q_3|N_3)}^{(x_{*\,3})}$) of the same type as the $\,D_A, A\in\{1,2\}$.

In what follows,\ we switch freely between the sheaf-theoretic and $\cS$-point picture in the description of the relevant supermanifolds.

\subsubsection{Superaligned bi-supersymmetric orbits}

Consider,\ first,\ two embeddings $\,\ep_{(1,1|16)}^{(\widehat x{}_{*\,A})}\ :\ D_{(1,1|16)}^{(\widehat x{}_{*\,A})}\emb\bT,\ A\in\{1,2\}\,$ which are {\bf superaligned} in the sense expressed by the identities
\qq\nn
\ep_{(1,1|16)}^{(\widehat x{}_{*\,A})}\bigl(\breve\th{}^{\unl\a},\breve x{}^0,\breve x{}^1\bigr)=\bigl(\breve{\unl\th}{}^\a,\breve x{}^0,\breve x{}^1,\widehat x{}^{\widehat a}_{*\,A}\bigr)\,,\qquad\qquad\breve{\unl\th}\in\im\,\sfP^{(1)}_{01}\,,\qquad\sfP^{(1)}_{01}\equiv\tfrac{1}{2}\,\bigl(\bd1_{32}+\G^0\,\G^1\bigr)\,.
\qqq
Take the respective ($\cS$-)points $\,(\breve{\unl\th}{}_A^\a,\breve x{}_A^0,\breve x{}_A^1,\widehat x{}_{*\,A}^{\widehat a})\in\ep_{(1,1|16)}^{(\widehat x{}_{*\,A})}(D_{(1,1|16)}^{(\widehat x{}_{*\,A})})\,$ and compute their product
\qq
\bigl(\breve{\unl\th}{}_1^\a,\breve x{}_1^0,\breve x{}_1^1,\widehat x{}_{*\,1}^{\widehat a}\bigr)\cdot\bigl(\breve{\unl\th}{}_2^\a,\breve x{}_2^0,\breve x{}_2^1,\widehat x{}_{*\,2}^{\widehat a}\bigr)=\bigl(\breve{\unl\th}{}_1^\a+\breve{\unl\th}{}_2^\a,\breve x{}_1^0+\breve x{}_2^0-\tfrac{1}{2}\,\breve{\unl\th}{}_1\,\ovl\G{}^0\,\breve{\unl\th}{}_2,\breve x{}_1^1+\breve x{}_2^1-\tfrac{1}{2}\,\breve{\unl\th}{}_1\,\ovl\G{}^1\,\breve{\unl\th}{}_2,\widehat x{}_{*\,1}^{\widehat a}+\widehat x{}_{*\,2}^{\widehat a}\bigr)\,,\cr
\label{eq:alignproD}
\qqq
in $\,\bT(\cS)$,\ using the equality $\,\breve{\unl\th}{}_1\,\ovl\G{}^{\widehat a}\,\breve{\unl\th}{}_2=0\,$ along the way.\ Clearly,\ $\,\breve{\unl\th}{}_1+\breve{\unl\th}{}_2\in\im\,\sfP^{(1)}_{01}$,\ and so we conclude that 
\qq\nn
\txm\circ\bigl(\ep_{(1,1|16)}^{(\widehat x{}_{*\,1})}\x\ep_{(1,1|16)}^{(\widehat x{}_{*\,2})}\bigr)\bigl(D_{(1,1|16)}^{(\widehat x{}_{*\,1})}\x D_{(1,1|16)}^{(\widehat x{}_{*\,2})}\bigr)=\ep_{(1,1|16)}^{(\widehat x{}_{*\,1}+\widehat x{}_{*\,2})}\bigl(D_{(1,1|16)}^{(\widehat x{}_{*\,1}+\widehat x{}_{*\,2})}\bigr)\,.
\qqq
We thus obtain a partial localisation of the trivalent-fusion supermanifold:
\qq\nn
\D_{(1/1,1/1|16/16)}^{(\widehat x{}_{*\,1}/\widehat x{}_{*\,2})}\bigl(1/1(1),1/1|16/16(16)\bigr)\equiv D_{(1,1|16)}^{(\widehat x{}_{*\,1})}\x D_{(1,1|16)}^{(\widehat x{}_{*\,2})}\,.
\qqq
Of course,\ the result does \emph{not} depend on the choice of the embeddings within the class under consideration,\ {\it i.e.},\ we may vary,\ {\it e.g.},\ the second Lorentz index of the embedding as in
\qq\nn
&\ep_{(1,1|16);i}^{(\widehat x{}_*)}\ :\ D_{(1,1|16)}^{(\widehat x{}_*)}\emb\bT\,,\qquad\qquad i\in\ovl{1,9}\,,&\cr\cr\cr
&\ep_{(1,1|16);i}^{(\widehat x{}_*)}\bigl(\breve\th{}^{\unl\a},\breve x{}^0,\breve x{}^1\bigr)=\bigl(\breve{\unl\th}{}^\a,\breve x{}^0,\widehat x{}^1_*,\widehat x{}^2_*,\ldots,\widehat x{}^{i-1}_*,\breve x{}^1,\widehat x{}^{i+1}_*,\widehat x{}^{i+2}_*,\ldots,\widehat x{}^9_*\bigr)\,,&\cr\cr
&\breve{\unl\th}\in\im\,\sfP^{(1)}_{0i}\,,\qquad\sfP^{(1)}_{0i}\equiv\tfrac{1}{2}\,\bigl(\bd1_{32}+\G^0\,\G^i\bigr)&
\qqq
(or take linear combinations of the space-like directions {\it etc.}),\ as long as we keep the two embeddings under `fusion' superaligned as above.

\subsubsection{Spatially transverse bi-supersymmetric orbits}

Next,\ we take a look at two embeddings $\,\ep_{(1(i_A),1|16)}^{(\widehat x{}_{*\,A})}\ :\ D_{(1(i_A),1|16)}^{(\widehat x{}_{*\,A})}\emb\bT,\ A\in\{1,2\}\,$ that are {\bf spatially transverse} in the sense expressed by the identities
\qq\nn
&\ep_{(1(i_A),1|16)}^{(\widehat x{}_{*\,A})}\bigl(\breve\th{}^\a,\breve x{}^0,\breve x{}^1\bigr)=\bigl(\breve{\unl\th}{}^\a_A,\breve x{}^0,\widehat x{}^1_{*\,A},\widehat x{}^2_{*\,A},\ldots,\widehat x{}^{i_A-1}_{*\,A},\breve x{}^1,\widehat x{}^{i_A+1}_{*\,A},\widehat x{}^{i_A+2}_{*\,A},\ldots,\widehat x{}^9_{*\,A}\bigr)\,,&\cr\cr
&\breve{\unl\th}{}_A\in\im\,\sfP^{(1)}_{0i_A}\,,\qquad\sfP^{(1)}_{0i_A}\equiv\tfrac{1}{2}\,\bigl(\bd1_{32}+\G^0\,\G^{i_A}\bigr)\,,\qquad\qquad i_1\neq i_2\in\ovl{1,9}\,.&
\qqq
Without any loss of generality,\ we shall assume that $\,i_1=1=i_2-1\,$ from now onwards.\ Given arbitrary ($\cS$-)points $\,(\breve{\unl\th}{}^\a_1,\breve x{}^0_1,\breve x{}^1_1,\widehat x{}^2_{*\,1},\widehat x{}^3_{*\,1},\ldots,\widehat x{}^9_{*\,1})\in\ep_{(1(1),1|16)}^{(\widehat x{}_{*\,1})}(D_{(1(1),1|16)}^{(\widehat x{}_{*\,1})})\,$ and $\,(\breve{\unl\th}{}^\a_2,\breve x{}^0_2,\widehat x{}^1_{*\,2},\breve x{}^2_2,\widehat x{}^3_{*\,2},\widehat x{}^4_{*\,2},\ldots,\widehat x{}^9_{*\,2})\in\ep_{(1(2),1|16)}^{(\widehat x{}_{*\,2})}(D_{(1(2),1|16)}^{(\widehat x{}_{*\,2})})$,\ we read off from their product
\qq\nn
&&\bigl(\breve{\unl\th}{}^\a_1,\breve x{}^0_1,\breve x{}^1_1,\widehat x{}^2_{*\,1},\widehat x{}^3_{*\,1},\ldots,\widehat x{}^9_{*\,1}\bigr)\cdot\bigl(\breve{\unl\th}{}^\a_2,\breve x{}^0_2,\widehat x{}^1_{*\,2},\breve x{}^2_2,\widehat x{}^3_{*\,2},\widehat x{}^4_{*\,2},\ldots,\widehat x{}^9_{*\,2}\bigr)\cr\cr
&=&\bigl(\breve{\unl\th}{}^\a_1+\breve{\unl\th}{}^\a_2,\breve x{}^0_1+\breve x{}^0_2-\tfrac{1}{2}\,\breve{\unl\th}{}_1\,\ovl\G{}^0\,\breve{\unl\th}{}_2,\breve x{}^1_1+\widehat x{}^1_{*\,2}-\tfrac{1}{2}\,\breve{\unl\th}{}_1\,\ovl\G{}^1\,\breve{\unl\th}{}_2,\widehat x{}^2_{*\,1}+\breve x{}^2_2-\tfrac{1}{2}\,\breve{\unl\th}{}_1\,\ovl\G{}^2\,\breve{\unl\th}{}_2,\widehat x{}^3_{*\,1}+\widehat x{}^3_{*\,2}-\tfrac{1}{2}\,\breve{\unl\th}{}_1\,\ovl\G{}^3\,\breve{\unl\th}{}_2,\cr\cr
&&\widehat x{}^4_{*\,1}+\widehat x{}^4_{*\,2}-\tfrac{1}{2}\,\breve{\unl\th}{}_1\,\ovl\G{}^4\,\breve{\unl\th}{}_2,\ldots,\widehat x{}^9_{*\,1}+\widehat x{}^9_{*\,2}-\tfrac{1}{2}\,\breve{\unl\th}{}_1\,\ovl\G{}^9\,\breve{\unl\th}{}_2\bigr)
\qqq
that unless we impose the constraints
\qq\nn
\breve{\unl\th}{}_1\,\ovl\G{}^{\widehat{\widehat a}}\,\breve{\unl\th}{}_2\must 0\,,\qquad\qquad\widehat{\widehat a}\in\ovl{3,9}\,,
\qqq
the Gra\ss mann-even coordinates with the Lorentz indices $\,\widehat{\widehat a}\,$ all cease to be constant,\ but rather than being shifted arbitrarily in the body \emph{and} the soul as in the ($\cS$-point) definition of the orbit $\,\ep_{(1,1|16)}^{(\widehat x{}_*)}(D_{(1,1|16)}^{(\widehat x{}_*)})\equiv\bD_{1,1|16}\lact(0,\widehat x{}_*)$,\ they are shifted in the soul exclusively,\ so that there is no chance to recover an orbit extending in the direction(s) labelled by (the) $\,\widehat{\widehat a}$.\ The seemingly most natural way to ensure that the constraints are obeyed is to assume either of the following conditions to hold true:
\qq\nn
\left\{ \barr{l} \breve{\unl\th}_1\in\im\,\sfP^{(1)}_{01} \\ \breve{\unl\th}_2\in\im\,\sfP^{(1)}_{02}\cap\im\,\sfP^{(1)}_{01} \earr\right.\qquad\qquad\lor\qquad\qquad\left\{ \barr{l} \breve{\unl\th}_1\in\im\,\sfP^{(1)}_{01}\cap\im\,\sfP^{(1)}_{02} \\ \breve{\unl\th}_2\in\im\,\sfP^{(1)}_{02} \earr\right.\,.
\qqq
Indeed,\ we then obtain the desired equalities $\,\breve{\unl\th}{}_1\,\ovl\G{}^{\widehat{\widehat a}}\,\breve{\unl\th}{}_2=0=\breve{\unl\th}{}_1\,\ovl\G{}^{\widehat{\widehat a}}\,\breve{\unl\th}{}_2$,\ respectively.\ In order to convince oneself that the space of solutions to the above conditions is non-trivial,\ note that $\,[\sfP^{(1)}_{01},\sfP^{(1)}_{02}]=\tfrac{1}{2}\,\G^1\,\G^2\neq 0\,$ (in fact the right-hand side is even invertible),\ and so the decomposition of the 32-dimensional Majorana-spinor module of $\,\Cliff(\bR^{9,1})\,$ that we work with all along into the two 16-dimensional eigenspaces of $\,\sfP^{(1)}_{02}\,$ (associated with the eigenvalues $0$ and $1$,\ respectively) is \emph{different} from the one induced by $\,\sfP^{(1)}_{01}$.\ We shall denote the intersection of the two eigenspaces associated with the common eigenvalue $1$ as
\qq\nn
\D_{12}:=\im\,\sfP^{(1)}_{01}\cap\im\,\sfP^{(1)}_{02}\,,\qquad\qquad D_{12}\equiv\dim\,\D_{12}\neq 0\,.
\qqq
Once either of the above conditions is satisfied,\ further reductions take place.\ Indeed,\ for a pair $\,(\breve{\unl\th}{}_1,\breve{\unl\th}{}_2)\in\im\,\sfP^{(1)}_{01}\x\D_{12}$,\ we have $\,\breve{\unl\th}{}_1\,\ovl\G{}^2\,\breve{\unl\th}{}_2\equiv\bigl(\sfP^{(1)}_{01}\,\breve{\unl\th}{}_1\bigr)\,\ovl\G{}^2\,\breve{\unl\th}{}_2=\breve{\unl\th}{}_1\,\ovl\G{}^2\,\bigl(\bigl(\bd1_{32}-\sfP^{(1)}_{01}\bigr)\,\breve{\unl\th}{}_2\bigr)=0$,\ and,\ similarly,\ for a pair $\,(\breve{\unl\th}{}_1,\breve{\unl\th}{}_2)\in\D_{12}\x\im\,\sfP^{(1)}_{02}$,\ we find $\,\breve{\unl\th}{}_1\,\ovl\G{}^1\,\breve{\unl\th}{}_2\equiv\breve{\unl\th}{}_1\,\ovl\G{}^1\,\bigl(\sfP^{(1)}_{02}\,\breve{\unl\th}{}_2\bigr)=\bigl(\bigl(\bd1_{32}-\sfP^{(1)}_{02}\bigr)\,\breve{\unl\th}{}_1\bigr)\,\ovl\G{}^1\,\breve{\unl\th}{}_2=0$.\ Consequently,\ in the former case,\ we arrive at
\qq\nn
&&\bigl(\breve{\unl\th}{}^\a_1,\breve x{}^0_1,\breve x{}^1_1,\widehat x{}^2_{*\,1},\widehat x{}^3_{*\,1},\ldots,\widehat x{}^9_{*\,1}\bigr)\cdot\bigl(\breve{\unl\th}{}^\a_2,\breve x{}^0_2,\widehat x{}^1_{*\,2},\breve x{}^2_2,\widehat x{}^3_{*\,2},\widehat x{}^4_{*\,2},\ldots,\widehat x{}^9_{*\,2}\bigr)\cr\cr
&=&\bigl(\breve{\unl\th}{}^\a_1+\breve{\unl\th}{}^\a_2,\breve x{}^0_1+\breve x{}^0_2-\tfrac{1}{2}\,\breve{\unl\th}{}_1\,\ovl\G{}^0\,\breve{\unl\th}{}_2,\breve x{}^1_1+\widehat x{}^1_{*\,2}-\tfrac{1}{2}\,\breve{\unl\th}{}_1\,\ovl\G{}^1\,\breve{\unl\th}{}_2,\widehat x{}^2_{*\,1}+\breve x{}^2_2,\widehat x{}^3_{*\,1}+\widehat x{}^3_{*\,2},\widehat x{}^4_{*\,1}+\widehat x{}^4_{*\,2},\ldots,\widehat x{}^9_{*\,1}+\widehat x{}^9_{*\,2}\bigr)\,,
\qqq
whereas in the latter case,\ we get
\qq\nn
&&\bigl(\breve{\unl\th}{}^\a_1,\breve x{}^0_1,\breve x{}^1_1,\widehat x{}^2_{*\,1},\widehat x{}^3_{*\,1},\ldots,\widehat x{}^9_{*\,1}\bigr)\cdot\bigl(\breve{\unl\th}{}^\a_2,\breve x{}^0_2,\widehat x{}^1_{*\,2},\breve x{}^2_2,\widehat x{}^3_{*\,2},\widehat x{}^4_{*\,2},\ldots,\widehat x{}^9_{*\,2}\bigr)\cr\cr
&=&\bigl(\breve{\unl\th}{}^\a_1+\breve{\unl\th}{}^\a_2,\breve x{}^0_1+\breve x{}^0_2-\tfrac{1}{2}\,\breve{\unl\th}{}_1\,\ovl\G{}^0\,\breve{\unl\th}{}_2,\breve x{}^1_1+\widehat x{}^1_{*\,2},\widehat x{}^2_{*\,1}+\breve x{}^2_2-\tfrac{1}{2}\,\breve{\unl\th}{}_1\,\ovl\G{}^2\,\breve{\unl\th}{}_2,\widehat x{}^3_{*\,1}+\widehat x{}^3_{*\,2},\widehat x{}^4_{*\,1}+\widehat x{}^4_{*\,2},\ldots,\widehat x{}^9_{*\,1}+\widehat x{}^9_{*\,2}\bigr)\,.
\qqq
It is now clear that upon freezing the Gra\ss mann-even coordinates $\,\breve x{}^2_2\,$ and $\,\breve x{}^1_1$,\ respectively,\ at constant (pure-body) values:\ $\,\breve x{}^2_2\equiv\widehat x{}^2_{*\,2}\,$ and $\,\breve x{}^1_1\equiv\widehat x{}^1_{*\,1}$,\ we recover the embedded orbits (other scenarios are possible {\it a priori})
\qq\nn
&\ep_{(1(1),1|16)}^{(\widehat x{}_{*\,1}+\widehat x{}_{*\,2})}\bigl(D_{(1(1),1|16)}^{(\widehat x{}_{*\,1}+\widehat x{}_{*\,2})}\bigr)\quad{\rm with}\quad\widehat x{}_{*\,A}\equiv(0,0,\widehat x{}^2_{*\,A},\widehat x{}^3_{*\,A},\ldots,\widehat x{}^9_{*\,A})\quad{\rm for}\quad\bigl(\breve{\unl\th}{}_1,\breve{\unl\th}{}_2\bigr)\in\im\,\sfP^{(1)}_{01}\x\D_{12}\,,&\cr\cr
&\ep_{(1(2),1|16)}^{(\widehat x{}_{*\,1}+\widehat x{}_{*\,2})}\bigl(D_{(1(2),1|16)}^{(\widehat x{}_{*\,1}+\widehat x{}_{*\,2})}\bigr)\quad{\rm with}\quad\widehat x{}_{*\,A}\equiv(0,\widehat x{}^1_{*\,A},0,\widehat x{}^3_{*\,A},\widehat x{}^4_{*\,A},\ldots,\widehat x{}^9_{*\,A})\quad{\rm for}\quad\bigl(\breve{\unl\th}{}_1,\breve{\unl\th}{}_2\bigr)\in\D_{12}\x\im\,\sfP^{(1)}_{02}\,.&
\qqq

Denote the relevant sub-supermanifolds determined by the projection and the freeze-out as
\qq\nn
D_{(0(2),1|D_{12})}^{(\widehat x{}_*)}\equiv D_{(1(2),1|16)}^{(\widehat x{}_*)}\rstr_{\breve{\unl\th}\in\D_{12},\ \breve x{}^2\equiv\widehat x{}^2_*}\subset D_{(1(2),1|16)}^{(\widehat x{}_*)}
\qqq
and
\qq\nn
D_{(0(1),1|D_{12})}^{(\widehat x{}_*)}\equiv D_{(1(1),1|16)}^{(\widehat x{}_*)}\rstr_{\breve{\unl\th}\in\D_{12},\ \breve x{}^1\equiv\widehat x{}^1_*}\subset D_{(1(1),1|16)}^{(\widehat x{}_*)}
\qqq
to be able to partially localise the trivalent-fusion supermanifolds as
\qq\nn
\D_{(1/1,1/1|16/16)}^{(\widehat x{}_{*\,1}/\widehat x{}_{*\,2})}\bigl(1(1)/0(2),1/1|16/D_{12}(16)\bigr)\equiv D_{(1(1),1|16)}^{(\widehat x{}_{*\,1})}\x D_{(0(2),1|D_{12})}^{(\widehat x{}_{*\,2})}
\qqq
and
\qq\nn
\D_{(1/1,1/1|16/16)}^{(\widehat x{}_{*\,1}/\widehat x{}_{*\,2})}\bigl(0(1)/1(2),1/1|D_{12}/16(16)\bigr)\equiv D_{(0(1),1|D_{12})}^{(\widehat x{}_{*\,1})}\x D_{(1(2),1|16)}^{(\widehat x{}_{*\,2})}\,.
\qqq

\subsubsection{Superpoints of opposite chirality}

Finally,\ we consider the fusion of a pair of superpoints $\,\d_{(0|N)}^{(x_{*\,A})}\ :\ D_{(0|N)}^{(x_{*\,A})}\too\bT,\ A\in\{1,2\}$.\ The product of the respective ($\cS$-)points $\,(\unl\th{}_A,x_{*\,A})\in\d_{(0|N)}^{(x_{*\,A})}(D_{(0|N)}^{(x_{*\,A})})
\,$ reads 
\qq\nn
\bigl(\unl\th{}_1,x_{*\,1}\bigr)\cdot\bigl(\unl\th{}_2,x_{*\,2}\bigr)=\bigl(\unl\th{}_1+\unl\th{}_2,x_{*\,1}+x_{*\,2}-\tfrac{1}{2}\,\unl\th{}_1\,\ovl\G{}^\cdot\,\unl\th{}_2\bigr)\,,
\qqq
and so we are led to impose the constraints
\qq\label{eq:LRspinorFus}
\unl\th{}_1\,\ovl\G{}^a\,\unl\th{}_2\must 0\,,\qquad a\in\ovl{0,9}
\qqq
on the Gra\ss mann-odd coordinates of the superpoints under fusion in order to end up with another superpoint.\ There is a natural way to satisfy the above constraints in dimension $\,9+1\,$ while recovering the full Majorana-spinor module in the product,\ to wit,\ it suffices to take the two spinors $\,\unl\th{}_1\,$ and $\,\unl\th{}_2\,$ of opposite chiralities.\ Indeed,\ the canonical (volume) element $\,\G_{11}\equiv\G^0\cdot\G^1\cdot\cdots\cdot\G^9\in\Cliff(\bR^{9,1})\,$ belongs to the anti-centre,\ $\,\{\G_{11},\G^a\}=0,\ a\in\ovl{0,9}$,\ and has the properties $\,C\cdot\G_{11}=-\G_{11}^{\rm T}\cdot C\,$ and $\,\G_{11}^2=\bd1_{32}$,\ therefore eigenvectors $\,\th_\pm\in\im\,\sfP_\pm\equiv\ker\,\sfP_\mp\,$ of the associated projectors $\,\sfP_\pm=\tfrac{1}{2}\,(\bd1_{32}\pm\G_{11})\,$ satisfy $\,\unl\th{}_+\,\ovl\G{}^a\,\unl\th{}_-=0$,\ and any Majorana spinor can be represented as the sum of spinors of opposite chiralities.\ Denote the relevant sub-supermanifolds determined by the projection as
\qq\nn
\D_{(0|16\pm)}^{(x_*)}\equiv D_{(0|N)}^{(x_*)}\rstr_{\unl\th\in\im\,\sfP_\pm}\,.
\qqq
Altogether,\ we have the two natural ways to localise the trivalent-fusion supermanifolds:
\qq\nn
\D_{(0/0|32/32)}^{(x_{*\,1}/x_{*\,2})}\bigl(0/0(0)|16+/16-(32)\bigr)=D_{(0|16+)}^{(x_{*\,1})}\x D_{(0|16-)}^{(x_{*\,2})}
\qqq
and
\qq\nn
\D_{(0/0|32/32)}^{(x_{*\,1}/x_{*\,2})}\bigl(0/0(0)|16-/16+(32)\bigr)=D_{(0|16-)}^{(x_{*\,1})}\x D_{(0|16+)}^{(x_{*\,2})}\,.
\qqq

\subsection{Further constraints from the Defect-Junction Identity}

The hitherto (super)algebraic considerations lead to a partial localisation of the $\cG_{\rm GS}$-inter-bi-brane worldvolume within $\,\bT\x D_1\x D_2$.\ Below,\ we take into account further constraints,\ of a tensorial nature,\ encoded by the DJI,\ which can now,\ upon invoking \Reqref{eq:varhothGS},\ be rewritten as
\qq\label{eq:sDJI}\hspace{2cm}
\pr_2^*\iota_{\cF_{D_1,D_2}^{D_3}}^*\bigl(\bigl(\iota_{D_3}^{-1}\circ\txm_{D_1,D_2}\bigr)^*\om_{D_3}-\pr_1^*\om_{D_1}-\pr_2^*\om_{D_2}\bigr)=-\pr_2^*\iota_{\cF_{D_1,D_2}^{D_3}}^*\iota_{D_1,D_2}^*\varrho_{\rm GS}+\jmath_{D_1,D_2}^{D_3\,*}\sfd\vartheta_{\rm GS}\,,
\qqq
with
\qq\nn
\om_{D_3}=\left\{ \barr{cl} \om_{(1,1|16)} & \tx{if}\quad D_3=D_{(1,1|16)}^{(\widehat x{}_{*\,3})} \cr\cr 0 & \tx{if}\quad D_3=D_{(0|N_3)}^{(x_{*\,3})}\earr \right.\,.
\qqq
This is to be satisfied on the (sub-)supermanifold $\,\jmath_{D_1,D_2}^{D_3}\ :\ T_{D_1,D_2}^{D_3}\equiv\bT\x\cF_{D_1,D_2}^{D_3}\emb\bT^{\x 3}\,$ introduced in the opening paragraph of the present section.\ Just as in the un-graded setting,\ the logic of our approach is to use the above DJI to \emph{identify} the admissible sub-supermanifolds $\,\cF_{D_1,D_2}^{D_3}$.

\subsubsection{Superaligned bi-supersymmetric worldvolumes}\label{subsub:salign-ibb}

Taking into account the explicit form \eqref{eq:alignproD} of the product of ($\cS$-)points in a pair of superaligned $\cG_{\rm GS}$-branes,\ we readily calculate the left-hand side of \eqref{eq:sDJI} over $\,\bT\x\D_{(1/1,1/1|16/16)}^{(\widehat x{}_{*\,1}/\widehat x{}_{*\,2})}(1/1(1),1|16/16(16))\ni((\th,x),(\breve{\unl\th}{}_1^\a,\breve x{}_1^0,\breve x{}_1^1,\widehat x{}_{*\,1}^{\widehat a}),(\breve{\unl\th}{}_2^\a,\breve x{}_2^0,\breve x{}_2^1,\widehat x{}_{*\,2}^{\widehat a}))\equiv(m,\breve m{}_1,\breve m{}_2)\,$ as
\qq\nn
&&\om_{D_{(1,1|16)}^{(\widehat x{}_{*\,1}+\widehat x{}_{*\,2})}}\bigl(\breve m{}_1\cdot\breve m{}_2\bigr)-\om_{D_{(1,1|16)}^{(\widehat x{}_{*\,1})}}\bigl(\breve m{}_1\bigr)-\om_{D_{(1,1|16)}^{(\widehat x{}_{*\,2})}}\bigl(\breve m{}_2\bigr)\cr\cr
&=&\bigl(\breve{\unl\th}{}_1+2\breve{\unl\th}{}_2\bigr)\,\ovl\g{}_0\,\sfd\breve{\unl\th}{}_1\wedge\sfd\breve x{}_{+\,2}+\breve{\unl\th}{}_2\,\ovl\g{}_0\,\sfd\bigl(2\breve{\unl\th}{}_1+\breve{\unl\th}{}_2\bigr)\wedge\sfd\breve x{}_{+\,1}+\sfd\breve x{}_{+\,1}\wedge\sfd\breve x{}_{-\,2}-\sfd\breve x{}_{-\,1}\wedge\sfd\breve x{}_{+\,2}\,.
\qqq
Upon comparison with the right-hand side,
\qq\nn
&&\sfd\vartheta_{\rm GS}\bigl(m,\breve m{}_1,\breve m{}_2\bigr)-\varrho_{\rm GS}\bigl(\breve m{}_1,\breve m{}_2\bigr)\cr\cr
&=&\bigl(\breve{\unl\th}{}_1+2\breve{\unl\th}{}_2\bigr)\,\ovl\g{}_0\,\sfd\breve{\unl\th}{}_1\wedge\sfd\breve x{}_{+\,2}+\breve{\unl\th}{}_2\,\ovl\g{}_0\,\sfd\bigl(2\breve{\unl\th}{}_1+\breve{\unl\th}{}_2\bigr)\wedge\sfd\breve x{}_{+\,1}+\sfd\breve x{}_{+\,2}\wedge\sfd\breve x{}_{-\,1}+\sfd\breve x{}_{-\,2}\wedge\sfd\breve x{}_{+\,1}\cr\cr
&&+\sfd\bigl(-\tfrac{2}{3}\,\breve{\unl\th}{}_1\,\ovl\G{}_{\unl a}\,\breve{\unl\th}{}_2\,\bigl(3e^{\unl a}_{\rm L}(m)+2\breve{\unl\th}{}_1\,\ovl\G{}^{\unl a}\,\sfd\th+\breve{\unl\th}{}_2\,\ovl\G{}^{\unl a}\,\sfd\th\bigr)\bigr)\,,
\qqq
this yields the condition 
\qq\nn
3\sfd\breve x{}_{-\,2}\wedge\sfd\breve x{}_{+\,1}-\sfd\bigl(\breve{\unl\th}{}_1\,\ovl\G{}_{\unl a}\,\breve{\unl\th}{}_2\,\bigl(3e^{\unl a}_{\rm L}(m)+2\breve{\unl\th}{}_1\,\ovl\G{}^{\unl a}\,\sfd\th+\breve{\unl\th}{}_2\,\ovl\G{}^{\unl a}\,\sfd\th\bigr)\bigr)=0
\qqq
to be satisfied identically on the sub-supermanifold sought after.\ In view of the arbitrariness of $\,m$,\ we are thus led to the pair of constraints
\qq\label{eq:fusconstrsal}
\breve{\unl\th}{}_1\,\ovl\g{}_0\,\breve{\unl\th}{}_2\must 0\qquad\land\qquad\sfd\breve x{}_{+\,1}\wedge\sfd\breve x{}_{-\,2}\must 0\,.
\qqq
These can be solved naturally by taking a pair 
\qq\nn
[a_1:a_2:b_{1,2}]\in\bR\bP^2\setminus\{[0:0:1]\}\,,\qquad\qquad[c_1:c_2]\in\bR\bP^1
\qqq 
and setting
\qq\nn
a_1\,\breve x{}_{+\,1}+a_2\,\breve x{}_{-\,2}=b_{1,2}\qquad\land\qquad c_1\,\breve{\unl\th}{}_1+c_2\,\breve{\unl\th}{}_2=0\,,
\qqq
which defines a sub-supermanifold within $\,\D_{(1/1,1/1|16/16)}^{(\widehat x{}_{*\,1}/\widehat x{}_{*\,2})}(1/1(1),1|16/16(16))$.\ As we want to recover the entire $\bD_{1,1|16}$-orbit through multiplication of ($\cS$-)points of the latter,\ we have to assume
\qq\nn
[a_1:a_2:b_{1,2}]\in\bR\bP^2\setminus\{[0:0:1]\}\,,\qquad\qquad[c_1:c_2]\in\bR\bP^1\setminus\{[1:1]\}\,.
\qqq
We shall denote the sub-supermanifold thus determined as
\qq\nn
\cF_{D^{(\widehat x{}_{*\,1})}_{(1,1|16)},D^{(\widehat x{}_{*\,2})}_{(1,1|16)}}^{D^{(\widehat x{}_{*\,1}+\widehat x{}_{*\,2})}_{(1,1|16)}}\bigl([a_1:a_2:b_{1,2}]\,|\,[c_1:c_2]\bigr)\emb\D_{(1/1,1/1|16/16)}^{(\widehat x{}_{*\,1}/\widehat x{}_{*\,2})}\bigl(1/1(1),1|16/16(16)\bigr)\,,
\qqq
and define the ensuing inter-$\cG_{\rm GS}$-bi-brane worldvolume
\qq\nn
T_{D^{(\widehat x{}_{*\,1})}_{(1,1|16)},D^{(\widehat x{}_{*\,2})}_{(1,1|16)}}^{D^{(\widehat x{}_{*\,1}+\widehat x{}_{*\,2})}_{(1,1|16)}}\bigl([a_1:a_2:b_{1,2}]\,|\,[c_1:c_2]\bigr):=\bT\x\cF_{D^{(\widehat x{}_{*\,1})}_{(1,1|16)},D^{(\widehat x{}_{*\,2})}_{(1,1|16)}}^{D^{(\widehat x{}_{*\,1}+\widehat x{}_{*\,2})}_{(1,1|16)}}\bigl([a_1:a_2:b_{1,2}]\,|\,[c_1:c_2]\bigr)\,.
\qqq

\subsubsection{Spatially transverse bi-supersymmetric worldvolumes}\label{subsub:sptransfuswv}

The reasoning and calculations are completely analogous to those from the previous paragraph,\ and lead us to the constraints
\qq\nn
\breve{\unl\th}{}_1\,\ovl\g{}_0\,\breve{\unl\th}{}_2\must 0\qquad\land\qquad\sfd\bigl(\breve x{}^0_1+\breve x{}^1_1\bigr)\wedge\sfd\breve x{}^0_2\must 0
\qqq
for $\,(\breve{\unl\th}{}_1,\breve{\unl\th}{}_2)\in\im\,\sfP^{(1)}_{01}\x\D_{12}$,\ and to
\qq\nn
\breve{\unl\th}{}_1\,\ovl\g{}_0\,\breve{\unl\th}{}_2\must 0\qquad\land\qquad\sfd\breve x{}^0_1\wedge\sfd\bigl(\breve x{}^0_2-\breve x{}^2_2\bigr)\must 0
\qqq
for $\,(\breve{\unl\th}{}_1,\breve{\unl\th}{}_2)\in\D_{12}\x\im\,\sfP^{(1)}_{02}$,\ and as before,\ the solution can be taken in the form
\qq\nn
a_1\,\bigl(\breve x{}^0_1+\breve x{}^1_1\bigr)+a_2\,\breve x{}^0_2=b_{1,2}\qquad\land\qquad c_1\,\breve{\unl\th}{}_1+c_2\,\breve{\unl\th}{}_2=0
\qqq 
or 
\qq\nn
a_1\,\breve x^0_1+a_2\,\bigl(\breve x{}^0_2-\breve x{}^2_2\bigr)=b_{1,2}\qquad\land\qquad c_1\,\breve{\unl\th}{}_1+c_2\,\breve{\unl\th}{}_2=0\,,
\qqq
respectively,\ with the previous restrictions for the constant coefficients.\ There is one subtle difference with respect to the previous case,\ though.\ Indeed,\ whenever $\,[c_1:c_2]\neq[0:1]\,$ (resp.\ $\,[c_1:c_2]\neq[1:0]$),\ we have $\,(\breve{\unl\th}{}_1,\breve{\unl\th}{}_2)\in\D_{12}\x\D_{12}$,\ and so we do \emph{not} retrieve the entire orbit through multiplication as the odd components both come from the subspace $\,\D_{12}\,$ (and therefore so does their sum).\ Consequently,\ we must require
\qq\nn
[a_1:a_2:b_{1,2}]\in\bR\bP^2\setminus\{[0:0:1]\}\,,\qquad\qquad[c_1:c_2]\in\{[1:0],[0:1]\}\,,
\qqq
and,\ accordingly,\ arrive at a pair of sub-supermanifolds:
\qq\nn
\cF_{D^{(\widehat x{}_{*\,1})}_{(1(1),1|16)},D^{(\widehat x{}_{*\,2})}_{(1(2),1|16)}}^{D^{(\widehat x{}_{*\,1}+\widehat x{}_{*\,2})}_{(1(1),1|16)}}\bigl([a_1:a_2:b_{1,2}]\,|\,[0:1]\bigr)\emb\D_{(1/1,1/1|16/16)}^{(\widehat x{}_{*\,1}/\widehat x{}_{*\,2})}\bigl(1(1)/0(2),1/1|16/D_{12}(16)\bigr)
\qqq
and
\qq\nn
\cF_{D^{(\widehat x{}_{*\,1})}_{(1(1),1|16)},D^{(\widehat x{}_{*\,2})}_{(1(2),1|16)}}^{D^{(\widehat x{}_{*\,1}+\widehat x{}_{*\,2})}_{(1(2),1|16)}}\bigl([a_1:a_2:b_{1,2}]\,|\,[1:0]\bigr)\emb\D_{(1/1,1/1|16/16)}^{(\widehat x{}_{*\,1}/\widehat x{}_{*\,2})}\bigl(1(1)/0(2),1/1|16/D_{12}(16)\bigr)\,,
\qqq
respectively.\ The corresponding inter-$\cG_{\rm GS}$-bi-brane worldvolumes shall be denoted as
\qq\nn
T_{D^{(\widehat x{}_{*\,1})}_{(1(1),1|16)},D^{(\widehat x{}_{*\,2})}_{(1(2),1|16)}}^{D^{(\widehat x{}_{*\,1}+\widehat x{}_{*\,2})}_{(1(1),1|16)}}\bigl([a_1:a_2:b_{1,2}]\,|\,[c_1:c_2]\bigr):=\bT\x\cF_{D^{(\widehat x{}_{*\,1})}_{(1(1),1|16)},D^{(\widehat x{}_{*\,2})}_{(1(2),1|16)}}^{D^{(\widehat x{}_{*\,1}+\widehat x{}_{*\,2})}_{(1(1),1|16)}}\bigl([a_1:a_2:b_{1,2}]\,|\,[c_1:c_2]\bigr)\,,
\qqq
and
\qq\nn
T_{D^{(\widehat x{}_{*\,1})}_{(1(1),1|16)},D^{(\widehat x{}_{*\,2})}_{(1(2),1|16)}}^{D^{(\widehat x{}_{*\,1}+\widehat x{}_{*\,2})}_{(1(2),1|16)}}\bigl([a_1:a_2:b_{1,2}]\,|\,[c_1:c_2]\bigr):=\bT\x\cF_{D^{(\widehat x{}_{*\,1})}_{(1(1),1|16)},D^{(\widehat x{}_{*\,2})}_{(1(2),1|16)}}^{D^{(\widehat x{}_{*\,1}+\widehat x{}_{*\,2})}_{(1(2),1|16)}}\bigl([a_1:a_2:b_{1,2}]\,|\,[c_1:c_2]\bigr)\,,
\qqq
with $\,[c_1:c_2]=[0:1]\,$ (resp.\ $\,[c_1:c_2]=[1:0]$),\ respectively.

\subsubsection{Superpoint worldvolumes of opposite chirality}\label{subsub:spoint-ibb}

In the last example,\ the situation simplifies dramatically as $\,\om_{D^{(x_*)}_{(0|N)}}\equiv 0$,\ which means that we have to identify a sub-supermanifold of 
\qq\nn
&\bT\x\D_{(0/0|32/32)}^{(x_{*\,1}/x_{*\,2})}\bigl(0/0(0)|16\pm/16\mp(32)\bigr)\ni\bigl((\th,x),\bigl(\unl\th{}_1,x_{*\,1}\bigr),\bigl(\unl\th{}_2,x_{*\,2}\bigr)\bigr)\equiv\bigl(m,\breve m{}_1,\breve m{}_2\bigr)\,,&\cr\cr
&\bigl(\unl\th{}_1,\unl\th{}_2\bigr)\in\im\,\sfP_\pm\x\im\,\sfP_\mp&
\qqq
on which $\,\varrho_{\rm GS}(\breve m{}_1,\breve m{}_2)=\sfd\vartheta_{\rm GS}(m,\breve m{}_1,\breve m{}_2)$.\ Using \eqref{eq:LRspinorFus},\ we readily establish that the right-hand side vanishes,\ $\,\sfd\vartheta_{\rm GS}(m,\breve m{}_1,\breve m{}_2)=0$,\ and so it suffices to identify those $\,(\unl\th{}_1,\unl\th{}_2)\,$ for which so does the left-hand side which reduces to the expression $\,\varrho_{\rm GS}\bigl(\breve m{}_1,\breve m{}_2\bigr)=\tfrac{1}{2}\,\unl\th{}_1\,\ovl\G{}_a\,\unl\th{}_2\cdot\sfd\unl\th{}_1\wedge\ovl\G{}^a\,\sfd\unl\th{}_2-\tfrac{1}{2}\,\unl\th{}_1\,\ovl\G{}_a\,\sfd\unl\th{}_2\wedge\unl\th{}_2\,\ovl\G{}^a\,\sfd\unl\th{}_1\,$ upon invoking the Fierz identity \eqref{eq:Fierz},\ and hence vanishes in consequence of \Reqref{eq:LRspinorFus}.\ As a result of the above straightforward check,\ we obtain the supermanifolds 
\qq\nn
\cF_{D^{(x_{*\,1})}_{(0|16\pm)},D^{(x_{*\,2})}_{(0|16\mp)}}^{D^{(x_{*\,1}+x_{*\,2})}_{(0|32)}}\equiv\D_{(0/0|32/32)}^{(x_{*\,1}/x_{*\,2})}\bigl(0/0(0)|16\pm/16\mp(32)\bigr)\,,
\qqq
and the candidate corresponding inter-$\cG_{\rm GS}$-bi-brane worldvolumes
\qq\nn
T_{D^{(x_{*\,1})}_{(0|16\pm)},D^{(x_{*\,2})}_{(0|16\mp)}}^{D^{(x_{*\,1}+x_{*\,2})}_{(0|32)}}:=\bT\x\cF_{D^{(x_{*\,1})}_{(0|16\pm)},D^{(x_{*\,2})}_{(0|16\mp)}}^{D^{(x_{*\,1}+x_{*\,2})}_{(0|32)}}\,.
\qqq

\subsection{The fusion 2-isomorphism}

The previous sections have provided us with candidate $\cG_{\rm GS}$-inter-bi-brane worldvolumes $\,T_{D_1,D_2}^{D_3}\equiv\bT\x\cF_{D_1,D_2}^{D_3}\,$ on which the corresponding DJI's \eqref{eq:sDJI} hold identically.\ In general,\ an ultimate localisation of the worldvolume would result only form a construction of the relevant fusion 2-isomorphism \eqref{eq:gen-fus-2-iso-gen},\ however,\ in the (de Rham-)cohomologically trivial environment of our current interest,\ no extra constraints arise as we look for sub-supermanifolds on which not only the DJI obtains but also the underlying higher-supergeometric structure exists.\ Accordingly,\ we merely have to decipher the data of the latter over the $\cG_{\rm GS}$-inter-bi-brane worldvolumes found.\ Before we do that,\ let us note,\ in the passing,\ that in all the cases of elementary fusion considered above  the tensorial constraints lead to the nullification of the contribution from the trivial 1-isomorphism $\,\cJ_{-\vartheta_{\rm GS}}$,\ and so we descend to the faithful image $\,D_{1,2}^3\equiv\iota_{\cF_{1,2}^3}(\cF_{1,2}^3)\,$ of $\,\cF_{1,2}^3\,$ in $\,D_1\x D_2$,\ with the associated mapping $\,\mu_{1,2}^3\ :\ D_{1,2}^3\too D_3\,$ and end up with the standard (non-generalised) fusion 2-isomorphism
\qq\nn
\alxydim{@C=3cm@R=2cm}{\tx{$\barr{c} \iota_{1,2}^*\bigl(\pr_1^*\cG_{\rm GS}\ox\pr_2^*\cG_{\rm GS}\bigr) \\ \cong_1 \iota_{1,2}^*\pr_1^*\cG_{\rm GS}\ox\iota_{1,2}^*\pr_2^*\cG_{\rm GS} \\ \cong_2 \pr_1^*\iota_1^*\cG_{\rm GS}\ox\pr_2^*\iota_2^*\cG_{\rm GS}\earr$} \ar[d]_{\pr_1^*\cT_1\ox\pr_2^*\cT_2}
\ar[r]^{\iota_{1,2}^*\cM_{\rm GS}} & \tx{$\barr{c} \iota_{1,2}^*\bigl(\txm^*\cG_{\rm GS}\ox\cI_{\varrho_{\rm GS}}\bigr) \\ \cong_3 \mu_{1,2}^{3\,*}\iota_3^*\cG_{\rm GS}\ox\cI_{\iota_{1,2}^*\varrho_{\rm GS}}\earr$} \ar[d]^{\mu_{1,2}^{3\,*}\cT_3\ox\id} \\ \cI_{\pr_1^*\om_1+\pr_2^*\om_2} \ar@{=>}[ur]|{\check\phi[\cF_{1,2}^3]} \ar@{=}[r]_{\id} & \cI_{\mu_{1,2}^{3\,*}\om_3+\iota_{1,2}^*\varrho_{\rm GS}}}\,,
\qqq
written in the shorthand notation:\ $\,X_A\equiv X_{{D^{(x_{*\,A})}_{(p_A,q_A|N_A)}\,*}},\ X\in\{\cT,\om,\iota\},\ A\in\{1,2,3\}\,$ and $\,\iota_{1,2}\equiv\iota_{D_1,D_2}\rstr_{D_{1,2}^3}$.\ Its predecessor was encountered previously in the non-graded setting in \Rcite{Runkel:2009sp} ({\it cf.}\ also \Rcite{Runkel:2008gr}).\ After these general remarks,\ we may summarise the detailed results in

\bethe\label{thm:sfusion}
Over each of the supermanifolds $\,T_{D_1,D_2}^{D_3}\equiv\bT\x\cF_{D_1,D_2}^{D_3}\,$ identified in Secs.\,\ref{subsub:salign-ibb}, \ref{subsub:sptransfuswv} and \ref{subsub:spoint-ibb},\ there exists a corresponding fusion 2-isomorphism $\,\check\phi[\cF_{1,2}^3]$.
\ethe
\beroof
{\it Cf.}\ App.\,\ref{app:sfusion}.
\eroof

\noindent This concludes our preliminary analysis of the elementary fusion of the maximally supersymmetric bi-branes.

\part*{Conclusions \& Outlook}\label{part:CandO}

In the present paper,\ the general framework for the lagrangean description of $\si$-model-type loop dynamics with Wess--Zumino-type terms in the presence of self-intersecting worldsheet defects,\ originally formulated for un-graded target spaces in \Rcite{Runkel:2008gr},\ has been extended to the $\bZ/2\bZ$-graded geometric category,\ with the two-dimensional spacetime of the ensuing superfield theory with defects kept Gra\ss mann-even consistently with the functorial paradigm laid out in \Rcite{Freed:1999}.\ The framework has two layers:\ the worldsheet-defect layer and the target (super)space layer,\ with the structure of the former,\ essentially homological in its nature,\ reflected faithfully in the quasi-simplicial relations between bicategorial components of the latter.\ The relations become particularly rich and prominent in the case of the so-called topological defects amenable to homotopy manipulations -- this simple observation,\ implicit already in \Rcite{Runkel:2008gr},\ and soon afterwards formalised and emphasised in \Rcite{Suszek:2012ddg},\ has led us to consider a rather natural class of (super)targets,\ to wit,\ the simplicial target (super-)spaces,\ introduced in Def.\,\ref{def:simpl-target-sspace},\ in which target inter-bi-brane worldvolumes associated with defect junctions of arbitrary valence form simplicial hierarchies whose face maps furnish a realisation of the induction scheme for topological defects discussed in the papers cited.\ The (super)geometry is augmented with the data of a simplicial (1-)gerbe over it,\ altogether giving rise to the simplicial (super)string background of Def.\,\ref{simplssbckgrndcompl}.\ Simpliciality has subsequently been coherently combined with the principle of (super)symmetry (or,\ indeed,\ that of full reducibility with respect to the action of a (super)symmetry group),\ in its categorified form suggested by the findings of Refs.\,\cite{Fuchs:2007fw,Runkel:2008gr,Runkel:2009sp} and,\ in particular,\ those of Refs.\,\cite{Suszek:2011hg,Suszek:2012ddg},\ whereby the notion of a maximally supersymmetric background has been introduced in Sec.\,\ref{sec:susygrb}.\ The abstract concept has been illustrated amply on the `canonical' examples,\ both Gra\ss mann-even and $\bZ/2\bZ$-graded,\ of the Lie background of Sec.\,\ref{sec:smultcat},\ defined over the disjoint union of those full orbits of the simplicial action of the (super)symmetry group on the nerve of the right-regular action groupoid of a target Lie (super)group on itself which support the bi-categorial data of the maximally (super)symmetric WZW defect of Refs.\,\cite{Fuchs:2007fw,Runkel:2008gr,Runkel:2009sp} resp.\ its $\bZ/2\bZ$-graded counterpart.\ In the un-graded Lie-group setting,\ the combined simplicial/symmetry analysis has been used to demonstrate -- upon a physically justified loop-group enrichment -- uniqueness of the tensorial data (the curvature) of the maximally symmetric WZW brane and bi-brane of \Rcite{Fuchs:2007fw} ({\it cf.}\ Props.\,\ref{prop:maxym-curv-bdry-existuniq} and \ref{prop:maxym-curv-existuniq}),\ and yielded a novel proposal,\ explicited in Def.\,\ref{def:fus-2-iso},\ for the target geometry carrying the data of the so-called fusion 2-isomorphisms (or inter-bi-brane) of Refs.\,\cite{Runkel:2008gr,Runkel:2009sp},\ based on and partially corroborated by the findings of the detailed case study of \Rcite{Runkel:2009sp}.\ The latter derivation has revealed a deep structural relationship between the (un-graded) maximally symmetric WZW background and the three-dimensional Chern--Simons theory with timelike Wilson lines,\ consistent with and elaborating substantially that established in Refs.\,\cite{Witten:1988hf,Gawedzki:1989rr,Witten:1991mm,Gawedzki:1999bq, Gawedzki:2001rm} and the findings of \Rcite{Suszek:2012ddg} -- the relationship is at the core of the fundamental Thm.\,\ref{thm:ker-Om}.\ Furthermore,\ it has motivated a conjectural identification,\ in Sec.\,\ref{sub:fus-2iso},\ of the (higher-)geometric structure that is anticipated to encode the Verlinde fusion rules and (a part of) the Moore--Seiberg data of the (chiral) bulk WZW $\si$-model,\ in an extension of and in conformity with the partial results of Refs.\,\cite{Runkel:2008gr,Runkel:2009sp}.\ In the supergeometric setting of the Green--Schwarz super-$\si$-model of the superstring in $\,{\rm sMink}(d,1|D_{d,1})$,\ on the other hand,\ the analysis has produced three classes of maximally supersymmetric (bi-)branes (Sec.\,\ref{sub:sbib}) and partial results on their (elementary) fusion (Sec.\,\ref{sec:sibib}).\ In full analogy with the un-graded setting,\ central r\^ole in the analysis has been played by the categorification of the binary operation on the (supersymmetry) Lie supergroup known as the multiplicative structure ({\it cf.}\ \Rcite{Carey:2004xt}).\ The latter notion has been generalised suitably in Sec.\,\ref{sub:homocat},\ and subsequently an instantiation of such a generalised multiplicative structure,\ manifestly compatible with the supersymmetry present,\ has been identified on the Green--Schwarz super-1-gerbe of \Rcite{Suszek:2017xlw} in Sec.\,\ref{sec:multGSs1g}.\ These results are,\ to the best of the Author's knowledge,\ the first examples of maximally (super)symmetric bi-branes and inter-bi-branes studied in the literature.\\

The work reported in the present paper leaves several important open questions:\ The first of them concerns the validity of the WZW conjectures of Sec.\,\ref{sub:fus-2iso}.\ It is worth pointing out that Conjecture \ref{conj:2-iso-vs-Verlinde},\ preliminarily verified in the case of $\,\txG={\rm SU}(2)\,$ (with simple fusion rules) in \Rcite{Runkel:2009sp},\ could readily be tested in the largely tractable yet nontrivial case of $\,\txG={\rm SU}(3)$,\ {\it cf.},\ {\it e.g.},\ \Rcite{Hayashi:1999},\ and Conjecture \ref{conj:indu} together with the field-theoretic content of Def.\,\ref{def:maxym-WZW-ibi-mat} could be investigated for the simplest non-abelian Lie group $\,\txG={\rm SU}(2)$.\ The second issue that ought to be explored at length is the intricate structural relationship between the WZW fusion 2-isomorphisms and the Chern--Simons theory -- here,\ one might start by performing a symplectic analysis of the WZW model in the presence of a self-intersecting maximally symmetric defect with reference to the known canonical description of the topological field theory (in close analogy with the original works by Gaw\c{e}dzki) and subsequently analyse the transgression of the higher-categorial data carried by the defect to the phase space of the $\si$-model under consideration (in keeping with Refs.\,\cite{Suszek:2011hg,Suszek:2012ddg}).\ In the $\bZ/2\bZ$-graded setting,\ the most pending question is that of the status of $\k$-symmetry,\ as understood -- also in the higher-geometric terms -- in the dual Hughes--Polchinsky formulation of the super-$\si$-model of \Rcite{Hughes:1986dn} in Refs.\,\cite{Suszek:2020xcu,Suszek:2020rev,Suszek:2021hjh},\ in the presence of the maximally supersymmetric (bi-)branes studied in the present paper.\ The answer,\ in conjunction with a positive outcome of the verification of the WZW conjectures,\ might pave the way towards the extraction of non-perturbative data of the super-WZW-model of Green and Schwarz from the higher-geometric analysis performed.\ Of course,\ a more systematic approach to the construction of maximally supersymmetric branes and bi-branes would most certainly also be welcome,\ and it is clear that it might go {\it via} analysis of Lie-superalgebraic skeleta of would-be travialisations of $\,\cG_{\rm GS}\,$ resp.\ $\cG_{\rm GS}$-bimodules akin to those considered in \Rcite{Suszek:2021hjh}.\ Another important direction in which the higher-supergeometric research initiated in the present paper might be developed is the study of maximally supersymmetric defects (boundary and non-boundary) in the curved supertargets reviewed recently in Refs.\,\cite{Suszek:2020xcu},\ with emphasis on the {\.I}n{\"o}n{\"u}--Wigner contractibility of Refs.\,\cite{Suszek:2018bvx,Suszek:2018ugf} as an additional organising principle (in the spirit of Fronsdal).\ Last but not least,\ and more fundamentally,\ it seems reasonable to think of the work reported hereabove as a natural springboard for a systematic investigation of cross-defect loop dynamics over nerves of (Lie) (super)groupoids of a more general kind than the ones encountered.\ We are hoping to return to all these issues in a future study.

\part*{Appendices}

\appendix

\stoptocwriting

\section{Conventions on mappings and pullbacks}\label{app:convs}

Let $\,\cM\,$ and $\,\cN_A,\ A\in\{1,2\}\,$ be supermanifolds,\ and let $\,f_A\ :\ \cM\too\cN_A\,$ be supermanifold mappings (morphisms).\ We then define a supermanifold mapping
\qq\nn
(f_1,f_2)\ :\ \cM\too\cN_1\x\cN_2
\qqq
as the unique one fixed by the conditions
\qq\nn
\pr_A\circ(f_1,f_2)=f_A\,,\qquad A\in\{1,2\}\,,
\qqq
written in terms of the canonical projections $\,\pr_A\ :\ \cN_1\x\cN_2\too\cN_A$.

Similarly,\ given supermanifolds $\,\cM_A,\ A\in\{1,2\}\,$ and $\,\cN_B,\ B\in\{1,2\}\,$ alongside supermanifold mappings $\,F_C\ :\ \cM_C\too\cN_C,\ C\in\{1,2\}$,\ we define a supermanifold mapping
\qq\nn
F_1\x F_2\ :\ \cM_1\x\cM_2\too\cN_1\x\cN_2
\qqq
as the unique one fixed by the conditions
\qq\nn
\pr_A\circ(F_1\x F_2)=F_A\circ\pr_A\,,\qquad A\in\{1,2\}\,.
\qqq

Let,\ next,\ $\,\cM_A,\ A\in\{1,2\}\,$ be supermanifolds,\ and let 
\qq\nn
\pi_{\sfY\cM_2}\ :\ \sfY\cM_2\too\cM_2
\qqq 
be a surjective submersion.\ Given a map
\qq\nn
f\ :\ \cM_1\too\cM_2\,,
\qqq
we shall employ,\ unless expressly stated otherwise,\ the model of the pullback of $\,\sfY\cM_2\,$ over $\,\cM_1\,$ along $\,f\,$ provided by the fibred product 
\qq\nn
f^*\sfY\cM_2=\cM_1{}_{f}\hspace{-1pt}\x_{\pi_{\sfY\cM_2}}\hspace{-1pt}\sfY\cM_2
\qqq
that is defined by the commutative diagram
\qq\nn
\alxydim{@C=2.5cm@R=1.5cm}{ \cM_1{}_{f}\hspace{-1pt}\x_{\pi_{\sfY\cM_2}}\hspace{-1pt}\sfY\cM_2 \ar[r]^{\qquad\widehat f\equiv\pr_2} \ar[d]_{\pi_{f^*\sfY\cM_2}\equiv\pr_1} & \sfY\cM_2 \ar[d]^{\pi_{\sfY\cM_2}} \\ \cM_1 \ar[r]_{f} & \cM_2 }\,.
\qqq

Furthermore,\ given a supermanifold $\,\cM\,$ and two surjective submersions $\,\pi_{\sfY_A\cM}\ :\ \sfY_A\cM\too\cM,\ A\in\{1,2\}\,$ with $\,\cM\,$ as the common base,\ we shall write the pullback of one of them along the other one as
\qq\nn
\sfY_1\cM\x_\cM\hspace{-1pt}\sfY_2\cM=\pi_{\sfY_1\cM}^*\sfY_2\cM\,.
\qqq
When fibring a given surjective submersion with itself,\ we shall use the by now standard notation 
\qq\nn
\sfY^{[n]}\cM\equiv\underbrace{\sfY\cM\x_\cM\sfY\cM\x_\cM\cdots\x_\cM\sfY\cM}_{n\ {\rm times}}\,.
\qqq
~\smallskip 

In the setting of (simplicial) Lie (super)group (super)geometry of $\,\sfN_\bullet\txG^{\rm op}$,\ we also use the following shorthand notation:\ Let $\,\txG\,$ be a Lie supergroup and fix $\,n\in\bN_{>1}$,\ writing 
\qq\nn
&\txm_1\equiv\id_\txG\,,\qquad\qquad\txm_2\equiv\txm\,,&\cr\cr
&\txm_k\equiv\txm\circ(\txm\x\id_\txG)\circ\cdots\circ(\txm\x\id_{\txG^{\x k-3}})\circ(\txm\x\id_{\txG^{\x k-2}})\ :\ \txG^{\x k}\too\txG\,,\qquad 2<k\leq n\,.&
\qqq
Given any subset $\,\{n_i\}_{i\in\ovl{1,m+1}}\,$ of integers satisfying the relations $\,1\leq n_1<n_2<\ldots<n_{m+1}\leq n+1$,\ we define supermanifold morphisms
\qq\nn
&&\txm^{(n)}_{n_1 n_1+1 \ldots n_2-1,n_2 n_2+1\ldots n_3-1,\ldots,n_m n_m+1 \ldots n_{m+1}-1}\cr\cr
&\equiv&\bigl(\txm_{n_2-n_1}\circ\pr^{(n)}_{n_1,n_1+1,\ldots,n_2-1},\txm_{n_3-n_2}\circ\pr^{(n)}_{n_2,n_2+1,\ldots,n_3-1},\ldots,\txm_{n_{m+1}-n_m}\circ\pr^{(n)}_{n_m n_m+1 \ldots n_{m+1}-1}\bigr)\cr\cr
&:&\ \txG^{\x n}\too\txG^{\x m}
\qqq
and write, for any (super-)differential-geometric object $\,\cO\,$ defined over $\,\txG^{\x m}\,$ for which the pullback along the above morphism is well-defined,
\qq\nn
\cO^{(n)}_{n_1 n_1+1 \ldots n_2-1,n_2 n_2+1\ldots n_3-1,\ldots,n_m n_m+1 \ldots n_{m+1}-1}\equiv\txm^{(n)\,*}_{n_1 n_1+1 \ldots n_2-1,n_2 n_2+1\ldots n_3-1,\ldots,n_m n_m+1 \ldots n_{m+1}-1}\cO\,.
\qqq

\section{A proof of Proposition \ref{prop:simplicial-target}}\label{app:proof-simpl}

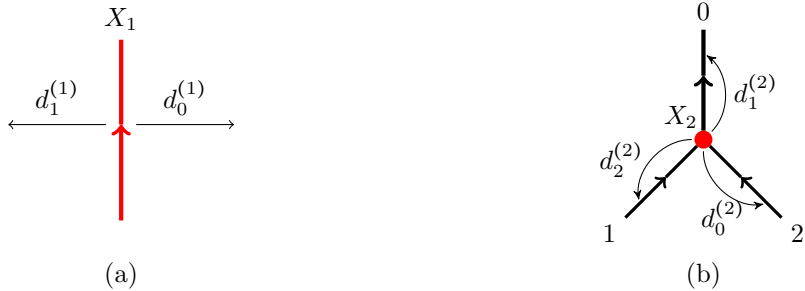
\begin{figure}[hb!]
\begin{tikzpicture}
\node (B) at (0,-1.2) {};
\node (T) at (0,1.6) {$X_1$};
\node at (0,-1.8) {(a)};

\draw[->,red,ultra thick] (B) -- (0,0.2);
\draw[-,red,ultra thick] (0,0.2) -- (T);
\draw[->] (0.2,0.2) to node [above] {$d^{(1)}_0$} (1.5,0.2);
\draw[->] (-0.2,0.2) to node [above] {$d^{(1)}_1$} (-1.5,0.2);
\end{tikzpicture}
\hspace{4.5cm}
\begin{tikzpicture}
[junction/.style={circle,draw=red!100,fill=red!100,thick,inner sep=0pt,minimum size=6pt}]

\node[junction] (vertex) at (0,0) {};
\node (X2) at (-0.3,0.3) {$X_2$};
\node (L1) at (-1.25,-1.25) {1};
\node (L2) at (1.25,-1.25) {2};
\node (L12) at (0,1.7) {0};
\node at (0,-1.8) {(b)};

\draw[->,very thick] (L1) -- (-0.5,-0.5); 
\draw[-,very thick] (-0.5,-0.5) -- (vertex.south west);
\draw[->,ultra thick] (vertex.north) -- (0,0.85);
\draw[-,ultra thick] (0,0.85) -- (L12);
\draw[->,very thick] (L2) -- (0.5,-0.5);
\draw[-,very thick] (0.5,-0.5) -- (vertex.south east);

\draw[->,bend right=50,shorten <=1pt,shorten >=1.5pt,>=stealth'] (vertex.north east) to node [right] {$d^{(2)}_1$} (0,1.15);
\draw[->,bend right=50,shorten <=1pt,shorten >=1.5pt,>=stealth'] (vertex.south) to node [below] {$d^{(2)}_0$} (0.85,-0.85);
\draw[->,bend right=50,shorten <=1pt,shorten >=1.5pt,>=stealth'] (vertex.west) to node [left] {$d^{(2)}_2$} (-0.85,-0.85);
\end{tikzpicture}
\caption{A pictorial representation of $\,X_1\,$ (a) and $\,X_2\,$ (b),\ with the respective face maps.} \label{fig:lineandtrinion}
\end{figure}

\begin{figure}[hb!]
\begin{tikzpicture}
[junction/.style={circle,draw=red!100,fill=red!100,thick,inner sep=0pt,minimum size=6pt},
fusion/.style={circle,draw=black!100,fill=black!100,thick,inner sep=0pt,minimum size=3pt}]

\node[junction] (vertex) at (0,0) {};
\node (Xn) at (-0.3,0.3) {$X_n$};
\node (L1) at (-2.0,-1.5) {1};
\node (L2) at (-1.0,-1.5) {2};
\node (Ln) at (2.0,-1.5) {n};
\node (L12) at (0,1.7) {0};
\node at (0.25,-0.75) {$\cdots$};

\draw[very thick] (L1) -- (vertex.south west);
\draw[very thick] (L2) -- (vertex.south west);
\draw[ultra thick] (vertex.north) -- (L12);
\draw[very thick] (Ln) -- (vertex.south east);

\node[fusion] (vertex03) at (-5.6,-1.3) {};
\node[junction] (vertex0) at (-5,-1.5) {};
\node (Xn0) at (-4.9,-1.2) {$X_{n-1}$};
\node (L10) at (-7.0,-3.0) {1};
\node (L20) at (-6.0,-3.0) {2};
\node (L30) at (-5.25,-3.0) {3};
\node (Ln0) at (-3.0,-3.0) {n};
\node (L120) at (-5.6,0.2) {0};
\node at (-4.65,-2.25) {$\cdots$};

\draw[very thick] (L10) -- (vertex03.south west);
\draw[very thick] (L20) -- (vertex0.south west);
\draw[very thick] (L30) -- (vertex0.south);
\draw[ultra thick] (vertex03.north) -- (L120);
\draw[very thick] (Ln0) -- (vertex0.south east);
\draw[densely dotted,very thick] (vertex0) -- (vertex03);

\node[fusion] (vertexn3) at (5.6,-1.3) {};
\node[junction] (vertexn) at (5,-1.5) {};
\node (Xnn) at (4.9,-1.2) {$X_{n-1}$};
\node (L1n) at (3.0,-3.0) {1};
\node (L2n) at (4.0,-3.0) {2};
\node (Lnn) at (7.0,-3.0) {n};
\node (Ln1n) at (6.0,-3.0) {n-1};
\node (L12n) at (5.6,0.2) {0};
\node at (5,-2.25) {$\cdots$};

\draw[very thick] (L1n) -- (vertexn.south west);
\draw[very thick] (L2n) -- (vertexn.south west);
\draw[ultra thick] (vertexn3.north) -- (L12n);
\draw[very thick] (Lnn) -- (vertexn3.south east);
\draw[very thick] (Ln1n) -- (vertexn.south east);
\draw[densely dotted,very thick] (vertexn) -- (vertexn3);

\node[fusion] (vertexk3) at (0,-5.0) {};
\node[junction] (vertexk) at (0,-4.2) {};
\node (Xnk) at (-0.45,-3.9) {$X_{n-1}$};
\node (L1k) at (-3.5,-5.7) {1};
\node (L2k) at (-2.5,-5.7) {2};
\node (Lkm1k) at (-1,-5.7) {k-1};
\node (Lkk) at (-0.35,-6.5) {k};
\node (Lk1k) at (0.35,-6.5) {k+1};
\node (Lk2k) at (1.0,-5.7) {k+2};
\node (Ln1k) at (2.5,-5.7) {n-1};
\node (Lnk) at (3.5,-5.7) {n};
\node (L12k) at (0,-2.5) {0};
\node at (-0.75,-4.95) {$\cdots$};
\node at (0.75,-4.95) {$\cdots$};

\draw[very thick] (L1k) -- (vertexk.south west);
\draw[very thick] (L2k) -- (vertexk.south west);
\draw[very thick] (Lkm1k) -- (vertexk.south west);
\draw[very thick] (Lkk) -- (vertexk3.south west);
\draw[very thick] (Lk1k) -- (vertexk3.south east);
\draw[ultra thick] (vertexk.north) -- (L12k);
\draw[very thick] (Lk2k) -- (vertexk.south east);
\draw[very thick] (Lnk) -- (vertexk.south east);
\draw[very thick] (Ln1k) -- (vertexk.south east);
\draw[densely dotted,very thick] (vertexk3) -- (vertexk);

\draw[->,bend left=55,shorten <=2.5pt,shorten >=2.5pt,>=stealth'] (vertex.south) to node [right] {$d^{(n)}_k$} (vertexk.north east);
\draw[->,bend right=10,shorten <=2.5pt,shorten >=2.5pt,>=stealth'] (vertex.east) to node [above] {$d^{(n)}_n$} (vertexn.west);
\draw[->,bend left=10,shorten <=2.5pt,shorten >=2.5pt,>=stealth'] (vertex.west) to node [above] {$d^{(n)}_0$} (vertex0.east);
\end{tikzpicture}
\caption{A pictorial representation of the face maps $\,d^{(n)}_i\ :\ X_n\too X_{n-1},\ i\in\ovl{0,n}$.} \label{fig:higherface}
\end{figure}
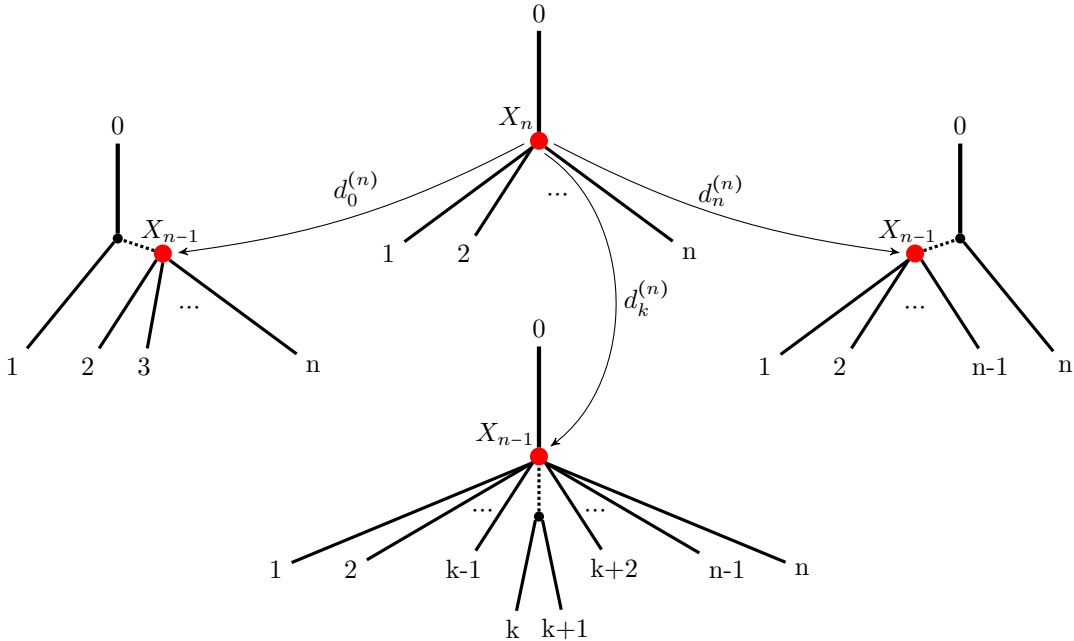

We prove the first part of the thesis by induction with respect to the value $\,n\geq 3\,$ of the family index.\ In so doing,\ we employ a convenient pictorial representation of the face maps suggested by the concrete realisation of the simplicial identities for the face maps in the nerve of a small category and by the field-theoretic intuition -- the representation helps to keep track of the admissible applications of the face maps:\ In it,\ $\,X_1\,$ is drawn as an oriented line and $\,X_n,\ n>1\,$ is drawn as a junction (graph) at which $n$ incoming lines and a single outgoing line come together.\ The face maps $\,d^{(1)}_\cdot\,$ assign to the line regions\footnote{The regions carry labels which are kept implicit in order to lighten the presentation.} to the left ($d^{(1)}_1$) and right ($d^{(1)}_0$) of the line,\ in the direction of its orientation,\ and the face maps $\,d^{(2)}_\cdot\,$ assign to the junction the lines that converge at it,\ in an appropriate order,\ {\it cf.}\ Fig.\,\ref{fig:lineandtrinion}.\ 
In fact,\ we may further assume that all lines are oriented `upwards',\ which enables us to drop the arrows in all graphs.\ Now,\ the application of a higher face map $\,d^{(n)}_i\,$ to a junction of valence $\,(n,1)\,$ representing $\,X_n\,$ yields a junction of valence $\,(n-1,1)\,$ at which $n-1$ lines of the original graph meet a single `virtual' line leaving an elementary junction (of valence $\,(2,1)$) at which the pair of adjacent lines of the original graph with indices $\,i\,$ and $\,i+1\,$ come together with the `virtual' line.\ The ensuing {\bf elementary resolutions} of the graph of valence $\,n+1$,' each of which produces a graph of valence $\,n\,$ and a trinion attached to it,\ are depicted in Fig.\,\ref{fig:higherface}.\ They can be used recursively to the subgraphs obtained by their application (in what we shall refer to as the {\bf recursive ternary resolution}),\ whereby descent to to an arbitrarily low level $\,l\in\ovl{0,n-1}\,$ can be attained.\ 
While there is no reason to view the latter geometric picture,\ inspired by the admissible operations on defects in a 2d CFT,\ as a reflection of an actual existence of some binary operation on `elements' of the `squares' of $\,X_1\,$ `embedded' in $\,X_n$,\ the adequacy of the representation adopted (beyond a trivial count of the \emph{number} of face maps) is attested by the fact that it does provide us with the correct identification of the \emph{equivalent} concatenations of the face maps $\,d^{(k)}_i,\ i\in\ovl{0,k},\ k\in\ovl{2,n}\,$ sending $\,X_n\,$ all the way to $\,X_1$.\ These we readily find below.

\begin{figure}[h!]
\begin{tikzpicture}
[junction/.style={circle,draw=red!100,fill=red!100,thick,inner sep=0pt,minimum size=6pt},
fusion/.style={circle,draw=black!100,fill=black!100,thick,inner sep=0pt,minimum size=3pt}]

\node[junction] (vertex) at (0,0) {};
\node (X3) at (-0.4,0.2) {$X_3$};
\node (L1) at (-1.5,-1.25) {1};
\node (L2) at (0,-1.25) {2};
\node (L3) at (1.5,-1.25) {3};
\node (L12) at (0,1.3) {0};

\draw[ultra thick] (vertex.north) -- (L12);
\draw[very thick] (L1) -- (vertex.south west);
\draw[ultra thick] (L2) -- (vertex.south);
\draw[very thick] (L3) -- (vertex.south east);

\node[fusion] (vertex03) at (-5.1,2.5) {};
\node[junction] (vertex0) at (-4.5,2.25) {};
\node (X20) at (-4.55,2.55) {$X_2$};
\node (L10) at (-6.6,1.35) {1};
\node (L20) at (-4.5,1.0) {2};
\node (L30) at (-3,1.0) {3};
\node (L120) at (-5.1,3.8) {0};

\draw[ultra thick] (vertex03.north) -- (L120);
\draw[very thick] (L10) -- (vertex03.south west);
\draw[ultra thick] (L20) -- (vertex0.south);
\draw[very thick] (L30) -- (vertex0.south east);
\draw[densely dotted,very thick] (vertex03) -- (vertex0);

\node[fusion] (vertex13) at (-5.1,-2.55) {};
\node[junction] (vertex1) at (-4.5,-2.3) {};
\node (X21) at (-4.8,-2.0) {$X_2$};
\node (L11) at (-6.6,-3.8) {1};
\node (L21) at (-5.1,-3.8) {2};
\node (L31) at (-3,-3.55) {3};
\node (L121) at (-4.5,-1.0) {0};

\draw[ultra thick] (vertex1.north) -- (L121);
\draw[very thick] (L11) -- (vertex13.south west);
\draw[ultra thick] (L21) -- (vertex13.south);
\draw[very thick] (L31) -- (vertex1.south east);
\draw[densely dotted,very thick] (vertex1) -- (vertex13);

\node[fusion] (vertex23) at (5.1,-2.5) {};
\node[junction] (vertex2) at (4.5,-2.3) {};
\node (X22) at (4.8,-2.0) {$X_2$};
\node (L12) at (3,-3.55) {1};
\node (L22) at (5.1,-3.8) {2};
\node (L32) at (6.6,-3.8) {3};
\node (L122) at (4.5,-1.0) {0};

\draw[ultra thick] (vertex2.north) -- (L122);
\draw[very thick] (L12) -- (vertex2.south west);
\draw[ultra thick] (L22) -- (vertex23.south);
\draw[very thick] (L32) -- (vertex23.south east);
\draw[densely dotted,very thick] (vertex23) -- (vertex2);

\node[fusion] (vertex33) at (5.1,2.5) {};
\node[junction] (vertex3) at (4.5,2.25) {};
\node (X23) at (4.55,2.55) {$X_2$};
\node (L13) at (3.0,1.0) {1};
\node (L23) at (4.5,1.0) {2};
\node (L33) at (6.6,1.25) {3};
\node (L123) at (5.1,3.8) {0};

\draw[ultra thick] (vertex33.north) -- (L123);
\draw[very thick] (L13) -- (vertex3.south west);
\draw[ultra thick] (L23) -- (vertex3.south);
\draw[very thick] (L33) -- (vertex33.south east);
\draw[densely dotted,very thick] (vertex33) -- (vertex3);

\draw[->,bend right=45,shorten <=2.5pt,shorten >=2.5pt,>=stealth'] (vertex.north west) to node [above] {$d^{(3)}_0$} (vertex0.east);
\draw[->,bend left=45,shorten <=2.5pt,shorten >=2.5pt,>=stealth'] (vertex.north east) to node [above] {$d^{(3)}_3$} (vertex3.west);
\draw[->,bend right=25,shorten <=2.5pt,shorten >=2.5pt,>=stealth'] (vertex.west) to node [above] {$d^{(3)}_1$} (vertex1.north east);
\draw[->,bend left=25,shorten <=2.5pt,shorten >=2.5pt,>=stealth'] (vertex.east) to node [above] {$d^{(3)}_2$} (vertex2.north west);

\node (L3t) at (-6.6,0.63) {};
\node (L3b) at (-5.1,-0.63) {3};

\draw[very thick] (L3b) -- (L3t);

\node (L0t) at (0,-2.5) {0};
\node (L0b) at (0,-3.8) {};

\draw[ultra thick] (L0b) -- (L0t);

\node (L1t) at (6.6,0.63) {};
\node (L1b) at (5.1,-0.63) {1};

\draw[very thick] (L1b) -- (L1t);

\node (L2t) at (0,3.8) {};
\node (L2b) at (0,2.55) {2};

\draw[ultra thick] (L2b) -- (L2t);

\draw[->,bend right=40,shorten <=2.5pt,shorten >=2.5pt,>=stealth'] (vertex0.south west) to node [right] {$d^{(2)}_0$} (-5.85,0);
\draw[->,bend left=45,shorten <=2.5pt,shorten >=2.5pt,>=stealth'] (vertex1.west) to node [left] {$d^{(2)}_0$} (-5.85,0);
\draw[->,bend left=30,shorten <=2.5pt,shorten >=2.5pt,>=stealth'] (vertex1.east) to node [below] {$d^{(2)}_1$} (0,-3.25);
\draw[->,bend right=30,shorten <=2.5pt,shorten >=2.5pt,>=stealth'] (vertex2.west) to node [below] {$d^{(2)}_1$} (0,-3.25);
\draw[->,bend right=45,shorten <=2.5pt,shorten >=2.5pt,>=stealth'] (vertex2.east) to node [right] {$d^{(2)}_2$} (5.85,0);
\draw[->,bend left=40,shorten <=2.5pt,shorten >=2.5pt,>=stealth'] (vertex3.south east) to node [left] {$d^{(2)}_2$} (5.85,0);
\draw[->,bend right=30,shorten <=2.5pt,shorten >=2.5pt,>=stealth'] (vertex3.north west) to node [above] {$d^{(2)}_0$} (0,3.25);
\draw[->,bend left=30,shorten <=2.5pt,shorten >=2.5pt,>=stealth'] (vertex0.north east) to node [above] {$d^{(2)}_2$} (0,3.25);
\end{tikzpicture}
\caption{The four inequivalent paths from $\,X_3\,$ to $\,X_1$.\ Every two paths with a given indexed edge at the terminal node are equivalent.} \label{fig:quadrinion}
\end{figure}
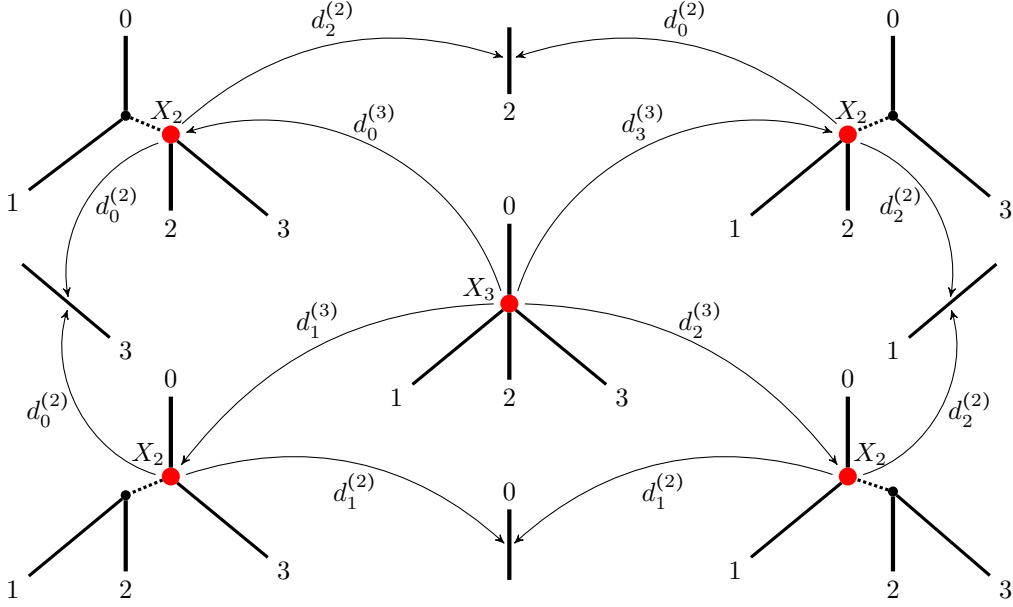

We begin by demonstrating that the two sets of face maps:\ $\,\{d^{(3)}_i\}_{i\in\{0,1,2,3\}}\,$ and $\,\{d^{(2)}_j\}_{j\in\{0,1,2\}}\,$ give rise to precisely four distinct morphisms $\,X_3\too X_1\,$ through superposition.\ The eight possible ways to descend from $\,X_3\,$ to $\,X_1\,$ are demonstrated in Fig.\,\ref{fig:quadrinion}.\ 
For each pair of paths leading to the same indexed edge in the figure,\ we establish equivalence of the two paths with the help of the simplicial identities (for the face maps) at level $\,n=3$.\ The four inequivalent morphisms admit the presentation
\qq\nn
\pi^{(4)}_{1,2}=d^{(2)}_2\circ d^{(3)}_2\,,\qquad\qquad\pi^{(4)}_{2,3}=d^{(2)}_2\circ d^{(3)}_0\,,\qquad\qquad\pi^{(4)}_{3,4}=d^{(2)}_0\circ d^{(3)}_0\,,\qquad\qquad\pi^{(4)}_{1,4}=d^{(2)}_1\circ d^{(3)}_1\,.
\qqq

Having verified the validity of the statement of the proposition on the lowest rung $\,n=3\,$ of the induction ladder,\ we proceed to assume its validity on the rung $\,n=N\geq 3$,\ and use this induction hypothesis to prove the induction step.\ First,\ we choose a convenient presentation of the morphisms $\,\pi^{(N+1)}_{1,N+1},\pi^{(N+1)}_{k,k+1}\ :\ X_N\too X_1,\ k\in\ovl{1,N}$,\ which we are at the liberty to do in virtue of the induction hypothesis.\ The straightforward `layered-exfoliation' procedures that yield these presentations for $\,\pi^{(N+1)}_{k,k+1},\ k\in\ovl{1,N}\,$ and for $\,\pi^{(N+1)}_{1,N+1}\,$ are illustrated in Fig.\,\ref{fig:facepeel} (a) and (b),\ respectively.\ 
\begin{figure}[hb!]
\begin{tikzpicture}
[junction/.style={circle,draw=red!100,fill=red!100,thick,inner sep=0pt,minimum size=6pt},
fusion/.style={circle,draw=black!100,fill=black!100,thick,inner sep=0pt,minimum size=3pt}]

\node[junction] (vertexk) at (0,-4.2) {};
\node[fusion] (vertexk30) at (-4,-3.2) {};
\node[fusion] (vertexk300) at (-3,-3.45) {};
\node[fusion] (vertexk3000) at (-1,-3.95) {};
\node (XNk) at (-4.35,-3) {$X_N$};
\node (XNkLred) at (0.8,-4.1) {$X_{N-k+2}$};
\node (L1k) at (-7.5,-4.45) {1};
\node (L2k) at (-5.5,-4.95) {2};
\node (Lkm2k) at (-2,-5.45) {k-2};
\node (Lkm1k) at (-0.35,-5.7) {k-1};
\node (Lkk) at (0.35,-5.7) {k};
\node (LNk) at (3.5,-5.7) {N};
\node (L12k) at (-4,-1.25) {0};
\node at (-2.75,-4.45) {$\cdots$};
\node at (0.75,-4.95) {$\cdots$};

\draw[very thick] (L1k) -- (vertexk30.south west);
\draw[very thick] (L2k) -- (vertexk300.south west);
\draw[very thick] (Lkm2k) -- (vertexk3000.south west);
\draw[very thick] (Lkm1k) -- (vertexk.south);
\draw[very thick] (Lkk) -- (vertexk.south);
\draw[ultra thick] (vertexk30.north) -- (L12k);
\draw[very thick] (LNk) -- (vertexk.south east);
\draw[densely dotted,very thick] (vertexk3000) --  (vertexk);
\draw[densely dotted,very thick] (vertexk3000) -- (vertexk300);
\draw[densely dotted,very thick] (vertexk300) --(vertexk30);

\draw[->,bend left=45,shorten <=2.5pt,shorten >=2.5pt,>=stealth'] (vertexk30.north east) to node [above] {$\quad d^{(N)}_0$} (vertexk300.north west);
\draw[->,bend left=45,shorten <=2.5pt,shorten >=2.5pt,>=stealth'] (vertexk300.north east) to node [above] {$\qquad\quad d^{(N-1)}_0$} (-2.25,-3.5);
\draw[->,bend left=45,shorten <=2.5pt,shorten >=2.5pt,>=stealth'] (-1.6,-3.6) to node [above] {$\qquad d^{(N-k+4)}_0$} (vertexk3000.north west);
\draw[->,bend left=45,shorten <=2.5pt,shorten >=2.5pt,>=stealth'] (vertexk3000.north east) to node [above] {$\qquad\qquad d^{(N-k+3)}_0$} (vertexk.north west);

\node[junction] (kvertex) at (-8.5,-7) {};
\node (kXN) at (-9.2,-6.7) {$X_{N-k+1}$};
\node (kL1) at (-10.5,-8.5) {k};
\node (kL2) at (-9.5,-8.5) {k+1};
\node (kLN) at (-6.5,-8.5) {N};
\node (kL12) at (-8.5,-6) {};
\node at (-8.35,-7.75) {$\cdots$};

\draw[very thick] (kL1) -- (kvertex.south west);
\draw[very thick] (kL2) -- (kvertex.south west);
\draw[densely dotted,ultra thick] (kvertex.north) -- (kL12);
\draw[very thick] (kLN) -- (kvertex.south east);

\draw[->,bend left=25,shorten <=2.5pt,shorten >=2.5pt,>=stealth'] (vertexk.south west) to node [left] {$d^{(N-k+2)}_0$} (kvertex.east);

\node[junction] (kvertex2) at (-4,-7.75) {};
\node[fusion] (kvertex23) at (-4.35,-8.45) {};
\node (kX2N) at (-4.55,-7.5) {$X_{N-k}$};
\node (kL21) at (-6,-9.25) {k};
\node (kL22) at (-5,-10.05) {k+1};
\node (kL23) at (-4,-10.05) {k+2};
\node (kL24) at (-3.5,-9.25) {k+3};
\node (kL2N) at (-2,-9.25) {N};
\node (kL212) at (-4,-6.75) {};
\node at (-3.4,-8.5) {$\cdots$};

\draw[very thick] (kL21) -- (kvertex2.south west);
\draw[very thick] (kL22) -- (kvertex23.south west);
\draw[very thick] (kL23) -- (kvertex23.south east);
\draw[densely dotted,ultra thick] (kvertex2.north) -- (kL212);
\draw[densely dotted,ultra thick] (kvertex23) -- (kvertex2);
\draw[very thick] (kL24) -- (kvertex2.south east);
\draw[very thick] (kL2N) -- (kvertex2.south east);

\draw[->,bend right=25,shorten <=2.5pt,shorten >=2.5pt,>=stealth'] (kvertex.south east) to node [above] {$\qquad d^{(N-k+1)}_2$} (kvertex2.west);

\node[junction] (kvertex22) at (1,-6.75) {};
\node[fusion] (kvertex223) at (-0.4,-9.55) {};
\node[fusion] (kvertex224) at (-0.05,-8.85) {};
\node[fusion] (kvertex225) at (0.65,-7.45) {};
\node (kX22N) at (1.35,-6.5) {$X_3$};
\node (kL221) at (-1.0,-8.25) {k};
\node (kL222) at (-1.05,-11.15) {k+1};
\node (kL223) at (-0.05,-11.15) {k+2};
\node (kL224) at (0.3,-10.45) {k+3};
\node at (0.45,-8.85) {$\cdots$};
\node (kL225) at (1.0,-9.05) {N-1};
\node (kL22N) at (3.0,-8.25) {N};
\node (kL2212) at (1.0,-5.75) {};

\draw[very thick] (kL221) -- (kvertex22.south west);
\draw[very thick] (kL222) -- (kvertex223.south west);
\draw[very thick] (kL223) -- (kvertex223.south east);
\draw[very thick] (kL224) -- (kvertex224.south east);
\draw[densely dotted,ultra thick] (kvertex22.north) -- (kL2212);
\draw[densely dotted,ultra thick] (kvertex223) -- (kvertex224);
\draw[densely dotted,ultra thick] (kvertex224) -- (kvertex225);
\draw[densely dotted,ultra thick] (kvertex225) -- (kvertex22);
\draw[very thick] (kL225) -- (kvertex225.south east);
\draw[very thick] (kL22N) -- (kvertex22.south east);

\node (cdots) at (-1.25,-7.25) {$\cdots$};

\draw[->,bend right=10,shorten <=2.5pt,shorten >=1.0pt,>=stealth'] (kvertex2.east) to node [above] {$\quad d^{(N-k)}_2\quad $} (cdots);
\draw[->,bend left=10,shorten <=1.0pt,shorten >=2.5pt,>=stealth'] (cdots) to node [above] {$\quad d^{(4)}_2$} (kvertex22.west);

\node (kT) at (-8.5,-9.65) {};
\node (kB) at (-10.5,-11.15) {k};

\draw[very thick] (kB) -- (kT);

\draw[->,bend left=33,shorten <=4.5pt,shorten >=2.5pt,>=stealth'] (kvertex22.south west) to node [below] {$\qquad\quad d^{(2)}_2\circ d^{(3)}_2$} (-9.5,-10.4);

\node at (-4,-12.0) {(a)};
\end{tikzpicture}\newline\newline\newline
\begin{tikzpicture}
[junction/.style={circle,draw=red!100,fill=red!100,thick,inner sep=0pt,minimum size=6pt},
fusion/.style={circle,draw=black!100,fill=black!100,thick,inner sep=0pt,minimum size=3pt}]

\node[junction] (vertex) at (-1,0) {};
\node[fusion] (vertex1) at (-5,-1) {};
\node[fusion] (vertex11) at (-4,-0.75) {};
\node[fusion] (vertex111) at (-2,-0.25) {};
\node (Xn) at (-5.4,-0.7) {$X_N$};
\node (X2) at (-0.6,0.3) {$X_2$};
\node (L1) at (-7.0,-2.5) {1};
\node (L2) at (-6.0,-2.5) {2};
\node (L3) at (-5.0,-2.25) {3};
\node (L4) at (-3.0,-1.75) {N-1};
\node (LN) at (1.0,-1.5) {N};
\node (L12) at (-1,1.7) {0};
\node at (-3.5,-1.2) {$\cdots$};

\draw[very thick] (L1) -- (vertex1.south west);
\draw[very thick] (L2) -- (vertex1.south west);
\draw[very thick] (L3) -- (vertex11.south west);
\draw[very thick] (L4) -- (vertex111.south west);
\draw[ultra thick] (vertex.north) -- (L12);
\draw[very thick] (LN) -- (vertex.south east);
\draw[densely dotted,ultra thick] (vertex1) -- (vertex11);
\draw[densely dotted,ultra thick] (vertex11) -- (vertex111);
\draw[densely dotted,ultra thick] (vertex111) -- (vertex);

\node (0B) at (3,0) {};
\node (0T) at (3,1.7) {0};

\draw[ultra thick] (0B) -- (0T);

\draw[->,bend left=45,shorten <=2.5pt,shorten >=2.5pt,>=stealth'] (vertex1.north east) to node [above] {$d^{(N)}_1$} (vertex11.north west);
\draw[->,bend left=45,shorten <=2.5pt,shorten >=2.5pt,>=stealth'] (vertex11.north east) to node [above] {$\qquad d^{(N-1)}_1$} (-3.25,-0.4);
\draw[->,bend left=45,shorten <=2.5pt,shorten >=2.5pt,>=stealth'] (-2.75,-0.2) to node [above] {$d^{(4)}_1$} (vertex111.north west);
\draw[->,bend left=45,shorten <=2.5pt,shorten >=2.5pt,>=stealth'] (vertex111.north east) to node [above] {$d^{(3)}_1$} (vertex.north west);
\draw[->,bend right=33,shorten <=2.5pt,shorten >=2.5pt,>=stealth'] (vertex.east) to node [above] {$d^{(2)}_1$} (3,0.8);

\node at (-2.5,-3.1) {(b)};
\end{tikzpicture}
\caption{A pictorial representation of the derivation of the morphisms $\,X_N\too X_{N-1}\too\cdots\too X_1\,$ through `layered exfoliation' for $\,\pi^{(N+1)}_{k,k+1},\ k\in\ovl{1,N}\,$ (a) and for $\,\pi^{(N+1)}_{1,N+1}\,$ (b).} \label{fig:facepeel}
\end{figure}
From these,\ we read off the expressions
\qq\nn
\pi^{(N+1)}_{k,k+1}&=&d^{(2)}_2\circ d^{(3)}_2\circ\cdots\circ d^{(N-k+1)}_2\circ d^{(N-k+2)}_0\circ d^{(N-k+3)}_0\circ\cdots\circ d^{(N)}_0\,,\qquad k\in\ovl{1,N}\,,\cr\cr
\pi^{(N+1)}_{1,N+1}&=&d^{(2)}_1\circ d^{(3)}_1\circ\cdots\circ d^{(N)}_1\,,
\qqq
which coincide with those given in \Reqref{eq:ind-ibb-maps}.\ They are instrumental in proving the induction step.\ Indeed,\ inspection of the face maps depicted in Fig.\,\ref{fig:higherface} for $\,n=N+1\,$ in conjunction with the induction hypothesis reveals that there are precisely $\,N\,$ ways to arrive at the $k$-th edge from the original graph of valence $\,(N+1,1)\,$ with the help of the face maps $\,d^{(N+1)}_i,\ i\in\ovl{0,N+1}$,\ to wit through all possible fusions of adjacent edges which do \emph{not} involve the $k$-th edge,\ each fusion being followed by the application of the \emph{unique} map $\,X_N\too X_1\,$ that maps the graph with a single `virtual' edge to its `real' edge with index $\,k$.\ Thus,\ we obtain the morphisms $\,\pi^{(N+1)}_{k-1,k}\circ d^{(N+1)}_l,\ l\in\ovl{0,k-2}$,\ represented by fusion of edges with indices smaller than $\,k$,\ and the morphisms $\,\pi^{(N+1)}_{k,k+1}\circ d^{(N+1)}_m,\ m\in\ovl{k+1,N+1}$,\ represented by fusion of edges with indices greater than $\,k$.\ At this stage,\ it suffices to check that all these morphisms are pairwise equal,\ which follows straightforwardly from the simplicial identities of Def.\,\ref{def:simplicial}.\ Indeed,\ on the one hand,
\qq\nn
\pi^{(N+1)}_{k-1,k}\circ d^{(N+1)}_l&\equiv&d^{(2)}_2\circ d^{(3)}_2\circ\cdots\circ d^{(N-(k-1)+1)}_2\circ d^{(N-(k-1)+2)}_0\circ d^{(N-(k-1)+3)}_0\circ\cdots\circ d^{(N)}_0\circ d^{(N+1)}_l\cr\cr
&=&d^{(2)}_2\circ d^{(3)}_2\circ\cdots\circ d^{(N-k+2)}_2\circ d^{(N-k+3)}_0\circ d^{(N-k+4)}_0\circ\cdots\circ d^{(N)}_0\circ d^{(N+1)}_0
\qqq
as $\,l-(N-(N-(k-1)+1))=l-(k-2)\leq 0$,\ and,\ on the other hand, 
\qq\nn
\pi^{(N+1)}_{k,k+1}\circ d^{(N+1)}_m&\equiv&d^{(2)}_2\circ d^{(3)}_2\circ\cdots\circ d^{(N-k+1)}_2\circ d^{(N-k+2)}_0\circ d^{(N-k+3)}_0\circ\cdots\circ d^{(N)}_0\circ d^{(N+1)}_m\cr\cr
&=&d^{(2)}_2\circ d^{(3)}_2\circ\cdots\circ d^{(N-k+1)}_2\circ d^{(N-k+2)}_{m-(k-1)}\circ d^{(N-k+3)}_0\circ d^{(N-k+4)}_0\circ\cdots\circ d^{(N+1)}_0\cr\cr
&=&d^{(2)}_2\circ d^{(3)}_2\circ\cdots\circ d^{(N-k+2)}_2\circ d^{(N-k+3)}_0\circ d^{(N-k+4)}_0\circ\cdots\circ d^{(N)}_0\circ d^{(N+1)}_0
\qqq
as $\,m-(k-1)-(N-k+1-1)=m+1-N\leq 2$.\ We treat the remaining composite morphisms that send $\,X_{N+1}\,$ to $\,X_1\,$ in a way represented by the assignment of the distinguished `out-going' edge with the index $\,0\,$ to the junction of valence $\,(N+1,1)\,$ analogously:\ Inspecting Fig.\,\ref{fig:higherface} once more,\ we infer that the $\,N\,$ paths to the edge begin with the moves representing the face maps $\,d^{(N+1)}_n,\ n\in\ovl{1,N}$,\ and so it remains to verify the equality of all morphisms of the form
\qq\nn
\pi^{(N+1)}_{1,N+1}\circ d^{(N+1)}_n\equiv d^{(2)}_1\circ d^{(3)}_1\circ\cdots\circ d^{(N)}_1\circ d^{(N+1)}_n=d^{(2)}_1\circ d^{(3)}_1\circ\cdots\circ d^{(N+1)}_1\,,
\qqq
which follows from $\,n-(N-1)\leq N-(N-1)=1$.\ This concludes the proof of the first part of the proposition.

The second part is proved through direct calculation using the simplicial identities for the face maps and the convenient presentation \eqref{eq:ind-ibb-maps},
\qq\nn
d^{(1)}_0\circ\pi^{(n+1)}_{k-1,k}&\equiv&d^{(1)}_0\circ d^{(2)}_2\circ d^{(3)}_2\circ\cdots\circ d^{(N-(k-1)+1)}_2\circ d^{(N-(k-1)+2)}_0\circ d^{(N-(k-1)+3)}_0\circ\cdots\circ d^{(N)}_0\cr\cr
&=&d^{(1)}_1\circ d^{(2)}_0\circ d^{(3)}_2\circ\cdots\circ d^{(N-(k-1)+1)}_2\circ d^{(N-(k-1)+2)}_0\circ d^{(N-(k-1)+3)}_0\circ\cdots\circ d^{(N)}_0\cr\cr
&=&\ldots=d^{(1)}_1\circ d^{(2)}_1\circ d^{(3)}_1\circ\cdots\circ d^{(N-(k-1))}_1\circ d^{(N-(k-1)+1)}_0\circ d^{(N-(k-1)+2)}_0\circ\cdots\circ d^{(N)}_0\cr\cr
&=&d^{(1)}_1\circ d^{(2)}_1\circ d^{(3)}_1\circ\cdots\circ d^{(N-(k-1)-1)}_1\circ d^{(N-(k-1))}_2\circ d^{(N-(k-1)+1)}_0\circ d^{(N-(k-1)+2)}_0\circ\cdots\circ d^{(N)}_0\cr\cr
&=&\ldots=d^{(1)}_1\circ d^{(2)}_2\circ d^{(3)}_2\circ\cdots\circ d^{(N-(k-1))}_2\circ d^{(N-(k-1)+2)}_0\circ d^{(N-(k-1)+3)}_0\circ\cdots\circ d^{(N)}_0\cr\cr
&\equiv&d^{(1)}_1\circ\pi^{(n+1)}_{k,k+1}\,,\qquad k\in\ovl{2,n}\,,\cr\cr\cr
d^{(1)}_1\circ\pi^{(n+1)}_{1,2}&\equiv&d^{(1)}_1\circ d^{(2)}_2\circ d^{(3)}_2\circ\cdots\circ d^{(n)}_2=d^{(1)}_1\circ d^{(2)}_1\circ d^{(3)}_2\circ d^{(4)}_2\circ\cdots\circ d^{(n)}_2=\ldots=d^{(1)}_1\circ d^{(2)}_1\circ\cdots\circ d^{(n)}_1\cr\cr
&\equiv&d^{(1)}_1\circ\pi^{(n+1)}_{1,n+1}\,,\cr\cr\cr
d^{(1)}_0\circ\pi^{(n+1)}_{1,n+1}&\equiv&d^{(1)}_0\circ d^{(2)}_1\circ d^{(3)}_1\circ\cdots\circ d^{(n)}_1=d^{(1)}_0\circ d^{(2)}_0\circ d^{(3)}_1\circ d^{(4)}_1\circ\cdots\circ d^{(n)}_1=\ldots=d^{(1)}_0\circ d^{(2)}_0\circ\cdots\circ d^{(n)}_0\cr\cr
&\equiv&d^{(1)}_0\circ\pi^{(n+1)}_{n,n+1}\,.
\qqq
The identities involving the degeneracy maps are proven similarly.
\qed

\section{A proof of Proposition \ref{prop:maxym-curv-bdry-existuniq}}\label{app:proof1}

In this appendix,\ we prove the uniqueness of the $\ell\wp^\p$-invariant curvature $\,\om_\p^{(\sfk)}\,$ of the maximally symmetric WZW D-brane.\ In view of the $\sfL\txG\x\sfL\txG$-covariance of the equations of motion,\ it suffices to ensure that the DGC \eqref{eq:DGC-b} written for the (component) D-brane worldvolume $\,\xcC_\la\,$ (and a 2-form $\,\om_{\p,\la}^{(\sfk)}\equiv\om_\p^{(\sfk)}\vert_{\xcC_\la}\in\Om^2(\xcC_\la)$,\ the latter assumed $\Ad_\cdot$-invariant) is satisfied by the $\sfL\txG$-translate $\,\ups{h}g\,$ of an arbitrary field configuration $\,g\in\xcF(\Si,d;\eta)$.\ To this end,\ we cast the DGC in a manageable form.\ First of all,\ we readily establish -- for $\,\ee_\la=\ee^{\frac{2\pi\sfi\,\la}{\sfk}}\,$ and an arbitrary element $\,x\in\txG\,$ -- the identity
\qq\label{eq:MConconjcl}
\th_{\rm L}\bigl(\Ad_x(\ee_\la)\bigr)=\bigl(\id_\ggt-\sfT_e\Ad_{\Ad_x(\ee_\la)}\bigr)\circ\sfT_e\Ad_{\Ad_x(\ee_\la)^{-1}}\circ\th_{\rm R}(x)\,,
\qqq
written in terms of the right-invariant ($\ggt$-valued) Maurer--Cartan 1-form $\,\theta_{\rm R}$,\ 
which shows that it is meaningful to invert the operator 
\qq\nn
O:=\sfT_e\Ad_\cdot\circ\Inv-\id_\ggt\in C^\infty\bigl(\txG,\End\,\ggt\bigr)
\qqq
on the $\theta_{\rm L}$-image of $\,\sfT\xcC_\la$,\ and so also on all of $\,\Om^1(\xcC_\la)$.\ Accordingly,\ we may write the D-brane curvature in the form
\qq\nn
\om_{\p,\la}^{(\sfk)}=\tfrac{\sfk}{8\pi}\,\tr_\ggt\,\bigl(\D\circ\theta_{\rm L}\wedge O^{-1}\circ\theta_{\rm L}\bigr)
\qqq
for some $\,\D\in C^\infty(\xcC_\la,\End\,\ggt)\,$ to be established in what follows,\ and we may further assume that the operators satisfy the identity 
\qq\label{eq:aSymmb}
\D_g^\dagger\circ O_g^{-1}=-O_{g^{-1}}^{-1}\circ\D_g\,,
\qqq
written in terms of the conjugate $\,\D_g^\dagger\,$ of $\,\D_g\,$ with respect to the scalar product $\,(X|Y)=\tr_\ggt\,(X\,Y)\,$ on $\,\ggt\,$ and encoding the skew symmetry of $\,\om_{\p,\la}$.\ Consider the vector fields
\qq\label{eq:left-min-right}
V_A(\cdot)=R_A(\cdot)-L_A(\cdot)\equiv O_{\cdot\,A}^{\ \ B}\,L_B(\cdot)\,,\qquad A\in\ovl{1,\dim\,\ggt}\,,
\qqq
written in terms of the left- ($L_A$) and right-invariant vector fields on $\,\txG\,$ dual to the left- and right-invariant component Maurer--Cartan 1-forms,\ respectively,\ $\,L_A\con\th_{\rm L}^B=\d_A^{\ B}=R_A\con\th_{\rm R}^B$,.\ The above are the fundamental vector fields for the adjoint action of the group on $\,\xcC_\la$,\ and so they generate the tangent bundle over $\,\xcC_\la$.\ For the choice
\qq\label{eq:tn-choice}
t=\p_2\,,\qquad\qquad\widetilde Nt=\p_1\,,
\qqq
and in the notation \eqref{eq:Convgdg},\ we now obtain the result (in which we have dropped the obvious dependence on $\,p\in\p\Si$)
\qq\nn
\txg_g^{(\sfk)}\left(\iota_{1\,*}V_A,g_{|1\,*}\widetilde Nt\right)&=&-\tfrac{\sfk}{4\pi}\,\tr_\ggt\left(t_A\,O_{g^{-1}}\left(g^{-1}\p_1 g\right)\right)\,,\cr\cr 
\om_{\p,\la\,g}^{(\sfk)}\left(V_A,g_*t\right)&=&-\tfrac{\sfk}{4\pi}\,\tr_\ggt\left(t_A\,\D_g\left(g^{-1}\,\p_2 g\right)\right)\,.
\qqq
Thus,\ in the end,\ the DGC reduces to
\qq\nn
\D_g\left(g^{-1}\,\p_2 g\right)=-O_{g^{-1}}\left(g^{-1}\p_1 g\right)\,.
\qqq

Next,\ we consider a loop-group transformation \eqref{eq:LG-trafo-bdry} of the $\si$-model field and restrict it to the boundary of $\,\Si$,\ that is to the locus of the equation $\,\si^1=0$.\ We have (in the obvious shorthand notation)
\qq\nn
{}^{\tx{\tiny $h$}}\hspace{-2pt}g^{-1}\p_1{}^{\tx{\tiny $h$}}
\hspace{-2pt}g(0,\si^2)&=&\sfT_e\Ad_h\left(g^{-1}\p_1 g\right)(\si^2)+
\left(\sfT_e\Ad_{h\cdot g^{-1}}+\sfT_e\Ad_h\right)\left(h^{-1}\,\p_2 h\right)(
\si^2)\,,\cr\cr {}^{\tx{\tiny $h$}}\hspace{-2pt}g^{-1}\p_2
{}^{\tx{\tiny $h$}}\hspace{-2pt}g(0,\si^2)&=&\sfT_e\Ad_h\left(g^{-1}\p_2
g\right)(\si^2)+ \left(\sfT_e\Ad_{h\cdot g^{-1}}-\sfT_e\Ad_h\right)\left(h^{-1}
\,\p_2 h\right)(\si^2)\,,
\qqq
and so the DGC for the transformed field (whose validity we assume as a constraint on $\,\om_{\p,\la}$) reads
\qq\nn
&&\D_{{}^{\tx{\tiny $h$}}\hspace{-2pt}g}\circ\sfT_e\Ad_h\left(g^{-1}\, \p_2 g\right)+\D_{{}^{\tx{\tiny $h$}}\hspace{-2pt}g}\circ\left(\sfT_e\Ad_{h\cdot g^{-1}}-\sfT_e\Ad_h\right)\left(h^{-1}\,\p_2 h\right)\cr\cr
&=&\sfT_e\Ad_h\circ(\id_\ggt-\sfT_e\Ad_g)\left(g^{-1}\p_1 g\right)+\left(\id_\ggt-\sfT_e\Ad_{\Ad_h(g)}\right)\circ\left(\sfT_e\Ad_{h\cdot g^{-1}}+\sfT_e\Ad_h\right)\left(h^{-1}\,\p_2 h\right)\,.
\qqq
The first terms on either side of the above equation cancel out in virtue of the assumed $\Ad$-invariance of $\,\om_{\p,\la}$,\ taken in conjunction with the DGC for $\,g$.\ Indeed,\ for a \emph{constant} loop $\,h$,\ the said $\Ad$-invariance yields the equality
\qq\nn
\tfrac{\sfk}{8\pi}\,\tr_\ggt\bigl(\sfT_e\Ad_{h^{-1}}\circ\D_{{}^{\tx{\tiny $h$}}\hspace{-2pt}g}\circ\sfT_e\Ad_h\circ\theta_{\rm L}(g)\wedge O^{-1}_g\circ\theta_{\rm L}(g)\bigr)\equiv\om_{\p,\la\,{}^{\tx{\tiny $h$}}\hspace{-2pt}g}\must\om_{\p,\la\,g}\equiv&\tfrac{\sfk}{8\pi}\,\tr_\ggt\,\bigl(\D_g\circ\theta_{\rm L}(g)\wedge O^{-1}_g\circ\theta_{\rm L}(g)\bigr)\,,
\qqq
and so upon contracting both sides of the above with $\,V_A$,\ we end up with the condition 
\qq\nn
\sfT_e\Ad_{h^{-1}}\circ\D_{{}^{\tx{\tiny $h$}}\hspace{-2pt}g}\circ\sfT_e\Ad_h\circ\theta_{\rm L}(g)=\D_g\circ\theta_{\rm L}(g)\,.
\qqq
The said cancellation now follows from the last relation upon evaluating both sides on the vector $\,g(0,\cdot)_*\p_2(\si^2)$,\ tangent to $\,\xcC_\la\,$ at $\,g(0,\si^2)$.

Thus,\ we are left with the condition
\qq\nn
\D_{{}^{\tx{\tiny $h$}}\hspace{-2pt}g}\circ\left(\sfT_e\Ad_{h\cdot
g^{-1}}-\sfT_e\Ad_h\right)\left(h^{-1}\,\p_2 h\right)=\left(\sfT_e\Ad_{h\cdot
g^{-1}}-\sfT_e\Ad_{h\cdot g}\right)\left(h^{-1}\,\p_2 h\right)\,,
\qqq
which can be further rewritten as
\qq\nn
\D_{{}^{\tx{\tiny $h$}}\hspace{-2pt}g}\left({}^{\tx{\tiny $h$}}
\hspace{-2pt}g^{-1}\,\p_2{}^{\tx{\tiny $h$}}\hspace{-2pt}g-\sfT_e\Ad_h
\left(g^{-1}\,\p_2 g\right)\right)={}^{\tx{\tiny $h$}}\hspace{-2pt}
g^{-1}\,\p_2{}^{\tx{\tiny $h$}}\hspace{-2pt}g+\left(\p_2
{}^{\tx{\tiny $h$}}\hspace{-2pt}g \right)\,{}^{\tx{\tiny $h$}}
\hspace{-2pt}g^{-1}-\sfT_e\Ad_h\left(g^{-1}\,\p_2 g+\left(\p_2 g\right)\,
g^{-1}\right)\,.
\qqq

In the last step of the argument,\ we restrict our attention to constant boundary loops $\,g(0,\cdot)=:g_*\in\xcC_\la$.\ These are,\ clearly,\ always admissible as boundary conditions for classical field configurations of the $\si$-model in hand\footnote{Take,\ {\it e.g.},\ the constant map $\,g(\si^1,\si^2)=g_*\,$ as a classical field configuration.}.\ For these,\ the last condition takes the simple form
\qq\nn
\D_{{}^{\tx{\tiny $h$}}\hspace{-2pt}g_*}\left({}^{\tx{\tiny
$h$}}\hspace{-2pt}g_*^{-1}\,\p_2{}^{\tx{\tiny $h$}}\hspace{-2pt}g_*
\right)={}^{\tx{\tiny $h$}}\hspace{-2pt}g_*^{-1}\,\p_2{}^{\tx{\tiny
$h$}}\hspace{-2pt}g_*+\left(\p_2 {}^{\tx{\tiny $h$}}\hspace{-2pt}
g_*\right)\,{}^{\tx{\tiny $h$}}\hspace{-2pt}g_*^{-1}\,.
\qqq
In view of the arbitrariness of both $\,h\,$ and $\,g_*$,\ we recognise in $\,{}^{\tx{\tiny $h$}}\hspace{-2pt}g_*^{-1}\,\p_2 {}^{\tx{\tiny $h$}}\hspace{-2pt}g_*\,$ a generic vector tangent to the conjugacy class $\,\xcC_\la\,$ at its arbitrary point $\,g_*$.\ Therefore,\ we may eventually cast the condition in the form
\qq\nn
\D_y\left(y^{-1}\,\p y\right)=y^{-1}\,\p y+(\p y)\,y^{-1}\,,
\qqq
with,\ now,\ an arbitrary loop $\,y\in\sfL\xcC_\la$.\ We conclude that $\,\D\,$ and,\ hence,\ also $\,\om_{\p,\la}^{(\sfk)}\,$ are unique if they exist.\ The latter fact can be verified explicitly by checking that the 2-form $\,\om_{\p,\la}^{(\sfk)}\,$ of \Reqref{eq:om-WZW-b} is $\Ad_\cdot$-invariant,\ satisfies \eqref{eq:aSymmb} and the condition
\qq\label{eq:omp-as-prim-H}
\txH_{\rm C}^{(\sfk)}\vert_{\xcC_\la}=\sfd\om_{\p,\la}^{(\sfk)}\,.
\qqq
Indeed,\ using \Reqref{eq:MConconjcl} alongside the obvious identities $\,\tr_\ggt\bigl(\th_{\rm L}\wedge\th_{\rm L}\bigr)=0=\tr_\ggt\bigl(\th_{\rm R}\wedge\th_{\rm R}\bigr)\,$ we readily verify the following chain of equalities:
\qq\nn
&&\tr_\ggt\bigg(\th_{\rm L}\bigl(\Ad_x(\ee_\la)\bigr)\wedge\tfrac{\sfT_e\Ad_{\Ad_x(\ee_\la)}}{\id_\ggt-\sfT_e\Ad_{\Ad_x(\ee_\la)}}\circ\th_{\rm L}\bigl(\Ad_x(\ee_\la)\bigr)\bigg)= \tr_\ggt\bigl(\th_{\rm R}(x)\wedge\sfT_e\Ad_{\Ad_x(\ee_\la)}\circ\th_{\rm R}(x)\bigr)\cr\cr
&=&\tr_\ggt\bigg(\th_{\rm L}\bigl(\Ad_x(\ee_\la)\bigr)\wedge\tfrac{\id_\ggt}{\id_\ggt-\sfT_e\Ad_{\Ad_x(\ee_\la)}}\circ\th_{\rm L}\bigl(\Ad_x(\ee_\la)\bigr)\bigg)\,,
\qqq
and so we find
\qq\label{eq:WZW-brane-curv-param-Cart}
\om_\la\bigl(\Ad_x(\ee_\la)\bigr)=\tfrac{\sfk}{4\pi}\,
\tr_\ggt\bigl(\th_{\rm L}(x)\wedge\sfT_e\Ad_{\ee_\la}\circ\th_{\rm L}(x)\bigr)\,.
\qqq
The $\Ad_\cdot$-invariance of $\,\om_{\p,\la}\,$ is now apparent as an immediate consequence of the invariance of the Maurer--Cartan 1-form under left translations.

More generally,\ we could write,\ for any fixed element $\,g_0\in\xcC_\la$,
\qq\label{eq:WZW-brane-curv-param}
\om_{\p,\la}^{(\sfk)}\bigl(\Ad_x(g_0)\bigr)=\tfrac{\sfk}{4\pi}\,\tr_\ggt\bigl(\th_{\rm L}(x)\wedge\sfT_e\Ad_{g_0}\circ\th_{\rm L}(x)\bigr)\,.
\qqq
Upon recalling the Maurer--Cartan equation $\,\sfd\th_{\rm L}=-\th_{\rm L}\wedge\th_{\rm L}$,\ it becomes completely straightforward to check the identity
\qq\nn
\sfd\om_{\p,\la}^{(\sfk)}\bigl(\Ad_x(g_0)\bigr)=\tfrac{\sfk}{4\pi}\,\tr_\ggt\bigl(\th_{\rm L}(x)\wedge\th_{\rm L}(x)\wedge\bigl(\sfT_e\Ad_{g_0^{-1}}-\sfT_e\Ad_{g_0}\bigr)\circ\th_{\rm L}(x)\bigr)=\txH_{\rm C}^{(\sfk)}\bigl(\Ad_x(g_0)\bigr)\,,
\qqq
which proves \Reqref{eq:omp-as-prim-H}.

Finally,\ note that under the identification $\,\D=\id_\ggt+\sfT_e\Ad_\cdot$,\ the DGC translates into the desired gluing condition \eqref{eq:bchirglueWZW} for the chiral Ka\v c--Moody currents.

We have just proved that imposition of the symmetry constraints determines the D-brane curvature uniquely.\ The converse implication of the proposition has long been known to hold true,\ {\it cf.},\ {\it e.g.},\ \Rcite{Gawedzki:2001rm}.\qed

\section{A proof of Proposition \ref{prop:maxym-curv-existuniq}}\label{app:proof2}

Reasoning in analogy with App.\,\ref{app:proof1},\ we verify the uniqueness of the curvature 2-form by exploiting the constraint that an $\sfL\txG\x\sfL\txG$-translate of (an extension of) an arbitrary element $\,\left(g_{\rm b},(g,h)\right)\in\cF_{({\rm eom})}(\Si,d;\eta)\,$ should satisfy the DGC \eqref{eq:DGC-WZW} and that the chiral Ka\v c--Moody currents should be continuous across the defect.\ We begin by decomposing,\ for the sake of the simplicity of the ensuing analysis,\ the restriction $\,\om_\la^{(\sfk)}\equiv\om^{(\sfk)}\vert_{Q_\la}\,$ of the sought-after 2-form as
\qq\nn
\om_\la^{(\sfk)}=\tfrac{\sfk}{8\pi}\,\tr_\ggt\left(\widetilde\om{}^1\circ\pr_1^*\theta_{\rm R}\wedge\pr_1^*\theta_{\rm R}+2\widetilde\om{}^2\circ\pr_1^*\theta_{\rm L}\wedge\pr_2^*\theta_{\rm R}+\widetilde\D\circ\pr_2^*\theta_{\rm L}\wedge\pr_2^*\bigl(O^{-1}\circ\theta_{\rm L}\bigr)\right)
\qqq
in terms of some linear operators $\,\widetilde\D,\widetilde\om{}^k\equiv\om^k_{AB}\,\tau^A\ox t_B\in C^\infty\bigl(Q_\la,\End\,\ggt\bigr),\ k\in\{1,2\}\,$ to be determined ($\{\tau^A\}^{A\in\ovl{1,\dim\,\ggt}}\,$ is the basis of $\,\ggt\,$ dual to $\,\{t_A\}_{A\in\ovl{1,\dim\,\ggt}}$),\ further assuming $\,\widetilde\om{}_{(g,h)}^{1\,\dagger}=-\widetilde\om{}_{(g,h)}^1\,$ and $\,\widetilde\D{}_{(g,h)}^\dagger\circ O_h^{-1}=-O_{h^{-1}}^{-1}\circ\widetilde\D{}_{(g,h)}\,$ ({\it cf.}\ App.\,\ref{app:proof1}).\ In writing the last term,\ we implicitly invoke the same argument as in the boundary case.\ The tangent space of the $\txG\x\txG$-homogeneous space $\,Q_\la\equiv\txG\x\xcC_\la\,$ at a given point $\,(g,h)\,$ is spanned by two types of vectors:
\qq\label{eq:V-tan-GCla}
X_A(g,h)=R_A(g)\,,\qquad\qquad Y_A(g,h)=-L_A(g)+V_A(h)\,,
\qqq
the latter being written in terms of the fundamental vector fields \eqref{eq:left-min-right} and jointly generating the action $\,\ell\wp^{(1)}$.\ This results in the emergence of two independent ($\ggt$-valued) DGCs that constrain the bi-brane curvature.\ Indeed,\ having made the choice \eqref{eq:tn-choice} once more,\ we find the DGC:
\qq\nn
\widetilde\om{}^1_{(g,h)}\left((\p_2 g)\,g^{-1}\right)-\sfT_e\Ad_g\circ\widetilde\om{}^{2\,\dagger}_{(g,h)}\left((\p_2 h)\,h^{-1}\right)=\bigl(\p_1(g\cdot h)\bigr)\,(g\cdot h)^{-1}-(\p_1 g)\,g^{-1}
\qqq
for $\,V=X_A$,\ and another one:
\qq\nn
&&\sfT_e\Ad_{g^{-1}}\circ\widetilde\om{}^1_{(g,h)}\left((\p_2 g)\,g^{-1}\right)-\widetilde\om{}^{2\,\dagger}_{(g,h)}\left((\p_2 h)\,h^{-1}\right)+O_h\circ\widetilde\om{}^2_{(g,h)}\left(g^{-1}\p_2 g\right)-\widetilde\D{}_{(g,h)}\left(h^{-1}\p_2 h\right)\cr\cr
&=&(g\cdot h)^{-1}\p_1(g\cdot h)-g^{-1}\,\p_1 g
\qqq
for $\,V=Y_A$.

Let us,\ to begin with,\ consider loop-group transformations of the type $\,(g,h)(0,\si^2)\longmapsto(x(\si^+)\cdot g(0,\si^2),h(0,\si^2))\vert_{\si^1=0}$.\ Under these,\ the two DGCs map to
\qq\nn
\widetilde\om{}^1_{(x\cdot g,h)}\circ\sfT_e\Ad_x\left((\p_2 g)\,g^{-1}\right)-\sfT_e\Ad_{x\cdot g}\circ\widetilde\om{}^{2\,\dagger}_{(x\cdot g,h)}\left((\p_2 h)\,h^{-1}\right)+\widetilde\om{}^1_{(x\cdot g,h)}\left((\p_2 x)\,x^{-1}\right)=\sfT_e\Ad_{x\cdot g}\bigl((\p_1 h)\,h^{-1}\bigr)
\qqq
and
\qq\nn
&&\sfT_e\Ad_{(x\cdot g)^{-1}}\circ\widetilde\om{}^1_{(x\cdot g,h)}\circ\sfT_e\Ad_x\left((\p_2 g)\,g^{-1}\right)-\widetilde\om{}^{2\,\dagger}_{(x\cdot g,h)}\left((\p_2 h)\,h^{-1}\right)+O_h\circ\widetilde\om{}^2_{(x \cdot g,h)}\left(g^{-1}\p_2 g\right)\cr\cr
&&-\widetilde\D{}_{(x\cdot g,h)}\left(h^{- 1} \p_2 h\right)+\sfT_e\Ad_{(x\cdot g)^{-1}}\circ\widetilde\om{}^1_{(x\cdot g,h)} \left((\p_2 x)\,x^{-1}\right)+O_h\circ\widetilde\om{}^2_{(x\cdot g,h)}\circ\sfT_e\Ad_{g^{-1}}\left(x^{-1}\,\p_2 x\right)\cr\cr
&=&(g\cdot h)^{-1}\p_1(g\cdot h)-g^{-1}\,\p_1 g+O_h\circ\sfT_e\Ad_{g^{-1}}\bigl(x^{-1}\,\p_1 x\bigr)=(g\cdot h)^{-1}\p_1(g\cdot h)-g^{-1}\,\p_1 g+O_h\circ\sfT_e\Ad_{g^{-1}}\bigl(x^{-1}\,\p_2 x\bigr)\,,
\qqq
respectively.\ We can subsequently use the assumed $\widetilde\ell$-invariance of the bi-brane curvature,\ in the form $\,\ell\wp_{(x,e)}^{(1)\,*}\om_\la^{(\sfk)}=\om_\la^{(\sfk)}$,\ implying (on contraction with appropriate vectors from $\,\sfT_{(g,h)}Q_\la$) the relations 
\qq\nn
&\sfT_e\Ad_{x^{-1}}\circ\widetilde\om{}^1_{(x\cdot g,h)}\circ\sfT_e\Ad_x\circ\theta_{\rm R}(g)=\widetilde\om{}^1_{(g,h)}\circ\theta_{\rm R}(g)\,,\qquad\qquad\bigl(\widetilde\om{}^{2\,\dagger}_{(x\cdot g,h)}-\widetilde\om{}^{2\,\dagger}_{(g,h)}\bigr)\circ\theta_{\rm R}(h)=0\,,&\cr\cr
&\bigl(\widetilde\D{}_{(x\cdot g,h)}-\widetilde\D{}_{(g,h)}\bigr)\circ\theta_{\rm L}(h)=0\,.&
\qqq
These,\ in conjunction with the original DGCs,\ enable us to drop all terms with derivatives of $\,g\,$ and $\,h\,$ from the transformed DGCs,\ whereby we obtain
\qq\nn
\widetilde\om{}^1_{(x\cdot g,h)}\left((\p_2 x)\,x^{-1}\right)&=&0\,,\cr\cr
\sfT_e\Ad_{(x\cdot g)^{-1}}\circ\widetilde\om{}^1_{(x\cdot g,h)}\left((\p_2 x)\,x^{-1}\right)+O_h\circ\widetilde\om{}^2_{(x\cdot g,h)}\circ\sfT_e\Ad_{g^{-1}}\left(x^{-1}\,\p_2 x\right)&=&O_h\circ\sfT_e\Ad_{g^{-1}}\bigl(x^{-1}\p_2 x\bigr)\,.
\qqq
Let us focus on the first relation.\ Setting $\,{}^{\tx{\tiny $x$}}\hspace{-2pt}g:=x\cdot g$,\ we rewrite it as
\qq\nn
\widetilde\om{}^1_{\left({}^{\tx{\tiny $x$}}\hspace{-2pt}g,h\right)}\left(\left(\p_2{}^{\tx{\tiny $x$}}\hspace{-2pt}g\right)\,{}^{\tx{\tiny$x$}}\hspace{-2pt}g^{-1}-\sfT_e\Ad_x\left((\p_2 g)\,g^{-1}\right)\right)=0
\,,
\qqq
and so,\ for an arbitrary constant loop $\,g(0,\cdot)=:g_*\in\txG$,\ we have
\qq\nn
\widetilde\om{}^1_{\left({}^{\tx{\tiny $x$}}\hspace{-2pt}g_*,h\right)}\left(\left(\p_2{}^{\tx{\tiny $x$}}\hspace{-2pt}g_*\right)\,
{}^{\tx{\tiny $x$}}\hspace{-2pt}g_*^{-1}\right)=0\,,
\qqq
from which we deduce
\qq\nn
\widetilde\om{}^1=0\,.
\qqq
This simplifies the other relation,\ giving
\qq\nn
O_h\circ\widetilde\om{}^2_{(x\cdot g,h)}\circ\sfT_e\Ad_{g^{-1}}\left(x^{-1}\,\p_2 x\right)=O_h\circ\sfT_e\Ad_{g^{-1}}\bigl(x^{-1}\p_2 x\bigr)\,,
\qqq
which we next subject to the same operations,\ to the following effect:
\qq\nn
O_h\circ\left(\widetilde\om{}^2_{\left({}^{\tx{\tiny $x$}}\hspace{-2pt}g_*,h\right)}-\id_\ggt\right)\left({}^{\tx{\tiny $x$}}\hspace{-2pt}g_*^{-1}\,\p_2{}^{\tx{\tiny $x$}}\hspace{-2pt}g_*\right)=0\,,
\qqq
whence\footnote{Recall that the tangent space $\,\sfT_h\xcC_\la\,$ is spanned by the vectors $\,V_A(h)$.} also
\qq\nn
\om_\la^{(\sfk)}=\tfrac{\sfk}{4\pi}\,\tr_\ggt\left(\pr_1^*\theta_{\rm L}\wedge\pr_2^*\theta_{\rm R}\right)+\tfrac{\sfk}{8\pi}\,\tr_\ggt\left(\widetilde\D\circ\pr_2^*\theta_{\rm L}\wedge\pr_2^*\bigl(O^{-1}\circ\theta_{\rm L}\bigr)\right)\,.
\qqq
The DGCs reduce accordingly,
\qq\nn
\bigl(\p_+(g\cdot h)\bigr)\,(g\cdot h)^{-1}-(\p_+ g)\,g^{-1}&=&0\,,\cr\cr
(g\cdot h)^{-1}\p_-(g\cdot h)-g^{-1}\,\p_- g&=&\bigl(\sfT_e\Ad_h+\id_\ggt\bigr)\bigl(h^{-1}\,\p_2 h\bigr)+\widetilde\D{}_{(g,h)}\left(h^{-1}\p_2 h\right)\,,
\qqq
and we readily recognise the continuity equation for the left-handed chiral current in the first of them,\ and that for the right-handed current in the first summand on the left-hand side of the second one.\ Taking both into account,\ we end up with the single constraint
\qq\nn
\widetilde\D{}_{(g,h)}\left(h^{-1}\p_- h\right)=-\bigl(\sfT_e\Ad_h+\id_\ggt\bigr)\bigl(h^{-1}\,\p_- h\bigr)\,.
\qqq
In view of the arbitrariness of $\,h$,\ this determines $\,\widetilde\D$,\ and so also $\,\om_\la^{(\sfk)}$. 

Having fixed all components of $\,\om$,\ we proceed to verify that no additional consequences arise from imposition of the requirement of invariance of the DGCs under loop-group transformations of the second type:\ $\,(g,h)(0,\si^2)\longmapsto(g(0,\si^2)\cdot y(\si^-)^{-1},y(\si^-)\cdot h(0,\si^2)\cdot y(\si^-)^{-1})\vert_{\si^1=0}$.\ This boils down to checking invariance of both continuity equations for the chiral Ka\v c--Moody currents under these loop-group transformations,\ a task readily accomplished through direct calculation.

In the next step,\ we check that the 2-form $\,\om_\la^{(\sfk)}\,$ given in \Reqref{eq:om-WZW-nb} (and derived above) has all the required properties.\ Thus,\ we first consider the identity $\,\pr_1^*\txH_{\rm C}^{(\sfk)}-\txm^*\txH_{\rm C}^{(\sfk)}=\sfd\om^{(\sfk)}$.\ That it holds true is a simple corollary of the previous result \eqref{eq:omp-as-prim-H} and of the relation $\,\txm^*\txH_{\rm C}^{(\sfk)}=\pr_1^*\txH_{\rm C}^{(\sfk)}+\pr_2^*\txH_{\rm C}^{(\sfk)}-\sfd\varrho^{(\sfk)}\,$ between 3-forms on $\,\txG\x\txG$.\ The last relation,\ known as the Polyakov--Wiegmann formula ({\it cf.}\ \Rcite{Polyakov:1984et}),\ is easily checked in a direct computation.

Finally,\ we convince ourselves that $\,\om_\la^{(\sfk)}\,$ is also $\ell\wp^{(1)}$-invariant as claimed.\ As $\,\ell\wp^{(1)}\,$ restricts to the adjoint action on the cartesian factor $\,\xcC_\la\,$ of $\,Q_\la\,$ from which the component $\,\pr_2^*\om_\la^{(\sfk)}\,$ is pulled back,\ and the latter was proven $\Ad$-invariant in Appendix \ref{app:proof1},\ it remains to show that the 2-form $\,\varrho^{(\sfk)}\,$ is $\ell\wp^{(1)}$-invariant.\ We leave this elementary exercise to the reader.\qed

\section{A proof of Proposition \ref{prop:Omn-in-kern}}\label{app:Omn-in-kern}

We want to show that the 2-form
\qq
\Om_n^{(\sfk)}=\sum_{i=1}^{n-1}\,\bigl(\pr_i,\txm_{i+1\,i+2\,\cdots\,n}^{(n)}\bigr)^*\varrho^{(\sfk)}
+\txm_n^*\om_\p^{(\sfk)}-\sum_{j=1}^n\,\pr_j^*\om_\p^{(\sfk)}\,,
\qqq
is annihilated by vector fields on the manifold $\,\cT_{\overrightarrow\la}^{[\overrightarrow w]}\,$ defined in
\Reqref{eq:Tlawe}.\ Clearly,\ the tangent space $\,\sfT_{(h_1,h_2,\ldots,h_n)}\cT_{\overrightarrow\la}^{[\overrightarrow w]}=\corr{\xcV_A(h_1,h_2, \ldots,h_n)\ \vert\ A\in{1,\dim\,\txG}}_\bR\,$ is spanned by the fundamental vector fields $\,\xcV_A(h_1,h_2,\ldots,h_n)=\sum_{j=1}^n\,V_A(h_j)\,$ ({\it cf.}\ \Reqref{eq:left-min-right}) for the (diagonal) adjoint action of $\,\txG\,$ on $\,\cT_{\overrightarrow\la}^{[\overrightarrow w]}$,\ and so we have to verify the identities $\,\xcV_A\con\Om_n^{(\sfk)}(h_1,h_2,\ldots,h_n)=0,\ A\in\ovl{1,\dim\,\txG}$.\ Denote $\,h_{k,k+1,\ldots,l}:=h_k\cdot h_{k+1}\cdots h_l,\ 1\leq k\leq l\leq n$.\ Upon substituting the defining expressions for $\,\varrho^{(\sfk)}\,$ and $\,\om_\p^{(\sfk)}$,\ we readily obtain
\qq\nn
\tfrac{4\pi}{\sfk}\,\xcV_A\con\Om_n^{(\sfk)}(h_1,h_2,\ldots,h_n)\cr\cr
=\sum_{i=1}^{n-1}\,\bigl((\sfT_e\Ad_{h_i}-\id_\ggt)\circ\th_{\rm R}(h_{i+1,i+2,\ldots,n})+(\sfT_e\Ad_{h_{i+1,i+2,\ldots,n}^{-1}}-\id_\ggt)\circ\th_{\rm L}(h_i)\bigr)\cr\cr
-(\id_\ggt+\sfT_e\Ad_{h_{1,2,\ldots,n}})\circ\th_{\rm L}(h_{1,2,\ldots,n})+\sum_{j=1}^n\,(\id_\ggt+\sfT_e\Ad_{h_j})\circ\th_{\rm L}(h_j)\cr\cr
=\sum_{j=2}^n\,\bigl(\sfT_e\Ad_{h_{1,2,\ldots,j-1}}-\id_\ggt\bigr)\circ\th_{\rm R}(h_j)+\sum_{i=1}^{n-1}\,\sfT_e\Ad_{h_{i+1,i+2,\ldots,n}^{-1}}\circ\th_{\rm L}(h_i)\cr\cr
-\th_{\rm L}(h_{1,2,\ldots,n})-\th_{\rm R}(h_{1,2,\ldots,n})+\th_{\rm L}(h_n)+\sum_{j=1}^n\,\th_{\rm R}(h_j)\,.
\qqq
The claim now follows from the identities
\qq\nn
\th_{\rm L}(h_{1,2,\ldots,n})=\sum_{i=1}^n\,\sfT_e\Ad_{h_{i+1,i+2,\ldots,n}^{-1}}\circ\th_{\rm L}(h_i)\,,\qquad\qquad\th_{\rm R}(h_{1,2,\ldots,n})=\sum_{i=1}^n\,\sfT_e\Ad_{h_{1,2,\ldots,i-1}}\circ\th_{\rm R}(h_i)\,,
\qqq
easily checked through direct inspection.\qed

\section{A proof of Proposition \ref{prop:Om-AM-vs-RS}}\label{app:Om-AM-vs-RS}

We shall be concerned with 2-forms on the manifold $\,\cT_{\overrightarrow\la}^\la$.\ Let $\,\Om_n^{(\sfk)}\,$ and $\,\Om_{\xcF_{0,n+1}}^{(\sfk)}\,$ be the 2-forms defined in Eqs.\,\eqref{eq:Omn-def} and \eqref{eq:Om-part-red},\ respectively.
Our goal in this appendix is to verify identity \eqref{eq:OmF-as-n} for $\,\hbar\,$ the isomorphism of \Reqref{eq:hbar}.\ We begin by actually identifying the 2-form $\,\Om_{\xcF_{0,n+1}}$,\ so far defined only implicitly.\ Using \Reqref{eq:WZW-brane-curv-param-Cart},\ we find
\qq\nn
\Om_{\xcF_{0,n+1}}^{(\sfk)}=\sum_{i=1}^n\,\pr_i^*\om_\p-\pr_{n+1}^*\om_\p+\hbar^{-1\,*}\sum_{j=1}^{n-1}\,\tfrac{\sfk}{4\pi}\,\tr_\ggt\bigl(\txm_{n+1-j\,n+2-j\,\cdots\,n}^{(n)\,*}\th_{\rm L}\wedge\txm_{n-j\,n+1-j\,\cdots\,n}^{(n)\,*}\th_{\rm L}\bigr)\,,\cr
\qqq
and so,\ upon consulting Eqs.\,\eqref{eq:Omn-def} and \eqref{eq:varrho},\ we conclude that identity
\eqref{eq:OmF-as-n} holds true iff
\qq\nn
\sum_{j=1}^{n-1}\,\tr_\ggt\bigl(\txm_{n+1-j\,n+2-j\,\cdots\,n}^{(n)\,*}\th_{\rm L}\wedge\txm_{n-j\,n+1-j\,\cdots\,n}^{(n)\,*}\th_{\rm L}\bigr)=-\sum_{j=1}^{n-1}\,\tr_\ggt\bigl(\pr_j^*\th_{\rm L}\wedge\txm_{j+1\,j+2\,\cdots\,n}^{(n)\,*}\th_{\rm R}\bigr)\,.\cr
\qqq
The latter readily follows from the identity $\,\tr_\ggt(\th_{\rm L}\wedge\th_{\rm L})=0\,$ on noting that
\qq\nn
\tr_\ggt\bigl(\txm_{n+1-j\,n+2-j\,\cdots\,n}^{(n)\,*}\th_{\rm L}\wedge\txm_{n-j\,n+1-j\,\cdots\,n}^{(n)\,*}\th_{\rm L}\bigr)=-\tr_\ggt\bigl(\pr_{n-j}^*\th_{\rm L}\wedge\txm_{n+1-j\,n+2-j\,\cdots\,n}^*\th_{\rm R}\bigr)\,.
\qqq
Indeed,\ we thus obtain
\qq\nn
-\sum_{j=1}^{n-1}\,\tr_\ggt\bigl(\pr_{n-j}^*\th_{\rm L}\wedge\txm_{n+1-j\,n+2-j\,\cdots\,n}^{(n)\,*}\th_{\rm R}\bigr)\equiv-\sum_{j=1}^{n-1}\,\tr_\ggt\bigl(\pr_{n-j}^*\th_{\rm L}\wedge\txm_{n+1-j\,n+2-j\,\cdots\,n}^*\th_{\rm R}\bigr)\,,
\qqq
which yields the desired result.\qed

\section{A proof of Theorem \ref{thm:mult-str-GS}}\label{app:mult-str-GS}

Below,\ we provide an explicit geometrisation of the super-4-cocycle $\,\cZ_{\rm GS}$.\ In so doing,\ we use the conventions of App.\,\ref{app:convs}.\ Thus,\ upon taking the surjective submersion of the pullback $\,f^*\cG_{\rm GS}\,$ of the GS super-1-gerbe along $\,f\ :\ \bT\x\bT\too\bT,\ f\in\{\txm,\pr_1,\pr_2\}\,$ in the form $\,\sfY_f\bT^{\x 2}\equiv f^*\sfY\bT$,\ and with coordinates $\,(m_1,m_2,(f(m_1,m_2),\xi))\in f^*\sfY\bT$,\ we readily obtain the surjective submersion of the tensor-product 1-gerbe $\,\pr_1^*\cG_{\rm GS}\ox\pr_2^*\cG_{\rm GS}\,$ given by $\,\sfY_{1,2}\bT^{\x 2}=\pr_1^*\sfY\bT\x_{\bT^{\x 2}}\hspace{-1pt}\pr_2^*\sfY\bT$,\ with the projection to the base $\,\pi_{\sfY_{1,2}\bT^{\x 2}}=\pi_{\pr_1^*\sfY\bT}\circ\pr_1\equiv\pi_{\pr_2^*\sfY\bT}\circ\pr_2\,$ and coordinates (here,\ as before,\ $\,y_A\equiv(m_A,\xi_A)\equiv(\theta_A,x_A,\xi_A),\ A\in\{1,2\}$) $\,((m_1,m_2,y_1),(m_1,m_2,y_2))\equiv(\widetilde y{}_1,\widetilde y{}_2)\in\sfY_{1,2}\bT^{\x 2}$.\ The tensor-product 1-gerbe $\,\txm^*\cG_{\rm GS}\ox\cI_{\varrho_{\rm GS}}$,\ on the other hand,\ has the surjective submersion $\,\pi_{\sfY_\txm\bT^{\x 2}}\equiv\pi_{\txm^*\sfY\bT}\ :\ \sfY_\txm\bT^{\x 2}\equiv\txm^*\sfY\bT\too\bT^{\x 2}$,\ and so the reconstruction of the 1-isomorphism $\,\cM\,$ starts with the erection of a principal $\bC^\x$-bundle $\,\pi_E\ :\ E\too\sfY_{1,2,\txm}\bT^{\x 2}\,$ over the fibred product $\,\sfY_{1,2,\txm}\bT^{\x 2}=\pi_{\sfY_{1,2}\bT^{\x 2}}^*\sfY_\txm\bT^{\x 2}$,\ with coordinates $\,((m_1,m_2,y_1),(m_1,m_2,y_2),(m_1,m_2,(m_1\cdot m_2,\xi_3)))\equiv(\widetilde y{}_1,\widetilde y{}_2,\widetilde y{}_\txm)\equiv\widetilde y{}_{1,2,\txm}\in\sfY_{1,2,\txm}\bT^{\x 2}\,$ and the surjective submersion $\,\pi_{\sfY_{1,2,\txm}\bT^{\x 2}}\equiv\pi_{\sfY_{1,2}\bT^{\x 2}}\circ\pr_1\ :\ \sfY_{1,2,\txm}\bT^{\x 2}\too\bT^{\x 2}$.\ The principal $\bC^\x$-connection on that bundle is to have the curvature
\qq\nn
\txF_E(\widetilde y{}_{1,2,\txm})&=&\txB_{\rm GS}(\widetilde y{}_\txm)+\pi_{\sfY_\txm\bT^{\x 2}}\varrho_{\rm GS}(\widetilde y{}_\txm)-\txB_{\rm GS}(\widetilde y{}_1)-\txB_{\rm GS}(\widetilde y{}_2)\,.
\qqq
The above is de Rham-exact by construction,\ and so the cohomological triviality of the underlying supergeometry enables us to employ the homotopy retraction $\,(\t,(\th_1,\th_2))\longmapsto(\t\,\th_1,\t\,\th_2),\ \t\in[0,1]\,$ of the Gra\ss mann-odd fibre to recast it in the desired exact form
\qq\nn
\txF_E(\widetilde y{}_{1,2,\txm})&=&\sfd\bigl[\th_1^\a\,\sfd\xi^{31}_\a+\th_2^\a\,\sfd\xi^{32}_\a-\eta_{ab}\,\bigl(x_1^a\,\sfd x_2^b-x_2^a\,\sfd x_1^b\bigr)\cr\cr
&&+\th_1\,\ovl\G{}_a\,\th_2\,\bigl(e^a_{\rm L}(m_1)+\sfd x_2^a+\tfrac{1}{6}\,\th_2\,\ovl\G{}^a\,\sfd\th_2+\tfrac{1}{2}\,\th_2\,\ovl\G{}^a\,\si_{\rm L}(m_1)\bigr)\bigr]\,.
\qqq
Upon setting 
\qq\nn
E=\sfY_{1,2,\txm}\bT^{\x 2}\x\bC^\x\ni(\widetilde y{}_{1,2,\txm},z)\,,
\qqq
we subsequently take the corresponding principal $\bC^\x$-connection super-1-form to be
\qq\nn
\cA_E(\widetilde y{}_{1,2,\txm},z)=\vartheta(z)+\txA_E(\widetilde y{}_{1,2,\txm})\,,
\qqq
with the base component
\qq\nn
&\txA_E(\widetilde y{}_{1,2,\txm})=\th_1^\a\,\sfd\xi^{31}_\a+\th_2^\a\,\sfd\xi^{32}_\a+\pi_{\sfY_{1,2,\txm}\bT^{\x 2}}^*\sfa_E(y_{1,2,\txm})\,,&\cr\cr
&\sfa_E(m_1,m_2)=-\eta_{ab}\,\bigl(x_1^a\,\sfd x_2^b-x_2^a\,\sfd x_1^b\bigr)+\th_1\,\ovl\G{}_a\,\th_2\,\bigl(e^a_{\rm L}(m_1)+\sfd x_2^a+\tfrac{1}{6}\,\th_2\,\ovl\G{}^a\,\sfd\th_2+\tfrac{1}{2}\,\th_2\,\ovl\G{}^a\,\si_{\rm L}(m_1)\bigr)\,.&
\qqq

In the next step,\ we extend the maps $\,\ell_{(\vep,t)}^{(2)}:=\ell_{(\vep,t)}\x\id_\bT\ :\ \bT^{\x 2}\circlearrowleft,\ (\vep,t)\in\bT\,$ to connection-preserving automorphisms of the principal $\bC^\x$-bundle $\,E\,$ of $\,\cM$.\ To this end,\ we introduce maps $\,(\sfY_{\pr_1}\ell_{(\vep,t)},\sfY_{\pr_2}\ell_{(\vep,t)},\sfY_\txm\ell_{(\vep,t)}):=(\sfY\ell_{(\vep,t,0)},\id_{\sfY\bT},\sfY\ell_{(\vep,t,0)})\,$ and use them to write the maps $\,f^*\ell^{(2)}_{(\vep,t)}:=\ell^{(2)}_{(\vep,t)}\x\sfY_f\ell_{(\vep,t)}\ :\ f^*\sfY\bT\circlearrowleft,\ f\in\{\pr_1,\pr_2,\txm\}\,$ that we combine in the definition $\,\sfY_{1,2,\txm}\ell_{(\vep,t)}:=\pr_1^*\ell^{(2)}_{(\vep,t)}\x\pr_2^*\ell^{(2)}_{(\vep,t)}\x\txm^*\ell^{(2)}_{(\vep,t)}\ :\ \sfY_{1,2,\txm}\bT\circlearrowleft$.\ Using the latter,\ we subsequently postulate
\qq\nn
E\ell^{(2)}_{(\vep,t)}(\widetilde y{}_{1,2,\txm},z):=\bigl(\sfY_{1,2,\txm}\ell_{(\vep,t)}(\widetilde y{}_{1,2,\txm}),\ee^{\sfi\,\chi_{(\vep,t)}(\widetilde y{}_{1,2,\txm})}\cdot z\bigr)
\qqq
and fix $\,\chi_{(\vep,t)}(\widetilde y{}_{1,2,\txm})\,$ through imposition of the requirement $\,E\ell_{(\vep,t)}^{(2)\,*}\cA_E\must\cA_E\,$ for all $\,(\vep,t)\in\bT$.\ That the postulate can be realised follows directly from the identity $\,\sfd(\sfY_{1,2,\txm}\ell_{(\vep,t)}^*\txA_E-\txA_E)=0$,\ readily proven directly with the help of Eqs.\,\eqref{eq:rhoLI} and \eqref{eq:curvLI},\ and implying $\,\sfY_{1,2,\txm}\ell_{(\vep,t)}^*\txA_E-\txA_E\in{\rm Ker}\,\sfd_{\rm dR}^{(1)}\equiv{\rm Im}\,\sfd_{\rm dR}^{(0)}$.\ With the help of the same homotopy retraction as before,\ we find
\qq\nn
\chi_{(\vep,t)}(\widetilde y{}_{1,2,\txm})=\vep^\a\,\xi^{31}_\a-\eta_{ab}\,\bigl(t^a+\tfrac{1}{2}\,\vep\,\ovl\G{}^a\,\theta_1\bigr)\,x_2^b+\tfrac{1}{6}\,\vep\,\ovl\G{}_a\,(2\theta_1+\theta_2)\cdot\theta_1\,\ovl\G{}^a\,\theta_2\,.
\qqq

In the next step,\ we look for a coherent principal-$\bC^\x$-bundle isomorphism of $\,\cM$.\ To this end,\ we first erect over the fibred square $\,\sfY_{1,2,\txm}\bT^{\x 2}\x_{\bT^{\x 2}}\sfY_{1,2,\txm}\bT^{\x 2}\,$ with coordinates $\,(\widetilde y{}_{1,2,\txm}^1,\widetilde y{}_{1,2,\txm}^2$,\ written in terms of the $\,\widetilde y{}_{1,2,\txm}^A\equiv((m_1,m_2,(m_1,\xi_1^A)),(m_1,m_2,(m_2,\xi_2^A)),(m_1,m_2,(m_1\cdot m_2,\xi_3^A)))\equiv((m_1,m_2,y_1^A),(m_1,m_2,$ $y_2^A),(m_1,m_2,(m_1\cdot m_2,\xi_3^A)))\equiv(\widetilde y{}_1^A,\widetilde y{}_2^A,\widetilde y{}_\txm^A),\ A\in\{1,2\}$,\ the two principal $\bC^\x$-bundles:\ $\,\pr_{1,4}^*\widehat\pr{}_1^{[2]*}L\ox\pr_{2,5}^*\widehat\pr{}_2^{[2]*}L\ox\pr_{4,5,6}^*E\,$ and $\,\pr_{1,2,3}^*E\ox\pr_{3,6}^*\widehat\txm{}^{[2]*}L$,\ defined in terms of the maps $\,\widehat f{}^{[2]}\equiv\widehat f\x\widehat f\ :\ \sfY_f^{[2]}\bT^{\x 2}\equiv\sfY_f\bT^{\x 2}\x_{\bT^{\x 2}}\hspace{-1pt}\sfY_f\bT^{\x 2}\too\sfY^{[2]}\bT,\ f\in\{\txm,\pr_1,\pr_2\}$,\ and subsequently compare the base components of the corresponding principal $\bC^\x$-connection super-1-forms,\ only to find the equality $\,\pr_{1,4}^*\widehat\pr{}_1^{[2]*}\txA_L+\pr_{2,5}^*\widehat\pr{}_2^{[2]*}\txA_L+\pr_{4,5,6}^*\txA_E=\pr_{1,2,3}^*\txA_E+\pr_{3,6}^*\widehat\txm{}^{[2]*}\txA_L\,$ and hence to conclude that the two tensor-product principal $\bC^\x$-bundles are,\ in fact,\ trivially isomorphic,
\qq\label{eq:alEmult}\qquad\qquad
\a_E\equiv\bd1\ :\ \pr_{1,4}^*\widehat\pr{}_1^{[2]*}L\ox\pr_{2,5}^*\widehat\pr{}_2^{[2]*}L\ox\pr_{4,5,6}^*E\xrightarrow{\ \cong\ }\pr_{1,2,3}^*E\ox\pr_{3,6}^*\widehat\txm{}^{[2]*}L\,,
\qqq
the isomorphism being manifestly coherent with the trivial groupoid structure $\,\mu_L\equiv\bd1$.\ We still need to check that it is also $\bT$-equivariant,\ {\it i.e.},\ that we have the equality 
\qq\nn
&&\a_E\circ\bigl(\pr_{1,4}^*\widehat\pr{}_1^{[2]*}L\ell_{((\vep,t,0),(\vep,t,0),1)}\ox\pr_{2,5}^*\widehat\pr{}_2^{[2]*}L\ell_{((0,0,0),(0,0,0),1)}\ox\pr_{4,5,6}^*E\ell^{(2)}_{(\vep,t)}\bigr)\cr\cr
&=&\bigl(\pr_{1,2,3}^*E\ell^{(2)}_{(\vep,t)}\ox\pr_{3,6}^*\widehat\txm{}^{[2]*}L\ell_{((\vep,t,0),(\vep,t,0),1)}\bigr)\circ\a_E
\qqq
for
\qq\nn
&&\bigl(\pr_{1,4}^*\widehat\pr{}_1^{[2]*}L\ell_{((\vep,t,0),(\vep,t,0),1)}\ox\pr_{2,5}^*\widehat\pr{}_2^{[2]*}L\ell_{((0,0,0),(0,0,0),1)}\ox\pr_{4,5,6}^*E\ell^{(2)}_{(\vep,t)}\bigr)\bigl(\widetilde y{}_{1,2,\txm}^1,\widetilde y{}_{1,2,\txm}^2,z\bigr)\cr\cr
&=&\bigl(\sfY_{1,2,\txm}\ell_{(\vep,t)}\bigl(\widetilde y{}_{1,2,\txm}^1\bigr),\sfY_{1,2,\txm}\ell_{(\vep,t)}\bigl(\widetilde y{}_{1,2,\txm}^2\bigr),\ee^{\sfi\,(\la((\vep,t,0),(\vep,t,0),y_1^1,y_1^2)+\chi_{(\vep,t)}(\widetilde y{}_{1,2,\txm}^2))}\cdot z\bigr)
\qqq
and
\qq\nn
&&\bigl(\pr_{1,2,3}^*E\ell^{(2)}_{(\vep,t)}\ox\pr_{3,6}^*\widehat\txm{}^{[2]*}L\ell_{((\vep,t,0),(\vep,t,0),1)}\bigr)\bigl(\widetilde y{}_{1,2,\txm}^1,\widetilde y{}_{1,2,\txm}^2,z\bigr)=\bigl(\sfY_{1,2,\txm}\ell_{(\vep,t)}\bigl(\widetilde y{}_{1,2,\txm}^1\bigr),\sfY_{1,2,\txm}\ell_{(\vep,t)}\bigl(\widetilde y{}_{1,2,\txm}^2\bigr),\cr\cr
&&\ee^{\sfi\,(\chi_{(\vep,t)}(\widetilde y{}_{1,2,\txm}^1)+\la((\vep,t,0),(\vep,t,0),(m_1\cdot m_2,\xi_3^1),(m_1\cdot m_2,\xi_3^2)))}\cdot z\bigr)\,.
\qqq
This follows directly from the identity
\qq\nn
\la\bigl((\vep,t,0),(\vep,t,0),y_1^1,y_1^2\bigr)+\chi_{(\vep,t)}\bigl(\widetilde y{}_{1,2,\txm}^2\bigr)=\chi_{(\vep,t)}\bigl(\widetilde y{}_{1,2,\txm}^1\bigr)+\la\bigl((\vep,t,0),(\vep,t,0),\bigl(m_1\cdot m_2,\xi_3^1\bigr),\bigl(m_1\cdot m_2,\xi_3^2\bigr)\bigr)\,.
\qqq
Thus,\ we obtain the ($\ell^{(2)}$-)supersymmetric 1-gerbe 1-isomorphism
\qq\nn
\cM_{\rm GS}\equiv\bigl(\sfY_{1,2,\txm}\bT^{\x 2},\pi_{\sfY_{1,2,\txm}\bT^{\x 2}},\sfY_{1,2,\txm}\bT^{\x 2}\x\bC^\x,\pr_1,\cA_E,\bd1\bigr)\ :\  d_2^{(2)\,*}\cG_{\rm GS}\ox d_0^{(2)\,*}\cG_{\rm GS}\xrightarrow{\ \cong\ } d_1^{(2)\,*}\cG_{\rm GS}\ox\cI_{\varrho_{\rm GS}}\,,
\qqq
with the lift of the action $\,\ell^{(2)}\,$ of the supersymmetry group $\,\bT\,$ to the total space $\,\sfY_{1,2,\txm}\bT^{\x 2}\x\bC^\x\,$ of its principal $\bC^\x$-bundle given by $\,E\ell^{(2)}_\cdot$.

With the 1-isomorphism in hand,\ we now look for a ($\ell^{(3)}$-)supersymmetric 1-gerbe 2-isomorphism $\,\a_{\rm GS}\,$ defined (just like $\,\a$) in Diagram \eqref{diag:aleph} but for the quintuple $\,(\txG,\cG,\cM,\varrho,\vartheta)=(\bT,\cG_{\rm GS},\cM_{\rm GS},\varrho_{\rm GS},\vartheta_{\rm GS})$.\ To this end,\ we first establish the (super)differential-geometric data of the two composite 1-gerbe 1-isomorphisms over $\,\bT^{\x 3}\,$ to be related by $\,\a_{\rm GS}$,\ {\it i.e.},\ $\,(\id_{\cG^{(3)}_{{\rm GS}\,123}}\ox\cJ_{-\vartheta_{\rm GS}})\circ( d_2^{(3)\,*}\cM_{\rm GS}\ox\id_{\cI_{ d_0^{(3)\,*}\varrho_{\rm GS}}})\circ(\id_{\cG^{(3)}_{{\rm GS}\,1}}\ox d_0^{(3)\,*}\cM_{\rm GS})\,$ and $\,(d_1^{(3)\,*}\cM_{\rm GS}\ox\id_{\cI_{ d_3^{(3)\,*}\varrho_{\rm GS}}})\circ(d_3^{(3)\,*}\cM_{\rm GS}\ox\id_{\cG^{(3)}_{{\rm GS}\,3}})$.\ In so doing,\ we employ (implicitly) some obvious and fairly straightforward identifications among the supergeometries involved which enable us to truncate and simplify the otherwise lengthy and tedious construction.\ Upon denoting,\ over $\,\bT^3$,\ the various relevant pullbacks as $\,\sfY_i\bT^{\x 3}=\pr_i^*\sfY\bT,\ i\in\{1,2,3\}\,$ and $\,\sfY_{123}\bT^{\x 3}=\txm_3^*\sfY\bT$,\ as well as $\,\sfY_{jk}\bT^{\x 3}=\pr_{j,k}^{(3)\,*}\sfY_\txm\bT^{\x 2},\ (j,k)\in\{(1,2),(2,3)\}$,\ and forming their $\bT^{\x 3}$-fibred multiple products:\ $\,\sfY_{1,23}\bT^{\x 3}\equiv(\sfY_1\bT^{\x 3}\x_{\bT^{\x 3}}\sfY_2\bT^{\x 3}\x_{\bT^{\x 3}}\sfY_3\bT^{\x 3})\x_{\bT^{\x 3}}(\sfY_1\bT^{\x 3}\x_{\bT^{\x 3}}\sfY_{23}\bT^{\x 3})\x_{\bT^{\x 3}}\sfY_{123}\bT^{\x 3}\x_{\bT^{\x 3}}\sfY_{123}\bT^{\x 3}\ni(\widehat y{}_1,\widehat y{}_2,\widehat y{}_3,\widehat y{}^2_1,\widehat y{}_{23},\widehat y{}_{123}^1,\widehat y{}_{123}^2)\,$ and $\,\sfY_{12,3}\bT^{\x 3}\equiv(\sfY_1\bT^{\x 3}\x_{\bT^{\x 3}}\sfY_2\bT^{\x 3}\x_{\bT^{\x 3}}\sfY_3\bT^{\x 3})\x_{\bT^{\x 3}}(\sfY_{12}\bT^{\x 3}\x_{\bT^{\x 3}}\sfY_3\bT^{\x 3})\x_{\bT^{\x 3}}\sfY_{123}\bT^{\x 3}\ni(\widehat y{}_1,\widehat y{}_2,\widehat y{}_3,\widehat y{}_{12},\widehat y{}^2_3,\widehat y{}_{123}^2)\,$ with the respective coordinates written in terms of $\,\widehat y{}_A=(m_1,m_2,m_3,(m_A,\xi_A)),\ A\in\{1,2,3\};\ \widehat y{}^2_B=(m_1,m_2,m_3,(m_B,\xi^2_B)),\ B\in\{1,3\};\ \widehat y{}_{23}\equiv(m_1,m_2,m_3,(m_2,m_3,(m_2\cdot m_3,\xi_4)));\ \widehat y{}_{12}\equiv(m_1,m_2,m_3,(m_1,m_2,(m_1\cdot m_2,\xi_5)))\,$ and $\,\widehat y{}_{123}^C\equiv(m_1,m_2,m_3,(m_1\cdot m_2\cdot m_3,\xi_{5+C})),\ C\in\{1,2\}\,$ and the respective projections to the base:\
$\,\pi_{\sfY_{1,23}\bT^{\x 3}}\equiv\pi_{\sfY_{123}\bT^{\x 3}}\circ\pr_7\ :\ \sfY_{1,23}\bT^{\x 3}\too\bT^{\x 3}\,$ and $\,\pi_{\sfY_{12,3}\bT^{\x 3}}\equiv\pi_{\sfY_{123}\bT^{\x 3}}\circ\pr_6\ :\ \sfY_{12,3}\bT^{\x 3}\too\bT^{\x 3}$,\ we find as the principal $\bC^\x$-bundles of the above two 1-isomorphisms 
\qq\nn
\pi_{LE_{1,23}}\ :\ LE_{1,23}=\pr_{1,4}^*\widehat\pr{}_1^{(3)\,*}L\ox\pr_{6,7}^*\widehat m{}_3^*L\ox\pr_{2,3,5}^*\widehat\pr{}_{2,3}^{(3)\,*}E\ox\pr_{4,5,6}^*\widehat\txm{}^{(3)\,*}_{1,23}E\ox(\sfY_{1,23}\bT^{\x 3}\x\bC^\x)\too\sfY_{1,23}\bT^{\x 3}
\qqq
and
\qq\nn
\pi_{LE_{12,3}}\ :\ LE_{12,3}=\pr_{3,5}^*\widehat\pr{}_3^{(3)\,*}L\ox\pr_{1,2,4}^*\widehat\pr{}_{1,2}^{(3)\,*}E\ox\pr_{4,5,6}^*\widehat\txm{}^{(3)\,*}_{12,3}E\too\sfY_{12,3}\bT^{\x 3}\,,
\qqq
respectively,\ where $\,\pr_{1,4}^*\widehat\pr{}_1^{(3)\,*}L\,$ (resp.\ $\,\pr_{3,5}^*\widehat\pr{}_3^{(3)\,*}L$) is the (obvious) pullback of the principal $\bC^\x$-bundle $\,L\,$ of $\,\cG_{\rm GS}\,$ to the factor $\,\sfY_1\bT^{\x 3}\x_{\bT^{\x 3}}\sfY_1\bT^{\x 3}\,$ within $\,\sfY_{1,23}\bT^{\x 3}\,$ (resp.\ to $\,\sfY_3\bT^{\x 3}\x_{\bT^{\x 3}}\sfY_3\bT^{\x 3}\,$ within $\,\sfY_{12,3}\bT^{\x 3}$),\ $\,\pr_{6,7}^*\widehat m{}_3^*L\,$ is the pullback of the same bundle to 
$\,\sfY_{123}\bT^{\x 3}\x_{\bT^{\x 3}}\sfY_{123}\bT^{\x 3}$,\ $\,\pr_{2,3,5}^*\widehat\pr{}_{2,3}^{(3)\,*}E\,$ (resp.\ $\,\pr_{1,2,4}^*\widehat\pr{}_{1,2}^{(3)\,*}E$) is the pullback of the principal $\bC^\x$-bundle $\,E\,$ of $\,\cM_{\rm GS}\,$ to the factor $\,\sfY_2\bT^{\x 3}\x_{\bT^{\x 3}}\sfY_3\bT^{\x 3}\x_{\bT^{\x 3}}\sfY_{23}\bT^{\x 3}\,$ within $\,\sfY_{1,23}\bT^{\x 3}\,$ (resp.\ to $\,\sfY_1\bT^{\x 3}\x_{\bT^{\x 3}}\sfY_2\bT^{\x 3}\x_{\bT^{\x 3}}\sfY_{12}\bT^{\x 3}\,$ within $\,\sfY_{12,3}\bT^{\x 3}$) and similarly for $\,\pr_{4,5,6}^*\widehat\txm{}^{(3)\,*}_{1,23}E\,$ (resp.\ $\,\pr_{4,5,6}^*\widehat\txm{}^{(3)\,*}_{12,3}E$),\ and where the last tensor factor $\,\sfY_{1,23}\bT^{\x 3}\x\bC^\x\,$ in $\,LE_{1,23}\,$ contributes the LI correction $\,-\pi_{LE_{1,23}}^*\pi_{\sfY_{1,23}\bT^{\x 3}}^*\vartheta_{\rm GS}\,$ to the tensor-product principal $\bC^\x$-connection (super-)1-form.

Equivalently,\ and conveniently,\ we may represent the two composite bundles as the (trivial) principal $\bC^\x$-bundles (just as all the tensor factors above) $\,LE_{1,23}\cong\sfY_{1,23}\bT^{\x 3}\x\bC^\x\ni((\widehat y{}_1,\widehat y{}_2,\widehat y{}_3,\widehat y{}^2_1,\widehat y{}_{23},\widehat y{}_{123}^1,\widehat y{}_{123}^2),$ $z)\equiv(\widehat y{}_{1,23},z)\,$ and $\,LE_{12,3}\cong\sfY_{12,3}\bT^{\x 3}\x\bC^\x\ni((\widehat y{}_1,\widehat y{}_2,\widehat y{}_3,\widehat y{}_{12},\widehat y{}^2_3,\widehat y{}_{123}^2),z)\equiv(\widehat y{}_{12,3},z)\,$ with the respective principal $\bC^\x$-connection (super-)1-forms given by the formul\ae
\qq\nn
&\cA_{LE_{1,23}}\bigl(\bigl(\widehat y{}_1,\widehat y{}_2,\widehat y{}_3,\widehat y{}^2_1,\widehat y{}_{23},\widehat y{}_{123}^1,\widehat y{}_{123}^2\bigr),z\bigr)=\vartheta(z)+\sfa_{1,23}(m_1,m_2,m_3)\,,&\cr\cr
&\sfa_{1,23}(m_1,m_2,m_3)=\th_1^\a\,\sfd\xi_{71\,\a}+\th_2^\a\,\sfd\xi_{72\,\a}+\th_3^\a\,\sfd\xi_{73\,\a}+\sfa_E(m_2,m_3)+\sfa_E(m_1,m_2\cdot m_3)-\vartheta_{\rm GS}(m_1,m_2,m_3)\,,&\cr\cr\cr
&\cA_{LE_{12,3}}\bigl(\bigl(\widehat y{}_1,\widehat y{}_2,\widehat y{}_3,\widehat y{}_{12},\widehat y{}^2_3,\widehat y{}_{123}^2\bigr),z\bigr)=\vartheta(z)+\sfa_{12,3}(m_1,m_2,m_3)\,,&\cr\cr
&\sfa_{12,3}(m_1,m_2,m_3)=\th_1^\a\,\sfd\xi_{71\,\a}+\th_2^\a\,\sfd\xi_{72\,\a}+\th_3^\a\,\sfd\xi_{73\,\a}+\sfa_E(m_1,m_2)+\sfa_E(m_1\cdot m_2,m_3)
\qqq
(in which $\,\xi_{7k}\equiv\xi_7-\xi_k,\ k\in\{1,2,3\}$) and endowed with the (elementwise) $\bT$-actions that are inherited from the natural ones on the respective bases,\ to wit,\ $\,\sfY_{1,23}\ell_{(\vep,t)}:=\sfY_1\ell^{(3)}_{(\vep,t)}\x\sfY_2\ell^{(3)}_{(\vep,t)}\x\sfY_3\ell^{(3)}_{(\vep,t)}\x\sfY_1\ell^{(3)}_{(\vep,t)}\x\sfY_{23}\ell^{(3)}_{(\vep,t)}\x\sfY_{123}\ell^{(3)}_{(\vep,t)}\x\sfY_{123}\ell^{(3)}_{(\vep,t)}\,$ and $\,\sfY_{12,3}\ell_{(\vep,t)}:=\sfY_1\ell^{(3)}_{(\vep,t)}\x\sfY_2\ell^{(3)}_{(\vep,t)}\x\sfY_3\ell^{(3)}_{(\vep,t)}\x\sfY_{12}\ell^{(3)}_{(\vep,t)}\x\sfY_3\ell^{(3)}_{(\vep,t)}\x\sfY_{123}\ell^{(3)}_{(\vep,t)}$,\ where $\,\sfY_X\ell^{(3)}_{(\vep,t)}:=\ell^{(3)}_{(\vep,t)}\x\sfY_X\ell_{(\vep,t)},\ X\in\{1,2,3,12,23,123\}\,$ and where $\,(\sfY_1\ell_{(\vep,t)},\sfY_2\ell_{(\vep,t)},\sfY_3\ell_{(\vep,t)}):=(\sfY\ell_{(\vep,t,0)},\id_{\sfY\bT},\id_{\sfY\bT}),\ (\sfY_{12}\ell_{(\vep,t)},\sfY_{23}\ell_{(\vep,t)}):=(\txm^*\ell^{(2)}_{(\vep,t)},\id_{\sfY_\txm\bT^{\x 2}}),\ \sfY_{123}\ell_{(\vep,t)}:=\sfY\ell_{(\vep,t,0)}$,\  
and those on the component bundles found earlier,\ and read
\qq\nn
LE_{1,23}\ell^{(3)}_{(\vep,t)}\ :\ \sfY_{1,23}\bT^{\x 3}\x\bC^\x\circlearrowleft\ :\ \bigl(\widehat y{}_{1,23},z\bigr)\longmapsto\bigl(\sfY_{1,23}\ell_{(\vep,t)}\bigl(\widehat y{}_{1,23}\bigr),\ee^{\sfi\,\Phi^{1,23}_{(\vep,t)}(\widehat y{}_{1,23})}\cdot z\bigr)
\qqq
and 
\qq\nn
LE_{12,3}\ell^{(3)}_{(\vep,t)}\ :\ \sfY_{12,3}\bT^{\x 3}\x\bC^\x\circlearrowleft\ :\ \bigl(\widehat y{}_{12,3},z\bigr)\longmapsto\bigl(\sfY_{12,3}\ell_{(\vep,t)}\bigl(\widehat y{}_{12,3}\bigr),\ee^{\sfi\,\Phi^{12,3}_{(\vep,t)}(\widehat y{}_{12,3})}\cdot z\bigr)\,,
\qqq
where
\qq\nn
\Phi^{1,23}_{(\vep,t)}\bigl(\widehat y{}_{1,23}\bigr)=\vep^\a\,\xi_{71\,\a}-\eta_{ab}\,\bigl(t^a+\tfrac{1}{2}\,\vep\,\ovl\G{}^a\,\th_1\bigr)\,\bigl(x_2^b+x_3^b-\tfrac{1}{2}\,\th_2\,\ovl\G{}^b\,\th_3\bigr)+\tfrac{1}{6}\,\vep\,\ovl\G{}_a\,\bigl(2\th_1+\th_2+\th_3\bigr)\cdot\th_1\,\ovl\G{}^a\,\bigl(\th_2+\th_3\bigr)
\qqq
and
\qq\nn
\Phi^{12,3}_{(\vep,t)}\bigl(\widehat y{}_{12,3}\bigr)&=&\vep^\a\,\xi_{71\,\a}-\eta_{ab}\,\bigl(t^a+\tfrac{1}{2}\,\vep\,\ovl\G{}^a\,\th_1\bigr)\,x_2^b-\eta_{ab}\,\bigl(t^a+\tfrac{1}{2}\,\vep\,\ovl\G{}^a\,(\th_1+\th_2)\bigr)\,x_3^b+\tfrac{1}{6}\,\vep\,\ovl\G{}_a\,\bigl(2\th_1+\th_2\bigr)\cdot\th_1\,\ovl\G{}^a\,\th_2\cr\cr
&&+\tfrac{1}{6}\,\vep\,\ovl\G{}_a\,\bigl(2\th_1+2\th_2+\th_3\bigr)\cdot(\th_1+\th_2)\,\ovl\G{}^a\,\th_3\,.
\qqq

Now,\ the point of departure in the construction of the 1-gerbe 2-isomorphism $\,\a\,$ is the identification of the bases of $\,LE_{12,3}\,$ and $\,LE_{1,23}\,$ as total spaces of the respective surjective submersions:\ $\,\pr_{1,2,3,7}\ :\ \sfY_{12,3}\bT^{\x 3}\too(\sfY_1\bT^{\x 3}\x_{\bT^{\x 3}}\sfY_2\bT^{\x 3}\x_{\bT^{\x 3}}\sfY_3\bT^{\x 3})\x_{\bT^{\x 3}}\sfY_{123}\bT^{\x 3}\equiv\sfY_{1,2,3}\bT^{\x 3}\,$ and $\,\pr_{1,2,3,6}\ :\ \sfY_{1,23}\bT^{\x 3}\too\sfY_{1,2,3}\bT^{\x 3}\,$ over the surjective submersion $\,\pi_{\sfY_{1,2,3}\bT^{\x 3}}\equiv\pi_{\sfY_{123}\bT^{\x 3}}\circ\pr_4\ :\ \sfY_{1,2,3}\bT^{\x 3}\too\bT^{\x 3}$,\ the latter being a fibre product of the surjective submersions of the 1-gerbes $\,\cG^{(3)}_{{\rm GS}\,1}\ox\cG^{(3)}_{{\rm GS}\,2}\ox\cG^{(3)}_{{\rm GS}\,3}\,$ and $\,\cG^{(3)}_{123}\ox\cI_{ d_3^{(3)\,*}\varrho_{\rm GS}+ d_1^{(3)\,*}\varrho_{\rm GS}}\,$ related by the two 1-gerbe 1-isomorphisms that we consider above.\ Thus,\ in order to establish $\,\a$,\ we take the fibre product of the two surjective submersions over their common base,\ $\,\sfY_{12,3\x1,23}\bT^{\x 3}=\sfY_{12,3}\bT^{\x 3}\x_{\sfY_{1,2,3}\bT^{\x 3}}\hspace{-1pt}\sfY_{1,23}\bT^{\x 3}$,\ and,\ over it,\ look for an isomorphism of the pullback principal $\bC^\x$-bundles (with connection) $\,\pr_1^*LE_{12,3}=\pr_1^*(\sfY_{12,3}\bT^{\x 3}\x\bC^\x)\equiv\sfY_{12,3\x1,23}\bT^{\x 3}{}_{\pr_1}\hspace{-3pt}\x_{\pr_1}\hspace{-1pt}(\sfY_{12,3}\bT^{\x 3}\x\bC^\x)\,$ and $\,\pr_2^*LE_{1,23}=\pr_2^*(\sfY_{1,23}\bT^{\x 3}\x\bC^\x)\equiv\sfY_{12,3\x1,23}\bT^{\x 3}{}_{\pr_2}\hspace{-3pt}\x_{\pr_1}\hspace{-1pt}(\sfY_{1,23}\bT^{\x 3}\x\bC^\x)$.\ In the light of the (de Rham-)cohomological triviality of the supergeometries involved,\ the existence of the isomorphism sought after is ensured by the vanishing of the exterior derivative of the difference of the base components of the two pullback connection super-1-forms,\ $\,\sfd(\sfa_{12,3}-\sfa_{1,23})=0$.\ This we check in a direct computation using the definitions of the curvatures of $\,L\,$ and $\,E\,$ alongside that of $\,\vartheta_{\rm GS}$.\ At this stage,\ we merely have to extract the data of the isomorphism from the comparison between $\,\sfa_{1,23}\,$ and $\,\sfa_{12,3}$.\ Assuming the isomorphism to have the form
\qq\nn
\b\ &:&\ \pr_1^*LE_{12,3}\too\pr_2^*LE_{1,23}\cr\cr 
&:&\ \bigl(\bigl(\widehat y{}_1,\widehat y{}_2,\widehat y{}_3,\widehat y{}_{123}^2\bigr),\bigl(\widehat y{}_{12,3},z\bigr)\bigr)\longmapsto(\bigl(\widehat y{}_1,\widehat y{}_2,\widehat y{}_3,\widehat y{}_{123}^2\bigr),\bigl(\widehat y{}_{1,23},\ee^{\sfi\,\Phi(\widehat y{}_1,\widehat y{}_2,\widehat y{}_3,\widehat y{}_{12},\widehat y{}_3^2,\widehat y{}_{123}^2)}\cdot z\bigr)\bigr)\,,
\qqq
we arrive at
\qq\nn
\sfd\Phi\bigl(\widehat y{}_1,\widehat y{}_2,\widehat y{}_3,\widehat y{}_{12},\widehat y{}_3^2,\widehat y{}_{123}^2\bigr)=(\sfa_{1,23}-\sfa_{12,3})(m_1,m_2,m_3)\equiv\bigl(\D^{(2;2)}_\bT\sfa_E-\vartheta_{\rm GS}\bigr)(m_1,m_2,m_3)\,,
\qqq
and so 
\qq\nn
\Phi\bigl(\widehat y{}_1,\widehat y{}_2,\widehat y{}_3,\widehat y{}_{12},\widehat y{}_3^2,\widehat y{}_{123}^2\bigr)\equiv\widetilde\Phi(m_1,m_2,m_3)\,,
\qqq
with the latter derived with the help of the usual homotopy retraction $\,(\t,(\th_1,\th_2,\th_3))\longmapsto(\t\,\th_1,\t\,\th_2,\t\,\th_3),$ $\t\in[0,1]\,$ in the form
\qq\nn
\widetilde\Phi(m_1,m_2,m_3)=\tfrac{1}{2}\,\bigl(x_1^a\,\th_2\,\ovl\G{}_a\,\th_3+x_3^a\,\th_1\,\ovl\G{}_a\,\th_2\bigr)-\tfrac{1}{6}\,\th_1\,\ovl\G{}_a\,\bigl(2\th_2+\th_3\bigr)\cdot\th_2\,\ovl\G{}^a\,\th_3\,.
\qqq
Coherence of the thus reconstructed isomorphism $\,\b\,$ with the trivial isomorphism $\,\a_E\equiv\bd1\,$ is a straightforward consequence of the dependence of the data $\,\Phi\,$ of the former on the base coordinates $\,(m_1,m_2,m_3)\,$ exclusively.

We are finally in a position to check the supersymmetry of the 1-gerbe 2-isomorphism $\,\a_{\rm GS}$.\ To this end,\ define the composite actions $\,\pr_1^*LE_{12,3}\ell^{(3)}_{(\vep,t)}\equiv\sfY_1\ell_{(\vep,t)}\x\sfY_2\ell_{(\vep,t)}\x\sfY_3\ell_{(\vep,t)}\x\sfY_{123}\ell_{(\vep,t)}\x LE_{12,3}\ell^{(3)}_{(\vep,t)}\ :\ \pr_1^*LE_{12,3}\circlearrowleft\,$ and $\,\pr_2^*LE_{1,23}\ell^{(3)}_{(\vep,t)}\equiv\sfY_1\ell_{(\vep,t)}\x\sfY_2\ell_{(\vep,t)}\x\sfY_3\ell_{(\vep,t)}\x\sfY_{123}\ell_{(\vep,t)}\x LE_{1,23}\ell^{(3)}_{(\vep,t)}\ :\ \pr_2^*LE_{1,23}\circlearrowleft$.\ The requirement of supersymmetry(-equivariance) now reads 
\qq\nn
\b\circ\pr_1^*LE_{12,3}\ell^{(3)}_{(\vep,t)}\must\pr_2^*LE_{1,23}\ell^{(3)}_{(\vep,t)}\circ\b\,,
\qqq
or,\ equivalently,
\qq\nn
\Phi^{12,3}_{(\vep,t)}\bigl(\widehat y{}_{12,3}\bigr)+\widetilde\Phi\bigl(\ell_{(\vep,t)}(m_1),m_2,m_3\bigr)\must\widetilde\Phi(m_1,m_2,m_3)+\Phi^{1,23}_{(\vep,t)}\bigl(\widehat y{}_{1,23}\bigr)\,,
\qqq
and is readily verified to follow from the Fierz identity \eqref{eq:Fierz}.\ Thus,\ the existence of the anticipated supersymmetric 1-gerbe 2-isomorphism 
\qq\nn
\a_{\rm GS}\equiv\bigl(\sfY_{12,3\x1,23}\bT^{\x 3},\pr_{1,2,3,6}\circ\pr_1,\b\bigr)
\qqq
is demonstrated.

We conclude the reconstruction of the supersymmetric multiplicative structure on the Green--Schwarz super-1-gerbe by convincing ourselves that the 2-isomorphism just found,\ $\,\a_{\rm GS}$,\ satisfies the coherence condition expressed by the commutativity of Diag.\,\eqref{diag:cohmult}.\ As the data $\,\widetilde\Phi\,$ of the 2-isomorphism depend only on the projection all the way to $\,\bT^{\x 3}$,\ the latter condition takes the simple form $\,\D^{(3;0)}_\bT\widetilde\Phi=0$,\ and is readily shown to be equivalent to the Fierz identity \eqref{eq:Fierz}.

\section{A proof of Proposition \ref{prop:sstring-GS-brane}}\label{app:sstring-GS-brane}

In what follows,\ we construct a 1-gerbe 1-isomorphism \eqref{eq:hyper-triv} invariant,\ in the previously discussed sense,\ with respect to the left action of the residual supersymmetry group 
\qq\nn
\bD_{(1,1|16)}\equiv{\rm sMink}(1,1|16)\,.
\qqq
As usual,\ we begin by setting up a convenient hands-on description of the pullback super-1-gerbe $\,\ep_{(1,1|16)}^{(\widehat x{}_*)\,*}\cG_{\rm GS}\,$ to the $(1+1|16)$-superdimensional worldvolume of a leaf $\,D_{(1,1|16)}^{(\widehat x{}_*)}\,$ `through' a fixed $\,(\widehat x{}^{\widehat a}),\ \widehat a\in\ovl{2,9}$,\ with the structure sheaf generated by the global coordinates $\,(\breve\th{}^{\unl\a},\breve x{}^{\unl a}),\ (\unl\a,\unl a)\in\ovl{1,16}\x\{0,1\}$ as introduced above.\ Its surjective submersion shall be taken in the form $\,\pi_{\sfY D_{(1,1|16)}^{(\widehat x{}_*)}}\equiv\pr_1\ :\ \sfY D_{(1,1|16)}^{(\widehat x{}_*)}=\ep_{(1,1|16)}^{(\widehat x{}_*)\,*}\sfY\bT\too D_{(1,1|16)}^{(\widehat x{}_*)}$,\ with the coordinates $\,((\breve\th{}^{\unl\a},\breve x{}^{\unl a}),(\breve{\unl\th}{}^\b,\breve x{}^{\unl a},\widehat x{}^{\widehat b}_*,\breve\xi{}_{\unl\d},\widehat\xi_{\widehat\ep}))\equiv(\breve m,(\ep_{(1,1|16)}^{(\widehat x{}_*)}(\breve m),\xi))\equiv\breve y\in\sfY D_{(1,1|16)}^{(\widehat x{}_*)}$.\ On it,\ we find the pullback curving
\qq\nn
\widehat\ep{}_{(1,1|16)}^{(\widehat x{}_*)\,*}\txB_{\rm GS}\bigl(\breve y\bigr)=\breve\th\,\ovl\g{}_0\,\sfd\breve\th\wedge\sfd\breve x{}_++\sfd\breve\xi{}_{\unl\a}\wedge\sfd\breve\th{}^{\unl\a}\equiv\breve e{}^{(2)}_{\unl\a}\bigl(\breve\th,\breve x,\breve\xi\bigr)\wedge\breve\si{}_{\rm L}^{\unl\a}\bigl(\breve\th,\breve x\bigr)\,,
\qqq
written in terms of the LI super-1-forms:\ $\,\breve\si{}_{\rm L}^{\unl\a}\,$ (introduced earlier) and 
\qq\nn
\breve e{}^{(2)}_{\unl\a}\bigl(\breve\th,\breve x,\breve\xi\bigr)=\sfd\breve\xi{}_{\unl\a}-\ovl\G{}_{\unl a\,\unl\a\unl\b}\,\breve\th{}^{\unl\b}\,\bigl(\sfd\breve x{}^{\unl a}+\tfrac{1}{6}\,\breve\theta\,\ovl\G{}^{\unl a}\,\sfd\breve\th\bigr)\,,
\qqq
{\it cf.}\ \Reqref{eq:LIBGS},\ and of the light-cone coordinates $\,\breve x{}_+\equiv\breve x{}^0+\breve x{}^1$.\
Next,\ over the $D^{(\widehat x{}_*)}_1$-fibred square of the pullback surjective submersion,\ $\,\sfY^{[2]}D_{(1,1|16)}^{(\widehat x{}_*)}=\widehat\ep{}_{(1,1|16)}^{(\widehat x{}_*)\,[2]\,*}\sfY^{[2]}\bT$,\ written for $\,\pi_{\sfY^{[2]}\bT}=\pi_{\sfY\bT}\circ\pr_1\equiv\pi_{\sfY\bT}\circ\pr_2$,\ with coordinates $\,(((\breve\th{}^{\unl\a},\breve x{}^{\unl a}),(\breve{\unl\th}{}^\b,\breve x{}^{\unl a},\widehat x{}^{\widehat b}_*,\breve\xi{}^1_{\unl\d},\widehat\xi^1_{\widehat\ep})),((\breve\th{}^{\unl\a},\breve x{}^{\unl a}),(\breve{\unl\th}{}^\b,\breve x{}^{\unl a},\widehat x{}^{\widehat b}_*,\breve\xi{}^2_{\unl\d},\widehat\xi^2_{\widehat\ep})))\equiv((\breve m,(\ep_{(1,1|16)}^{(\widehat x{}_*)}(\breve m),\xi^1)),(\breve m,$ $(\ep_{(1,1|16)}^{(\widehat x{}_*)}(\breve m),\xi^2)))\equiv(\breve y{}_1,\breve y{}_2)$,\ 
we erect the pullback principal $\bC^\x$-bundle $\,\widehat\ep{}_{(1,1|16)}^{(\widehat x{}_*)\,[2]\,*}L\,$ which may equivalently,\ and conveniently,\ be presented as the (trivial) principal $\bC^\x$-bundle 
\qq\label{eq:LD1triv}
\widehat\ep{}_{(1,1|16)}^{(\widehat x{}_*)\,[2]\,*}L\cong\sfY^{[2]}D_{(1,1|16)}^{(\widehat x{}_*)}\x\bC^\x
\qqq
with the (global) coordinates $\,(\breve y{}_1,\breve y{}_2,z)$,\ and endowed with the principal $\bC^\x$-connection
\qq\nn
\cA_{\widehat\ep{}_{(1,1|16)}^{(\widehat x{}_*)\,[2]\,*}L}\bigl(\breve y{}_1,\breve y{}_2,z\bigr)=\tfrac{\sfi\,\sfd z}{z}+\breve\th{}^{\unl\a}\,\sfd\breve\xi{}^{21}_{\unl\a}\equiv\vartheta(z)+\txA_{\widehat\ep{}_{(1,1|16)}^{(\widehat x{}_*)\,[2]\,*}L}\bigl(\breve y{}_1,\breve y{}_2\bigr)
\qqq
and the trivial pullback groupoid structure on the fibres,\ $\,\mu_{\widehat\ep{}_{(1,1|16)}^{(\widehat x{}_*)\,[2]\,*}L}\equiv\widehat\ep{}_{(1,1|16)}^{(\widehat x{}_*)\,[3]\,*}\mu_L=\bd1$.\ The various supergeometries appearing above carry induced left actions of the residual supersymmetry group $\,\bD_{(1,1|16)}\,$ with respect to which the tensorial data are invariant (by construction).\ Thus,\ on the pullback surjective submersion $\,\sfY D_{(1,1|16)}^{(\widehat x{}_*)}$,\ we have the induced action $\,\sfY\breve\ell{}^{(1,1|16)}\ :\ \bD_{(1,1|16)}\x\sfY D_{(1,1|16)}^{(\widehat x{}_*)}\too\sfY D_{(1,1|16)}^{(\widehat x{}_*)}\,$ with the coordinate presentation $\,\sfY\breve\ell{}_{\breve c}^{(1,1|16)}(\breve m,(\ep_{(1,1|16)}^{(\widehat x{}_*)}(\breve m),\xi))=(\breve\ell_{\breve c}\x\sfY\ell_{(\jmath_{(1,1|16)}(\breve c),0)})(\breve m,$ $(\ep_{(1,1|16)}^{(\widehat x{}_*)}(\breve m),\xi))$,\ {\it cf.}\ \Reqref{eq:inD1act},\ whereas on the principal $\bC^\x$-bundle $\,\widehat\ep{}_{(1,1|16)}^{(\widehat x{}_*)\,[2]\,*}L\cong\sfY^{[2]}D_{(1,1|16)}^{(\widehat x{}_*)}\x\bC^\x$,\ we obtain the action $\,L\breve\ell{}^{(1,1|16)}\ :\ \bD_{(1,1|16)}\x\widehat\ep{}_{(1,1|16)}^{(\widehat x{}_*)\,[2]\,*}L\too\widehat\ep{}_{(1,1|16)}^{(\widehat x{}_*)\,[2]\,*}L\,$ with the coordinate presentation $\,L\breve\ell{}_{\breve c}^{(1,1|16)}(\breve y{}_1,\breve y{}_2,z)=(\sfY\breve\ell{}_{\breve c}^{(1,1|16)}(\breve y{}_1),\sfY\breve\ell{}_{\breve c}^{(1,1|16)}(\breve y{}_2),\ee^{\sfi\,\la((\jmath_{(1,1|16)}(\breve c),0),(\jmath_{(1,1|16)}(\breve c),0),(\ep_{(1,1|16)}^{(\widehat x{}_*)}(\breve m),\xi^1),(\ep_{(1,1|16)}^{(\widehat x{}_*)}(\breve m),\xi^2))}\cdot z)$.\ By now,\ we have all the requisites to address the question of existence of the trivialisation $\,\cT_{(1,1|16)}^{(\widehat x{}_*)}$.

We begin by comparing the two curvings over $\,\sfY D_{(1,1|16)}^{(\widehat x{}_*)}\x_{D_{(1,1|16)}^{(\widehat x{}_*)}}\hspace{-1pt}\sfY_0 D_{(1,1|16)}^{(\widehat x{}_*)}\equiv\sfY D_{(1,1|16)}^{(\widehat x{}_*)}$,\ using identity \eqref{eq:Cgam0-Cgam1} along the way,\ whereby we obtain
\qq\nn
\bigl(\pr_2^*\om_{(1,1|16)}-\widehat\ep{}_{(1,1|16)}^{(\widehat x{}_*)\,*}\txB_{\rm GS}\bigr)\bigl(\breve y\bigr)=-2\breve e{}^0_{\rm L}\wedge\breve e{}^1_{\rm L}\bigl(\breve m\bigr)-\breve\th\,\ovl\g{}_0\,\sfd\breve\th\wedge\sfd\breve x{}_+-\sfd\breve\xi{}_{\unl\a}\wedge\sfd\breve\th{}^{\unl\a}=\sfd\bigl(\tfrac{1}{2}\,\bigl(\breve x{}_+\,\sfd\breve x{}_--\breve x{}_-\,\sfd\breve x{}_+\bigr)-\breve\th{}^{\unl\a}\,\sfd\breve\xi{}_{\unl\a}\bigr)\,,
\qqq
written in terms of the light-cone coordinates $\,\breve x{}_\pm=\breve x{}^0\pm\breve x{}^1$.\ From the latter result,\ we read off the structure of the principal $\bC^\x$-connection super-1-form on the principal $\bC^\x$-bundle
\qq\nn
\pi_{T_{(1,1|16)}^{(\widehat x{}_*)}}=\pr_1\ :\ T_{(1,1|16)}^{(\widehat x{}_*)}=\sfY D_{(1,1|16)}^{(\widehat x{}_*)}\x\bC^\x\too\sfY D_{(1,1|16)}^{(\widehat x{}_*)}
\qqq
of the trivialisation,\ to wit,
\qq\nn
&\cA_{T_{(1,1|16)}^{(\widehat x{}_*)}}\bigl(\breve y,z\bigr)=\vartheta(z)+\txA_{T_{(1,1|16)}^{(\widehat x{}_*)}}\bigl(\breve y\bigr)\,,&\cr\cr
&\txA_{T_{(1,1|16)}^{(\widehat x{}_*)}}\bigl(\breve y\bigr)\equiv-\breve\th{}^{\unl\a}\,\sfd\breve\xi{}_{\unl\a}+\tfrac{1}{2}\,\bigl(\breve x{}_+\,\sfd\breve x{}_--\breve x{}_-\,\sfd\breve x{}_+\bigr)\equiv-\breve\th{}^{\unl\a}\,\sfd\breve\xi{}_{\unl\a}+\pi_{\sfY D_{(1,1|16)}^{(\widehat x{}_*)}}^*\sfa_{T_{(1,1|16)}^{(\widehat x{}_*)}}\bigl(\breve y\bigr)\,.&
\qqq
Direct inspection of the variance of its base component $\,\txA_{T_{(1,1|16)}^{(\widehat x{}_*)}}\,$ under a supersymmetry transformation,\ with $\,\breve c=(\breve\vep,\breve t)$,\
\qq\nn
&&\bigl(\sfY\breve\ell{}_{\breve c}^{(1,1|16)\,*}-\id_{\sfY D_{(1,1|16)}^{(\widehat x{}_*)}}^*\bigr)\txA_{T_{(1,1|16)}^{(\widehat x{}_*)}}\bigl(\breve m,\bigl(\ep_{(1,1|16)}^{(\widehat x{}_*)}\bigl(\breve m\bigr),\xi\bigr)\bigr)\cr\cr
&=&\sfd\bigl[-\breve\vep{}^{\unl\a}\,\breve\xi{}_{\unl\a}+\tfrac{1}{2}\,\bigl(\breve t{}_+\,\breve x{}_--\breve t{}_-\,\breve x{}_+\bigr)+\tfrac{1}{2}\,\bigl(\breve x{}_++\breve t{}_+\bigr)\,\breve\vep\,\ovl\g{}_0\,\breve\th+\tfrac{1}{6}\,\bigl(\breve\vep\,\ovl\g{}_{\unl a}\,\breve\th\bigr)\,\bigl(\breve\vep\,\ovl\g{}^{\unl a}\,\breve\th\bigr)\bigr]+\tfrac{1}{6}\,\breve\th{}^{\unl\a}\,\sfd\bigl(\breve\vep\,\ovl\g{}_{\unl a}\,\breve\th\cdot\ovl\g{}^{\unl a}_{\unl\a\unl\b}\,\bigl(2\breve\vep+\breve\th\bigr)^{\unl\b}\bigr)\,,
\qqq
in which $\,\breve t{}_\pm\equiv\breve t{}^0\pm\breve t{}^1\,$ and which,\ upon invoking the identity
\qq\label{eq:Cgamtens0r}
\ovl\g{}_{\unl a}\ox\ovl\g{}^{\unl a}=\ovl\g{}_0\ox\ovl\g{}^0+\ovl\g{}_1\ox\ovl\g{}^1=-\ovl\g{}_0\ox\ovl\g{}_0+\ovl\g{}_1\ox\ovl\g{}_1=-\ovl\g{}_1\ox\ovl\g{}_1+\ovl\g{}_1\ox\ovl\g{}_1=0\,,
\qqq
yields the final result
\qq\nn
\bigl(\sfY\breve\ell{}_{\breve c}^{(1,1|16)\,*}-\id_{\sfY D_{(1,1|16)}^{(\widehat x{}_*)}}^*\bigr)\txA_{T_{(1,1|16)}^{(\widehat x{}_*)}}\bigl(\breve m,\bigl(\ep_{(1,1|16)}^{(\widehat x{}_*)}\bigl(\breve m\bigr),\xi\bigr)\bigr)=\sfd\bigl(-\breve\vep{}^{\unl\a}\,\breve\xi{}_{\unl\a}+\tfrac{1}{2}\,\bigl(\breve t{}_+\,\breve x{}_--\breve t{}_-\,\breve x{}_+\bigr)+\tfrac{1}{2}\,\bigl(\breve x{}_++\breve t{}_+\bigr)\,\breve\vep\,\ovl\g{}_0\,\breve\th\bigr)\,,
\qqq
leads to the definition of the desired lift $\,T\breve\ell{}^{(1,1|16)}\ :\ \bD_{(1,1|16)}\x T_{(1,1|16)}^{(\widehat x{}_*)}\too T_{(1,1|16)}^{(\widehat x{}_*)}\,$ of the action $\,\sfY\breve\ell{}^{(1,1|16)}\,$ to the total space $\,T_{(1,1|16)}^{(\widehat x{}_*)}\,$ of the principal $\bC^\x$-bundle,\ $\,T\breve\ell{}^{(1,1|16)}_{\breve c}(\breve y,z)=(\sfY\breve\ell{}_{\breve c}^{(1,1|16)}(\breve y),\ee^{\sfi\,\D_{\breve c}^{(1,1|16)}(\breve m,\breve\xi)}\cdot z\bigr)\,$ with 
$\,\D_{\breve c}^{(1,1|16)}(\breve m,\breve\xi)=-\breve\vep{}^{\unl\a}\,\breve\xi{}_{\unl\a}+\tfrac{1}{2}\,(\breve t{}_+\,\breve x{}_--\breve t{}_-\,\breve x{}_+)+\tfrac{1}{2}\,(\breve x{}_++\breve t{}_+)\,\breve\vep\,\ovl\g{}_0\,\breve\th$,\ 
that preserves the principal connection super-1-form $\,\cA_{T_{(1,1|16)}^{(\widehat x{}_*)}}$.

At this stage,\ it remains to verify the existence,\ over $\,\sfY^{[2]}D_{(1,1|16)}^{(\widehat x{}_*)}$,\ of a principal $\bC^\x$-bundle isomorphism $\,\a_{T_{(1,1|16)}^{(\widehat x{}_*)}}\ :\ \widehat\ep{}_{(1,1|16)}^{(\widehat x{}_*)\,[2]\,*}L\ox\pr_2^*T_{(1,1|16)}^{(\widehat x{}_*)}\xrightarrow{\ \cong\ }\pr_1^*T_{(1,1|16)}^{(\widehat x{}_*)}$,\ and check its equivariance with respect to the $\bD_{(1,1|16)}$-actions on its domain and codomain,\ {\it i.e.},\ under the identification \eqref{eq:LD1triv},
\qq\label{eq:alT1equiv}
\a_{T_{(1,1|16)}^{(\widehat x{}_*)}}\circ\bigl(L\breve\ell{}_{\breve c}^{(1,1|16)}\ox\pr_2^*T\breve\ell{}^{(1,1|16)}_{\breve c}\bigr)=\pr_1^*T\breve\ell{}^{(1,1|16)}_{\breve c}\circ\a_{T_{(1,1|16)}^{(\widehat x{}_*)}}\,,
\qqq
where on the pullback bundles
\qq\nn
\alxydim{@C=2.5cm@R=1.5cm}{ \pr_A^*T_{(1,1|16)}^{(\widehat x{}_*)}\cong\sfY^{[2]}D_{(1,1|16)}^{(\widehat x{}_*)}\x\bC^\x \ar[r]^{\qquad\qquad\widehat\pr{}_A\equiv\pr_A\x\id_{\bC^\x}} \ar[d]_{\pi_{\pr_A^*T_{(1,1|16)}^{(\widehat x{}_*)}}\equiv\pr_1} & T_{(1,1|16)}^{(\widehat x{}_*)} \ar[d]^{\pi_{T_{(1,1|16)}^{(\widehat x{}_*)}}} \\ \sfY^{[2]}D_{(1,1|16)}^{(\widehat x{}_*)} \ar[r]_{\pr_A} & \sfY D_{(1,1|16)}^{(\widehat x{}_*)} }\,,\qquad A\in\{1,2\}\,,
\qqq
we induce actions $\,\pr_A^*T_{(1,1|16)}^{(\widehat x{}_*)}\breve\ell\ :\ \bD_{(1,1|16)}\x\pr_A^*T_{(1,1|16)}^{(\widehat x{}_*)}\too\pr_A^*T_{(1,1|16)}^{(\widehat x{}_*)}\,$ with the respective coordinate presentations $\,\pr_A^*T\breve\ell{}^{(1,1|16)}_{\breve c}(\breve y{}_1,\breve y{}_2,z)=(\sfY\breve\ell{}_{\breve c}^{(1,1|16)}(\breve y{}_1),\sfY\breve\ell{}_{\breve c}^{(1,1|16)}(\breve y{}_2),\ee^{\sfi\,\D_{\breve c}^{(1,1|16)}(\breve m,\breve\xi{}^A)}\cdot z)$.\ For the sake of simplicity of the subsequent analysis,\ we present the tensor-product bundle as the principal $\bC^\x$-bundle $\,\widehat\ep{}_{(1,1|16)}^{(\widehat x{}_*)\,[2]\,*}L\ox\pr_2^*T_{(1,1|16)}^{(\widehat x{}_*)}\cong\sfY^{[2]}D_{(1,1|16)}^{(\widehat x{}_*)}\x\bC^{\x }\,$ equipped with the principal $\bC^\x$-connection super-1-form
\qq\nn
\cA_{\widehat\ep{}_{(1,1|16)}^{(\widehat x{}_*)\,[2]\,*}L\ox\pr_2^*T_{(1,1|16)}^{(\widehat x{}_*)}}\bigl(\breve y{}_1,\breve y{}_2,z\bigr)\equiv\vartheta(z)+\bigl(\pr_2^*\txA_{T_{(1,1|16)}^{(\widehat x{}_*)}}+\txA_{\widehat\ep{}_{(1,1|16)}^{(\widehat x{}_*)\,[2]\,*}L}\bigr)\bigl(\breve y{}_1,\breve y{}_2\bigr)\,.
\qqq
In search of $\,\a_{T_{(1,1|16)}^{(\widehat x{}_*)}}$,\ we consider the difference of the base components of the relevant principal $\bC^\x$-connection super-1-forms,\ $\,(\pr_1^*\txA_{T_{(1,1|16)}^{(\widehat x{}_*)}}-\pr_2^*\txA_{T_{(1,1|16)}^{(\widehat x{}_*)}}-\txA_{\widehat\ep{}_{(1,1|16)}^{(\widehat x{}_*)\,[2]\,*}L}\bigr)(\breve y{}_1,\breve y{}_2)=0$,\ and derive therefrom 
\qq\label{eq:alTbi}
\a_{T_{(1,1|16)}^{(\widehat x{}_*)}}\equiv\bd1\,,
\qqq
manifestly coherent with the trivial groupoid structure on (the pullback of) $\,L$.\ We complete our analysis with a check of the equivariance of $\,\a_{T_{(1,1|16)}^{(\widehat x{}_*)}}$.\ The condition of equivariance,\ \Reqref{eq:alT1equiv},\ rewrites as
\qq\nn
\la\bigl(\bigl(\jmath_{(1,1|16)}\bigl(\breve c\bigr),0\bigr),\bigl(\jmath_{(1,1|16)}(\breve c\bigr),0\bigr),\bigl(\ep_{(1,1|16)}^{(\widehat x{}_*)}\bigl(\breve m\bigr),\xi^1\bigr),\bigl(\ep_{(1,1|16)}^{(\widehat x{}_*)}\bigl(\breve m\bigr),\xi^2\bigr)\bigr)+\D_{\breve c}^{(1,1|16)}\bigl(\breve m,\breve\xi{}^2\bigr)=\D_{\breve c}^{(1,1|16)}\bigl(\breve m,\breve\xi{}^1\bigr)\,,
\qqq
and so holds true trivially.

Thus,\ we obtain a manifestly (left-)supersymmetric trivialisation
\qq\nn
\cT_{(1,1|16)}^{(\widehat x{}_*)}=\bigl(\sfY D_{(1,1|16)}^{(\widehat x{}_*)},\pi_{\sfY D_{(1,1|16)}^{(\widehat x{}_*)}},T_{(1,1|16)}^{(\widehat x{}_*)},\pi_{T_{(1,1|16)}^{(\widehat x{}_*)}},\cA_{T_{(1,1|16)}^{(\widehat x{}_*)}},\a_{T_{(1,1|16)}^{(\widehat x{}_*)}}\bigr)
\qqq
on the restricted super-1-gerbe $\,\ep_{(1,1|16)}^{(\widehat x{}_*)\,*}\cG_{\rm GS}$.

\section{A proof of Proposition \ref{prop:Rsusy-GS-grb}}\label{app:Rsusy-GS-grb}

We begin our constructive proof by lifting the maps $\,\breve\wp{}_{\breve c}\,$ to $\,\sfY D_{(1,1|16)}^{(\widehat x{}_*)}\,$ in an obvious manner,\ {\it i.e.},\ as mappings $\,\sfY\breve\wp{}_{\breve c}^{(1,1|16)}\ :\ \sfY D_{(1,1|16)}^{(\widehat x{}_*)}\too\sfY D_{(1,1|16)}^{(\widehat x{}_*)}\,$ with the coordinate presentation $\,\sfY\breve\wp{}_{\breve c}^{(1,1|16)}(\breve y)\equiv\sfY\breve\wp{}_{\breve c}^{(1,1|16)}(\breve m,(\ep_{(1,1|16)}^{(\widehat x{}_*)}(\breve m),\xi))$ $=(\breve\wp{}_{\breve c}(\breve m),\sfY\wp_{(\jmath_{(1,1|16)}(\breve c),0)}(\ep_{(1,1|16)}^{(\widehat x{}_*)}(\breve m),\xi))$,\ written in terms of the mappings $\,\sfY\wp_{y_2}(y_1)=\sfY\txm(y_1,y_2)\,$ (for $\,(y_1,y_2)\in\sfY\bT\x\sfY\bT$),\ and using the lift to define the suitable mappings on the fibred product $\,\sfY^{[2]}\breve\wp{}_{\breve c}^{(1,1|16)}\ :\ \sfY^{[2]}D_{(1,1|16)}^{(\widehat x{}_*)}\too\sfY^{[2]}D_{(1,1|16)}^{(\widehat x{}_*)}\,$ as {\it per} $\,\sfY^{[2]}\breve\wp{}_{\breve c}^{(1,1|16)}(\breve y{}_1,\breve y{}_2)=(\sfY\breve\wp{}_{\breve c}^{(1,1|16)}(\breve y{}_1),\sfY\breve\wp{}_{\breve c}^{(1,1|16)}(\breve y{}_2))$,\ we find the transformations $\,\breve\xi{}_{\unl\a}\longmapsto\breve\xi{}_{\unl\a}+\breve t{}^{\unl a}\,(\ovl\g{}_{\unl a}\,\breve\th){}_{\unl\a}+\tfrac{1}{6}\,\breve\vep\,\ovl\g{}_{\unl a}\,\breve\th\cdot(\ovl\g{}^{\unl a}\,(2\breve\th+\breve\vep))_{\unl\a}=\breve\xi{}_{\unl\a}+\breve t{}_+\,(\ovl\g{}_0\,\breve\th){}_{\unl\a}\,$ in whose simplification we used identity \eqref{eq:Cgamtens0r}.\ These yield
\qq\nn
\sfY\breve\wp{}_{\breve c}^{(1,1|16)\,*}\widehat\ep{}_{(1,1|16)}^{(\widehat x{}_*)\,*}\txB_{\rm GS}\bigl(\breve y\bigr)=\widehat\ep{}_{(1,1|16)}^{(\widehat x{}_*)\,*}\txB_{\rm GS}\bigl(\breve y\bigr)+\sfd\bigl(\breve\vep\,\ovl\g{}_0\,\breve\th\,\sfd\breve x{}_++\breve t{}_+\,\breve\th\,\ovl\g{}_0\,\sfd\breve\th\bigr)\,,
\qqq
and 
\qq\nn
\sfY^{[2]}\breve\wp{}_{\breve c}^{(1,1|16)\,*}\txA_{\widehat\ep{}_{(1,1|16)}^{(\widehat x{}_*)\,[2]\,*}L}\bigl(\breve y{}_1,\breve y{}_2\bigr)=\txA_{\widehat\ep{}_{(1,1|16)}^{(\widehat x{}_*)\,[2]\,*}L}\bigl(\breve y{}_1,\breve y{}_2\bigr)+\sfd\bigl(\breve\vep{}^{\unl\a}\,\breve\xi{}^{21}_{\unl\a}\bigr)\,.
\qqq
From the above\footnote{As usual in the (de Rham-)cohomologically trivial setting,\ the conclusion follows directly from the top-level identity \eqref{eq:rinvHGSrestr},\ the latter being the nontrivial one.},\ we deduce the existence of a $\bD_{(1,1|16)}$-indexed family of 1-isomorphisms
\qq\nn
\Phi_{\breve c}\ :\ \wp_{\breve c}^*\ep_{(1,1|16)}^{(\widehat x{}_*)\,*}\cG_{\rm GS}\xrightarrow{\ \cong\ }\ep_{(1,1|16)}^{(\widehat x{}_*)\,*}\cG_{\rm GS}\,,\qquad\qquad\breve c\in\bD_{(1,1|16)}\,,
\qqq
which is the standard gerbe-theoretic marker of a global supersymmetry.\ We choose the surjective submersion of the former 1-gerbe in the form
\qq\nn
\alxydim{@C=2.5cm@R=1.5cm}{ \sfY_{\breve c}D_{(1,1|16)}^{(\widehat x{}_*)}\equiv\wp_{\breve c}^*\sfY D_{(1,1|16)}^{(\widehat x{}_*)}\equiv \sfY D_{(1,1|16)}^{(\widehat x{}_*)} \ar[r]^{\qquad\qquad\qquad\widehat{\breve\wp}_{\breve c}\equiv\sfY\breve\wp{}_{\breve c}^{(1,1|16)}} \ar[d]_{\pi_{\sfY_{\breve c}D_{(1,1|16)}^{(\widehat x{}_*)}}\equiv\pi_{\sfY D_{(1,1|16)}^{(\widehat x{}_*)}}} & \sfY D_{(1,1|16)}^{(\widehat x{}_*)} \ar[d]^{\pi_{\sfY D_{(1,1|16)}^{(\widehat x{}_*)}}} \\ D_{(1,1|16)}^{(\widehat x{}_*)} \ar[r]_{\breve\wp{}_{\breve c}} & D_{(1,1|16)}^{(\widehat x{}_*)} }\,,
\qqq
and,\ then,\ erect over the $D_{(1,1|16)}^{(\widehat x{}_*)}$-fibred product $\,\sfY_{\breve c,0}D_{(1,1|16)}^{(\widehat x{}_*)}=\sfY_{\breve c}D_{(1,1|16)}^{(\widehat x{}_*)}\x_{D_{(1,1|16)}^{(\widehat x{}_*)}}\hspace{-1pt}\sfY D_{(1,1|16)}^{(\widehat x{}_*)}\,$ the principal $\bC^\x$-bundle 
\qq\nn
\pi_{P_{\breve c}}=\pr_1\ :\ P_{\breve c}=\sfY_{\breve c,0}D_{(1,1|16)}^{(\widehat x{}_*)}\x\bC^\x\too\sfY_{\breve c,0}D_{(1,1|16)}^{(\widehat x{}_*)}
\qqq
equipped with the principal $\bC^\x$-connection super-1-form with a coordinate presentation
\qq\nn
\cA_{P_{\breve c}}\bigl(\breve y{}_1,\breve y{}_2,z\bigr)=\vartheta(z)+\breve\th{}^{\unl\a}\,\sfd\breve\xi{}^{21}_{\unl\a}-\breve\vep\,\ovl\g{}_0\,\breve\th\,\sfd\breve x{}_+-\breve t{}_+\,\breve\th\,\ovl\g{}_0\,\sfd\breve\th\equiv\vartheta(z)+\txA_{P_{\breve c}}\bigl(\breve y{}_1,\breve y{}_2\bigr)\,.
\qqq
Next,\ we pull back the principal $\bC^\x$-bundle $\,\widehat\ep{}_{(1,1|16)}^{(\widehat x{}_*)\,[2]\,*}L\,$ of $\,\ep{}_{(1,1|16)}^{(\widehat x{}_*)\,*}\cG_{\rm GS}\,$ (in the presentation \eqref{eq:LD1triv}) to the $D_{(1,1|16)}^{(\widehat x{}_*)}$-fibred square $\,\sfY^{[2]}_{\breve c,0}D_{(1,1|16)}^{(\widehat x{}_*)}\equiv\sfY_{\breve c,0}D_{(1,1|16)}^{(\widehat x{}_*)}\x_{D_{(1,1|16)}^{(\widehat x{}_*)}}\sfY_{\breve c,0}D_{(1,1|16)}^{(\widehat x{}_*)}\,$ along $\,\widehat{\breve\wp}{}^{[2]}_{\breve c}\equiv\sfY^{[2]}\breve\wp{}^{(1,1|16)}_{\breve c}\,$ as
\qq\nn
\alxydim{@C=3.cm@R=1.5cm}{ L_{\breve c}\equiv\widehat{\breve\wp}{}^{[2]\,*}_{\breve c}\widehat\ep{}_{(1,1|16)}^{(\widehat x{}_*)\,[2]\,*}L=\widehat\ep{}_{(1,1|16)}^{(\widehat x{}_*)\,[2]\,*}L \ar[r]^{\qquad\qquad\quad\widehat{\widehat{\breve\wp}}{}^{[2]}_{\breve c}\ \equiv L\breve\wp{}_{\breve c}^{(1,1|16)}} \ar[d]_{\pi_{L_{\breve c}}\equiv\pi_{\widehat\ep{}_{(1,1|16)}^{(\widehat x{}_*)\,[2]\,*}L}} & \widehat\ep{}_{(1,1|16)}^{(\widehat x{}_*)\,[2]\,*}L \ar[d]^{\pi_{\widehat\ep{}_{(1,1|16)}^{(\widehat x{}_*)\,[2]\,*}L}} \\ \sfY^{[2]}_{\breve c}D_{(1,1|16)}^{(\widehat x{}_*)} \ar[r]_{\widehat{\breve\wp}{}^{[2]}_{\breve c}} & \sfY^{[2]}D_{(1,1|16)}^{(\widehat x{}_*)} }
\qqq
in terms of the natural lifts $\,L\breve\wp{}^{(1,1|16)}_{\breve c}\ :\ \widehat\ep{}_{(1,1|16)}^{(\widehat x{}_*)\,[2]\,*}L\too\widehat\ep{}_{(1,1|16)}^{(\widehat x{}_*)\,[2]\,*}L\,$ of the $\,\breve\wp{}_{\breve c}\,$ with the coordinate presentations $\,L\breve\wp{}_{\breve c}^{(1,1|16)}(\breve y{}_1,\breve y{}_2,z)=(\sfY^{[2]}\breve\wp{}^{(1,1|16)}_{\breve c}(\breve y{}_1,\breve y{}_2),\ee^{\sfi\,\la((\ep_{(1,1|16)}(\breve m),\xi^1),(\ep_{(1,1|16)}(\breve m),\xi^2),(\jmath_{(1,1|16)}(\breve c),0),(\jmath_{(1,1|16)}(\breve c),0))}\cdot z)$,\ and subsequently compare,\ over the $D_{(1,1|16)}^{(\widehat x{}_*)}$-fibred square $\,\sfY^{[2]}_{\breve c,0}D_{(1,1|16)}^{(\widehat x{}_*)}\equiv\sfY_{\breve c,0}D_{(1,1|16)}^{(\widehat x{}_*)}\x_{D_{(1,1|16)}^{(\widehat x{}_*)}}\sfY_{\breve c,0}D_{(1,1|16)}^{(\widehat x{}_*)}\ni(\breve y{}_1,\breve y{}_2,\breve y{}_3,\breve y{}_4)$,\ the base components of the relevant principal $\bC^\x$-connection super-1-forms
\qq\nn
\bigl(\pr_{1,2}^*\txA_{P_{\breve c}}+\pr_{2,4}^*\txA_{\widehat\ep{}_{(1,1|16)}^{(\widehat x{}_*)\,[2]\,*}L}-\pr_{1,3}^*\widehat{\breve\wp}{}^{[2]\,*}_{\breve c}\txA_{\widehat\ep{}_{(1,1|16)}^{(\widehat x{}_*)\,[2]\,*}L}-\pr_{3,4}^*\txA_{P_{\breve c}}\bigr)\bigl(\breve y{}_1,\breve y{}_2,\breve y{}_3,\breve y{}_4\bigr)=\sfd\bigl(-\breve\vep{}^{\unl\a}\,\breve\xi{}^{31}_{\unl\a}\bigr)\,.
\qqq
From this comparison,\ we read off the coordinate presentation
\qq\nn
\a_{P_{\breve c}}\bigl(\breve y{}_1,\breve y{}_2,\breve y{}_3,\breve y{}_4,z\bigr)=\bigl(\breve y{}_1,\breve y{}_2,\breve y{}_3,\breve y{}_4,\ee^{\sfi\,\Pi^{(1,1|16)}_{\breve c}(\breve y{}_1,\breve y{}_2,\breve y{}_3,\breve y{}_4)}\cdot z\bigr)\,,\qquad\qquad\Pi^{(1,1|16)}_{\breve c}\bigl(\breve y{}_1,\breve y{}_2,\breve y{}_3,\breve y{}_4\bigr)=-\breve\vep{}^{\unl\a}\,\breve\xi{}^{31}_{\unl\a}
\qqq
of the principal $\bC^\x$-bundle isomorphism
\qq\nn
\a_{P_{\breve c}}\ :\ \pr_{1,3}^*L_{\breve c}\ox\pr_{3,4}^*P_{\breve c}\xrightarrow{\ \cong\ }\pr_{1,2}^*P_{\breve c}\ox\pr_{2,4}^*\widehat\ep{}_{(1,1|16)}^{(\widehat x{}_*)\,[2]\,*}L\,,
\qqq
in whose definition we adopted the convenient presentations:\ $\,\pr_{1,3}^*L_{\breve c}\ox\pr_{3,4}^*P_{\breve c}\cong\sfY^{[2]}_{\breve c,0}D_{(1,1|16)}^{(\widehat x{}_*)}\x\bC^\x\,$ and $\,\pr_{1,2}^*P_{\breve c}\ox\pr_{2,4}^*\widehat\ep{}_{(1,1|16)}^{(\widehat x{}_*)\,[2]\,*}L\cong\sfY^{[2]}_{\breve c,0}D_{(1,1|16)}^{(\widehat x{}_*)}\x\bC^\x$.\ Compatibility of the latter with the trivial groupoid structure $\,\mu_{\widehat\ep{}_{(1,1|16)}^{(\widehat x{}_*)\,[2]\,*}L}=\bd1$,\ as measured over $\,\sfY_{\breve c,0}D_{(1,1|16)}^{(\widehat x{}_*)}\x_{D_{(1,1|16)}^{(\widehat x{}_*)}}\hspace{-1pt}\sfY_{\breve c,0}D_{(1,1|16)}^{(\widehat x{}_*)}\x_{D_{(1,1|16)}^{(\widehat x{}_*)}}\hspace{-1pt}\sfY_{\breve c,0}D_{(1,1|16)}^{(\widehat x{}_*)}\ni(\breve y{}_1,\breve y{}_2,\breve y{}_3,\breve y{}_4,\breve y{}_5,\breve y{}_6)$,\ is ensured by the trivial identity
\qq\nn
\bigl(\pr_{1,2,3,4}^*\Pi^{(1,1|16)}_{\breve c}+\pr_{3,4,5,6}^*\Pi^{(1,1|16)}_{\breve c}\bigr)\bigl(\breve y{}_1,\breve y{}_2,\breve y{}_3,\breve y{}_4,\breve y{}_5,\breve y{}_6\bigr)=\pr_{1,2,5,6}^*\Pi^{(1,1|16)}_{\breve c}\bigl(\breve y{}_1,\breve y{}_2,\breve y{}_3,\breve y{}_4,\breve y{}_5,\breve y{}_6\bigr)\,.
\qqq
This completes the reconstruction of the right supersymmetric structure
\qq\nn
\Phi_{\breve c}\equiv\bigl(\sfY_{\breve c,0}D_{(1,1|16)}^{(\widehat x{}_*)},\id_{\sfY_{\breve c,0}D_{(1,1|16)}^{(\widehat x{}_*)}},P_{\breve c},\pi_{P_{\breve c}},\cA_{P_{\breve c}},\a_{P_{\breve c}}\bigr)\ :\ \breve\wp{}_{\breve c}^*\ep_{(1,1|16)}^{(\widehat x{}_*)\,*}\cG_{\rm GS}\xrightarrow{\ \cong\ }\ep_{(1,1|16)}^{(\widehat x{}_*)\,*}\cG_{\rm GS}\,,\qquad\quad\breve c\in\bD_{(1,1|16)}
\qqq
on the restricted super-1-gerbe $\,\ep_{(1,1|16)}^{(\widehat x{}_*)\,*}\cG_{\rm GS}$.

\section{A proof of Proposition \ref{prop:Rsusy-inv-GS-1-brane}}\label{app:Rsusy-inv-GS-1-brane}

Having established the right supersymmetry of the restriction $\,\ep_{(1,1|16)}^{(\widehat x{}_*)\,*}\cG_{\rm GS}\,$ of the GS super-1-gerbe in the proof of Prop.\,\ref{prop:Rsusy-GS-grb},\ we may,\ now,\ proceed with the examination of its trivialisation $\,\cT_{(1,1|16)}^{(\widehat x{}_*)}$.\ The first step on this path consists in checking the behaviour of the curving $\,\om_{(1,1|16)}\bigl(\breve m\bigr)=\sfd\breve x{}_+\wedge\sfd\breve x{}_-+\breve\th\,\ovl\g{}_0\,\sfd\breve\th\wedge\sfd\breve x{}_+\,$ of the trivial super-1-gerbe $\,\cI_{\om_{(1,1|16)}}\,$ under a right translation $\,\breve\wp{}_{\breve c}$.\ The latter is established in a direct calculation using the transformation formul\ae ~\eqref{eq:rsusybr} augmented with $\,\breve x{}_-\longmapsto\breve x{}_-+\breve t{}_--\tfrac{1}{2}\,\breve\th\,(\ovl\g{}^0-\ovl\g{}^1)\,\breve\vep=\breve x{}_-+\breve t{}_--\breve\vep\,\ovl\g{}_0\,\breve\th$,\ which yields $\,\breve\wp{}_{\breve c}^*\om_{(1,1|16)}(\breve m)=\om_{(1,1|16)}(\breve m)+\sfd(2\breve\vep\,\ovl\g{}_0\,\breve\th\,\sfd\breve x{}_+)$.\ Thus,\ the \emph{left} trivial super-1-gerbe $\,\cI_{\om_{(1,1|16)}}\,$ is \emph{not} a \emph{right} trivial super-1-gerbe.\ Nevertheless,\ invariance of its curvature $\,\sfd\om_{(1,1|16)}\,$ under the right supersymmetry ensures its right supersymmetry as a 1-gerbe.\ Indeed,\ reasoning along the lines drawn previously for $\,\ep_{(1,1|16)}^{(\widehat x{}_*)\,*}\cG_{\rm GS}$,\ we arrive at the (trivial) principal $\bC^\x$-bundle 
\qq\nn
\pi_{J_{\breve c}}=\pr_1\ :\ J_{\breve c}=D_{(1,1|16)}^{(\widehat x{}_*)}\x\bC^\x\too D_{(1,1|16)}^{(\widehat x{}_*)} 
\qqq 
with,\ in the global coordinates $\,(\breve m,z)\in J_{\breve c}$,\ the principal $\bC^\x$-connection super-1-form
\qq\nn
\cA_{J_{\breve c}}\bigl(\breve m,z\bigr)=\vartheta(z)-2\breve\vep\,\ovl\g{}_0\,\breve\th\,\sfd\breve x{}_+\equiv\vartheta(z)+\txA_{J_{\breve c}}(\breve m)
\qqq
and the trivial isomorphism 
\qq\nn
\a_{J_{\breve c}}\equiv\bd1\ :\ J_{\breve c}\circlearrowleft\,,
\qqq
manifestly coherent with the trivial groupoid structure of $\,\cI_{\om_{(1,1|16)}}$.\ Hence,\ altogether,\ the right supersymmetry is implemented on the latter by the $\bD_{(1,1|16)}$-indexed family of 1-gerbe 1-isomorphisms
\qq\nn
\Psi_{\breve c}\equiv\bigl(D_{(1,1|16)}^{(\widehat x{}_*)},\id_{D_{(1,1|16)}^{(\widehat x{}_*)}},J_{\breve c},\pi_{J_{\breve c}},\cA_{J_{\breve c}},\a_{J_{\breve c}}\bigr)\ :\ \breve\wp{}_{\breve c}^*\cI_{\om_{(1,1|16)}}\xrightarrow{\ \cong\ }\cI_{\om_{(1,1|16)}}\,,\qquad\qquad\breve c\in\bD_{(1,1|16)}\,.
\qqq

The next,\ and final,\ step towards geometrisation of the extended supersymmetry over $\,D_{(1,1|16)}\,$ is the confirmation (or,\ indeed,\ description) of the compatibility of the trivialisation with the right-supersymmetric structures on the two super-1-gerbes involved:\ $\,\{\Phi_{\breve c}\}_{\breve c\in\bD_{(1,1|16)}}\,$ on $\,\ep_{(1,1|16)}^{(\widehat x{}_*)\,*}\cG_{\rm GS}\,$ and $\,\{\Psi_{\breve c}\}_{\breve c\in\bD_{(1,1|16)}}\,$ on $\,\cI_{\om_{(1,1|16)}}$.\ This we achieve by identifying the family of 2-isomorphisms of Diag.\,\eqref{diag:R-on-L-GS-1-brane}.\ To this end,\ we consider the two surjective submersions:\ $\,\pr_1\ :\ \sfY_{\breve c,0}D_{(1,1|16)}^{(\widehat x{}_*)}\too\sfY_{\breve c}D_{(1,1|16)}^{(\widehat x{}_*)}\,$ and $\,\id_{\sfY_{\breve c}D_{(1,1|16)}^{(\widehat x{}_*)}}\ :\ \sfY_{\breve c}D_{(1,1|16)}^{(\widehat x{}_*)}\too\sfY_{\breve c}D_{(1,1|16)}^{(\widehat x{}_*)}\,$ over the surjective submersion of either of the composite 1-isomorphisms $\,\wp_{\breve c}^*\ep_{(1,1|16)}^{(\widehat x{}_*)\,*}\cG_{\rm GS}\xrightarrow{\ \cong\ }\cI_{\om_{(1,1|16)}}$,\ and take their product fibred over the common base,\ to the effect $\,\sfY_{\breve c,0}D_{(1,1|16)}^{(\widehat x{}_*)}\equiv\sfY_{\breve c,0}D_{(1,1|16)}^{(\widehat x{}_*)}\x_{\sfY_{\breve c}D_{(1,1|16)}^{(\widehat x{}_*)}}\hspace{-1pt}\sfY_{\breve c}D_{(1,1|16)}^{(\widehat x{}_*)}$,\ whereupon we present the principal $\bC^\x$-bundles of the two composite 1-gerbe 1-isomorphisms over the latter conveniently as the trivial principal $\bC^\x$-bundles $\,\pr_2^*T_{(1,1|16)}^{(\widehat x{}_*)}\ox P_{\breve c}\cong\sfY_{\breve c,0}D_{(1,1|16)}^{(\widehat x{}_*)}\x\bC^\x\,$ and $\,\widehat{\breve\wp}{}^*_{\breve c}T_{(1,1|16)}^{(\widehat x{}_*)}\ox(\pi_{\sfY_{\breve c}D_{(1,1|16)}^{(\widehat x{}_*)}}\circ\pr_1)^*J_{\breve c}\cong\sfY_{\breve c,0}D_{(1,1|16)}^{(\widehat x{}_*)}\x\bC^\x$.\ A direct comparison of the base components of the respective principal $\bC^\x$-connection super-1-forms\,,
\qq\nn
&&\bigl(\pr_1^*\sfY\breve\wp{}_{\breve c}^{(1,1|16)\,*}\txA_{T_{(1,1|16)}^{(\widehat x{}_*)}}+\pr_1^*\pi_{\sfY_{\breve c}D_{(1,1|16)}^{(\widehat x{}_*)}}^*\txA_{J_{\breve c}}-\pr_2^*\txA_{T_{(1,1|16)}^{(\widehat x{}_*)}}-\txA_{P_{\breve c}}\bigr)\bigl(\breve y{}_1,\breve y{}_2\bigr)\cr\cr
&=&\sfd\bigl(\tfrac{1}{2}\,\bigl(\breve t{}_+\,\breve x{}_--\breve t{}_-\,\breve x{}_+\bigr)-\tfrac{1}{2}\,\bigl(\breve x{}_++3\breve t{}_+\bigr)\,\breve\vep\,\ovl\g{}_0\,\breve\th-\breve\vep{}^{\unl\a}\,\breve\xi{}^1_{\unl\a}\bigr)\,,
\qqq
gives us the coordinate presentation of the isomorphism sought-after,
\qq\nn
\b_{\breve c}\ :\ \pr_2^*T_{(1,1|16)}^{(\widehat x{}_*)}\ox P_{\breve c}\xrightarrow{\ \cong\ }\widehat{\breve\wp}{}^*_{\breve c}T_{(1,1|16)}^{(\widehat x{}_*)}\ox\bigl(\pi_{\sfY_{\breve c}D_{(1,1|16)}^{(\widehat x{}_*)}}\circ\pr_1\bigr)^*J_{\breve c}\,,
\qqq
to wit
\qq\nn
\b_{\breve c}\bigl(\breve y{}_1,\breve y{}_2,z\bigr)=\bigl(\breve y{}_1,\breve y{}_2,\ee^{\sfi\,\La^{(1,1|16)}_{\breve c}(\breve y{}_1,\breve y{}_2)}\cdot z\bigr)\,,\qquad\quad\La^{(1,1|16)}_{\breve c}\bigl(\breve y{}_1,\breve y{}_2\bigr)=\tfrac{1}{2}\,\bigl(\breve t{}_+\,\breve x{}_--\breve t{}_-\,\breve x{}_+\bigr)-\tfrac{1}{2}\,\bigl(\breve x{}_++3\breve t{}_+\bigr)\,\breve\vep\,\ovl\g{}_0\,\breve\th-\breve\vep{}^{\unl\a}\,\breve\xi{}^1_{\unl\a}\,.
\qqq
Coherence of the above isomorphism with those of the composite 1-isomorphisms over $\,\sfY_{\breve c,0}D_{(1,1|16)}^{(\widehat x{}_*)}\x_{D_{(1,1|16)}^{(\widehat x{}_*)}}\sfY_{\breve c,0}D_{(1,1|16)}^{(\widehat x{}_*)}\ni(\breve y{}_1,\breve y{}_2,\breve y{}_3,\breve y{}_4)\,$ is neatly reflected by the identity
\qq\nn
\Pi^{(1,1|16)}_{\breve c}\bigl(\breve y{}_1,\breve y{}_2,\breve y{}_3,\breve y{}_4\bigr)+\La^{(1,1|16)}_{\breve c}\bigl(\breve y{}_1,\breve y{}_2\bigr)=\La^{(1,1|16)}_{\breve c}\bigl(\breve y{}_3,\breve y{}_4\bigr)\,.
\qqq

Altogether,\ we obtain the desired family of 2-isomorphisms
\qq\nn
\varphi_{\breve c}=\bigl(\sfY_{\breve c,0}D_{(1,1|16)}^{(\widehat x{}_*)},\b_{\breve c}\bigr)\,,\qquad\qquad\breve c\in\bD_{(1,1|16)}\,.
\qqq

\section{A proof of Proposition \ref{prop:spoint-GS-brane}}\label{app:spoint-GS-brane}

The first step of the reconstruction of the trvialisation consists in defining the surjective submersion $\,\sfY D_{(0|N)}^{(x_*)}=\d_{(0|N)}^{(x_*)\,*}\sfY\bT\,$ of the pullback 1-gerbe,\ with the coordinates $\,(\th{}^{\unl\a},(\unl\th^\a,x_*^a,\xi_\b))\equiv(\th,(\d_{(0|N)}^{(x_*)}(\th),\xi))\equiv\widetilde y$.\ On it,\ we have $\,\widehat\d{}_{(0|N)}^{(x_*)\,*}\txB_{\rm GS}(\widetilde y)=\sfd(\unl\th{}^\a\,\sfd\xi_\a)$,\ and so we erect a trivial principal $\bC^\x$-bundle 
\qq\nn
\pi_{T_{(0|N)}^{(x_*)}}=\pr_1\ :\ T_{(0|N)}^{(x_*)}=\sfY D_{(0|N)}^{(x_*)}\x\bC^\x\too\sfY D_{(0|N)}^{(x_*)}
\qqq
with coordinates $\,(\widetilde y,z)$,\ and with the principal $\bC^\x$-connection super-1-form
\qq\nn
\cA_{T_{(0|N)}^{(x_*)}}\bigl(\widetilde y,z\bigr)=\vartheta(z)-\unl\th{}^\a\,\sfd\xi_\a\equiv\vartheta(z)+\txA_{T_{(0|N)}^{(x_*)}}\bigl(\widetilde y\bigr)
\qqq
as the principal $\bC^\x$-bundle of the trivialisation sought after.\ Next,\ on the $D_{(0|N)}^{(x_*)}$-fibred square 
$\,\sfY^{[2]}D_{(0|N)}^{(x_*)}\equiv\sfY D_{(0|N)}^{(x_*)}\x_{D_{(0|N)}^{(x_*)}}\sfY D_{(0|N)}^{(x_*)}\,$ of the surjective submersion,\ with coordinates $\,((\th,(\d_{(0|N)}^{(x_*)}(\th),\xi^1)),(\th,(\d_{(0|N)}^{(x_*)}(\th),$ $\xi^2)))\equiv(\widetilde y{}_1,\widetilde y{}_2)$,\ we obtain -- for $\,\widehat\d{}_{(0|N)}^{(x_*)\,[2]}\equiv\widehat\d{}_{(0|N)}^{(x_*)}\x\widehat\d{}_{(0|N)}^{(x_*)}\,$ -- the identity $\,\widehat\d{}_{(0|N)}^{(x_*)\,[2]\,*}\txA_L(\widetilde y{}_1,\widetilde y{}_2)=\unl\th{}^\a\,\sfd\xi^{21}_\a$,\ and so also $\,\widehat\d{}_{(0|N)}^{(x_*)\,[2]\,*}\txA_L+\pr_2^*\txA_{T_{(0|N)}^{(x_*)}}=\pr_1^*\txA_{T_{(0|N)}^{(x_*)}}$,\ from which we read off,\ over $\,\sfY^{(x_*)\,[2]}D_{(0|N)}^{(x_*)}$,\ the trivial isomorphism
\qq\label{eq:alTsspt}
\a_{T_{(0|N)}^{(x_*)}}\equiv\bd1\ :\ \widehat\d{}_{(0|N)}^{(x_*)\,[2]\,*}L\ox\pr_2^*T_{(0|N)}^{(x_*)}\xrightarrow{\ \cong\ }\pr_1^*T_{(0|N)}^{(x_*)}\,,
\qqq
manifestly coherent with the trivial groupoid structures of $\,\d_{(0|N)}^{(x_*)\,*}\cG_{\rm GS}\,$ and $\,\cI_0$.\ All in all,\ we have the trivialisation
\qq\nn
\cT_{(0|N)}^{(x_*)}\equiv\bigl(\sfY D_{(0|N)}^{(x_*)},\pi_{\sfY D_{(0|N)}^{(x_*)}},T_{(0|N)}^{(x_*)},\pi_{T_{(0|N)}^{(x_*)}},\cA_{T_{(0|N)}^{(x_*)}},\a_{T_{(0|N)}^{(x_*)}}\bigr)\,.
\qqq

At this stage,\ it remains to verify the ($\bD_{(0|N)}^{(x_*)}$-)supersymmetry of the trivialisation just derived.\ Proceeding as in the case of the right supersymmetry of the trivialisation $\,\cT_{(1,1|16)}^{(\widehat x{}_*)}$,\ we begin by lifting the maps $\,\t_\vep\,$ to $\,\sfY D_{(0|N)}^{(\widehat x{}_*)}\,$ as mappings $\,\sfY\t_\vep^{(0|N)}\ :\ \sfY D_{(0|N)}^{(x_*)}\too\sfY D_{(0|N)}^{(x_*)}\,$ with the coordinate presentation $\,\sfY\t_\vep^{(0|N)}(\widetilde y)\equiv\sfY\t_\vep^{(0|N)}(\th,(\d_{(0|N)}^{(x_*)}(\th),\xi))=(\t_\vep(\th),\sfY\widetilde\Ad{}_\vep(\d_{(0|N)}^{(x_*)}(\th),\xi))$,\ written in terms of the mappings $\,\sfY\widetilde\Ad{}_\vep(y)=\sfY\wp_{(\frac{1}{2}\,\unl\vep,0,0)}\circ\sfY\ell_{(\frac{1}{2}\,\unl\vep,0,0)}(y)\,$ (for $\,y\in\sfY\bT$),\ and subsequently employing them to induce mappings on the fibred product $\,\sfY^{[2]}\t_\vep^{(0|N)}\ :\ \sfY^{[2]}D_{(0|N)}^{(\widehat x{}_*)}\too\sfY^{[2]}D_{(0|N)}^{(\widehat x{}_*)}\,$ as {\it per} $\,\sfY^{[2]}\t_\vep^{(0|N)}(\widetilde y{}_1,\widetilde y{}_2)=(\sfY\t_\vep^{(0|N)}(\widetilde y{}_1),\sfY\t_\vep^{(0|N)}(\widetilde y{}_2))$,\ whence the coordinate transformations $\,\xi_\a\longmapsto\xi_\a+\tfrac{1}{2}\,x_*^a\,(\ovl\G{}^a\,\unl\vep)_\a+\tfrac{1}{24}\,\unl\vep\,\ovl\G{}_a\,\unl\th\cdot(\ovl\G{}^a\,(2\unl\th+\unl\vep))_\a$.\ These imply
\qq\nn
\sfY\t_\vep^{(0|N)\,*}\widehat\d{}_{(0|N)}^{(x_*)\,*}\txB_{\rm GS}\bigl(\widetilde y\bigr)=\widehat\d{}_{(0|N)}^{(x_*)\,*}\txB_{\rm GS}\bigl(\widetilde y\bigr)+\sfd\bigl(\tfrac{1}{12}\,\unl\vep\,\ovl\G{}_a\,\unl\th\cdot\unl\th\,\ovl\G{}^a\,\sfd\unl\th\bigr)\,,
\qqq
and 
\qq\nn
\sfY^{[2]}\t_\vep^{(0|N)\,*}\widehat\d{}_{(0|N)}^{(x_*)\,[2]\,*}\txA_L\bigl(\widetilde y{}_1,\widetilde y{}_2\bigr)=\widehat\d{}_{(0|N)}^{(x_*)\,[2]\,*}\txA_L\bigl(\widetilde y{}_1,\widetilde y{}_2\bigr)+\sfd\bigl(\unl\vep{}^\a\,\xi{}^{21}_\a\bigr)\,,
\qqq
leading to the following description of a $\bD_{(0|N)}$-indexed family of 1-isomorphisms
\qq\nn
\Xi_\vep\ :\ \t_\vep^*\d_{(0|N)}^{(x_*)\,*}\cG_{\rm GS}\xrightarrow{\ \cong\ }\d_{(0|N)}^{(x_*)\,*}\cG_{\rm GS}\,,\qquad\qquad\vep\in\bD_{(0|N)}\,.
\qqq
The surjective submersion of the domain 1-gerbe is taken in the form
\qq\nn
\alxydim{@C=2.5cm@R=1.5cm}{ \sfY_\vep D_{(0|N)}^{(x_*)}\equiv\t_\vep ^*\sfY D_{(0|N)}^{(x_*)}=\sfY D_{(0|N)}^{(x_*)} \ar[r]^{\qquad\qquad\qquad\widehat\t{}_\vep \equiv\sfY\t_\vep^{(0|N)}} \ar[d]_{\pi_{\sfY_\vep D_{(0|N)}^{(x_*)}}\equiv\pi_{\sfY D_{(0|N)}^{(x_*)}}} & \sfY D_{(0|N)}^{(x_*)} \ar[d]^{\pi_{\sfY D_{(0|N)}^{(x_*)}}} \\ D_{(0|N)}^{(x_*)} \ar[r]_{\t_\vep} & D_{(0|N)}^{(x_*)} }\,,
\qqq
whereupon,\ over the $D_{(0|N)}^{(x_*)}$-fibred product $\,\sfY_{\vep,0}D_{(0|N)}^{(x_*)}\equiv\sfY_\vep D_{(0|N)}^{(x_*)}\x_{D_{(0|N)}^{(x_*)}}\sfY D_{(0|N)}^{(x_*)}$,\ the principal $\bC^\x$-bundle 
\qq\nn
\pi_{Q_\vep}=\pr_1\ :\ Q_\vep=\sfY_{\vep,0}D_{(0|N)}^{(x_*)}\x\bC^\x\too\sfY_{\vep,0}D_{(0|N)}^{(x_*)}
\qqq
is erected and equipped with the principal $\bC^\x$-connection super-1-form with a coordinate presentation
\qq\nn
\cA_{Q_\vep }\bigl(\widetilde y{}_1,\widetilde y{}_2,z\bigr)=\vartheta(z)+\unl\th{}^\a\,\sfd\xi{}^{21}_\a-\tfrac{1}{12}\,\unl\vep\,\ovl\G{}_a\,\unl\th\cdot\unl\th\,\ovl\G{}^a\,\sfd\unl\th\equiv\vartheta(z)+\txA_{Q_\vep }\bigl(\widetilde y{}_1,\widetilde y{}_2\bigr)\,.
\qqq
Next,\ we pull back the principal $\bC^\x$-bundle $\,\widehat\d{}_{(0|N)}^{(x_*)\,[2]\,*}L\,$ of $\,\d_{(0|N)}^{(x_*)\,*}\cG_{\rm GS}\,$ to the $D_{(0|N)}^{(x_*)}$-fibred square $\,\sfY^{[2]}_{\vep,0}D_{(0|N)}^{(x_*)}\equiv\sfY_{\vep,0}D_{(0|N)}^{(x_*)}\x_{D_{(0|N)}^{(x_*)}}\sfY_{\vep,0}D_{(0|N)}^{(x_*)}\,$ along $\,\widehat\t{}^{[2]}_\vep \equiv\sfY^{[2]}\t^{(0|N)}_\vep \,$ as 
\qq\nn
\alxydim{@C=3.cm@R=1.5cm}{ \widetilde L_\vep \equiv\widehat\t{}^{[2]\,*}_\vep \widehat\d{}_{(0|N)}^{(x_*)\,[2]\,*}L=\widehat\d{}_{(0|N)}^{[2]\,*}L \ar[r]^{\qquad\qquad\quad \widehat{\widehat\t}{}^{[2]}_\vep\ \equiv L\t_\vep^{(0|N)}} \ar[d]_{\pi_{L_\vep }\equiv\pi_{\widehat\d{}_{(0|N)}^{(x_*)\,[2]\,*}L}} & \widehat\d{}_{(0|N)}^{(x_*)\,[2]\,*}L \ar[d]^{\pi_{\widehat\d{}_{(0|N)}^{(x_*)\,[2]\,*}L}} \\ \sfY^{[2]}_\vep D_{(0|N)}^{(x_*)} \ar[r]_{\widehat{\t}{}^{[2]}_\vep } & \sfY^{[2]}D_{(0|N)}^{(x_*)} }
\qqq
using the natural lifts $\,L\t^{(0|N)}_\vep \ :\ \widehat\d{}_{(0|N)}^{(x_*)\,[2]\,*}L\too\widehat\d{}_{(0|N)}^{(x_*)\,[2]\,*}L\,$ of the $\,\t_\vep\,$ with the coordinate presentations $\,L\t_\vep^{(0|N)}(\widetilde y{}_1,\widetilde y{}_2,z)=(\sfY^{[2]}\t{}^{(0|N)}_\vep(\widetilde y{}_1,\widetilde y{}_2),\ee^{\sfi\,\nu_\vep(\th;x_*)}\cdot z)$,\ where $\,
\nu_\vep(\th;x_*)=\la((\tfrac{1}{2}\,\unl\vep,0,0),(\tfrac{1}{2}\,\unl\vep,0,0),(\d_{(0|N)}^{(x_*)}(\th),$ $\xi^1),(\d_{(0|N)}^{(x_*)}(\th),\xi^2))+\la(\sfY\ell_{(\frac{1}{2}\,\unl\vep,0,0)}(\d_{(0|N)}^{(x_*)}(\th),\xi^1),\sfY\ell_{(\frac{1}{2}\,\unl\vep,0,0)}(\d_{(0|N)}^{(x_*)}(\th),\xi^2),(\tfrac{1}{2}\,\unl\vep,0,0),(\tfrac{1}{2}\,\unl\vep,0,0))\equiv\tfrac{1}{2}\,\unl\vep{}^\a\,\xi^{21}_\a$.\ Over the $D_{(0|N)}^{(x_*)}$-fibred square $\,\sfY^{[2]}_{\vep,0}D_{(0|N)}^{(x_*)}\equiv\sfY_{\vep,0}D_{(0|N)}^{(x_*)}\x_{D_{(0|N)}^{(x_*)}}\sfY_{\vep,0}D_{(0|N)}^{(x_*)}\ni(\widetilde y{}_1,\widetilde y{}_2,\widetilde y{}_3,\widetilde y{}_4)$,\ we then compare 
\qq\nn
\bigl(\pr_{1,2}^*\txA_{Q_\vep }+\pr_{2,4}^*\txA_{\widehat\d{}_{(0|N)}^{(x_*)\,[2]\,*}L}-\pr_{1,3}^*\widehat\t{}^{[2]\,*}_\vep \txA_{\widehat\d{}_{(0|N)}^{(x_*)\,[2]\,*}L}-\pr_{3,4}^*\txA_{Q_\vep }\bigr)\bigl(\widetilde y{}_1,\widetilde y{}_2,\widetilde y{}_3,\widetilde y{}_4\bigr)=\sfd\bigl(-\unl\vep{}^\a\,\xi^{31}_\a\bigr)\,,
\qqq
and infer that the principal $\bC^\x$-bundle isomorphism
\qq\nn
\a_{Q_\vep }\ :\ \pr_{1,3}^*\widetilde L_\vep \ox\pr_{3,4}^*Q_\vep \xrightarrow{\ \cong\ }\pr_{1,2}^*Q_\vep \ox\pr_{2,4}^*\widehat\d{}_{(0|N)}^{(x_*)\,[2]\,*}L
\qqq
admits,\ for $\,\pr_{1,3}^*L_\vep \ox\pr_{3,4}^*P_\vep \cong\sfY^{[2]}_{\vep,0}D_{(0|N)}^{(x_*)}\x\bC^\x\,$ and $\,\pr_{1,2}^*P_\vep \ox\pr_{2,4}^*\widehat\d{}_{(0|N)}^{(x_*)\,[2]\,*}L\cong\sfY^{[2]}_{\vep,0}D_{(0|N)}^{(x_*)}\x\bC^\x$,\ the coordinate presentation
\qq\nn
\a_{Q_\vep }\bigl(\widetilde y{}_1,\widetilde y{}_2,\widetilde y{}_3,\widetilde y{}_4,z\bigr)=\bigl(\widetilde y{}_1,\widetilde y{}_2,\widetilde y{}_3,\widetilde y{}_4,\ee^{\sfi\,\Upsilon_\vep (\widetilde y{}_1,\widetilde y{}_2,\widetilde y{}_3,\widetilde y{}_4)}\cdot z\bigr)\,,\qquad\qquad\Upsilon_\vep \bigl(\widetilde y{}_1,\widetilde y{}_2,\widetilde y{}_3,\widetilde y{}_4\bigr)=-\unl\vep{}^\a\,\xi^{31}_\a\,.
\qqq
Its compatibility with the trivial groupoid structure $\,\mu_{\widehat\d{}_{(0|N)}^{(x_*)\,[2]\,*}L}=\bd1\,$ over $\,\sfY_{\vep,0}D_{(0|N)}^{(x_*)}\x_{D_{(0|N)}^{(x_*)}}\sfY_{\vep,0}D_{(0|N)}^{(x_*)}\x_{D_{(0|N)}^{(x_*)}}\sfY_{\vep,0}D_{(0|N)}^{(x_*)}\ni(\widetilde y{}_1,\widetilde y{}_2,\widetilde y{}_3,\widetilde y{}_4,\widetilde y{}_5,\widetilde y{}_6)$,\ follows from the identity
\qq\nn
\bigl(\pr_{1,2,3,4}^*\Upsilon_\vep +\pr_{3,4,5,6}^*\Upsilon_\vep \bigr)\bigl(\widetilde y{}_1,\widetilde y{}_2,\widetilde y{}_3,\widetilde y{}_4,\widetilde y{}_5,\widetilde y{}_6\bigr)=\pr_{1,2,5,6}^*\Upsilon_\vep \bigl(\widetilde y{}_1,\widetilde y{}_2,\widetilde y{}_3,\widetilde y{}_4,\widetilde y{}_5,\widetilde y{}_6\bigr)\,.
\qqq
We thus arrive at the $\bD_{(0|N)}$-supersymmetric structure
\qq\nn
\Xi_\vep \equiv\bigl(\sfY_{\vep,0}D_{(0|N)}^{(x_*)},\id_{\sfY_{\vep,0}D_{(0|N)}^{(x_*)}},Q_\vep ,\pi_{Q_\vep },\cA_{Q_\vep },\a_{Q_\vep }\bigr)\ :\ \t_\vep^*\d_{(0|N)}^{(x_*)\,*}\cG_{\rm GS}\xrightarrow{\ \cong\ }\d_{(0|N)}^{(x_*)\,*}\cG_{\rm GS}\,,\qquad\qquad\vep\in\bD_{(0|N)}
\qqq
on the restricted super-1-gerbe $\,\d_{(0|N)}^{(x_*)\,*}\cG_{\rm GS}$.

Once again,\ upon identifying the above,\ we pass to inspect the behaviour of the previously established trivialisation $\,\cT_{(0|N)}^{(x_*)}\,$ under (suitable lifts of) the maps $\,\t_\vep\,$ in relation to the the $\bD_{(0|N)}$-supersymmetric structure $\,\Xi_\vep$.\ What simplifies the situation considerably is the manifest $\bD_{(0|N)}$-supersymmetry of the trivial 1-gerbe $\,\cI_0$,\ which enables us to pass directly to the implementation of the supersymmetry present on the trivialisation given by the $\bD_{(0|N)}$-indexed family of 1-gerbe 2-isomorphisms
\qq\nn
\alxydim{@C=3.cm@R=2cm}{\d_{(0|N)}^{(x_*)\,*}\cG_{\rm GS} \ar[r]^{\cT_{(0|N)}^{(x_*)}} \ar@{=>}[dr]|{\ \varphi_\vep \ }  & \cI_0 \\ \t_\vep ^*\d_{(0|N)}^{(x_*)\,*}\cG_{\rm GS} \ar[u]^{\Xi_\vep } \ar[r]_{\t_\vep^*\cT_{(0|N)}^{(x_*)}} & \t_\vep^*\cI_0 \ar@{=}[u] }\,,\qquad\qquad\vep\in\bD_{(0|N)}\,,
\qqq
As before,\ we consider the two surjective submersions:\ $\,\pr_1\ :\ \sfY_{\vep,0}D_{(0|N)}^{(x_*)}\too\sfY_\vep D_{(0|N)}^{(x_*)}\,$ and $\,\id_{\sfY_\vep D_{(0|N)}^{(x_*)}}\ :\ \sfY_\vep D_{(0|N)}^{(x_*)}\too\sfY_\vep D_{(0|N)}^{(x_*)}\,$ over the surjective submersion of the composite 1-isomorphisms,\ $\,\cT_{(0|N)}^{(x_*)}\circ\Xi_\vep\,$ and $\,\t_\vep^*\cT_{(0|N)}^{(x_*)}$,\ and take their product fibred over the common base,\ whereby we obtain $\,\sfY_{\vep,0}D_{(0|N)}^{(x_*)}\equiv\sfY_{\vep,0}D_{(0|N)}^{(x_*)}\x_{\sfY_\vep D_{(0|N)}^{(x_*)}}\hspace{-1pt}\sfY_\vep D_{(0|N)}^{(x_*)}$.\ Inspection of the difference of the base components of the relevant principal $\bC^\x$-bundles $\,\pr_2^*T_{(0|N)}^{(x_*)}\ox Q_\vep \cong\sfY_{\vep,0}D_{(0|N)}^{(x_*)}\x\bC^\x\,$ and $\,\widehat\t{}^*_\vep T_{(0|N)}^{(x_*)}\cong\sfY_{\vep,0}D_{(0|N)}^{(x_*)}\x\bC^\x\,$ now yields
\qq\nn
\bigl(\pr_1^*\sfY\t_\vep^{(0|N)\,*}\txA_{T_{(0|N)}^{(x_*)}}-\pr_2^*\txA_{T_{(0|N)}^{(x_*)}}-\txA_{Q_\vep }\bigr)\bigl(\widetilde y{}_1,\widetilde y{}_2\bigr)=\sfd\bigl(-\unl\vep{}^\a\,\xi^1_\a-\tfrac{1}{16}\,\unl\vep\,\ovl\G{}_a\,\unl\th\cdot\unl\vep\,\ovl\G{}^a\,\unl\th\bigr)\,,
\qqq
whence the coordinate presentation of the isomorphism $\,\widetilde\b{}_\vep \ :\ \pr_2^*T_{(0|N)}^{(x_*)}\ox Q_\vep \xrightarrow{\ \cong\ }\widehat\t{}^*_\vep T_{(0|N)}^{(x_*)}\,$ given by $\,\widetilde\b_\vep(\widetilde y{}_1,\widetilde y{}_2,z)=(\widetilde y{}_1,\widetilde y{}_2,\ee^{\sfi\,\widetilde\La{}_\vep (\widetilde y{}_1,\widetilde y{}_2)}\cdot z)$,\ where $\,\widetilde\La{}_\vep(\widetilde y{}_1,\widetilde y{}_2)=-\unl\vep{}^\a\,\xi^1_\a-\tfrac{1}{16}\,\unl\vep\,\ovl\G{}_a\,\unl\th\cdot\unl\vep\,\ovl\G{}^a\,\unl\th$.\ Its coherence with the isomorphisms of the composite 1-isomorphisms over $\,\sfY_{\vep,0}D_{(0|N)}^{(x_*)}\x_{D_{(0|N)}^{(x_*)}}\sfY_{\vep,0}D_{(0|N)}^{(x_*)}\ni(\widetilde y{}_1,\widetilde y{}_2,\widetilde y{}_3,\widetilde y{}_4)\,$ is ensured by the identity
\qq\nn
\Upsilon_\vep \bigl(\widetilde y{}_1,\widetilde y{}_2,\widetilde y{}_3,\widetilde y{}_4\bigr)+\widetilde\La{}_\vep \bigl(\widetilde y{}_1,\widetilde y{}_2\bigr)=\widetilde\La{}_\vep \bigl(\widetilde y{}_3,\widetilde y{}_4\bigr)\,.
\qqq

\section{A proof of Theorem \ref{thm:sfusion}}\label{app:sfusion}

In all three cases,\ the reconstruction of $\,\check\phi[\cF_{1,2}^3]\,$ develops along the same lines,\ which we draw next.\ Thus,\ with the surjective submersions of the 1-gerbes in the top line given by 
\qq\nn
\iota_{1,2}^*\bigl(\pr_1^*\cG_{\rm GS}\ox\pr_2^*\cG_{\rm GS}\bigr)\quad&\rightsquigarrow&\quad \pi_{\sfY_{(1,2)}D_{1,2}^3}=\pr_1\ :\ \sfY_{(1,2)}D_{1,2}^3=\iota_{1,2}^*\sfY_{1,2}\bT^{\x 2}\too D_{1,2}^3\,,\cr\cr
\iota_{1,2}^*\pr_1^*\cG_{\rm GS}\ox\iota_{1,2}^*\pr_2^*\cG_{\rm GS}\quad&\rightsquigarrow&\quad \pi_{\sfY_{1;2}D_{1,2}^3}=\pr_1\circ\pr_1\ :\ \sfY_{1;2}D_{1,2}^3=\iota_{1,2}^*\sfY_1\bT^{\x 2}\x_{D_{1,2}^3}\hspace{-1pt}\iota_{1,2}^*\sfY_2\bT^{\x 2}\too D_{1,2}^3\,,\cr\cr
\pr_1^*\iota_1^*\cG_{\rm GS}\ox\pr_2^*\iota_2^*\cG_{\rm GS}\quad&\rightsquigarrow&\quad \pi_{\sfY_{1|2}D_{1,2}^3}=\pr_1\circ\pr_1\ :\ \sfY_{1|2}D_{1,2}^3=\pr_1^*\sfY D_1\x_{D_{1,2}^3}\hspace{-1pt}\pr_2^*\sfY D_2\too D_{1,2}^3\,,\cr\cr
\iota_{1,2}^*\bigl(\txm^*\cG_{\rm GS}\ox\cI_{\varrho_{\rm GS}}\bigr)\quad&\rightsquigarrow&\quad \pi_{\sfY_{\txm(1,2)}D_{1,2}^3}=\pr_1\ :\ \sfY_{\txm(1,2)}D_{1,2}^3=\iota_{1,2}^*\sfY_\txm\bT^{\x 2}\too D_{1,2}^3\,,\cr\cr
\mu_{1,2}^{3\,*}\iota_3^*\cG_{\rm GS}\ox\cI_{\iota_{1,2}^*\varrho_{\rm GS}}\quad&\rightsquigarrow&\quad \pi_{\sfY_{3(1*2)}D_{1,2}^3}=\pr_1\ :\ \sfY_{3(1*2)}D_{1,2}^3=\mu_{1,2}^{3\,*}\sfY D_3\too D_{1,2}^3\,,
\qqq
respectively,\ and written in terms of the surjective submersions $\,\sfY D_A=\iota_A^*\sfY\bT,\ A\in\{1,2,3\}\,$ and with the principal $\bC^\x$-bundles for the various 1-isomorphisms given by (we are abusing the notation by overloading symbols of lifts)
\qq\nn
\cong_1\quad&\rightsquigarrow&\quad\bigl(\pr_1\circ\widehat\iota{}_{1,2}\x\widehat\iota{}_{1,2}\circ\pr_1\bigr)^*L\ox\bigl(\pr_2\circ\widehat\iota{}_{1,2}\x\widehat\iota{}_{1,2}\circ\pr_2\bigr)^*L\too\sfY_{(1,2)}D_{1,2}^3\x_{D_{1,2}^3}\hspace{-1pt}\sfY_{1;2}D_{1,2}^3\,,\cr\cr
\cong_2\quad&\rightsquigarrow&\quad\bigl(\widehat\pr{}_1\circ\widehat\iota{}_{1,2}\circ\pr_1\x\widehat\iota{}_1\circ\widehat\pr{}_1\circ\pr_1\bigr)^*L\ox\bigl(\widehat\pr{}_2\circ\widehat\iota{}_{1,2}\circ\pr_2\x\widehat\iota{}_2\circ\widehat\pr{}_2\circ\pr_2\bigr)^*L\cr\cr
&&\hspace{2cm}\too\sfY_{1;2}D_{1,2}^3\x_{D_{1,2}^3}\hspace{-1pt}\sfY_{1|2}D_{1,2}^3\,,\cr\cr
\cong_3\quad&\rightsquigarrow&\quad\bigl(\widehat\txm\circ\widehat\iota{}_{1,2}\x\widehat\iota{}_3\circ\widehat\mu{}_{1,2}^3\bigr)^*L\too\sfY_{\txm(1,2)}D_{1,2}^3\x_{D_{1,2}^3}\hspace{-1pt}\sfY_{3(1*2)}D_{1,2}^3\,,\cr\cr
\iota_{1,2}^*\cM_{\rm GS}\quad&\rightsquigarrow&\quad\bigl(\widehat\iota{}_{1,2}\x\widehat\iota{}_{1,2}\bigr)^*E\too\sfY_{(1,2)}D_{1,2}^3\x_{D_{1,2}^3}\hspace{-1pt}\sfY_{\txm(1,2)}D_{1,2}^3\,,\cr\cr
\pr_1^*\cT_1\ox\pr_2^*\cT_2\quad&\rightsquigarrow&\quad\bigl(\widehat\pr{}_1\circ\pr_1\bigr)^*T_1\ox\bigl(\widehat\pr{}_2\circ\pr_2\bigr)^*T_2\too\sfY_{1|2}D_{1,2}^3\,,\cr\cr
\mu_{1,2}^{3\,*}\cT_3\ox\id\quad&\rightsquigarrow&\quad\widehat\mu{}_{1,2}^{3\,*}T_3\too\sfY_{3(1*2)}D_{1,2}^3\,,
\qqq
and written in terms of the principal $\bC^\x$-bundles $\,T_A\,$ of the trivialisations $\,\cT_A$,\ we may present the principal $\bC^\x$-bundles of the composite 1-isomorphisms
\qq\nn
\bigl(\mu_{1,2}^{3\,*}\cT_3\ox\id\bigr)\circ\cong_3\circ\iota_{1,2}^*\cM_{\rm GS}\quad&\rightsquigarrow&\quad P_{1,2}=\pr_{1,2}^*\bigl(\widehat\iota{}_{1,2}\x\widehat\iota{}_{1,2}\bigr)^*E\ox\pr_{2,3}^*\bigl(\widehat\txm\circ\widehat\iota{}_{1,2}\x\widehat\iota{}_3\circ\widehat\mu{}_{1,2}^3\bigr)^*L\ox\pr_3^*\widehat\mu{}_{1,2}^{3\,*}T_3\cr\cr
&&\hspace{2cm}\too\sfY_{(1,2)}D_{1,2}^3\x_{D_{1,2}^3}\hspace{-1pt}\sfY_{\txm(1,2)}D_{1,2}^3\x_{D_{1,2}^3}\hspace{-1pt}\sfY_{3(1*2)}D_{1,2}^3\,,\cr\cr\cr
\bigl(\pr_1^*\cT_1\ox\pr_2^*\cT_2\bigr)\circ\cong_2\circ\cong_1\quad&\rightsquigarrow&\quad P_3=\pr_{1,2}^*\bigl(\bigl(\pr_1\circ\widehat\iota{}_{1,2}\x\widehat\iota{}_{1,2}\circ\pr_1\bigr)^*L\ox\bigl(\pr_2\circ\widehat\iota{}_{1,2}\x\widehat\iota{}_{1,2}\circ\pr_2\bigr)^*L\bigr)\cr\cr
&&\hspace{-1.25cm}\ox\pr_{2,3}^*\bigl(\bigl(\widehat\pr{}_1\circ\widehat\iota{}_{1,2}\circ\pr_1\x\widehat\iota{}_1\circ\widehat\pr{}_1\circ\pr_1\bigr)^*L\ox\bigl(\widehat\pr{}_2\circ\widehat\iota{}_{1,2}\circ\pr_2\x\widehat\iota{}_2\circ\widehat\pr{}_2\circ\pr_2\bigr)^*L\bigr)\cr\cr
&&\hspace{-1.75cm}\ox\pr_3^*\bigl(\bigl(\widehat\pr{}_1\circ\pr_1\bigr)^*T_1\ox\bigl(\widehat\pr{}_2\circ\pr_2\bigr)^*T_2\bigr)\too\sfY_{(1,2)}D_{1,2}^3\x_{D_{1,2}^3}\hspace{-1pt}\sfY_{1;2}D_{1,2}^3\x_{D_{1,2}^3}\hspace{-1pt}\sfY_{1|2}D_{1,2}^3
\qqq
simply as the trivial ones:
\qq\nn
P_{1,2}&\cong&\bigl(\sfY_{(1,2)}D_{1,2}^3\x_{D_{1,2}^3}\hspace{-1pt}\sfY_{\txm(1,2)}D_{1,2}^3\x_{D_{1,2}^3}\hspace{-1pt}\sfY_{3(1*2)}D_{1,2}^3\bigr)\x\bC^\x\in\bigl(y_{(1,2)},y_{\txm(1,2)},y_{3(1*2)},z\bigr)\equiv\bigl(\widehat y{}_{1,2},z\bigr)\,,\cr\cr
P_3&\cong&\bigl(\sfY_{(1,2)}D_{1,2}^3\x_{D_{1,2}^3}\hspace{-1pt}\sfY_{1;2}D_{1,2}^3\x_{D_{1,2}^3}\hspace{-1pt}\sfY_{1|2}D_{1,2}^3\bigr)\x\bC^\x\in\bigl(y_{(1,2)},y_{1;2},y_{1|2},z\bigr)\equiv\bigl(\widehat y{}_3,z\bigr)
\qqq
endowed with the respective principal $\bC^\x$-connection super-1-forms
\qq\nn
\cA_{P_{1,2}}&=&\pr_2^*\vartheta+\pr_1^*\txA_{P_{1,2}}\,,\cr\cr
\txA_{P_{1,2}}&=&\pr_{1,2}^*\bigl(\widehat\iota{}_{1,2}\x\widehat\iota{}_{1,2}\bigr)^*\txA_E+\pr_{2,3}^*\bigl(\widehat\txm\circ\widehat\iota{}_{1,2}\x\widehat\iota{}_3\circ\widehat\mu{}_{1,2}^3\bigr)^*\txA_L+\pr_3^*\widehat\mu{}_{1,2}^{3\,*}\txA_{T_3}\,,
\qqq
and
\qq\nn
\cA_{P_3}&=&\pr_2^*\vartheta+\pr_1^*\txA_{P_3}\,,\cr\cr
\txA_{P_3}&=&\pr_{1,2}^*\bigl(\bigl(\pr_1\circ\widehat\iota{}_{1,2}\x\widehat\iota{}_{1,2}\circ\pr_1\bigr)^*\txA_L+\bigl(\pr_2\circ\widehat\iota{}_{1,2}\x\widehat\iota{}_{1,2}\circ\pr_2\bigr)^*\txA_L\bigr)\cr\cr
&&+\pr_{2,3}^*\bigl(\bigl(\widehat\pr{}_1\circ\widehat\iota{}_{1,2}\circ\pr_1\x\widehat\iota{}_1\circ\widehat\pr{}_1\circ\pr_1\bigr)^*\txA_L+\bigl(\widehat\pr{}_2\circ\widehat\iota{}_{1,2}\circ\pr_2\x\widehat\iota{}_2\circ\widehat\pr{}_2\circ\pr_2\bigr)^*\txA_L\bigr)\cr\cr
&&+\pr_3^*\bigl(\bigl(\widehat\pr{}_1\circ\pr_1\bigr)^*\txA_{T_1}+\bigl(\widehat\pr{}_2\circ\pr_2\bigr)^*\txA_{T_2}\bigr)\,.
\qqq
The coordinate presentation of the principal $\bC^\x$-bundle isomorphism 
\qq\nn
\b_{1,2}^3\ :\ \pr_1^*P_{1,2}\xrightarrow{\ \cong\ }\pr_2^*P_3
\qqq
of $\,\check\phi[\cF_{1,2}^3]\,$ over the common base $\,(\sfY_{(1,2)}D_{1,2}^3\x_{D_{1,2}^3}\hspace{-1pt}\sfY_{\txm(1,2)}D_{1,2}^3\x_{D_{1,2}^3}\hspace{-1pt}\sfY_{3(1*2)}D_{1,2}^3)\x_{\sfY_{(1,2)}D_{1,2}^3}\hspace{-1pt}(\sfY_{(1,2)}D_{1,2}^3\x_{D_{1,2}^3}\hspace{-1pt}\sfY_{1;2}D_{1,2}^3$ $\x_{D_{1,2}^3}\hspace{-1pt}\sfY_{1|2}D_{1,2}^3)\,$ can now be read off from direct comparison of the base components of the respective principal $\bC^\x$-connection super-1-forms.\ Indeed,\ denote
\qq\nn
\b_{1,2}^3\bigl(\bigl(\widehat y{}_{1,2},\widehat y{}_3\bigr)\bigl(\widehat y{}_{1,2},z\bigr)\bigr)=\bigl(\bigl(\widehat y{}_{1,2},\widehat y{}_3\bigr)\bigl(\widehat y{}_3,\ee^{\sfi\,\nu_{1,2}^3(\widehat y{}_{1,2},\widehat y{}_3)}\cdot z\bigr)\bigr)
\qqq
to obtain
\qq\nn
\sfd\nu_{1,2}^3\bigl(\widehat y{}_{1,2},\widehat y{}_3\bigr)&=&\txA_E\bigl(\widehat\iota{}_{1,2}(y_{(1,2)}),\widehat\iota{}_{1,2}(y_{\txm(1,2)})\bigr)+\txA_L\bigl(\widehat\txm\circ\widehat\iota{}_{1,2}(y_{\txm(1,2)}),\widehat\iota{}_3\circ\widehat\mu{}_{1,2}^3(y_{3(1*2)})\bigr)+\txA_{T_3}\bigl(\widehat\mu{}_{1,2}^3(y_{3(1*2)})\bigr)\cr\cr
&&-\txA_L\bigl(\pr_1\circ\widehat\iota{}_{1,2}(y_{(1,2)}),\widehat\iota{}_{1,2}\circ\pr_1(y_{1;2})\bigr)-\txA_L\bigl(\pr_2\circ\widehat\iota{}_{1,2}(y_{(1,2)}),\widehat\iota{}_{1,2}\circ\pr_2(y_{1;2})\bigr)\cr\cr
&&-\txA_L\bigl(\widehat\pr{}_1\circ\widehat\iota{}_{1,2}\circ\pr_1(y_{1;2}),\widehat\iota{}_1\circ\widehat\pr{}_1\circ\pr_1(y_{1|2})\bigr)-\txA_L\bigl(\widehat\pr{}_2\circ\widehat\iota{}_{1,2}\circ\pr_2(y_{1;2}),\widehat\iota{}_2\circ\widehat\pr{}_2\circ\pr_2(y_{1|2})\bigr)\cr\cr
&&-\txA_{T_1}\bigl(\widehat\pr{}_1\circ\pr_1(y_{1|2})\bigr)-\txA_{T_2}\bigl(\widehat\pr{}_2\circ\pr_2(y_{1|2})\bigr)\,,
\qqq
the right-hand side being de Rham-exact due to (de Rham-)cohomological triviality of the supergeometries involved.\ In fact,\ it is easy to see that in all the cases considered the components of the above super-1-forms that depend on the coordinates in the fibre over the underlying product worldvolume $\,D_{1,2}^3\,$ cancel out,\ and so by the end of the long day,\ we are left with
\qq\label{eq:betbase}
\nu_{1,2}^3\bigl(\widehat y{}_{1,2},\widehat y{}_3\bigr)\equiv\unl\nu{}_{1,2}^3\bigl(\breve m{}_1,\breve m{}_2\bigr)\,,
\qqq
where
\qq\nn
\sfd\unl\nu{}_{1,2}^3\bigl(\breve m{}_1,\breve m{}_2\bigr)=\sfa_E\bigl(\breve m{}_1,\breve m{}_2,\breve m{}_1\cdot\breve m{}_2\bigr)+\sfa_{T_3}\bigl(\mu_{1,2}^3\bigl(\breve m{}_1,\breve m{}_2\bigr)\bigr)-\sfa_{T_1}\bigl(\breve m{}_1\bigr)-\sfa_{T_2}\bigl(\breve m{}_2\bigr)\,,
\qqq
with
\qq\nn
\sfa_{T_A}=\left\{ \barr{cl} \sfa_{T_{(1,1|16)}^{(\widehat x{}_{*\,A})}} & \tx{if}\quad D_A=D_{(1,1|16)}^{(\widehat x{}_{*\,A})} \cr\cr 0 & \tx{if}\quad D_A=D_{(0|N_A)}^{(x_{*\,A})}\earr \right.\,.
\qqq
Coherence of the thus established 2-isomorphism is a straightforward consequence of the triviality of $\,\a_E\,$ ({\it cf.}\ \Reqref{eq:alEmult}) and of the $\,\a_{T_A}\,$ ({\it cf.} Eqs.\,\eqref{eq:alTbi} and \eqref{eq:alTsspt}),\ as well as that of the multiplicative structure $\,\mu_L\,$ of the GS super-1-gerbe ({\it cf.}\ \Reqref{eq:muL}),\ and of the dependence of (the coordinate presentation of) $\,\b_{1,2}^3\,$ solely on the base coordinates ({\it cf.} \Reqref{eq:betbase}).\smallskip

In the final step of our analysis,\ we inspect the various cases one by one.

\subsubsection{Fusion of superaligned bi-supersymmetric branes} 

Given $\,((\breve{\unl\th}{}_1^\a,\breve x{}_1^0,\breve x{}_1^1,\widehat x{}_{*\,1}^{\widehat a}),(\breve{\unl\th}{}_2^\a,\breve x{}_2^0,\breve x{}_2^1,\widehat x{}_{*\,2}^{\widehat a}))\equiv(\breve m{}_1,\breve m{}_2)\in\iota_{D^{(\widehat x{}_{*\,1})}_{(1,1|16)},D^{(\widehat x{}_{*\,2})}_{(1,1|16)}}(\cF_{D^{(\widehat x{}_{*\,1})}_{(1,1|16)},D^{(\widehat x{}_{*\,2})}_{(1,1|16)}}^{D^{(\widehat x{}_{*\,1}+\widehat x{}_{*\,2})}_{(1,1|16)}}([a_1:a_2:b_{1,2}]\,|\,[c_1:c_2]))$,\ we calculate
\qq\nn
\sfa_E\bigl(\breve m{}_1,\breve m{}_2,\breve m{}_1\cdot\breve m{}_2\bigr)+\sfa_{T^{(\widehat x{}_{*\,1}+\widehat x{}_{*\,2})}_{(1,1|16)}}\bigl(\mu_{1,2}^3\bigl(\breve m{}_1,\breve m{}_2\bigr)\bigr)-\sfa_{T^{(\widehat x{}_{*\,1})}_{(1,1|16)}}\bigl(\breve m{}_1\bigr)-\sfa_{T^{(\widehat x{}_{*\,2})}_{(1,1|16)}}\bigl(\breve m{}_2\bigr)=\breve x{}_{+\,1}\,\sfd\breve x{}_{-\,2}-\breve x{}_{-\,2}\,\sfd\breve x{}_{+\,1}\,,
\qqq
and the exactness of the result is ensured by the second constraints in \Reqref{eq:fusconstrsal}.\ Indeed,\ whenever 
$\,a_1=0\,$ and $\,\breve x{}_{-\,2}=d_{-\,2}\in\bR$,\ we obtain $\,\unl\nu{}_{1,2}^3\bigl(\breve m{}_1,\breve m{}_2\bigr)=-d_{-\,2}\,\breve x{}_{+\,1}$,\ whereas if $\,a_2=0\,$ and $\,\breve x{}_{+\,1}=d_{+\,1}\in\bR$,\ we have
$\,\unl\nu{}_{1,2}^3\bigl(\breve m{}_1,\breve m{}_2\bigr)=d_{+\,1}\,\breve x{}_{-\,2}$.\ Finally,\ for $\,a_1,a_2\neq 0$,\ we find $\,\unl\nu{}_{1,2}^3(\breve m{}_1,\breve m{}_2)=-\tfrac{b_{1,2}}{a_2}\,\breve x{}_{+\,1}$.

\subsubsection{Fusion of spatially transverse bi-supersymmetric branes} 

Here,\ we have two subcases of Sec.\,\ref{subsub:sptransfuswv}.\ In both subcases,\ we may simply adapt the results of the analysis carried out in the previous case.

\subsubsection{Fusion of superpoint branes of opposite chirality} 

Given $\,((\unl\th{}_1,x_{*\,1}),(\unl\th{}_2,x_{*\,2}))\equiv(\breve m{}_1,\breve m{}_2)\in\iota_{D^{(x_{*\,1})}_{(0|16\pm)},D^{(x_{*\,2})}_{(0|16\mp)}}(\cF_{D^{(x_{*\,1})}_{(0|16\pm)},D^{(x_{*\,2})}_{(0|16\mp)}}^{D^{(x_{*\,1}+x_{*\,2})}_{(0|32)}})$,\ we readily establish
\qq\nn
\sfa_E\bigl(\breve m{}_1,\breve m{}_2,\breve m{}_1\cdot\breve m{}_2\bigr)+\sfa_{T^{(x_{*\,1}+x_{*\,2})}_{(0|32)}}\bigl(\mu_{1,2}^3\bigl(\breve m{}_1,\breve m{}_2\bigr)\bigr)-\sfa_{T^{(x_{*\,1})}_{(0|16\pm)}}\bigl(\breve m{}_1\bigr)-\sfa_{T^{(x_{*\,2})}_{(0|16\mp)}}\bigl(\breve m{}_2\bigr)=0\,,
\qqq
and so we may consistently set $\,\unl\nu{}_{1,2}^3(\breve m{}_1,\breve m{}_2)=0$. 

\resumetocwriting

\end{document}